\documentclass[fleqn,usenatbib]{rasti}

\usepackage{newtxtext,newtxmath}

\usepackage[T1]{fontenc}

\DeclareRobustCommand{\VAN}[3]{#2}
\let\VANthebibliography\thebibliography
\def\thebibliography{\DeclareRobustCommand{\VAN}[3]{##3}\VANthebibliography}

\usepackage{graphicx}	
\usepackage{amsmath}	
\usepackage[dvipsnames]{xcolor}

\newcommand\revised[1]{#1}

\newcommand\HST{\textit{HST}}
\newcommand\JWST{\textit{JWST}}
\newcommand\Kepler{\textit{Kepler}}
\newcommand\Ktwo{\textit{K2}}
\newcommand\Magellan{\textit{Magellan}}
\newcommand\Spitzer{\textit{Spitzer}}
\newcommand\TESS{\textit{TESS}}

\title[Stellar Contamination in Transmission Spectra]{The Effect of Stellar Contamination on \revised{Low-resolution} Transmission Spectroscopy: Needs Identified by NASA's Exoplanet Exploration Program Study Analysis Group 21}

\author[B. V. Rackham, N. Espinoza et al.]{Benjamin V. Rackham,$^{1,2}$\thanks{Email: brackham@mit.edu}\thanks{51 Pegasi b Fellow}
N\'estor Espinoza,$^{3}$\thanks{Email: nespinoza@stsci.edu}
Svetlana V. Berdyugina,$^{4,5}$
Heidi Korhonen,$^{6}$
\newauthor
Ryan J. MacDonald,$^{7}$
Benjamin T. Montet,$^{8}$
Brett M. Morris,$^{9}$
Mahmoudreza Oshagh,$^{10}$
\newauthor
Alexander I. Shapiro,$^{11}$
Yvonne C. Unruh,$^{12}$
Elisa V. Quintana,$^{13}$
Robert T. Zellem,$^{14}$
D\'aniel Apai,$^{15,16}$
\newauthor
Thomas Barclay,$^{17,13}$
Joanna K. Barstow,$^{18}$
Giovanni Bruno,$^{19}$
Ludmila Carone,$^{20}$
Sarah L. Casewell,$^{21}$
\newauthor
Heather M. Cegla,$^{22}$
Serena Criscuoli,$^{23}$
Catherine Fischer,$^{4}$
Damien Fournier,$^{11}$
Mark S. Giampapa,$^{15}$
\newauthor
Helen Giles,$^{24}$
Aishwarya Iyer,$^{25}$
Greg Kopp,$^{26}$
Nadiia M. Kostogryz,$^{11}$
Natalie Krivova,$^{11}$
\newauthor
Matthias Mallonn,$^{27}$
Chima McGruder,$^{28}$
Karan Molaverdikhani,$^{29,30,31}$
Elisabeth R. Newton,$^{32}$
\newauthor
Mayukh Panja,$^{11}$
Sarah Peacock,$^{13}$
Kevin Reardon,$^{33,34}$
Rachael M. Roettenbacher,$^{35,36}$
\newauthor
Gaetano Scandariato,$^{19}$
Sami Solanki,$^{11}$
Keivan G. Stassun,$^{37}$
Oskar Steiner,$^{4}$
Kevin B. Stevenson,$^{38}$
\newauthor
Jeremy Tregloan-Reed,$^{39}$
Adriana Valio,$^{40}$
Sven Wedemeyer,$^{41,42}$
Luis Welbanks,$^{25}$
Jie Yu,$^{11}$
\newauthor
Munazza K. Alam,$^{43}$
James R. A. Davenport,$^{44}$
Drake Deming,$^{45}$
Chuanfei Dong,$^{46,47}$
Elsa Ducrot,$^{48}$
\newauthor
Chloe Fisher,$^{9}$
Emily Gilbert,$^{49}$
Veselin Kostov,$^{13,50}$
Mercedes L\'opez-Morales,$^{28}$
Mike Line,$^{25}$
\newauthor
Teo Mo{\v{c}}nik,$^{51}$
Susan Mullally,$^{3}$
Rishi R. Paudel,$^{52}$
Ignasi Ribas,$^{53,54}$
Jeff A. Valenti$^{3}$
\\
\\
\textit{A list of affiliations is given at the end of the paper}
}

\date{Accepted 13/03/2023. Received 03/03/2023; in original form 20/09/2022}

\pubyear{2023}

\begin{document}
\label{firstpage}
\pagerange{\pageref{firstpage}--\pageref{lastpage}}
\maketitle

\begin{abstract}
Study Analysis Group 21 (SAG21) of NASA's Exoplanet Exploration Program Analysis Group (ExoPAG) was organized to study the effect of stellar contamination on space-based transmission spectroscopy, a method for studying exoplanetary atmospheres by measuring the wavelength-dependent radius of a planet as it transits its star.
Transmission spectroscopy relies on a precise understanding of the spectrum of the star being occulted.
However, stars are not homogeneous, constant light sources but have temporally evolving photospheres and chromospheres with inhomogeneities like spots, faculae, plages\revised{, granules, and flares}.
This SAG brought together an interdisciplinary team of more than 100 scientists, with observers and theorists from the heliophysics, stellar astrophysics, planetary science, and exoplanetary atmosphere research communities, to study the current research needs that can be addressed in this context to make the most of transit studies from current NASA facilities like HST and JWST.
The analysis produced 14 findings, which fall into three Science Themes encompassing (1) how the Sun is used as our best laboratory to calibrate our understanding of stellar heterogeneities ("The Sun as the Stellar Benchmark"), (2) how stars other than the Sun extend our knowledge of heterogeneities ("Surface Heterogeneities of Other Stars") and (3) how to incorporate information gathered for the Sun and other stars into transit studies ("Mapping Stellar Knowledge to Transit Studies").
In this invited review, we \revised{largely} reproduce the final report of SAG21 as a contribution to the peer-reviewed literature.
\end{abstract}




\section{Introduction}
\label{S:Introduction}

\label{sec:introduction}

Study Analysis Group 21 (SAG21) of \revised{NASA's} Exoplanet Exploration Program Analysis Group (ExoPAG) was organized to study the effect of stellar contamination on space-based transmission spectroscopy.
\revised{This invited review summarizes the work of SAG21 and largely reproduces the final report submitted to NASA \citep{Rackham2022} as a contribution to the peer-reviewed literature.}
We begin with a brief description of the motivation for SAG21, the goals of the group, the methods by which we conducted our analysis, and the structure of this \revised{review}.

\subsection{Motivation}
Transmission spectroscopy is a method for identifying the atmospheric composition of an exoplanet by measuring a wavelength-dependent radius of a planet as it transits its star \citep{Seager2000, Brown-01}. 
This technique has been successfully applied to many exoplanets, primarily using \revised{the Hubble Space Telescope (\HST{})} and \revised{Spitzer Space Telescope} \citep[\Spitzer{}; e.g.,][]{Charbonneau2002, Tinetti2007, Agol2010, Desert2011, Knutson2011, Deming2013, Fraine2014, Kreidberg2014, Sing2016}, but also using ground-based telescopes \citep[e.g.,][]{Redfield2008, Snellen2008, Snellen2010, Bean2010, Jordan2013, Danielski2014, Nikolov2016}. 
Most transmission spectroscopy studies to date have targeted gas giant worlds due to their larger expected signatures, which scale with the planet’s atmospheric scale height. 
During the next two decades, and thanks to the next generation of ground- and space-based observatories, the technique is expected to also be a major source of information on the atmospheres of rocky exoplanets.
 
Transmission spectroscopy relies on a precise understanding of the wavelength-dependent brightness of the star being occulted. 
However, stars are not homogeneous, constant light sources but have temporally evolving photospheres and chromospheres with inhomogeneities like spots, faculae, and plages \citep[e.g.,][]{Berdyugina2005}. 
The surface features of the star change both intrinsically as active regions evolve and from the perspective of an observer as the star rotates. 
Spots and faculae have different temperatures from the disk-averaged photosphere, and for cooler stars, can have molecular features distinct from the star itself but similar to those in a planet’s atmosphere, including H$_2$O \citep{Wohl1971}, TiO \citep{Neff1995}, and other molecules \citep{Berdyugina2003mol}.
Some studies have found that the signal from stellar inhomogeneities can exceed the signal from the planetary spectral features \citep[e.g.,][]{Rackham2018, Rackham2019}. 
In particular, small exoplanets---mini-Neptunes and rocky planets---around M-dwarf host stars are susceptible to this ``transit light source effect.'' 
To make the most of transit studies from current NASA facilities like \HST{} and \JWST{}, it is essential we quantify the impact of stellar contamination on transmission spectroscopy and develop methods to mitigate for it.

\subsubsection{Relevance to Astro2020}

This topic is directly relevant to the first science theme identified in the Astro2020 Decadal Survey, \textit{Worlds and Suns in Context}, and its priority area, \textit{Pathways to Habitable Worlds}.
Before the launch in the 2040s of \revised{the Habitable Worlds Observatory, a large space telescope} capable of directly imaging Earth-like worlds, transit observations will provide our best pathway to identifying and studying rocky exoplanets, including potentially habitable ones.
These will necessarily transit cool dwarfs, which tend to remain active long after their formation and possess heterogeneous photospheres.
Even with the advanced direct-imaging capabilities enabled by \revised{the Habitable Worlds Observatory}, transit observations will remain a critical tool in the study of exoplanetary atmospheres, as emphasized by the LUVOIR and HabEx Mission Concept Study Final Reports \citep{TheLUVOIRTeam2019, Gaudi2020}.
Thus, the topic of SAG21 is especially relevant to NASA's science priorities in the next two decades and beyond.

\subsection{Goals of SAG21}

This SAG brought together an interdisciplinary team of \revised{more than 100} scientists, with observers and theorists from the heliophysics, stellar astrophysics, planetary science, and exoplanetary atmosphere research communities to address both the impact of stellar contamination on transmission spectra and constraints on stellar photospheric heterogeneity enabled by transiting exoplanets.

SAG21 has six goals:
\begin{enumerate}
    \item Report on what effect stellar contamination could have on future space-based transmission spectroscopy measurements;
    
    \item Identify regions of the parameter space in which care should be exercised with respect to stellar contamination in the context of transmission spectroscopy studies;
    
    \item Identify measures that can be taken to understand the magnitude of stellar contamination;
    
    \item Identify what modeling efforts can further our understanding of stellar contamination;
    
    \item Develop methods to identify measurements that might be contaminated; and
    
    \item Pinpoint complementary observations that can be combined with transmission spectroscopy to mitigate or correct for stellar contamination.
\end{enumerate}
 
\subsection{Methods}

\subsubsection{Soliciting contributions}

The SAG21 co-chairs solicited members for the group via open calls for members to the ExoPAG newsletter and other email listserves, social media sites, presentations at ExoPAG meetings, and direct communications to researchers with relevant expertise.
In total, 122 people joined the dedicated email list for SAG21, and the group ultimately included more than 60 active participants that contributed to the discussion in some form.

\subsubsection{Subgroups}

Given the large interest from the community, we divided the analysis of SAG21 into five thematic areas and formed subgroups with relevant expertise to study each area.
The subgroups studied the topics of:
\begin{enumerate}
    \item Stellar Photospheric \& Chromospheric Heterogeneity,
    \item Unocculted Active Regions,
    \item Occulted Active Regions,
    \item Stellar \& Planetary Retrievals, and
    \item Future Complementary Observations.
\end{enumerate}
Each subgroup held roughly a dozen virtual meetings over the timespan of the SAG to refine the questions to be considered, assign member duties, discuss updates on the analysis, and decide on its final findings.
Subgroup leads met with the SAG21 co-chairs monthly to share and discuss progress.
Significant conversations also took place on public channels of SAG21's dedicated Slack Workspace\revised{\footnote{\url{https://slack.com/}}}.
The detailed scopes of these groups are defined in their corresponding \revised{sections of this review}.

\subsubsection{SAG21 Community Symposium}

SAG21 also hosted a two-day virtual community symposium over Zoom on 8--9\,Mar\,2021 to share preliminary results from the analysis and solicit feedback from the wider community.
The meeting included five overview presentations on the analysis from subgroup leads as well as 21 contributed community talks on subjects relevant to the SAG.
In total, the symposium brought together roughly 110 attendees, including 46 active participants in question-and-answer sessions.
The recordings of all talks are available on the SAG21 Community Symposium Website\footnote{\url{https://sites.google.com/view/sag21symposium}}.

\subsubsection{Final report}

Considering the analysis of the subgroups and inputs from the community symposium, each subgroup prepared a draft \revised{chapter of the} report in mid-2021 outlining their top-level findings and the analysis that lead to them.
Each chapter was then reviewed by at least three members from other subgroups, who provided detailed feedback that the original subgroup leads later addressed.
The entire report was then thoroughly edited by the SAG21 co-chairs with an eye towards consistency, logical ordering of ideas between subgroups, and relevance to the scope of SAG21.
The report was then shared again among all members for a final round of feedback and, after minor revisions, submitted to NASA and posted on the arXiv \revised{\citep{Rackham2022}}.

\revised{The combined analysis produced 14 findings, which are contextualized statements of what we understand to be the current needs that can be addressed to further our understanding of the photospheres of exoplanetary host stars and make the best use of precise space-based transmission spectra of exoplanets.
While SAG21 was focused on space-based observations specifically, the findings are generally applicable to low-to-medium-resolution transmission spectroscopy, both ground- and space-based.
Following the format of the Exoplanet Exploration Program Science Gap List, each finding includes a summary statement along with a statement of the capability needed, capability in progress, and mitigation in progress to address each need.}

\subsection{The \revised{s}tructure of this \revised{review}}

\revised{Like the SAG21 report submitted to NASA, the} structure of this \revised{invited review} follows that of SAG21 itself, with \revised{sections} dedicated to the work of each subgroup.
In each \revised{section}, we introduce the scope of the analysis and the individual topics we studied.
We then present the information on each topic we considered in reaching a \revised{finding}.
\revised{We summarize the 14 findings and conclude in \autoref{S:Conclusions}.}
\revised{\autoref{tab:acronyms} lists the abbreviations and acronyms used throughout this review.}

\section{Stellar Photospheric \& Chromospheric Heterogeneity}
\label{S:PhotosphericHeterogeneity}

\subsubsection*{Essential \revised{q}uestions:}
\begin{enumerate}
    \item What do we know about solar and stellar photospheric and chromospheric heterogeneities (PCH)?  
    \item Which properties of photospheric and chromospheric heterogeneities are of relevance to exoplanet transit spectroscopy?  
    \item How can the knowledge of the Sun help to constrain such PCH properties?
    \item Which of these PCH properties can be constrained by stellar observations? Which observations are the most useful (photometry, spectroscopy, X-ray, etc.)?
    \item Which of these PCH properties can be constrained using advanced MHD and HD simulations (magnetic and hydrodynamic)?
    \item How do stellar fundamental parameters influence PCH?
    \item What is known about PCH of high-priority exoplanet host stars?
    \item Which PCH properties of the known host stars are critical for the characterization of exoplanetary atmospheres?
\end{enumerate}


\subsection{Introduction}

From high-spatial-resolution observations of the Sun we know that the solar atmosphere is anything but homogeneous .
The main sources of the heterogeneities in the solar photosphere and chromosphere are due to the plasma flows and magnetic fields as well as the interplay between them \revised{\citep[see, e.g., the review by][]{solanki2006}}. Observations suggest that such heterogeneities are also present in the atmospheres of other stars \revised{\citep{Berdyugina2005, Reiners2012LRSP}}. 
They manifest themselves via, e.g., photometric and spectral variability \revised{\citep[see, e.g., recent  books by][]{Envgold2019, Basri2021}}. 

When a planet transits a star, such heterogeneities lead to a background spectral signal and temporal modulation on the transit light curve, contaminating the transmission spectrum \revised{\citep[e.g.,][]{Pont2008}}. 
The magnitude of the contamination is defined by the brightness contrasts and the fractional areas of various magnetic features and their fine structure \revised{\citep[e.g.,][]{Sing2011}}. 
From solar observations, we know that the contrast between different solar structures is strongly wavelength dependent \revised{\citep[e.g.,][]{Wohl1970}}.
Consequently, the underlying stellar transmission spectrum depends on the wavelength and can, thus, hide or mimic a genuine planetary atmospheric signal \revised{\citep[e.g.,][]{Ballerini-12, Rackham2018, Rackham2019}}. 

To distinguish between planetary and stellar signals we must observe and model the contribution of stellar heterogeneities to transmission spectra. 
In particular, we need to understand the dependence of the stellar signal as a function of wavelength \revised{and time}. Modeling requires accurate knowledge of properties of various stellar features on stars with different fundamental parameters and magnetic activity levels.

The main goal of this \revised{section} is to summarize current knowledge, ongoing work, and limits in our understanding of stellar heterogeneities that is of relevance for transmission spectroscopy. 
In particular, the focus here is given to the optical and near infrared (NIR), given the relevance of these wavelengths to the topic of this \revised{review}: the ``stellar contamination" of transmission spectra \citep{Rackham2018, Rackham2019}. 
We begin in \autoref{sec:thesun} by introducing observations and modeling efforts to understand the star for which we can obtain arguably the most detailed information about these heterogeneities: the Sun, together with examples of \revised{S}olar \revised{S}ystem planetary transits of the Sun-as-a-star. We then introduce simulations of surface structures on other stars in \autoref{sec:stellarmodels}. 
We conclude this \revised{section} by discussing observations of stars other than our Sun in \autoref{sec:otherstars}.

\subsection{The Sun: a prime window to understand stellar heterogeneities}
\label{sec:thesun}

\begin{table*}
    \caption{
    Continuum contrasts of a solar quiet region granulation at disk center ($c_{\rm rms}$) measured with two space-based telescopes of different aperture.
    }
    \label{tab:crms_space}
     \setlength{\tabcolsep}{3pt} 
    \centering
    \begin{tabular}{llccccp{6cm}}
    \hline
    \hline
    Satellite & Instrument & Aperture & Wavelength & $c_{\rm rms}$ & Deconvolved & Reference\\
    \hline
    SDO     & HMI      & 14\,cm   & 617.3\,nm  &  4.0\,\%  & 12.2\,\%  & \citet{yeo+al2014}\\
    Hinode  & SOT/SP   & 50\,cm   & 630.0\,nm  &  7.0\,\%  & 14.4\,\%  & \citet{danilovic+al2008}\\
    Hinode  & SOT/BFI  & 50\,cm   & 450.5\,nm  & 12.8\,\%  & 26.7\,\%  & \citet{wed2009} \\
         &          &          & 555.0\,nm  &  8.3\,\%  & 19.4\,\%  &  \\
         &          &          & 668.4\,nm  &  6.2\,\%  & 16.6\,\%  & \\
    \hline
    
    \end{tabular}
 \end{table*}

Stellar heterogeneities in our Sun have been observed by the unaided eye at least since 800 BC, when dark spots---today known as ``sunspots"---were seen to sporadically appear on the solar disk as reported by East Asian sunspot sightings \citep{Stephenson:1990}.
The advent of telescopes in the 1600s allowed for a much more detailed and systematic analysis of 
these features, which Galileo Galilei interpreted as surface features 
\textit{on the (spherical) Sun} \citep[see, e.g.,][]{casanovas:1997}. 
Later observations confirmed these early explanations and, 
in fact, extended the diversity of solar surface heterogeneities 
beyond these first noticed dark spots.

\begin{figure}
    \centering
    \includegraphics[height=\textheight, width=\columnwidth, keepaspectratio]{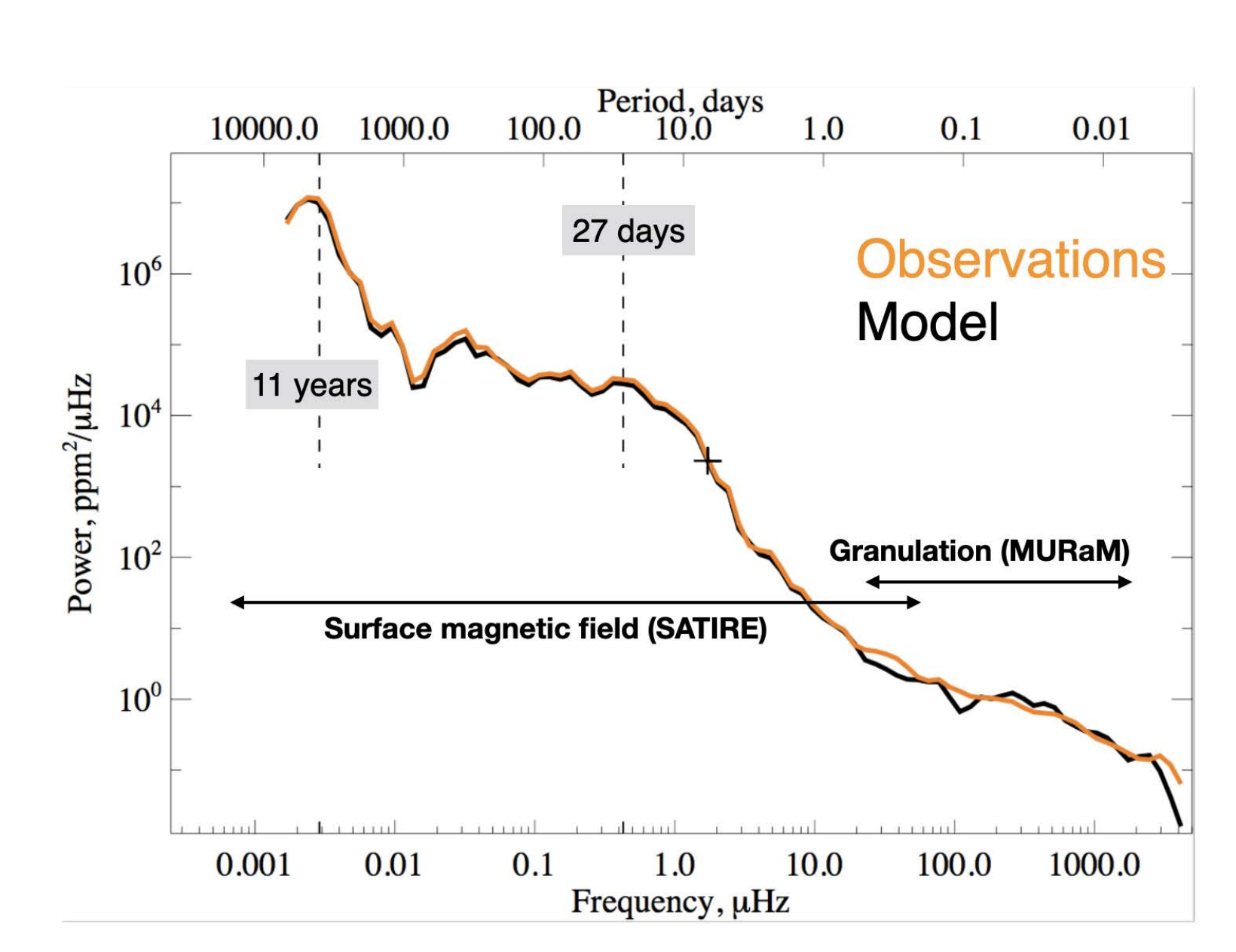}
    \caption{
        Global wavelet power spectrum of modeled (black) and measured (orange) TSI variations. 
        Double arrows indicate frequency ranges where variability is dominated by surface magnetic field and granulation, modeled with the SATIRE \citep{yeo+al2014}  and MURaM \citep{Vogler2005} codes, respectively.
        From \citet{2017NatAs...1..612S}.
    }
    \label{fig:solar_astero}
\end{figure}

Solar (and, by extension, stellar) atmospheric heterogeneities are now 
understood, in general, \revised{to} arise from both 
plasma-motion\revised{-}driven processes and magnetic\revised{-}field processes. The former features are generated by a constantly evolving cellular granulation pattern on 
the visible solar surface, accompanied by oscillations dominating within a rather narrow frequency interval. Surface magnetic structures, on the other hand, are driven by evolution of small-scale magnetic (a.k.a. ``bright") features and spot-like magnetic (a.k.a. ``dark") features. Both plasma-motion and magnetic\revised{-}driven processes have distinct properties, e.g., the time-scales on which they are observed to act. While 
plasma-motion\revised{-}driven features seem to impact various solar observables (e.g., the total solar irradiance\revised{,} TSI) on timescales of minutes to hours \citep{2013AN....334..145S}, solar magnetic features 
give rise to longer time-scale variations, ranging from hours to years \citep[][see \autoref{fig:solar_astero}]{2017NatAs...1..612S}. Even though some properties of these two main processes, such as timescales and flux variations at certain wavelengths, might not 
seem to depend on each other \citep[see, e.g.,][]{Bastien:2013}, they are 
physically related to one another as plasma-driven motions are inherently impacted by magnetic processes and vice versa \citep[e.g.,][]{Stein:2012}. Distinct spatial and wavelength-dependant signatures of these processes and their modeling are reviewed in this \revised{section}.  

In the following sub-subsections, we summarize observed and modeled properties of solar surface plasma-motion and magnetic field structures and processes relevant for transmission spectroscopy, together with examples of Mercury and Venus transits of the Sun-as-a-star. 

\subsubsection{Solar granulation and oscillations}

Brightness variations due to granulation and oscillations in the Sun are important for understanding their influence on transit spectroscopy studies, since the time-scales on which they occur are similar to those of transit events (minutes to hours). 
Also, to date, the Sun is the only star on which we can spatially resolve these features in detail, so our host star is key to understanding the properties and main mechanisms driving them.
Solar granulation and oscillations are believed to be reasonably well understood for the Sun. 

The granulation brightness fluctuations are driven by intensity variations of hot, bright (up-welling) granules and cool, dark (down-flowing) inter-granular lanes \revised{\citep[see, e.g.,][]{Nordlund2009, Stein:2012}}. 
The contrast between these features is typically measured through the mean-normalized RMS of the intensity fluctuations within an area/image of the Sun taken at a given wavelength or a bandpass. 

The true RMS contrast of granulation, as seen in broad-band or continuum images of very quiet regions of the Sun, has been the subject of debates for decades, mainly because of the great disparity of results from early 3D hydrodynamic simulations and observations.
Observations with the space-borne Hinode Solar Optical Telescope \citep[SOT,][]{tsuneta+al2008} show 7.0\,\% intensity contrasts at 630\,nm \citep{danilovic+al2008}.
Employing the highest spatial resolution simulations and convolving with SOT's point spread function, which is believed to be well known, \citet{danilovic+al2008} showed that the simulated 14.4\,\% at 630\,nm was degraded to close to the observed 7.0\,\%.
\citet{wed2009} measured \revised{RMS} contrast values of $12.8 \pm 0.5$\,\%, $8.3 \pm 0.4$\,\%, and $6.2 \pm 0.2$\,\% at disc-center for the blue, green, and red continuum, respectively using the Broadband Filter Imager (BFI) of SOT. 
These values translate to $26.7 \pm 1.3$\,\%, $19.4 \pm 1.4$\,\%, and $16.6 \pm 0.7$\,\% for the blue, green, and red continuum, respectively, when deconvolving the images with the appropriate \revised{point spread function}. 
The filters were centered at 450.45\,nm, 555.05\,nm, and 668.40\,nm, respectively. 
Consequently, there is now good agreement between simulations and deconvolved resolved granulation observations, which both can be considered to yield the true intensity contrast of the granular structure on the surface of the Sun. 
\autoref{tab:crms_space} gives a summary of these values.

These measurements show a strong decrease of the granule contrast towards longer wavelengths, which translates into smaller amplitudes of the time-dependent RMS brightness variability at longer wavelengths. 
Such a contrast decrease was also observed to occur towards infrared wavelengths and was shown to strongly depend on the solar limb distance and location, with the smallest contrast near the limb \citep[see, e.g.,][and references therein]{SC:2003}. 
This occurs because these intensity variations track temperature fluctuations: when assuming a blackbody spectrum for the granulation, its contrast and variation amplitudes are reduced at longer wavelengths. 
A similar argument can be made for oscillations, whose brightness fluctuation amplitudes are expected to strongly decrease with wavelength as well \citep[$\sim 1/\lambda$; see, e.g., ][]{kjeldsen2011a, samadi2012a}.

In contrast to granulation, solar oscillations demonstrate much smaller brightness variability amplitudes on their time-scales (on the order of minutes). 
A full-disk RMS variability measurement over eight-hour time-scales of about $\sim 15$ parts-per-million (ppm) were measured by the Solar and Heliospheric Observatory (SOHO) using the Variability of solar IRradiance and Gravity Oscillations \citep[VIRGO;][]{VIRGO} instrument \citep{Bastien:2013}. 
Also, \revised{the Kepler Space Telescope \citep[\Kepler{};][]{Borucki2010} was used to measure} variability of the solar reflected light from Neptune of the order of several ppm \citep[e.g.,][and references therein]{Gaulme:2016}.

Given the level of brightness fluctuations and their dependence on the wavelength discussed above, it might follow that if other stars behaved like our Sun, granulation and oscillations may not be an important part of the budget for stellar contamination in the transmission spectrum of planets orbiting Sun-like stars in the infrared. 
Rather, the impact of these processes could be a simple noise contribution (a.k.a. ``jitter") to the transit light-curves themselves. 
However, we still lack multi-wavelength and precise center-to-limb variation measurements of this effect on the Sun in order to make a concrete assessment of how important this effect could be as a contamination source in transmission spectra from UV (${\sim}0.3\,\micron$) to near-infrared wavelengths (${\sim}5\,\micron$), which are relevant for transit studies with present and future space-based observatories. 
In particular, some studies \citep[e.g.\revised{,}][]{Chiavassa-17} do suggest that granulation could become an important effect for Earth\revised{--}Sun transiting exoplanet analogues. 
We discuss this in the light of current knowledge on granulation and oscillations for other stars in \autoref{sec:otherstars}.

\subsubsection{Small-scale magnetic heterogeneities: \revised{f}aculae, network, inter-network and mixed-polarity magnetic fields (bright features)}

Faculae, from the Latin ``torch'', are small-scale concentrations of bright patchy features with diameters of about 100--400\,km, often found in the vicinity of sunspots in inter-granular spaces. 
The term faculae refers to the brightening observed in the photosphere; at disc center they appear as magnetic bright points. 
Their counterparts in the chromosphere (e.g., observed in Ca\,\textsc{ii}\,H and K lines) are called plages. 
On the Sun, faculae and plages are typically found in magnetic active regions emerging within the latitudinal activity belts. 
A larger pattern of bright small-scale magnetic fields appears at borders of supergranular cells, as magnetic flux is swept toward the cell boundaries by convective horizontal flows.
These concentrations form the magnetic network which is distributed almost uniformly across the solar disk, at any latitude and longitude.
Thus, faculae, plage, and network heterogeneities are manifestations of photospheric and chromospheric small-scale magnetic fields which are not strong enough to form sunspots \citep{Babcock1955,Kiepenheuer1953, solanki1993}.
Here we summarize the main characteristics of these magnetic bright features, focusing on the aspects most relevant to exoplanet transit spectroscopy.  

\begin{figure}
    \centering
    \includegraphics[width=\columnwidth, height=\textheight, keepaspectratio]{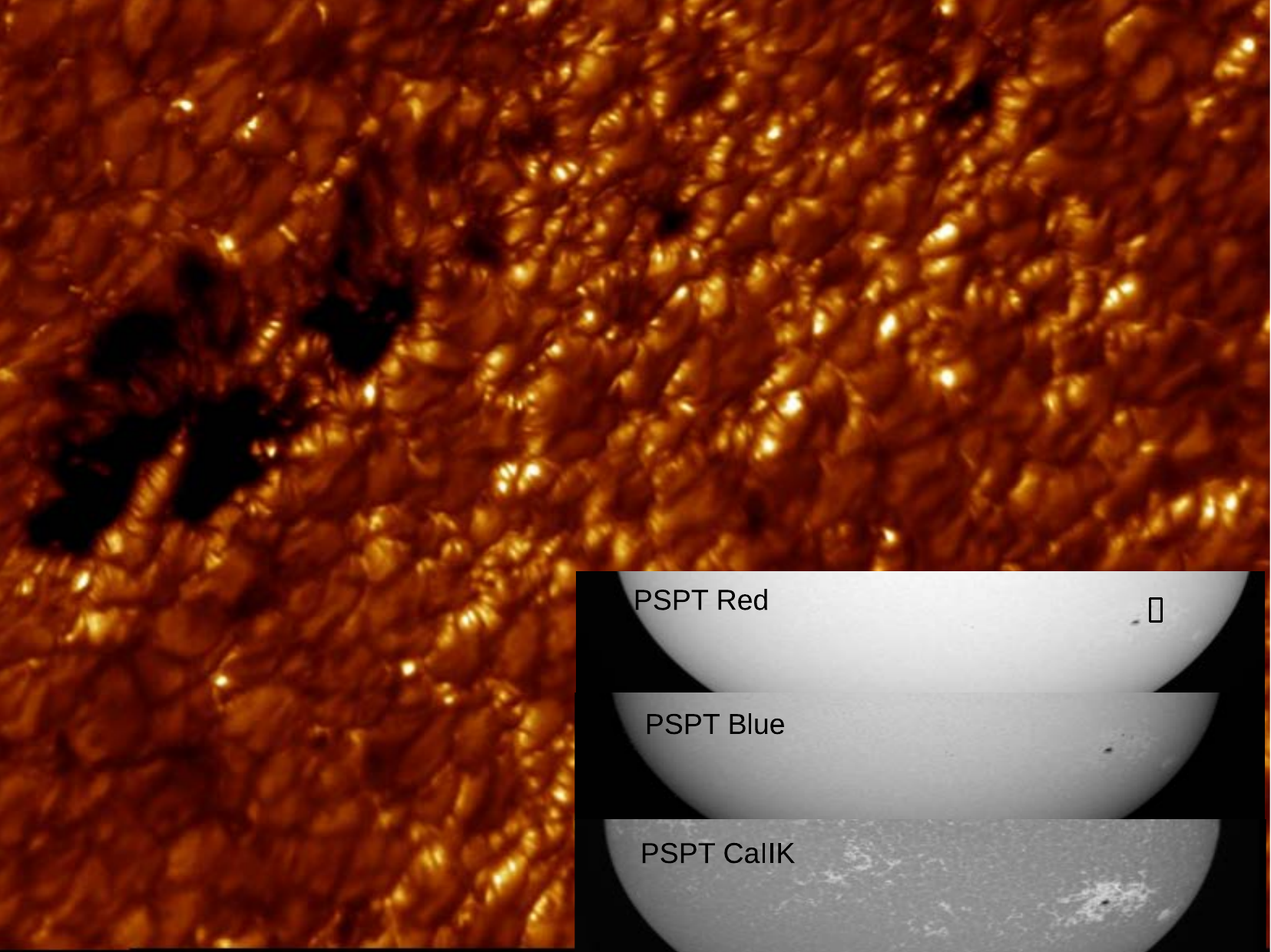}
    \caption{
    A portion of an active region observed at high-spatial resolution in the G-band (430.4\,nm) with the Swedish Solar Tower (SST). 
    Faculae appear as small (sub-arcsec), bright, elongated features. 
    The same region, observed at medium resolution with the Precision Solar Photometric Telescope (PSPT) at Mauna Loa, appears as a bright patch, whose contrast is strongly dependent on the wavelength.
    The images were taken in the passbands 607.1$\pm$0.46 nm (PSPT Red), 409.4$\pm$0.26 nm (PSPT Blue), and 393.4$\pm$ 0.27 (PSPT Ca\,\textsc{ii}\,K).
    The black box in the PSPT Red image shows roughly the field-of-view of the SST G-band image.}
    \label{fig:facula_example}
\end{figure}

Solar observations at high spatial resolution (sub-arcsecond, corresponding to a few hundred kilometers on the solar surface) reveal that faculae are conglomerates of small features associated with kilogauss magnetic field concentrations \revised{\citep[\autoref{fig:facula_example};][]{berger2001,dewijn2009,blanco2010,keys2019}}.
Both high- (sub-arcsec) \citep[e.g.][]{Berger2007, kobel2011, romano2012, kahil2017, buehler2019} and medium- (arcsec) \citep[e.g.][]{ortiz2002, yeo2013, criscuoli2017,  Chatzistergos:2019a} resolution observations indicate that the radiative properties of faculae depend on their size, magnetic flux, degree of aggregation, wavelength, and position on the solar disk.

In the visible continuum, faculae are about 5--10\% brighter than the quiet Sun when observed close to the limb, while they are almost indistinguishable from the quiet photosphere, if not slightly darker, when observed close to the disk center \citep{unruh1999,ahern2000,ermolli2007,yeo2013}.
In the infrared, faculae appear darker than the surroundings \revised{\citep{sanchez2002,Norris2017,Witzke2022}}. 

Facular contrast reaches several tens of percent in the cores of strong Fraunhofer lines, e.g., in the chromospheric Ca\,\textsc{ii}\,H \& K lines \citep[e.g.,][]{Mehltretter1974,walton2003, rutten2007,Chatzistergos:2019a} and H-alpha wings \citep{Dunn-Zirker1973}.
A similar facular contrast is observed in certain molecular bands: 
the CN-band at 388\,nm \citep{Chapman1970,Sheeley1971,zakharov2005}, the G-band at 430\,nm \citep{Kitai-Muller1984,berger2001,langhans2004,hirzberger2005,Berger2007} and the UV OH band at 313\,nm \citep{Berdyugina2003mol,Hirzberger2010}. 
The dependence of the facular contrast on the line-of-sight magnetic field was found to be logarithmic \citep{kahil2017}.
The center-to-limb variation (CLV) of the quiescent photosphere brightness is less steep than that of faculae \citep{walton2003, ermolli2010}. 

\revised{ Magnetic n}etwork is also a manifestation of the photospheric magnetic field \citep[e.g.][]{ortiz2002, yeo2013,criscuoli2017}, and, unlike faculae, which either are located in active regions or result from the decay of active regions, network is present at any latitude and longitude.
Network also results from the aggregation of small-size magnetic elements, which are transported by surface convective motion to the edges of super-granules, thus forming a typical lattice structure particularly clear in chromospheric images (e.g., in near-UV images \revised{from the Solar Dynamics Obsevatory, SDO}).

However, solar photometric measurements currently present two main difficulties. First, these measurements are affected by both finite spatial resolution and scattered light, which reduce the observed contrasts and affect the relation between physical and photometric properties. 
This problem affects both ground-\revised{based} \citep{toner1997,criscuoli2008,viticchie2010} and space-based observations  \citep{mathew2007, wed2009,yeo2013}. 
Second, measurements are made relative to the quiet Sun background, whose definition is somewhat arbitrary, thus creating ambiguity and discrepancies between results presented in the literature \citep{peck2015}. 

Moreover, the quiet Sun is not void of magnetic fields \citep[see a review by][]{BellotRubio-OrozcoSuarez2019}.
Weaker, predominantly horizontal fields occur as inter-network features flowing towards the network at supergranular borders \citep{Livingston-Harvey1971,Lites2008}. 
Also, a large volume of mixed-polarity magnetic fields occur at scales which are unresolved with current instrumentation, but fortunately this rather hidden solar magnetism can be detected via the Hanle effect \citep{Stenflo1982,Faurobert-Scholl1993,Stenflo1998,Berdyugina-Fluri2004,TrujilloBueno2004,Kleint2011,Shapiro2011}. 
Whether these quiet-Sun magnetic features vary with the solar cycle \citep{Kleint2010} or affect variations of the solar irradiance \citep{faurobert2020, rempel2020, Yeo2020} is still not fully understood.
High spatial resolution as well as high photometric and polarimetric sensitivity are therefore essential to improve our understanding of how solar magnetism affects irradiance variability. 

Future observations with the upcoming Daniel K.\ Inouye Solar Telescope \citep[DKIST,][]{DKIST} will have unprecedented spatial and temporal resolution, complemented by observations with space missions, e.g., the Interface Region Imaging Spectrograph (IRIS) and the Atacama Large Millimeter/submillimeter Array (ALMA).
These data are expected to improve our understanding of the physics that underlies the radiative emission of quiet and magnetic regions, especially those occurring in the chromosphere and corona. 
Coordinated observations of DKIST with the recently launched Solar Orbiter are expected to improve our knowledge on polar faculae, whose properties, due to the ecliptic position of the Earth, are still poorly understood \citep{petrie2021}.
These features have negligible effects on the solar irradiance received on the Earth, but they can play a major role for planets with orbits strongly inclined with respect to the rotation axis of the host star \citep{shapiro2016} or for stars with more abundant or stronger polar faculae. 
Observations with the upcoming Solar Ultraviolet Imaging Telescope (SUIT) onboard Aditya-L1 \citep{Ghosh:2016}, \revised{planned for launch in 2023}, will provide critical information on the contrasts and the CLV of the magnetic features in the UV between 200 and 400\,nm.

\subsubsection{Spot-like magnetic heterogeneities: \revised{u}mbra, penumbra, pores (dark features)}

Sunspots are the highest-contrast features in the solar photosphere visible in the optical and near-infrared. 
The largest sunspot in recent history was observed in April 1947 with an area of more than 6000 millionths of the solar hemisphere area (MHS). 
Spots ten times smaller are visible by the protected naked eye from the Earth. 
Spots larger than about 2000\,MHS are regularly observed near solar activity maxima. These are several times larger than the Earth’s area.  
There are several reviews of the literature related to physical properties of sunspots  \citep[e.g.,][]{Solanki-03,Borrero:2011}.
Here, we will only consider sunspot properties relevant to exoplanet transmission spectroscopy, and will refer to only a few important results.

Because of a strong magnetic field, up to 6\,\revised{kG} \revised{\citep[e.g., ][]{Hale:1908,Harvey:1969,Livingston:2006,delaCruzRodriguez2013,vanNoort:2013,Siu-Tapia:2017,Okamoto:2018}}, the bolometric intensity of sunspot umbra is 0.2--0.3 of that of the surrounding quiet (non-magnetic) photosphere, implying effective temperatures of 4000--4500\,K. 
The intensity decreases with the umbral size \citep{mathew2007}.
An umbra larger than about 4000\,km in diameter is usually (but not always) surrounded by a filamentary penumbra, which include both bright filaments and dark gaps and can be symmetric or partially asymmetric, depending on the spot complexity.
The total spot area (including the penumbra) may exceed that of the umbra by up to a factor of five \citep[e.g.,][]{Solanki-03}. 
The penumbra’s bolometric contrast to the photosphere is about 0.8, i.e., its effective temperature is about 5500\,K (average for bright filaments and dark gaps). 
A strongly inclined 0.5--1\,kG magnetic field is responsible for the filamentary structure of the penumbra \citep{Siu-Tapia2019}. 
An umbra without a penumbra is known as a pore. 
Pores, which generally have relatively weak magnetic field strengths \citep{Suetterlin1998}, are often observed to merge together and form larger spots  with penumbra and structures called light-bridges, outlining the previous pores. 
The brightness of light-bridges is similar to that of the penumbra. 

The sunspot size distribution is log-normal \citep{Bogdan:1988}, with smaller spots occurring more frequently.
This implies that the sunspot emergence is a fragmentation process, possibly of a large flux-tube anchored in the convection zone.
The largest sunspot in a group occupies more than half the total area of all spots in a group, while the number of individual umbrae within a given group, $N$, increases with the group size as a power law $N\propto A_\mathrm{group}^{0.58}$ \citep{Mandal2021}. 
By extrapolating their results to significantly bigger spot groups, like those observed on very active stars, \citet{Mandal2021} conclude that such groups may be composed of multiple spots, with the biggest spot in a group occupying 55--75\% of the total group area. 
This, however, does not apply to spots formed by coalescence.

Sunspots exhibit various dynamic phenomena, such as magneto-convection, flows, oscillations, and short-term events. 
The umbral fine structure is best revealed in images taken in the passband centered at the TiO 706\,nm molecular band \citep{Berger-Berdyugina2003}.
Remnants of convection within the umbra are observed as umbral dots, bright features of 100\,km or less in diameter.
In the photosphere, the intensity of \revised{umbral dots} is significantly higher than that of the dark umbra, and their lifetimes span from minutes up to 2 hours  \citep{Beckers-Schroeter1968,Adjabshirzadeh-Koutchmy1983,Sobotka1997a,Riethmueller:2008,Watanabe2014,Kilcik2020}. 
Longer-lasting \revised{umbral dots} tend to be larger and brighter.
As was found by \citet{Sobotka-Hanslmeier2005}, observed intensities of \revised{umbral dots} correlate with local intensities of the umbral background. 
In terms of temperature, \revised{umbral dots} were found on average hotter than the coolest area in the umbra by about 1000\,K and cooler than the undisturbed photosphere by 500--1000\,K, while individual \revised{umbral dots} reached or even exceeded the average photospheric brightness and temperature.
The density and brightness of umbral dots increase towards the umbral edge \citep{Sobotka1997a,Yadav:2018}, and the brightness may vary quasi-periodically on time-scales of 3--30 minutes \citep{Sobotka1997b,Kilcik2020}.
Umbral dots extend up to the chromosphere where they are larger and more numerous \citep{Kitai1986}.
The horizontal outflow observed in penumbra (the Evershed effect) is reversed in the chromosphere, becoming an inflow in the penumbra and downflow in the umbra.
In addition, umbral flashes---sudden brightenings of the dark umbra---are observed in the upper atmosphere and reveal a further complexity of the dark umbra near the temperature minimum and lower chromosphere \citep{Turova1984,Henriques2020}.
Short-lived brightenings in the penumbra are associated with chromospheric micro-jets \citep{Katsukawa2007}.

Sunspots constitute active regions that emerge with an 11-year solar cycle (22 years if accounting for a magnetic reversal).
A typical active region consists of larger leading and smaller trailing spots, pores, and faculae. 
The lifetime of sunspots linearly increases with their maximum size (the Gnevyshev\revised{--}Waldmeier rule). 
The decay rate is about 11\,MHS/day \citep{Petrovay1997}.
This implies that most spots live less than a day, but larger ones can persist for months. 
It is not clear whether the sunspot lifetime and/or decay rate vary with the cycle.
In the context of stellar and exoplanetary research, it is important to investigate cycle-dependent characteristics of sunspots and small-scale magnetic fields. 
This includes the contrast and relative areas of magnetic features and their spectral variations.

In view of this, a full-disk, synoptic capability is essential to achieve a comprehensive view of solar variability in all interlinked space and time domains.
These domains include:
global and temporal scales ranging from transient flare outbursts to cycle timescales; 
spatial structures we refer to as active regions as delineated by spots, faculae and plages; 
and field extrapolations into the corona and heliosphere that influence solar/stellar wind properties and, in turn, their potential impact on planetary atmospheres.
The SOLIS (Synoptic Long-term Investigations of the Sun) facility of the \revised{National Solar Observatory} is an example of an advanced instrument for global solar synoptic vector magnetic field investigations \citep{Keller2003}.
A solar synoptic capability of this kind yields the partition of magnetic energy among the fundamental constituents of solar heterogeneity for the Sun-as-a-star on all time scales and, in conjunction with DKIST, at all relevant spatial scales.
Full-disk synoptic observations, uniquely characterized by long-term stability and continuity, provide a fundamental database for the understanding of magnetic activity that is encountered in our \revised{S}olar \revised{S}ystem, the planetary systems around Sun-like stars and the extrapolation of the solar example to other kinds of stellar hosts.

\subsubsection{Modeling solar observables}
\label{sec:solarmodels}

Modeling solar observables requires, on one hand, a solid knowledge on physics of radiation, plasma and magnetic processes and, on the other hand, accounting for instrumental degradation of the observed phenomena, especially for high-resolution imaging data.

From \autoref{tab:crms_space} and other observations, one can see that the uncorrected RMS contrast of granulation measured with space telescopes increases with increasing the telescope aperture, i.e., with increasing the spatial resolution. 
Thus, the distribution of the intensities is resolution dependent.
At low spatial resolution, the intensities show a distinct bi-modal intensity distribution, reflecting the different intensities of (bright) granular and (dark) intergranular space. 
With increasing spatial resolution of simulations, the bi-modality of the intensity distribution becomes less pronounced. 
For ground-based observations, achieving true contrasts is even more challenging because of the constantly varying terrestrial atmosphere causing image distortions on time-scales shorter than the data acquisition rates. 
In addition to adaptive optics corrections during observations, post-processing of imaging data is carried out with various image restoration techniques to achieve the diffraction limit of the instrument. 
After that, deconvolution similar to that applied to space data may allow achieving true contrasts of observed features. 
Alternatively, simulated and modeled observables can be convoluted with known instrumentation effects to be compared with the data. 
Both approaches have uncertainties and removing observing biases are to be improved. 
Modeling solar irradiance variability modulated by surface magnetism and comparing this with measurements are important for (1) reconstructing the solar irradiance variability \citep{chapman2013,chat2020}; (2) validating atmosphere models employed in solar irradiance reconstruction \citep{Fontenla:1993, Fontenla:1999, Fontenla:2006, Fontenla:2009, unruh1999, penza2004, ermolli2010}; and (3) understanding relations observed between various magnetic-activity indices \citep{choud2020, ayres2020, clette2021}.

Understanding the physics of faculae in terms of their magnetic structure, radiative transfer, and dynamic substructure is important for predicting their effects at different wavelengths. 
Here both realistic 3D simulations and a detailed modeling of opacities are of great importance. 
For example, \citet{Schuessler2003} showed why solar magnetic flux concentrations are bright in molecular bands using radiative magnetoconvection simulations and proper molecular opacities in the presence of magnetic fields by \citet{Berdyugina2003mol}. 
This was followed by similar work by \citet{keller+al2004}, \citet{steiner2005}, and \citet{de_pontieu+al2006}.

\begin{figure}
    \centering
    \includegraphics[height=\textheight, width=\columnwidth, keepaspectratio]{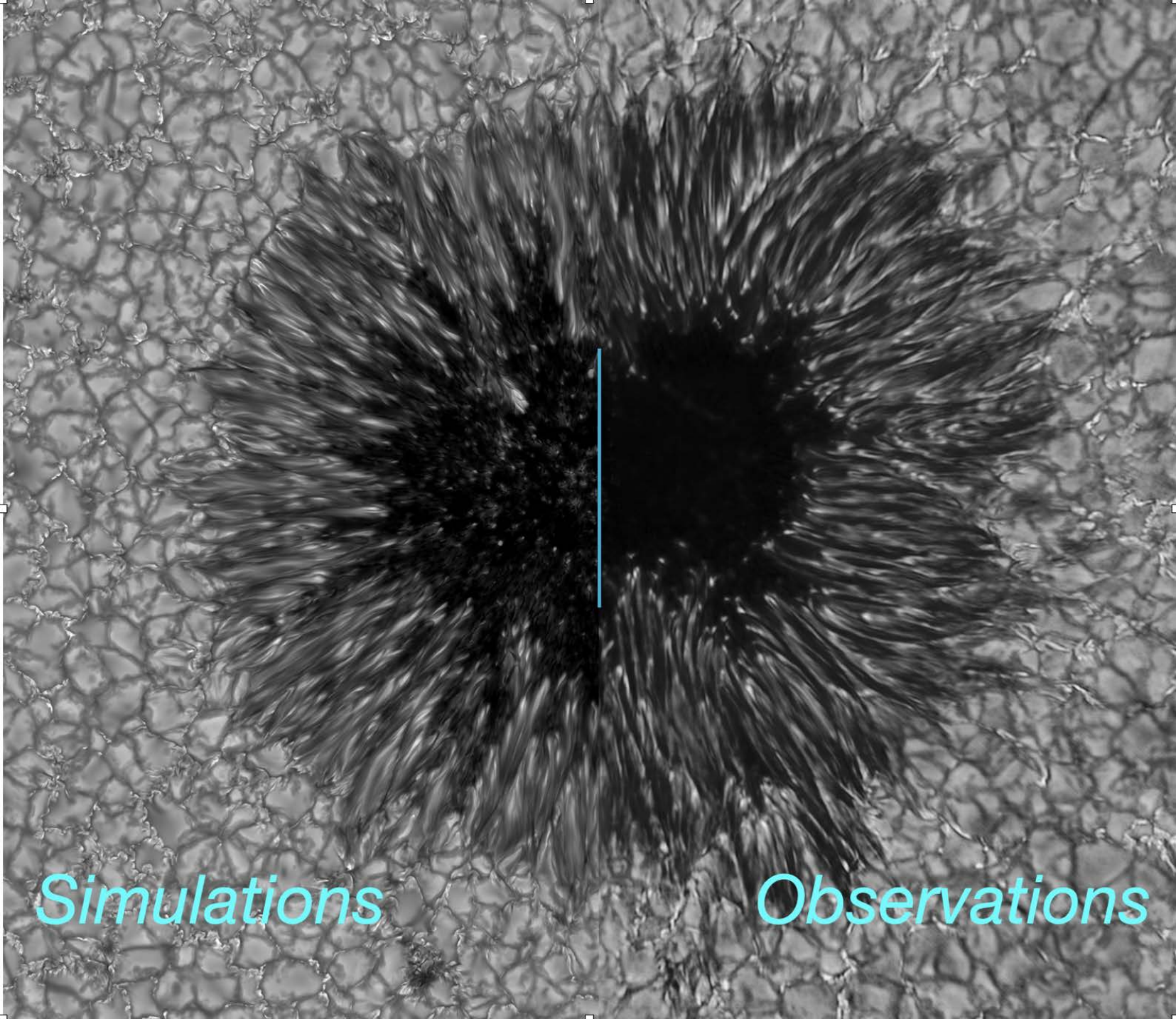}
    \caption{
    MURaM simulations of a sunspot by Matthias Rempel (left half of the plot) vs. Dunn Solar Telescope observations by F.\ W{\"o}ger (right half of the plot). 
    \revised{The vertical blue line separates the simulated image from the observed one.}
    Credit: F.\ W{\"o}ger.
    }
    \label{fig:MURAM_vs_DST}
\end{figure}

Similarly, understanding physics of sunspots and obtaining their realistic models is helpful for modeling starspot effects on other stars at different wavelengths.
\cite{rempel09a} discussed the magnetoconvective processes that occur in a sunspot and provided the first insights into penumbral filament formation. 
Subsequently, \cite{Rempel:2009Sci} simulated  circular sunspots with full-fledged penumbrae that reached such levels of realism that the simulated intensity images were almost indistinguishable from observed structures (see \autoref{fig:MURAM_vs_DST}).
However, physical drivers for forming sunspot penumbra are still not fully understood \citep{Rempel2012}.

The fine structure and large range of temperatures in sunspots results in a dramatic difference of their spectra as compared to that of the quiet photosphere. 
In particular, spectra of the dark umbra are similar to those of M dwarf's quiet photospheres, with strong molecular bands from metal hydrides and oxides (e.g., MgH, CaH, FeH, CrH, SH, TiO, VO). 
In addition, CNO-based molecules (e.g., CH, NH, OH, CO, CN, C2) form in both the quiet photosphere and chromosphere as well as in sunspots \citep[see an overview by][]{Berdyugina2011}. 
Magnetic effects in these molecules are currently well understood and implemented in the HotMol numerical library. 
They are employed for measuring and modeling magnetic fields on the Sun and other stars.
Other molecules have been detected in sunspots in the near infrared, e.g., H$_2$O \citep{Wohl1971,Polyansky1997}. 
These molecular bands, along with strong atomic (resonance) lines, are spectral features that are being detected in exoplanetary atmospheres. 
Thus, modeling spectra of sunspot substructures is important for the decontamination of composite star-planet spectra.

\subsubsection{Observations of Venus and Mercury transits of the Sun as a star}
\label{sec:solartransits}

Photometric observations of exoplanets across their host stars contend with the many stellar-variability issues discussed here, including global oscillations, convection, granulation, magnetically active regions, and center-to-limb variations.
In the case of the Sun, the spatially averaged variations due to oscillations, granulation, and convection cause ${\sim}10^{-4}$ variations in spatially and spectrally integrated light \citep{2016JSWSC...6A..30K}. 
For comparison, the transit of the Earth across the solar disk viewed from a distant system would have a maximum transit depth of ${\sim}8 \times 10^{-5}$, and can thus be masked by the background solar variability.

Fortunately, the timescales of solar variability and planetary transits differ sufficiently that transits of Earth-sized planets across Sun-like stars would, in fact, be detectable since they affect different portions of the stellar light-curve power spectrum. 
Solar variability due to oscillations, convection, and granulation primarily occurs on timescales of several minutes, while active regions, such as sunspots and faculae, vary on multi-day timescales. 
In contrast, planetary transits occur on ${\sim}$1--10\,hr timescales. 
This timescale difference makes the exoplanet detection of Earth-like planets across solar-like stars feasible via photometric observations.

Observations of Mercury transits across the solar disk as viewed from the Earth demonstrate the difficulties in discovering small exoplanets orbiting G-type stars via photometric measurements of the stellar light curves during planetary transits. 
Four Mercury and two Venus transits have been observed by the Total Irradiance Monitor (TIM) on NASA's Earth-orbiting \revised{Solar Radiation and Climate Experiment (SORCE) mission} \citep{Kopp2021}. 
This instrument measures the TSI, the spatially and spectrally integrated radiant solar energy flux incident at the top of the Earth's atmosphere. 
This essentially views the Sun as a star. 
Instrument sensitivities are ${\sim}4 \times 10^{-6}$, making the TIM inherently capable of detecting both intrinsic solar variabilities and solar-disk transits by inner planets. 

\begin{figure}
     \centering
     \includegraphics[height=\textheight, width=\columnwidth, keepaspectratio]{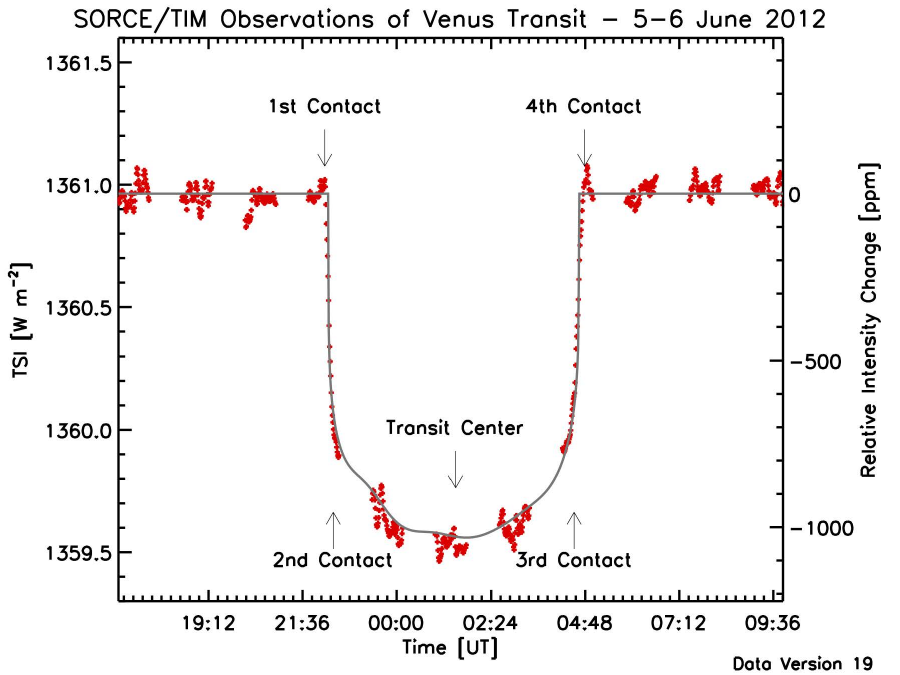}
     \caption{
     The \mbox{SORCE/TIM} measured a decrease in the TSI (red dots) as Venus transited the Sun on both 5--6\,Jun\,2012 (shown) and 8\,Jun\,2004. 
     In agreement with predictions (grey curve) accounting for solar limb-darkening and the SORCE position, the incident sunlight decreased by ${\sim}10^{-3}$ during the transits, which is comparable to the effect of a medium-sized sunspot in TSI. 
     The gaps in the plotted data are from times when the SORCE spacecraft was in the Earth's shadow and could not view the Sun.
     All four \revised{transit contact points}, covering both ingress and egress, were observed in the shown Venus transit. 
     The increases in brightness near ingress and egress during the transit are due to solar limb-darkening, which makes the center of the solar disk brighter than the edges and hence the transit-depth greater when Venus is nearer to disk-center. 
     The small fluctuations in brightness on short timescales are from normal solar convection and oscillations, and can be seen in the unocculted times both before and after the transits. 
     From \citet{Kopp2021}.
     }
     \label{fig:Venus_Transit}
 \end{figure}

\begin{figure}
     \centering
     \includegraphics[height=\textheight, width=\columnwidth, keepaspectratio]{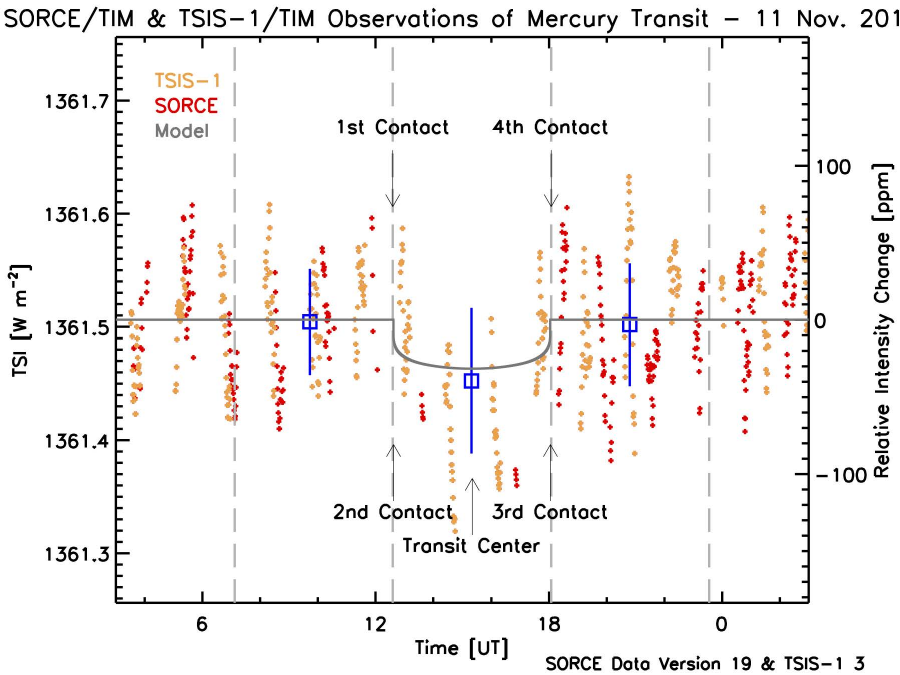}
     \caption{
     Both the \mbox{SORCE/TIM} and the \mbox{TSIS-1/TIM} observed the 11\,Nov\,2019 Mercury transit. 
     The high-cadence TSIS-1/TIM TSI values are plotted in orange and SORCE (scaled to the \mbox{TSIS-1} values) in red. 
     The scatter in these high-cadence TSI values is almost all due to actual solar variability. 
     The blue boxes and whiskers are, respectively, the averages and standard deviations of equal-length time regions before, during, and after the transit from both instruments. 
     The grey curve is the predicted signal for a limb-darkened solar disk viewed from the TSIS-1. 
     As with three prior Mercury transits observed by the SORCE, the transit signal is likely there but would not be readily apparent without knowing where to look, since the ${\sim}10^{-4}$ background solar variations mask the $3.7 \times 10^{-5}$ expected transit-depth signal. 
     From \citet{Kopp2021}.
     }
     \label{fig:Mercury_Transit}
 \end{figure}

Viewed from the proximity of the Earth, Venus-transit light curves have large transit depths ($\sim$~$10^{-3}$), making such transits unambiguous from the smaller effects of background solar oscillations and convection. \revised{While detections of exoplanet signals typically involve several transits, these snapshots of stellar variability allow for their detailed study.} 
The TIM radiometric measurements of the 2012 Venus transit are shown in \autoref{fig:Venus_Transit}. 
Even viewed from the Earth, Mercurial transits are much smaller. 
These have ${\sim}4 \times 10^{-5}$ transit depths in integrated light, or approximately half the light-curve depth that an Earth-sized exoplanet would have when transiting a G-type star in a distant planetary system. 
These transits are largely masked by the ever-present background solar variability
(see \autoref{fig:Mercury_Transit}).

With such solar-variability masking, multiple transit observations are required for the unambiguous detection of an Earth-sized exoplanet across a solar-like star.
Knowledge of the amplitudes and timescales of the host star's intrinsic variability is also necessary for detecting small exoplanets via transits. 
Measurements of the Sun provide such knowledge for the most ``Sun-like'' star known, and actual radiometric transit observations of the Sun's innermost planet, Mercury, exemplify the issues expected for exoplanet discoveries of \revised{transits of} Earth-sized planets across host G-type stars.

\subsubsection{Our finding}
From this analysis, we draw the following finding.
\\
\\
\noindent \underline{\textbf{Finding 1.1}}
\\
\\
\noindent \textit{Summary}: 
The Sun provides the benchmark for stellar studies.
Studying the spatial, spectral, and temporal variations of stellar surface structures is necessary for understanding the impact of analogous structures on transmission spectroscopy of exoplanets.
\\
\\
\noindent \textit{Capability Needed}: 
More work is needed to study spatially resolved properties \revised{of the Sun and their variations} on timescales from minutes to years \revised{at optical and infrared (0.3--5\,$\micron$) wavelengths}. 
\revised{Theoretical and observational perspectives are needed to further our understanding of both granules and active regions.
For both granules and active regions, key parameters to understand better are sizes and contrasts at optical and near-infrared wavelengths.
For active regions, key parameters to study also include locations, lifetimes, and vertical structure variations from the photosphere to the chromosphere.}
\\
\\
\noindent \textit{Capability Today}:
High-resolution observations of the Sun are available for fine-structure studies of photospheric and chromospheric inhomogeneities, but their wavelength range is limited. 
Advanced codes \revised{necessary solar and stellar magnetic studies are available}, such as the HotMol numerical library \revised{for spectral synthesis using realistic molecular opacities in the presence of magnetic fields \citep{Berdyugina2002mol,Berdyugina2000mol,Berdyugina2003mol,Berdyugina2005mol}}.
The total solar irradiance of the Sun can be modeled accurately with codes such as SATIRE \revised{\citep[Spectral And Total Irradiance REconstruction;][]{Fligge2000, Krivova2003, yeo+al2014}} and MURaM \revised{\citep[Max-Planck-Institute for Aeronomy/University of Chicago Radiation Magneto-hydrodynamics code;][]{Vogler2005,rempel09a,Beeck1,Beeck2}}.
Three-dimensional (3D) magnetohydrodynamic (MHD) codes, such as MURaM \revised{\citep{Vogler2005}}, CO5BOLD \revised{\citep[COnservative COde for the COmputation of COmpressible COnvection in a BOx of L Dimensions, L=2,3;][]{freytag2002, ludwig2009cfist,Co5bold_freytag12,bonifacio2018cfist,co5bold_salhab}}, and Bifrost \revised{\citep{gudiksen2011bifrost},} can be used to produce realistic simulations of the solar photosphere and chromosphere, including active regions and sunspots. 
\\
\\
\noindent \textit{Mitigation in Progress}: 
When fully operational, DKIST will have unprecedented spatial and temporal resolution for observations of the solar photosphere and chromosphere at wavelengths of 0.4--5\,$\micron$ \revised{\citep{DKIST}}.
The augmentation of DKIST observations with full-disk, synoptic observations, such as those from SOLIS \revised{\citep{Keller2003}}, will provide a global context that is continuous in space and time and, with DKIST observations, will extend across all relevant spatial scales.
This complementary approach will yield a comprehensive view of solar activity and, by extension, the magnetic activity in \revised{S}un-like hosts of planetary systems.

\begin{table*}
    \caption{Available \revised{non-magnetic 3D} model grids.}
    \label{tab:hydro_codes}
    \begin{center} 
    \begin{tabular}{ c | c | c | c | c}
        \hline  
        \hline
        &  & & &\\
        Code & \large T$_\mathrm{eff}$ [K]& \large log g & \large [Fe/H] & \large Papers\\ 
        \hline
        & & & &\\
        STAGGER & 4185--6901  & 2.2--4.74 & 0.0  & \cite{Stein_Nordlund98,Trampedach_STAGGER}\\
                & 4000--7000  & 1.5--5.00 & -4.0--0.5 & \cite{magic+al2013a}\\
        & & & &\\
        C\revised{O}5BOLD & 3600--6750 & 1.0--5.0 & -3,-2,-1,0 &  \cite{ludwig2009cfist}
        \cite{Co5bold_freytag12}\\ 
         & & & &\\
        MURaM & 3690--6893  & 4.3--4.826 & 0.0 & \cite{Vogler2005,rempel09a,Beeck_2011,Beeck1,Beeck2}\\ 
         & & & &\\
         \hline
         
    \end{tabular}
      
    \end{center}
\end{table*}


\subsection{3D \revised{s}tellar modeling: from the Sun to other stars}
\label{sec:stellarmodels}

Since the first time-dependent 3D simulations of the Sun \citep[e.g.,][]{1982nordlund,1984nordlund,1985SoPh..100..209N} and other stars \citep[e.g.,][]{atroshchenko+al1989a,atroshchenko+al1989b,Nord_Dravins_90A,dravins+nordlund1990a,dravins+nordlund1990b}, stellar 3D models have been improved in terms of the spatial resolution of the numerical grid, the extent of the computational box, and the included physical processes. 
As all these improvements compete for limited computational resources, some models are geared towards either high resolution or covering large spatial scales \citep[even up to global simulations of giant stars,][]{freytag2002}. 
Others aim for a high degree of realism by including as many of the relevant physical processes as possible.  
The extension from pure hydrodynamics to ideal magnetohydrodynamics for 3D models of the lower solar atmosphere was a costly but necessary step in the early 2000s. 
Today, \revised{3D MHD models are} considered \revised{the} standard for solar simulations and \revised{are} increasingly applied to simulations of other stars. 

The selected compromise between number of grid cells and included physics is typically determined by the intended scientific application. 
For instance, simulations of sunspots or even active regions require large computational boxes, but the minimum spatial resolution of the computational grid is constrained by the spatial scales that need to be resolved adequately, e.g., for realistically reproducing the photospheric flow field. 
The resulting number of grid cells then limits the detail and extent to which physical processes can be modeled. 
Another example is the extension of models in height into the chromosphere or even the corona. 
In these layers, many assumptions that can be made for photospheric conditions are no longer valid. 
For instance, the ionisation of hydrogen is no longer in statistical equilibrium and instead requires detailed calculations that are computationally expensive \citep{carlsson2002hion,leenaarts+wedemeyer2006,leenaarts+al2007} despite using simplified model atoms. 
Nonetheless, these improvements are necessary for realistic simulations of the upper atmosphere and thus for the interpretation of spectral lines and phenomena originating in these layers. 
An example is the widely used 3D model of an enhanced network region by \citet{carlsson2016sim} that has been produced with the Bifrost code. 
Depending on the exact application, even otherwise seemingly quiescent conditions in the solar photosphere can require models with a high degree of realism. 
This is because small-scale variations of the atmospheric structure in a line-forming layer and its dynamical (and magnetic) properties can have a notable impact on the resulting spectral line shapes. 
For instance, 3D simulations have resulted in the more precise determination of chemical abundances \citep[see, e.g.,][]{steffen2002, Asplund1, Asplund2} and insights into the formation of the H$\alpha$ line in the solar chromosphere \citep{Leenaarts2012}.

In the context of observing the atmospheres of exoplanets, numerical simulations of the atmospheres of host stars can support the interpretation of such observations and can also aid the design of adequate observing and data reduction strategies. 
The first step towards producing numerical simulations for this purpose is thus to define the wanted spectral diagnostics and the required precision of the anticipated measurements. 
These requirements then determine the needed setup and degree of realism of the simulations. 
The second step is to see how to produce these models based on already existing models (discussion in the following sections). 
It should be emphasised that the production of advanced 3D numerical simulations of stellar atmospheres is a computationally very expensive endeavour that requires much computational time on high-performance-computing infrastructure. 
Substantial \revised{high-performance-computing} resources are thus needed as an integral part of the preparatory phase of the scientific strategy outlined here. 

In the following sections, we detail the available state-of-the-art non-magnetic and magnetohydrodynamic 3D model grids (\autoref{sec:stellarmodels_grids}) and modeling considerations as they relate to heterogeneities due to plasma motion (granulation and oscillations, \autoref{sec:stellarmodels_largescale}), small-scale bright magnetic features (faculae and network, \autoref{sec:stellarmodels_smallscale}), and dark spot-like magnetic heterogeneities (umbra, penumbra, and pores, \autoref{sec:stellarmodels_spots}).

\subsubsection{Available non-magnetic model grids}
\label{sec:stellarmodels_grids}

Systematic grids of local 3D model stellar atmospheres have been produced with several radiation (magneto-)hydrodynamics (MHD) codes.
The main codes used to produce grids of non-magnetic models for studying stellar near-surface convection are STAGGER \citep{Stein_Nordlund98,magic+al2013a,Trampedach_STAGGER}, C\revised{O}5BOLD, and MURaM.
\autoref{tab:hydro_codes} lists the range of spectral types, surface gravities, and metallicities covered by some of the prominent model grids. 
We refer the reader to \citet{Beeck2012codecomparison} for a detailed comparison of the codes. 
Naturally, all three listed codes have been also used in numerous studies, sometimes simulating stellar atmospheres outside the listed grids. 
For example, C\revised{O}5BOLD has been used to simulate M stars with effective temperatures down to 2500\,K \citep{Wende2009,Wedemeyer2013}.
These model grids are complemented by small grids of local, three-dimensional stellar atmospheres including magnetic fields \citep{Beeck3,Beeck4, co5bold_salhab} and 3D MHD models of M-dwarf atmospheres with a chromosphere \citep{Wedemeyer2013}.

In sum, there exists today a large range of 3D stellar non-magnetic models ready for future analysis.
The grids cover main-sequence stars with $3600 < T_\mathrm{eff} < 7000$ (\autoref{tab:hydro_codes}), spanning the parameters of \revised{more than 90\% of known} exoplanet host stars\footnote{\revised{NASA Exoplanet Archive, accessed 20~Feb~2023 at \url{https://exoplanetarchive.ipac.caltech.edu/}.}}.

However, the models include varying degrees of realism in terms of setup, metallicity, the inclusion of an upper atmosphere, and more. 
Unfortunately, these models do not yet include the effect of magnetic fields and simulations of stellar magnetic features are currently only available for several values of effective temperature (see \autoref{sec:stellarmodels_smallscale}--\ref{sec:stellarmodels_spots}). 
The simulations of magnetic features on stars with a wide range of effective temperatures, metallicities, and surface gravities \revised{(in particular, covering the fundamental parameters of known cool stars hosting exoplanets, i.e., from early F- to late M-dwarfs and metallicity values from at least $-1$ to $+0.5$)}  are urgently needed for assessing the magnetic component of stellar contamination of transmission spectra. 
\revised{While the above mentioned simulations typically cover only a small part of a stellar atmosphere, they} are ready to serve as the basis for the production of models in the context of exoplanet observations, given sufficient investments in the computational and human resources required for these efforts. 

\begin{figure*}
    \centering
    \includegraphics[height=0.9\textheight, width=\textwidth, keepaspectratio]{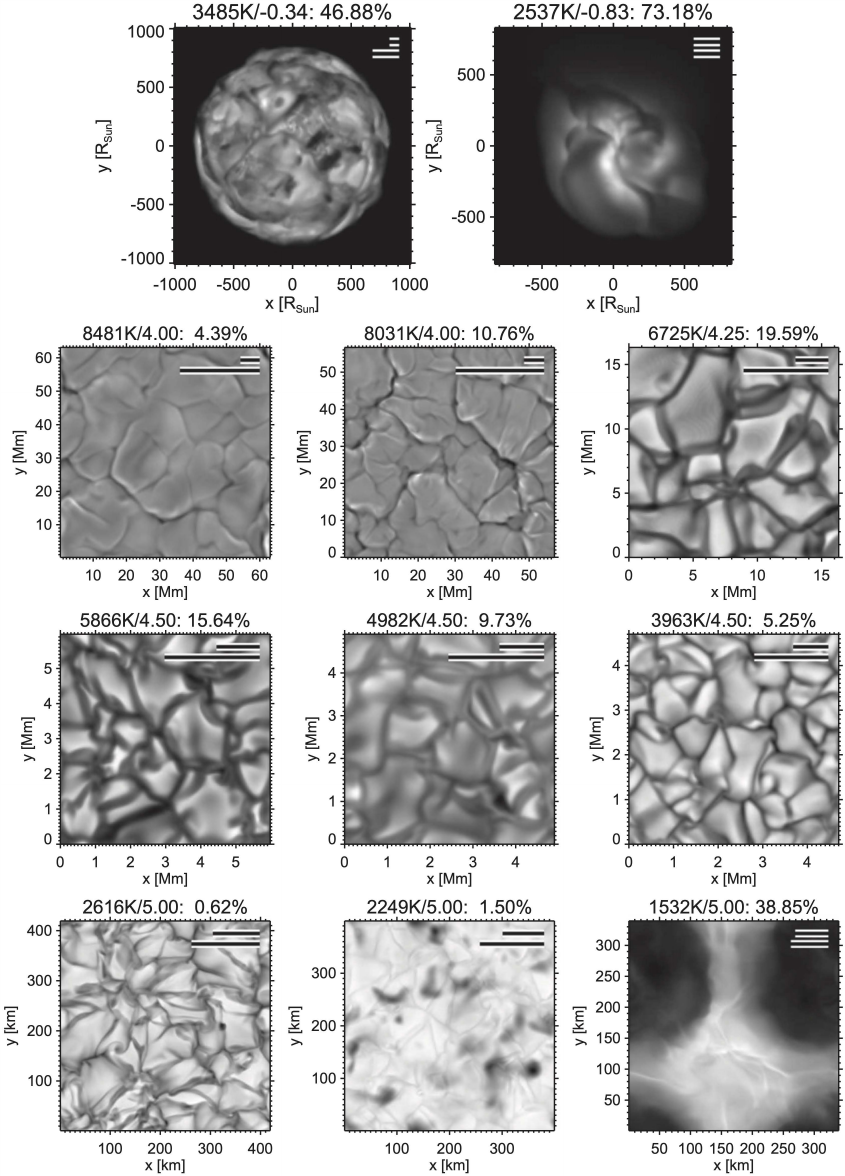}
    \caption{
    Instances of the \revised{normalised, time-averaged, and} frequency-integrated (bolometric) intensity \revised{(grey scale)} from various simulated \revised{solar-metallicity} stellar surfaces \revised{with CO5BOLD}. 
    \textit{Top row:} Global models of a red supergiant and an AGB star of low surface gravity. 
    \textit{Following rows:} Local models of \revised{main-sequence} stars \revised{and brown dwarfs}.
    The title lines show the effective temperature, the decadic logarithm of the surface gravity in cm\,s$^{-2}$, and the relative bolometric intensity contrast averaged over a representative time span. 
    The length of the upper bar in the top right of each panel is $10\times$ the surface pressure scale height.
    The bar below is 10$\times$ the pressure scale height but measured 3~pressure scale heights below the surface. 
    From \citet{Co5bold_freytag12}.
    }
    \label{fig:stellar_gran_contrast}
\end{figure*}

\subsubsection{Granulation and oscillations in stellar models}
\label{sec:stellarmodels_largescale}

\paragraph{Granulation.} 
The intensity contrast of granules defines the amplitude of the granulation-driven contamination of the transmission spectrum and, thus, its accurate modeling for a broad class of stars is of special importance. 

The first 3D simulations of granulation on the Sun were done by \cite{1982nordlund,1984nordlund,1985SoPh..100..209N}. 
Soon after that, \cite{Nord_Dravins_90A} extended the solar simulations to other Sun-like stars: Procyon F5 IV\revised{--}V, $\alpha$ Cen A G2V, $\beta$ Hyi G2IV, and $\beta$ Cen B  K1V (T$_\mathrm{eff}$ between 5200K and 6600 K), and further computed synthetic Fe line profiles and their bisectors to compare with observations.
These early simulations established the strong dependence of granular velocity, lifetimes, and size on the spectral type. 

\autoref{fig:stellar_gran_contrast} shows examples of granulation on stellar surfaces as bolometric intensity maps with the RMS intensity contrast given in each frame.
\revised{
The upper panel shows simulations of a red supergiant and an AGB star. 
The granulation scale in such stars is comparable to the size of a star so that only several very prominent (note the very high values of the RMS intensity contrast) granules are present on their surfaces. 
Therefore giant stars can be simulated using a star-in-a-box approach, in which the entire stellar atmosphere is included in the simulation box. 
The remaining panels in \autoref{fig:stellar_gran_contrast} present simulations of dwarfs for which the surfaces contain many granules. 
For such stars, the star-in-a-box approach would demand a very fine spatial mesh to resolve the granulation, which is presently not feasible. 
Thus, they are presently simulated using a box-in-a-star approach, in which the simulation box covers only a small part of stellar atmosphere.  
The RMS contrast generally increases with increasing effective temperature; 
in other words, for hotter stars there is generally a larger contrast variation between the bright (up-welling) granules and dark (down-flowing) inter-granular lanes. 
The exception from this rule is ultracool stars and substellar objects, e.g., brown dwarfs (see middle and right panels in the bottom row) as well as hot A-stars (see left and middle panels of the second row)}.

For a solar effective temperature and surface gravity, the RMS contrast increases with decreasing metallicity \citep{magic+al2013a, Witzke2023}. 
The RMS granular contrast over a wide range of effective temperatures, surface gravities, and metallicities is given by \citet{tremblay+al2013}. 
They find that the intensity contrast correlates well with the Mach number\revised{, the ratio of flow and sound speeds}, at the stellar surface over the full range of the HR diagram.
This reflects the fact that both the Mach number and the intensity contrast are a measure of the vigor of convection.

\begin{figure*}
    \centering
    \includegraphics[height=0.9\textheight, width=\textwidth, keepaspectratio]{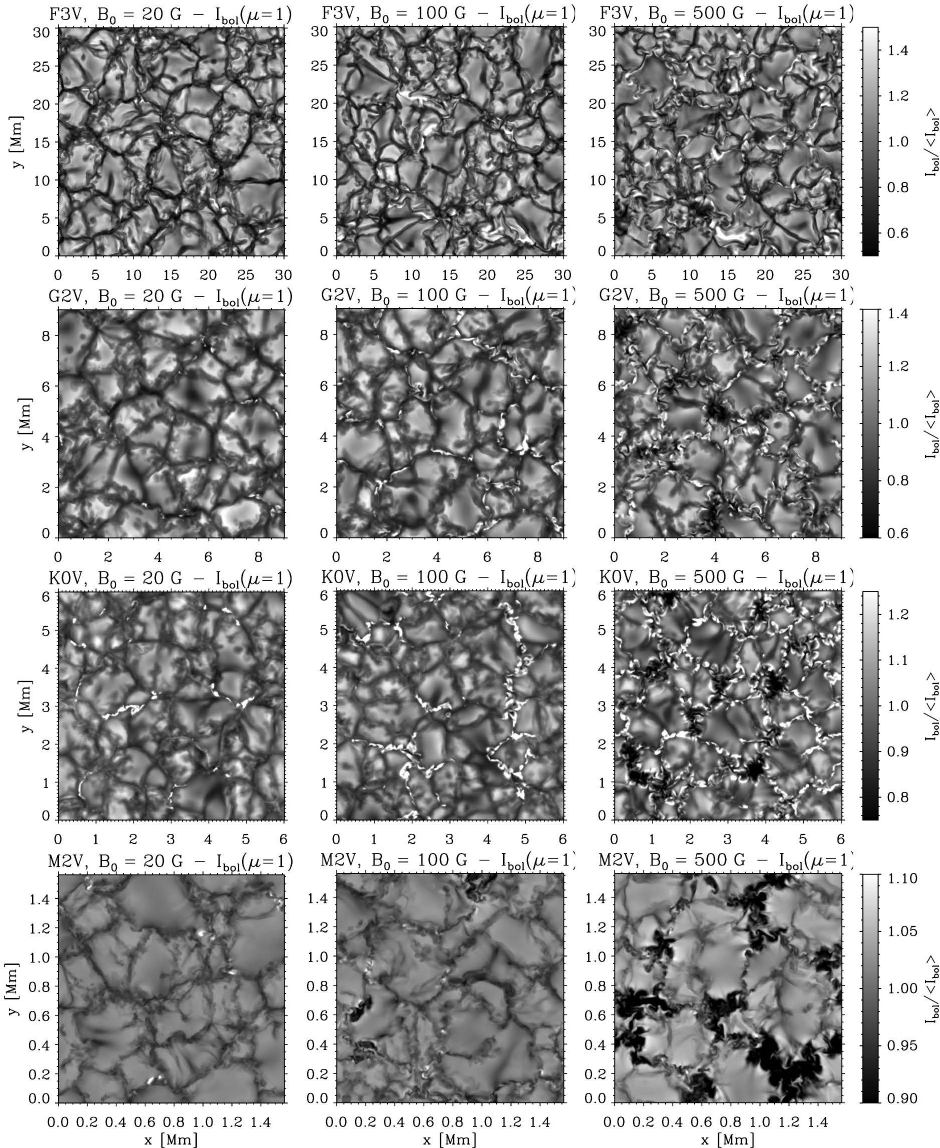}
    \caption{
    Maps of the vertically directed bolometric intensity for 12 of the 18 magnetic simulation runs. 
    Note that the scale of the normalized intensity is adjusted for each spectral type for improved image contrast, and the grayscales saturate at the limits indicated for each row. 
    As the average magnetic field strength increases, more \revised{magnetic bright points} are visible within the inter-granular lanes; 
    the quantity of \revised{magnetic bright points} also increases with effective temperature, but decreases again at temperatures greater than solar. 
    From \citet{Beeck3}.
    }
    \label{fig:stellar_filigree}
\end{figure*}

The key points from these studies are as follows.

\begin{enumerate}
    \item The temperature contrast, and therefore intensity contrast, between granules and intergranular lanes depends strongly on the stellar surface temperature. 
    Hotter stars have higher temperature contrast. 
    The temperature contrast is weakly dependent on metallicity.
    \revised{For a solar effective temperature and gravity, the contrast decreases with metallicity.}
    
    \item Granule size increases with temperature and decreases with surface gravity. 
    Thus, hotter stars with lower surface gravities have larger granules than cooler stars with higher surface gravities. 
    In cooler stars with very low surface gravities (e.g., supergiants), granule sizes can be comparable to the stellar radius. 
    Like the temperature contrast, granule size also shows a weak dependence on metallicity. 
    At the same surface temperature, decreasing the metallicity results in smaller granules.  
\end{enumerate}

\paragraph{Oscillations.} 
Predicting the amplitudes of oscillations on other stars remains challenging \citep[see, e.g.,][for a recent review]{Aerts2021}.
Instead, a number of scaling relations using fundamental stellar properties have been proposed \citep{kjeldsen2011a, samadi2012a}. 
For example, \cite{kjeldsen2011a} suggested 
\begin{equation}
    A_{\rm{vel}} \propto \frac{L \tau_{\rm{osc}}^{0.5}}{M^{1.5}T_{\rm{eff}}^{2.25}}
\end{equation}
for velocity amplitudes, and 
\begin{equation}
    A_{\lambda} \propto \frac{L \tau_{\rm{osc}}^{0.5}}{\lambda M^{1.5}T_{\rm{eff}}^{2.25+r}}
\end{equation}
for intensity amplitudes observed at a typical wavelength $\lambda$. 
Here $L, M, T_{\rm{eff}}$ and $\tau_{\rm{osc}}$ refer to luminosity, mass, effective temperature, and mode lifetime, respectively.
The exponent $r$ takes slightly different values in literature \citep[see][and references therein]{kjeldsen2011a}. 
The second scaling relation has been extensively tested and modified based on \Kepler{} data \citep[e.g.,][]{Stello2011, Huber2011, Mosser2012, Corsaro2013, Kallinger2014}. 
The oscillation amplitudes were later found to be also metallicity dependent by \cite{Yu2018a} and \cite{Vrard2018a}. 
These studies showed that, all else being equal, metal-rich stars tend to have larger oscillation amplitudes, in agreement with theoretical predictions \citep{houdek1999a, samadi2010b, samadi2010a}.

These scaling relations have been used to estimate the significance of oscillation signals against granulation background and noise to prioritise the asteroseismic target selection for the \Kepler{}, \Ktwo{}, and \revised{Transiting Exoplanet Survey Satellite \citep[\TESS{};][]{Ricker2015}} space missions \citep{Chaplin2011, Chaplin2015, Schofield2019}. 
These efforts have shown that the photometric oscillation amplitudes are very low compared to granulation and noise levels in the Fourier spectra of late K- and M-type main-sequence stars.

\subsubsection{Faculae and network in 3D stellar models}
\label{sec:stellarmodels_smallscale}

Although hydrodynamic simulations of near-surface granulation have been achieved by a number of groups, only two codes have been used to simulate magnetic fields on stars other than the Sun---MURaM \citep{Beeck_2011,Beeck3,Beeck4} and CO5BOLD \citep{Steiner2014,co5bold_salhab}. 
The MURaM papers covered the spectral types F3V, G2V, K0V, K5V, M0V, and M2V, while the CO5BOLD ones covered the types F5V, G2V, K2V and K8V.
In both cases, the authors computed network or facular fields, starting with vertical homogeneous fields as an initial condition. 
\cite{co5bold_salhab} used an initial strength of 50\,Gauss for all of the runs, while \cite{Beeck_2011,Beeck3, Beeck4} used initial strengths of 20, 100 and 500\,Gauss. 

\autoref{fig:stellar_filigree} shows the bolometric intensity maps of models computed by \citet{Beeck3}. 
The F-, G- and K-type models show the typical bright magnetic elements familiar from solar network and plage regions. 
Only when the initial magnetic field strength is taken to be a copious 500\,G do a few dark features appear among the mostly bright magnetic structures. 
In contrast, in the M2V star (and in the M0V star---not shown in \autoref{fig:stellar_filigree}), dark magnetic structures frequently form in the 100\,G runs.
In the 500\,G runs, however, large dark structures dominate, and there are essentially no bright magnetic structures left.

One remarkable behavior of the magnetic flux concentrations that form in these simulations is that their magnetic field strength\revised{,} measured at the surface with mean visible optical depth of unity \revised{(}$\langle \tau_{500}\rangle = 1$)\revised{,} is only weakly dependent on stellar spectral type and initial magnetic field strength (for fixed solar surface gravity and metallicity). 
It increases towards later spectral types, from 1250\,G for the F5V to 1440\,G for K8V \citep{co5bold_salhab}, and increases with increasingly available magnetic flux \citep{Beeck3} but remains in the range of kG field strength.

\begin{figure*}
     \includegraphics[height=\textheight, width=\textwidth, keepaspectratio]{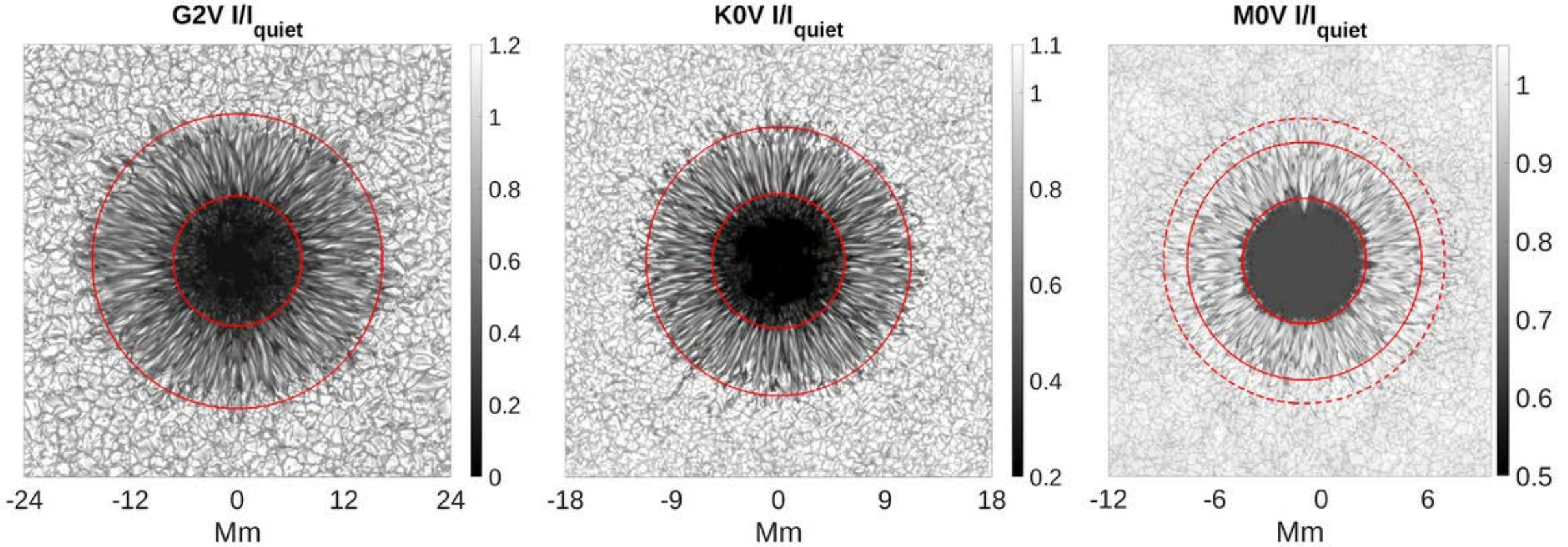}
     \caption{
     Simulated bolometric intensity maps of complete circular spots for the spectral types G2V, K0V and M0V.
     \revised{From Panja et al. (in prep.).}
     }
     \label{fig:temp_tau_spots_circ}
 \end{figure*}

\citet{Steiner2014} and \citet{co5bold_salhab} found that the presence of small-scale magnetism (e.g., \revised{magnetic bright points}) increases the bolometric intensity and flux in all their investigated models compared with equivalent magnetic field-free models. 
The surplus in radiative flux of the magnetic over the magnetic field-free atmosphere increases with increasing effective temperature, $T_\mathrm{eff}$, from 0.47\% for spectral type K8V to 1.05\% for the solar model, but decreases again for effective temperatures greater than solar. 
This agrees with the results of \citet{Beeck3}, as shown in \autoref{fig:stellar_filigree}. 
Thus, for mean magnetic flux densities of approximately 50\,G, we expect the small-scale magnetism of stars with spectral types F5V---K8V to produce a positive contribution to their bolometric luminosity \citep[see also][for the first simulations of small-scale dynamo in stars of various spectral classes]{Bhatia2021}. 
While the overall effective temperature changes in the presence of small-scale magnetic features can be small, the flux increases are strongly wavelength dependent, with particularly strong enhancement in the UV. 
The enhancement also depends strongly on the disk position so that magnetic activity is expected to noticeably affect center-to-limb variations of emergent intensity.

All the simulations described above have been performed for the solar value of the metallicity. 
However, metallicity has a strong effect on the opacity so that one can expect that  the visibility of facular features strongly depends on the metallicity \citep{witzke2018, Witzke2020}. 
Furthermore, most hot Jupiters are discovered around stars with a higher metallicity than the Sun \citep[e.g.,][]{Osborn2020}. 
All in all, accounting for a metallicity effect on magnetic features is crucial for a proper characterisation of magnetic contamination and, thus, 3D MHD simulations of magnetised stellar atmospheres for a broad range of metallicities are urgently needed.

The key points from these studies are as follows.

\begin{enumerate}
    \item The presence of small-scale magnetic fields on stellar surfaces leads to the formation of network and faculae, which can lead to strong \revised{contamination} signals in transmission spectra.
    
    \item  Small-scale magnetic features typical of network and faculae (plage) may be both bright and dark, depending on the spectral type of the star. 
    In M stars, very few bright \revised{features} form and the change in radiative flux is largely dominated by the darker magnetic features. 
   
    \item Currently very little is known about the effect of metallicity on the facular contrast. 
    While simplified 1D modeling indicates that facular contrasts might strongly depend on the metallicity \citep{witzke2018, Witzke2020}, a more realistic 3D MHD modeling of faculae on stars with different metallicities is urgently needed.
\end{enumerate}

\subsubsection{Umbra, penumbra and pores in stellar models}
\label{sec:stellarmodels_spots}

Models featuring spots have attained increasing levels of realism in the last decade \citep[e.g.,][]{Rempel:2009Sci, rempel09a, Rempel2012}.
Building upon the first 3D sunspot simulation in a slab geometry using the MURaM code \citep{rempel09a}, the current state-of-the-art MHD simulations with the MURaM code produce sunspot intensity images, complete with expansive penumbral filaments that are often indistinguishable from observations (see, e.g., \autoref{fig:MURAM_vs_DST}). 

Turning to stars other than the Sun, \cite{Panja-2020} recently conducted the first radiative magnetohydrodynamic simulations of spots, focusing on their umbral properties. 
They modeled G2V, K0V, and M0V stars and found the temperature contrast between the umbra and the surrounding photosphere to be around 1400\,K for the G2V star, 650\,K for the K0V star and 350\,K for the M0V star. 
The umbral bolometric intensities relative to the quiet star intensities are 0.3, 0.5 and 0.7 for the G2V, K0V and M0V stars, respectively. 
The umbral magnetic field strengths are all in the range 3 to 4.5\,kG. 
These simulations focused on the umbra; 
however, spots on the Sun are known to be dominated by the penumbra, with a typical umbra:penumbra areal ratio on the Sun of 1:4 \citep{Solanki-03}. 
Due to the slab geometry chosen by \cite{Panja-2020} to keep computing costs manageable, the penumbra is underrepresented in these simulations. 
Future studies will expand on this work to understand the properties of spots with expansive penumbral filaments (\autoref{fig:temp_tau_spots_circ} shows such simulated spots for the spectral types G2V, K0V and M0V), including intensity contrasts of spots at different wavelengths and different disk positions (Panja et.\ al, in prep.).

The key points from these studies are as follows.

\begin{enumerate}
    \item Modern 3D MHD simulations of sunspots excel at producing sunspot intensity images that essential\revised{ly} mirror solar observations.
    
    \item Only recently have studies produced 3D MHD simulations of spots on other stars. 
    Much work remains to be done in this arena, but the existing results reveal a trend of decreasing spot--photosphere temperature contrast for later spectral types, matching observations \citep[e.g.,][]{Berdyugina2005}.
\end{enumerate}

\subsubsection{Our finding}
From this analysis, we draw the following finding.
\\
\\
\noindent \underline{\textbf{Finding 1.2}}
\\
\\
\noindent \textit{Summary}: 
More modeling work is needed to understand how the fundamental parameters of stars other than the Sun govern their magnetic fields and the associated properties of their surface inhomogeneities.
In accordance with observations, simulations show that many lower-activity stars, such as exoplanet hosts that \revised{might be deemed as ``low risk''} for transmission spectroscopy \revised{in terms of stellar activity}, are \revised{facula}-dominated \revised{\citep{Shapiro2014, Nemec2022}.}
\revised{However}, little is known on how facular contrasts and coverages depend on stellar activity levels and fundamental parameters, such as metallicity and surface gravity. 
Both observing and modeling the fine structure of spots on stellar surfaces remain a challenge. 
\\
\\
\noindent \textit{Capability Needed}: 
MHD simulations of magnetic features and spectral synthesis for a larger range of stellar parameters, such as lower and higher metallicity, cooler temperatures, various stellar ages and/or activity levels are needed. 
In particular, we need to understand how facular and spot contrasts and coverages depend on stellar fundamental parameters and activity.
We need to be able to infer the observational properties of photospheric and chromospheric features from easily accessible grids of simulations or, ideally, from simulations tailored to particular high-priority exoplanet host stars.
Importantly, models of cool stars ($T_{\textrm{eq}} < 3700$\,K) are necessary to understand the hosts of many exoplanets that are high priority for follow-up.
\\
\\
\noindent \textit{Capability Today}: 
A limited amount of MHD \revised{radiative} simulations exist now. 
\revised{Thus, a}tmospheric structures of various magnetic features are often represented by 1D radiative equilibrium models \revised{\citep[e.g., PHOENIX;][]{Husser2013}}, but they fail dramatically for bright features \revised{\citep[e.g., faculae;][]{Norris2017, Witzke2022}}.
3D \revised{radiative} MHD codes like MURaM \revised{\citep{Vogler2005}} and CO5BOLD \revised{\citep{ludwig2009cfist}} have been used to simulate photospheric magnetic fields and surface features for a few (${<}$10) FGK and early-M main-sequence spectral types \revised{\citep{Beeck2012codecomparison,Beeck1,Bhatia2022, Co5bold_freytag12}}, but physical constraints on the fine structure of spots (umbra and penumbra) are still not understood.
\\
\\
\noindent \textit{Mitigation in Progress}: 
\revised{
Groups have recently used the 3D radiative MHD code MURam to simulate starspots on G2V, K0V, and M0V stars \citep{Panja-2020}.
Others have also used MURaM to study facular contrasts of G2V stars with varying magnetic field strengths \citep{Witzke2022, Witzke2023}.
}


\subsection{Observations of stars other than the Sun}
\label{sec:otherstars}

\subsubsection{Granulation and oscillations in other stars}

\begin{figure}
    \centering
    \includegraphics[width=1\linewidth]{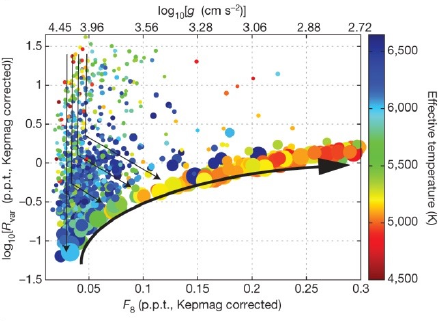}
    \includegraphics[width=1\linewidth,trim=30 375 520 108,clip]{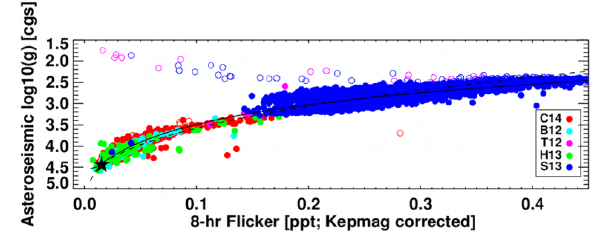}
    \caption{
    \textit{Top:} Light-curve ``flicker" on timescales of less than 8~hr ($F_8$; horizontal axis) in units of parts-per-thousand (ppt) is shown for \Kepler{} stars as a function of their evolutionary state ($\log g$ scale at top) and as a function of overall photometric variability amplitude (vertical axis).
    Arrows depict approximate evolution of stars for visualization, beginning as highly magnetically variable dwarf stars (upper left), evolving straight down as they become magnetically quieter at constant $\log g$, then evolving toward the right as they become subgiants and eventually red giants. 
    The thick arrow from the lower left toward the right represents the ``flicker floor'' set by granulation alone in the absence of magnetic activity. 
    The photometric signal of surface granulation in solar-type stars ranges from $\sim$0.02~ppt as main-sequence dwarfs up to $\gtrsim$0.3\,ppt as red giants. 
    From \citet{Bastien:2013}. 
    \textit{Bottom:} Expanded sample from various \Kepler{} asteroseismic studies, from Sun-like dwarfs (star symbol at lower left) to red giants at upper right, with their $F_8$ photometric amplitudes ranging from $\sim$0.02\,ppt to $\sim$0.45\,ppt, respectively. 
    From \citet{Bastien:2016}.
    }
    \label{fig:flicker}
\end{figure}

While plasma-driven processes like granulation and oscillations cannot yet be spatially resolved in stars other than the Sun, their impact in other stars can still be detected in unresolved observations.
These include extremely precise radial velocity and spectral line profile measurements (not reviewed here) as well as photometric time-series measurements (i.e., light curves), particularly from space-based observatories such as the Convection Rotation and Planetary Transits \citep[CoRoT;][]{Corot} Space Telescope, the \revised{\Kepler{}/\Ktwo{}} Space Telescope, and \revised{\TESS{}}. 
For example, \Kepler{} observations of thousands of red giant stars have yielded detailed asteroseismic analyses of the oscillation frequencies of these stars, oscillations which are driven largely by surface convection. 
This enables accurate determination of bulk stellar properties, such as mass and radius, as well as detailed internal stellar structure and rotation profiles \citep[see, e.g.,][and references therein]{Chaplin:2011, Huber2011, Mathur:2011, Stello:2013}. 
In terms of granulation, \citet{Bastien:2013, Bastien:2016} demonstrated a technique to measure the properties of the surface convective motions of solar-type main-sequence dwarfs, known as ``flicker" ($F_8$). 
$F_8$ is the amplitude of stochastic variations in the light curve on timescales shorter than 8\,hr and can be directly tied to the stars' surface granulation properties (\autoref{fig:flicker}).  

Comparisons with the predicted granulation properties (i.e., the $F_8$ amplitude) from 3D stellar models show good agreement in general. 
However, for stars hotter than $\sim$6200\,K, the models predict amplitudes significantly larger (a factor of ${\sim}2$) than what is observed. 
The smaller observed granulation amplitudes may imply some suppression mechanism (e.g., surface magnetic fields) that reduces the granulation amplitudes relative to expectations \citep{Cranmer:2014}. 
In addition, comparisons of observations and models in order to understand granulation in ensemble of stars have typically not used the full array of information that exists on the power spectrum of the time-series where they are observed---a technique that studies of stellar oscillations do regularly use. 
Recent work, however, has begun to observationally characterize the full granulation power spectrum via data-driven inference in \TESS{} light curves \citep[e.g.,][]{Sayeed:2021}. 

While an evident source of noise to be accounted for when analyzing transit light curves, the impact of granulation and oscillations as a contamination source to the transit spectrum has been only explored using simulations. 
\cite{Chiavassa-17}, simulating transits in front of stars using 3D stellar models, suggest that indeed biases on the transit depth can occur due to the fact that exoplanets could transit on chords in the stellar surface that have a significantly different brightness due to the granulation pattern. 
The effect could be of the order of a few percent on the transit depth of a terrestrial exoplanet orbiting a Sun-like star in the red-end of the optical (${\sim}0.7$\,\micron), amounting to a few ppm in the case of an Earth\revised{--}Sun 
analogue exoplanetary system---within the same order of magnitude as atmospheric features in a transiting Earth\revised{--}Sun analogue exoplanetary system \citep{BK:2013}.
This percentage seems to rapidly decline at longer wavelengths, which suggests could make it a negligible effect in practice for most up-coming space-based observatories focusing on the infrared. \
However, it could become a problem for transit observations in the optical range, such as the ones that missions proposed by, e.g., the Astro2020 Decadal Survey could perform. 
Validating these granulation simulations with wavelength-dependant data of both the Sun and stars other than the Sun seems to be fundamental to understand the true impact of this effect on future high-precision observations of transiting exoplanetary systems. 
Although likely having a smaller impact due to the smaller associated amplitudes, performing the same type of simulations and validations for oscillations would be highly beneficial to understand their impact as a contamination source.

\subsubsection{Observing starspots \& faculae: techniques and their interplay}
\label{sec:observing-starspots-and-faculae}

Similar to granulation and oscillations, photometric monitoring of stars other than the Sun has enabled the study of the hours-to-multi-year variations produced by 
surface inhomogeneities for many decades \citep{Kron1947}. 
With the advent of stellar high-resolution spectroscopy and spectropolarimetry, these were found to be caused by magnetic regions similar to those on the Sun---starspots \citep{Berdyugina2005}.

Early attempts to interpret the observed photometric variability in other stars involved extrapolating properties from synthetic light-curves of model stars with differing spot parameters: circular uniform spots \citep[e.g.,][]{Budding1977}; spots with defined umbral and penumbral areas \citep[e.g.,][]{Dorren1987}; and active latitudes and/or longitudes \citep[e.g.,][]{BoppEvans1973}. 
A single-starspot model is rarely a solution for observed photometric variations.
Also, it is unlikely that a light-curve analysis results in 
a unique solution. 
Instead, a set of various solutions usually satisfies the same data. 
\revised{Moreover, intrinsic degeneracies exist in the inversion of rotational light curves, and sets of surface features with zero impact are impossible to infer from single-band photometry \citep{Cowan2013, Luger2021}.}
\revised{Nonetheless, a}s the volume of photometric data has dramatically increased, more advanced tools have been developed to analyze and interpret \revised{these observations} \citep[e.g.\revised{,}][]{Lanza1998,Berdyugina2002lqhya,Ribarik2003,Maxted-16, Luger2021b, Luger2021}.

\begin{figure}
    \centering
    \includegraphics[width=\columnwidth, height=\textheight, keepaspectratio, trim={190 80 90 90}, clip]{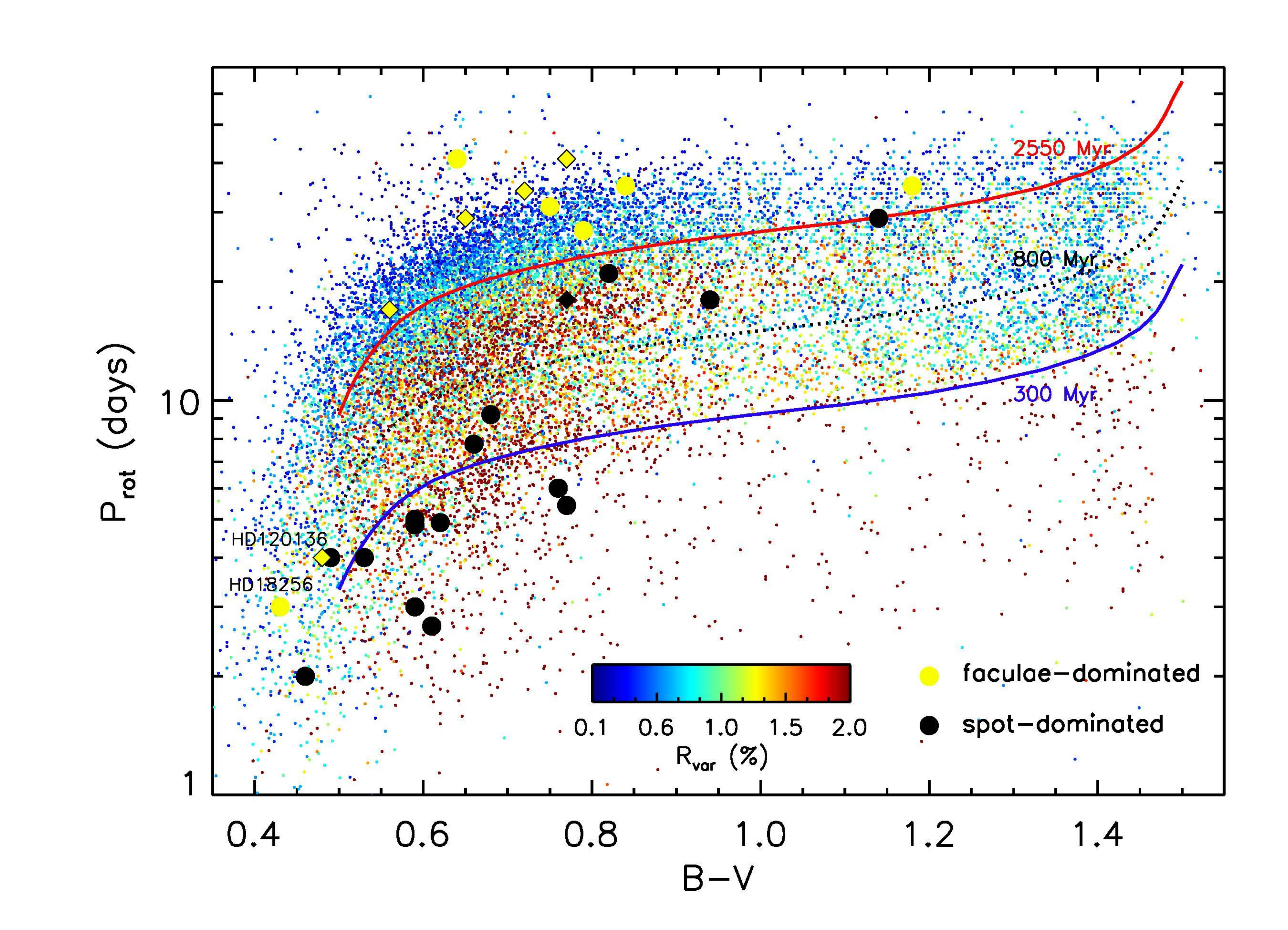}
    \caption{
    \Kepler{} rotation periods \revised{$P_\mathrm{rot}$} against B-V color. 
    The data points are color-coded with the variability range \revised{$R_\mathrm{var}$}. 
    Black and yellow symbols show spot- and \revised{facula}-dominated stars, respectively. 
    Dots show dwarf stars (luminosity class V) and diamond show (sub-)giants (luminosity class III, IV, IV--V). 
    The solid blue, dotted black, and solid red lines show 300\,Myr, 800\,Myr, and 2550\,Myr isochrones, respectively \citep{reinhold2019transition}.
    }
    \label{fig:kepler_rot_dreath}
\end{figure}

Understanding solar variability has been very helpful in solving 
problems posed by stellar observations---in particular, the brightness variations as detected by ground-based efforts and space-based missions 
such as CoRoT, \Kepler{} and \TESS{} \revised{\citep[e.g.,][]{Michel2008, Southworth2011, Reinhold-13, witzke2018, shapiro2020inflection}}. 
For instance, inspired by solar observations, \citet{Radick1998} and 
\citet{reinhold2019transition} demonstrated that active stars are dominated by dark spots, whereas less-active stars are dominated by bright faculae and possibly networks. 
\citet{reinhold2019transition} found the transition from spot- to \revised{facula}-domination to happen at a Rossby number of about one (i.e., when the rotation period is roughly equal to the convective overturn time). 
This corresponds to the Vaughan–Preston gap \revised{\citep{Vaughan1980}}, where a dearth of F-, G-, and K-type stars with intermediate levels of magnetic activity has been observed  \revised{\citep{Noyes-84, Baliunas-95, Henry1996, Fossati2015}}.
They proposed an age of around 800\,Myr for this transition, in accordance with the conclusions of \citet{Radick1998}. 
This also explains the apparent lack of stellar rotation periods between 15--25\,days in the \Kepler{} field, which would result from the contributions of dark spots and bright faculae brightness cancelling at these periods \revised{\citep{Montet2017}}. 
This result suggests that the observations do not necessarily imply an under-representation in the real period distribution but rather the lack of photometric modulation in the light curves. 
\revised{\citet{reinhold2019transition}} also concluded that stars at ages $\geq$2.55\,Gyr should be \revised{facula}-dominated (see \autoref{fig:kepler_rot_dreath}), although further investigations with a larger sample size might be required to robustly extend these results to a wider range of stellar types and properties. 
While good empirical models based on the Sun exist for solar-like stars, similar 
models for other spectral types need more investigation and data to support them 
\citep{Shapiro2014, yeo2020faculae}.
For example, \citet{see2021photometric} show a positive correlation between 
the variability amplitude and metallicity of stars and suggest rotation period detection might be biased toward the periods of metal-rich stars for stars of a given mass \citep[see also][]{Reinhold2021}.

In addition to significant photometric effects, starspots and faculae 
demonstrate characteristic spectroscopic and spectropolarimetric signatures. 
Modeling of chromospheric features in the integrated spectra of 
stars can yield meaningful measurements of plage-like fractional area 
coverages \citep[see, e.g.,][]{Andretta:2017}; line-depth ratios are also 
a reliable source for starspot temperatures \citep[e.g.,][]{TonerGray1988,GrayJohanson1991,Catalano2002}. 
Furthermore, starspots on stars with higher effective temperatures can result in molecular lines that would not otherwise be present in their spectra \citep{Berdyugina2003mol}. 
These molecular features can be used to measure starspot properties \citep[e.g.,][]{Vogt1979,Huenemoerder1989,Neff1995,ONeal1998,Berdyugina-02,Afram-Berdyugina2015}. 
In general, hotter stars have a larger difference between the unspotted surface temperature and the spot temperature than cooler stars do \citep[e.g.,][]{Berdyugina2005,AndersenKorhonen2015}. 
This temperature difference can be as large as 2000\,K for spectral type G0 and only 200\,K for spectral type M4.
\revised{While the trend with spectral type is similar, temperature differences derived from spectral observations can be both smaller and larger than those derived from observations \citep{Panja-2020}.}
Spectropolarimetric analysis of many atomic and molecular lines reveals an internal, height-dependent temperature and magnetic structure of starspots with strong magnetic fields, which were found to be smaller and warmer in the lower atmospheres of earlier M-dwarfs than later M-dwarfs, where starspots with strong fields are bigger and cooler throughout their entire atmosphere \citep{Berdyugina2011,Afram-Berdyugina2019}.

The distribution of starspots on stellar surfaces can be recovered in some cases using the Doppler \revised{i}maging (DI) technique, as surface features induce perturbations in Doppler-broadened spectral line profiles \citep[e.g.,][]{Deutsch1958, VogtPenrod1983, Piskunov1990di, Berdyugina1998di}. 
This method is reliant on stars being rapid rotators and starspots being relatively large. 
Such large active regions have been detected on young, rapidly-rotating solar-type stars \citep[e.g.,][]{Jeffers2007abdor,Jarvinen2008v889her,Jarvinen2018ekdra} as well as on evolved, rotationally synchronized binary components, such as RS CVn-type stars \citep[e.g.,][]{Vogt1999hr1099,Berdyugina1998iipeg,Berdyugina1999impeg,Strassmeier-09}. 
Time series of DI maps and photometric light-curves reveal stellar differential rotation and latitudinal activity belts \revised{\citep[e.g.,][]{Berdyugina-Henry2007hr1099, Donati-CollierCameron1997abdor, Vida2007},} longitudinal migration of active regions with respect to the stellar rotation and various activity cycles \citep[e.g.,][]{Berdyugina1998al,Jarvinen2005ekdra,Jarvinen2005abdor,Olah2009}\revised{, and meridional motions and poleward migration of active regions \citep{Kovari2007}}.
Such long-term starspot phenomena are important for monitoring transits of Earth-like planets around solar-type stars with orbital periods more than a hundred days or so. 
High spectral resolution is key for resolving spots and their substructure, e.g., a possible penumbra signature on a young solar analog EK Dra \citep{Jarvinen2018ekdra}. 

DI also shows that, for active stars, starspots can occur at much higher latitudes than within the $\pm 30 \degr$ from the equator that is typically seen on the Sun. 
Large spots have been found to cover even the visible pole of the star \citep[e.g.\revised{,}][]{StrassmeierRice1998,Donati1999,Korhonen2021}. 
Recently, large starspots on giant stars were directly imaged using long-baseline interferometric imaging \citep{Roettenbacher2016,Roettenbacher2017}.
This method confirmed the existence of polar starspots on active stars \citep{Roettenbacher2016}. 
It also allows obtaining accurate information on spot locations with respect to the stellar equator that is not possible with DI \citep{Roettenbacher2017}.
A polar spot that is perfectly symmetrical around the pole does not cause rotational modulation in the light curve.
On the other hand, it would still cause effects on the observed spectrum of the star and also lead to the wrong estimation of the transit depth \citep[e.g.,][]{Rackham2018, Zhang2018}. 
However, typically stars showing polar spots would also show large spots on other latitudes, causing significant variation in their light curves. 
Therefore, as with other active stars, the presence of polar spots must be taken into account when assessing targets for transit spectroscopy and analyzing observations from these systems. 
Here molecular spectroscopy can also help detecting such spots, even if they do not modulate photometric and spectroscopic signals.

Similar to DI, a Zeeman\revised{--}Doppler \revised{i}maging (ZDI) technique based on spectropolametric measurements can help to recover the distribution of magnetic fields on stellar surfaces \revised{\citep{Semel1989zdi,Donati-97,Donati-Landstreet2009, Shulyak2019, Fuhrmeister2022}}. 
However, surface features detected photometrically and in DI maps seem not directly related to features in ZDI maps. 
Overcoming this discrepancy and improving underlying physical assumptions \citep[e.g.,][]{Berdyugina2009maghrd} may allow in the future relate ZDI magnetic features with transit spectrophotometry.

In terms of activity level and its relation with age, a clear relation has been found between the two for G, K and M-stars. 
Stars are born as rapid rotators, and their rotation slows down with age due to magnetic braking \citep{Skumanich1972}. 
On the other hand, the dynamo action is enhanced with more rapid rotation until a saturation level \citep{Pallavicini1981}. 
Somewhat surprisingly, both fully convective and partially convective stars follow similar activity\revised{--}rotation relations \citep{2007AcA....57..149K, 2017ApJ...834...85N, 2018MNRAS.479.2351W}. 
There is evidence for the age\revised{--}rotation relations breaking down at the halfway point of the main sequence lifetime of the star, i.e., close to the current age of the Sun, suggested to be due weakened magnetic braking \citep[e.g.][]{vanSaders2016,NicholsFleming2020,Hall2021}. 
\revised{Additionally, gyrochronological age estimates are significantly lower than isochronal age estimates for planet-hosting stars, which may be due to tidal or magnetic interactions between the star and planet \citep{Brown2014, Maxted2015}.}

The study of stars other than the Sun has also been fundamental to understand the time-scales on which active regions live on stellar surfaces. 
From the Sun we know that the lifetime of small sunspots is proportional to their size \citep{Gnevyshev1938,Waldmeier1955}. 
Similar behavior for starspots has been indicated from relatively sparse photometric observations \citep[e.g.\revised{,}][]{HallHenry1994}. 
Precise, high-cadence space photometry from \Kepler{} and \TESS{} has recently been used to further investigate starspot lifetimes as well \citep{Giles:2017, Namekate2020}, which can be compared to theoretical and empirical models \citep{BradshawHartigan2014, Namekate2019}. 
This is also an active area of research in the radial-velocity community, which is also interested in constraining these time-scales 
\citep[see, e.g.,][]{Haywood-14, Rajpaul-15, Faria-16}. 
It seems that for other stars, starspot lifetimes fall slightly below the solar trend, but are within an order of magnitude of those expected from sunspots studies. 
Very large starspots on active stars can live a very long time, maybe even years \citep{Hussain2002}. 
On the other hand, it is not clear if these huge spots are really single spots or an active-region-like collection of individual spots. 
On rapidly rotating M dwarfs, starspot modulation, observed in photometry and spectral features, can remain coherent for months to years, suggesting long-lived features \citep{2016ApJ...821...93N, 2020ApJ...897..125R}.
\revised{The rapidly rotating ($P_\mathrm{rot} = 0.44$\,d), fully convective, single M4 dwarf V374\,Peg, for example, exhibits a large-scale magnetic topology that is consistent for years \citep{Morin2008} and even decades \citep{Vida2016}.}
\revised{Rather than originating from shearing in the tachocline region as is thought to be the case for stars with convective outer envelopes \citep{Parker1955}, it has been proposed for fully convective stars that a distributed dynamo generates the surface magnetic field, which interacts with turbulent convection in the outer layers to produce the observed small-scale fields \citep{Yadav2015dynamo}.}

\subsubsection{Atmospheric heterogeneity in ultra-cool dwarfs}

Stellar heterogeneity imposes special challenges for transiting planets orbiting ultracool dwarfs---stars and brown dwarfs with $T_\mathrm{eff} < 3000$\,K.
With the planet-to-star radius ratio being a key parameter underpinning the feasibility of observational studies of transiting exoplanets, ultracool dwarfs---including the smallest dwarf stars and most massive brown dwarfs---are set to play key roles in the future of exoplanet science. 
These stars, however, are not just smaller versions of the Sun but, instead, are a fundamentally different class of objects. 

A spectral-type- and thus temperature-dependent pattern of rotational variability has been observed for the mainly temperature-driven sequence from late M-dwarfs through early L-dwarfs to T-type brown dwarfs.
Time-resolved, multi-color or spectrally-resolved observations of such rotational variability provide powerful probes of the nature of the atmospheric heterogeneity in these objects. 
For the coolest stars and hottest brown dwarfs (at the M/L spectral type transition), it is very likely that rotational variability is caused by multi-pole magnetic fields, resulting in large starspots. 
Such spots, possibly anchored by large-scale magnetic fields, have been found (at least for some early L-type dwarfs) to lead to periodic variability that is stable over multiple years \citep[e.g.,][]{Gizis2015}. 
Indeed, a strong surface magnetic field of 5\revised{\,kG} was successfully detected on one of such young brown dwarfs displaying transient rotationally-modulated radio bursts \citep{Berdyugina2017bdw,Kuzmychov2017}. 
Also, the rotational modulation of the Balmer H\,\textsc{i}, blue Ca\,\textsc{ii} and Na\,\textsc{i}~D line emission as well as radio bursts were found to be related to the detected magnetic region and magnetic loops anchored in it. 
Since rotation periods of such objects (and thus the emission modulation) are 2--3\,hr, this variability may affect transit spectra of exoplanets.

In contrast, later spectral type brown dwarfs (from mid-L to mid-T types) often display rotational variability \citep[e.g.,][]{Buenzli2014,Radigan2014,Metchev2015} that are different in nature: 
their light-curve shapes are constantly evolving \citep[e.g.,][]{Artigau2009,Biller2013,Apai2017}, likely caused by cloud thickness variations \citep[e.g.,][]{Radigan2012,Apai2013} (although chemical disequilibrium may also play a role, \citealt[][]{Tremblin2020}).
Due to the low atmospheric temperatures and ionization rates, the cloud thickness modulations are very likely independent of the magnetic field strengths \citep[][]{Miles-Paez2017}. 
Instead, cloud thickness variations are driven by atmospheric circulation and lead to atmospheric jets and zonal circulation \cite[][]{Zhang2014,Apai2017,Millar-Blanchaer2020,TanShowman2021,Apai2021}.

\subsubsection{Limb-darkening \& its interplay with stellar heterogeneities}

As has been discussed in this \revised{section}, understanding center-to-limb variations on stars (limb-darkening or brightening profiles) is fundamental to constraining stellar heterogeneities. 
They are not only critical for constraining the amplitude and contrast of granulation in stars and the observed contrast variations of faculae and spots as a function of their position on stellar surfaces, but they also impact transit light-curve modeling itself and hence the transit spectrum that is derived from this modeling. 
Through simulations, it has been shown that inaccurate modeling of limb darkening may lead to a 1--10\,\% discrepancy in the determination of exoplanetary radii obtained from transit photometry observed at UV or visible wavelengths \citep[see, e.g.,][]{Espinoza:2015, Csizmadia-13}. 
Studies trying to constrain how well theoretical limb-darkening profiles extracted from stellar model atmospheres compare against empirically determined limb-darkening profiles from precise photometry have typically found disagreement between the two \citep{Cabrera2010A, Claret2009, Espinoza:2015, Maxted2018}. 

Limb darkening has an important interplay with stellar heterogeneities too. 
\cite{Csizmadia-13} showed that if stellar heterogeneities are present in the star, this can effectively modify the limb-darkening profile. 
Using theoretical limb-darkening laws to model this effect, which do not account for the presence of those features, will create a bias on the model that would translate to a bias in the retrieved transit depth. 
This could itself be an important source of contamination on the transit modeling. 

Taking into account polarization due to scattering in stellar atmospheres was also found important for theoretical limb-darkening models.
This is especially true closer to the limb and for cooler atmospheres, where scattering becomes significant as compared to absorption \citep{Kostogryz-Berdyugina2015,Kostogryz2016,Kostogryz2017}. 
Hence near-limb transits of cooler stars will be most sensitive to (and, thus, biased if not accounted for) polarization effects.  

\subsubsection{Our finding}
From this analysis, we draw the following finding.
\\
\\
\noindent \underline{\textbf{Finding 1.3}}
\\
\\
\noindent \textit{Summary}:
Simultaneous multi-wavelength (multi-instrument) stellar observations are needed to provide feedback to modeling efforts and improve our understanding of the photospheres and chromospheres of other stars, including high-priority exoplanet host stars.
This is particularly critical for K, M, and L dwarfs, for which models are relatively poorly constrained.
\\
\\
\noindent \textit{Capability Needed}:
Time-resolved multiband photometry, spectroscopy, and spectropolarimetry of high-priority exoplanet host stars are \revised{needed} to provide feedback for simulations of these stars.
In particular, we need to understand the variability of exoplanet hosts within the wavelength range of interest for current and upcoming missions (0.3--5\,$\mu$m) at timescales relevant to granulation (minutes), transits (minutes to hours), rotation periods (days to months), and magnetic cycles (years).
\revised{More work is needed to understand the precisions at which variabilities should be constrained to be useful in this respect and the dependence of the required precision on the stellar parameters and the transit science case.}
Limb darkening as a function of wavelength and stellar activity should be constrained with future observations in order to better understand its impact as a contamination source on the transit spectrum.
\\
\\
\noindent \textit{Capability Today}:
Precise broad-band photometric light curves from CoRoT, \Kepler{}/\Ktwo{}, and \TESS{} probe surface convective motions \revised{\citep[or ``flicker'';][]{Michel2008, Bastien:2013, Bastien:2016}} and spot distribution and evolution \revised{\citep{Southworth2011, Reinhold-13, Morris-17}}, providing feedback for 3D models of stellar surface features.
Multi-decade monitoring of stellar chromospheric emission \revised{\citep{reinhold2019transition}} as well as recent population studies relying on photometric light curves \revised{\citep{Montet2017}} have been used to infer that more-active stars are dominated by dark spots while less-active stars are dominated by bright faculae.
Ground-based spectroscopy and spectropolarimetry provide constraints on physical properties of stellar magnetic features \revised{\citep[e.g.,][]{TonerGray1988, GrayJohanson1991, Neff1995, Catalano2002, Berdyugina2011, Andretta:2017, Afram-Berdyugina2019}}. 
(Zeeman) Doppler imaging can probe structures in photospheres of rapidly rotating stars \revised{\citep[e.g.,][]{Semel1989zdi, Donati-97, Berdyugina1998al, Berdyugina-Henry2007hr1099, Jeffers2007abdor, Shulyak2019, Fuhrmeister2022}}, but it is unclear how well these findings may extend to the general population of exoplanet host stars, which tend to rotate more slowly \revised{\citep[e.g.,][]{Gaidos2023}}.
\\
\\
\noindent \textit{Mitigation in Progress}:
\revised{Recent work has explored the limits of light-curve inversion \citep{Luger2021} and presented an interpretable Gaussian process model that can be used to infer physical parameters from light curves \citep{Luger2021b}.}
\revised{Meanwhile, other work has shown that combining} high-resolution spectroscopy and spectropolarimetry, precise space photometry, and interferometric imaging helps to construct stellar surface maps of cool stars \revised{\citep{Roettenbacher2016, Roettenbacher2017}}, more recently also including \revised{a main-sequence} exoplanet host \revised{\citep{Roettenbacher2022}}.

\section{Occulted Active Regions}
\label{S:Occulted Active Regions}

\subsubsection*{Essential \revised{q}uestions:}

\begin{enumerate}
    \item What is the state of the art for modeling and observing occulted active regions?
    
    \item What are the outstanding theoretical challenges for this work?
    
    \item What are the available modeling tools for active-region occultations? Are they sufficient for the precise transit datasets that are expected in the next decade?
    
\end{enumerate}

\subsection{Introduction}
As described in \autoref{S:Introduction}, stellar surfaces are very inhomogeneous. 
In particular, active regions, including dark (spots and pores) and bright (faculae, plage, and network) magnetic features, are of concern to transmission spectroscopy due to their effects on transmission spectra. 
If the transit chord of a planet crosses a spot (facula), this will appear as a temporal brightening (darkening) in the resulting light curve. 
The prime effect of unocculted active regions, which are discussed in \autoref{S:UnoccultedActiveRegions}, is a change in apparent transit depth.
Here we focus on the effects of occulted active regions on the observed transit light curves.
We consider three topics in this context: 
the state of the art for observations and models of occulted active regions (\autoref{sec:occulted-state-of-the-art}),
outstanding theoretical challenges in this context (\autoref{sec:occulted-challenges}), and
the availability of well-maintained tools for conducting these analyses (\autoref{sec:occulted-models}).
In the following sections, we summarize the analysis that leads to our primary findings in these areas.

\subsection{State of the art for observations and models}
\label{sec:occulted-state-of-the-art}

\subsubsection{Observations of spot occultations}

\begin{figure}
    \centering
    \includegraphics[width=\columnwidth]{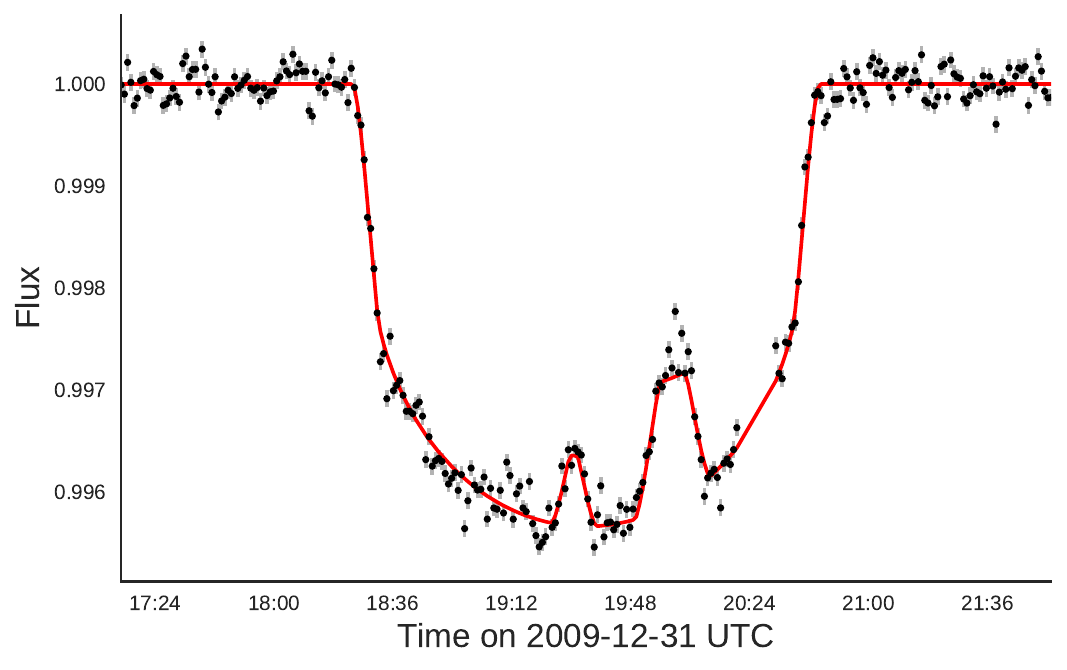}
    \includegraphics[width=\columnwidth]{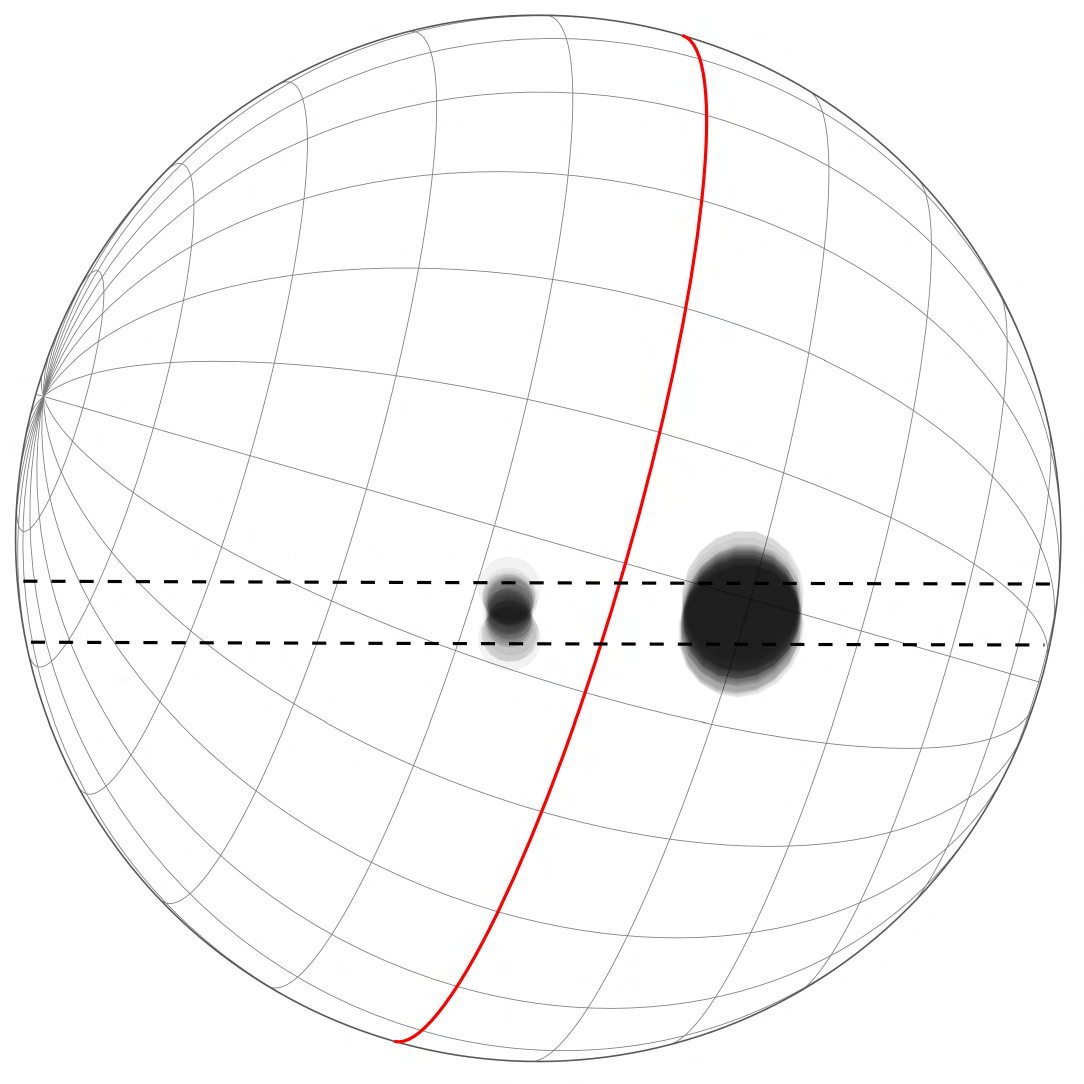}
    \caption{
    \label{fig:STSP}
    {\it \revised{Top}}: Transit light curve of HAT-P-11 b observed by the \Kepler{} mission (black) fit with the {\tt STSP} forward model (red), which models occultations by two starspots \citep{Morris-17}. 
    {\it \revised{Bottom}}: Inferred starspot map produced by the MCMC fit of the {\tt STSP} model to the \Kepler{} observations, shown with the stellar rotational pole to the right and into the page. 
    The stellar equator is marked in red and the transit chord is given by the dotted black horizontal lines. 
    The two clusters of overlapping, low-opacity spots represent draws from the posterior distributions for the spot positions and radii.
}
\end{figure}

Due to spots' high contrast with respect to the quiet stellar surface and their more distinct structure as compared to faculae, spot occultations are significantly easier to identify, manifesting as temporal bumps in the transit light curves (see \revised{top} panel of \autoref{fig:STSP} for an example). 
Having lower contrast, faculae are much patchier in their surface structure and thus are more difficult to separate from noise and stellar variability.
The following paragraphs summarize some notable studies of spot crossings from space- and ground-based observations.

\paragraph{Space-based.}
The first known transiting exoplanet, HD\,209458b \citep{Charbonneau-00, Henry-00}, also provided the first observation of a exoplanet occulting a starspot, both from space with \HST{} \citep{Brown-01} and from the ground \citep{Deeg-01}.
The occulted spots on HD\,209458 were first modeled by \citet{Silva-03}, yielding their sizes and temperatures.
The next detection was with the CoRoT telescope of CoRoT-2 \citep{Silva-10}. 
The number of observed starspot occultation events has increased since \Kepler{} and the extended \Ktwo{} mission (e.g., Kepler-30; \citealt{Sanchis-Ojeda-12, Netto-20}: Kepler-63;  \citealt{Sanchis-Ojeda-13}: Kepler-539; \citealt{Mancini-16}; Kepler-17: \citealt{Desert-11,Valio-17}; HAT-P-11: \citealt{Sanchis-Ojeda-11b, Oshagh-13a,Morris-17, Murgas2019}; Kepler-71: \citealt{Zaleski-19}; Kepler-45: \citealt{Zaleski-20}; Kepler-411: \citealt{Araujo-21}), although with low-cadence temporal sampling (30-min cadence), the \Kepler{} mission is neither ideal nor well suited to such detections. 
The cadence has a strong impact on the detection limits of starspot anomalies for planetary systems with small planet-to-star radius ratios \citep{Tregloan-Reed-19}. 
This effect is mostly evident in the detection limits of the smallest and/or hottest (low contrast, $\rho>0.9$, where $\rho$ is the intensity of the spot relative to the photosphere) dark starspots, when the timescale of the occultation event is equivalent to or shorter than the observational cadence. 
Starspot occultations are more easily detected with the 1-min-cadence \Ktwo{} observations. 
Such detection examples include the aligned WASP-85 \citep{Mocnik-2016} and Qatar-2 \citep{Mocnik-2017a} systems with recurring occultation events, and the misaligned WASP-107 system, where the occultations were not found to be recurring \citep{Dai-2017,Mocnik-2017b}.

\paragraph{Ground-based.}

\begin{table*}
\centering
\caption{
\label{tab:spots} Selected planetary systems found to show a starspot occultation event in their light curves from ground-based photometry.
}
\setlength{\tabcolsep}{8pt} 
\begin{tabular}{lll} 
\hline\hline
 Planetary system & Telescope/Instrument & Reference \\    
\hline
HD\,209458 & Observatorio de Sierra Nevada 90\,cm/Str\"omgren Photometer & \citet{Deeg-01} \\
WASP-4     & 1.54-m Danish Telescope/DFOSC & \citet{Southworth-09} \\
           & Magellan Baade 6.5\,m/MagIC & \citet{Sanchis-Ojeda-11a} \\
HAT-P-11   & Astrophysical Research Consortium 3.5\,m/ARCTIC & \citet{Morris-2018}\\
WASP-19    & 1.54-m Danish Telescope/DFOSC & \citet{Mancini-13} \\
           & MPG/ESO 2.2\,m/GROND & \citet{Mancini-13} \\
           & 3.6-m New Technology Telescope/EFOSC2 & \citet{Tregloan-Reed-13} \\
           & 8.2-m VLT/FORS2 & \citet{Sedaghati-15}  \\
           & Magellan Baade 6.5\,m/IMACS & \citet{Espinoza2019} \\
HATS-2     & MPG/ESO 2.2\,m/GROND & \citet{Mohler-Fischer13} \\
WASP-21    & Cassini 1.52\,m/BFOSC & \citet{Ciceri-13} \\
           & Calar Alto 1.23-m/DLR-MKIII & \citet{Ciceri-13} \\
Qatar-2    & MPG/ESO 2.2\,m/GROND & \citet{Mancini-14b} \\
           & CAHA 2.2\,m/BUSCA & \citet{Mancini-14b} \\
           & Cassini 1.52\,m/BFOSC & \citet{Mancini-14b} \\
           & Calar Alto 1.23-m/DLR-MKIII & \citet{Mancini-14b} \\
WASP-89    & TRAPPIST 0.6\,m/TRAPPIST photometer & \citet{Hellier-15} \\
           & Euler 1.2\,m/EulerCAM & \citet{Hellier-15} \\
WASP-6     & 1.54-m Danish Telescope/DFOSC & \citet{Tregloan-Reed-15} \\
HAT-P-36   & Cassini 1.52\,m/BFOSC & \citet{Mancini-15} \\
           & Calar Alto 1.23-m/DLR-MKIII & \citet{Mancini-15} \\
WASP-41    & 1.54-m Danish Telescope/DFOSC & \citet{Southworth-16} \\
           & TRAPPIST 0.6\,m/TRAPPIST photometer & \citet{Neveu-VanMalle-16} \\ 
           & 1.54-m Danish Telescope/DFOSC & \citet{Neveu-VanMalle-16} \\ 
WASP-52    & William Herschel Telescope 4.2\,m/ULTRACAM & \citet{Kirk-16} \\       
           & MPG/ESO 2.2\,m/GROND & \citet{Mancini-17} \\
           & 1.54-m Danish Telescope/DFOSC & \citet{Mancini-17} \\
           & Cassini 1.52\,m/BFOSC & \citet{Mancini-17} \\
           & Calar Alto 1.23-m/DLR-MKIII & \citet{Mancini-17} \\
           & Euler 1.2\,m/EulerCAM & \citet{Mancini-17} \\
HAT-P-20   & YO 1\,m/Andor $2\mathrm{K} \times 2\mathrm{K}$ CCD Camera & \citet{Sun-17} \\
GJ\,3470   & 10.4-m Gran Telescopio Canarias/OSIRIS & \citet{Chen-17} \\
GJ\,1214   & 8.4-m Large Binocular Telescope/Large Binocular Camera & \citet{Nascimbeni2015} \\
\hline 
\end{tabular} 
\end{table*}

There is a substantially greater number of observed starspot occultation events in transit light curves from ground-based telescopes compared to those from space-based telescopes. 
This is principally due to the many observations using high-cadence ($<2$\,min) observations, which allows for smaller starspots to be detected.
A sample of systems with starspot detections from ground-based photometry is given in \autoref{tab:spots}.

\subsubsection{Influences on the retrieved transmission spectra} 

\begin{figure}
\centering 
\includegraphics[width=\columnwidth]{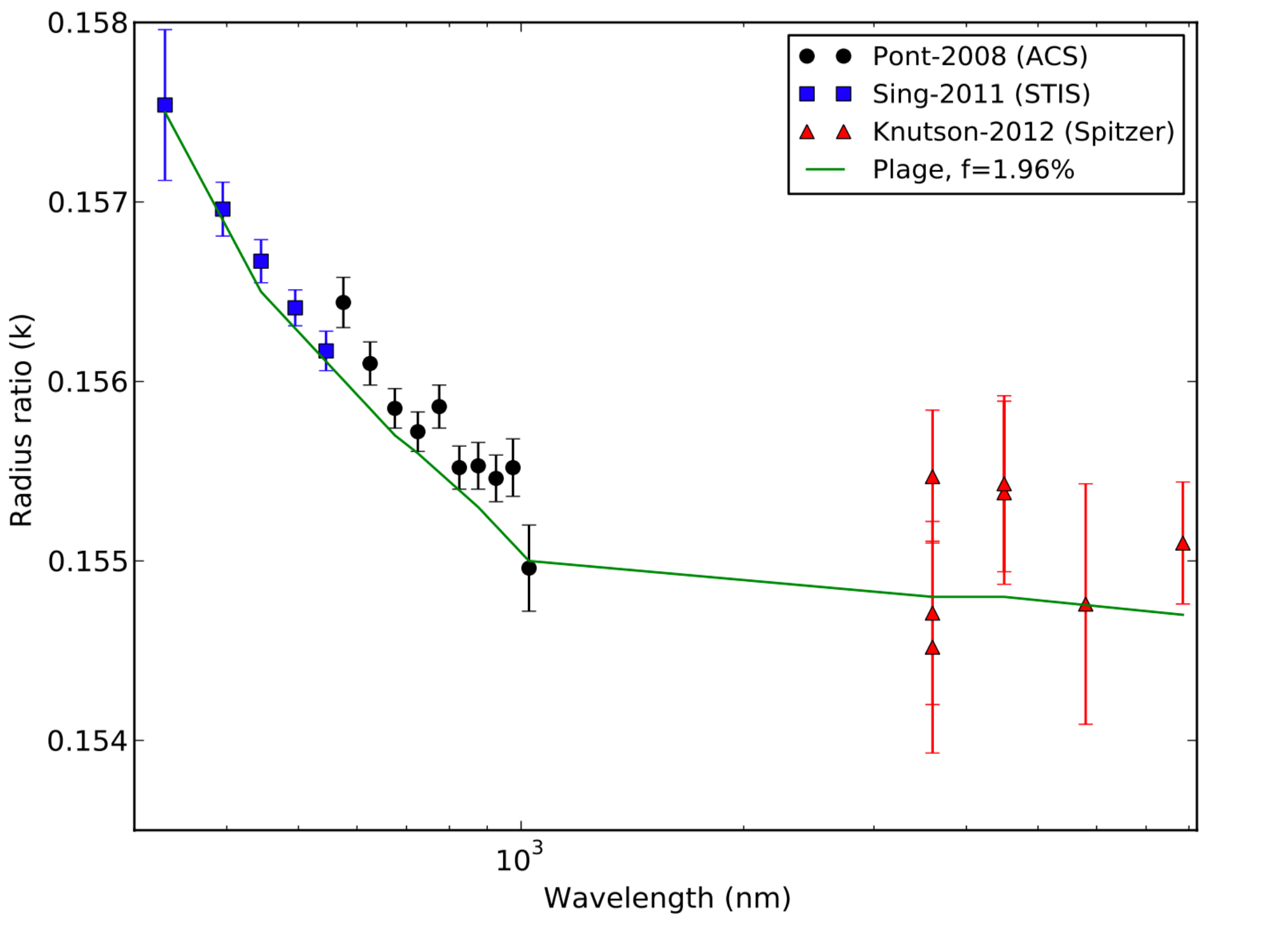} 
\includegraphics[width=\columnwidth]{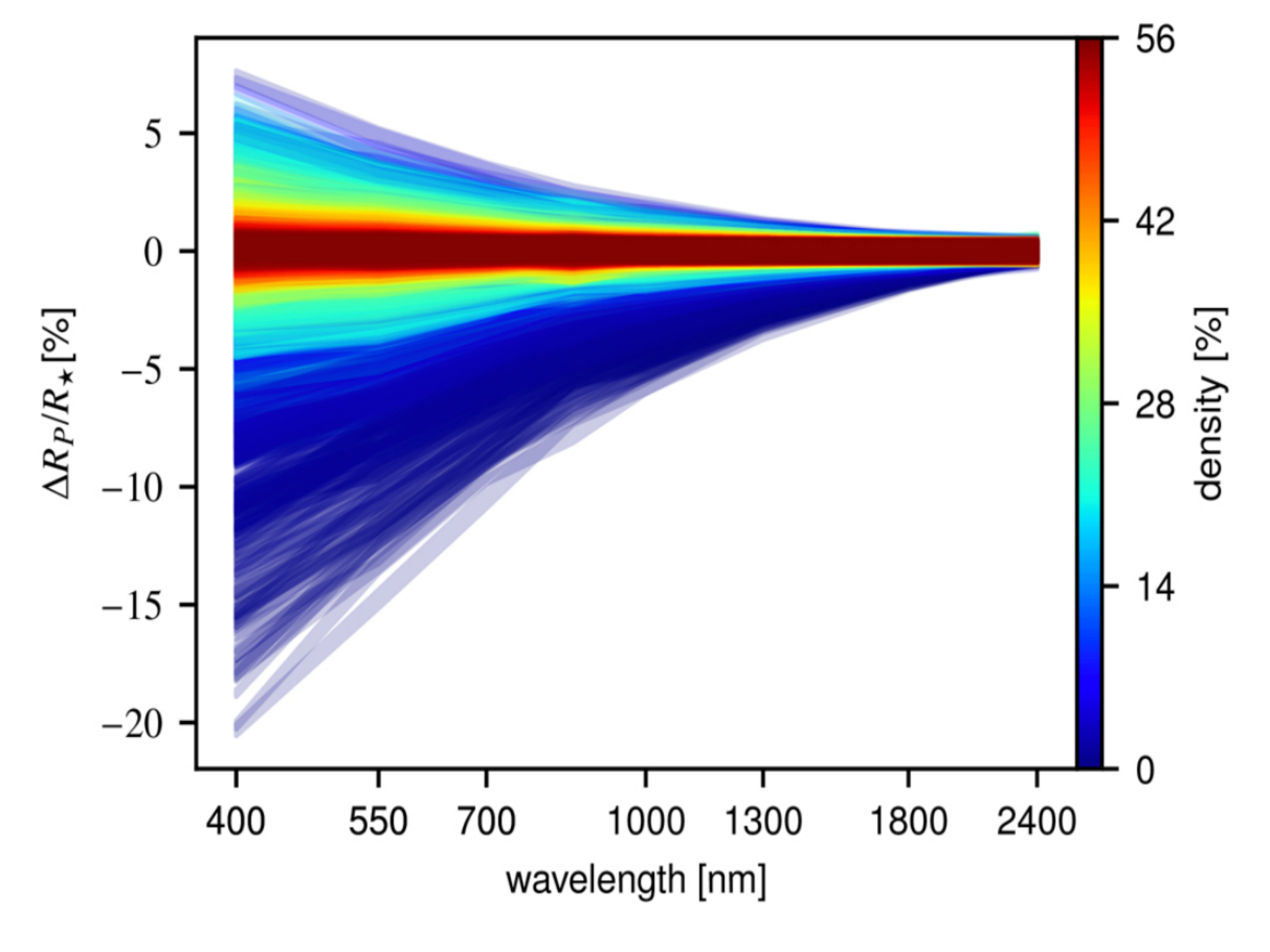} 
\caption{
    \label{Fig.3} 
    \textit{\revised{Top}:} Impact of stellar facula occultation on the retrieved transmission spectra of HD\,189733b, which is similar to what one would expect from a hazy atmosphere in a planet that causes a Rayleigh scattering slope \citep{Oshagh2014}. 
    \textit{\revised{Bottom}:} \revised{Simulated transit depth changes for WASP-19b in chromatic Rossiter-McLaughlin observations due to occultations of randomly distributed active regions \citep{Boldt-20}.
    Results from 10\,000 simulations are shown, highlighting a wide range of broadband features produced.
    The colour scale indicates the percentage of other simulated results that fall within $\Delta R_p/R_s \le 0.0005$ of the current result across the studied wavelength range, highlighting that most random distributions of active regions do not produce strong changes that mimic broadband features. 
    }
}
\end{figure}

When occulted during planetary transits, starspots and faculae decrease and increase the apparent transit depth, respectively, \citep[e.g.][]{Czesla-09}, especially in the visible (\autoref{Fig.3}). 
As their effect is wavelength-dependent, uncorrected active region crossings produce effects opposing those of unocculted active regions.
In broadband data, occultations of stellar active regions (spots/faculae) during planetary transits can mimic broadband characteristics of planetary atmospheres in transmission spectra, as has been shown with both observations and simulations \citep{Sing-10, Pont-11, Oshagh2014, Herrero-16}. 
In the context of narrow-band features, several studies have explored the possibility that realistic stellar active regions (both occulted and unocculted) on various types of stellar hosts could also have a significant impact on the strength of atomic and molecular features \citep{Rackham2017, Rackham2018, Rackham2019, Tinetti-18, Apai-18, Chachan2019}.

If the occultation is isolated and its signal large compared to the noise level, it is usually removed from the transit. 
However, \cite{Barros-13} showed that the occulted feature can impact multiple transit parameters other than the transit depth, such as mid-transit time, scaled orbital semi-major axis, and orbital inclination. 
In particular, impact parameter variations due to the effect of active regions on limb-darkening parameters can introduce positive or negative slopes from the visible to the infrared transmission spectrum \citep{alexoudi2020}. 
Such effects can reach up to one or more scale heights.

Hence, simply removing active region occultations from the transit profile likely does not completely negate their effect on the determination of transit parameters, particularly with high-precision spectrophotometry. 
The problem becomes trickier still when multiple occultations are observed in the same transit, as has been found in many cases in the visible \citep[e.g.,][]{Czesla-09,Desert-11,Morris-17}. 
In these cases, removing the occultations might simply not leave enough data points for the fit of the transit profile, such that using a transit-starspot model becomes a necessity.
In addition, modeling active-region occultations, instead of removing them, allows us to leverage their wavelength-dependent impact on planetary transits to derive active-region parameters \citep{Sing2011, Bruno2022}---valuable information for correcting transmission spectra for the effects of stellar activity (see \autoref{S:Retrievals}). 
Finally, even in cases without obvious active-region occultations, occultation models can be used to place upper limits on the impact of low signal-to-noise occulted active regions, which introduce an uncertainty in the level of stellar-activity contamination similar to that of unocculted active regions but with an opposite sign \citep{Ballerini-12}.

\subsubsection{Our finding}

From this analysis, we draw the following finding.
\\
\\
\noindent \underline{\textbf{Finding 2.1}}
\\
\\
\noindent \textit{Summary}: 
Precise transit observations increasingly reveal occultations of stellar active regions.
Rather than flagging and removing active region occultations, which results in decreased observing efficiency and possibly biased transit depth measurements, future observations should move towards joint inference of the active region and planetary properties. 
\\
\\
\noindent \textit{Capability Needed}: 
Detections of occulted active regions are more common in precise datasets.
From a practical perspective, discarding data impacted by active region occultations in many interesting systems will result in continuously lower observing efficiencies as precisions improve.
Additionally, more study is needed to understand the extent to which transit depths may be biased by active regions that are occulted but undetected.
\\
\\
\noindent \textit{Capability Today}:
Many recent studies have included joint inferences of transit and active region properties using a variety of spot modeling codes \revised{(see, e.g., \autoref{tab:spots})}.
However, the active regions we can detect in precise transit light curves, both ground- and space-based, are much larger than those that have been observed on the Sun \revised{\citep{Mandal2021}}.
At the same time, data impacted by active region occultations are commonly discarded in other studies.
\\
\\
\noindent \textit{Mitigation in Progress}:
Joint analyses of active region and planetary properties from impacted light curves have been presented recently \revised{\citep[e.g.,][]{Mancini-17, Espinoza2019}}.
Additionally, some groups are beginning to look into the impact of heavily spotted transit chords on inferences \revised{\citep[e.g.,][]{Cauley2018, Morris2018b}}.

\subsection{Theoretical challenges for active region occultations}
\label{sec:occulted-challenges}

\subsubsection{Solar spots and comparison to models of stellar spots}

In the context of studying transiting planets occulting starspots, the most relevant property is the relative brightness of the starspot with respect to its surroundings. 
For spots on the Sun, there is a clear connection between the size of the spot and its umbral intensity: larger spots tend to have darker umbrae than smaller spots \citep{mathew2007,2014-Schad,Valio2020_sunspots}. 
In the study by \cite{mathew2007}, spots that have an umbral radius between 5\,$\arcsec$ (1\,$\arcsec \approx 720$\,km) and 15\,$\arcsec$, the mean umbral relative intensity drops from 0.5 to 0.3 relative to the quiet-Sun intensity at the continuum near the Ni\,$\textsc{i}$ line at 676.8\,nm. 
They also found a very weak dependence of penumbral intensity on spot size, between 0.84 and 0.81 at the same wavelength. 

Another property determining the effective brightness of a spot on a star is the ratio of penumbral to umbral area. 
For the Sun, this factor is roughly 4 and generally independent of spot area above a minimal spot area \citep[although there is some scatter in the literature; e.g.,][]{Solanki-03}. 

For starspots, there is a strong connection between the surface temperature of the star and the spot contrast \citep{Berdyugina2005}. 
The spot temperature contrast is larger for hotter stars, and this seems to be valid for stars of different surface gravities. 
Updated summaries of starspot contrasts as a function of stellar surface temperature can be found in \cite{Mancini-14} and \cite{herbst2021}. 
Interestingly, starspot temperatures inferred from planetary transits seem to be warmer than temperatures retrieved by other methods. 
This may be because the transit method returns the average temperature of the entire spot while methods like modeling of molecular bands are mainly sensitive to the umbral temperatures (due to the non-linear response of molecular lines to temperature). 

A particularly useful example for drawing comparisons to solar activity is that of the active G-dwarf Kepler-71.
Both spots and plages have been detected in its photosphere via occultations by its transiting hot Jupiter \citep{Zaleski-19}.
Despite the higher contrast of Kepler-71 faculae than the solar counterparts, these plages showed similar properties, such as increasing contrast towards the limb and larger sizes than sunspots.
Assuming a solar-type differential-rotation profile, the results obtained from independent modeling of starspots and faculae indicated similar rotational periods at the transit latitude and the same rotational shear.

\subsubsection{The importance of MHD simulations for modeling occulted active regions}

\citet{Johnson2021} have used quiet-Sun limb darkening and limb-dependent facular contrasts derived from MURaM 3D MHD simulations by \citet{Beeck3}, together with spot contrasts computed from 1D model atmospheres with effective temperatures in agreement with 3D MHD results by \citet{Panja-2020}, to model rotational light curves of active G2, K0, M0, and M2 stars in the  \Kepler{} band.
They found that the spot temperature together with spot coverage determines the amplitude of the variability, while faculae have a strong influence on mean brightness levels and the shape of light curves. 
The latter might have significant implications for occultation modeling. 

Using 3D radiative magnetohydrodynamic simulations, \citet{Panja-2020} have recently confirmed the dependence of spot contrast on stellar surface temperature for dwarfs within the spectral range G2V--M0V.
These authors simulated the spectral types G2V, K0V and M0V and obtained umbral bolometric intensities of 0.3, 0.5, and 0.7 relative to the stellar photospheric intensities. 
Ongoing work includes simulations of spots with expansive penumbral filaments, similar to those observed in sunspots for the simulated G2V starspot, and simulations that allow the calculation of spot contrasts as a function of wavelength and disk position (Panja et al., in prep).

\subsubsection{Common degeneracies in the retrieved spot and \revised{facula} properties}

The main difficulty in accounting for surface inhomogeneities is that these are essentially never single, isolated features. 
Thus, dark and bright features are likely present simultaneously, both occulted and unocculted.
Simultaneous ground-based spectroscopic and space-based photometric observations of G and K stars show that starspots and plage are often, although not always, spatially associated, as is true on the Sun \citep{Morris2018c}.
Moreover, it is not always clear whether a bright or dark feature is being occulted, as models with either type of feature at different times may offer comparable fits to transit light curves \citep{Kirk-16, Bruno-16}.
The resulting effect from all features depends on their coverage (size), location, and the wavelength of observations, and is generally impossible to separate.

When observing stellar rotation in a single photometric bandpass, both faculae and starspots can produce rotational modulation signatures.
Observing in more than one bandpass nearly simultaneously can break this degeneracy and suggest whether spots or faculae dominate. 
This technique has been applied to photometry of the planet-hosting M8 dwarf TRAPPIST-1, which shows (1) no clear spot occultations in transit photometry, \revised{though a few candidate events have been identified \citep{Burdanov2019, Ducrot2020},} (2) strong rotational modulation in the \Kepler{} bandpass \revised{\citep{Luger-17, Vida2017}}, and (3) very little modulation in \Spitzer{} photometry \revised{\citep{RoettenbacherKane2017, Morris2018e, Morris2018d}}.

\subsubsection{Our finding}

From this analysis, we draw the following finding.
\\
\\
\noindent \underline{\textbf{Finding 2.2}}
\\
\\
\noindent \textit{Summary}: 
Theoretical advances are needed to understand the limits of what we can infer about active-region properties from transit light curves and how best to incorporate prior knowledge from magnetohydrodynamic models into transit studies.
\\
\\
\noindent \textit{Capability Needed}: 
More study is needed to understand the limits of inferences from occultations and how to break degeneracies in retrieved parameters of occulted active regions, such as size and contrast, with additional observations, such as spectroscopic transit depths or multiwavelength monitoring.
Additionally, more study is needed to understand best approaches for propagating information from stellar magnetic activity models into priors on spot properties, such as typical size and contrast, as a function of stellar spectral type and activity level.
\\
\\
\noindent \textit{Capability Today}: 
\revised{Studies generally use uninformative} priors on active-region properties \revised{when studying occultations of active regions \citep[e.g.,][]{Morris-17}}.
Retrieved sizes are generally large and contrasts are generally small relative to those inferred from other methods \revised{\citep[e.g.,][]{Mancini-13, Espinoza2019}}, which may point to biases with the technique or the combined impact of umbral, penumbral, and facular areas.
\\
\\
\noindent \textit{Mitigation in Progress}: 
Multiple groups are developing magnetohydrodynamic models for active regions on stars with differing spectral types and activity levels \revised{\citep{Panja-2020, Johnson2021, Witzke2022}, which can be used to provide informative priors in future studies}.

\subsection{The availability of well-maintained, open-source tools} 
\label{sec:occulted-models}

Over the years, there have been multiple starspot occultation models developed both by the eclipsing binary star community (e.g., {\tt Wilson-Devinney} code: \citealt{Wilson-71,Wilson-79,Wilson-90,Wilson-12}; \revised{PHysics Of Eclipsing BinariEs ({\tt PHEOBE})}: \citealt{Prsa-05,Prsa-16}) and the exoplanet community (e.g., {\tt ECLIPSE}: \citealt{Silva-03}; {\tt SOAP-T}: \citealt{Oshagh-13a}; {\tt PRISM}: \citealt{Tregloan-Reed-13,Tregloan-Reed-15,Tregloan-Reed-18}; {\tt SPOTROD}: \citealt{Beky-14}; {\tt KSint}: \citealt{Montalto-14}; {\tt ellc}: \citealt{Maxted-16}; {\tt StarSim}: \citealt{Herrero-16}; {\tt PyTranSpot}: \citealt{Juven-18}); {\tt STSP} \citealt{Morris-17}; {\tt TOSC}: \citealt{Scandariato2017}).
Such models require integration over a 2D grid on the stellar surface or in the projection plane, which is computationally expensive, with computational time growing as the square of the grid resolution.
The only exceptions to our knowledge are the semi-analytical {\tt SPOTROD} and {\tt STSP} models.
At the same time, full 2D integration (pixelation) has the advantage that it allows individual intensities (accounting for surface inhomogeneities) to be assigned to every individual element at specific coordinates. 

We summarize some of the most prominent models below.

\paragraph{ECLIPSE.}

\begin{figure}
\centering 
\includegraphics[width=\columnwidth]{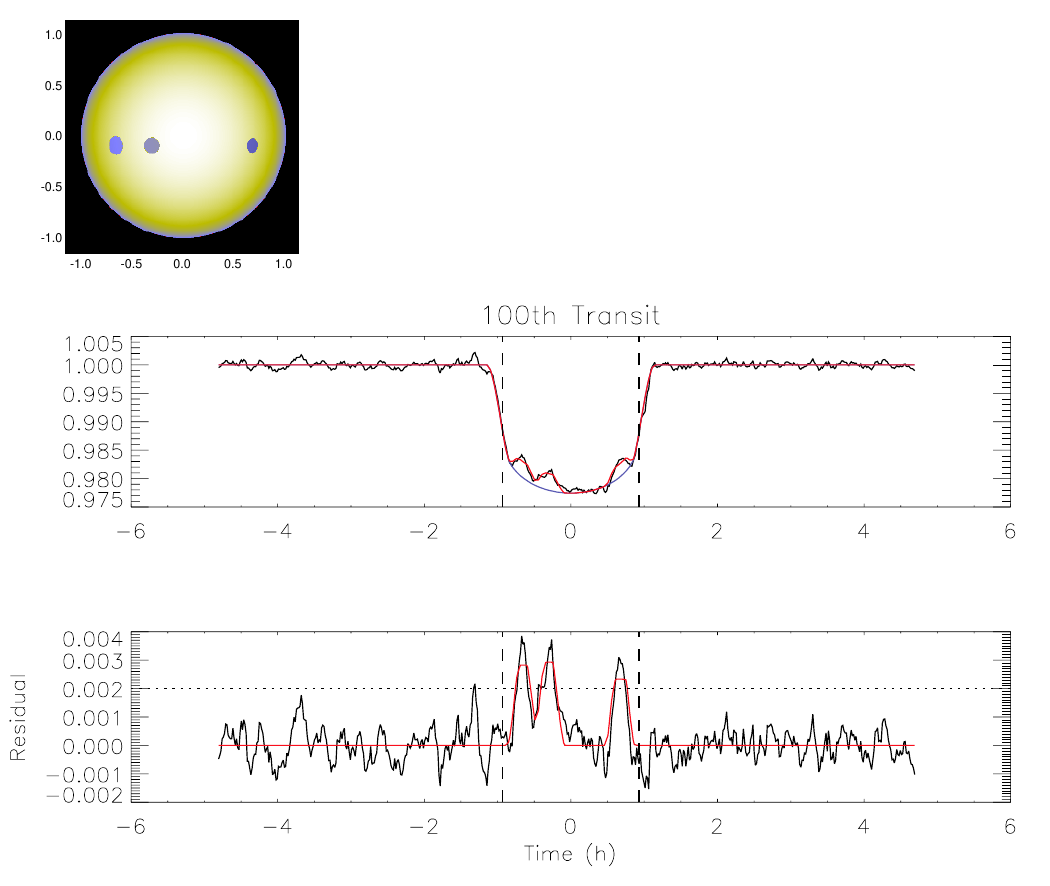}  
\caption{
    \label{fig:Kepler-17} 
    \textit{Top:} Simulated star with three spots. 
    \textit{Middle:} The 100th transit light curve with a spotless model (blue) and the three spot model (red). 
    \textit{Bottom:} Residuals after subtraction of the spotless model from the transit light curve with the modeled three spots (red curve). 
    Adapted from \citet{Valio-17}.
}
\end{figure}

\texttt{ECLIPSE} \citep{Silva-03}, a code first written in IDL and then translated to Python, models active-region occultations using a 2D image of a synthetic star with spots (dark features) or plages (bright features) on the surface of the limb-darkened stellar disk.
The planet, modeled as a circular dark disk, is positioned in its elliptical orbit every two minutes (or the desired time interval) and the intensity of all pixels in the image is summed to yield the light curve intensity at a given time. 
The spots (or plages) are simulated as dark (or bright) circles positioned along the transit chord. 
Spot intensities vary from 0--1, whereas the plage intensities vary from 1--1.5, where 1 is the brightness of the activity-free stellar disk center. 
The temperature of the spots can be estimated assuming blackbody emission for both the spots/plages and the stellar photosphere \citep{Valio-17}. 
Foreshortening of the spots is taken into account when they are close to the stellar limb.
This model was first applied to HD\,209458 \citep{Silva-03} and has since been applied to many targets with high-cadence \Kepler{} data \citep[e.g.,][]{Silva-10, Valio-17, Netto-20, Zaleski-20, Araujo-21}. 
An example of such modeling is shown in \autoref{fig:Kepler-17} for Kepler-17 with three spots.
\revised{\texttt{ECLIPSE}} can be found at GitHub\footnote{\url{https://github.com/biaduque/astronomy}} and was last updated in 2021.

\paragraph{SOAP-T.}

Spot Oscillation and Planet Transit (\texttt{SOAP-T}) \citep{Oshagh-13a} is a \texttt{SOAP} \citep{Boisse-12} adaptation that can generate light curves and radial velocity variations for systems containing a rotating spotted star and a transiting planet.
The stellar disk, active regions on it, and a transiting exoplanet are simulated numerically using a pixelation approach.
This tool has been used to investigate spot-crossing anomalies within transit light curves \citep{Oshagh-13b}, as well as to assess their impacts on the estimation of the transit duration, depth, and timing. 
\revised{\texttt{SOAP-T}} was used to study, for the first time, the impact of the occultation of a stellar spot and plage on the transmission spectra of transiting exoplanets \citep{Oshagh2014}.
\revised{\texttt{SOAP-T}} was also used to study the impact of stellar active region occultation on the Rossiter-McLaughlin effect signal \citep{Oshagh-16, Boldt-20}, as well as on the estimation of orbital configuration and properties of planetary systems \citep{Oshagh-18} and active regions on their host stars \citep{Oshagh-15, Serrano-20}.
\texttt{SOAP-T} is available via an online interface\footnote{\url{http://www.astro.up.pt/resources/soap-t/}}.

\paragraph{PRISM.}

\begin{figure}
\centering 
\includegraphics[width=\columnwidth]{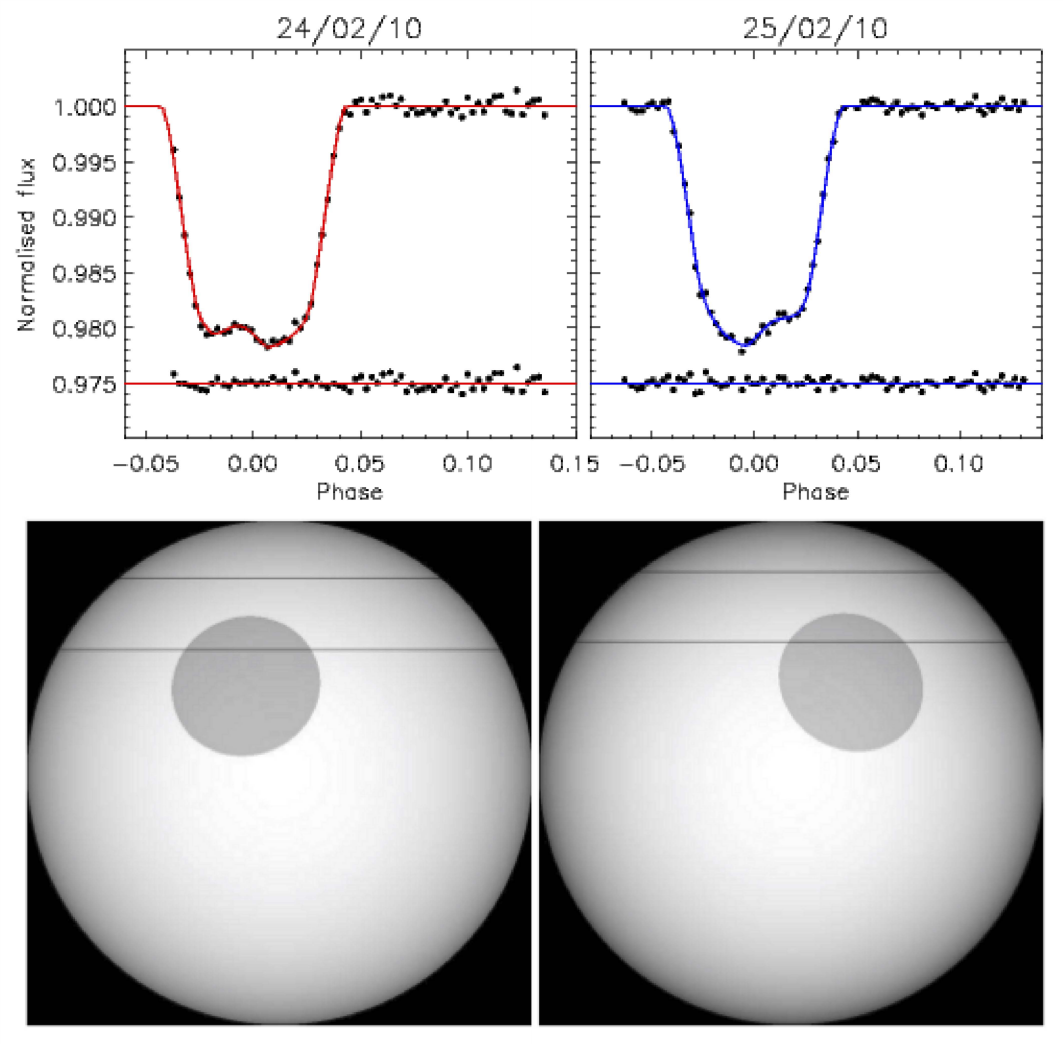}  
\caption{
    \label{fig:wasp19_tregloan-reed13} 
    \textit{Top:} Two consecutive transits of G8\,V WASP-19 by the 0.8-d, hot-Jupiter WASP-19b from ESO NTT, along with the best-fitting \texttt{PRISM} model. 
    The spot anomalies are clearly visible as upward blips around the midpoints of the two transits.
    \textit{Bottom:} Representation of the stellar disc, starspot and transit chord for the two datasets, showing the movement of the starspot between the consecutive nights. 
    The stellar rotation axis lies in the plane of the page and points upwards.
    From \citet{Tregloan-Reed-13}.
} 
\end{figure}

The Planetary Retrospective Integrated Starspot Model \revised{\citep[\texttt{PRISM},][]{Tregloan-Reed-13}} is written in IDL and uses a pixelation approach to model the stellar disc in a 2D array by subdividing the star into many individual elements. 
These elements can then be described by a 2D vector in Cartesian coordinates. 
Each element is then assigned an intensity value based on whether a stellar feature is present at that location and then the quadratic limb darkening law is applied over the entire stellar disc. 
The planet is then set to transit the star. 
For each data point in the transit light curve, the total received intensity is calculated based on which elements of the star are visible.
The {\tt PRISM} model has been used primarily to model starspots found in ground-based transit exoplanetary light curves \citep[e.g.][]{Tregloan-Reed-13, Tregloan-Reed-15, Chen-17, Mancini-17}. 
More recently, {\tt PRISM} has also been used to simulate spotted transits observed with 2-min-cadence \TESS{} data \citep[e.g.][]{Tregloan-Reed-19, Tregloan-Reed-21}.
An example of the use of {\tt PRISM} is provided by \autoref{fig:wasp19_tregloan-reed13}.
\texttt{PRISM} is available on GitHub\footnote{\url{https://github.com/JTregloanReed/PRISM_GEMC}} and was last updated in 2019.

\paragraph{SPOTROD.}

\texttt{SPOTROD} \citep{Beky-14} is a semi-analytic tool for modeling planetary transits of stars having an arbitrary limb-darkening law (being identical for spots and the stellar photosphere) along with a number of homogeneous, circular spots on the stellar surface. 
It can account for both eclipsed and uneclipsed starspots. 
The program is written in C and available in Python.
The homogeneous spots are represented by their flux ratio, $f$, which is the ratio of the flux from a spot (as viewed by the observer) to the flux from an unocculted stellar surface at the same disc location.
Umbra-penumbra structures of spots can be accounted for by superimposing two concentric spots of differing intensities, and bright features can be introduced by using $f>1$.
\texttt{SPOTROD} works in polar coordinates, with the advantage being that the integral over the polar angle can be computed analytically and only the integral over the radial coordinate requires numerical integration.
The significant reduction of the required computation time is the main advantage of this model compared to other, typically 2D, models.
Thus, \texttt{SPOTROD} can be used for fitting large numbers of transits with occulted active regions and also for efficient statistical investigations using MCMC approaches.
\texttt{SPOTROD} is available on GitHub\footnote{\url{https://github.com/bencebeky/spotrod}} and was last updated in 2015.
It has been wrapped into a nested-sampling retrieval framework, \texttt{SpotNest} \citep{Espinoza2019}, which is also available on GitHub\footnote{\url{https://github.com/nespinoza/spotnest}} and was last updated in 2018.

\paragraph{StarSim2.}

\texttt{StarSim2} \citep{Rosich-20}, an update of the \texttt{StarSim} tool \citep{Herrero-16}, uses a fine grid of surface elements to model integrated spectral contributions of a spotted rotating photosphere.
It is written in Python and Fortran 90.
The flux intensities and wavelengths of spectral features produced by active regions are included in the model. 
The original \texttt{StarSim} tool has been used to derive stellar surface parameters from light curve inversion as well as to mitigate their impacts on the transiting planet's transmission spectrum \citep{Mallonn-18}.
\texttt{StarSim2} is available on GitHub\footnote{\url{https://github.com/rosich/Starsim-2}} and was last updated in 2020.

\paragraph{STSP.}

\texttt{STSP} is a fast forward model for starspot occultations by transiting exoplanets and unocculted spot modulation, written in \texttt{C} by L.~Hebb and G.~Rohn \citep{Davenport2015, Morris-17, Schutte2022}.
It computes the overlap between the planet, circular starspots with geometric foreshortening, and concentric nested disks of varying intensity to approximate stellar limb darkening. 
The code is highly efficient and coupled to a Markov Chain Monte Carlo sampler also written in \texttt{C}, making it ideal for studies on high performance computing facilities.
\texttt{STSP} is available on GitHub\footnote{\url{https://github.com/lesliehebb/stsp}} and was last updated in 2021.

\paragraph{ellc.}

\texttt{ellc} \citep{Maxted-16} is a semi-analytical and semi-numerical model to analyse the light curves of detached eclipsing binary stars and transiting exoplanet systems.
It can include the effects of both occulted and unocculted starspots simultaneously. 
It is implemented as Fortran subroutines called directly from a user interface written in Python, and is fast enough to enable the use of modern Monte Carlo methods for data analysis and model testing.
The source code is available at the \revised{Strasbourg Astronomical Data Center} via anonymous \revised{file transfer protocol}\footnote{\url{http://cdsarc.u-strasbg.fr/viz-bin/qcat?J/A+A/591/A111}}.

\subsubsection{Our finding}

From this analysis, we draw the following finding.
\\
\\
\noindent \underline{\textbf{Finding 2.3}}
\\
\\
\noindent \textit{Summary}: 
\revised{Several} publicly available tools \revised{have been developed} for forward-modeling active-region occultations.
Their maintenance and further development is necessary to ensure their utility for analyses of precise transit observations. 
\\
\\
\noindent \textit{Capability Needed}: 
Publicly accessible, well-maintained, and up-to-date tools for modeling active-region occultations are necessary for analyses of precise transit observations from current and future facilities.
\revised{A comparative understanding of the performance of these tools is also needed.}
\\
\\
\noindent \textit{Capability Today}: 
\revised{Several} publicly available tools \revised{exist for modeling} for active-region occultations in broad photometric bandpasses today\revised{, including \texttt{ECLIPSE} \citep{Silva-03}, \texttt{SOAP-T} \citep{Oshagh-13a}, \texttt{PRISM} \citep{Tregloan-Reed-13}, \texttt{SPOTROD} \citep{Beky-14}, \texttt{StarSim2} \citep{Rosich-20}, \texttt{STSP} \citep{Davenport2015, Morris-17, Schutte2022}, and \texttt{ellc} \citep{Maxted-16}}.
However, many are not actively developed or maintained.
\\
\\
\noindent \textit{Mitigation in Progress}:
Multiple groups have developed \revised{tools for modeling active-region occultations and made their source code publicly available.}
\revised{Continued development of some tools has been spurred on by recent analyses \citep[e.g.,][]{Schutte2022}.}

\section{Unocculted Active Regions}
\label{S:UnoccultedActiveRegions}

\subsubsection*{Essential \revised{q}uestions:}

\begin{enumerate}
    \item Can we learn about unocculted active regions through the study of transit depth variations in large photometric planet search surveys?
    \item How well can other signatures of stellar activity in these datasets such as flares provide information about \revised{unocculted active regions}?
    \item What are the limitations of what we can learn about \revised{unocculted active regions} from transit photometry?
    \item Can additional photometric or spectroscopic data taken simultaneously with transit data help to better understand the information encoded in light curves?
    \item What data can best inform our knowledge of active region filling factors?
    \item How can we best distinguish between starspots and faculae?
    \item Can we quantify the noise floor due to granulation as a function of stellar parameters?
\end{enumerate}

\subsection{Introduction}

\subsubsection{Motivation}

Unocculted active regions on the heterogeneous surface of a star can impact measured transmission spectra of planets \citep[e.g.,][]{Pont2008, Czesla-09, Rackham2017, Rackham2018, Rackham2019, Zellem2017}. 
In transit photometry, the particular configuration of active regions produces a degenerate signal with an infinite number of different configurations, making unique identification of unocculted regions challenging. 
Moreover, while the most obvious signature of rotational modulation in broadband stellar time-series photometry is often the signal from starspots, facular regions are expected to induce a more significant impact on transmission spectra due to their complicated spectral signatures \citep[e.g.,][]{Cauley2018}. 
In this \revised{section}, we examine our understanding of unocculted regions and our ability to infer their presence before and during transmission spectroscopy of active stars.

\subsubsection{Scope}

We focus on three areas of interest \revised{here}, primarily centered around the measurement and understanding of unocculted active regions rather than their impacts on atmospheric retrievals, which is instead discussed in \revised{\autoref{S:Retrievals}}. 
First, in \autoref{sec:transits} we investigate what information can be obtained about unocculted active regions from high-cadence transit photometry. 
In \autoref{sec:otherdata} we explore the usefulness of other datasets, current or planned, in understanding these regions.
Finally, in \autoref{sec:uar_physics} we discuss the effects of stellar granulation on transmission spectra.

\subsection{Information from high-cadence transit photometry}
\label{sec:transits}

High-cadence and high-precision transit photometry, such as that enabled by the \Kepler{} and \TESS{} missions \citep{Borucki2010, Ricker2015}, also encodes information about the underlying stellar surfaces. 
Here we investigate opportunities from these datasets to directly infer properties of unocculted active regions.

\subsubsection{Transit depth variations}

Transit shapes can be affected by dynamical interactions between multiple planets or between a planet and its host star, which affects the relative geometry of the system along our line of sight. 
Transit shapes can also be affected by variability on the stellar surface with no change in the geometry \citep{Agol-05, Ragozzine-09, Barros-13}. 
In this section, we explore how to disambiguate between the two. 

Changes in the observed parameters of transiting planets have provided significant opportunities to understand planet-planet and star-planet dynamical interactions.
The most dramatic result of a perturbation to a planet's orbital parameters is in the measured transit timing, which has been used to determine orbital configurations and planet masses, as well as to identify tidal inspiral of hot Jupiter systems \citep{Maciejewski:2016,Patra:2020}.
In other systems, transit duration variations have provided information about orbital precession. 

Transit depth variations, in which the magnitude of the depth of the transit changes over time, are at their most fundamental a change in the relative brightness of the occulted stellar chord compared to the unocculted starlight in the atmosphere. 
This change can be the result of a few different physical variations. 
The transit chord geometry itself could be changing as a result of dynamical interactions \citep{Carter2010, Hamann2019}.
The unocculted stellar surface could also change in brightness due to a change in the rotational phase of active regions or their emergence or decay. 
The occulted stellar surface could change in brightness for the same reason, producing an equivalent effect in the opposite direction.

The last of these possibilities is discussed more fully in \autoref{S:Occulted Active Regions}. 
Here we note active regions on the occulted surface produce additional structure in observed transit photometry \citep{Sanchis-Ojeda-11b}, while unocculted active regions do not: 
sufficiently high-precision photometry can disambiguate between these cases.

Separating depth variations from dynamical interactions compared to those from unocculted active regions is also possible. 
One straightforward consideration is that a change in the transit chord should also produce a change in the impact parameter, and thus the transit duration, while brightness variations should not significantly affect the transit duration as observed in broadband photometry.
Transit duration variations from long-\revised{term} photometry can be detected at high confidence, with variations of $\approx$1 minute/year expected to be common \citep{Boley2020}. 
In most cases, duration variations should be more significant than depth variations \citep{Holczer2016}. 
Careful observations over a long \revised{temporal} baseline, whether from the same telescope or others, can provide the evidence needed to separate these cases.

If a system is observed to have transit depth variations and not timing or duration variations, and there is no evidence for these variations to be predominantly caused by the growth of active regions along the transit chord, then it is plausible that this is the result of changes in the unocculted stellar surface. 
However, for \Kepler{} and \TESS{} photometry, the primary hindrance in inferring these changes is in identifying which stellar surface---that of the target star or another nearby---is changing. 
The \Kepler{} telescope point spread function is approximately 6\,$\arcsec$ in size; 
for \TESS{}, it is typically 30--50\,$\arcsec$.\footnote{For both telescopes, the \revised{point spread function} is a strong function of location on the focal plane and, more weakly, a function of magnitude and color.} 
A typical photometric aperture used in the \Kepler{} mission has a diameter of 10--20\,pixels; 
for \TESS{} the typical diameter is more than 1\,$\arcmin$.  
As a result, each aperture is contaminated by the light of multiple nearby stars, and it is not uncommon for the Exoplanet Follow-up Observing Program (ExoFOP) database\footnote{\url{https://exofop.ipac.caltech.edu/tess/}} to report 50 or more stars within one arcminute of a \TESS{} Object of Interest (TOI). 
In addition, at least 25\% of TOIs have unresolved stellar binary companions \citep{Ziegler2020}, which can equally plausibly be the root of stellar brightness variations even when all known nearby stars can be ruled out.
It is therefore challenging from transit photometry alone to determine whether a depth variation is due to a change in brightness of the target star, or a change in brightness of a background star, which would have an equivalent effect on the light curve. 

This issue can be mitigated at some level by obtaining high-resolution imaging of these systems to limit the parameter space in which relatively bright nearby stars can evade seeing-limited detection \citep[e.g.,][]{Law2014, Baranec2016, Ziegler2017}. 
\revised{In order to remove neighboring stars from the observed aperture, occasional} monitoring of potential candidate transits with ground- or space-based telescopes with considerably higher pixel scales \revised{provides} a benefit over large photometric surveys themselves.
\revised{This sort of analysis (from the ground) is a common component of \TESS{} candidate validation efforts \citep[e.g.,][]{Wells2021, Schanche2022, Delrez2022, Gan2022_TOI530b, Gan2022_TOI2136b}.}
An even more promising route would be to use the existence of transit depth variations in concert with other proxies of stellar activity indicators. 
For example, \revised{measurements of} chromospheric activity indicators \revised{\citep[e.g.,][]{Baliunas-95} could shed light on} whether potential transit duration variations are due to the target star or \revised{some} chance brightening of \revised{another} star in the aperture.

\subsubsection{Active region distributions}
\label{s:active_distributions}

In the context of transmission spectroscopy observations, not all active regions may be equally important to consider.
\citet{Cauley2018} demonstrate that spectral line contrast absorption decreases dramatically as the separation between a planet and an active latitude increases, with the most significant events occurring when the two are within five degrees of each other.
Although not all planets have a low obliquity between their orbital axis and the stellar spin axis \citep{Winn2010}, for the systems where the obliquity can be measured, this consideration can impact our interpretation significantly if we can determine the distribution of active regions on the surface of the star.

Inferring the two-dimensional distribution of active regions on a stellar surface from one-dimensional transit photometry is an ill-posed and degenerate problem, with different stellar surface models able to produce identical light curves \citep{Vogt1987,Basri2020,Luger2021}. 
This is the case even if one is trying to recover information on dark (or bright) features alone. 
However, the problem is compounded as active regions typically contain bright (facular or plage-like) features and dark spots.
The contributions from these can add or partially cancel, depending on the location of the features, the wavelength at which the star is observed, and the feature size (or magnetic field strength) and the stellar spectral type. 
The contribution of faculae is particularly poorly characterised (see \autoref{sec:stellarmodels_smallscale} and \autoref{subsec:retrieval_limitations}). 

In general, longitudinal information is more easily recovered than information about the latitudes of unocculted active regions \citep{Luger2021}, although this is complicated further by the potential of active regions that manifest at all longitudes, such as a banded region or a polar spot. 
While there is some asymmetry even in banded features, their photometric signal is only representative of the asymmetry rather than the total filling factor, requiring other techniques to infer their structure. 

As discussed in \autoref{S:Occulted Active Regions}, significantly more information can be learned about the size and structure of active regions when they are occulted by a planet. 
In the context of transmission spectroscopy, the most promising avenue to use transit photometry to understand active regions outside of a narrow transit chord is to consider multi-planet systems with strong dynamical interactions. 
Over years to decades, multi-body perturbations can affect neighboring planets' inclinations by a few degrees; 
in some cases, planets can move from transiting the stellar equator to not transiting.

\begin{figure}
    \centering
    \includegraphics[width=\columnwidth, height=0.5\textheight, keepaspectratio]{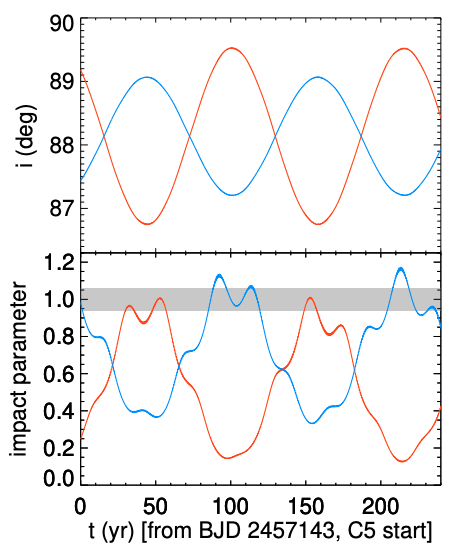}
    \caption{
    Long-term evolution of the (top) inclination and (bottom) observed impact parameter of the two planets in the K2-146 system. 
    Over timescales of a decade, the impact parameters from both planets can vary by as much as 10\%, causing previously occulted active regions to become unocculted and vice versa.
    From \citet{Hamann2019}.
    \label{fig:hamann}
    }
\end{figure}

The K2-146 system, described by \citet{Hamann2019} and \citet{Lam2020}, provides an example of these perturbations in action (\autoref{fig:hamann}).
Over 10 years, the impact parameter for both planets can change by 10\%, leading to a $\sim$10-degree change in the projected latitude of the transit chord on the surface of the star. 
Regular monitoring of occulted active regions over many years can then give a broader understanding of unocculted surfaces at the time of transmission spectroscopy observations, especially those surfaces within a few degrees of the transiting planet that are most critical for interpreting the transmission spectrum.

It is also helpful to know the inclination and rotation period of the star, and to observe a large number of rotations \citep{Basri2020}. 
This could, in particular, help to distinguish between the effects of the active regions on the light curves due to the stellar rotation (change in the projected area and contrasts) and the evolution of active regions (e.g., redistribution of the magnetic flux between the dark and bright components leading to a change in their relative contributions as well as flux decay).
Furthermore, a promising tool to distinguish between the signatures of planets, starspots and faculae in the light curves is offered by multiband photometry, due to the different center-to-limb contrasts of the different occulters (see \autoref{subsec:multiband} for further details).

\subsubsection{Stellar flares}

The presence of unocculted active regions can also be traced in principle by measuring the occurrence rate and energy of stellar flares. 
\Kepler{} and \TESS{} have been effective flare-finding missions \citep[e.g.,][]{Davenport2016, Guenther2020}.
Flares appear in these light curves as stochastic spikes, which can be separated from slower and usually weaker variations or sources of noise \citep{Feinstein2020}. 
Since flares are distinct events, it can be easier to unambiguously measure their specific occurrence rate for stars, compared with the degeneracies in estimating the true filling factor of, e.g., starspots, as discussed above.

Detailed statistical analyses of flares on Sun-like stars \citep[e.g.][]{Maehara2015,Notsu:2019,Okamoto2021} are limited in their ability to compare directly to the Sun itself: 
the smallest flares on Sun-like stars observed by \Kepler{} are still more energetic than the largest flares observed on the Sun or even than the upper limit on the possible solar flare energy \citep{Schrijver:2012,Aulanier:2013}. 
Yet these so-called ``superflares'' seem to follow the same relationship between occurrence and emitted energy as smaller flares, suggesting a common physical origin. 
There is clear and convincing evidence that the energies of solar flares are correlated with the magnetic energy in active regions on the Sun \citep{Mayfield1985}. 
For example, flare rates on the Sun are ${\sim}10 \times$ higher during activity cycle maximum, when the largest number of sunspots and active regions are present \citep{aschwanden2012}.
Three potential extreme solar particle events, identified by \citet{Brehm:2021} in the newest radiocarbon data with annual resolution, all occurred close to solar activity maxima \citep{Usoskin:2021}.
\citet{Notsu2013} similarly demonstrate that the energies of superflares observed on solar-type stars trace the total starspot coverage on these stars. 
Yet the occurrence of strongest flares or flare candidates on the Sun does not seem to be closely correlated with the strength of the cycle, and strong events can happen even during relatively weak cycles \citep{Bazilevskaya:2014, Omodei2018, Usoskin:2021}.
From an analysis of main-sequence stars, \citet{Roettenbacher2018} found that only low-amplitude flares were correlated with observed starspot phase.

For stars that are different from the Sun, there are more uncertainties about the relationship between active region coverage and flare frequency, making \revised{flare frequency} a more difficult proxy to interpret for transmission spectroscopy analyses. 
\revised{On the other hand, tracing} flare energies on stars provides us with key information about the overall active region coverage on these stars. 
\revised{Studies of the spectral behavior of flares on M dwarfs are useful for constraining radiative-hydrodynamic flare models and thus areal coverages of active regions associated with flares \citep{Kowalski2013}.}
However, locating these active regions on the stellar surface is not possible from observing the flares alone, somewhat limiting our ability to infer the impact of these active regions on exoplanet transmission spectra.

\citet{Doyle2018} find no correlation with rotational phase for any flares on M dwarfs, an interpretation matched by \citet{Feinstein2020} in an analysis of stars in young moving groups. 
These observations may suggest very large ($\approx$80\%) active-region filling factors are typical for M dwarfs and young stars, consistent with values inferred via spectroscopic methods of the young star LkCa4 \citep{Gully-Santiago2017}, or may suggest variations in the underlying relations between active-region size and flares on non-Sun-like stars. 
For M dwarfs, flares may primarily occur from smaller active regions spread across the entire photosphere, while the starspot modulations observed by \Kepler{} and \TESS{} may be tracing larger and more stable active regions or ``spot caps'' \citep{Hawley2014}. 
There is an opportunity for continued theoretical modeling of active regions on these stars to be combined with the additional photometry for young and low-mass stars that will be collected over the coming years with \TESS{} to more fully understand the relations between flares and active regions on stars different from the Sun.

\subsubsection{The potential for improved data analysis methods}

While some features of active regions, such as the overall latitude distribution, can demonstrably not be uniquely inferred from transit photometry, here we consider what limitations may be due to current data processing practices. 

The \Kepler{} and \TESS{} data analysis pipelines are motivated by the search for exoplanets. 
While they do try to preserve stellar signals when possible, attempting to maintain those signals is sometimes at odds with the goal of planet detection.

Planet transits are relatively small signals (hundreds of ppm) with sharp features, few-hour timescales, and characteristic shapes, and are mostly achromatic. 
Therefore, relative precision is only needed on timescales of hours to a few days, while slow variability can be effectively removed by data processing pipelines. 
Since the signals are nearly achromatic, single-bandpass broadband photometry is acceptable for finding transits.  

Significant effort has been made towards increasing the photometric precision achievable by analyzing these data for transits, though there is clear evidence that additional information can be recovered to better understand the observed stars themselves. For example, \citet{Montet2017} used the \Kepler{} full-frame images, a set of calibration data obtained approximately monthly over the \Kepler{} mission, to calibrate photometry from that telescope and measure absolute brightness variations over multi-year timescales. 
That data product provides the opportunity to probe the relative contributions of facular and spot regions on the surface of the star as they grow and decay due to changing activity levels.

Similarly, \citet{Hedges2021} characterized the color dependence of the \Kepler{} point spread function, enabling identification of changing average stellar temperatures (and thus colors) over the course of the mission. 
This technique has the potential to further our knowledge of large active regions, possibly breaking some degeneracies between starspot size and temperature by providing effective multi-wavelength photometry over a four-year baseline. 
It also could enable improved inference of temperatures of stellar flares observed by \Kepler{}. 

Further development of new data analysis methods for \Kepler{} and \TESS{} data can help understand unocculted active regions in spaces where information may be quashed by the instrument pipeline or otherwise overwhelmed by systematic effects. 
Possible opportunities where significant progress may be both achievable and particularly beneficial include the improved recoverability of low-amplitude or long-period rotational modulation, the characterization of the smallest flares, and the development of machine learning methods to separate astrophysical and instrumental signals. 

\begin{figure*}
    \centering
    \includegraphics[width=\textwidth, height=0.5\textheight, keepaspectratio]{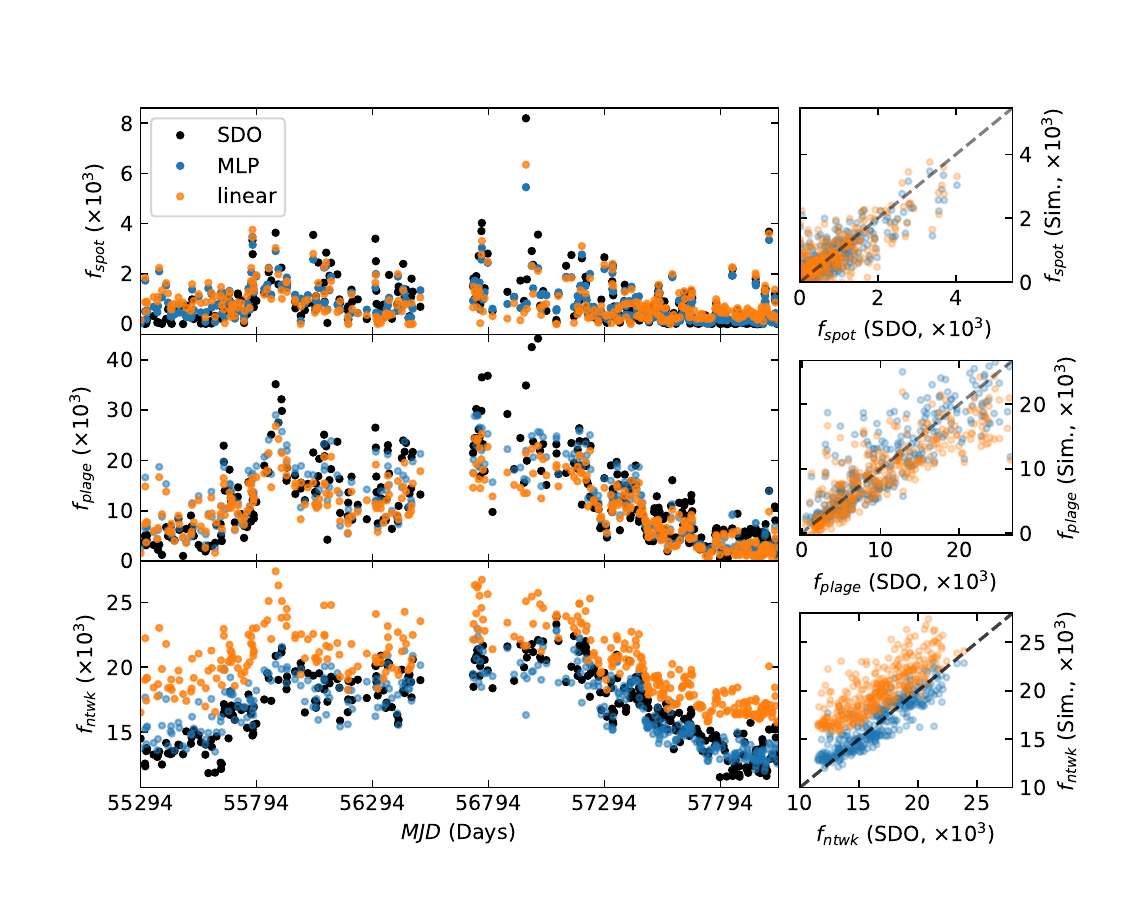}
    \caption{Spot (top), facular (middle), and network (bottom) filling factors as obtained from SDO (black dots; obtained using segmentation) and reconstructed using TSI and Ca S-index time series. 
    Blue symbols are for a machine-learning algorithm, orange symbols are from linear fitting. 
    From \citet{Milbourne2021}.
    \label{fig:milbourne}
    }
\end{figure*}

\subsubsection{Our finding}

From this analysis, we draw the following finding.
\\
\\
\noindent \underline{\textbf{Finding 3.1}}
\\
\\
\noindent \textit{Summary}: 
High-cadence light curves provide the potential to understand unocculted active regions, but the information is not comprehensive enough to make unambiguous measurements at present. 
Theoretical advances are needed to make full use of these light curves.
\\
\\
\noindent \textit{Capability Needed}: 
Significant theoretical work is needed to understand the relations between observational signatures of stellar activity in light curves\revised{.}
\revised{Continued} advances in data analysis methods developed to re-analyze archival data sets can provide major improvements in our ability to leverage information from \revised{precise light curves}. 
\\
\\
\noindent \textit{Capability Today}:
\revised{The large data sets provided by missions such as \Kepler{} \citep{Borucki2010} and \TESS{} \citep{Ricker2015} have enabled the development of data-driven methodologies for studying unocculted active regions through photometry \citep[e.g.,][]{Montet2017, Hedges2021, Luger2021b}.
These same data sets} are producing opportunities for improved data analysis tools \revised{\citep[e.g.,][]{Lightkurve, Feinstein2019}}. 
\revised{Efforts like these} can be supported by programs such as \revised{the} NASA \revised{Astrophysics Data Analysis Program}.
\\
\\
\noindent \textit{Mitigation in Progress}: 
The NASA\revised{--}NSF \revised{Extreme Precision Radial Velocity Initiative\footnote{\url{https://exoplanets.nasa.gov/exep/NNExplore/EPRV/}}, a strategic effort to develop methods and facilities for measuring the masses of temperate terrestrial planets orbiting Sun-like stars, has} enabled opportunities for the development of theoretical work to improve our understanding of the effects of stellar activity on time-series spectroscopy \revised{\citep[e.g.,][]{Newman2022}}, which is a parallel need to the theoretical work outlined here.

\subsection{Information from other datasets}
\label{sec:otherdata}

In characterizing the effects of unocculted active regions on transmission spectra, the high-cadence photometry obtained to discover these systems does not exist in a vacuum. 
Here we consider other observations, whether or not they are obtained simultaneously with the transmission spectrum, and their ability to understand active regions. 
Solar observations show that the absolute as well as the relative surface coverage of spots and faculae (or plage) varies as a function of time. 
As the transmission spectrum ``contamination'' depends on the feature contrasts in the given spectral band, a single active-region tracer is generally insufficient and ideally two active region tracers should be combined, ideally with very distinct sensitivity levels to bright and dark active region components.

\subsubsection{Multiband photometry}
\label{subsec:multiband}

Both \Kepler{} and \TESS{} observe with a single broadband filter. 
With simultaneous observations at multiple wavelengths, the changing color of an active stellar surface over time can be inferred. 
This strategy has been used to determine temperatures of stellar flares \citep{Howard2020}, and there is a long tradition of using multiband photometric light curve modeling to infer spot temperatures and covering fractions on active stars (see \autoref{sec:observing-starspots-and-faculae} and \citet{Berdyugina2005} for a review on starspot properties and observations). 
While multiband photometry can lift some of the degeneracies between filling factors and active region temperatures \citep[e.g.][]{Rosich-20}, spot contrasts only vary slowly as a function of wavelength, meaning that spot temperature determinations remain very uncertain. 
Recovering active-region filling factors becomes even more of a problem when trying to map spots and faculae simultaneously using broad-band photometry. 
However, as facular regions have a much stronger spectral signature than starspot regions \citep{Shapiro2014}, narrow-band photometry focused on particular spectral regions (e.g., Ca\,\textsc{ii} H \& K and molecular bands such as CN bands, the CH G-band, or the TiO band in the red optical; \citealt{Zellem2010}) could provide an outsized benefit in understanding the distributions of faculae, motivating future theoretical modeling. 
These spectral regions are also discussed more fully in \autoref{subsec:unocculted_lowres}.

Work by \citet{Milbourne2021} suggests that for solar twins, where spot and facular contrasts are relatively well constrained by theory and solar observations, it should be possible to recover spot and facular filling factors using a combination of high-fidelity broadband photometry and a magnetic-activity tracer such as the Ca S-index (which can either be obtained using high-resolution spectroscopy or narrow-band photometry). 
The first results are promising (\autoref{fig:milbourne}), though they rely on accurate knowledge of the spot and facular contrasts. 
There is also both theoretical and observational evidence that spot and facular contrasts are affected by stellar metallicity, but the effects have not been fully explored in detail for a large population \citep{Karoff18, witzke2018}.
Thus, more information on active-region contrasts for different spectral types and metallicities is needed before this can be extended to stars with non-solar spectral types. 
Further work will also be needed to explore how the location and relative importance of spots and faculae affects such reconstructions. 

To map and understand the distribution of faculae, their spectral signatures should be further explored. 
Faculae have strong UV contrasts, though this has as yet been not fully explored, mainly due to a lack of time-resolved UV photometry.
Broadly, optical photometry has been traditionally considered to retrieve stellar heterogeneities from long-term photometry, with the Str\"{o}mgren $b$ and $y$ filters commonly considered \citep[e.g.\revised{,}][]{Gray-05}. 
\revised{Far ultraviolet} photometry may provide a promising addition to this set, though current capabilities may enable only a limited benefit compared to monitoring of stellar calcium emission features \citep{Findeisen11}. 
In the \revised{NIR}, there is some evidence that signatures of stellar activity may be even smaller than anticipated for some young stars \citep{Miyakawa21}, making them a sub-optimal choice to measure long-term photometric variability. 
However, \citet{Stefansson18} highlight a promising custom filter, \revised{300-\AA} wide and centered at \revised{8570\,\AA}. 
This filter spans the Ca\,\textsc{ii} infrared triplet \revised{(8498\,\AA, 8542\,\AA, 8662\,\AA)} and minimises the effects of telluric absorption. 
These authors suggest this filter offers an inexpensive opportunity to study stellar activity in the \revised{NIR}; 
further work should be undertaken to determine the usefulness of this or other potential bandpasses\revised{, spanning features such as the Na D doublet (5890\,\AA, 5896\,\AA) or the metastable He 10830\,\AA{} line \citep[e.g.,][]{Vaughan1968},} in the context of identifying and quantifying the extent of stellar surface heterogeneities.

\subsubsection{\revised{Long-term} photometry}
\label{sec:sg4-vlb_photometry}

For many bright stars, additional photometry exists from ground-based surveys with \revised{long temporal} baselines\revised{, i.e.,} extending ten years or more before the present day. 
These data are often leveraged by the exoplanet community to great effect in the confirmation and characterization of planetary systems, demonstrating their utility in combination with high-cadence transit photometry. 
In particular, the \revised{Wide Angle Search for Planets (WASP)} \citep{Pollacco2006}, \revised{Kilodegree Extremely Little Telescope (KELT)} \citep{Pepper2007}, and \revised{All Sky Automated Survey for SuperNovae (ASAS-SN)} \citep{Kochanek2017} surveys often have the precision to separate planet candidates from background eclipsing binaries and infer stellar rotation periods through starspot modulation.
On much longer timescales, the \revised{Digital Access to a Sky Century @ Harvard (DASCH)} survey \revised{\citep{Grindlay2009, Grindlay2012}} provides photometry with 0.1 magnitude precision over 100 years through the digitization of a photographic survey that began in 1885.
These data have been used to study long-term variability in the \Kepler{} field \citep{Tang2013}.

These datasets offer a tradeoff between time baseline and photometric precision, with a century of data available at $\approx 10$\% precision or a decade at $\approx 1$\% precision. 
\revised{They can be used to measure photometric variability at these precisions as well as changes in variability over time \citep[e.g.,][]{Lanza1998, Gully-Santiago2017}, owing to activity cycle evolution} or the growth and decay of large polar spots.

While these data have information about unocculted active regions that can be used to understand the overall activity level of the star, they are collected years or decades before the transmission spectra are obtained. 
Long-\revised{term} observations do provide data on the relative overall activity of the star at the current time, through spectroscopic tracers such as a single measurement of chromospheric activity from the Ca\,\textsc{ii} H and K lines can provide a similar level of information contemporaneously with the transmission spectrum. 
There is currently a paucity of theoretical motivation suggesting that these observations can provide useful information in interpreting a transmission spectrum. 
The converse has even been suggested: \citet{Iyer2020} suggest that prior knowledge of stellar heterogeneities does not improve precision in planetary parameters achievable from an atmospheric retrieval if the heterogeneities are appropriately marginalized.
Expanded theoretical work highlighting benefits of this long-term photometry specifically in the context of transmission spectroscopy would help to motivate the future utility of these data in this context.

\subsubsection{Interferometry}

The nearby active star $\zeta$~And is among the so far few stars for which a global surface temperature map has been successfully constructed via aperture synthesis imaging \citep{Roettenbacher2016}. 
The data, which were obtained with the interferometric \revised{Center for High Angular Resolution Astronomy (CHARA)} Array at a wavelength of 1600\,nm and an angular resolution of 0.5\,mas, reveal variations in the surface temperature that are related to starspots.  

Interferometric observations are possible with present technology only for bright stars with large angular sizes; 
these are typically red giants. 
Recent work has enabled constraints on the spot locations for the K2V exoplanet host $\epsilon$ Eridani through a combination of RV spectroscopy, \TESS{} photometry, and \revised{interferometry with the Michigan InfraRed Combiner-eXeter (MIRC-X) six-telescope interferometric imager at the CHARA array} \citep{Roettenbacher2022}, though the spot detections are marginal with this dataset, underscoring the challenge of imaging exoplanet hosts.
Thus, the primary utility of optical interferometry at present is in enabling a better understanding of possible configurations of active regions on stellar surfaces, which can be compared to theoretical models to provide constraints on the stellar dynamo.
In the context of transmission spectroscopy, an important step is theoretical modeling of the differences between the dynamo and resultant magnetic fields between typical planet-hosting stars and the giants that are observable with optical interferometry.

\subsubsection{Low- and medium-resolution spectroscopy}
\label{subsec:unocculted_lowres}

\begin{figure}
    \centering
    \includegraphics[width=\columnwidth]{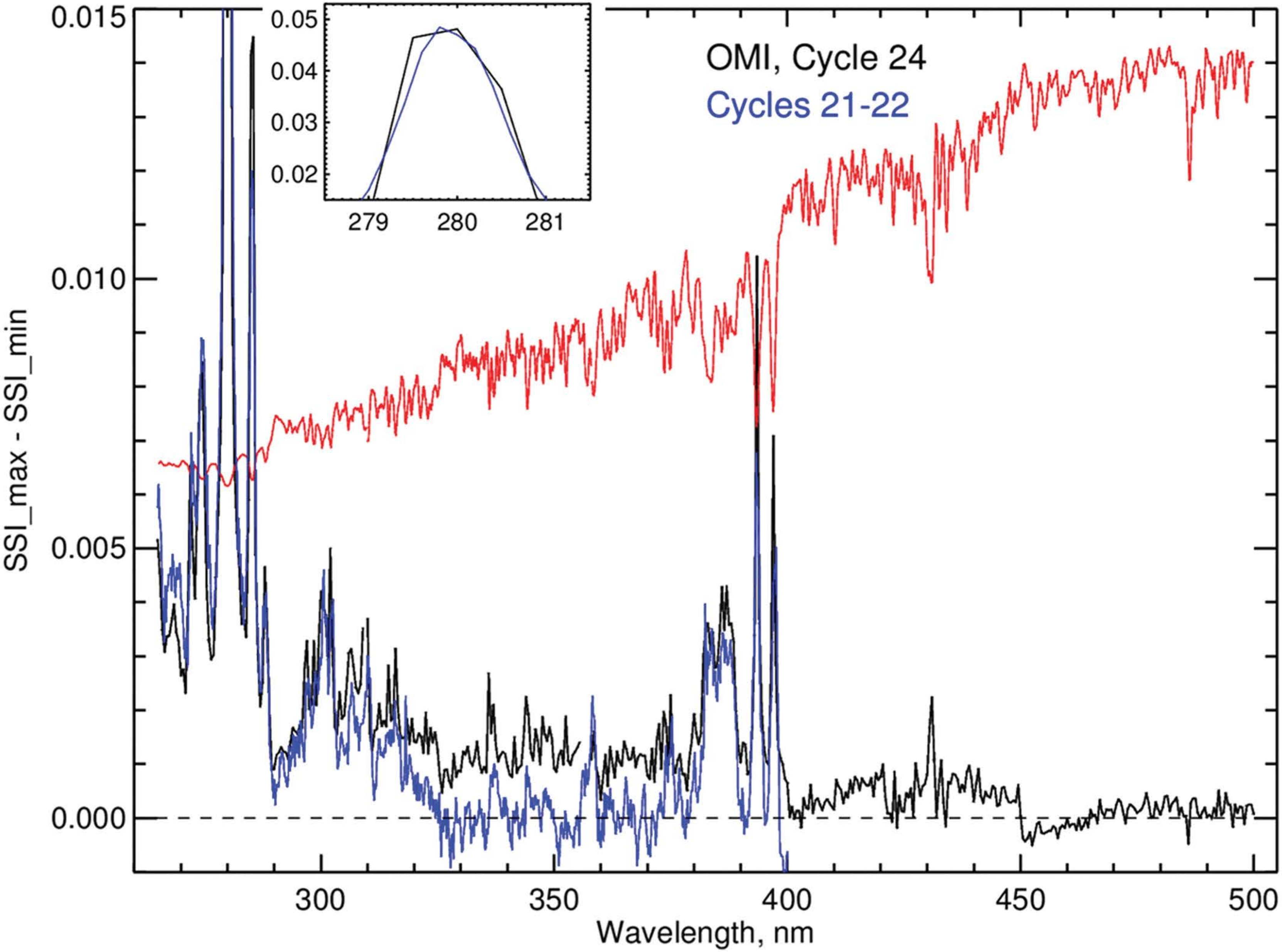}
   \caption{Spectral solar-cycle variability for cycle 24 as observed by \revised{Aura/OMI} (black line; the blue line shows the cycle 21 and 22 variability below 400\,nm scaled to match the cycle 24 variability for Mg~\textsc{ii}).
   Enhanced variability is clearly seen in the Ca\,\textsc{ii} H \& K lines as well as the CN violet band (between $\sim$ 380\,nm and 390\,nm), and the CH G-band (at 430\,nm). 
   From \citet{Marchenko2019}.
   \label{fig:Marchenko}
   }
\end{figure}

Photometric observations of active stars can be used to identify active regions, especially starspots. 
However, they typically produce a degeneracy between filling factors and active region temperature: 
a given amount of photometric modulation could be produced by a relatively small but very dark starspot or a larger spot with a relatively small differential between its temperature and the unocculted stellar surface. 
Low- and medium-resolution spectroscopy can offer a more detailed view than photometry, provided the instrumental stability is sufficient to monitor relatively subtle spectral changes. 
An example of (facular-dominated) solar cycle variability observed at mid-resolution ($R{\sim}1000$) with the \revised{Ozone Monitoring Instrument (OMI) onboard the NASA Aura Mission} is shown in \autoref{fig:Marchenko} \citep{Marchenko2019}. 
The blue and UV wavelength regions are of particular use here as facular contrasts are much larger at shorter wavelengths. 
A number of relatively broad spectral features stand out, such as the Ca~\textsc{ii} H \& K lines, the CH G-band, and the CN violet system (see \autoref{subsec:multiband}).

\revised{Despite recent advances \citep{Panja-2020}, generally} little is known about contrasts of active regions on stars of spectral types other than the Sun\revised{,} and more theoretical work is needed to understand and model bright features (see \autoref{subsec:retrieval_limitations}). 
Magnetoconvection simulations show that faculae will become less bright for later spectral types, in particular for M stars \citep{Beeck3}. 
Preliminary calculations suggest variability in the CN and CH bands will decrease, while TiO bands become more important; 
at the same time, facular UV contrasts remain comparable to those observed for Sun-like stars \citep{Norris2018}.

\subsubsection{High-resolution spectroscopy}

Much as in the low/medium resolution spectroscopy discussed in the previous section, high-resolution spectroscopy can be used to better understand active-region filling factors.
With sufficiently high resolution, molecular bands can be detected at high significance. 
Spectroscopic signatures of molecular features are strong functions of temperature and pressure, so these features provide an ideal laboratory to understand the stellar surface as the sum of multiple components at different temperatures \citep[e.g.,][]{Berdyugina2003mol}.
For example, \citet{ONeal-98} demonstrated that the active K star II Peg had a coverage fraction of ${\sim}$50\% from observations of TiO molecular bands.
\citet{Gully-Santiago2017} showed an even higher filling factor for the young star Lk\,Ca\,4 using the same molecular features. 
In fact, the most studied active K stars II Peg and IM Peg were never observed in their unspotted states, as evidenced by the continued presence of variable TiO bands in their spectra \citep{Neff1995,Berdyugina1998iipeg,Berdyugina1999impeg}. 
Thus, the disappearance of the TiO bands forming only in cool spots can provide the reference for the unspotted stellar atmosphere and brightness \citep{Berdyugina1998iipeg}.
Similar observations of other stars at different levels of activity remain critical to understanding overall active-region filling factors and thus the underlying sizes, atmospheric parameters, and distributions of active regions.

Additionally, molecular features trace the stellar magnetic field through their polarization \citep{Berdyugina2011}. 
High-resolution spectropolarimetry of these lines provide estimates of magnetic fields on active stars, also directly within spatially unresolved spots \citep{Berdyugina-02,Berdyugina2006tio}. 
In particular, \citet{Afram-Berdyugina2015} highlight that high-resolution spectropolarimetry of TiO on M dwarfs, CaH on K dwarfs, and MgH and FeH on G dwarfs provide particularly promising avenues to understand stellar magnetic fields. 
This has been demonstrated for M dwarfs by studying atomic and molecular spectropolarimetric signatures, revealing variations of the magnetic field complexity on stellar surfaces, with the height in the atmosphere and depending on the stellar effective temperature \citep{Berdyugina2011,Afram-Berdyugina2019}.

Extreme precision radial velocity (EPRV) observations to measure planet masses and orbits also require care to mitigate the effects of active regions on stellar spectra \revised{\citep[e.g.,][]{Robertson-14, Robertson2015}}. 
Considerable effort has been expended to attempt to separate the effects of stellar active regions from planetary signals.
These efforts have been successful in developing methods that can effectively predict the RV signal from active regions from observed photometric variability \citep[e.g.][]{Aigrain2012} and to model the surfaces of very active stars to measure the masses of their planets \citep[e.g.][]{Barragan2019}.
Observations of the Sun with the \revised{High Accuracy Radial velocity Planet Searcher for the Northern hemisphere (HARPS-N)} spectrograph also demonstrate that RVs can effectively be modeled in the face of known active region distributions \citep{Milbourne2021}.

A common finding in EPRV studies of active regions is the importance of simultaneous photometry and spectroscopy to condition data-driven models. 
Having both data sets is critical in many cases to measure planet masses \citep{Oshagh2017}; 
it is likely that in many cases simultaneous observations will also provide a benefit to uniquely characterise unocculted active regions in transmission spectroscopy data.

\subsubsection{Conclusion}

Each of the methods described in this section holds the potential to improve our understanding of unocculted active regions for transmission spectroscopy.
While there is evidence from EPRV observations that simultaneous photometry is highly important for active stars, it is not as obvious for stars with lower activity levels. 
An important step forward is not just understanding what data additional sets are important, but to what level the answer is a function of spectral type, age, and/or activity level. 
Theoretical work that can effectively predict the effects of stellar activity on transmission spectra would be beneficial in answering this question.

\subsubsection{Our finding}

From this analysis, we draw the following finding.
\\
\\
\noindent \underline{\textbf{Finding 3.2}}
\\
\\
\noindent \textit{Summary}: 
Simultaneous multi-band photometry and contemporaneous spectroscopy provide critical information towards understanding the potential effects of active regions on transmission spectroscopy observations. 
While other data sets can provide information on filling factors, theoretical work is needed to maximise the utility of these data for transmission spectroscopy. 
\\
\\
\noindent \textit{Capability Needed}: 
A better understanding of the emergent flux from stellar active regions, including both photometric (\revised{i.e., spectral energy distribution, }SED) and spectral signatures (e.g., Na D \revised{doublet, Ca\,\textsc{ii} infrared triplet, He\,10830\,\AA}, etc.), especially for spectral types and metallicities other than the Sun, is essential to robustly interpret transmission spectra. 
\\
\\
\noindent \textit{Capability Today}: 
There is a relatively good understanding of active-region contrasts for solar twins from solar observations and modeling. 
\revised{
Models including calculated spot and facula contrasts as a function of wavelength and limb angle can accurately reconstruct solar irradiance variations on timescale of the solar rotation \citep{Fligge2000} and solar magnetic cycle \citep{Krivova2003}.
Extended to other stars, these models can reproduce the observed photometric variability of Sun-like stars using different coverages of magnetic features \citep{Shapiro2014}.
Using a combination of spectroscopy and photometry, the coverage of both spots and faculae on Sun-like stars can similarly be constrained \citep{Milbourne2021}.
}
\\
\\
\noindent \textit{Mitigation in Progress}: 
Ab-initio modelling of emergent fluxes from spots and faculae for different spectral types is being developed by multiple research groups \revised{\citep[e.g.,][]{Norris2017, Panja-2020, Witzke2022}}.

\subsection{Stellar granulation}
\label{sec:uar_physics}

One of the ultimate sources of an astrophysical ``noise floor'' in stellar light curves is stellar surface granulation. 
Because it is a fundamental feature of stars with convective envelopes (i.e., $T_{\rm eff} \lesssim 6700$~K), it represents a true lower limit to the photometric variability that any star can attain. 
This is summarized in \autoref{fig:flicker}, where the lower ``granulation flicker floor" shows the lowest possible photometric variability amplitude for otherwise inactive stars, as a function principally of the stellar surface gravity, $\log g$. 
On the timescale of typical planet transits (i.e., $\lesssim$8~hr), in the \Kepler{} bandpass (i.e., visible light), the ``granulation flicker'' noise ranges from $\sim$0.02\,ppt for the least active dwarfs to $\gtrsim$0.4\,ppt for the least active giants. 

The amplitude of the ``granulation flicker'' noise scales with the granule/inter-granule brightness contrast, which is minimized at longer wavelengths. 
For example, the granular brightness contrast for a solar-type star is ${\sim}100 \times$ lower at 10\,$\micron$ than in visible light. 
Thus, for inactive dwarfs, the ``granulation flicker'' noise at 10\,$\micron$ should be as low as $\sim$0.2\,ppm on planet-transit timescales.

While these are relatively small contributions to the overall stellar photometric noise, from the standpoint of a planet transit spectrum, the impact can become the dominant source of uncertainty, particularly for transits around larger stars and at shorter wavelengths. 
That is because a larger star of a given mass will have a lower surface gravity, hence a stronger granulation signal, and as noted above, that signal is strongest in the visible. 
\revised{\citet{Morris2020_plato} examined the impact of granulation on observations from the ESO PLAnetary Transit and Oscillations (PLATO) mission \citep{Rauer-14}, and \citet{Sulis:2020} performed a similar analysis for both PLATO and CHaracterising ExOPlanet Satellite \citep[CHEOPS,][]{Benz2021} observations.} 
\revised{These simulations show that} for planets transiting Sun-like stars observed in the visible, the absolute error on the retrieved planet radius due to granulation noise can be as large as $\sim$2\% and $\sim$10\% for super-Earth and Earth-sized planets, respectively.
\revised{Indeed, CHEOPS commissioning observations of the bright, G8/K0 subgiant KELT-11 show granulation-induced variability with an amplitude of 200\,ppm \citep{Benz2021}, in line with these predictions.}

\begin{figure}
    \centering
    \includegraphics[width=\columnwidth, height=\textheight, keepaspectratio]{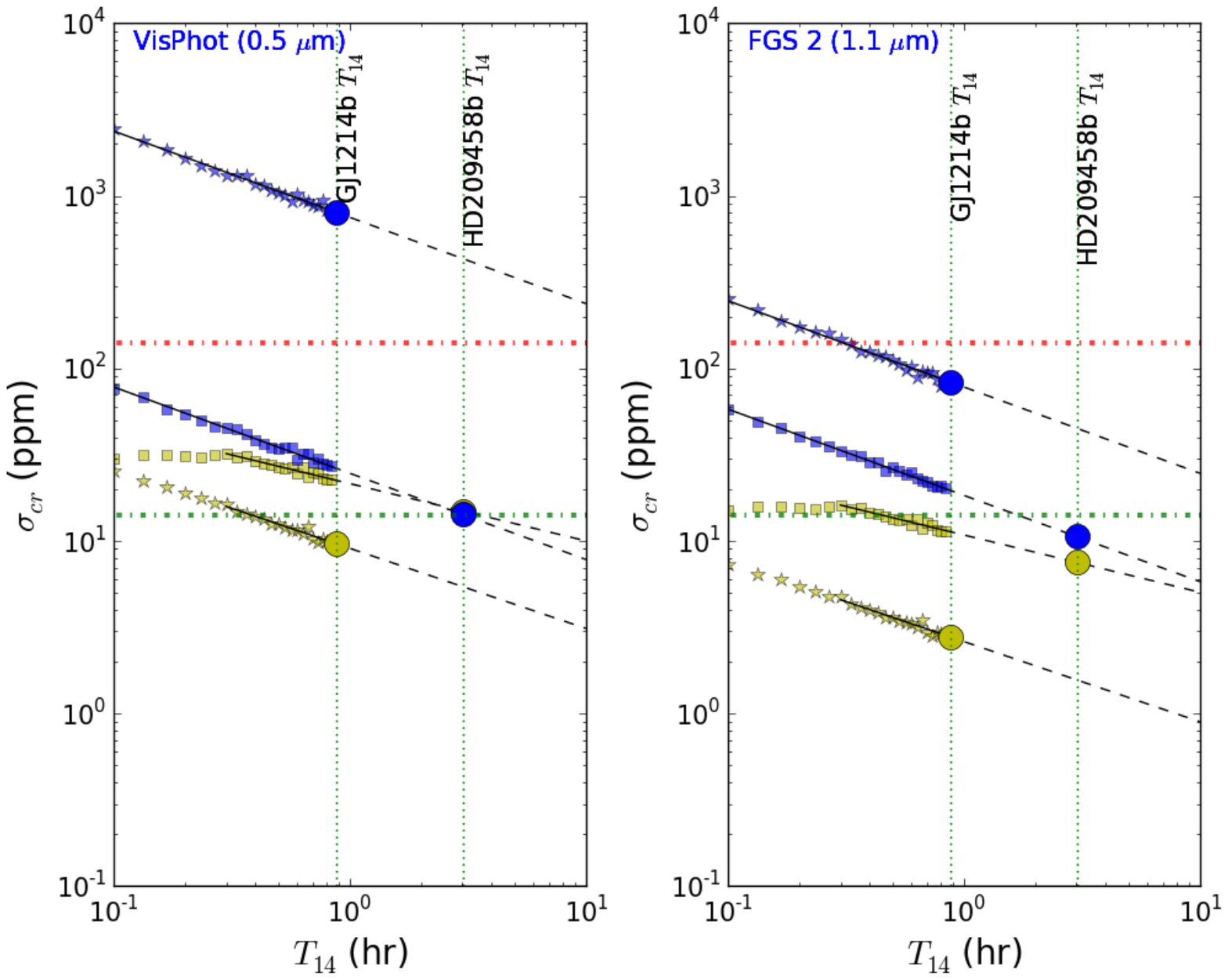}
    \caption{Simulations of the relative noise contributions to planet-transit depth \revised{fractional uncertainty ($\sigma_{cr}$)} due to stellar \revised{pulsation and} granulation (yellow) and photon noise (blue) as a function of transit duration\revised{.}
    \revised{The} cases of a hot Jupiter orbiting a G0V-type star (HD\,209458b\revised{; squares}) and a super-Earth orbiting an M4.5V-type star (GJ\,1214b\revised{; stars}) \revised{are shown for two filters of the Ariel Space Mission, VisPhot (visible; left) and FGS 2 (near-infrared; right).}
    \revised{The black dashed lines give power-law fits to the noise levels, and the upper red and lower green dash-dotted lines show the Ariel noise floor requirement and goal, respectively.}
    For each planet, compare the large yellow dot to the large blue dot; 
    this is the relative contribution of stellar granulation noise to photon noise for atmospheric retrieval. 
    Note that in the left panel (visible light), the yellow and blue dots overlap for HD\,209458b, meaning that the stellar granulation noise begins to dominate the error budget. 
    From \citet{Sarkar:2018}.
    \label{fig:ariel_sim}
    }
\end{figure}

In practice, stellar granulation can impact the transit spectrum thus in two ways. 
The first and most obvious one is the one related to the extra noise this granulation adds to light curves, effectively setting a lower 
limit to the possible precision attainable on the retrieved transit depths as a function of wavelength. 
This source of noise, however, should be minor in most cases. 
While studying this, \citet{Sarkar:2018} considered the relative and absolute contributions to the error budget for atmosphere retrieval of the planets HD\,209458b (a hot Jupiter orbiting a G0V-type star) and GJ\,1214b (a super-Earth orbiting an M4.5V-type star). 
The results are summarized in \autoref{fig:ariel_sim} and \autoref{tab:ariel_sim}, where we see that in the worst case (HD\,209458b in the visible), where the stellar noise from granulation dominates over the photon noise, the impact on the transit-depth measurement for atmospheric retrieval is at most $\sim$0.05\%.

\begin{table}
    \centering
    \caption{
        Tabular summary of \autoref{fig:ariel_sim}.
        Relative uncertainty in planet transit depth for the two representative cases (in ppm), and the resulting absolute uncertainty on the planet radius for atmospheric retrieval (in percent) for different wavelength regimes of observation.
        From \citet{Sarkar:2018}.
        \revised{Details of the listed spectral channels can be found in \citet{Tinetti-18}.}
        \label{tab:ariel_sim}
    }
    \begin{tabular}{lccccc}
    \hline\hline
    \multicolumn{1}{c}{Channel} &
    \multicolumn{2}{c}{GJ\,1214b} &
    \multicolumn{2}{c}{HD\,209458b} \\
    
    \multicolumn{1}{c}{(\textmu m)} &
    \multicolumn{1}{c}{$\sigma_{cr, sn}$} &
    \multicolumn{1}{c}{$\frac{\sigma_{R_p, sn}}{R_p}$} &
    \multicolumn{1}{c}{$\sigma_{cr, sn}$} &
    \multicolumn{1}{c}{$\frac{\sigma_{R_p, sn}}{R_p}$} \\
    
    \multicolumn{1}{c}{} &
    \multicolumn{1}{c}{ppm} &
    \multicolumn{1}{c}{\%} &
    \multicolumn{1}{c}{ppm} &
    \multicolumn{1}{c}{\%} \\
    
    \hline
    AIRS Ch1 (3.9--7.8) &	1.5 & 0.01 & 3.6 & 0.01 \\
    AIRS Ch0 (1.95--3.9) & 2.3 & 0.01 &  3.5 & 0.01\\
    NIRSpec (1.25--1.9) & 3.1 &  0.01  & 4.5 &  0.02\\
    FGS 2 (1.1) &	 2.8  & 0.01 & 7.5 & 0.03 \\
    FGS 1 (0.9) & 4.4 &  0.02 & 9.0 & 0.03\\
    VisPhot (0.5) & 9.6 & 0.04 & 14.9 & 0.05\\
    \hline
    \end{tabular}
\end{table}

The second way in which granulation could impact on the transit spectrum, however, is through the transit light source effect, 
similar to the impact caused by unnoc\revised{c}ulted spots and faculae described by \cite{Rackham2018, Rackham2019}, but 
produced by the effectively different granulation pattern in different parts of the stellar surface. 
This is an effect that \cite{Chiavassa-17} studied through 3D simulations using the STAGGER grid and synthetic images computed with the radiative transfer code OPTIM3D \revised{\citep{Chiavassa-09}}. 
The study concluded that the impact of this contamination could be as large as a few percent in the transit depth of small, terrestrial exoplanets orbiting Sun-like stars at $0.7\mu$m. 
While \cite{Chiavassa-17} did this experiment with simulations for both a Sun-like star and a K-dwarf star, it is unclear how big this impact is in reality, and how realistic those simulations are both as a function of wavelength and granulation amplitudes. 
This latter effect is one that should be studied in detail, especially given missions such as the 
ones proposed in the Astro2020 Decadal Survey, which could perform precise optical spectrophotometry.

\subsubsection{Our finding}

From this analysis, we draw the following finding.
\\
\\
\noindent \underline{\textbf{Finding 3.3}}
\\
\\
\noindent \textit{Summary}: 
Stellar ``granulation flicker" constitutes a fundamental ``noise floor" on stellar light curves that increases with decreasing stellar surface gravity and at shorter wavelengths. 
The impact of this stellar granulation can enter as a source of noise and/or contamination for transmission spectroscopy in two ways. 
The first is through extra light curve scatter, \revised{the} impact \revised{of which} is minimal in most cases (at most $\sim$0.05\% error on the transit depth in the visible and even less at longer wavelengths).
The second is through a contamination source similar to that of spots and faculae, albeit at a much lower amplitude ($\sim 1\%$ of the transit depth at about 0.7\,$\mu$m).
This second source is significant at short wavelengths for smaller exoplanets around Sun-like stars, but decreases strongly as a function of wavelength.
\\
\\
\noindent \textit{Capability Needed}: 
3D granulation simulations need to be validated against observed light curves as a function of spectral type and wavelengths relevant for future space-based observatories ($\sim$0.3--5\,$\micron$). 
Simulations of exoplanets crossing those simulated stellar surfaces should be \revised{used further} to study the full impact of these features as a contamination source in the transmission spectrum of high-profile exoplanet systems. 
\\
\\
\noindent \textit{Capability Today}: 
Granulation is well understood in large optical bandpasses from space-based missions such as \Kepler{}\revised{,} \TESS{}\revised{, and CHEOPS}. 
Methodology is in place in order to understand and analyze its amplitude given an observed precise light curve. 
3D stellar models can be used to model granulation as a function of wavelength for different spectral types.
\revised{
CHEOPS observations of the G8/K0 subgiant KELT-11 show granulation-induced variability of 200\,ppm \citep{Benz2021} that is in line with predictions \citep{Sulis:2020}.
For future Ariel observations, simulations show that the impact of granulation can be comparable to photon noise uncertainties for a 3-hr transit of the G0V HD\,209458 \citep{Sarkar:2018}.
Similarly, simulated transits of the Sun show that solar oscillations and granulation produce signals with an amplitude of 100\,ppm, producing a planetary radius uncertainty that accounts for a significant fraction of the error budget of the PLATO mission \citep{Morris2020_plato}.
}
\\
\\
\noindent \textit{Mitigation in Progress}: 
All-sky missions such as \TESS{} can retrieve the properties of the granulation signal for high-profile exoplanet host stars \revised{\citep{Campante2016, Stassun:2017}}.
\revised{Studies of the impact of oscillations and granulation on Ariel \citep{Sarkar:2018}, CHEOPS \citep{Sulis:2020}, and PLATO \citep{Morris2020_plato} have recently been carried out.}

\section{Stellar \& Planetary Retrievals}
\label{S:Retrievals}

\begin{figure*}
    \centering
    \includegraphics[width=\textwidth, height=\textheight, keepaspectratio]{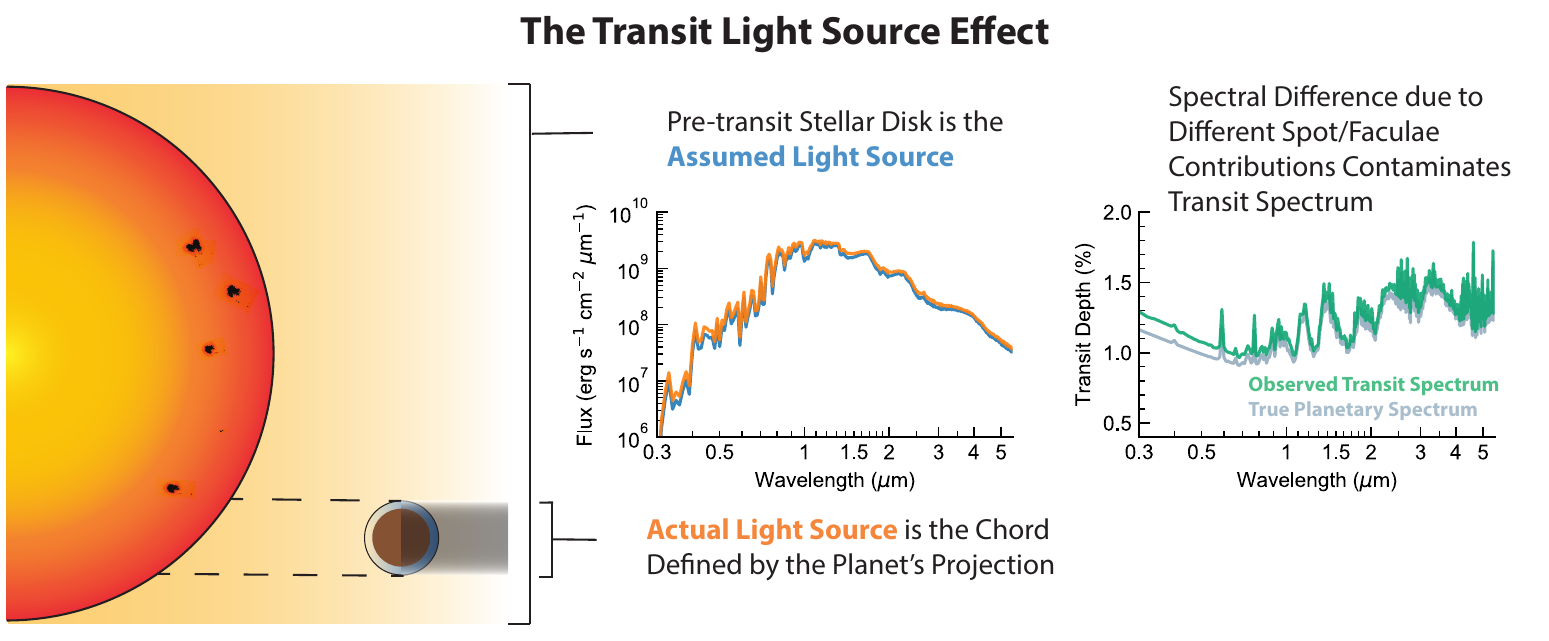}
    \caption{
    Illustration of the transit light source effect. 
    Unocculted heterogeneities (i.e., magnetic active regions) can mask or mimic spectral features in transmission spectra by introducing a difference between the spectrum of the light source illuminating an exoplanet's atmosphere---the emergent spectrum of the transit chord---and the light source necessarily used as a reference for the transit depth measurement---the disk-averaged spectrum of the star. 
    From \citet{Rackham2018}.
    \label{fig:tlse}
    }
\end{figure*}

\subsubsection*{Essential \revised{q}uestions:}

\begin{enumerate}
    \item What is the state of the art for stellar and planetary retrievals of transmission spectra?
    \item What wavelengths and resolutions are necessary for useful constraints on active-region contrasts and coverages? Do these requirements vary with spectral type?
    \item What are the best practices for propagating stellar spectral information into atmospheric retrievals of transit observations?
    \item What additional information is necessary to uniquely disentangle stellar and planetary signals via retrievals?
    \item What are the limitations of using retrievals to disentangle stellar and planetary features?
    \item What future work should be pursued in this context?
\end{enumerate}

\subsection{Introduction}

As noted in \autoref{sec:introduction}, a primary concern with transmission spectroscopy is the potential of unocculted active regions (including, e.g., spots or faculae) to alter transit depths through the transit light source effect (TLSE, see \autoref{fig:tlse}) and, thus, bias inferences of the properties of the exoplanetary atmosphere.
Here we examine the state of the art for Bayesian frameworks that jointly fit for the properties of a planetary atmosphere and a heterogeneous stellar photosphere, i.e., one with magnetic active regions.
We begin by defining more precisely these retrieval frameworks and outlining the scope of our analysis.

\subsubsection{Retrievals}
\label{subsec: sg2: Retrievals}

\begin{figure}
    \centering
    \includegraphics[width=\columnwidth, height=0.5\textheight, keepaspectratio]{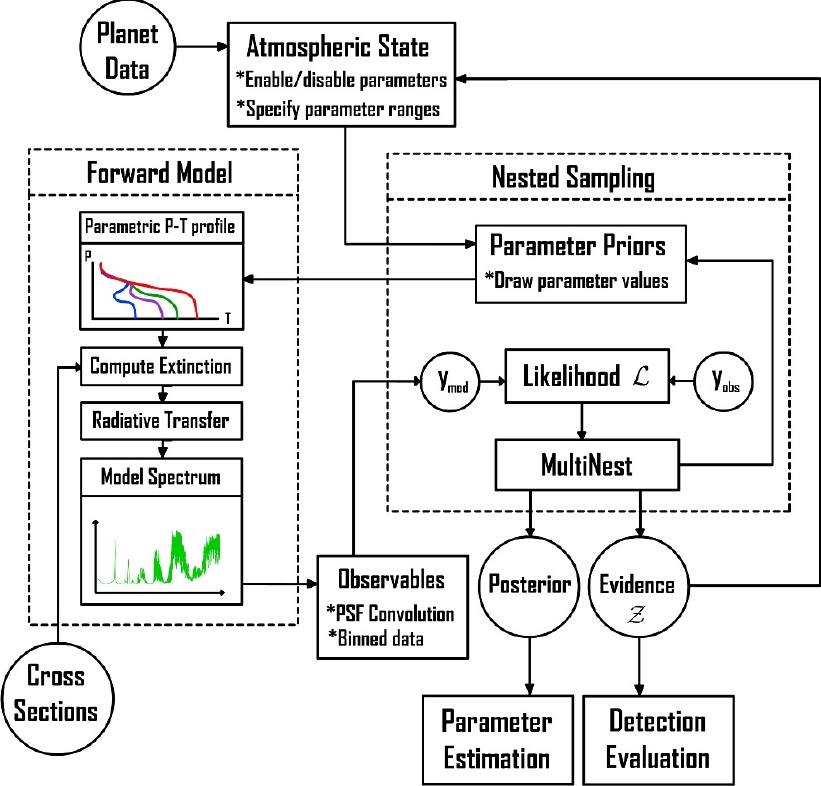}
    \caption{
    Schematic illustration of a retrieval framework. 
    Retrievals start with an observed (or simulated) spectrum. 
    A forward model is called repeatedly to compare model spectra for a wide range of atmospheric properties to the observations. 
    A Bayesian sampling algorithm (here, nested sampling) guides the parameter space exploration, producing the posterior distribution for each parameter and detection significances for each model component (e.g., molecular species, presence of stellar heterogeneities). 
    From \citet{MacDonald2017_hd209458b}.
    \label{fig:retrievals}
    }
\end{figure}

Atmospheric retrieval refers to a broad class of techniques used to extract constraints on properties of a planetary atmosphere (e.g., composition, temperature, clouds) from an observed spectrum. 
Retrieval codes start with a set of spectral observations, such as \HST{} transmission spectra, and compute the range of models consistent with the data, as illustrated in \autoref{fig:retrievals}. 
A retrieval model describes an atmosphere by a set of free parameters (e.g., molecular abundances, planetary radius) and uses a Bayesian sampling algorithm (typically MCMC or nested sampling) to explore the posterior distribution of the model.
A retrieval typically computes $\gtrsim 10^5$ spectra to map the parameter space defining a model, with nested models then compared to identify the statistical significance of each model component (e.g., whether a given molecule is supported by the data).

\subsubsection{The scope of this analysis}

Extracting the most information from transmission spectra requires inference frameworks that allow us to constrain the properties of stars and planets simultaneously.
In this context, the analysis presented here considers how we can best use the information we receive on both stars and planets in transit observations from space-based facilities.
We focus on three areas of interest: 
the utility of joint stellar and planetary retrievals generally (\autoref{subsec:retrievals_utility}); 
the known limitations of the approach, with a focus on the availability of suitable model spectra for stellar active regions (\autoref{subsec:retrieval_limitations}); 
and the complementarity of short-wavelength observations to \HST{} and \JWST{} NIR observations for inferring stellar photospheric properties (\autoref{subsec:retrievals_blue}).
In the following sections, we summarize the analysis that leads to our top-level findings in these three areas.

\subsection{The utility of joint stellar and planetary retrievals}
\label{subsec:retrievals_utility}

We first consider the general utility of employing joint stellar and planetary retrievals over other, simpler approaches.
In reaching our finding, we \revised{reviewed} how unocculted stellar heterogeneities modify transmission spectra, approaches for directly correcting for stellar spottedness without using retrievals, \revised{and} the current state of the art for modeling stellar photospheric heterogeneity in retrievals\revised{.}
\revised{We also reviewed} recent studies that suggest direct approaches for corrections may underestimate the actual level of stellar heterogeneity\revised{, underscoring} the utility of retrievals.
\revised{The following sections present each of these topics in turn.}

\subsubsection{Unocculted stellar heterogeneities modify transmission spectra}

\begin{figure*}
    \centering
    \includegraphics[width=\textwidth, height=0.5\textheight, keepaspectratio]{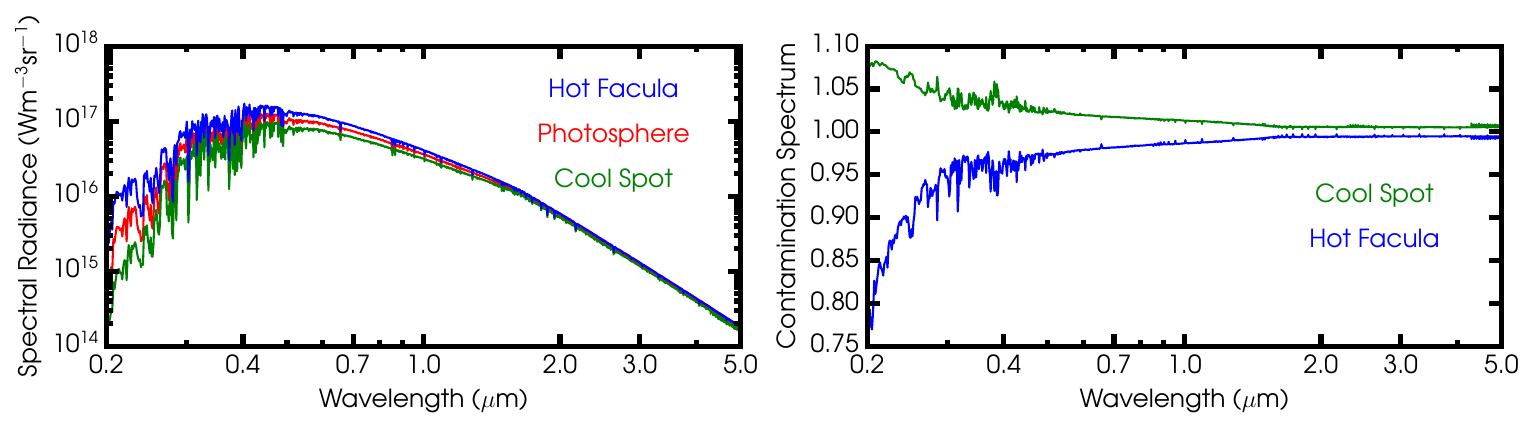}
    \caption{
        Stellar spectral components and their impacts on transmission spectra.
        The left panel shows example model spectra of three primary stellar photospheric components \revised{for the case of HAT-P-1, assuming $T_\mathrm{phot} = 5980$\,K and adding (subtracting) 300\,K for the hot facula (cool spot) spectrum.}.
        The right panel illustrates the multiplicative change in transit depth introduced by unocculted spots and faculae, assuming a 10\% coverage fraction.
        From \citet{Pinhas2018}.
        \label{fig:pinhas2018}
    }
\end{figure*}

When a stellar surface contains heterogeneities residing outside the transit chord (i.e., unocculted spots and faculae), the transmission spectrum of a transiting planet can accrue a wavelength dependence not caused by the planetary atmosphere.
The origin of this effect is a mismatch between the stellar intensity incident on the planetary atmosphere and the average intensity of the full stellar surface (see \autoref{fig:tlse}).
The TLSE can be expressed by the following functional form:
\begin{equation}
D_{\lambda, \, \rm{obs}} = D_{\lambda, \, \rm{atm}} \, \epsilon_{\lambda, \, \rm{het}}
\label{eqn:stellar_contam_1}   
\end{equation}
where $D_{\lambda, \, \rm{obs}}$ is the observed transmission spectrum, $D_{\lambda, \, \rm{atm}} = (R_{p, \lambda}/R_s )^2$ is the intrinsic transmission spectrum of the planetary atmosphere (i.e., the square of the wavelength-dependent planet-to-star radius ratio), and $\epsilon_{\lambda, \, \rm{het}}$ is the wavelength-dependent ``contamination factor'' introduced by the TLSE.
For a stellar surface with a single type of unocculted heterogeneity (i.e., a population of similar spots or, alternatively, faculae), the contamination factor can be written as \citep[e.g.,][]{Rackham2018}
\begin{equation}
\epsilon_{\lambda, \, \rm{het}} = \frac{1}{1 - f_{\rm het} \left(1 - \frac{S_{\lambda, \, \rm{het}} (T_{*, \, \rm{het}})}{S_{\lambda, \, \rm{phot}} (T_{*, \, \rm{phot}})} \right)}
\label{eqn:stellar_contam_2}   
\end{equation}
where $f_{\rm het}$ is the fractional coverage area of the heterogeneous regions, $S_{\lambda, \, \rm{het}}$ and $S_{\lambda, \, \rm{phot}}$ are the specific intensities of the heterogeneity and background photosphere, respectively, with $T_{*, \, \rm{het}}$ and $T_{*, \, \rm{phot}}$ being their corresponding temperatures.
\autoref{fig:pinhas2018} shows example model spectra of $S_{\lambda, \, \rm{het}}$ (left panel) and the contamination spectra $\epsilon_{\lambda, \, \rm{het}}$ (right panel).
We see that the contamination factor deviates from unity significantly at short wavelengths ($\lesssim 1\,\micron$), with transit depths enhanced by unocculted spots and lowered by unocculted faculae.

\subsubsection{Non-retrieval approaches for correcting for stellar spottedness}

Without even turning to retrievals, a rich body of work exists on mitigating the impact of stellar heterogeneities on transmission spectra \citep[see review by][]{Bruno2021}.
In the transmission spectroscopy literature, the primary non-retrieval approach used to account for a heterogeneous stellar photosphere is to apply a direct correction for stellar heterogeneity based on the photometric variability of the star.
Modeling the stellar flux as a linear combination of the flux from the quiescent photosphere $F_{\mathrm{phot}, \lambda}$ and spots $F_{\mathrm{spot}, \lambda}$, which cause a total dimming of $\Delta f(\lambda_0, t)$ at a nominal wavelength $\lambda_0$, \citet{Sing2011} derive an expression for the correction required for the transit depth $D_\lambda$, given by
\begin{equation}
    \frac{\Delta D_\lambda}{D_\lambda} = 
    \Delta f(\lambda_0, t)
    \left(1 - \frac{ F_{\mathrm{spot}, \lambda} }{ F_{\mathrm{phot}, \lambda} }\right) / 
    \left(1 - \frac{ F_{\mathrm{spot}, \lambda_0} }{ F_{\mathrm{phot}, \lambda_0} }\right).
\end{equation}
This implies a concomitant correction to the planet-to-star radius ratio of
\begin{equation}
    \Delta (R_{p, \lambda} / R_s) \simeq
    \frac{1}{2} 
    \frac{\Delta D_\lambda}{D_\lambda}
    (R_{p, \lambda} / R_s).
\end{equation}
\citet{Zellem2017} present a similar approach, relying on the variability of the out-of-transit stellar spectra, which are used to establish the baseline for transmission spectroscopy studies, to infer a per-visit stellar activity correction.

However, as noted by \citet{Berta2011}, who derived a correction similar to that of \citet{Sing2011}, this approach (whether photometric or spectroscopic) relies on the assumption that the maximum brightness in a light curve corresponds to an unspotted photosphere.
This is an increasingly poorer assumption for active stars, which can have many active regions that contribute to a persistent level of spottedness as the star rotates.

These highly spotted stars are those for which corrections for stellar heterogeneity are most important to consider.
\citet{McCullough2014}, for example, note that the strongly sloped transmission spectrum of HD\,189733\,b could be explained by spots alone (without any contribution from haze in the planetary atmosphere) if the total covering fraction of unocculted spots were 5.6\%, not the 1--1.7\% coverage that had been inferred through variability monitoring and previously taken into account \citep{Pont2008, Pont2013, Sing2009, Sing2011}.
From this example and others, we conclude that rotational variability provides a lower limit on the heterogeneity of a stellar photosphere.
Thus, applying a direct correction based on the stellar variability may underestimate the actual correction required.

\subsubsection{Including stellar heterogeneities in retrievals} \label{subsubsec:stellar_het_retrievals}

\begin{figure*}
    \centering
    \includegraphics[width=0.57\textwidth, height=0.42\textwidth]{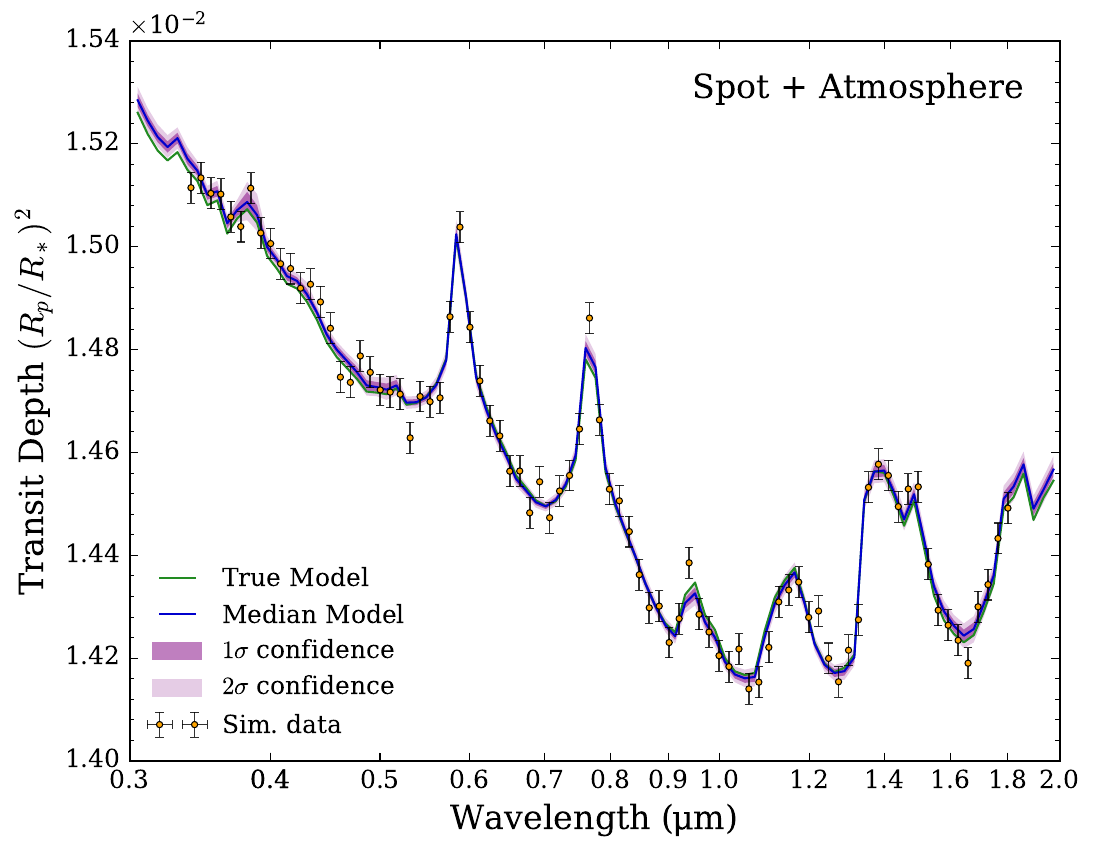}
    \includegraphics[width=0.41\textwidth, height=0.40\textwidth]{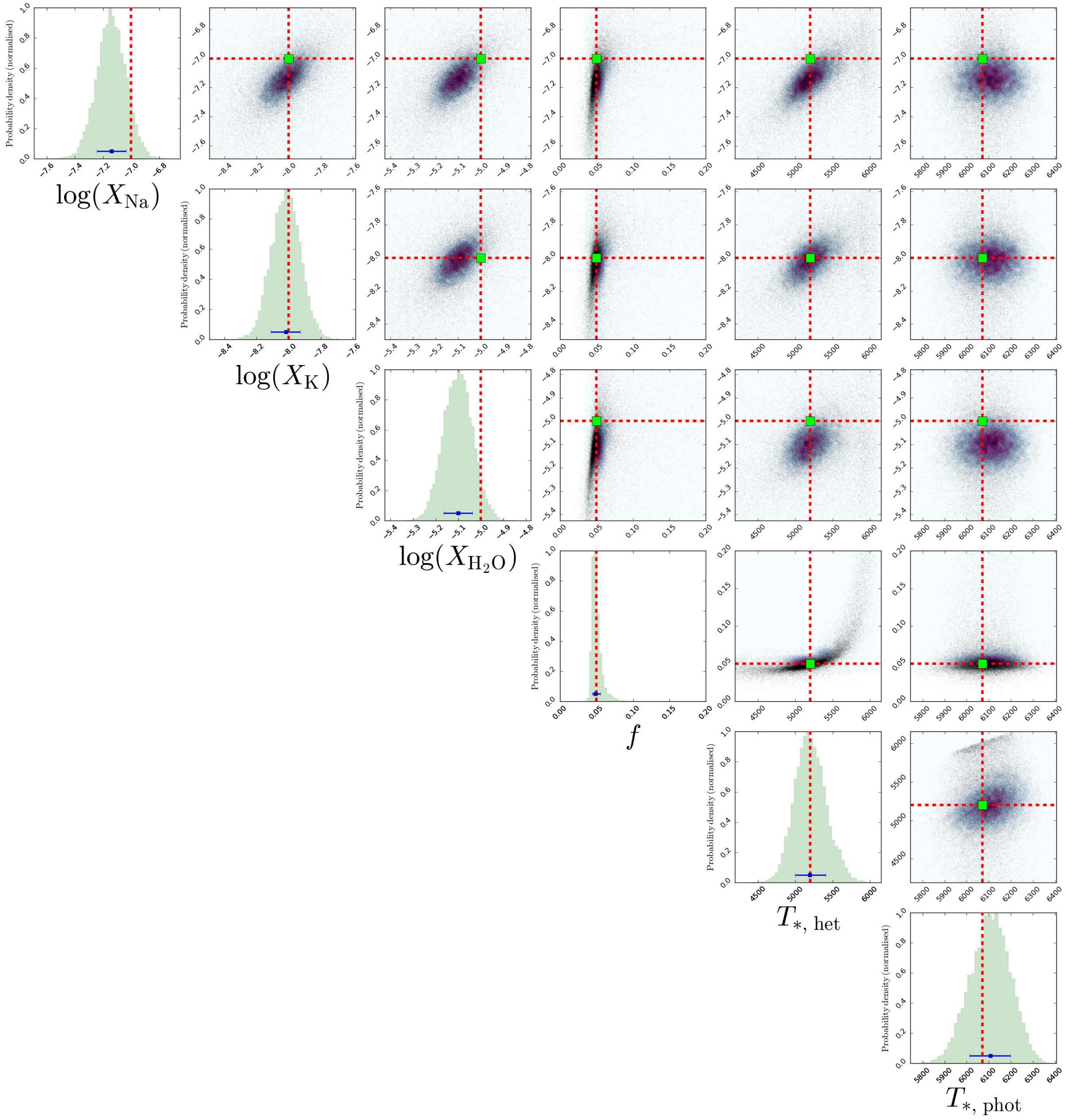}
    \includegraphics[width=0.57\textwidth, height=0.42\textwidth]{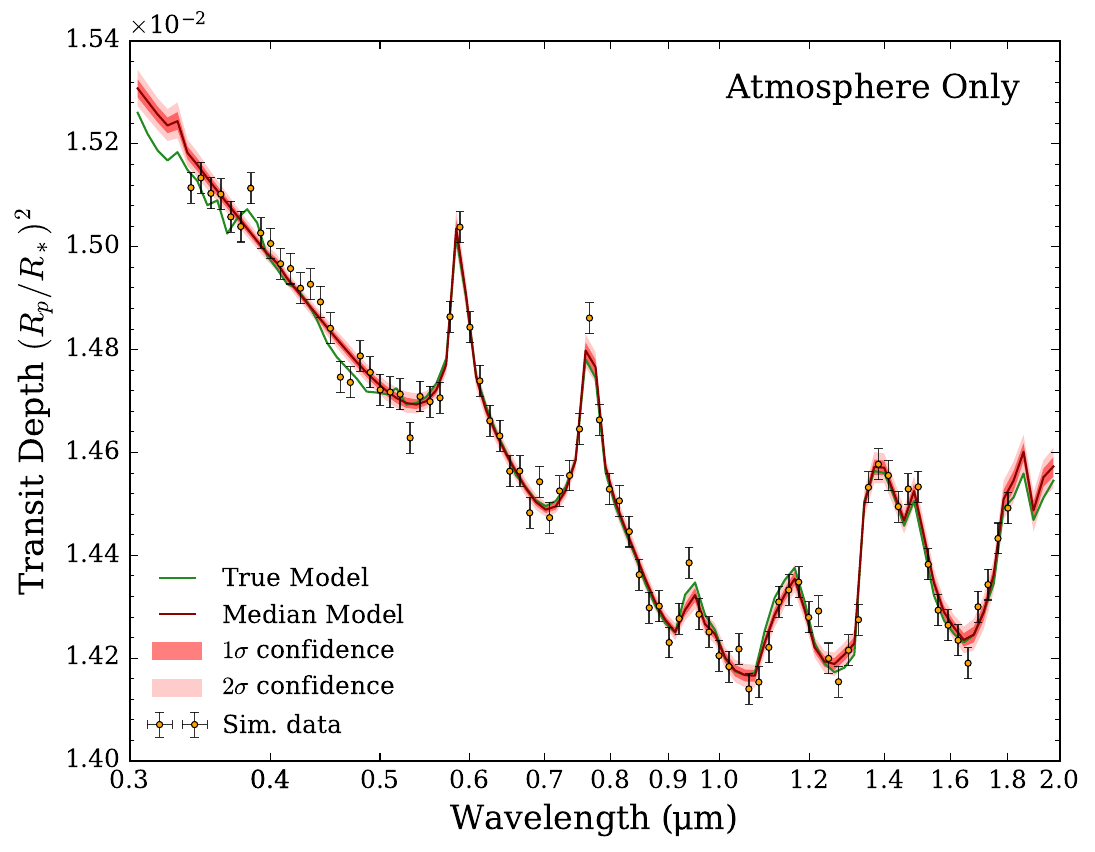}
    \includegraphics[width=0.41\textwidth, height=0.40\textwidth]{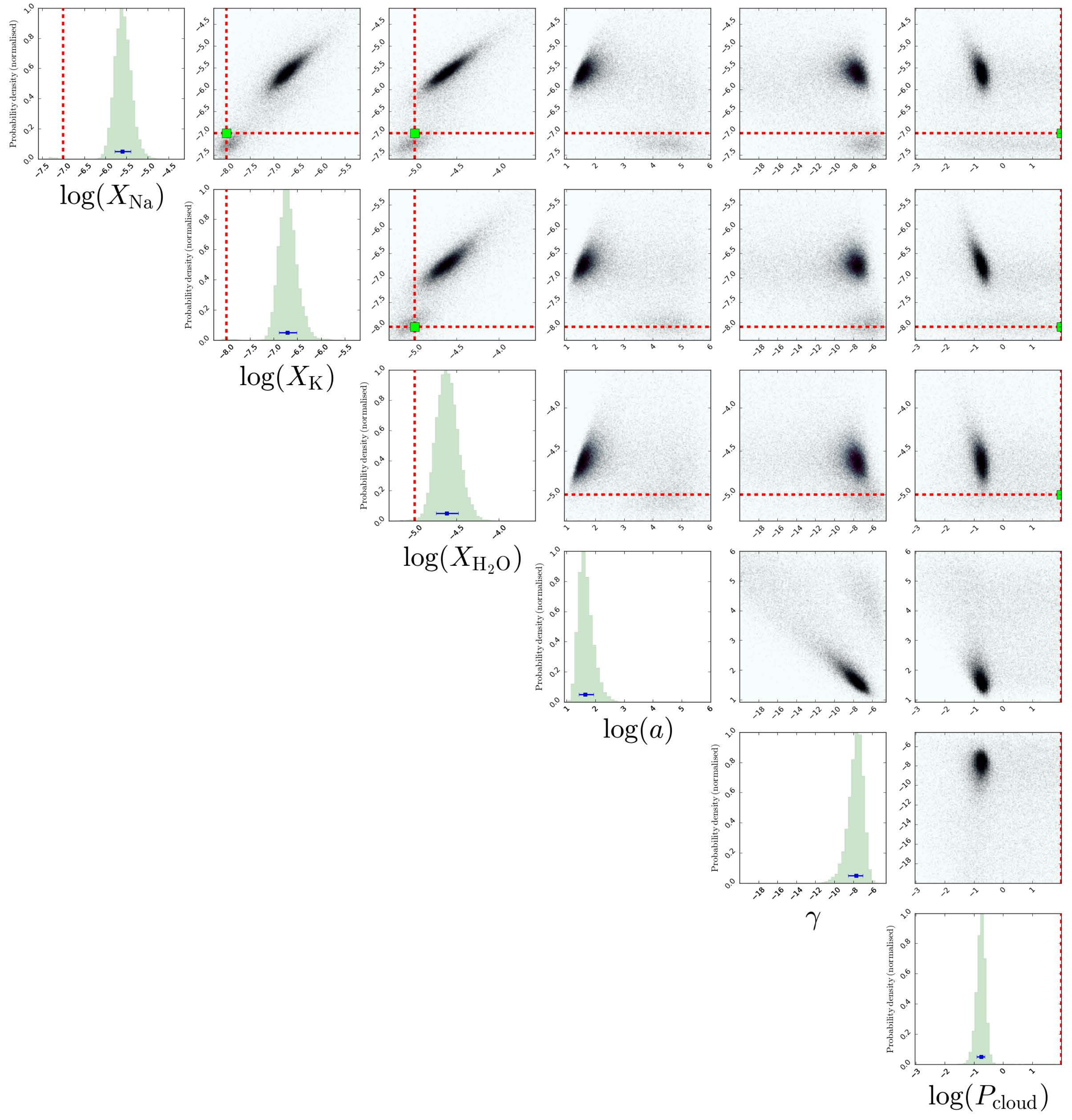}
    \caption{
        The impact of unocculted spots in a retrieval analysis. 
        A simulated hot Jupiter transmission spectrum contaminated by unocculted spots (orange data) is subjected to two retrievals: 
        (i) simultaneous modeling of the planetary atmosphere and unocculted spots (top panels); and
        (ii) the planetary atmosphere only (bottom panels). 
        The posterior distributions (right panels) compare the true planetary atmosphere and stellar properties (red dashed lines) to the $1\,\sigma$ confidence regions for each parameter (blue error bars). 
        The model without spots attempts to compensate by adding a haze slope and a cloud deck, resulting in a similar fit but significantly biased ($\gtrsim 3\,\sigma$) composition inferences. 
        The model including unocculted spots within the retrieval achieves a better fit to the data and successfully recovers the input planetary atmosphere and stellar properties.
        \label{fig:spot_retrieval}
    }
\end{figure*}

By constraining the impact of stellar contamination directly from transmission spectra, retrievals can identify photospheric heterogeneity that is not evident in variability monitoring.
These analyses can fit simultaneously for the properties of the planetary atmosphere and any heterogeneities and appropriately marginalize over the contamination signal while estimating planetary parameters.
Retrieval codes typically accomplish this by adding three additional free parameters encoding the heterogeneous regions \citep[e.g.,][]{Pinhas2018, Rathcke2021}: $f_{\rm het}$, $T_{*, \, \rm{het}}$, and $T_{*, \, \rm{phot}}$.
During parameter space exploration, each pair of $T_{*, \, \rm{het}}$ and $T_{*, \, \rm{phot}}$ are used to compute spectra of the heterogeneity and the photosphere by interpolating grids of stellar models, e.g., the PHOENIX \citep{Husser2013} or Castelli-Kurucz grids \citep{Castelli2003}. 
Given the interpolated stellar spectra and $f_{\rm het}$, the contamination factor is computed using \autoref{eqn:stellar_contam_2} and combined with the atmospheric spectrum as in \autoref{eqn:stellar_contam_1}. 
The combined transmission spectrum is then compared with the observations at each point in the parameter space. 

We provide an example retrieval analysis in \autoref{fig:spot_retrieval}.
We generated a simulated transmission spectrum of a hot Jupiter (with planet properties based on HD\,209458b) transiting a star with unocculted spots.
For this example, we assumed 5\% spot coverage with a spot temperature 900\,K colder than the photosphere.
The unocculted spots produce a steep slope at visible wavelengths and secondary features imprinted by stellar absorption. 
The simulated data cover 0.34--1.80\,$\micron$, roughly corresponding to the range of \HST{}/STIS+WFC3, at a precision of 30\,ppm.
We subjected this dataset to two retrievals with the POSEIDON code \citep{MacDonald2017_hd209458b} under different assumptions: (i) unocculted spots are included within the retrieval; and (ii) no spots are considered, only the planetary atmosphere shapes the spectrum.
We find that the atmosphere-only retrieval attempts to account for the lack of spots by adding a haze (producing a similar optical slope) and a cloud deck.
While both models result in an acceptable fit, the atmospheric properties are significantly biased when unocculted spots are not considered: the Na and K abundances are biased by $\sim$5$\sigma$, the H$_2$O abundance is biased by $\sim$3$\sigma$, and one would incorrectly conclude that the atmosphere has potent aerosols.
Conversely, the retrieval including unocculted spots correctly recovers all input parameters within the 1--2$\,\sigma$ expected from Gaussian scatter.
The Bayesian evidence also prefers the model including spots ($\Delta \ln Z = 11.6$), showing that a retrieval code can differentiate between these scenarios.
This exercise demonstrates that biases in atmospheric properties can be mitigated by including stellar heterogeneities in retrievals.

\begin{figure}
    \centering
    \includegraphics[width=\columnwidth, height=\textheight, keepaspectratio]{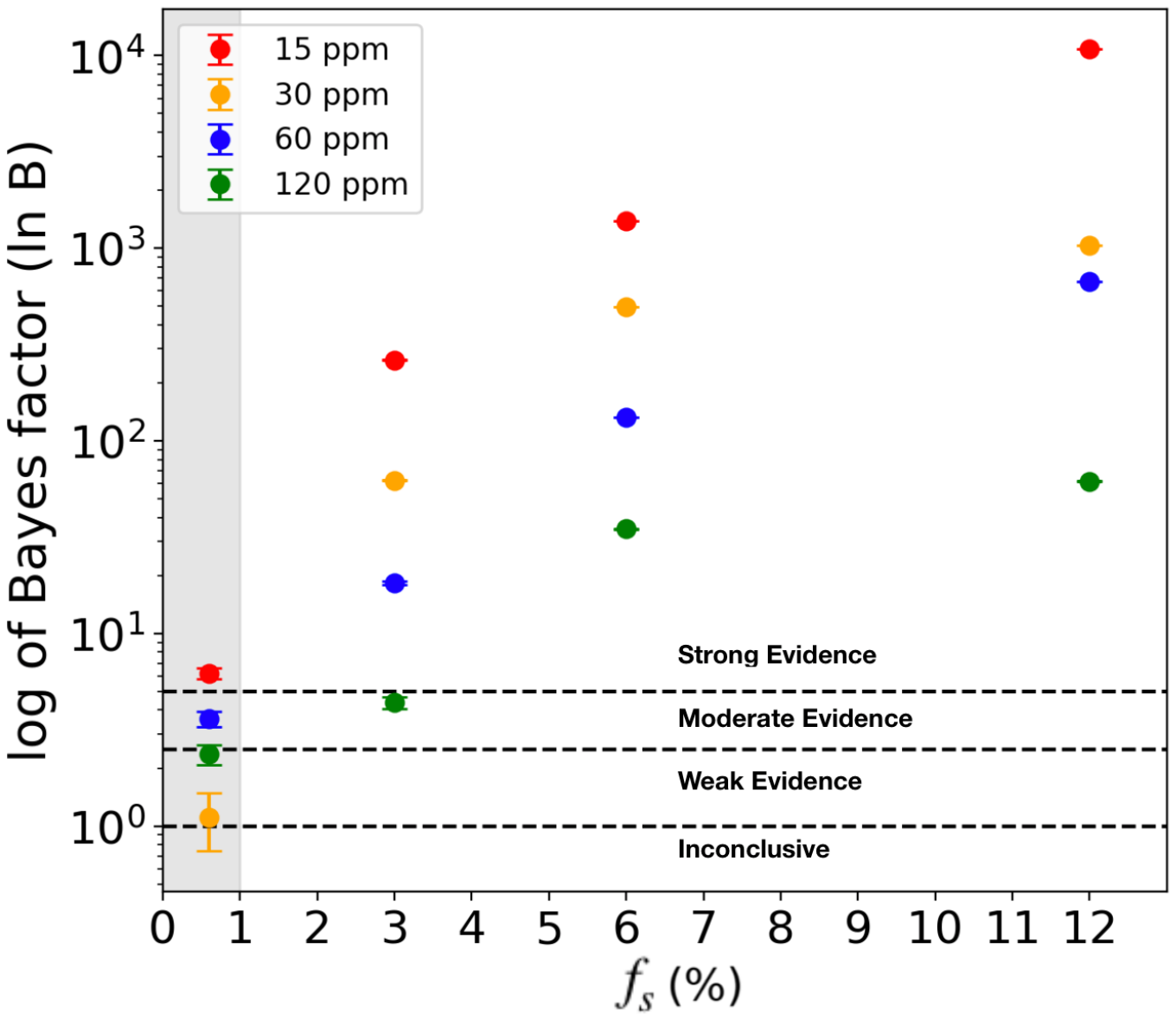}
    \caption{
        The statistical preference for including the TLSE in retrievals of synthetic \JWST{}/NIRISS data as a function of spot covering fraction. 
        In this exercise, \citet{Iyer2020} performed retrievals on the transmission spectrum of a sub-Neptune orbiting an M-dwarf with unocculted spots, varying the data precision from 15 to 120\,ppm (point colors) and the spot covering fractions from 0.6$\%$ to 12$\%$. 
        The black dashed lines indicate the degree of statistical preference for the TLSE-correction-included retrieval model \citep{Trotta2008}. 
        The overall trend is that the model including the TLSE is increasingly preferred for better precision and for spot covering fractions above 1$\%$. 
        Below 1$\%$ spot coverage, the biases from ignoring the correction are not as noteworthy (except for 15\,ppm precision, where the TLSE model is still strongly preferred). 
        Such small spot coverage (a low evidence regime) also incurs inaccuracies in computing Bayes factors, causing the crossing over of the 30\,ppm point \citep[e.g.,][]{Lupu14}.
        From \citet{Iyer2020}.
        \label{fig:iyer_2020_bayesfactor}
    }
\end{figure}

The confidence with which unocculted stellar spots or faculae can be identified is a function of spectral precision, coverage fraction, and wavelength range. 
Using simulated \JWST{}/NIRISS data and Bayesian model comparisons, \citet{Iyer2020} show how the level of preference for stellar contamination varies with these factors for typical sub-Neptunes orbiting M-dwarfs.
For large covering fractions of unocculted spots (above 1$\%$, see \autoref{fig:iyer_2020_bayesfactor}), the Bayes factor increasingly favors a model including the proper TLSE correction across spectrophotometric precisions from 15 to 120 ppm. 
For small spot covering fractions (below 1$\%$), the TLSE signal is evident in high-fidelity \JWST{}/NIRISS data (${\sim}$15 ppm precision). 
However, it may be harder to identify stellar contamination for small spot coverage at more typical precisions ($\gtrsim$ 30 ppm). 
\cite{Iyer2020} draw consistent conclusions for including the TLSE in retrievals---in their case, for \JWST{}/NIRISS data of sub-Neptunes transiting M-dwarfs---compared to those we draw from the analysis of simulated \HST{}/STIS+WFC3 data of a hot Jupiter transiting a G dwarf in \autoref{fig:spot_retrieval}.
Notwithstanding these analyses, there is considerable scope for future work to further quantify when the inclusion of stellar contamination within a retrieval framework is necessary, e.g., considering the sensitivity of such inferences to other instruments with different precisions, wavelength coverage, and spectral resolutions.  

The inclusion of unocculted stellar heterogeneities within retrievals is a relatively new addition to the literature.
Consequently, approaches can vary between different studies and a ``best practice'' consensus has yet to emerge.
Four of the main differences are as follows.

\begin{enumerate}
    \item \textbf{Free parameters}: A single heterogeneity can be described by three free parameters ($f_{\rm het}$, $T_{*, \, \rm{het}}$, and $T_{*, \, \rm{phot}}$, see \citealt{Pinhas2018}), or a subset can be fixed \citep[e.g.,][]{Bruno2021}. 
    Additional parameters could be added for multiple heterogeneities.
    When such models \citep[e.g., including both spots and faculae;][]{Zhang2018} are necessary is unclear and worthy of future study.
    \item \textbf{Out-of-transit stellar spectra}: Recent studies have leveraged baseline, out-of-transit stellar spectra in real and simulated datasets to fit for stellar heterogeneity parameters, either while simultaneously modeling their impact on transmission spectra or simply gathering prior information for analyses of transmission spectra \citep{Zhang2018, Wakeford2019, Iyer2020, Cracchiolo2021}.
    It is unclear whether this approach provides an additional advantage over analyses of transmission spectra alone.
    \item \textbf{Priors}: Some studies ascribe Gaussian priors to $T_{*, \, \rm{phot}}$ \citep[e.g.,][]{Pinhas2018,Rathcke2021}, encoding \textit{a priori} knowledge on the stellar $T_{\rm eff}$, while others use uniform priors \citep[e.g.,][]{Iyer2020}. 
    \item \textbf{Stellar grids}: Many model stellar grids are available, with different grids implemented in different retrieval codes. 
    We consider the impact of this further in \autoref{subsec:retrieval_limitations}. 
\end{enumerate}

Alongside these differences, other common assumptions include: fixing the metallicity and surface gravity of any heterogeneities to those of the photosphere (both set to \textit{a priori} values) and, as noted previously, the consideration of only one heterogeneity. 
The variety of approaches in the literature to date serves to illustrate that more work on retrievals including stellar heterogeneities will be beneficial.

\subsubsection{Existing spectra are impacted}

Recent studies employing retrievals have found mounting evidence for unocculted heterogeneities impacting transmission spectra, in many cases at a level beyond what was previously inferred via variability measurements.
\citet{Pinhas2018} reanalyzed nine hot Jupiter transmission spectra from \HST{} and \Spitzer{} \citep{Sing2016}, finding moderate evidence (3.1$\sigma$) for stellar heterogeneity in the spectrum of WASP-6b, beyond that previously estimated to be present from variability measurements \citep{Jordan2013, Nikolov2015}.
Another three hot Jupiters show moderate to weak suggestions of stellar heterogeneity: WASP-39b (2.5$\sigma$), HD\,209458b (1.6$\sigma$), and HAT-P-12b (1.4$\sigma$).
\citet{Rathcke2021} studied the optical and NIR transmission spectrum of the hot Jupiter WASP-79b and inferred the presence of unocculted faculae (4.7$\sigma$) covering $\sim$15\% of the disk of the F5V host star with a temperature contrast of $\sim$500\,K.
Similarly, \citet{Kirk2021} studied the optical and NIR transmission spectrum of the ultrahot Jupiter WASP-103b, notably using 11 optical transits that paint a consistent picture between epochs and across four instruments.
From this spectrum, which shows a marked decrease in transit depth at blue-optical wavelengths, they inferred the presence of unocculted faculae (4.1$\sigma$) covering $\sim$20\% of the stellar disk with a temperature contrast of $\sim$400\,K.
In these examples, the impact of stellar heterogeneity on the transmission spectra was either underestimated in previous studies applying direct corrections or may have been missed if not for the use of retrievals, which points to the utility of this approach for identifying and disentangling signals from stellar heterogeneity and planetary atmospheres in transmission spectra.

\begin{figure*}
    \centering
    \includegraphics[width=\textwidth, height=\textheight, keepaspectratio]{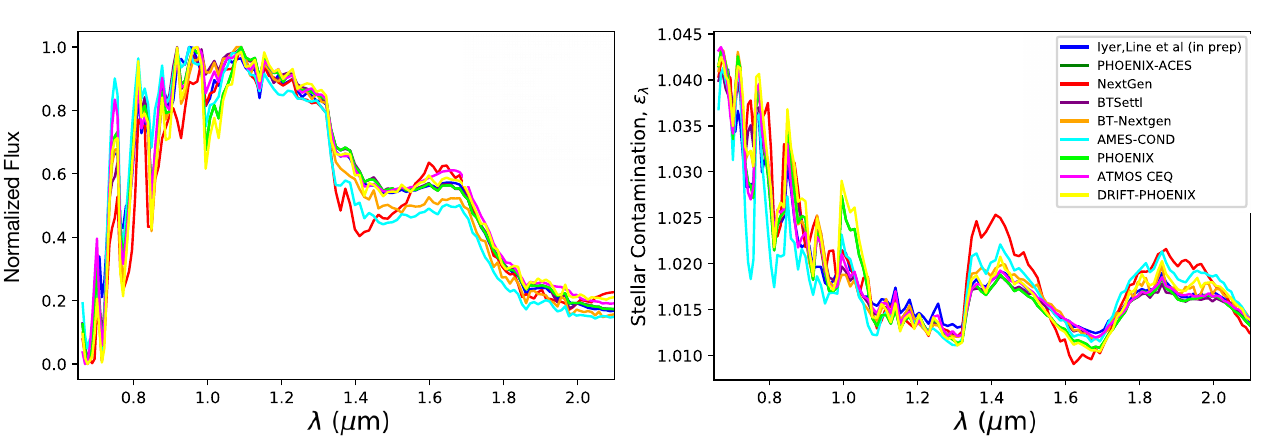}
    \caption{
        Disagreements among model stellar spectra and their resulting contamination factors. 
        \textit{Left}: Spectra from popular stellar model grids of an M dwarf with $T_\mathrm{eff} = 3000$\,K, $\log g=5.0$, and [Fe/H]$=0.0$. 
        \textit{Right}: The stellar contamination factor that influences planetary spectra via the TLSE (\autoref{eqn:stellar_contam_2}) constructed from these stellar models, assuming a fractional coverage of unocculted spots on the photosphere of $f_s = 5\%$ and a spot temperature of $T_\mathrm{s} = 2700$\,K. 
        \revised{Adapted from \citet{Iyer2023}}.
        \label{fig:iyer_spectra}
    }
\end{figure*}

\subsubsection{Our finding}

From this analysis, we draw the following finding.
\\
\\
\noindent \underline{\textbf{Finding 4.1}}
\\
\\
\noindent \textit{Summary}:
Retrievals of transmission spectra that include the effects of unocculted active regions can guard against biases.
More work is needed to understand when these retrievals are necessary and what are the limitations and best practices of this approach. \\
\\
\noindent \textit{Capability Needed}:
A detailed understanding is needed of when to marginalize over stellar heterogeneity in retrievals of planetary atmospheric parameters and the best practices for doing so.
\revised{A benchmarking of the available retrieval tools is also needed.}
\\
\\
\noindent \textit{Capability Today}:
Multiple groups have developed retrievals that use simple parameterizations of unocculted heterogeneities and rely on \revised{1D} stellar models \revised{for spectra of both the quiescent photosphere and active regions \citep[e.g.,][]{Rackham2017, Pinhas2018, Zhang2018, Zhang2019_platon, MacDonald2023}}.
\\
\\
\noindent \textit{Mitigation in Progress}:
\revised{
Retrieval frameworks that marginalize over stellar heterogeneity are commonly used in modern analyses of transmission spectra \citep[e.g.,][]{Espinoza2019, Kirk2021, Rathcke2021}.
Applications to simulated \JWST{} data show that marginalizing over potential stellar impacts is useful for avoiding biases in retrieved planetary properties introduced by photospheric heterogeneity \citep{Iyer2020}.
Ongoing work is focused on applying these frameworks to \JWST{} transmission spectra to understand the best practices when working with these precise observations (Rackham et al., in prep.).
}

\subsection{Limitations of the approach}
\label{subsec:retrieval_limitations}

We next consider the known limitations of the retrieval approach, focusing on the availability of suitable model spectra for stellar magnetic active regions.
In reaching our finding, we considered the disagreement among current model stellar spectra and the related impact of employing different stellar spectra on retrieved planetary properties as well as recent advances in modeling spectra of stellar magnetic active regions.

\subsubsection{Disagreement among model stellar spectra and its impact on retrievals}

\begin{figure*}
    \centering
    \includegraphics[width=\textwidth, height=0.5\textheight, keepaspectratio]{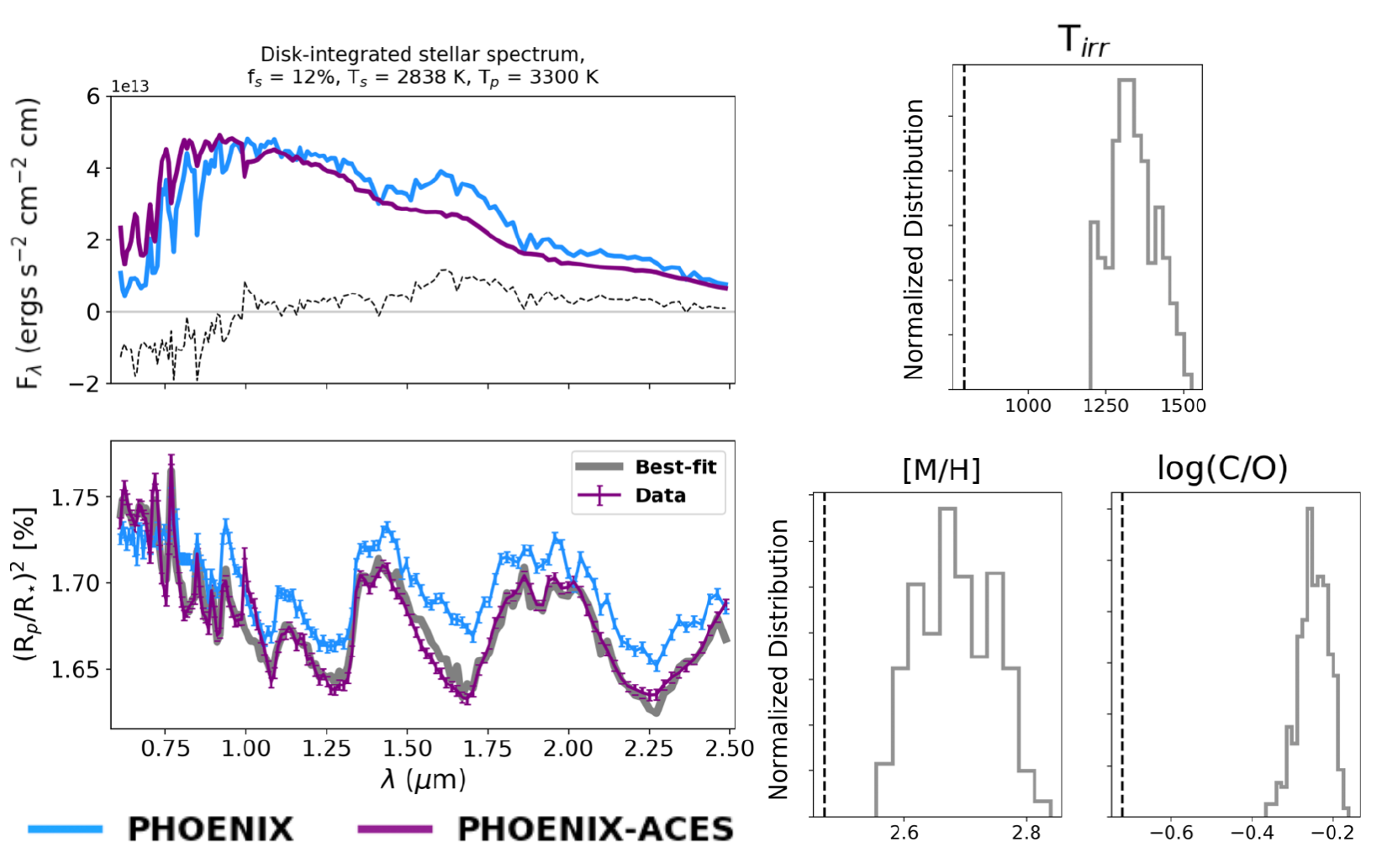}
    \caption{
        The impact of disagreements between stellar models on retrievals.
        Here \citet{Iyer2020} simulated a dataset contaminated by unocculted spots using the PHOENIX-ACES model grid \citep{Husser2013} but retrieved it using the PHOENIX model grid \citep{Allard2003}. 
        They find significant biases in the retrieved values of the atmospheric irradiation temperature $T_\mathrm{irr}$, metallicity [M/H], and carbon-to-oxygen ratio $\log$\,(C/O).
        \textit{Top Left}: disk-integrated stellar spectra from both model grids and their residuals.
        \textit{Bottom Left}: contaminated transmission spectra in both cases and the retrieved spectrum.
        \textit{Right}: posterior distributions of the atmospheric parameters compared to the true model inputs (dashed lines).
        From \citet{Iyer2020}.
        \label{fig:iyer_2020_stellar_spectra}
    }
\end{figure*}

A primary limitation of retrieval approaches for handling stellar contamination is their reliance on stellar models, which vary substantially in their predictions of the spectra of cool stars.
Substantial variations in spectral shape among widely used stellar models (\autoref{fig:iyer_spectra}, left panel) owe to a variety of factors, including the choice of opacity data included and spectral line list completeness; treatment of line-broadening effects; and physical assumptions, such as plane-parallel vs.\ spherical geometry and \revised{local thermodynamic equilibrium (LTE)} vs.\ non-LTE chemistry conditions. 
The variation is more pronounced for mid-to-late M stars (\revised{$T_\mathrm{eff}$}$<$3500\,K) than for FGK stars due to molecular opacities dominating such atmospheres and photospheric surface heterogeneities that are common in active cool stars \citep{Hawley1991, Hawley2014, Schmidt2014}.
As a result, the choice of stellar models used while correcting for TLSE signals can impact planetary properties retrieved from transmission spectra.

This impact has been explored quantitatively in a recent study by \citet[][\autoref{fig:iyer_2020_stellar_spectra}]{Iyer2020}.
These authors used the PHOENIX-ACES model grid \citep{Husser2013} to mimic a ``true'' disk-integrated stellar spectrum of an M dwarf ($T_\mathrm{phot} = 3300$\,K) contaminated by unocculted spots ($T_s = 2838$\,K) covering a fractional area f$_{s}$ of 12$\%$.
They performed a retrieval on a simulated transmission spectrum with this contamination, using the PHOENIX model grid \citep{Allard2003} to calculate the TLSE contribution to the transmission spectrum. 
They find the fit reasonably explains the shape of the transmission spectrum.
However, the induced variations in the best-fit spectrum are probably due to false opacity signatures or incorrect estimates of the thermal profile---evident in the form of bias in the posterior probability distributions of the planetary irradiation temperature T$_\mathrm{irr}$, metallicity [M/H], and carbon-to-oxygen ratio $\log$\,(C/O) (\autoref{fig:iyer_2020_stellar_spectra}, right panel). 
This exercise highlights that, despite an appropriate implementation of the TLSE, our ability to correct for the contribution of stellar contamination heavily depends on the accuracy of the stellar models used to represent the contaminated stellar photosphere---particularly for cooler host stars (K- and M-type)---thereby emphasizing the need for improved stellar atmosphere models.

\subsubsection{Advances in modeling spectra of active regions}

\begin{figure*}
    \centering
    \includegraphics[width=0.475\textwidth, height=0.5\textheight, keepaspectratio]{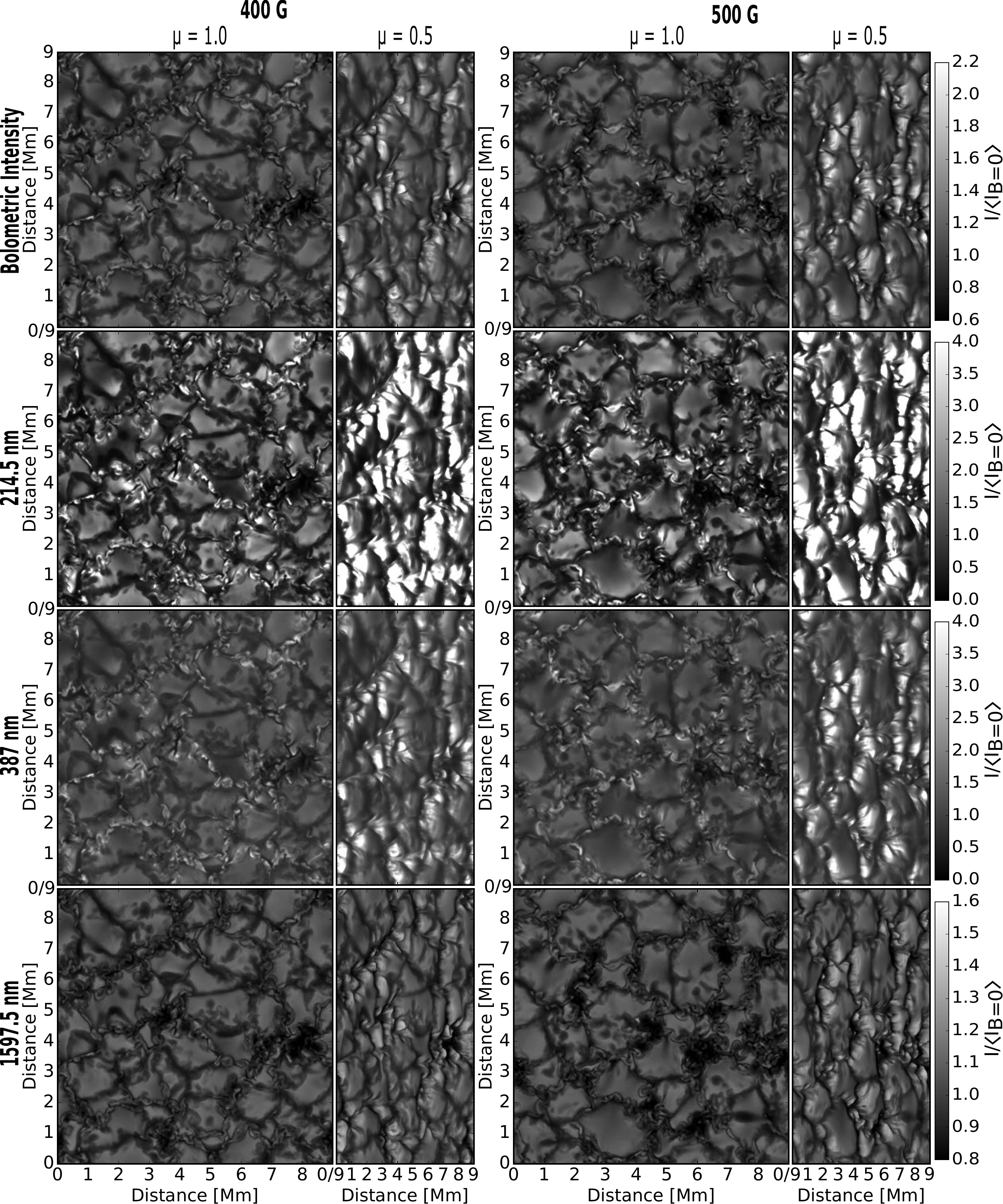}
    \hfill
    \includegraphics[width=0.475\textwidth, height=0.5\textheight, keepaspectratio]{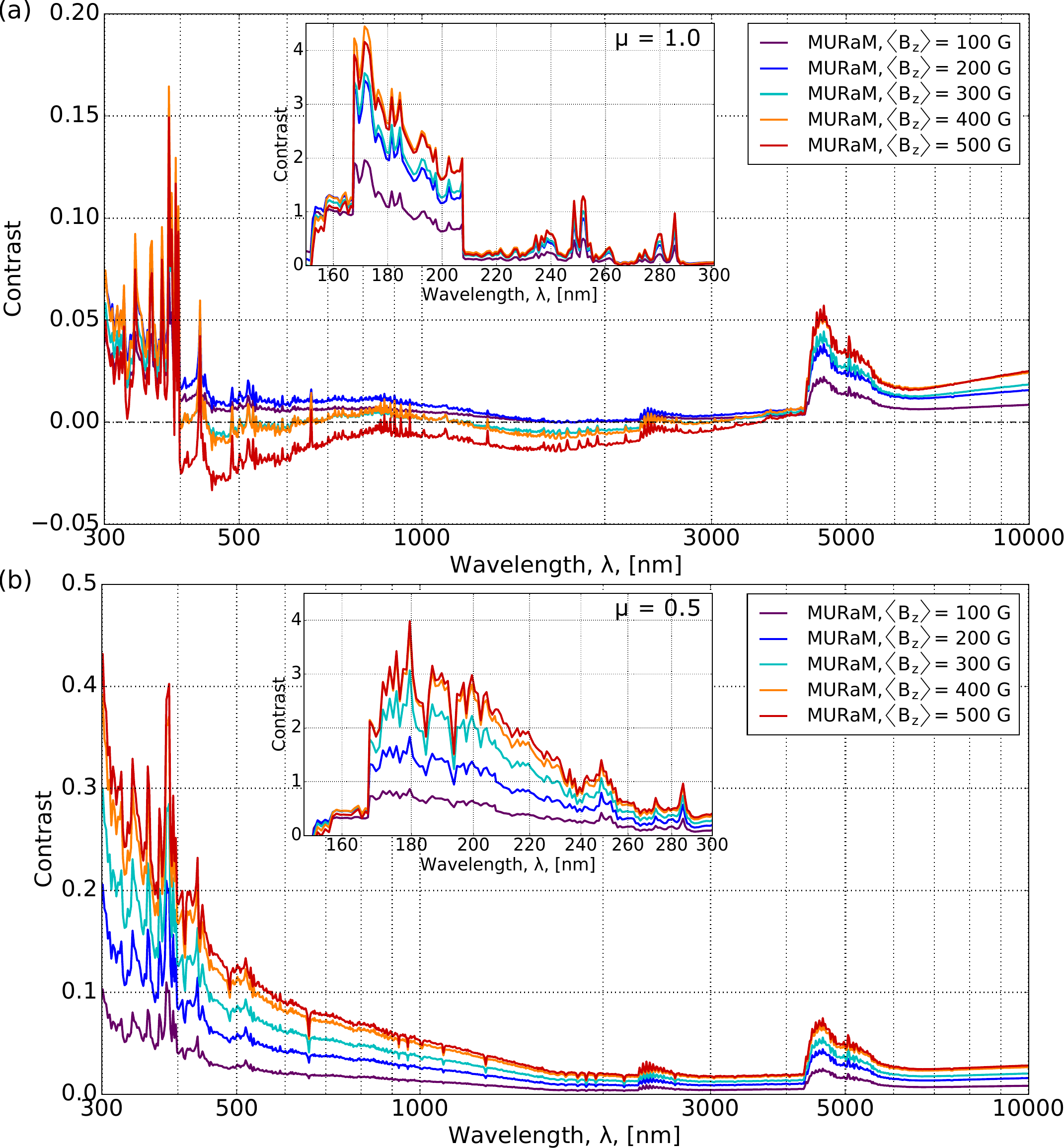}
    \caption{
        Simulations of faculae with the 3D radiation-magnetohydrodynamics code MURaM \citep{Vogler2005} and their derived contrast spectra.
        \textit{Left}: normalized emergent intensities from a simulated G2 atmosphere at different viewing angles, wavelengths, and for two average magnetic field strengths (400\,G and 500\,G).
        Faculae have markedly higher contrasts near the limb and at shorter wavelengths.
        \textit{Right}: average facular contrast spectra for G2 simulated atmospheres with differing field strengths, both at disk center (top right) and nearer to the limb (bottom right).
        The spectra show different structures than what are currently captured by approximations of facular spectra with disk-integrated stellar spectra.
        The center-to-limb variations shown here are also not captured by current approximations of faculae in retrievals.
        From \citet{Norris2017}.
        \label{fig:norris_2017_faculae}
    }
\end{figure*}

Facular and sunspot contrasts employed by retrieval models are typically derived from 1D, plane-parallel, semi-empirical model atmospheres \citep[e.g.,][]{Vernazza:1981,Kurucz:1992a,Kurucz:1992b,Kurucz:1992c,Fontenla:1993,Fontenla:1999,Fontenla:2002,Fontenla:2006}.
For faculae in particular, such models are adjusted to agree with solar observations \citep{Vernazza:1981}.
However, for stars other than the Sun, such observations are not available, and thus adjustment of the models is not possible. 
Furthermore, while such semi-empirical models capture the overall disc-integrated properties of faculae reasonably well, they do not recover the observed center-to-limb variation (CLV) of the facular contrast \citep[e.g.,][]{yeo2013}.

More realistic CLV of facular contrasts is provided by 3D \revised{MHD} simulations, as discussed in \autoref{S:PhotosphericHeterogeneity}.
The first 3D radiative \revised{HD} simulations for stars were reported by \citet{Nord_Dravins_90A}.
The magnetic field was introduced in such simulations more recently, and by now a series of models for stars of spectral types F to M with different mean vertical magnetic fields is available
\citep{Beeck_2011,Beeck_2014,Beeck3,Wedemeyer2013,Steiner2014,co5bold_salhab}.
\citet{Norris2017} and \citet{Johnson2021} used the 3D MURaM atmospheres from \citet{Beeck3} to calculate the CLV of the facular contrast for G2 (see \autoref{fig:norris_2017_faculae}), K0, M0, and M2 stars with the mean vertical magnetic field ranging between 100\,G and 500\,G. 
\citet{Norris2017} found the computed CLV profiles of a G2 star to be in good agreement with solar observations \citep{Neckel1994}.
More recently, simulations by \citet{Johnson2021} have clearly demonstrated a strong effect of faculae on the mean brightness and shape of stellar light curves, pointing to the need for more accurate facular models.
Using a self-consistent approach and 1D model atmospheres, \citet{witzke2018} have additionally shown that facular contrast increases with stellar metallicity, underscoring the importance of accurate fundamental stellar parameters when modeling stellar magnetic activity.

In the case of spots, 3D MHD simulations using the MURaM code have been largely successful in reproducing the fundamental observed features of spots, both on the Sun and other stars.
\citet{Rempel:2009Sci, rempel09a} have produced MURaM simulations of sunspots and found good agreement with observations.
Recently, \citet{Panja-2020} have used the MURaM code to perform the first \textit{ab initio} simulations of spots on cool main-sequence stars other than the Sun.
Their simulations for G2V, K0V, and M0V stars provide umbral and penumbral effective temperatures for these stars, along with other fundamental parameters of starspots, such as the brightness relative to the stellar surface and magnetic field strength.
They conclude that the trend of increasing temperature difference with photospheric temperature that they observe is consistent with observations of starspots.

Taken together, these studies highlight recent advances in modeling spectra of magnetic active regions and underscore the existing limitations.
While 1D models, such as those used to derive template spectra for state-of-the-art retrievals of transmission spectra, generally reproduce well the observed features of spots, they fail to capture aspects of facular contrasts that can impact transit observations.
In particular, the CLV of the facular contrast and the dependence of facular contrast on stellar metallicity are important in this context but not addressed in modern retrievals.
Recent work with 3D MHD simulations can be leveraged to improve template spectra for spots and faculae, though the output of these realistic stellar models has yet to be incorporated into exoplanetary retrieval analyses.

\subsubsection{Our finding}

From this analysis, we draw the following finding.
\\
\\
\noindent \underline{\textbf{Finding 4.2}}
\\
\\
\noindent \textit{Summary}:
Retrieval approaches rely on stellar models, and thus their accuracy is limited by model fidelity \revised{and, depending on the approach, the prior information on stellar parameters used in the retrieval}.
Further efforts to develop model spectra for spots and faculae and incorporate them into exoplanetary atmospheric retrievals are needed.
\\
\\
\noindent \textit{Capability Needed}:
More work is needed to (1) further test and develop models for cool stars,  (2) assess the impact of using stellar spectra to approximate active regions, and (3) develop model spectra for active regions, particularly faculae, for different spectral types.
\revised{Studies should also investigate the impact of simple parameterizations that neglect the viewing-angle dependence of spot and facula spectra.}
\\
\\
\noindent \textit{Capability Today}:
Retrieval analyses today rely on stellar spectra derived from 1D models \revised{\citep[e.g.,][]{Husser2013}} as templates for spot and \revised{facula} spectra.
\revised{
Recent retrieval frameworks incorporate the option using multiple stellar model grids for generating component spectra \citep[e.g., quiescent photosphere, spots, and faculae][]{MacDonald2023}.
Simple parameterizations of signals introduced by unocculted heterogeneities do not incorporate any viewing-angle dependence of spot and facula spectra \citep[e.g.,][]{Rackham2017}.
}
\\
\\
\noindent \textit{Mitigation in Progress}:
Recent studies have produced \textit{ab initio} spot and \revised{facula} spectra from 3D MHD models for a few FGKM spectral types \revised{\citep{Norris2017, Panja-2020}}.
These studies reveal spectral differences with respect to spectra from 1D stellar spectral models that underscore the limitation of this approach, particularly for faculae \revised{\citep{Witzke2022}}.

\subsection{The complementarity of short-wavelength observations}
\label{subsec:retrievals_blue}

\begin{figure*}
    \centering
    \includegraphics[width=0.475\textwidth, height=0.5\textheight, keepaspectratio]{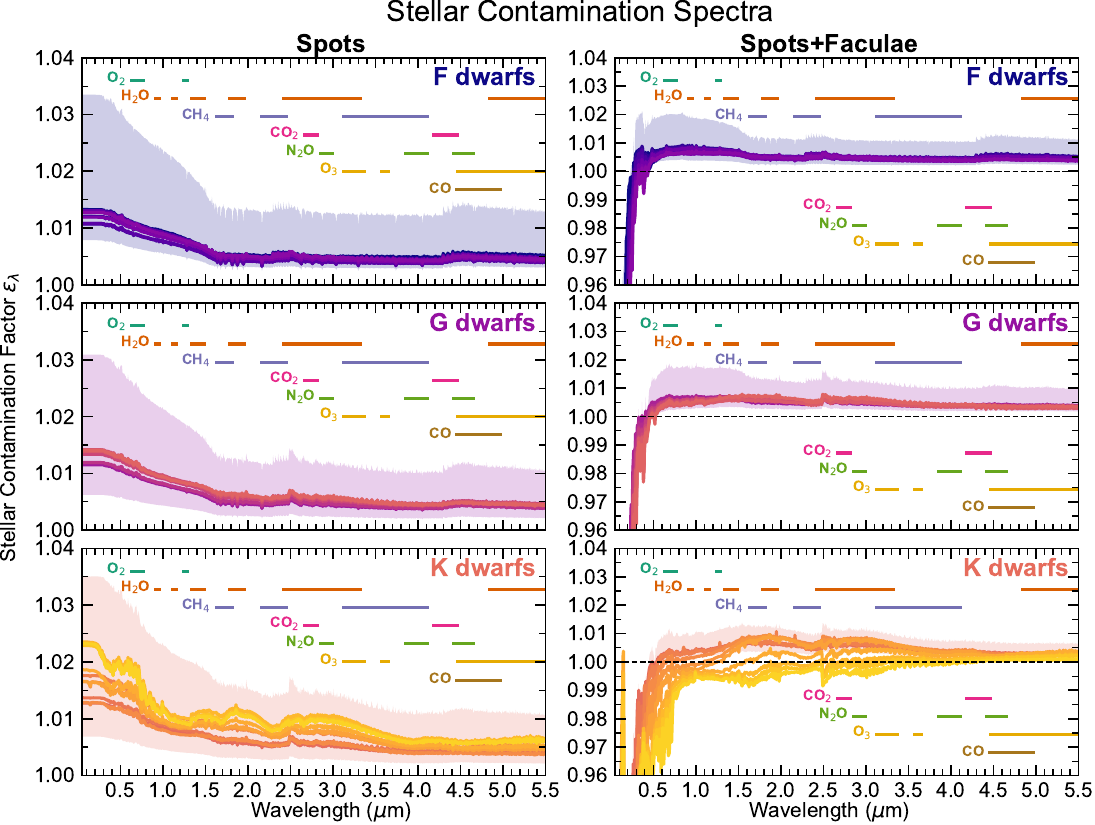}
    \hfill
    \includegraphics[width=0.475\textwidth, height=0.5\textheight, keepaspectratio]{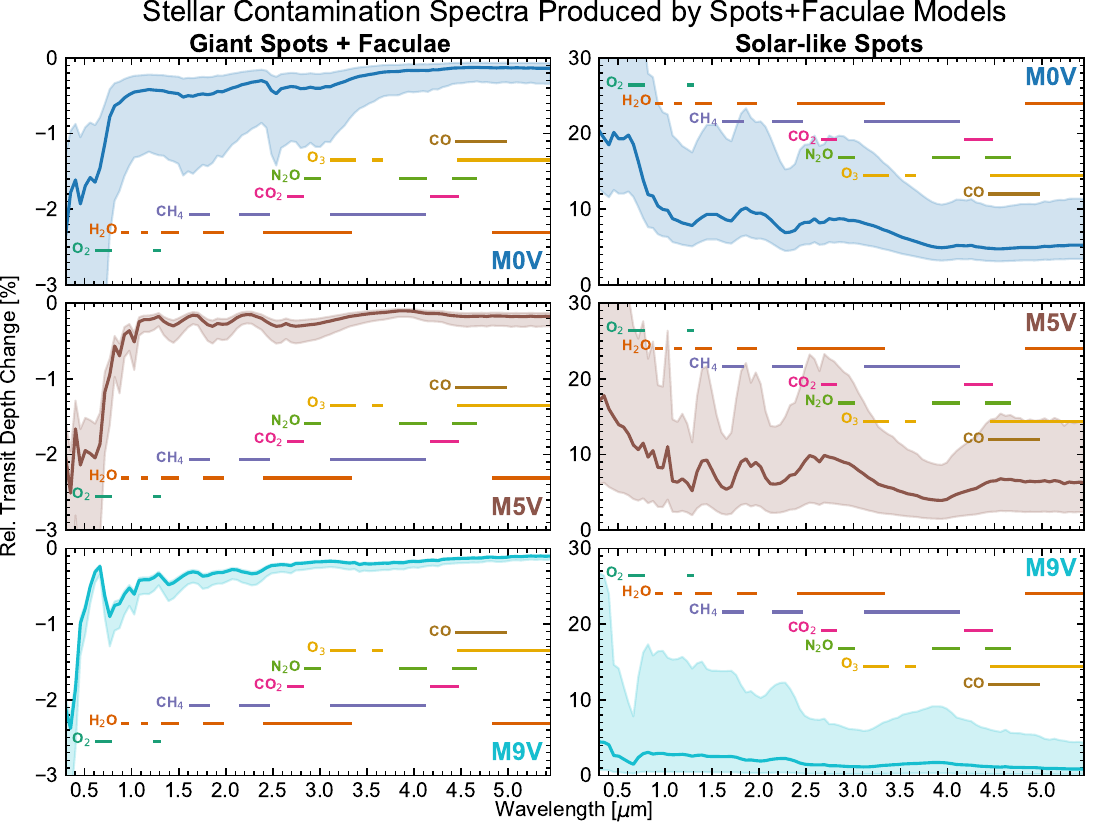}
    \caption{
        Estimated contamination spectra due to spots and faculae on FGKM dwarfs.
        Using spot and \revised{facula} covering fractions inferred from rotational variabilities, forward modeling suggests that typical impacts of stellar contamination will be larger for cooler stars, though considerable uncertainty exists as to the relative impacts of spots and faculae.
        In any case, spectral contamination due to stellar photospheric heterogeneity is predicted to be largest at short wavelengths (${<}0.6$\,\micron).
        From left to right, the columns show the spectral contamination from unocculted spots on FGK dwarfs, spots and faculae on FGK dwarfs, spots and faculae on M dwarfs, and spots on M dwarfs.
        From \citet{Rackham2018, Rackham2019}.
        \label{fig:est_cont_spectra}
    }
\end{figure*}

The third and final general aspect we considered is the complementarity of short-wavelength observations to \HST{} and \JWST{} NIR observations for inferring stellar photospheric properties.
In reaching our finding, we considered forward-modeling efforts that suggest the impacts of heterogeneous stellar photospheres are largest at UV and blue-optical wavelengths and recent studies that either point to stellar contamination in short-wavelength transit data or increase the model preference for a heterogeneous stellar photosphere by including optical data.

\begin{figure*}
    \centering
    \includegraphics[width=\textwidth, height=0.5\textheight, keepaspectratio]{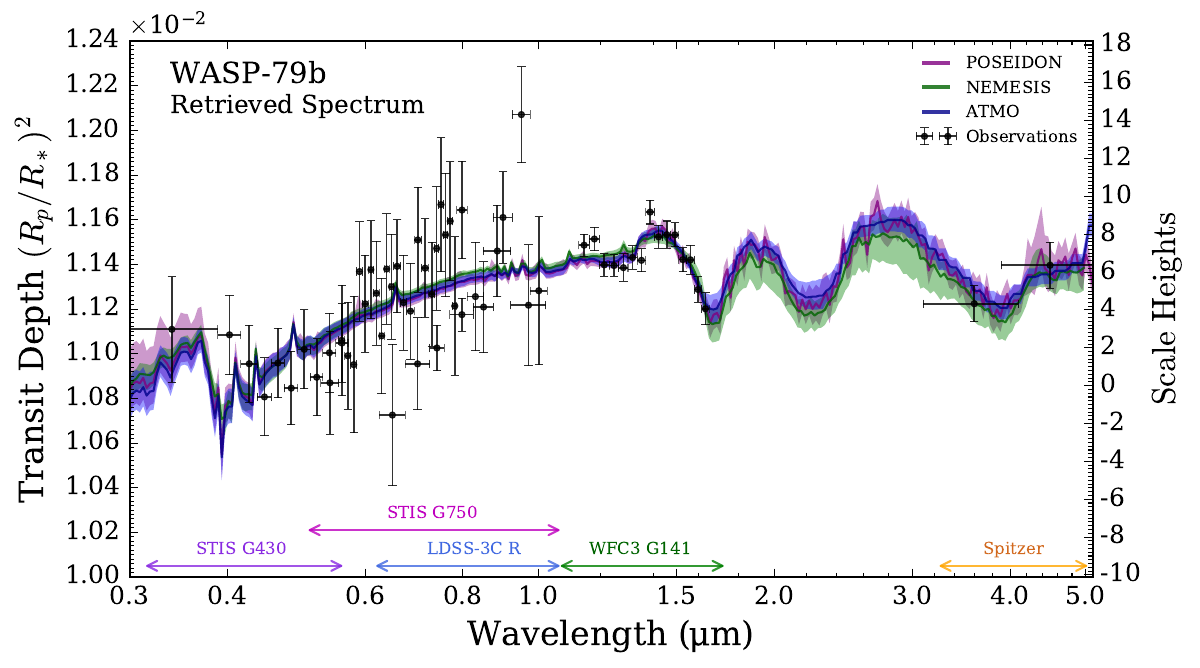}
    \includegraphics[width=\textwidth, height=0.5\textheight, keepaspectratio]{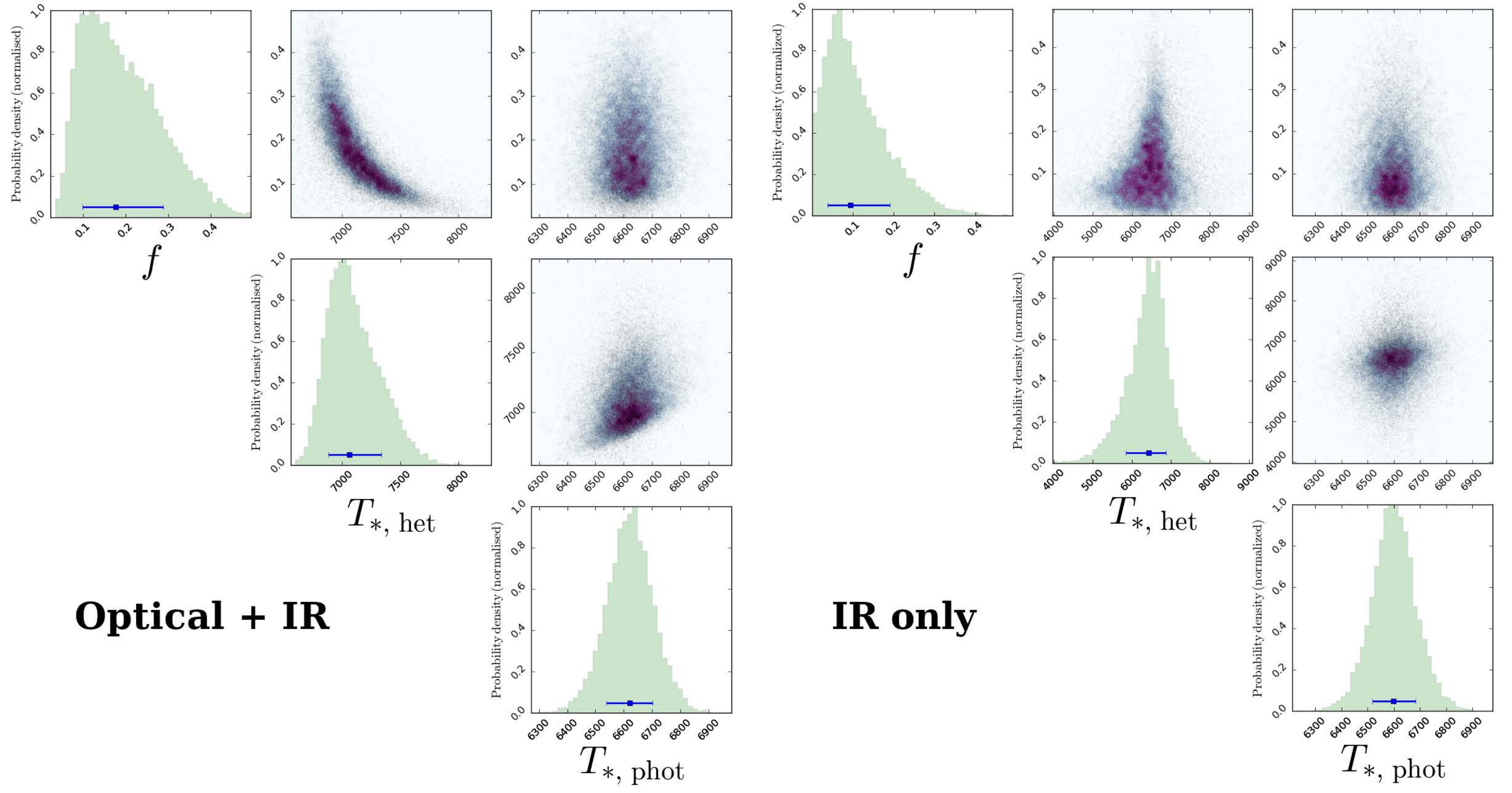}
    \caption{
        Top: the optical and near-infrared transmission spectrum of the hot Jupiter WASP-79b. 
        The decreasing transit depth towards blue wavelengths is a characteristic signature of unocculted faculae (detected at 4.7$\sigma$). 
        Three distinct retrieval codes independently inferred the necessity of faculae to explain WASP-79b's transmission spectrum, demonstrating the robustness of this interpretation.
        Bottom: retrieved stellar heterogeneity parameters from the full optical + near-infrared dataset (left panel) and from the near-infrared data alone (right panel).
        The identification of faculae for WASP-79b crucially hinges on data at wavelengths $< 1.0\,\micron$. 
        From \citet{Rathcke2021}.
        \label{fig:WASP79b_faculae}
    }
\end{figure*}

\begin{figure*}
    \centering
    \includegraphics[width=\textwidth, height=0.5\textheight, keepaspectratio]{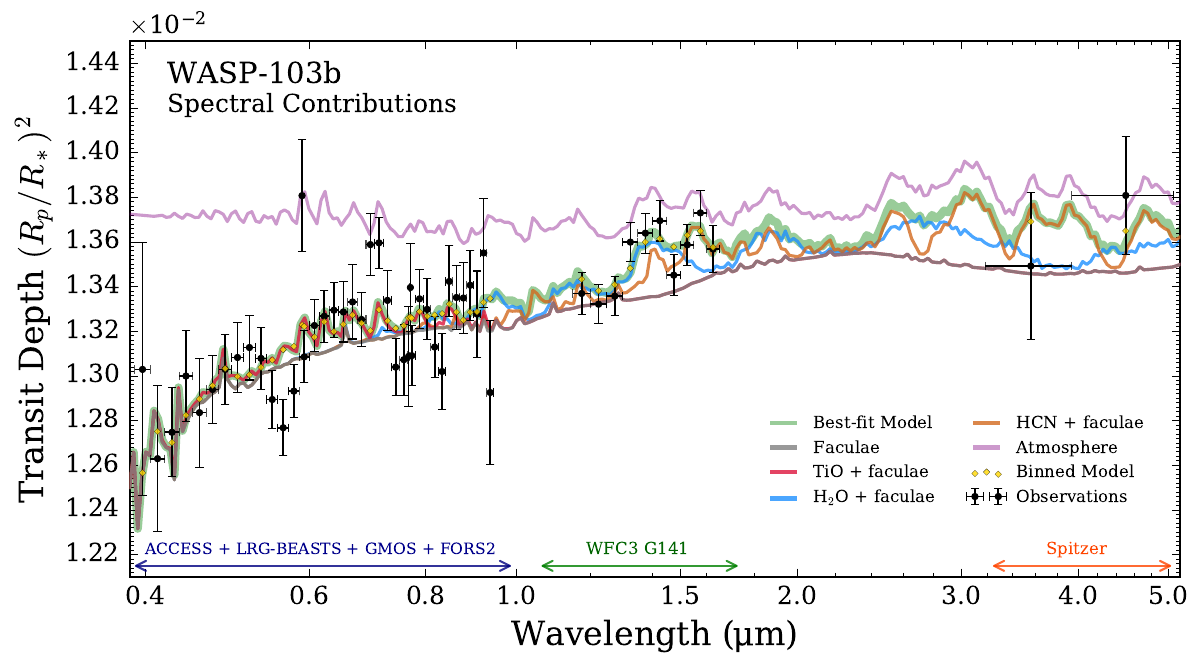}
    \caption{
        The panchromatic transmission spectrum of ultrahot Jupiter WASP-103b. 
        The spectral impact of unocculted faculae is more pronounced at shorter wavelengths.
        Retrievals on the NIR \HST{} and \Spitzer{} data favor the inclusion of faculae at 2.4$\sigma$.
        Including the optical data, combined from 11 ground-based transits with four different facilities, boosts the detection significance of unocculted faculae to 4.3$\sigma$.
        The optical data also boost the detection significance for a planetary atmosphere, inferred by suggestions of TiO, H$_2$O, and HCN, to 2.8$\sigma$ from 1.9$\sigma$ with the NIR data alone.
        From \citet{Kirk2021}.
        \label{fig:WASP103b_contributions}
    }
\end{figure*}

\subsubsection{Projected impacts are strongest at short wavelengths}

Many studies in the past decade have considered the impact of unocculted active regions on transit observations.
Usually, these studies focus on interpretations of transmission spectra from individual systems with active host stars, such as HD\,189733 \citep{Pont2008, Pont2013, Sing2009, Sing2011, McCullough2014, Oshagh2014},  WASP-6 \citep{Jordan2013, Nikolov2015}, and GJ\,1214 \citep{Berta2011, Berta2012, Rackham2017}.
Despite the different spectral types and activity levels of these host stars, studies of individual targets like these consistently find that the projected impacts of unocculted active regions are most evident at blue-optical and shorter wavelengths, where the contrasts between active regions (whether spots or faculae) and the quiescent photosphere are greatest.

This finding is borne out as well by studies of unocculted active regions for less-active stars, like the Sun.
\citet{Llama2015} studied the impact of solar-like photospheric heterogeneity on transits by injecting a hot Jupiter transit into resolved observations of Solar Cycle 24 from NASA's \textit{Solar Dynamics Observatory}.
While they recovered accurate planetary radii in the optical (4500\,\AA{}), they found that solar-like unocculted active regions biased their recovered $R_p/R_s$ values in $\sim$20\% of their simulated soft X-ray (94\,\AA{}) and UV (131--1700\,\AA{}) light curves.
At these wavelengths, unocculted active regions generally led to a mean underestimate of 10\% of the planetary radius, with up to 25\% underestimates occurring during high-activity periods
\revised{In the FUV (1600\,Å and 1700\,Å) simulations, the mean recovered value of $R_p/R_s$ was overestimated by $\sim$20\%.
Similarly, in simulated Ly\,$\alpha$ transits \citep{Llama2016}, recovered values of $R_p/R_s$ are biased by up to 50\% in 25\% of the simulated light curves.}

Turning to stars other than the Sun, \citet{Rackham2018, Rackham2019} estimated the impact of unocculted active regions for FGKM stars with typical activity levels.
Using a suite of Monte Carlo simulations, they estimated the spot and \revised{facula} covering fractions that correspond to the observed variabilities of FGKM dwarfs by spectral type.
They found that spot covering fractions, in particular, increase from $\sim$0.1\% for F dwarfs to 2--4\% for late-K dwarfs to $\sim$10\% for M dwarfs.
If present on the unocculted stellar disk, spots and faculae alter transit depths in the 0.3--5.5\,$\micron$ wavelength range by a few to tens of percent, depending on assumptions about spot size and spot-to-\revised{facula} areal ratio (\autoref{fig:est_cont_spectra}).
Along with the other studies summarized here, these studies of exoplanet host stars with typical activity levels agree that impacts should be most evident at blue-optical and UV wavelengths.

\revised{For this reason, the need to observe at short wavelengths is a key motivating factor in the design of upcoming transit-focused space missions.
The Ariel Space Mission, a medium-class ESA mission planned for launch in 2029, will survey a diverse sample of roughly 1\,000 exoplanets during its four-year mission.
Its design includes a 1-m-class elliptical ($1.1 \times 0.7$\,m) primary mirror and enables simultaneous visible and near-infrared observations spanning 0.5--7.8\,$\micron$.
The choice of the short wavelength limit of the Ariel bandpass was driven precisely by the need to correct for stellar activity \citep{Tinetti2021}.
Likewise, Pandora is a NASA Pioneers Mission, planned for launch in the mid-2020s, with the goal of disentangling stellar and planetary signals in transmission spectra and reliably determining the compositions of exoplanet atmospheres.
Pandora will study transits of at least 20 planets hosted by K and M dwarfs during its one-year primary mission.
Its design includes a 0.45-m primary mirror and, like Ariel, it will also observe simultaneously in the visible and near-infrared (${\sim}$0.4--1.6\,$\micron$) with a visible photometric channel and a low-resolution ($R{\ge}30$) near-infrared (1.0--1.6\,$\micron$) spectroscopic channel \citep{Quintana2021}.
}

\subsubsection{Short-wavelength observations reveal stellar contamination}

Many existing exoplanet transmission spectra are assembled piece-wise via observations with various instruments covering different wavelength ranges \citep[e.g., \HST{}/STIS, \HST{}/WFC3, \Spitzer{}/IRAC;][]{Sing2016}. 
Consequently, the atmospheric interpretations inferred from a given planet's spectrum can change over time as the spectral wavelength range becomes more complete. 
Given that the impact of unocculted stellar heterogeneities is most pronounced at short wavelengths, one may anticipate that initial atmospheric reconnaissance using only infrared observations can miss signatures of stellar contamination. 
Here we focus on one recent case study demonstrating the utility of short-wavelength transmission spectra to reveal unocculted stellar heterogeneities.

The hot Jupiter WASP-79b \citep{Smalley2012} has been considered a strong candidate for the \JWST{} Early Release Science program \citep{Bean2018}. 
\citet{Sotzen2020} presented initial observations of WASP-79b obtained with \HST{}/WFC3, \Spitzer{}/IRAC, \TESS{}, and \Magellan{}/LDSS3. They inferred the presence of H$_2$O absorption from the infrared observations and attributed a blue downward slope from the ground-based LDSS3 data to FeH. 
An independent analysis by \citet{Skaf2020}, using only \HST{}/WFC3 data, reached similar conclusions. 
However, neither of these studies considered unocculted stellar heterogeneities in their retrievals.
Recently, \citet{Rathcke2021} presented additional \HST{}/STIS observations of WASP-79b, which extended the transmission spectrum down to 0.3\,$\micron$ and confirmed the lower transit depths at short wavelengths suggested by LDSS3 (see \autoref{fig:WASP79b_faculae}). 
Their retrieval analysis identified a strong preference for unocculted faculae (4.7$\sigma$), covering $\sim$15\% of the stellar disk with a temperature contrast of $\sim$500\,K, but without the need for FeH. 
This provides another demonstration that, in the presence of informative data, atmospheric interpretations can be biased if stellar contamination is not considered (see \autoref{subsubsec:stellar_het_retrievals}).

Since \JWST{} observations will be confined to wavelengths longer than 0.6\,$\micron$, we assessed the information content provided by optical transmission spectra in the context of WASP-79b. 
\autoref{fig:WASP79b_faculae} shows how the inferred stellar properties alter when only the infrared ($> 1.0\,\micron$) observations are considered. 
We see that the retrieval does not identify unocculted faculae when the optical data is omitted ($T_{*, \, \rm{het}}$ is consistent with $T_{*, \, \rm{phot}}$ and $f_{\rm het}$ is consistent with 0). 
This suggests that complementary blue-UV observations, at shorter wavelengths than \JWST{} can observe, may aid the identification of unocculted stellar heterogeneities. 
However, additional work would be beneficial to explore whether the conclusions from this case study apply more broadly. 

\subsubsection{Combining visible and NIR spectra improves physical inferences}

The inclusion of visible wavelength observations provides an additional quantitative benefit: better inferences for both the planetary atmospheric properties and any unocculted stellar heterogeneities. 
\citet{Kirk2021} recently demonstrated this with a retrieval analysis of the ultrahot Jupiter WASP-103b. 
Similar to the case of WASP-79b, they found the transit depth of WASP-103b to decrease towards blue wavelengths (\autoref{fig:WASP103b_contributions}). 
They showed that retrievals including only infrared \HST{} and \Spitzer{} observations can infer unocculted faculae (2.4$\sigma$) from a slope in the WFC3 observations, but the addition of precise ground-based observations helps confirm this inference (4.3$\sigma$).
At the same time, the detection significance of the planetary atmosphere increased marginally from 1.9$\sigma$ to 2.8$\sigma$. 
This provides a quantitative example of how short-wavelength observations can increase our confidence in a given interpretation of an exoplanet's transmission spectrum.

\subsubsection{Our finding}

From this analysis, we draw the following finding.
\\
\\
\noindent \underline{\textbf{Finding 4.3}}
\\
\\
\noindent \textit{Summary}:
For low-resolution transmission spectra, the impact of unocculted active regions is larger at shorter wavelengths.
More work is needed to quantify the complementary nature of such spectra for \JWST{} observations.
\\
\\
\noindent \textit{Capability Needed}:
\revised{A detailed} understanding \revised{is needed} of when short-wavelength data are necessary to complement NIR transmission spectra from \HST{} and \JWST{}.
\revised{Studies should address the impact of short-wavelength data on retrievals as a function of the stellar spectral type and activity level and the scale of planetary atmospheric features.}
Similarly, an understanding \revised{is needed} of what biases may be introduced by studying NIR spectra alone---even in the context of joint stellar and planetary retrievals.
\\
\\
\noindent \textit{Capability Today}:
\revised{
Simulated observations of a hot Jupiter transiting the Sun show that Sun-like activity can bias $R_p/R_s$ measurements by $\sim$20\% in soft X-ray, EUV, and FUV observations \citep{Llama2015}.
Other simulations show that typical coverages of spots and faculae on K and M dwarfs can bias $R_p/R_s$ measurements by tens of percent in the 0.3--5.5\,$\micron$ range \citep{Rackham2018, Rackham2019}.
Such effects are evident in some precise visible transmission spectra, both from space \citep{McCullough2014} and the ground \citep{Rackham2017}.
\revised{Recent studies have shown that supplementing space-based NIR observations with optical observations (either ground- or space-based) boosts the detection significance of stellar contamination in transmission spectra and alters interpretations of planetary atmospheric properties \revised{\citep{Kirk2021, Rathcke2021}}}.
}
\\
\\
\noindent \textit{Mitigation in Progress}:
\revised{
The upcoming NASA Pandora Mission and Ariel Space Mission will collect simultaneous visible and NIR transmission spectra explicitly to account for stellar activity. 
Pandora, a 0.45-m space telescope planned for launch in the mid-2020s, will observe simultaneously across (${\sim}$0.4--1.6\,$\micron$) with a photometric channel and a low-resolution ($R{\ge}30$) NIR (1.0--1.6\,$\micron$) spectroscopic channel \citep{Quintana2021}.
Ariel, a 1-m-class space telescope planned for launch in 2029, will similarly observe simultaneously across 0.5--7.8\,$\micron$ with a combination of short-wavelength photometric and long-wavelength spectroscopic channels \citep{Tinetti2021}.
}

\section{Future Complementary Observations}
\label{S:ComplementaryObservations}

\subsubsection*{Essential \revised{q}uestions:}

\begin{enumerate}
    \item What datasets and/or techniques can help to unveil stellar photospheres?
    \item What strategies would optimize transit spectroscopy given the findings of the SAG?
\end{enumerate}

\subsection{Introduction}
Exoplanet transmission spectroscopy is a key science case for \revised{\JWST{}} \citep[e.g.,][]{Barstow_habitable_2016, greene2016, Morley_observing_2017} and the focus for the upcoming \revised{Ariel Space Mission \citep[e.g.,][]{Tinetti-18} and its related NASA Contribution to Ariel Spectroscopy of Exoplanets (CASE) Mission of Opportunity \citep[e.g.,][]{Zellem2019}.}
\revised{Both \JWST{} and Ariel/CASE} will have unprecedented precision to probe exoplanet atmospheres\revised{, the former at wavelengths of 0.6--14\,$\micron$ and the latter at 0.5--7.8\,$\micron$}. 
Guided by the findings from \autoref{S:PhotosphericHeterogeneity} through \autoref{S:Retrievals}, the goal of this \revised{section} is to identify current capabilities and future complementary observations that could provide datasets and techniques to constrain stellar heterogeneities in aid of exoplanet transmission spectroscopy studies. 
Here we describe the current landscape for ground- and space-based facilities, and identify gaps in observations and capabilities that are needed to mitigate or correct for stellar contamination.
We focus first on observations of the Sun and second on observations of exoplanetary host stars.

\subsection{Observations of the Sun}

Observations of our nearest star provide the best opportunity to resolve and characterize stellar active regions over time. 
Here we describe two lines of research that leverage solar observations to inform our understanding of exoplanetary transits:
studying the Sun as a resolved star, and studying the Sun as a resolved planetary host.
In each context, we discuss the parameters of useful observations and current or future facilities for gathering these data.

\subsubsection{Studying the Sun as a resolved star}

As described in \autoref{S:PhotosphericHeterogeneity}, the brightness of the Sun (total and over selected spectral ranges) has been monitored from space since 1978.
The Sun's brightness varies on all timescales accessible to observations, from minutes to decades \citep[e.g.,][]{2016JSWSC...6A..30K}. 
Despite the wealth of solar observations collected to date, resolved observations of the Sun in the visible and IR are still needed to study spectral variations of active regions on minute-to-year timescales, as noted in Finding~1.1.
These are the relevant timescales spanning single-transit observations and observational campaigns that stack ten or more transits to build up the necessary signal-to-noise ratio to probe atmospheric features for smaller, cooler exoplanets \citep[e.g.,][]{Kreidberg2014, Benneke2019_GJ3470b, Benneke2019_K2-18b}.

To understand and model the effect of stellar active regions on these timescales, it is important to combine information obtained from both long-term, full-disk (i.e., disk-integrated) observations and high-spatial-resolution (sub-arcsecond) observations. 
The former allows an estimate of the temporal evolution of the areal coverage by the different features and an analysis of their properties (both individual and statistical), such as typical sizes, magnetic field, and photometric contrast in specific spectral bands. 
These quantities are fundamental ingredients of irradiance models and are the main observables for the validation of global dynamo models. 

\paragraph{Full-disk observations.} 
A number of existing facilities provide systematic full-disk observations of the Sun. 
Ground-based facilities targeted to investigate solar variability are typically furnished with broadband Ca\,\textsc{ii}\,K filters (see \citealt{petrie2021} for a recent review on Ca\,\textsc{ii}\,K solar observations), and other broadband photospheric filters and/or H-alpha filters. 
Examples of current facilities include the Precision Solar Photometric Telescopes in Rome \citep{ermolli1998} and Hawaii \citep{rast1999}, and the Solar Observatories of Kanzelh\"ohe, Tenerife, San Fernando, Kodaikanal and Mitaka.
Furthermore, multiple historical Ca\,\textsc{ii}\,K archives have been compiled, together providing more than a century of solar full-disk observations \citep{Chatzistergos:2019b,Chatzistergos:2020}.
Full-disk observations in the UV and EUV, fundamental to understanding and modelling the variability at the shortest wavelengths, are currently provided by AIA \revised{onboard SDO} and \revised{the Solar Ultraviolet Imager (SUVI)} \revised{onboard} the \revised{Geostationary Operational Environmental Satellites (GOES)} 16 and 17.
These will be complemented in the near future with observations acquired in the UV by the SUIT instrument on the Indian Aditya-L1 mission \citep{Ghosh:2016}. 
Full-disk continuum images, such as those provided SOHO/MDI or SDO/HMI, have been used to study brightness and other properties of sunspots and their changes over the solar cycle \citep[e.g.,][]{mathew2007,Watson:2014,Kiess:2014}.
Full-disk magnetograms acquired in both the photosphere (e.g., SOHO/MDI, SDO/HMI, \revised{International Space Station [ISS]}) and chromosphere (e.g., \revised{Global Oscillations Network Group}, ISS) allow studies of the evolution of the magnetic field over the activity cycle.
A combination of full-disk magnetograms and broad-band imagery allows understanding of the relation between the magnetic field and the radiative properties of different features \citep[e.g.,][]{yeo2013, criscuoli2017}, and are assimilated in some irradiance reconstruction models \citep[e.g.,][]{Krivova2003}.

\paragraph{High-spatial-resolution observations.} 
High-spatial-resolution observations are fundamental to understanding the physical processes that generate, sustain, and dissipate magnetic structures, and that determine their radiative properties. 
These, in turn, are fundamental for the development and validation of models. 
Current ground-based facilities\revised{, including the} Dunn Solar Telescope\revised{\footnote{\url{https://nso.edu/telescopes/dunn-solar-telescope/dunn/ 
}}}, Swedish Solar Tower \revised{\citep{Scharmer2003}}, Goode Solar Telescope \revised{\citep{Goode2010}}, \revised{and} GREGOR Solar Telescope \revised{\citep{Schmidt2012},} and space-based facilities\revised{, including Hinode} \revised{\citep{Shimizu2009} and} IRIS \revised{\citep{DePontieu2014},} allow the Sun to be observed at sub-arcsecond spatial resolution (corresponding to a few hundred kilometers on the Solar surface), at high temporal cadence, and at moderate-to-high spectral resolution.
These facilities also typically support spectropolarimetric observations to infer properties of the magnetic fields. 
The five first-light instruments of \revised{DKIST} \citep{DKIST} provide solar observations at unprecedented spatial and temporal resolution, at high spectropolarimetric sensitivity, and in a wide spectral range (380--5000\,nm) that allows observations of the Sun from the base of the photosphere to the corona. 
Synergistic observations in the millimeter spectral range with ALMA \citep{Wootten2009} and in the UV with IRIS will provide essential information to validate and improve modeling, especially of the chromosphere and corona.
Finally, the recently launched Solar Orbiter \citep{SolarOrbiterMision} will observe for the first time polar magnetic fields at sub-arcsecond spatial resolution, thus providing essential information to improve our understanding of the global dynamo and of the radiative emission of faculae at high latitude.

Although breakthrough discoveries are expected in the understanding of the multi-scale, dynamical processes occurring in the solar atmosphere, it should be noted that, because detailed information about physical and magnetic properties of plasma are derived from the analysis of spectral and spectropolarimetric observations, current and future solar facilities typically acquire data at high spectral resolution in relatively narrow spectral ranges (usually not more than $\sim$1\,nm).
For the purposes of studying the impact of stellar variability on both planetary atmospheres themselves and our observations of them, however, we are often interested in variability over wider---and sometimes poorly studied---spectral ranges \citep[e.g.,][]{Ermolli:2013,Matthes2017}.
UV variability, in particular, is known to affect planetary atmospheres and their habitability \citep{Linsky2017}, but unfortunately, this is also the region in which models present the largest discrepancies.
Additionally, with the exception of IRIS, which has been providing solar spectra in three UV regions since 2012, spatially and spectrally resolved observations in the UV and shorter wavelength ranges, necessary to validate and improve models are scarce \citep{criscuoli2019}. 
Most of spatially and spectrally resolved observations in the UV/EUV ranges have been so far rocket- or balloon-born, such as the recently launched (2015) Chromospheric Lyman-Alpha Spectro-Polarimeter (CLASP) or the 2019 Chromospheric Layer Spectropolarimeter (CLASP2), which have collected spectropolarimetric data in the Mg\,\textsc{ii}\,h\,\&\,k range \citep[][]{CLASP2}.
The Sunrise UV Spectropolarimeter and Imager \citep[][]{SUSI} \revised{on}board the SUNRISE III balloon, \revised{planned for launch in 2023}, will provide for the first time sub-arcsecond, spectropolarimetric observations of the Sun in the near-UV (300--400\,nm) at a spectral resolution of $\sim$2\,nm. 
The SUIT instrument on the Indian Aditya-L1 space mission, \revised{also planned for launch in 2023}, will image the Sun \revised{for the first time} in both broad and narrow spectral bands in the UV range from 200 to 400\,nm \citep{Ghosh:2016}.

\subsubsection{Studying the Sun as a resolved planetary host}

In addition to using the Sun to learn about other stars, the planets in our Solar System provide opportunities to learn about exoplanets.
Several ``Earth as an exoplanet'' studies have discussed the potential of using observations of the Earth from past planetary science missions (fly-bys) or current Earth science missions (e.g., EPOXI/DISCOVR) to identify key features that are observable for exoplanets \citep[e.g.,][]{Cowan2009, Llama2015, Llama2016, Berdyugina2019}. 
Recently, \citet{Mayorga2021} proposed that a small spacecraft at Earth\revised{--}Sun L2 could potentially observe transits of the Earth and Moon across the Sun during periods of high activity.
These observations would provide several means towards understanding the biases imparted on transmission spectra from active regions. 
For example, knowing \textit{a priori} that a planet transits a star with numerous active regions and has a flat transmission spectrum (because the planet has no atmosphere, a high mean-molecular-weight atmosphere, or high gravity) would provide a testing ground for understanding the effects of occulted and unocculted active regions on the planet's measured transmission spectrum.
Such an experiment would benefit from our knowledge of the properties of active regions from ground-based solar telescopes, thus providing quantitative points of comparison with the ground truth for active regions of various temperatures and sizes.

\subsubsection{Our finding}

\noindent From this analysis, we draw the following finding.
\\
\\
\noindent \underline{\textbf{Finding 5.1}}
\\
\\
\noindent \textit{Summary}: 
Our understanding of stellar photospheric heterogeneity and its impact on observations of transiting exoplanets is informed greatly by studies of the Sun.
While many suitable facilities exist to study the Sun as a star, the field would benefit from additional capabilities to study planetary transits of the Sun.
\\
\\
\noindent \textit{Capability Needed}: 
Resolved observations of the Sun in the visible and IR are needed to study spectral variations of active regions on minute-to-year timescales.
Observations of planetary bodies transiting the Sun, either resolved or unresolved, are needed to provide the ground-truth for exoplanetary transits.
\\
\\
\noindent \textit{Capability Today}: 
Many ground- and space-based facilities exist to study the Sun as a resolved star.
\revised{Full-disk Ca\,\textsc{ii}\,K observations of the Sun are available spanning more than the last century \citep{Chatzistergos:2019a, Chatzistergos:2020}.}
\revised{Full-disk observations in the UV and EUV are provided by SDO/AIA and GOES/SUVI.}
\revised{Ground-based facilities, such as the Dunn Solar Telescope, Swedish Solar Telescope \citep{Scharmer2003}, Goode Solar Telescope \citep{Goode2010}, and GREGOR Solar Telescope \citep{Schmidt2012}, as well as space-based facilities, including Hinode \citep{Shimizu2009} and IRIS \citep{DePontieu2014}, provide high-cadence observations of the Sun at sub-arcsecond spatial resolution and moderate-to-high spectral resolution.}
No dedicated facilities exist to study planetary transits of the Sun.
\\
\\
\noindent \textit{Mitigation in Progress}: 
\revised{When fully operational,} DKIST \revised{\citep{DKIST}} will provide sensitive solar observations with unprecedented spatial and temporal resolution and across a wide spectral range of interest to exoplanetary transit observations (380--5000\,nm).
\revised{The Sunrise UV Spectropolarimeter and Imager \citep{SUSI}, planned for launch in 2023 onboard the SUNRISE III balloon, will provide sub-arcsecond, spectropolarimetric observations of the Sun in the NUV (300--400\,nm) at a spectral resolution of $\sim$2\,nm.}
\revised{Also planned for launch in 2023, the SUIT instrument \citep{Ghosh:2016} on the Indian Aditya-L1 space mission will image the Sun in broad and narrow spectral bands in the UV (200--400\,nm).}

\subsection{Observations of exoplanetary host stars}

Understanding stellar activity, both the observations and the theory behind what physical processes drive the observations, requires a multiwavelength, multi-facility approach. 
Here we describe useful avenues to better constrain exoplanet host stars in upcoming years in the context of photometry, UV observations, and advances in interferometry.

\subsubsection{Photometric approaches to constrain active-region parameters}

\paragraph{Long-baseline photometry.}
Precise, time-domain photometric surveys (e.g., CoRoT, \Kepler{}, \Ktwo{}, and \TESS{}) have revolutionized our understanding of stellar activity by enabling the measurements of stellar flares and stellar rotation for a wide range of spectral types. 
As discussed in \autoref{sec:sg4-vlb_photometry}, long-baseline photometry of exoplanet host stars can be used effectively to monitor for epoch-to-epoch stellar variability \citep[e.g.,][]{knutson12, Sing2015, Mallonn2016, Mallonn-18, Zellem2017, Mansfield2018, Kilpatrick2020, Rosich-20}. 
Complementary observations from the ground allow us to place comparatively temporally-limited observations with space-based observatories into a larger baseline to explore whether any observed variations can be attributed to the variability of the host star. 
These observations can be conducted on a variety of platforms, from automatic robotic telescopes \citep[e.g.,][]{dukes92} to even \revised{15-cm}-class telescopes, which have the potential to observe even a relatively dim 11.3~V-mag host star with a per-minute precision of 0.67\% \citep{zellem20}.
From space, the recently selected Pandora SmallSat Mission \citep{Quintana2021} will explore the potential of small telescopes (0.45\,m) to reliably constrain stellar photospheres and exoplanetary atmospheres by collecting multi-epoch, long-baseline transit observations of active K- and M-type exoplanet hosts with simultaneous visible photometry and NIR spectroscopy.
\revised{Likewise, the Ariel Space Mission \citep{Tinetti2021} will study planetary transits across 0.5--7.8\,$\micron$ with a combination of short-wavelength photometric channels (0.5--0.60\,$\micron$, 0.6--0.80\,$\micron$, 0.80--1.10\,$\micron$) and long-wavelength spectroscopic channels (1.10--1.95\,$\micron$ at $R{\ge}15$, 1.95--7.8\,$\micron$ at $R{\sim}$30--100).}

\paragraph{Multiband photometry.}
While photometric monitoring of exoplanet host stars can give insights into their epoch-to-epoch variability, detailed modeling of the stellar activity is generally limited by degeneracies between the spot temperature and spot size. 
For instance, a small, cool spot can produce the same amount of flux to first order as a large, comparatively warm/bright spot. 
As discussed in \autoref{subsec:multiband}, however, multiband photometry can provide the necessary constraints to resolve this degeneracy and better understand the impact of active regions on stellar spectral contamination in exoplanetary transmission spectra. 
For example, TiO forms in cooler spots but dissociates at hotter temperatures \citep{mirtorabi03,Berdyugina2003mol}. 
Therefore, one can perform simultaneous observations of the host star both inside the TiO band and just outside to provide measurements of the starspot temperature and, thus, its size \citep[e.g.,][]{Zellem2010}. 
Such a study was done of the chromospherically active star IM Peg using the relatively small 0.75-m Four College Consortium \revised{T}elescope \citep{dukes92} as well as spectroscopy with the 2.5-m Nordic Optical Telescope \citep{Berdyugina1999impeg}. 
This study and others like it show that, while these observations can be time-consuming, they potentially could be shifted to smaller telescopes, alleviating the need for larger telescopes \citep{zellem20}.
Here too the Pandora SmallSat Mission \citep{Quintana2021} will provide a relevant space-based counterpart with its simultaneous visible photometry and NIR spectroscopy designed to uniquely identify the properties of heterogeneous photospheres with absolutely calibrated, multi-epoch observations.

\paragraph{Existing datasets and underutilized avenues for photometry.}
Existing single- or multi-band photometry may be available for a given exoplanet host, given the abundance of existing time-domain photometric surveys, including those searching for transiting exoplanets, such as \Kepler{} \citep{Borucki2010}, \TESS{} \citep{TESS}, HAT-P \citep{hat-p_bakos2004}, WASP \citep{Pollacco2006}, HATS \citep{hats_bakos2013}, MEarth \citep{Mearth_nutzman_charbonneau2008}, NGTS \citep{NGTS_weathley2018}, SPECULOOS \citep{speculoos_Delrez2018}, EDEN \citep{eden_gibbs2020}, and more.
If data is not already available from a targeted or wide-field survey, monitoring the star over multiple rotation periods could be time-consuming. 
Spun-down M dwarfs, for example, can have periods as long as 140\,days \citep{Newton2018}.
However, while monitoring exoplanet host stars over multiple rotation periods is challenging due to the oversubscription of professional observatories, there is the opportunity for amateur astronomers to contribute. 
For example, the American Association of Variable Star Observers has been observing variable stars for $\sim$100 years and has over 2 million stars in its Variable Star Index \citep[VSX,][]{watson06}.

\paragraph{High-cadence photometry.}
Finally, obtaining high-cadence photometry will help in a number of areas.
\revised{Searches for flares with high-cadence photometry has a surprisingly long history, including detections of flares with sub-second time resolution in the 1980s \citep{Beskin1988}.}
\revised{\TESS{} has opened the latest frontier in this research area with its recently introduced} 20-sec cadence mode, which can be used to resolved short-duration activity such as flares, but the number of stars observed at short cadence is typically limited. 
\revised{CHEOPS can also collect photometry at a sub-minute cadence \citep{Benz2021}, and PLATO will include 32 cameras with a 25-s readout cadence and two with a 2.5-s cadence \citep{Rauer-14}.}
High-cadence observations which fully resolve the ingress and egress of the transit can independently constrain the  planetary radius even when the transit is contaminated by significant spot occultation events \citep{Morris2018b}.

\subsubsection{UV observations to inform models of host-star atmospheres}
With the wealth and quality of existing solar observations, highly sophisticated semi-empirical models of the Sun have been developed, including complex models of spots, flares, and other magnetic phenomena (See \autoref{S:PhotosphericHeterogeneity}). 
Extensive modeling efforts have furthered our understanding of the solar interior \citep[e.g.,][]{Vinyoles2017}, the detailed structure in the upper atmosphere \citep{Fontenla:2006}, and the solar activity cycle and associated energetic events \revised{\citep[e.g.,][]{Yeo2014_SATIRE, Hathaway2015, Desai2016}}. 
\revised{The} current state-of-the-art for stellar atmosphere models \revised{of lower-mass, planet-host\revised{ing} stars, including K and M dwarfs,} is not at the same level as solar models\revised{, due in part to the additional complexity involved in modelling these objects \citep[e.g.,][]{Iyer2023} and in part to the disparity in the quality and quantity of X-ray--NUV spectra available \citep[e.g.,][]{France2016}}.
\revised{In the context of exoplanet transmission spectroscopy, an accurate understanding of the high-energy radiation of host stars is crucially important, as it drives hydrodynamic escape of planetary atmospheres \citep[e.g.,][]{Lammer2003, Murray-Clay2009}, leading to observations of comet-like evaporation tails \citep[e.g.,][]{Vidal-Madjar2004, Kulow2014, Spake2018} and perhaps shaping the mass--radius distribution of close-in exoplanets \citep[e.g.,][]{Owen2012, Owen2013, Owen2017}.}
Stellar modeling efforts would greatly benefit from a future EUV/FUV/NUV dedicated mission with time-dependent measurements to improve our understanding of active-region coverage and temperature, and flare frequency, size, and color on these low-mass, planet-host\revised{ing} stars. 
With this knowledge, composite semi-empirical models of stellar atmospheres could be developed, complete with accurate \revised{and precise} modeling of photospheric and chromospheric heterogeneities that would, amongst other things, advance the general understanding of stellar atmospheres and improve our ability to disentangle the stellar and planetary contributions from UV transit spectra.

\subsubsection{Constraining host-star surface inhomogeneities through interferometric imaging}

\paragraph{Global surface temperature maps.}
Stellar surfaces have been indirectly imaged for decades using Doppler imaging \citep[e.g.,][]{Berdyugina2005,Strassmeier-09,Korhonen2021} and light-curve imaging \citep[e.g.,][]{Lanza1998, Berdyugina2002lqhya,Roettenbacher2013} in order to detect surface inhomogeneities, such as starspots and faculae (see also \autoref{S:PhotosphericHeterogeneity}). 
In these methods, either high-resolution spectra or photometric observations are used to obtain maps of the stellar surface with inversion methods.  
Light-curve inversion imaging reconstructs the stellar surface based upon the observed brightness fluctuations.  
Doppler imaging utilizes the evolution of distortions of absorption lines caused by surface features rotating in and out of view to reconstruct the features of the stellar surface.  
Both methods suffer from degeneracies, especially for the latitudinal information of the spots.  
The determination of starspot latitudes is constrained in light-curve inversion with limb-darkening laws and estimated coefficients, while in Doppler imaging with the location of the spot feature within absorption line profiles and the speed at which it crosses the profile as the star rotates.
While the light-curve method cannot provide starspot latitudes, the Doppler imaging can locate spot latitudes and longitudes within the stellar hemisphere that is best visible to the observer with respect to the equator, as was demonstrated by modeling various spots \citep[e.g.,][]{VogtPenrod1983,Piskunov1990di,Berdyugina1998di}.

In recent years, advances in long-baseline optical/near-infrared interferometry have enabled the direct-imaging of starspots on stellar surfaces \citep{Roettenbacher2016,Roettenbacher2017,Parks2021,Martinez2021}. 
This is the only method that can obtain reliable, unambiguous information of the spot locations in latitude and longitude within both stellar hemispheres relative to the equator. 
However, the targets currently accessible with present interferometric resources are limited, as this technique requires very bright targets with spots that appear large enough on the sky.  
Additionally, obtaining a complete image of the stellar surface requires repeated observations with phase coverage across a rotation.
This can require large amounts of dedicated observing time.  
The relative rate of evolution of the surface structure compared to the stellar rotation must also be considered.

Currently, the highest spatial resolution can be reached by \revised{MIRC-X} instrument \citep{Anugu2020} at the CHARA Array \citep{tenBrummelaar2005}, which can provide an angular resolution of about 0.4\,mas. MIRC-X is also the only instrument that can combine light from all the six 1-meter telescopes of the CHARA Array, making it the instrument that is best suited for interferometric imaging of stellar surfaces. 
Another facility that can be used for interferometric imaging is European Southern Observatory's Very Large Telescope Interferometer (VLTI). 
The best angular resolution of the VLTI can be reached using the Precision Integrated Optics Near Infrared ExpeRiment instrument \citep[PIONIER;][]{LeBouquin2011} that gives angular resolution of about 1.3\,mas. 
VLTI can combine four telescopes simultaneously and has changeable array configurations with the 1.8-meter auxiliary telescopes. 
VLTI can also combine the light from the four 8.2-meter Unit Telescopes telescopes, which gives the current faintest limiting magnitudes for infra-red interferometry \citep[GRAVITY instrument: $K{=}10.5$ for on-axis observations and $K{=}17$ when using a close-by bright fringe tracker star;][]{Gravity2017}.

The number of targets that can be accurately mapped using interferometric imaging is still limited to a handful of bright, nearby stars that have angular diameters of about 2.0\,mas or larger. 
The technique, however, importantly opens a new parameter space that is not accessible using Doppler or light-curve inversion imaging: cool, aged, main-sequence stars. 
Due to their slow rotation, Doppler imaging cannot be used to image these stars. 
Cool, nearby, main-sequence stars include some exoplanet hosts, like $\epsilon$\,Eridani.

New exciting discoveries are made with each new image obtained of stellar surfaces. 
Future advances in interferometric imaging can be made by: increasing the number of telescope that are combined to improve the image fidelity, moving towards shorter wavelengths to obtain higher angular resolution, and increasing the length of the longest baselines to obtain higher angular resolution. 
Many of these aspects will be addressed, for example, by the Visible Imaging System for Interferometric Observations \citep[VISION; ][]{Garcia2016} instrument at the upgraded Navy Precision Optical Interferometer \citep[NPOI;][]{vanBelle2020} and the Stellar Parameters and Images with a Cophased Array \citep[SPICA;][]{Pannetier2020} at \revised{the CHARA Array} \citep{tenBrummelaar2005}.

Because the light detected with long-baseline optical interferometers must be combined at the telescopes, there are physical limitations to the size of ground-based interferometers.  
While efforts are being made to investigate kilometer baselines \citep[e.g.,][]{Monnier2018}, space-based interferometers \citep[e.g.,][]{Monnier2018,vanBelle2020b} could alleviate some of the constraints on ground-based optical interferometers, allowing for increased resolution and increased targets.  

\paragraph{Empirical determination of fundamental stellar parameters.}
Precise stellar parameters, including stellar radius, surface temperature, and limb darkening, are crucial for correctly interpreting exoplanet observations.
\revised{High-resolution spectroscopy from the ground provides the main avenue for determining stellar parameters of exoplanet hosts, including large efforts focused on FGK stars by the SWEET-Cat \citep{Santos2013, Sousa2021}, Ariel \citep{Danielski2022, Magrini2022}, and GAPS teams \citep[e.g.,][]{Biazzo2022} and on M dwarfs by the CARMENES team \citep[e.g.,][]{Passegger2018, Marfil2021}}.
\revised{As a complement to these efforts, interferometry offers a powerful, direct, empirical means to determine the fundamental physical properties of stars.}
While most stars cannot be resolved with current single-dish, ground- and space-based telescopes, long-baseline optical and near-infrared interferometers consisting of individual telescopes connected as an array can provide accurate information on these parameters. 
When resolved stars are observed with interferometry the measured squared visibility changes with the baseline (distance between two telescopes) and wavelength. 
When limb-darkened disk models are fit to the observed squared visibilities, the apparent stellar diameter and limb-darkening profile can be determined. 
Additionally, the effective temperature can be measured when the stellar diameter is combined with observations of the bolometric flux.

Stellar diameters and limb-darkening laws have been determined interferometrically for many stars, both \revised{F to M subgiants and} giants and \revised{A to M nearby} main\revised{-}sequence stars \revised{\citep[e.g.,][]{Nordgren1999,DiFolco2004,Boyajian2012,Boyajian2012_KM,Boyajian2013,Mann2015}}.
\revised{Interferometric measurements of stellar diameters have enabled the calibration of surface--brightness relations of dwarfs, subgiants, giants, and more \citep{Kervella2004_cepheids, Kervella2004_dwarfs, Kervella2008}.}
Diameter measurements have also been carried out for exoplanet host stars \citep[see, e.g.,][]{Baines2008, Baines2009, Bazot2011, Crida2018, White2018}. 
These observations have shown the necessity of having sufficiently long baselines for accurately resolving targets with small angular diameters. 
Additionally, one has to take great care when selecting the calibration stars.  
Calibration stars are ideally nearby on the sky, of similar spectral type, single, and unresolved by the interferometer.  
Instrumentation developments that are ideally suited for studying smaller exoplanet host stars include VISION at NPOI \citep{Garcia2016,vanBelle2020} and SPICA at the CHARA Array \citep{tenBrummelaar2005, Pannetier2020}. 
With instruments working in the visible wavelength range and with the long baselines of 432\,m and 330\,m, respectively, NPOI and the CHARA Array are ideal facilities for studying fundamental stellar parameters of the exoplanet host stars in the future. 

Complementing these capabilities is the method of ``pseudo-interferometry" \citep{Stassun:2017,Stassun:2018}, in which the stellar angular radius is determined through an inversion of the Stefan-Boltzmann relation, $\Theta = \left(F_{\rm bol} / \sigma_{\rm SB} T_{\rm eff}^4 \right)^{1/2}$\revised{, in which $F_\mathrm{bol}$ is the bolometric flux, $\sigma_\mathrm{SB}$ is the Stefan-Boltzmann constant, and $T_\mathrm{eff}$ is the effective temperature}. 
The physical radius then follows directly via the distance as determined, e.g., with a precise Gaia parallax. 
This approach does require an independent determination of $T_{\rm eff}$, preferably via a high quality spectroscopic analysis. 
As demonstrated by \citet{Stassun:2017}, the availability of broadband apparent magnitudes spanning from the UV to the mid-IR allows $F_{\rm bol}$ to be determined with a typical precision of a few percent, and in the best cases $T_{\rm eff}$ determined to a precision of 1--2\%, such that angular radii for typical planet-host stars can be measured to 2--3\,$\micron$, rivaling that of interferometric imaging for the brightest stars (though without the benefit of resolving surface inhomogeneities, of course). 
Moreover, as ultra-precise parallaxes are now routinely available with Gaia, physical radii for typical stars via pseudo-interferometry can be determined to better than a few percent \citep{Stassun:2017}. 
Finally, such pseudo-interferometric stellar radii can be further leveraged to measure the masses of single stars empirically as well, if a precise surface gravity is available such as through the technique of granulation ``flicker" from a light curve \citep{Bastien:2013,Bastien:2016}. 
Masses of individual \revised{FGK dwarf, subgiant, and giant stars can be determined} via this pseudo-interferometric $+$ flicker gravity method to better than 10\% in many cases and as good as a few percent in the best cases \citep{Stassun:2018}.

\subsubsection{Our finding}

\noindent From this analysis, we draw the following finding.
\\
\\
\noindent \underline{\textbf{Finding 5.2}}
\\
\\
\noindent \textit{Summary}: 
Many existing facilities can be leveraged to study exoplanet host stars photometrically, though there is a need to scale observations to study a large number of exoplanet host stars with long-term, multi-band photometry.
UV observations are also essential to inform models of host-star atmospheres, and time-dependent measurements provide valuable information on active-region coverage and temperature as well as flare frequency and magnitude.
At the same time, advances in interferometric technique enable constraints on photospheric properties of nearby dwarfs, including some exoplanet hosts; 
of specific interest is the direct mapping of temperature inhomogeneities on stellar surfaces. 
\\
\\
\noindent \textit{Capability Needed}: 
Long-baseline, multi-band photometry of active exoplanet host stars is needed to uniquely constrain active region temperatures and sizes.
Stellar modeling efforts would greatly benefit from a future EUV/FUV/NUV dedicated mission with time-dependent measurements to improve our understanding of active-region coverage and temperature, and flare frequency, size, and color on these low-mass, planet-host stars. 
Much longer baseline interferometric imaging capabilities on the ground and/or in space would allow mapping of stellar surface inhomogeneities for a much larger sample of stars; 
currently this can only be done for a handful of bright, nearby stars with angular diameters larger than $\sim$2\,mas. 
\\
\\
\noindent \textit{Capability Today}: 
Photometric studies of exoplanet host stars benefit from many capable facilities, including dedicated transit-search networks and longstanding contributions from amateur astronomers \revised{\citep[e.g.,][]{zellem20}}.
\revised{Space-based platforms are being used to build panchromatic (X-ray--mid-IR) SEDs of nearby planet-hosting stars \citep{France2016}, providing inputs to models of stellar atmospheres and chemical evolution of associated planetary atmospheres.}
\revised{High-resolution stellar spectroscopy from the ground provides a strong avenue for characterizing the fundamental properties of exoplanet host stars, and large community efforts have pushed this work forward \citep[e.g.,][]{Santos2013, Passegger2018, Sousa2021, Marfil2021, Biazzo2022, Danielski2022,  1982nordlund}.}
\revised{Long-baseline optical/NIR interferometry provides diameter measurements of exoplanet host stars \citep[e.g.,][]{Baines2008, Baines2009, Bazot2011, Crida2018, White2018}.}
\revised{In tandem with photometric and RV measurements, interferometry also provides an opportunity to directly image starspots of nearby, main-sequence exoplanet hosts \citep{Roettenbacher2022}.}
\\
\\
\noindent \textit{Mitigation in Progress}: 
The extended \TESS{} mission \revised{\citep{Ricker2015}} continues to provide long-baseline photometry in a broad red-optical bandpass for targets over nearly the whole sky.
The recently selected Pandora SmallSat Mission \revised{\citep{Quintana2021}} will provide both long-baseline monitoring and transit observations simultaneously in a visible photometric bandpass and a NIR spectroscopic channel \revised{for at least 20 planets transiting K and M dwarfs over a one-year mission}.
\revised{Similarly, the Ariel Space Mission \citep{Tinetti2021} will provide simultaneous visible photometry and NIR spectroscopy of roughly 1\,000 exoplanets over a four-year mission.}

\section{Conclusions}
\label{S:Conclusions}

\begin{figure*}
    \centering
    \includegraphics[width=\textwidth, height=0.95\textheight, keepaspectratio]{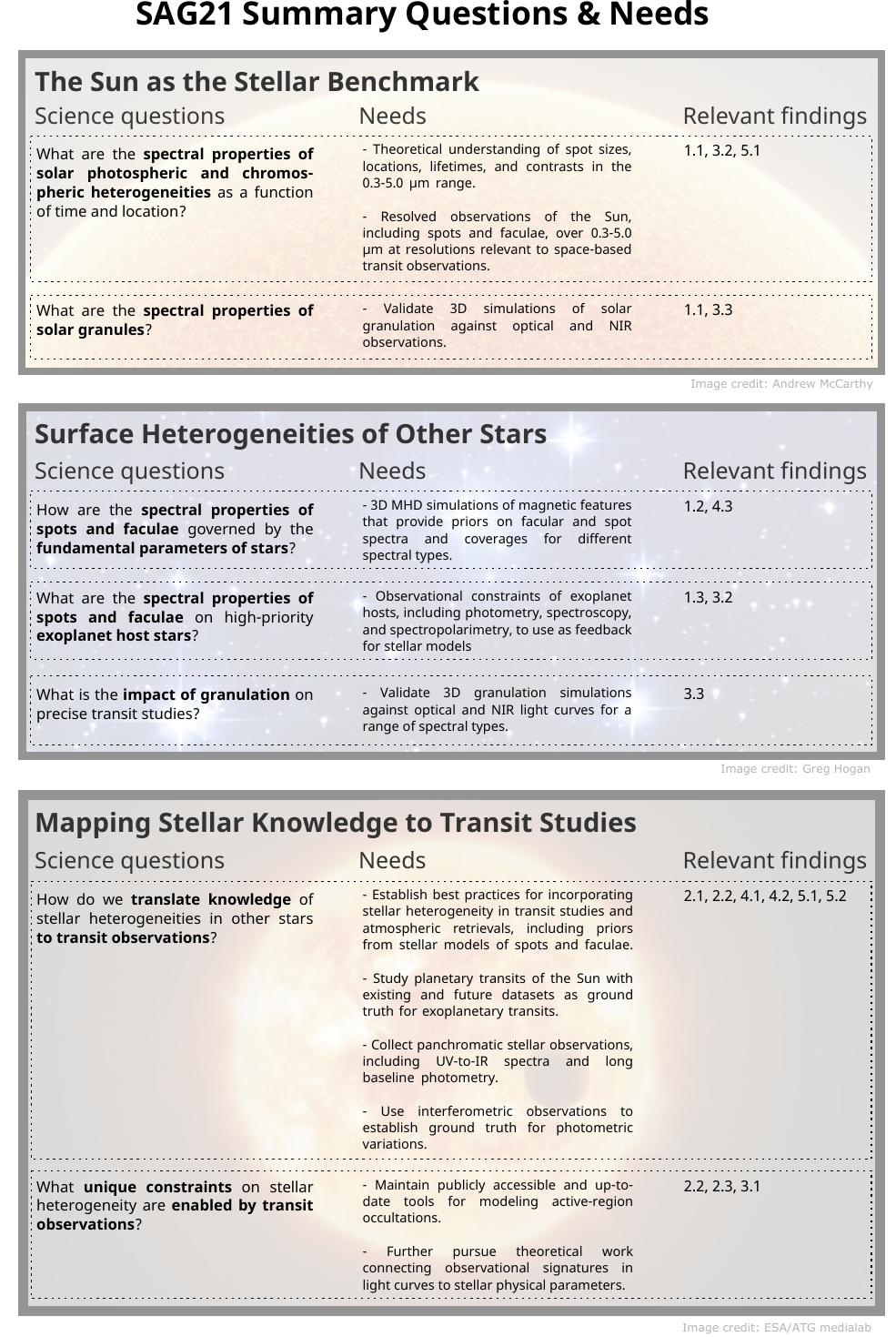}
    \caption{
        A summary of SAG21 in terms of its science questions, needs identified, and relevant findings.
        \label{fig:exsum}
    }
\end{figure*}

\revised{We have presented the analysis and findings of SAG21, an interdisciplinary group of more than 100 scientists that examined the impact of stellar contamination on space-based transmission spectra of exoplanets.
The analysis was expansive, with subgroups of SAG21 focused on stellar photospheric and chromospheric heterogeneity, unocculted active regions, occulted active regions, stellar and planetary retrievals, and future complementary observations.
In total, the analysis produced 14 findings, contextualized statements of what we understand to be the current research needs that can be addressed to further our understanding of stellar photospheres and make the best use of precise space-based transmission spectra of exoplanets.}

To summarize these findings further, \autoref{fig:exsum} presents a set of summary science questions and needs that are common to several findings at a time.
These fall into three Science Themes encompassing (1) how the Sun is used as our best laboratory to calibrate our understanding of stellar heterogeneities (``The Sun as the Stellar Benchmark"), (2) how stars other than the Sun extend our knowledge of heterogeneities (``Surface Heterogeneities of Other Stars") and (3) how to incorporate information gathered for the Sun and other stars into transit studies (``Mapping Stellar Knowledge to Transit Studies").

The following subsections summarize in turn the scope of the five subgroups and their findings.
For brevity, only the summary statement of each finding is reproduced here.
The full findings \revised{in the format of NASA's Exoplanet Exploration Program Science Gap List}, including descriptions of the capability needed, capability today, and mitigation in progress, are available in their respective \revised{sections of this review}.

\subsection{Stellar Photospheric \& Chromospheric Heterogeneity}

The Stellar Photospheric \& Chromospheric Heterogeneity subgroup was tasked with summarizing our knowledge of the heterogeneity of stellar photospheres and chromospheres and ongoing work to understand relevant properties for the purpose of transmission spectroscopy.
The members considered observational and modeling efforts to understand the heterogeneity of the Sun at wavelengths relevant to transit studies and comparable efforts for other stars, including exoplanet hosts.
Their analysis produced three findings.

\subsubsection*{Finding 1.1}
The Sun provides the benchmark for stellar studies.
Studying the spatial, spectral, and temporal variations of stellar surface structures is necessary for understanding the impact of analogous structures on transmission spectroscopy of exoplanets.

\subsubsection*{Finding 1.2}
More modeling work is needed to understand how the fundamental parameters of stars other than the Sun govern their magnetic fields and the associated properties of their surface inhomogeneities.
In accordance with observations, simulations show that many lower-activity stars, such as exoplanet hosts that \revised{might be deemed as ``low risk''} for transmission spectroscopy \revised{in terms of stellar activity}, are \revised{facula}-dominated \revised{\citep{Shapiro2014, Nemec2022}.}
\revised{However}, little is known on how facular contrasts and coverages depend on stellar activity levels and fundamental parameters, such as metallicity and surface gravity. 
Both observing and modeling the fine structure of spots on stellar surfaces remain a challenge. 

\subsubsection*{Finding 1.3}
Simultaneous multi-wavelength (multi-instrument) stellar observations are needed to provide feedback to modeling efforts and improve our understanding of the photospheres and chromospheres of other stars, including high-priority exoplanet host stars.
This is particularly critical for K, M, and L dwarfs, for which models are relatively poorly constrained.

\subsection{Occulted Active Regions}

The Occulted Active Regions subgroup was tasked with evaluating what we can learn from planetary occultations of stellar active regions during transits.
The members considered the state of the art for observations and models of active-region occultations, outstanding theoretical challenges for these studies, and the availability of well-maintained tools for conducting these analyses.
Their analysis produced three findings.

\subsubsection*{Finding 2.1}
Precise transit observations increasingly reveal occultations of stellar active regions.
Rather than flagging and removing active region occultations, which results in decreased observing efficiency and possibly biased transit depth measurements, future observations should move towards joint inference of the active region and planetary properties. 

\subsubsection*{Finding 2.2}
Theoretical advances are needed to understand the limits of what we can infer about active-region properties from transit light curves and how best to incorporate prior knowledge from magnetohydrodynamic models into transit studies.

\subsubsection*{Finding 2.3}
\revised{Several} publicly available tools \revised{have been developed} for forward-modeling active-region occultations.
Their maintenance and further development is necessary to ensure their utility for analyses of precise transit observations.

\subsection{Unocculted Active Regions}

The Unocculted Active Regions subgroup was tasked with examining our knowledge of host-star active regions that are not occulted by transiting exoplanets.
The members considered what information can be obtained about unocculted active regions from high-cadence transit photometry, the utility of other datasets---current or planned---for inferring the presence of active regions and their properties, and the effects of stellar granulation on transmission spectra.
Their analysis produced three findings.

\subsubsection*{Finding 3.1}
High-cadence light curves provide the potential to understand unocculted active regions, but the information is not comprehensive enough to make unambiguous measurements at present. 
Theoretical advances are needed to make full use of these light curves.

\subsubsection*{Finding 3.2}
Simultaneous multi-band photometry and contemporaneous spectroscopy provide critical information towards understanding the potential effects of active regions on transmission spectroscopy observations. 
While other data sets can provide information on filling factors, theoretical work is needed to maximise the utility of these data for transmission spectroscopy. 

\subsubsection*{Finding 3.3}
Stellar ``granulation flicker" constitutes a fundamental ``noise floor" on stellar light curves that increases with decreasing stellar surface gravity and at shorter wavelengths. 
The impact of this stellar granulation can enter as a source of noise and/or contamination for transmission spectroscopy in two ways. 
The first is through extra light curve scatter, \revised{the} impact \revised{of which} is minimal in most cases (at most $\sim$0.05\% error on the transit depth in the visible and even less at longer wavelengths).
The second is through a contamination source similar to that of spots and faculae, albeit at a much lower amplitude ($\sim 1\%$ of the transit depth at about 0.7\,$\mu$m).
This second source is significant at short wavelengths for smaller exoplanets around Sun-like stars, but decreases strongly as a function of wavelength.

\subsection{Stellar \& Planetary Retrievals}

The Stellar \& Planetary Retrievals subgroup was tasked with studying how we can best use inference frameworks to constrain the properties of stellar photospheres and exoplanetary atmospheres jointly from transmission spectra and other relevant inputs.
The members considered the utility of joint stellar and planetary retrievals generally, the known limitations of the approach, and the complementarity of short-wavelength observations to \HST{} and \JWST{} NIR observations for inferring stellar photospheric properties.
Their analysis produced three findings.

\subsubsection*{Finding 4.1}
Retrievals of transmission spectra that include the effects of unocculted active regions can guard against biases.
More work is needed to understand when these retrievals are necessary and what are the limitations and best practices of this approach.

\subsubsection*{Finding 4.2}
Retrieval approaches rely on stellar models, and thus their accuracy is limited by model fidelity \revised{and, depending on the approach, the prior information on stellar parameters used in the retrieval}.
Further efforts to develop model spectra for spots and faculae and incorporate them into exoplanetary atmospheric retrievals are needed.

\subsubsection*{Finding 4.3}
For low-resolution transmission spectra, the impact of unocculted active regions is larger at shorter wavelengths.
More work is needed to quantify the complementary nature of such spectra for JWST observations.

\subsection{Future Complementary Observations}

The Future Complementary Observations subgroup was tasked with assessing the current landscape for ground- and space-based facilities that can provide useful complements for \HST{} and \JWST{} studies of transiting exoplanets.
The subgroup was also tasked with identifying gaps in capabilities that are needed to mitigate or correct for stellar contamination.
The members considered observations of the Sun that are informative in this context as well as observations of exoplanetary host stars themselves.
Their analysis produced two findings.

\subsubsection*{Finding 5.1}
Our understanding of stellar photospheric heterogeneity and its impact on observations of transiting exoplanets is informed greatly by studies of the Sun.
While many suitable facilities exist to study the Sun as a star, the field would benefit from additional capabilities to study planetary transits of the Sun.

\subsubsection*{Finding 5.2}
Many existing facilities can be leveraged to study exoplanet host stars photometrically, though there is a need to scale observations to study a large number of exoplanet host stars with long-term, multi-band photometry.
UV observations are also essential to inform models of host-star atmospheres, and time-dependent measurements provide valuable information on active-region coverage and temperature as well as flare frequency and magnitude.
At the same time, advances in interferometric technique enable constraints on photospheric properties of nearby dwarfs, including some exoplanet hosts; 
of specific interest is the direct mapping of temperature inhomogeneities on stellar surfaces.

\section*{List of affiliations}
$^{1}$Department of Earth, Atmospheric and Planetary Sciences, Massachusetts Institute of Technology, 77 Massachusetts Avenue, Cambridge, MA 02139, USA\\
$^{2}$Kavli Institute for Astrophysics and Space Research, Massachusetts Institute of Technology, Cambridge, MA 02139, USA\\
$^{3}$Space Telescope Science Institute, 3700 San Martin Drive, Baltimore, MD 21218, USA\\
$^{4}$Leibniz-Institut f\"ur Sonnenphysik (KIS), Schöneckstrasse 6, D-79104 Freiburg, Germany\\
$^{5}$Istituto Ricerche Solari Aldo e Cele Daccò (IRSOL), Faculty of Informatics, Università della Svizzera italiana, Locarno, Switzerland\\
$^{6}$European Southern Observatory (ESO), Alonso de Córdova 3107, Vitacura, Santiago, Chile\\
$^{7}$Department of Astronomy, University of Michigan, 1085 S. University Ave, Ann Arbor, MI 48109, USA\\
$^{8}$School of Physics, University of New South Wales, Sydney, NSW 2052, Australia\\
$^{9}$Center for Space and Habitability, University of Bern, Gesellschaftsstrasse 6, 3012 Bern, Switzerland\\
$^{10}$Instituto de Astrofísica de Canarias, 38200 La Laguna, Tenerife, Spain\\
$^{11}$Max-Planck-Institut f\"ur Sonnensystemforschung, Justus-von-Liebig-Weg 3, 37077 Göttingen, Germany\\
$^{12}$Department of Physics, Imperial College London, Prince Consort Road, London SW7 2AZ, UK\\
$^{13}$NASA Goddard Space Flight Center, 8800 Greenbelt Road, Greenbelt, MD 20771, USA\\
$^{14}$Jet Propulsion Laboratory, California Institute of Technology, 4800 Oak Grove Drive, Pasadena, CA 91109, USA\\
$^{15}$Steward Observatory, The University of Arizona, 933 N. Cherry Avenue, Tucson, AZ 85721, USA\\
$^{16}$Lunar and Planetary Laboratory, The University of Arizona, 1629 E. University Blvd., Tucson, AZ 85721, USA\\
$^{17}$University of Maryland, Baltimore County, 1000 Hilltop Circle, Baltimore, MD 21250, USA\\
$^{18}$School of Physical Sciences, The Open University, Walton Hall, Milton Keynes, MK7 6AA, UK\\
$^{19}$INAF, Osservatorio Astrofisico di Catania, Via S. Sofia 78, 95123 Catania, Italy\\
$^{20}$Space Research Institute, Austrian Academy of Sciences, Schmiedlstrasse 6, A-8042 Graz, Austria\\
$^{21}$School of Physics and Astronomy, University of Leicester, University Road, Leicester LE1 7RH, UK\\
$^{22}$Department of Physics, University of Warwick, Coventry, CV4 7AL, UK\\
$^{23}$National Solar Observatory, 3665 Discovery Drive, Boulder, CO 80303, USA\\
$^{24}$Observatoire de Gen\`eve, Universit\'e de Gen\`eve, Chemin des Maillettes 51, 1290 Versoix, Switzerland\\
$^{25}$School of Earth and Space Exploration, Arizona State University, 525 E. University Dr., Tempe AZ 85281\\
$^{26}$Laboratory for Atmospheric and Space Physics, University of Colorado Boulder, Boulder, CO, USA\\
$^{27}$Leibniz-Institut f\"ur Astrophysik Potsdam (AIP), An der Sternwarte 16, 14482 Potsdam, Germany\\
$^{28}$Center for Astrophysics | Harvard \& Smithsonian, 60 Garden St, Cambridge, MA 02138, USA\\
$^{29}$Universitäts-Sternwarte, Ludwig-Maximilians-Universität München, Scheinerstrasse 1, 81679 München, Germany\\
$^{30}$Exzellenzcluster Origins, Boltzmannstraße 2, 85748 Garching, Germany\\
$^{31}$Landessternwarte, Zentrum für Astronomie der Universität Heidelberg, Königstuhl 12, 69117 Heidelberg, Germany\\
$^{32}$Department of Physics and Astronomy, Dartmouth College, Hanover, NH 03755, USA\\
$^{33}$National Solar Observatory, Boulder, CO 80303, USA\\
$^{34}$Department of Astrophysics and Planetary Sciences, University of Colorado, Boulder, CO 80303, USA\\
$^{35}$Yale Center for Astronomy and Astrophysics, Yale University, 46 Hillhouse Avenue, New Haven, CT 06511, USA\\
$^{36}$Department of Astronomy, Yale University, 52 Hillhouse Avenue, New Haven, CT 06511, USA\\
$^{37}$Department of Physics and Astronomy, Vanderbilt University, Nashville, TN 37235, USA\\
$^{38}$Johns Hopkins University Applied Physics Laboratory, 11100 Johns Hopkins Rd, Laurel, MD 20723, USA\\
$^{39}$Instituto de Investigación en Astronomia y Ciencias Planetarias, Universidad de Atacama, Copiapó, Atacama, Chile\\
$^{40}$Centro de R\'adio Astronomia e Astrofisica Mackenzie, Mackenzie Presbyterian University, Rua da Consolacao, 896, Sao Paulo, Brazil\\
$^{41}$Rosseland Centre for Solar Physics, University of Oslo, Postboks 1029 Blindern, 0315 Oslo, Norway\\
$^{42}$Institute of Theoretical Astrophysics, University of Oslo, Postboks 1029 Blindern, 0315 Oslo, Norway\\
$^{43}$Carnegie Earth \& Planets Laboratory, 5241 Broad Branch Road NW, Washington, DC 20015, USA\\
$^{44}$Astronomy Department, University of Washington, Box 951580, Seattle, WA 98195, USA\\
$^{45}$Department of Astronomy, University of Maryland, College Park, MD 20742, USA\\
$^{46}$Department of Astronomy, Boston University, Boston, MA 02215\\
$^{47}$Department of Astrophysical Sciences, Princeton University, Princeton, NJ, USA\\
$^{48}$Astrobiology Research Unit, Universit\'e de Li\`ege, 19C All\'ee du 6 Aout, 4000 Li\`ege, Belgium\\
$^{49}$Department of Astronomy and Astrophysics, University of Chicago, 5640 S. Ellis Avenue, Chicago, IL 60637, USA\\
$^{50}$SETI Institute, 189 Bernardo Avenue, Suite 200, Mountain View, CA 94043, USA\\
$^{51}$Gemini Observatory/NSF’s NOIRLab, 670 N. A’ohoku Place, Hilo, HI 96720, USA\\
$^{52}$NASA Goddard Space Flight Center / University of Maryland, Baltimore County\\
$^{53}$Institut de Ciències de l’Espai (ICE, CSIC), Campus UAB, C/Can Magrans s/n, 08193 Bellaterra, Spain\\
$^{54}$Institut d’Estudis Espacials de Catalunya (IEEC), 08034 Barcelona, Spain\\

\section*{Acknowledgements}

J.K.B. acknowledges support from an STFC Ernest Rutherford Fellowship. 
S.V.B. was supported by the European Research Council Advanced Grant HotMol ERC-2011-AdG291659.
G.B. acknowledges support from CHEOPS ASI-INAF agreement n.\ 2019-29-HH.0.
H.M.C. acknowledges support from a UKRI Future Leaders Fellowship fellowship MR/S035214/1.
C.D. acknowledges the support from NASA XRP grant 80NSSC21K0608. 
E.D. acknowledges the Paris Region fellowship programme which is supported by the Ile-de-France Region and has received funding under the Horizon 2020 innovation framework programme and the Marie Sklodowska-Curie grant agreement no.\ 945298.
A.I. and M.L. acknowledge support from NSF Award Number: 2009592. 
G.K. was supported by NASA's TSIS program under contract NNG09HP02C. 
B.V.R. and R.M.R. thank the Heising-Simons Foundation for support.
A.I.S. is supported by the by the European Research Council (ERC) under the European Union’s Horizon 2020 research and innovation programme (grant agreement No.\ 715947). 
K.G.S. acknowledges generous support from NASA XRP grant 80NSSC18K0445. 
Y.C.U. acknowledges support from STFC through grant ST/S000372/1. 
S.W. is supported by the European Research Council (ERC) under the European Union’s Horizon 2020 research and innovation programme (grant agreement No.\ 682462), and by the Research Council of Norway through its Centres of Excellence scheme (project number 262622) and FRIPRO scheme (project number 286853). 
\revised{We thank the three anonymous reviewers for their careful reading of the manuscript and their helpful comments.}
Part of the research was carried out at the Jet Propulsion Laboratory, California Institute of Technology, under a contract with the National Aeronautics and Space Administration (80NM0018D0004). 
\revised{This material is based upon work supported by the National Aeronautics and Space Administration under Agreement No.\ 80NSSC21K0593 for the program ``Alien Earths''.
The results reported herein benefited from collaborations and/or information exchange within} NASA’s Nexus for Exoplanet System Science (NExSS) research coordination network sponsored by NASA’s Science Mission Directorate.
This research was supported by the Excellence Cluster ORIGINS which is funded by the Deutsche Forschungsgemeinschaft (DFG, German Research Foundation) under Germany's Excellence Strategy - EXC-2094 - 390783311. 
The National Solar Observatory is operated by the Association of Universities for Research in Astronomy, Inc.\ (AURA), under cooperative agreement with the National Science Foundation. 
\revised{This research has made use of the NASA Exoplanet Archive, which is operated by the California Institute of Technology, under contract with the National Aeronautics and Space Administration under the Exoplanet Exploration Program.}
\revised{This research has made use of NASA’s Astrophysics Data System.}

\section*{Data Availability}

Most datasets presented in this report have been summarized from the scientific literature and can be found at the cited references.
Data relating to any new analyses presented here are available upon reasonable request.



\bibliographystyle{rasti}
\bibliography{references}

\begin{thebibliography}{659}
\expandafter\ifx\csname natexlab\endcsname\relax\def\natexlab#1{#1}\fi

\bibitem[{Adjabshirzadeh} \& {Koutchmy}(1983)]{Adjabshirzadeh-Koutchmy1983}
{Adjabshirzadeh}, A. \& {Koutchmy}, S., 1983.
\newblock {Photometric analysis of sunspot umbral dots. III Spectrophotometry
  and preliminary model of a 2-component umbra}, {\it \aap\/}, {\bf 122}(1-2),
  1--8.

\bibitem[{Aerts}(2021)]{Aerts2021}
{Aerts}, C., 2021.
\newblock {Probing the interior physics of stars through asteroseismology},
  {\it Reviews of Modern Physics\/}, {\bf 93}(1), 015001.

\bibitem[{Afram} \& {Berdyugina}(2015)]{Afram-Berdyugina2015}
{Afram}, N. \& {Berdyugina}, S.~V., 2015.
\newblock {Molecules as magnetic probes of starspots}, {\it \aap\/}, {\bf 576},
  A34.

\bibitem[{Afram} \& {Berdyugina}(2019)]{Afram-Berdyugina2019}
{Afram}, N. \& {Berdyugina}, S.~V., 2019.
\newblock {Complexity of magnetic fields on red dwarfs}, {\it \aap\/}, {\bf
  629}, A83.

\bibitem[{Agol} et~al.(2005){Agol}, {Steffen}, {Sari}, \& {Clarkson}]{Agol-05}
{Agol}, E., {Steffen}, J., {Sari}, R., \& {Clarkson}, W., 2005.
\newblock {On detecting terrestrial planets with timing of giant planet
  transits}, {\it \mnras\/}, {\bf 359}, 567--579.

\bibitem[{Agol} et~al.(2010){Agol}, {Cowan}, {Knutson}, {Deming}, {Steffen},
  {Henry}, \& {Charbonneau}]{Agol2010}
{Agol}, E., {Cowan}, N.~B., {Knutson}, H.~A., {Deming}, D., {Steffen}, J.~H.,
  {Henry}, G.~W., \& {Charbonneau}, D., 2010.
\newblock {The Climate of HD 189733b from Fourteen Transits and Eclipses
  Measured by Spitzer}, {\it \apj\/}, {\bf 721}(2), 1861--1877.

\bibitem[{Ahern} \& {Chapman}(2000)]{ahern2000}
{Ahern}, S. \& {Chapman}, G.~A., 2000.
\newblock {An analysis of full-disk observations of facular contrast in the
  blue and red}, {\it \solphys\/}, {\bf 191}(1), 71--84.

\bibitem[{Aigrain} et~al.(2012){Aigrain}, {Pont}, \& {Zucker}]{Aigrain2012}
{Aigrain}, S., {Pont}, F., \& {Zucker}, S., 2012.
\newblock {A simple method to estimate radial velocity variations due to
  stellar activity using photometry}, {\it \mnras\/}, {\bf 419}(4), 3147--3158.

\bibitem[{Alexoudi} et~al.(2020){Alexoudi}, {Mallonn}, {Keles},
  {Poppenh{\"a}ger}, {von Essen}, \& {Strassmeier}]{alexoudi2020}
{Alexoudi}, X., {Mallonn}, M., {Keles}, E., {Poppenh{\"a}ger}, K., {von Essen},
  C., \& {Strassmeier}, K.~G., 2020.
\newblock {Role of the impact parameter in exoplanet transmission
  spectroscopy}, {\it \aap\/}, {\bf 640}, A134.

\bibitem[{Allard} et~al.(2003){Allard}, {Guillot}, {Ludwig}, {Hauschildt},
  {Schweitzer}, {Alexander}, \& {Ferguson}]{Allard2003}
{Allard}, F., {Guillot}, T., {Ludwig}, H.-G., {Hauschildt}, P.~H.,
  {Schweitzer}, A., {Alexander}, D.~R., \& {Ferguson}, J.~W., 2003.
\newblock {Model Atmospheres and Spectra: The Role of Dust}, in {\em Brown
  Dwarfs\/}, vol. 211, p. 325.

\bibitem[{Andersen} \& {Korhonen}(2015)]{AndersenKorhonen2015}
{Andersen}, J.~M. \& {Korhonen}, H., 2015.
\newblock {Stellar activity as noise in exoplanet detection - II. Application
  to M dwarfs}, {\it \mnras\/}, {\bf 448}(4), 3053--3069.

\bibitem[{Andretta} et~al.(2017){Andretta}, {Giampapa}, {Covino}, {Reiners}, \&
  {Beeck}]{Andretta:2017}
{Andretta}, V., {Giampapa}, M.~S., {Covino}, E., {Reiners}, A., \& {Beeck}, B.,
  2017.
\newblock {Estimates of Active Region Area Coverage through Simultaneous
  Measurements of the He I {\ensuremath{\lambda}}{\ensuremath{\lambda}} 5876
  and 10830 Lines}, {\it \apj\/}, {\bf 839}(2), 97.

\bibitem[{Anugu} et~al.(2020){Anugu}, {Le Bouquin}, {Monnier}, {Kraus},
  {Schaefer}, {Setterholm}, {Davies}, {Gardner}, {Labdon}, {Lanthermann},
  {Ennis}, {ten Brummelaar}, {Sturmann}, {Anderson}, {Farrington}, {Vargas}, \&
  {Majoinen}]{Anugu2020}
{Anugu}, N., {Le Bouquin}, J.-B., {Monnier}, J.~D., {Kraus}, S., {Schaefer},
  G., {Setterholm}, B.~R., {Davies}, C.~L., {Gardner}, T., {Labdon}, A.,
  {Lanthermann}, C., {Ennis}, J., {ten Brummelaar}, T., {Sturmann}, J.,
  {Anderson}, M., {Farrington}, C., {Vargas}, N., \& {Majoinen}, O., 2020.
\newblock {CHARA/MIRC-X: a high-sensitive six telescope interferometric imager
  concept, commissioning and early science}, in {\em Society of Photo-Optical
  Instrumentation Engineers (SPIE) Conference Series\/}, vol. 11446 of {\bf
  Society of Photo-Optical Instrumentation Engineers (SPIE) Conference Series},
  p. 114460N.

\bibitem[{Apai} et~al.(2013){Apai}, {Radigan}, {Buenzli}, {Burrows}, {Reid}, \&
  {Jayawardhana}]{Apai2013}
{Apai}, D., {Radigan}, J., {Buenzli}, E., {Burrows}, A., {Reid}, I.~N., \&
  {Jayawardhana}, R., 2013.
\newblock {HST Spectral Mapping of L/T Transition Brown Dwarfs Reveals Cloud
  Thickness Variations}, {\it \apj\/}, {\bf 768}(2), 121.

\bibitem[{Apai} et~al.(2017){Apai}, {Karalidi}, {Marley}, {Yang}, {Flateau},
  {Metchev}, {Cowan}, {Buenzli}, {Burgasser}, {Radigan}, {Artigau}, \&
  {Lowrance}]{Apai2017}
{Apai}, D., {Karalidi}, T., {Marley}, M.~S., {Yang}, H., {Flateau}, D.,
  {Metchev}, S., {Cowan}, N.~B., {Buenzli}, E., {Burgasser}, A.~J., {Radigan},
  J., {Artigau}, E., \& {Lowrance}, P., 2017.
\newblock {Zones, spots, and planetary-scale waves beating in brown dwarf
  atmospheres}, {\it Science\/}, {\bf 357}(6352), 683--687.

\bibitem[{Apai} et~al.(2018){Apai}, {Rackham}, {Giampapa}, {Angerhausen},
  {Teske}, {Barstow}, {Carone}, {Cegla}, {Domagal-Goldman}, {Espinoza},
  {Giles}, {Gully-Santiago}, {Haywood}, {Hu}, {Jordan}, {Kreidberg}, {Line},
  {Llama}, {L{\'o}pez-Morales}, {Marley}, \& {de Wit}]{Apai-18}
{Apai}, D., {Rackham}, B.~V., {Giampapa}, M.~S., {Angerhausen}, D., {Teske},
  J., {Barstow}, J., {Carone}, L., {Cegla}, H., {Domagal-Goldman}, S.~D.,
  {Espinoza}, N., {Giles}, H., {Gully-Santiago}, M., {Haywood}, R., {Hu}, R.,
  {Jordan}, A., {Kreidberg}, L., {Line}, M., {Llama}, J., {L{\'o}pez-Morales},
  M., {Marley}, M.~S., \& {de Wit}, J., 2018.
\newblock {Understanding Stellar Contamination in Exoplanet Transmission
  Spectra as an Essential Step in Small Planet Characterization}, {\it arXiv
  e-prints\/}.

\bibitem[{Apai} et~al.(2021){Apai}, {Nardiello}, \& {Bedin}]{Apai2021}
{Apai}, D., {Nardiello}, D., \& {Bedin}, L.~R., 2021.
\newblock {TESS Observations of the Luhman 16 AB Brown Dwarf System: Rotational
  Periods, Lightcurve Evolution, and Zonal Circulation}, {\it \apj\/}, {\bf
  906}(1), 64.

\bibitem[{Ara{\'u}jo} \& {Valio}(2021)]{Araujo-21}
{Ara{\'u}jo}, A. \& {Valio}, A., 2021.
\newblock {Kepler-411 Differential Rotation from Three Transiting Planets},
  {\it \apjl\/}, {\bf 907}(1), L5.

\bibitem[{Artigau} et~al.(2009){Artigau}, {Bouchard}, {Doyon}, \&
  {Lafreni{\`e}re}]{Artigau2009}
{Artigau}, {\'E}., {Bouchard}, S., {Doyon}, R., \& {Lafreni{\`e}re}, D., 2009.
\newblock {Photometric Variability of the T2.5 Brown Dwarf SIMP
  J013656.5+093347: Evidence for Evolving Weather Patterns}, {\it \apj\/}, {\bf
  701}(2), 1534--1539.

\bibitem[{Aschwanden} \& {Freeland}(2012)]{aschwanden2012}
{Aschwanden}, M.~J. \& {Freeland}, S.~L., 2012.
\newblock {Automated Solar Flare Statistics in Soft X-Rays over 37 Years of
  GOES Observations: The Invariance of Self-organized Criticality during Three
  Solar Cycles}, {\it \apj\/}, {\bf 754}, 112.

\bibitem[{Asplund} et~al.(2005){Asplund}, {Grevesse}, \& {Sauval}]{Asplund2}
{Asplund}, M., {Grevesse}, N., \& {Sauval}, A.~J., 2005.
\newblock {The Solar Chemical Composition}, in {\em Cosmic Abundances as
  Records of Stellar Evolution and Nucleosynthesis\/}, vol. 336 of {\bf
  Astronomical Society of the Pacific Conference Series}, p.~25.

\bibitem[{Asplund} et~al.(2009){Asplund}, {Grevesse}, {Sauval}, \&
  {Scott}]{Asplund1}
{Asplund}, M., {Grevesse}, N., {Sauval}, A.~J., \& {Scott}, P., 2009.
\newblock {The Chemical Composition of the Sun}, {\it \araa\/}, {\bf 47}(1),
  481--522.

\bibitem[{Atroshchenko} et~al.(1989{\natexlab{a}}){Atroshchenko}, {Gadun}, \&
  {Kostyk}]{atroshchenko+al1989a}
{Atroshchenko}, I.~N., {Gadun}, A.~S., \& {Kostyk}, R.~I., 1989{\natexlab{a}}.
\newblock {The Simulation of the Convective Motions in the Procyon Envelope -
  Part One - Ideology and Analysis of the Three-Dimensional Inhomogeneous
  Models}, {\it Astrofizika\/}, {\bf 31}, 281.

\bibitem[{Atroshchenko} et~al.(1989{\natexlab{b}}){Atroshchenko}, {Gadun}, \&
  {Kostyk}]{atroshchenko+al1989b}
{Atroshchenko}, I.~N., {Gadun}, A.~S., \& {Kostyk}, R.~I., 1989{\natexlab{b}}.
\newblock {Three-Dimensional Simulation of Convective Motions in the Procyon
  Envelope}, in {\em Solar and Stellar Granulation\/}, vol. 263 of {\bf NATO
  ASIC Proc.}, p. 521.

\bibitem[{Aulanier} et~al.(2013){Aulanier}, {D{\'e}moulin}, {Schrijver},
  {Janvier}, {Pariat}, \& {Schmieder}]{Aulanier:2013}
{Aulanier}, G., {D{\'e}moulin}, P., {Schrijver}, C.~J., {Janvier}, M.,
  {Pariat}, E., \& {Schmieder}, B., 2013.
\newblock {The standard flare model in three dimensions. II. Upper limit on
  solar flare energy}, {\it \aap\/}, {\bf 549}, A66.

\bibitem[{Auvergne} et~al.(2009){Auvergne}, {Bodin}, {Boisnard}, {Buey},
  {Chaintreuil}, {Epstein}, {Jouret}, {Lam-Trong}, {Levacher}, {Magnan},
  {Perez}, {Plasson}, {Plesseria}, {Peter}, {Steller}, {Tiph{\`e}ne}, {Baglin},
  {Agogu{\'e}}, {Appourchaux}, {Barbet}, {Beaufort}, {Bellenger}, {Berlin},
  {Bernardi}, {Blouin}, {Boumier}, {Bonneau}, {Briet}, {Butler}, {Cautain},
  {Chiavassa}, {Costes}, {Cuvilho}, {Cunha-Parro}, {de Oliveira Fialho},
  {Decaudin}, {Defise}, {Djalal}, {Docclo}, {Drummond}, {Dupuis}, {Exil},
  {Faur{\'e}}, {Gaboriaud}, {Gamet}, {Gavalda}, {Grolleau}, {Gueguen},
  {Guivarc'h}, {Guterman}, {Hasiba}, {Huntzinger}, {Hustaix}, {Imbert},
  {Jeanville}, {Johlander}, {Jorda}, {Journoud}, {Karioty}, {Kerjean},
  {Lafond}, {Lapeyrere}, {Landiech}, {Larqu{\'e}}, {Laudet}, {Le Merrer},
  {Leporati}, {Leruyet}, {Levieuge}, {Llebaria}, {Martin}, {Mazy}, {Mesnager},
  {Michel}, {Moalic}, {Monjoin}, {Naudet}, {Neukirchner}, {Nguyen-Kim},
  {Ollivier}, {Orcesi}, {Ottacher}, {Oulali}, {Parisot}, {Perruchot},
  {Piacentino}, {Pinheiro da Silva}, {Platzer}, {Pontet}, {Pradines},
  {Quentin}, {Rohbeck}, {Rolland}, {Rollenhagen}, {Romagnan}, {Russ}, {Samadi},
  {Schmidt}, {Schwartz}, {Sebbag}, {Smit}, {Sunter}, {Tello}, {Toulouse},
  {Ulmer}, {Vandermarcq}, {Vergnault}, {Wallner}, {Waultier}, \&
  {Zanatta}]{Corot}
{Auvergne}, M., {Bodin}, P., {Boisnard}, L., {Buey}, J.~T., {Chaintreuil}, S.,
  {Epstein}, G., {Jouret}, M., {Lam-Trong}, T., {Levacher}, P., {Magnan}, A.,
  {Perez}, R., {Plasson}, P., {Plesseria}, J., {Peter}, G., {Steller}, M.,
  {Tiph{\`e}ne}, D., {Baglin}, A., {Agogu{\'e}}, P., {Appourchaux}, T.,
  {Barbet}, D., {Beaufort}, T., {Bellenger}, R., {Berlin}, R., {Bernardi}, P.,
  {Blouin}, D., {Boumier}, P., {Bonneau}, F., {Briet}, R., {Butler}, B.,
  {Cautain}, R., {Chiavassa}, F., {Costes}, V., {Cuvilho}, J., {Cunha-Parro},
  V., {de Oliveira Fialho}, F., {Decaudin}, M., {Defise}, J.~M., {Djalal}, S.,
  {Docclo}, A., {Drummond}, R., {Dupuis}, O., {Exil}, G., {Faur{\'e}}, C.,
  {Gaboriaud}, A., {Gamet}, P., {Gavalda}, P., {Grolleau}, E., {Gueguen}, L.,
  {Guivarc'h}, V., {Guterman}, P., {Hasiba}, J., {Huntzinger}, G., {Hustaix},
  H., {Imbert}, C., {Jeanville}, G., {Johlander}, B., {Jorda}, L., {Journoud},
  P., {Karioty}, F., {Kerjean}, L., {Lafond}, L., {Lapeyrere}, V., {Landiech},
  P., {Larqu{\'e}}, T., {Laudet}, P., {Le Merrer}, J., {Leporati}, L.,
  {Leruyet}, B., {Levieuge}, B., {Llebaria}, A., {Martin}, L., {Mazy}, E.,
  {Mesnager}, J.~M., {Michel}, J.~P., {Moalic}, J.~P., {Monjoin}, W., {Naudet},
  D., {Neukirchner}, S., {Nguyen-Kim}, K., {Ollivier}, M., {Orcesi}, J.~L.,
  {Ottacher}, H., {Oulali}, A., {Parisot}, J., {Perruchot}, S., {Piacentino},
  A., {Pinheiro da Silva}, L., {Platzer}, J., {Pontet}, B., {Pradines}, A.,
  {Quentin}, C., {Rohbeck}, U., {Rolland}, G., {Rollenhagen}, F., {Romagnan},
  R., {Russ}, N., {Samadi}, R., {Schmidt}, R., {Schwartz}, N., {Sebbag}, I.,
  {Smit}, H., {Sunter}, W., {Tello}, M., {Toulouse}, P., {Ulmer}, B.,
  {Vandermarcq}, O., {Vergnault}, E., {Wallner}, R., {Waultier}, G., \&
  {Zanatta}, P., 2009.
\newblock {The CoRoT satellite in flight: description and performance}, {\it
  \aap\/}, {\bf 506}(1), 411--424.

\bibitem[{Ayres}(2021)]{ayres2020}
{Ayres}, T.~R., 2021.
\newblock {In the Trenches of the Solar-Stellar Connection. II. Extreme
  Ultraviolet Flux-Flux Correlations across Solar Cycle 24}, {\it \apj\/}, {\bf
  908}(2), 205.

\bibitem[{Babcock} \& {Babcock}(1955)]{Babcock1955}
{Babcock}, H.~W. \& {Babcock}, H.~D., 1955.
\newblock {The Sun's Magnetic Field, 1952-1954.}, {\it \apj\/}, {\bf 121}, 349.

\bibitem[{Baines} et~al.(2008){Baines}, {McAlister}, {ten Brummelaar},
  {Turner}, {Sturmann}, {Sturmann}, {Goldfinger}, \& {Ridgway}]{Baines2008}
{Baines}, E.~K., {McAlister}, H.~A., {ten Brummelaar}, T.~A., {Turner}, N.~H.,
  {Sturmann}, J., {Sturmann}, L., {Goldfinger}, P.~J., \& {Ridgway}, S.~T.,
  2008.
\newblock {CHARA Array Measurements of the Angular Diameters of Exoplanet Host
  Stars}, {\it \apj\/}, {\bf 680}(1), 728--733.

\bibitem[{Baines} et~al.(2009){Baines}, {McAlister}, {ten Brummelaar},
  {Sturmann}, {Sturmann}, {Turner}, \& {Ridgway}]{Baines2009}
{Baines}, E.~K., {McAlister}, H.~A., {ten Brummelaar}, T.~A., {Sturmann}, J.,
  {Sturmann}, L., {Turner}, N.~H., \& {Ridgway}, S.~T., 2009.
\newblock {Eleven Exoplanet Host Star Angular Diameters from the Chara Array},
  {\it \apj\/}, {\bf 701}(1), 154--162.

\bibitem[{Bakos} et~al.(2004){Bakos}, {Noyes}, {Kov{\'a}cs}, {Stanek},
  {Sasselov}, \& {Domsa}]{hat-p_bakos2004}
{Bakos}, G., {Noyes}, R.~W., {Kov{\'a}cs}, G., {Stanek}, K.~Z., {Sasselov},
  D.~D., \& {Domsa}, I., 2004.
\newblock {Wide-Field Millimagnitude Photometry with the HAT: A Tool for
  Extrasolar Planet Detection}, {\it \pasp\/}, {\bf 116}(817), 266--277.

\bibitem[{Bakos} et~al.(2013){Bakos}, {Csubry}, {Penev}, {Bayliss},
  {Jord{\'a}n}, {Afonso}, {Hartman}, {Henning}, {Kov{\'a}cs}, {Noyes},
  {B{\'e}ky}, {Suc}, {Cs{\'a}k}, {Rabus}, {L{\'a}z{\'a}r}, {Papp}, {S{\'a}ri},
  {Conroy}, {Zhou}, {Sackett}, {Schmidt}, {Mancini}, {Sasselov}, \&
  {Ueltzhoeffer}]{hats_bakos2013}
{Bakos}, G.~{\'A}., {Csubry}, Z., {Penev}, K., {Bayliss}, D., {Jord{\'a}n}, A.,
  {Afonso}, C., {Hartman}, J.~D., {Henning}, T., {Kov{\'a}cs}, G., {Noyes},
  R.~W., {B{\'e}ky}, B., {Suc}, V., {Cs{\'a}k}, B., {Rabus}, M.,
  {L{\'a}z{\'a}r}, J., {Papp}, I., {S{\'a}ri}, P., {Conroy}, P., {Zhou}, G.,
  {Sackett}, P.~D., {Schmidt}, B., {Mancini}, L., {Sasselov}, D.~D., \&
  {Ueltzhoeffer}, K., 2013.
\newblock {HATSouth: A Global Network of Fully Automated Identical Wide-Field
  Telescopes}, {\it \pasp\/}, {\bf 125}(924), 154.

\bibitem[{Baliunas} et~al.(1995){Baliunas}, {Donahue}, {Soon}, {Horne},
  {Frazer}, {Woodard-Eklund}, {Bradford}, {Rao}, {Wilson}, {Zhang}, {Bennett},
  {Briggs}, {Carroll}, {Duncan}, {Figueroa}, {Lanning}, {Misch}, {Mueller},
  {Noyes}, {Poppe}, {Porter}, {Robinson}, {Russell}, {Shelton}, {Soyumer},
  {Vaughan}, \& {Whitney}]{Baliunas-95}
{Baliunas}, S.~L., {Donahue}, R.~A., {Soon}, W.~H., {Horne}, J.~H., {Frazer},
  J., {Woodard-Eklund}, L., {Bradford}, M., {Rao}, L.~M., {Wilson}, O.~C.,
  {Zhang}, Q., {Bennett}, W., {Briggs}, J., {Carroll}, S.~M., {Duncan}, D.~K.,
  {Figueroa}, D., {Lanning}, H.~H., {Misch}, T., {Mueller}, J., {Noyes}, R.~W.,
  {Poppe}, D., {Porter}, A.~C., {Robinson}, C.~R., {Russell}, J., {Shelton},
  J.~C., {Soyumer}, T., {Vaughan}, A.~H., \& {Whitney}, J.~H., 1995.
\newblock {Chromospheric variations in main-sequence stars}, {\it \apj\/}, {\bf
  438}, 269--287.

\bibitem[{Ballerini} et~al.(2012){Ballerini}, {Micela}, {Lanza}, \&
  {Pagano}]{Ballerini-12}
{Ballerini}, P., {Micela}, G., {Lanza}, A.~F., \& {Pagano}, I., 2012.
\newblock {Multiwavelength flux variations induced by stellar magnetic
  activity: effects on planetary transits}, {\it \aap\/}, {\bf 539}, A140.

\bibitem[{Baranec} et~al.(2016){Baranec}, {Ziegler}, {Law}, {Morton}, {Riddle},
  {Atkinson}, {Schonhut}, \& {Crepp}]{Baranec2016}
{Baranec}, C., {Ziegler}, C., {Law}, N.~M., {Morton}, T., {Riddle}, R.,
  {Atkinson}, D., {Schonhut}, J., \& {Crepp}, J., 2016.
\newblock {Robo-AO Kepler Planetary Candidate Survey. II. Adaptive Optics
  Imaging of 969 Kepler Exoplanet Candidate Host Stars}, {\it \aj\/}, {\bf
  152}(1), 18.

\bibitem[{Barrag{\'a}n} et~al.(2019){Barrag{\'a}n}, {Aigrain}, {Kubyshkina},
  {Gandolfi}, {Livingston}, {Fridlund}, {Fossati}, {Korth}, {Parviainen},
  {Malavolta}, {Palle}, {Deeg}, {Nowak}, {Rajpaul}, {Zicher}, {Antoniciello},
  {Narita}, {Albrecht}, {Bedin}, {Cabrera}, {Cochran}, {de Leon},
  {Eigm{\"u}ller}, {Fukui}, {Granata}, {Grziwa}, {Guenther}, {Hatzes},
  {Kusakabe}, {Latham}, {Libralato}, {Luque},
  {Monta{\~n}{\'e}s-Rodr{\'\i}guez}, {Murgas}, {Nardiello}, {Pagano}, {Piotto},
  {Persson}, {Redfield}, \& {Tamura}]{Barragan2019}
{Barrag{\'a}n}, O., {Aigrain}, S., {Kubyshkina}, D., {Gandolfi}, D.,
  {Livingston}, J., {Fridlund}, M.~C.~V., {Fossati}, L., {Korth}, J.,
  {Parviainen}, H., {Malavolta}, L., {Palle}, E., {Deeg}, H.~J., {Nowak}, G.,
  {Rajpaul}, V.~M., {Zicher}, N., {Antoniciello}, G., {Narita}, N., {Albrecht},
  S., {Bedin}, L.~R., {Cabrera}, J., {Cochran}, W.~D., {de Leon}, J.,
  {Eigm{\"u}ller}, P., {Fukui}, A., {Granata}, V., {Grziwa}, S., {Guenther},
  E., {Hatzes}, A.~P., {Kusakabe}, N., {Latham}, D.~W., {Libralato}, M.,
  {Luque}, R., {Monta{\~n}{\'e}s-Rodr{\'\i}guez}, P., {Murgas}, F.,
  {Nardiello}, D., {Pagano}, I., {Piotto}, G., {Persson}, C.~M., {Redfield},
  S., \& {Tamura}, M., 2019.
\newblock {Radial velocity confirmation of K2-100b: a young, highly irradiated,
  and low-density transiting hot Neptune}, {\it \mnras\/}, {\bf 490}(1),
  698--708.

\bibitem[{Barros} et~al.(2013){Barros}, {Bou{\'e}}, {Gibson}, {Pollacco},
  {Santerne}, {Keenan}, {Skillen}, \& {Street}]{Barros-13}
{Barros}, S.~C.~C., {Bou{\'e}}, G., {Gibson}, N.~P., {Pollacco}, D.~L.,
  {Santerne}, A., {Keenan}, F.~P., {Skillen}, I., \& {Street}, R.~A., 2013.
\newblock {Transit timing variations in WASP-10b induced by stellar activity},
  {\it \mnras\/}, {\bf 430}, 3032--3047.

\bibitem[{Barstow} \& {Irwin}(2016)]{Barstow_habitable_2016}
{Barstow}, J.~K. \& {Irwin}, P.~G.~J., 2016.
\newblock {Habitable worlds with JWST: transit spectroscopy of the TRAPPIST-1
  system?}, {\it \mnras\/}, {\bf 461}(1), L92--L96.

\bibitem[{Basri}(2021)]{Basri2021}
{Basri}, G., 2021.
\newblock {\it {An Introduction to Stellar Magnetic Activity}\/}.

\bibitem[{Basri} \& {Shah}(2020)]{Basri2020}
{Basri}, G. \& {Shah}, R., 2020.
\newblock {The Information Content in Analytic Spot Models of Broadband
  Precision Light Curves. II. Spot Distributions and Lifetimes and Global and
  Differential Rotation}, {\it \apj\/}, {\bf 901}(1), 14.

\bibitem[{Bastien} et~al.(2013){Bastien}, {Stassun}, {Basri}, \&
  {Pepper}]{Bastien:2013}
{Bastien}, F.~A., {Stassun}, K.~G., {Basri}, G., \& {Pepper}, J., 2013.
\newblock {An observational correlation between stellar brightness variations
  and surface gravity}, {\it \nat\/}, {\bf 500}(7463), 427--430.

\bibitem[{Bastien} et~al.(2016){Bastien}, {Stassun}, {Basri}, \&
  {Pepper}]{Bastien:2016}
{Bastien}, F.~A., {Stassun}, K.~G., {Basri}, G., \& {Pepper}, J., 2016.
\newblock {A Granulation ``Flicker''-based Measure of Stellar Surface Gravity},
  {\it \apj\/}, {\bf 818}(1), 43.

\bibitem[{Bazilevskaya} et~al.(2014){Bazilevskaya}, {Cliver}, {Kovaltsov},
  {Ling}, {Shea}, {Smart}, \& {Usoskin}]{Bazilevskaya:2014}
{Bazilevskaya}, G.~A., {Cliver}, E.~W., {Kovaltsov}, G.~A., {Ling}, A.~G.,
  {Shea}, M.~A., {Smart}, D.~F., \& {Usoskin}, I.~G., 2014.
\newblock {Solar Cycle in the Heliosphere and Cosmic Rays}, {\it \ssr\/}, {\bf
  186}(1-4), 409--435.

\bibitem[{Bazot} et~al.(2011){Bazot}, {Ireland}, {Huber}, {Bedding},
  {Broomhall}, {Campante}, {Carfantan}, {Chaplin}, {Elsworth}, {Mel{\'e}ndez},
  {Petit}, {Th{\'e}ado}, {Van Grootel}, {Arentoft}, {Asplund}, {Castro},
  {Christensen-Dalsgaard}, {Do Nascimento}, {Dintrans}, {Dumusque}, {Kjeldsen},
  {McAlister}, {Metcalfe}, {Monteiro}, {Santos}, {Sousa}, {Sturmann},
  {Sturmann}, {ten Brummelaar}, {Turner}, \& {Vauclair}]{Bazot2011}
{Bazot}, M., {Ireland}, M.~J., {Huber}, D., {Bedding}, T.~R., {Broomhall},
  A.~M., {Campante}, T.~L., {Carfantan}, H., {Chaplin}, W.~J., {Elsworth}, Y.,
  {Mel{\'e}ndez}, J., {Petit}, P., {Th{\'e}ado}, S., {Van Grootel}, V.,
  {Arentoft}, T., {Asplund}, M., {Castro}, M., {Christensen-Dalsgaard}, J., {Do
  Nascimento}, J.~D., {Dintrans}, B., {Dumusque}, X., {Kjeldsen}, H.,
  {McAlister}, H.~A., {Metcalfe}, T.~S., {Monteiro}, M.~J.~P.~F.~G., {Santos},
  N.~C., {Sousa}, S., {Sturmann}, J., {Sturmann}, L., {ten Brummelaar}, T.~A.,
  {Turner}, N., \& {Vauclair}, S., 2011.
\newblock {The radius and mass of the close solar twin 18 Scorpii derived from
  asteroseismology and interferometry}, {\it \aap\/}, {\bf 526}, L4.

\bibitem[{Bean} et~al.(2010){Bean}, {Miller-Ricci Kempton}, \&
  {Homeier}]{Bean2010}
{Bean}, J.~L., {Miller-Ricci Kempton}, E., \& {Homeier}, D., 2010.
\newblock {A ground-based transmission spectrum of the super-Earth exoplanet GJ
  1214b}, {\it \nat\/}, {\bf 468}(7324), 669--672.

\bibitem[{Bean} et~al.(2018){Bean}, {Stevenson}, {Batalha}, {Berta-Thompson},
  {Kreidberg}, {Crouzet}, {Benneke}, {Line}, {Sing}, {Wakeford}, {Knutson},
  {Kempton}, {D{\'e}sert}, {Crossfield}, {Batalha}, {de Wit}, {Parmentier},
  {Harrington}, {Moses}, {Lopez-Morales}, {Alam}, {Blecic}, {Bruno}, {Carter},
  {Chapman}, {Decin}, {Dragomir}, {Evans}, {Fortney}, {Fraine}, {Gao},
  {Garc{\'\i}a Mu{\~n}oz}, {Gibson}, {Goyal}, {Heng}, {Hu}, {Kendrew},
  {Kilpatrick}, {Krick}, {Lagage}, {Lendl}, {Louden}, {Madhusudhan}, {Mandell},
  {Mansfield}, {May}, {Morello}, {Morley}, {Nikolov}, {Redfield}, {Roberts},
  {Schlawin}, {Spake}, {Todorov}, {Tsiaras}, {Venot}, {Waalkes}, {Wheatley},
  {Zellem}, {Angerhausen}, {Barrado}, {Carone}, {Casewell}, {Cubillos},
  {Damiano}, {de Val-Borro}, {Drummond}, {Edwards}, {Endl}, {Espinoza},
  {France}, {Gizis}, {Greene}, {Henning}, {Hong}, {Ingalls}, {Iro}, {Irwin},
  {Kataria}, {Lahuis}, {Leconte}, {Lillo-Box}, {Lines}, {Lothringer},
  {Mancini}, {Marchis}, {Mayne}, {Palle}, {Rauscher}, {Roudier}, {Shkolnik},
  {Southworth}, {Swain}, {Taylor}, {Teske}, {Tinetti}, {Tremblin}, {Tucker},
  {van Boekel}, {Waldmann}, {Weaver}, \& {Zingales}]{Bean2018}
{Bean}, J.~L., {Stevenson}, K.~B., {Batalha}, N.~M., {Berta-Thompson}, Z.,
  {Kreidberg}, L., {Crouzet}, N., {Benneke}, B., {Line}, M.~R., {Sing}, D.~K.,
  {Wakeford}, H.~R., {Knutson}, H.~A., {Kempton}, E. M.~R., {D{\'e}sert},
  J.-M., {Crossfield}, I., {Batalha}, N.~E., {de Wit}, J., {Parmentier}, V.,
  {Harrington}, J., {Moses}, J.~I., {Lopez-Morales}, M., {Alam}, M.~K.,
  {Blecic}, J., {Bruno}, G., {Carter}, A.~L., {Chapman}, J.~W., {Decin}, L.,
  {Dragomir}, D., {Evans}, T.~M., {Fortney}, J.~J., {Fraine}, J.~D., {Gao}, P.,
  {Garc{\'\i}a Mu{\~n}oz}, A., {Gibson}, N.~P., {Goyal}, J.~M., {Heng}, K.,
  {Hu}, R., {Kendrew}, S., {Kilpatrick}, B.~M., {Krick}, J., {Lagage}, P.-O.,
  {Lendl}, M., {Louden}, T., {Madhusudhan}, N., {Mandell}, A.~M., {Mansfield},
  M., {May}, E.~M., {Morello}, G., {Morley}, C.~V., {Nikolov}, N., {Redfield},
  S., {Roberts}, J.~E., {Schlawin}, E., {Spake}, J.~J., {Todorov}, K.~O.,
  {Tsiaras}, A., {Venot}, O., {Waalkes}, W.~C., {Wheatley}, P.~J., {Zellem},
  R.~T., {Angerhausen}, D., {Barrado}, D., {Carone}, L., {Casewell}, S.~L.,
  {Cubillos}, P.~E., {Damiano}, M., {de Val-Borro}, M., {Drummond}, B.,
  {Edwards}, B., {Endl}, M., {Espinoza}, N., {France}, K., {Gizis}, J.~E.,
  {Greene}, T.~P., {Henning}, T.~K., {Hong}, Y., {Ingalls}, J.~G., {Iro}, N.,
  {Irwin}, P. G.~J., {Kataria}, T., {Lahuis}, F., {Leconte}, J., {Lillo-Box},
  J., {Lines}, S., {Lothringer}, J.~D., {Mancini}, L., {Marchis}, F., {Mayne},
  N., {Palle}, E., {Rauscher}, E., {Roudier}, G., {Shkolnik}, E.~L.,
  {Southworth}, J., {Swain}, M.~R., {Taylor}, J., {Teske}, J., {Tinetti}, G.,
  {Tremblin}, P., {Tucker}, G.~S., {van Boekel}, R., {Waldmann}, I.~P.,
  {Weaver}, I.~C., \& {Zingales}, T., 2018.
\newblock {The Transiting Exoplanet Community Early Release Science Program for
  JWST}, {\it Publications of the Astronomical Society of the Pacific\/}, {\bf
  130}(993), 114402.

\bibitem[{Beckers} \& {Schr{\"o}ter}(1968)]{Beckers-Schroeter1968}
{Beckers}, J.~M. \& {Schr{\"o}ter}, E.~H., 1968.
\newblock {The Intensity, Velocity and Magnetic Structure of a Sunspot Region.
  II: Some Properties of Umbral Dots}, {\it \solphys\/}, {\bf 4}(3), 303--314.

\bibitem[{Beeck}(2014)]{Beeck_2014}
{Beeck}, B., 2014.
\newblock {\it {Simulations of magnetoconvection in cool main-sequence
  stars}\/}, Ph.D. thesis, Georg-August-UniversitGoettingen.

\bibitem[{Beeck} et~al.(2011){Beeck}, {Sch{\"u}ssler}, \&
  {Reiners}]{Beeck_2011}
{Beeck}, B., {Sch{\"u}ssler}, M., \& {Reiners}, A., 2011.
\newblock {MHD Simulations Reveal Crucial Differences Between Solar and Very
  Cool Star Magnetic Structures}, in {\em 16th Cambridge Workshop on Cool
  Stars, Stellar Systems, and the Sun\/}, vol. 448 of {\bf Astronomical Society
  of the Pacific Conference Series}, p. 1071.

\bibitem[{Beeck} et~al.(2012){Beeck}, {Collet}, {Steffen}, {Asplund},
  {Cameron}, {Freytag}, {Hayek}, {Ludwig}, \&
  {Sch{\"u}ssler}]{Beeck2012codecomparison}
{Beeck}, B., {Collet}, R., {Steffen}, M., {Asplund}, M., {Cameron}, R.~H.,
  {Freytag}, B., {Hayek}, W., {Ludwig}, H.~G., \& {Sch{\"u}ssler}, M., 2012.
\newblock {Simulations of the solar near-surface layers with the CO5BOLD,
  MURaM, and Stagger codes}, {\it \aap\/}, {\bf 539}, A121.

\bibitem[{Beeck} et~al.(2013{\natexlab{a}}){Beeck}, {Cameron}, {Reiners}, \&
  {Sch{\"u}ssler}]{Beeck1}
{Beeck}, B., {Cameron}, R.~H., {Reiners}, A., \& {Sch{\"u}ssler}, M.,
  2013{\natexlab{a}}.
\newblock {Three-dimensional simulations of near-surface convection in
  main-sequence stars. I. Overall structure}, {\it \aap\/}, {\bf 558}, A48.

\bibitem[{Beeck} et~al.(2013{\natexlab{b}}){Beeck}, {Cameron}, {Reiners}, \&
  {Sch{\"u}ssler}]{Beeck2}
{Beeck}, B., {Cameron}, R.~H., {Reiners}, A., \& {Sch{\"u}ssler}, M.,
  2013{\natexlab{b}}.
\newblock {Three-dimensional simulations of near-surface convection in
  main-sequence stars. II. Properties of granulation and spectral lines}, {\it
  \aap\/}, {\bf 558}, A49.

\bibitem[{Beeck} et~al.(2015{\natexlab{a}}){Beeck}, {Sch{\"u}ssler}, {Cameron},
  \& {Reiners}]{Beeck3}
{Beeck}, B., {Sch{\"u}ssler}, M., {Cameron}, R.~H., \& {Reiners}, A.,
  2015{\natexlab{a}}.
\newblock {Three-dimensional simulations of near-surface convection in
  main-sequence stars. III. The structure of small-scale magnetic flux
  concentrations}, {\it \aap\/}, {\bf 581}, A42.

\bibitem[{Beeck} et~al.(2015{\natexlab{b}}){Beeck}, {Sch{\"u}ssler}, {Cameron},
  \& {Reiners}]{Beeck4}
{Beeck}, B., {Sch{\"u}ssler}, M., {Cameron}, R.~H., \& {Reiners}, A.,
  2015{\natexlab{b}}.
\newblock {Three-dimensional simulations of near-surface convection in
  main-sequence stars. IV. Effect of small-scale magnetic flux concentrations
  on centre-to-limb variation and spectral lines}, {\it \aap\/}, {\bf 581},
  A43.

\bibitem[{B{\'e}ky} et~al.(2014){B{\'e}ky}, {Kipping}, \& {Holman}]{Beky-14}
{B{\'e}ky}, B., {Kipping}, D.~M., \& {Holman}, M.~J., 2014.
\newblock {SPOTROD: a semi-analytic model for transits of spotted stars}, {\it
  MNRAS\/}, {\bf 442}, 3686--3699.

\bibitem[{Bellot Rubio} \& {Orozco
  Su{\'a}rez}(2019)]{BellotRubio-OrozcoSuarez2019}
{Bellot Rubio}, L. \& {Orozco Su{\'a}rez}, D., 2019.
\newblock {Quiet Sun magnetic fields: an observational view}, {\it Living
  Reviews in Solar Physics\/}, {\bf 16}(1), 1.

\bibitem[{Benneke} et~al.(2019{\natexlab{a}}){Benneke}, {Knutson},
  {Lothringer}, {Crossfield}, {Moses}, {Morley}, {Kreidberg}, {Fulton},
  {Dragomir}, {Howard}, {Wong}, {D{\'e}sert}, {McCullough}, {Kempton},
  {Fortney}, {Gilliland}, {Deming}, \& {Kammer}]{Benneke2019_GJ3470b}
{Benneke}, B., {Knutson}, H.~A., {Lothringer}, J., {Crossfield}, I. J.~M.,
  {Moses}, J.~I., {Morley}, C., {Kreidberg}, L., {Fulton}, B.~J., {Dragomir},
  D., {Howard}, A.~W., {Wong}, I., {D{\'e}sert}, J.-M., {McCullough}, P.~R.,
  {Kempton}, E. M.~R., {Fortney}, J., {Gilliland}, R., {Deming}, D., \&
  {Kammer}, J., 2019{\natexlab{a}}.
\newblock {A sub-Neptune exoplanet with a low-metallicity methane-depleted
  atmosphere and Mie-scattering clouds}, {\it Nature Astronomy\/}, {\bf 3},
  813--821.

\bibitem[{Benneke} et~al.(2019{\natexlab{b}}){Benneke}, {Wong}, {Piaulet},
  {Knutson}, {Lothringer}, {Morley}, {Crossfield}, {Gao}, {Greene}, {Dressing},
  {Dragomir}, {Howard}, {McCullough}, {Kempton}, {Fortney}, \&
  {Fraine}]{Benneke2019_K2-18b}
{Benneke}, B., {Wong}, I., {Piaulet}, C., {Knutson}, H.~A., {Lothringer}, J.,
  {Morley}, C.~V., {Crossfield}, I. J.~M., {Gao}, P., {Greene}, T.~P.,
  {Dressing}, C., {Dragomir}, D., {Howard}, A.~W., {McCullough}, P.~R.,
  {Kempton}, E. M.~R., {Fortney}, J.~J., \& {Fraine}, J., 2019{\natexlab{b}}.
\newblock {Water Vapor and Clouds on the Habitable-zone Sub-Neptune Exoplanet
  K2-18b}, {\it \apjl\/}, {\bf 887}(1), L14.

\bibitem[{Benz} et~al.(2021){Benz}, {Broeg}, {Fortier}, {Rando}, {Beck},
  {Beck}, {Queloz}, {Ehrenreich}, {Maxted}, {Isaak}, {Billot}, {Alibert},
  {Alonso}, {Ant{\'o}nio}, {Asquier}, {Bandy}, {B{\'a}rczy}, {Barrado},
  {Barros}, {Baumjohann}, {Bekkelien}, {Bergomi}, {Biondi}, {Bonfils},
  {Borsato}, {Brandeker}, {Busch}, {Cabrera}, {Cessa}, {Charnoz}, {Chazelas},
  {Collier Cameron}, {Corral Van Damme}, {Cortes}, {Davies}, {Deleuil},
  {Deline}, {Delrez}, {Demangeon}, {Demory}, {Erikson}, {Farinato}, {Fossati},
  {Fridlund}, {Futyan}, {Gandolfi}, {Garcia Munoz}, {Gillon}, {Guterman},
  {Gutierrez}, {Hasiba}, {Heng}, {Hernandez}, {Hoyer}, {Kiss}, {Kovacs},
  {Kuntzer}, {Laskar}, {Lecavelier des Etangs}, {Lendl}, {L{\'o}pez}, {Lora},
  {Lovis}, {L{\"u}ftinger}, {Magrin}, {Malvasio}, {Marafatto}, {Michaelis}, {de
  Miguel}, {Modrego}, {Munari}, {Nascimbeni}, {Olofsson}, {Ottacher},
  {Ottensamer}, {Pagano}, {Palacios}, {Pall{\'e}}, {Peter}, {Piazza}, {Piotto},
  {Pizarro}, {Pollaco}, {Ragazzoni}, {Ratti}, {Rauer}, {Ribas}, {Rieder},
  {Rohlfs}, {Safa}, {Salatti}, {Santos}, {Scandariato}, {S{\'e}gransan},
  {Simon}, {Smith}, {Sordet}, {Sousa}, {Steller}, {Szab{\'o}}, {Szoke},
  {Thomas}, {Tschentscher}, {Udry}, {Van Grootel}, {Viotto}, {Walter},
  {Walton}, {Wildi}, \& {Wolter}]{Benz2021}
{Benz}, W., {Broeg}, C., {Fortier}, A., {Rando}, N., {Beck}, T., {Beck}, M.,
  {Queloz}, D., {Ehrenreich}, D., {Maxted}, P.~F.~L., {Isaak}, K.~G., {Billot},
  N., {Alibert}, Y., {Alonso}, R., {Ant{\'o}nio}, C., {Asquier}, J., {Bandy},
  T., {B{\'a}rczy}, T., {Barrado}, D., {Barros}, S.~C.~C., {Baumjohann}, W.,
  {Bekkelien}, A., {Bergomi}, M., {Biondi}, F., {Bonfils}, X., {Borsato}, L.,
  {Brandeker}, A., {Busch}, M.~D., {Cabrera}, J., {Cessa}, V., {Charnoz}, S.,
  {Chazelas}, B., {Collier Cameron}, A., {Corral Van Damme}, C., {Cortes}, D.,
  {Davies}, M.~B., {Deleuil}, M., {Deline}, A., {Delrez}, L., {Demangeon}, O.,
  {Demory}, B.~O., {Erikson}, A., {Farinato}, J., {Fossati}, L., {Fridlund},
  M., {Futyan}, D., {Gandolfi}, D., {Garcia Munoz}, A., {Gillon}, M.,
  {Guterman}, P., {Gutierrez}, A., {Hasiba}, J., {Heng}, K., {Hernandez}, E.,
  {Hoyer}, S., {Kiss}, L.~L., {Kovacs}, Z., {Kuntzer}, T., {Laskar}, J.,
  {Lecavelier des Etangs}, A., {Lendl}, M., {L{\'o}pez}, A., {Lora}, I.,
  {Lovis}, C., {L{\"u}ftinger}, T., {Magrin}, D., {Malvasio}, L., {Marafatto},
  L., {Michaelis}, H., {de Miguel}, D., {Modrego}, D., {Munari}, M.,
  {Nascimbeni}, V., {Olofsson}, G., {Ottacher}, H., {Ottensamer}, R., {Pagano},
  I., {Palacios}, R., {Pall{\'e}}, E., {Peter}, G., {Piazza}, D., {Piotto}, G.,
  {Pizarro}, A., {Pollaco}, D., {Ragazzoni}, R., {Ratti}, F., {Rauer}, H.,
  {Ribas}, I., {Rieder}, M., {Rohlfs}, R., {Safa}, F., {Salatti}, M., {Santos},
  N.~C., {Scandariato}, G., {S{\'e}gransan}, D., {Simon}, A.~E., {Smith},
  A.~M.~S., {Sordet}, M., {Sousa}, S.~G., {Steller}, M., {Szab{\'o}}, G.~M.,
  {Szoke}, J., {Thomas}, N., {Tschentscher}, M., {Udry}, S., {Van Grootel}, V.,
  {Viotto}, V., {Walter}, I., {Walton}, N.~A., {Wildi}, F., \& {Wolter}, D.,
  2021.
\newblock {The CHEOPS mission}, {\it Experimental Astronomy\/}, {\bf 51}(1),
  109--151.

\bibitem[{Berdyugina}(1998)]{Berdyugina1998di}
{Berdyugina}, S.~V., 1998.
\newblock {Surface imaging by the Occamian approach. Basic principles,
  simulations, and tests}, {\it \aap\/}, {\bf 338}, 97--105.

\bibitem[{Berdyugina}(2002)]{Berdyugina-02}
{Berdyugina}, S.~V., 2002.
\newblock {Sunspot and starspot interiors as seen from molecular lines}, {\it
  Astronomische Nachrichten\/}, {\bf 323}, 192--195.

\bibitem[{Berdyugina}(2005)]{Berdyugina2005}
{Berdyugina}, S.~V., 2005.
\newblock {Starspots: A Key to the Stellar Dynamo}, {\it Living Reviews in
  Solar Physics\/}, {\bf 2}(1), 8.

\bibitem[{Berdyugina}(2009)]{Berdyugina2009maghrd}
{Berdyugina}, S.~V., 2009.
\newblock {Stellar magnetic fields across the H-R diagram: observational
  evidence}, in {\em Cosmic Magnetic Fields: From Planets, to Stars and
  Galaxies\/}, vol. 259, pp. 323--332.

\bibitem[{Berdyugina}(2011)]{Berdyugina2011}
{Berdyugina}, S.~V., 2011.
\newblock {Polarimetry of Cool Atmospheres: From the Sun to Exoplanets}, in
  {\em Solar Polarization 6\/}, vol. 437 of {\bf Astronomical Society of the
  Pacific Conference Series}, p. 219.

\bibitem[{Berdyugina} \& {Fluri}(2004)]{Berdyugina-Fluri2004}
{Berdyugina}, S.~V. \& {Fluri}, D.~M., 2004.
\newblock {Evidence for the Hanle effect in molecular lines}, {\it \aap\/},
  {\bf 417}, 775--784.

\bibitem[{Berdyugina} \& {Henry}(2007)]{Berdyugina-Henry2007hr1099}
{Berdyugina}, S.~V. \& {Henry}, G.~W., 2007.
\newblock {Butterfly Diagram and Activity Cycles in HR 1099}, {\it \apjl\/},
  {\bf 659}(2), L157--L160.

\bibitem[{Berdyugina} \& {Kuhn}(2019)]{Berdyugina2019}
{Berdyugina}, S.~V. \& {Kuhn}, J.~R., 2019.
\newblock {Surface Imaging of Proxima b and Other Exoplanets: Albedo Maps,
  Biosignatures, and Technosignatures}, {\it \aj\/}, {\bf 158}(6), 246.

\bibitem[{Berdyugina} \& {Solanki}(2002)]{Berdyugina2002mol}
{Berdyugina}, S.~V. \& {Solanki}, S.~K., 2002.
\newblock {The molecular Zeeman effect and diagnostics of solar and stellar
  magnetic fields. I. Theoretical spectral patterns in the Zeeman regime}, {\it
  \aap\/}, {\bf 385}, 701--715.

\bibitem[{Berdyugina} \& {Tuominen}(1998)]{Berdyugina1998al}
{Berdyugina}, S.~V. \& {Tuominen}, I., 1998.
\newblock {Permanent active longitudes and activity cycles on RS CVn stars},
  {\it \aap\/}, {\bf 336}, L25--L28.

\bibitem[{Berdyugina} et~al.(1998){Berdyugina}, {Jankov}, {Ilyin}, {Tuominen},
  \& {Fekel}]{Berdyugina1998iipeg}
{Berdyugina}, S.~V., {Jankov}, S., {Ilyin}, I., {Tuominen}, I., \& {Fekel},
  F.~C., 1998.
\newblock {The active RS Canum Venaticorum binary II Pegasi. I. Stellar and
  orbital parameters}, {\it \aap\/}, {\bf 334}, 863--872.

\bibitem[{Berdyugina} et~al.(1999){Berdyugina}, {Ilyin}, \&
  {Tuominen}]{Berdyugina1999impeg}
{Berdyugina}, S.~V., {Ilyin}, I., \& {Tuominen}, I., 1999.
\newblock {The long-period RS Canum Venaticorum binary IM Pegasi. I. Orbital
  and stellar parameters}, {\it \aap\/}, {\bf 347}, 932--936.

\bibitem[{Berdyugina} et~al.(2000){Berdyugina}, {Frutiger}, {Solanki}, \&
  {Livingstone}]{Berdyugina2000mol}
{Berdyugina}, S.~V., {Frutiger}, C., {Solanki}, S.~K., \& {Livingstone}, W.,
  2000.
\newblock {Successful spectral synthesis of Zeeman-split molecular bands in
  sunspot spectra}, {\it \aap\/}, {\bf 364}, L101--L104.

\bibitem[{Berdyugina} et~al.(2002){Berdyugina}, {Pelt}, \&
  {Tuominen}]{Berdyugina2002lqhya}
{Berdyugina}, S.~V., {Pelt}, J., \& {Tuominen}, I., 2002.
\newblock {Magnetic activity in the young solar analog LQ Hydrae. I. Active
  longitudes and cycles}, {\it \aap\/}, {\bf 394}, 505--515.

\bibitem[{Berdyugina} et~al.(2003){Berdyugina}, {Solanki}, \&
  {Frutiger}]{Berdyugina2003mol}
{Berdyugina}, S.~V., {Solanki}, S.~K., \& {Frutiger}, C., 2003.
\newblock {The molecular Zeeman effect and diagnostics of solar and stellar
  magnetic fields. II. Synthetic Stokes profiles in the Zeeman regime}, {\it
  \aap\/}, {\bf 412}, 513--527.

\bibitem[{Berdyugina} et~al.(2005){Berdyugina}, {Braun}, {Fluri}, \&
  {Solanki}]{Berdyugina2005mol}
{Berdyugina}, S.~V., {Braun}, P.~A., {Fluri}, D.~M., \& {Solanki}, S.~K., 2005.
\newblock {The molecular Zeeman effect and diagnostics of solar and stellar
  magnetic fields. III. Theoretical spectral patterns in the Paschen-Back
  regime}, {\it \aap\/}, {\bf 444}(3), 947--960.

\bibitem[{Berdyugina} et~al.(2006){Berdyugina}, {Petit}, {Fluri}, {Afram}, \&
  {Arnaud}]{Berdyugina2006tio}
{Berdyugina}, S.~V., {Petit}, P., {Fluri}, D.~M., {Afram}, N., \& {Arnaud}, J.,
  2006.
\newblock {Detection of the Molecular Zeeman Effect in Circular Polarization on
  Cool Active Stars}, in {\em Solar Polarization 4\/}, vol. 358 of {\bf
  Astronomical Society of the Pacific Conference Series}, p. 381.

\bibitem[{Berdyugina} et~al.(2017){Berdyugina}, {Harrington}, {Kuzmychov},
  {Kuhn}, {Hallinan}, {Kowalski}, \& {Hawley}]{Berdyugina2017bdw}
{Berdyugina}, S.~V., {Harrington}, D.~M., {Kuzmychov}, O., {Kuhn}, J.~R.,
  {Hallinan}, G., {Kowalski}, A.~F., \& {Hawley}, S.~L., 2017.
\newblock {First Detection of a Strong Magnetic Field on a Bursty Brown Dwarf:
  Puzzle Solved}, {\it \apj\/}, {\bf 847}(1), 61.

\bibitem[{Berger} \& {Berdyugina}(2003)]{Berger-Berdyugina2003}
{Berger}, T.~E. \& {Berdyugina}, S.~V., 2003.
\newblock {The Observation of Sunspot Light-Bridge Structure and Dynamics},
  {\it \apjl\/}, {\bf 589}(2), L117--L121.

\bibitem[{Berger} \& {Title}(2001)]{berger2001}
{Berger}, T.~E. \& {Title}, A.~M., 2001.
\newblock {On the Relation of G-Band Bright Points to the Photospheric Magnetic
  Field}, {\it \apj\/}, {\bf 553}(1), 449--469.

\bibitem[{Berger} et~al.(2007){Berger}, {Rouppe van der Voort}, \&
  {L{\"o}fdahl}]{Berger2007}
{Berger}, T.~E., {Rouppe van der Voort}, L., \& {L{\"o}fdahl}, M., 2007.
\newblock {Contrast Analysis of Solar Faculae and Magnetic Bright Points}, {\it
  \apj\/}, {\bf 661}(2), 1272--1288.

\bibitem[{Berta} et~al.(2011){Berta}, {Charbonneau}, {Bean}, {Irwin}, {Burke},
  {D{\'e}sert}, {Nutzman}, \& {Falco}]{Berta2011}
{Berta}, Z.~K., {Charbonneau}, D., {Bean}, J., {Irwin}, J., {Burke}, C.~J.,
  {D{\'e}sert}, J.-M., {Nutzman}, P., \& {Falco}, E.~E., 2011.
\newblock {The GJ1214 Super-Earth System: Stellar Variability, New Transits,
  and a Search for Additional Planets}, {\it \apj\/}, {\bf 736}(1), 12.

\bibitem[{Berta} et~al.(2012){Berta}, {Charbonneau}, {D{\'e}sert},
  {Miller-Ricci Kempton}, {McCullough}, {Burke}, {Fortney}, {Irwin}, {Nutzman},
  \& {Homeier}]{Berta2012}
{Berta}, Z.~K., {Charbonneau}, D., {D{\'e}sert}, J.-M., {Miller-Ricci Kempton},
  E., {McCullough}, P.~R., {Burke}, C.~J., {Fortney}, J.~J., {Irwin}, J.,
  {Nutzman}, P., \& {Homeier}, D., 2012.
\newblock {The Flat Transmission Spectrum of the Super-Earth GJ1214b from Wide
  Field Camera 3 on the Hubble Space Telescope}, {\it \apj\/}, {\bf 747}(1),
  35.

\bibitem[{Beskin} et~al.(1988){Beskin}, {Chekh}, {Gershberg}, {Mitronova},
  {Neizvestnyi}, {Plakhotnichenko}, {Pustilnik}, {Shvartsman}, \&
  {Zhurovkov}]{Beskin1988}
{Beskin}, G.~M., {Chekh}, S.~A., {Gershberg}, R.~E., {Mitronova}, S.~N.,
  {Neizvestnyi}, S.~I., {Plakhotnichenko}, V.~L., {Pustilnik}, L.~A.,
  {Shvartsman}, V.~F., \& {Zhurovkov}, A.~V., 1988.
\newblock {A Search at 0.3-MICROSECOND Resolution for Fine Structure in Uv-Ceti
  Type Flares}, {\it Soviet Astronomy Letters\/}, {\bf 14}, 65.

\bibitem[{B{\'e}tr{\'e}mieux} \& {Kaltenegger}(2013)]{BK:2013}
{B{\'e}tr{\'e}mieux}, Y. \& {Kaltenegger}, L., 2013.
\newblock {Transmission Spectrum of Earth as a Transiting Exoplanet from the
  Ultraviolet to the Near-infrared}, {\it \apjl\/}, {\bf 772}(2), L31.

\bibitem[{Bhatia} et~al.(2021){Bhatia}, {Cameron}, {Solanki}, {Peter},
  {Przybylski}, {Witzke}, \& {Shapiro}]{Bhatia2021}
{Bhatia}, T., {Cameron}, R., {Solanki}, S., {Peter}, H., {Przybylski}, D.,
  {Witzke}, V., \& {Shapiro}, A., 2021.
\newblock {Small-scale Dynamo in Cool Main-Sequence Stars: Effect on
  Stratification, Convection and Bolometric Intensity}, in {\em American
  Astronomical Society Meeting Abstracts\/}, vol.~53 of {\bf American
  Astronomical Society Meeting Abstracts}, p. 304.04.

\bibitem[{Bhatia} et~al.(2022){Bhatia}, {Cameron}, {Solanki}, {Peter},
  {Przybylski}, {Witzke}, \& {Shapiro}]{Bhatia2022}
{Bhatia}, T.~S., {Cameron}, R.~H., {Solanki}, S.~K., {Peter}, H., {Przybylski},
  D., {Witzke}, V., \& {Shapiro}, A., 2022.
\newblock {Small-scale dynamo in cool stars. I. Changes in stratification and
  near-surface convection for main-sequence spectral types}, {\it \aap\/}, {\bf
  663}, A166.

\bibitem[{Biazzo} et~al.(2022){Biazzo}, {D'Orazi}, {Desidera}, {Turrini},
  {Benatti}, {Gratton}, {Magrini}, {Sozzetti}, {Baratella}, {Bonomo}, {Borsa},
  {Claudi}, {Covino}, {Damasso}, {Di Mauro}, {Lanza}, {Maggio}, {Malavolta},
  {Maldonado}, {Marzari}, {Micela}, {Poretti}, {Vitello}, {Affer}, {Bignamini},
  {Carleo}, {Cosentino}, {Fiorenzano}, {Giacobbe}, {Harutyunyan}, {Leto},
  {Mancini}, {Molinari}, {Molinaro}, {Nardiello}, {Nascimbeni}, {Pagano},
  {Pedani}, {Piotto}, {Rainer}, \& {Scandariato}]{Biazzo2022}
{Biazzo}, K., {D'Orazi}, V., {Desidera}, S., {Turrini}, D., {Benatti}, S.,
  {Gratton}, R., {Magrini}, L., {Sozzetti}, A., {Baratella}, M., {Bonomo},
  A.~S., {Borsa}, F., {Claudi}, R., {Covino}, E., {Damasso}, M., {Di Mauro},
  M.~P., {Lanza}, A.~F., {Maggio}, A., {Malavolta}, L., {Maldonado}, J.,
  {Marzari}, F., {Micela}, G., {Poretti}, E., {Vitello}, F., {Affer}, L.,
  {Bignamini}, A., {Carleo}, I., {Cosentino}, R., {Fiorenzano}, A.~F.~M.,
  {Giacobbe}, P., {Harutyunyan}, A., {Leto}, G., {Mancini}, L., {Molinari}, E.,
  {Molinaro}, M., {Nardiello}, D., {Nascimbeni}, V., {Pagano}, I., {Pedani},
  M., {Piotto}, G., {Rainer}, M., \& {Scandariato}, G., 2022.
\newblock {The GAPS Programme at TNG. XXXV. Fundamental properties of
  transiting exoplanet host stars}, {\it \aap\/}, {\bf 664}, A161.

\bibitem[{Biller} et~al.(2013){Biller}, {Crossfield}, {Mancini}, {Ciceri},
  {Southworth}, {Kopytova}, {Bonnefoy}, {Deacon}, {Schlieder}, {Buenzli},
  {Brandner}, {Allard}, {Homeier}, {Freytag}, {Bailer-Jones}, {Greiner},
  {Henning}, \& {Goldman}]{Biller2013}
{Biller}, B.~A., {Crossfield}, I. J.~M., {Mancini}, L., {Ciceri}, S.,
  {Southworth}, J., {Kopytova}, T.~G., {Bonnefoy}, M., {Deacon}, N.~R.,
  {Schlieder}, J.~E., {Buenzli}, E., {Brandner}, W., {Allard}, F., {Homeier},
  D., {Freytag}, B., {Bailer-Jones}, C. A.~L., {Greiner}, J., {Henning}, T., \&
  {Goldman}, B., 2013.
\newblock {Weather on the Nearest Brown Dwarfs: Resolved Simultaneous
  Multi-wavelength Variability Monitoring of WISE J104915.57-531906.1AB}, {\it
  \apjl\/}, {\bf 778}(1), L10.

\bibitem[{Blanco Rodr{\'\i}guez} \& {Kneer}(2010)]{blanco2010}
{Blanco Rodr{\'\i}guez}, J. \& {Kneer}, F., 2010.
\newblock {Faculae at the poles of the Sun revisited: infrared observations},
  {\it \aap\/}, {\bf 509}, A92.

\bibitem[{Bogdan} et~al.(1988){Bogdan}, {Gilman}, {Lerche}, \&
  {Howard}]{Bogdan:1988}
{Bogdan}, T.~J., {Gilman}, P.~A., {Lerche}, I., \& {Howard}, R., 1988.
\newblock {Distribution of Sunspot Umbral Areas: 1917--1982}, {\it \apj\/},
  {\bf 327}, 451.

\bibitem[{Boisse} et~al.(2012){Boisse}, {Bonfils}, \& {Santos}]{Boisse-12}
{Boisse}, I., {Bonfils}, X., \& {Santos}, N.~C., 2012.
\newblock {SOAP. A tool for the fast computation of photometry and radial
  velocity induced by stellar spots}, {\it \aap\/}, {\bf 545}, A109.

\bibitem[{Boldt} et~al.(2020){Boldt}, {Oshagh}, {Dreizler}, {Mallonn},
  {Santos}, {Claret}, {Reiners}, \& {Sedaghati}]{Boldt-20}
{Boldt}, S., {Oshagh}, M., {Dreizler}, S., {Mallonn}, M., {Santos}, N.~C.,
  {Claret}, A., {Reiners}, A., \& {Sedaghati}, E., 2020.
\newblock {Stellar activity consequence on the retrieved transmission spectra
  through chromatic Rossiter-McLaughlin observations}, {\it \aap\/}, {\bf 635},
  A123.

\bibitem[{Boley} et~al.(2020){Boley}, {Van Laerhoven}, \& {Granados
  Contreras}]{Boley2020}
{Boley}, A.~C., {Van Laerhoven}, C., \& {Granados Contreras}, A.~P., 2020.
\newblock {Transit Duration Variations in Multiplanet Systems}, {\it \aj\/},
  {\bf 159}(5), 207.

\bibitem[{Bonifacio} et~al.(2018){Bonifacio}, {Caffau}, {Ludwig}, {Steffen},
  {Castelli}, {Gallagher}, {Ku{\v{c}}inskas}, {Prakapavi{\v{c}}ius}, {Cayrel},
  {Freytag}, {Plez}, \& {Homeier}]{bonifacio2018cfist}
{Bonifacio}, P., {Caffau}, E., {Ludwig}, H.~G., {Steffen}, M., {Castelli}, F.,
  {Gallagher}, A.~J., {Ku{\v{c}}inskas}, A., {Prakapavi{\v{c}}ius}, D.,
  {Cayrel}, R., {Freytag}, B., {Plez}, B., \& {Homeier}, D., 2018.
\newblock {Using the CIFIST grid of CO$^{5}$BOLD 3D model atmospheres to study
  the effects of stellar granulation on photometric colours. I. Grids of 3D
  corrections in the UBVRI, 2MASS, HIPPARCOS, Gaia, and SDSS systems}, {\it
  \aap\/}, {\bf 611}, A68.

\bibitem[{Bopp} \& {Evans}(1973)]{BoppEvans1973}
{Bopp}, B.~W. \& {Evans}, D.~S., 1973.
\newblock {The spotted flare stars BY Dra and CC Eri: a model for the spots and
  some astrophysical implications.}, {\it \mnras\/}, {\bf 164}, 343--356.

\bibitem[{Borrero} \& {Ichimoto}(2011)]{Borrero:2011}
{Borrero}, J.~M. \& {Ichimoto}, K., 2011.
\newblock {Magnetic Structure of Sunspots}, {\it Living Reviews in Solar
  Physics\/}, {\bf 8}(1), 4.

\bibitem[{Borucki} et~al.(2010){Borucki}, {Koch}, {Basri}, {Batalha}, {Brown},
  {Caldwell}, {Caldwell}, {Christensen-Dalsgaard}, {Cochran}, {DeVore},
  {Dunham}, {Dupree}, {Gautier}, {Geary}, {Gilliland}, {Gould}, {Howell},
  {Jenkins}, {Kondo}, {Latham}, {Marcy}, {Meibom}, {Kjeldsen}, {Lissauer},
  {Monet}, {Morrison}, {Sasselov}, {Tarter}, {Boss}, {Brownlee}, {Owen},
  {Buzasi}, {Charbonneau}, {Doyle}, {Fortney}, {Ford}, {Holman}, {Seager},
  {Steffen}, {Welsh}, {Rowe}, {Anderson}, {Buchhave}, {Ciardi}, {Walkowicz},
  {Sherry}, {Horch}, {Isaacson}, {Everett}, {Fischer}, {Torres}, {Johnson},
  {Endl}, {MacQueen}, {Bryson}, {Dotson}, {Haas}, {Kolodziejczak}, {Van Cleve},
  {Chandrasekaran}, {Twicken}, {Quintana}, {Clarke}, {Allen}, {Li}, {Wu},
  {Tenenbaum}, {Verner}, {Bruhweiler}, {Barnes}, \& {Prsa}]{Borucki2010}
{Borucki}, W.~J., {Koch}, D., {Basri}, G., {Batalha}, N., {Brown}, T.,
  {Caldwell}, D., {Caldwell}, J., {Christensen-Dalsgaard}, J., {Cochran},
  W.~D., {DeVore}, E., {Dunham}, E.~W., {Dupree}, A.~K., {Gautier}, T.~N.,
  {Geary}, J.~C., {Gilliland}, R., {Gould}, A., {Howell}, S.~B., {Jenkins},
  J.~M., {Kondo}, Y., {Latham}, D.~W., {Marcy}, G.~W., {Meibom}, S.,
  {Kjeldsen}, H., {Lissauer}, J.~J., {Monet}, D.~G., {Morrison}, D.,
  {Sasselov}, D., {Tarter}, J., {Boss}, A., {Brownlee}, D., {Owen}, T.,
  {Buzasi}, D., {Charbonneau}, D., {Doyle}, L., {Fortney}, J., {Ford}, E.~B.,
  {Holman}, M.~J., {Seager}, S., {Steffen}, J.~H., {Welsh}, W.~F., {Rowe}, J.,
  {Anderson}, H., {Buchhave}, L., {Ciardi}, D., {Walkowicz}, L., {Sherry}, W.,
  {Horch}, E., {Isaacson}, H., {Everett}, M.~E., {Fischer}, D., {Torres}, G.,
  {Johnson}, J.~A., {Endl}, M., {MacQueen}, P., {Bryson}, S.~T., {Dotson}, J.,
  {Haas}, M., {Kolodziejczak}, J., {Van Cleve}, J., {Chandrasekaran}, H.,
  {Twicken}, J.~D., {Quintana}, E.~V., {Clarke}, B.~D., {Allen}, C., {Li}, J.,
  {Wu}, H., {Tenenbaum}, P., {Verner}, E., {Bruhweiler}, F., {Barnes}, J., \&
  {Prsa}, A., 2010.
\newblock {Kepler Planet-Detection Mission: Introduction and First Results},
  {\it Science\/}, {\bf 327}, 977--.

\bibitem[{Boyajian} et~al.(2012{\natexlab{a}}){Boyajian}, {McAlister}, {van
  Belle}, {Gies}, {ten Brummelaar}, {von Braun}, {Farrington}, {Goldfinger},
  {O'Brien}, {Parks}, {Richardson}, {Ridgway}, {Schaefer}, {Sturmann},
  {Sturmann}, {Touhami}, {Turner}, \& {White}]{Boyajian2012}
{Boyajian}, T.~S., {McAlister}, H.~A., {van Belle}, G., {Gies}, D.~R., {ten
  Brummelaar}, T.~A., {von Braun}, K., {Farrington}, C., {Goldfinger}, P.~J.,
  {O'Brien}, D., {Parks}, J.~R., {Richardson}, N.~D., {Ridgway}, S.,
  {Schaefer}, G., {Sturmann}, L., {Sturmann}, J., {Touhami}, Y., {Turner},
  N.~H., \& {White}, R., 2012{\natexlab{a}}.
\newblock {Stellar Diameters and Temperatures. I. Main-sequence A, F, and G
  Stars}, {\it \apj\/}, {\bf 746}(1), 101.

\bibitem[{Boyajian} et~al.(2012{\natexlab{b}}){Boyajian}, {von Braun}, {van
  Belle}, {McAlister}, {ten Brummelaar}, {Kane}, {Muirhead}, {Jones}, {White},
  {Schaefer}, {Ciardi}, {Henry}, {L{\'o}pez-Morales}, {Ridgway}, {Gies}, {Jao},
  {Rojas-Ayala}, {Parks}, {Sturmann}, {Sturmann}, {Turner}, {Farrington},
  {Goldfinger}, \& {Berger}]{Boyajian2012_KM}
{Boyajian}, T.~S., {von Braun}, K., {van Belle}, G., {McAlister}, H.~A., {ten
  Brummelaar}, T.~A., {Kane}, S.~R., {Muirhead}, P.~S., {Jones}, J., {White},
  R., {Schaefer}, G., {Ciardi}, D., {Henry}, T., {L{\'o}pez-Morales}, M.,
  {Ridgway}, S., {Gies}, D., {Jao}, W.-C., {Rojas-Ayala}, B., {Parks}, J.~R.,
  {Sturmann}, L., {Sturmann}, J., {Turner}, N.~H., {Farrington}, C.,
  {Goldfinger}, P.~J., \& {Berger}, D.~H., 2012{\natexlab{b}}.
\newblock {Stellar Diameters and Temperatures. II. Main-sequence K- and
  M-stars}, {\it \apj\/}, {\bf 757}(2), 112.

\bibitem[{Boyajian} et~al.(2013){Boyajian}, {von Braun}, {van Belle},
  {Farrington}, {Schaefer}, {Jones}, {White}, {McAlister}, {ten Brummelaar},
  {Ridgway}, {Gies}, {Sturmann}, {Sturmann}, {Turner}, {Goldfinger}, \&
  {Vargas}]{Boyajian2013}
{Boyajian}, T.~S., {von Braun}, K., {van Belle}, G., {Farrington}, C.,
  {Schaefer}, G., {Jones}, J., {White}, R., {McAlister}, H.~A., {ten
  Brummelaar}, T.~A., {Ridgway}, S., {Gies}, D., {Sturmann}, L., {Sturmann},
  J., {Turner}, N.~H., {Goldfinger}, P.~J., \& {Vargas}, N., 2013.
\newblock {Stellar Diameters and Temperatures. III. Main-sequence A, F, G, and
  K Stars: Additional High-precision Measurements and Empirical Relations},
  {\it \apj\/}, {\bf 771}(1), 40.

\bibitem[{Bradshaw} \& {Hartigan}(2014)]{BradshawHartigan2014}
{Bradshaw}, S.~J. \& {Hartigan}, P., 2014.
\newblock {On Sunspot and Starspot Lifetimes}, {\it \apj\/}, {\bf 795}(1), 79.

\bibitem[{Brehm} et~al.(2021){Brehm}, {Bayliss}, {Christl}, {Synal}, {Adolphi},
  {Beer}, {Kromer}, {Muscheler}, {Solanki}, {Usoskin}, {Bleicher},
  {Bollhalder}, {Tyers}, \& {Wacker}]{Brehm:2021}
{Brehm}, N., {Bayliss}, A., {Christl}, M., {Synal}, H.-A., {Adolphi}, F.,
  {Beer}, J., {Kromer}, B., {Muscheler}, R., {Solanki}, S.~K., {Usoskin}, I.,
  {Bleicher}, N., {Bollhalder}, S., {Tyers}, C., \& {Wacker}, L., 2021.
\newblock {Eleven-year solar cycles over the last millennium revealed by
  radiocarbon in tree rings}, {\it Nature Geoscience\/}, {\bf 14}(1), 10--15.

\bibitem[{Brown}(2014)]{Brown2014}
{Brown}, D.~J.~A., 2014.
\newblock {Discrepancies between isochrone fitting and gyrochronology for
  exoplanet host stars?}, {\it \mnras\/}, {\bf 442}(2), 1844--1862.

\bibitem[{Brown} et~al.(2001){Brown}, {Charbonneau}, {Gilliland}, {Noyes}, \&
  {Burrows}]{Brown-01}
{Brown}, T.~M., {Charbonneau}, D., {Gilliland}, R.~L., {Noyes}, R.~W., \&
  {Burrows}, A., 2001.
\newblock {Hubble Space Telescope Time-Series Photometry of the Transiting
  Planet of HD 209458}, {\it \apj\/}, {\bf 552}, 699--709.

\bibitem[{Bruno} \& {Deleuil}(2021)]{Bruno2021}
{Bruno}, G. \& {Deleuil}, M., 2021.
\newblock {Stellar activity and transits}, {\it arXiv e-prints\/}, p.
  arXiv:2104.06173.

\bibitem[{Bruno} et~al.(2016){Bruno}, {Deleuil}, {Almenara}, {Barros}, {Lanza},
  {Montalto}, {Boisse}, {Santerne}, {Lagrange}, \& {Meunier}]{Bruno-16}
{Bruno}, G., {Deleuil}, M., {Almenara}, J.~M., {Barros}, S.~C.~C., {Lanza},
  A.~F., {Montalto}, M., {Boisse}, I., {Santerne}, A., {Lagrange}, A.~M., \&
  {Meunier}, N., 2016.
\newblock {Disentangling planetary and stellar activity features in the CoRoT-2
  light curve}, {\it \aap\/}, {\bf 595}, A89.

\bibitem[{Bruno} et~al.(2022){Bruno}, {Lewis}, {Valenti}, {Pagano}, {Wilson},
  {Schlawin}, {Lothringer}, {Lanza}, {Fraine}, {Scandariato}, {Micela}, \&
  {Cracchiolo}]{Bruno2022}
{Bruno}, G., {Lewis}, N.~K., {Valenti}, J.~A., {Pagano}, I., {Wilson}, T.~J.,
  {Schlawin}, E., {Lothringer}, J., {Lanza}, A.~F., {Fraine}, J.,
  {Scandariato}, G., {Micela}, G., \& {Cracchiolo}, G., 2022.
\newblock {Hiding in plain sight: observing planet-starspot crossings with the
  James Webb Space Telescope}, {\it \mnras\/}, {\bf 509}(4), 5030--5045.

\bibitem[{Budding}(1977)]{Budding1977}
{Budding}, E., 1977.
\newblock {The Interpretation of Cyclical Photometric Variations in Certain
  Dwarf ME-Type Stars}, {\it \apss\/}, {\bf 48}(1), 207--223.

\bibitem[{Buehler} et~al.(2019){Buehler}, {Lagg}, {van Noort}, \&
  {Solanki}]{buehler2019}
{Buehler}, D., {Lagg}, A., {van Noort}, M., \& {Solanki}, S.~K., 2019.
\newblock {A comparison between solar plage and network properties}, {\it
  \aap\/}, {\bf 630}, A86.

\bibitem[{Buenzli} et~al.(2014){Buenzli}, {Apai}, {Radigan}, {Reid}, \&
  {Flateau}]{Buenzli2014}
{Buenzli}, E., {Apai}, D., {Radigan}, J., {Reid}, I.~N., \& {Flateau}, D.,
  2014.
\newblock {Brown Dwarf Photospheres are Patchy: A Hubble Space Telescope
  Near-infrared Spectroscopic Survey Finds Frequent Low-level Variability},
  {\it \apj\/}, {\bf 782}(2), 77.

\bibitem[{Burdanov} et~al.(2019){Burdanov}, {Lederer}, {Gillon}, {Delrez},
  {Ducrot}, {de Wit}, {Jehin}, {Triaud}, {Lidman}, {Spitler}, {Demory},
  {Queloz}, \& {Van Grootel}]{Burdanov2019}
{Burdanov}, A.~Y., {Lederer}, S.~M., {Gillon}, M., {Delrez}, L., {Ducrot}, E.,
  {de Wit}, J., {Jehin}, E., {Triaud}, A.~H.~M.~J., {Lidman}, C., {Spitler},
  L., {Demory}, B.~O., {Queloz}, D., \& {Van Grootel}, V., 2019.
\newblock {Ground-based follow-up observations of TRAPPIST-1 transits in the
  near-infrared}, {\it \mnras\/}, {\bf 487}(2), 1634--1652.

\bibitem[{Cabrera} et~al.(2010){Cabrera}, {Bruntt}, {Ollivier}, {D{\'\i}az},
  {Csizmadia}, {Aigrain}, {Alonso}, {Almenara}, {Auvergne}, {Baglin}, {Barge},
  {Bonomo}, {Bord{\'e}}, {Bouchy}, {Carone}, {Carpano}, {Deleuil}, {Deeg},
  {Dvorak}, {Erikson}, {Ferraz-Mello}, {Fridlund}, {Gandolfi}, {Gazzano},
  {Gillon}, {Guenther}, {Guillot}, {Hatzes}, {Havel}, {H{\'e}brard}, {Jorda},
  {L{\'e}ger}, {Llebaria}, {Lammer}, {Lovis}, {Mazeh}, {Moutou}, {Ofir}, {von
  Paris}, {P{\"a}tzold}, {Queloz}, {Rauer}, {Rouan}, {Santerne}, {Schneider},
  {Tingley}, {Titz-Weider}, \& {Wuchterl}]{Cabrera2010A}
{Cabrera}, J., {Bruntt}, H., {Ollivier}, M., {D{\'\i}az}, R.~F., {Csizmadia},
  S., {Aigrain}, S., {Alonso}, R., {Almenara}, J.~M., {Auvergne}, M., {Baglin},
  A., {Barge}, P., {Bonomo}, A.~S., {Bord{\'e}}, P., {Bouchy}, F., {Carone},
  L., {Carpano}, S., {Deleuil}, M., {Deeg}, H.~J., {Dvorak}, R., {Erikson}, A.,
  {Ferraz-Mello}, S., {Fridlund}, M., {Gandolfi}, D., {Gazzano}, J.~C.,
  {Gillon}, M., {Guenther}, E.~W., {Guillot}, T., {Hatzes}, A., {Havel}, M.,
  {H{\'e}brard}, G., {Jorda}, L., {L{\'e}ger}, A., {Llebaria}, A., {Lammer},
  H., {Lovis}, C., {Mazeh}, T., {Moutou}, C., {Ofir}, A., {von Paris}, P.,
  {P{\"a}tzold}, M., {Queloz}, D., {Rauer}, H., {Rouan}, D., {Santerne}, A.,
  {Schneider}, J., {Tingley}, B., {Titz-Weider}, R., \& {Wuchterl}, G., 2010.
\newblock {Transiting exoplanets from the CoRoT space mission . XIII.
  CoRoT-13b: a dense hot Jupiter in transit around a star with solar
  metallicity and super-solar lithium content}, {\it \aap\/}, {\bf 522}, A110.

\bibitem[{Campante} et~al.(2016){Campante}, {Schofield}, {Kuszlewicz}, {Bouma},
  {Chaplin}, {Huber}, {Christensen-Dalsgaard}, {Kjeldsen}, {Bossini}, {North},
  {Appourchaux}, {Latham}, {Pepper}, {Ricker}, {Stassun}, {Vanderspek}, \&
  {Winn}]{Campante2016}
{Campante}, T.~L., {Schofield}, M., {Kuszlewicz}, J.~S., {Bouma}, L.,
  {Chaplin}, W.~J., {Huber}, D., {Christensen-Dalsgaard}, J., {Kjeldsen}, H.,
  {Bossini}, D., {North}, T.~S.~H., {Appourchaux}, T., {Latham}, D.~W.,
  {Pepper}, J., {Ricker}, G.~R., {Stassun}, K.~G., {Vanderspek}, R., \& {Winn},
  J.~N., 2016.
\newblock {The Asteroseismic Potential of TESS: Exoplanet-host Stars}, {\it
  \apj\/}, {\bf 830}(2), 138.

\bibitem[{Carlsson} \& {Stein}(2002)]{carlsson2002hion}
{Carlsson}, M. \& {Stein}, R.~F., 2002.
\newblock {Dynamic Hydrogen Ionization}, {\it \apj\/}, {\bf 572}(1), 626--635.

\bibitem[{Carlsson} et~al.(2016){Carlsson}, {Hansteen}, {Gudiksen},
  {Leenaarts}, \& {De Pontieu}]{carlsson2016sim}
{Carlsson}, M., {Hansteen}, V.~H., {Gudiksen}, B.~V., {Leenaarts}, J., \& {De
  Pontieu}, B., 2016.
\newblock {A publicly available simulation of an enhanced network region of the
  Sun}, {\it \aap\/}, {\bf 585}, A4.

\bibitem[{Carter} \& {Winn}(2010)]{Carter2010}
{Carter}, J.~A. \& {Winn}, J.~N., 2010.
\newblock {The Detectability of Transit Depth Variations Due to Exoplanetary
  Oblateness and Spin Precession}, {\it \apj\/}, {\bf 716}(1), 850--856.

\bibitem[{Casanovas}(1997)]{casanovas:1997}
{Casanovas}, J., 1997.
\newblock {Early Observations of Sunspots: Scheiner and Galileo}, in {\em 1st
  Advances in Solar Physics Euroconference. Advances in Physics of Sunspots\/},
  vol. 118 of {\bf Astronomical Society of the Pacific Conference Series},
  p.~3.

\bibitem[{Castelli} \& {Kurucz}(2003)]{Castelli2003}
{Castelli}, F. \& {Kurucz}, R.~L., 2003.
\newblock {New Grids of ATLAS9 Model Atmospheres}, in {\em Modelling of Stellar
  Atmospheres\/}, vol. 210 of {\bf IAU Symposium}, p. A20.

\bibitem[{Catalano} et~al.(2002){Catalano}, {Biazzo}, {Frasca}, \&
  {Marilli}]{Catalano2002}
{Catalano}, S., {Biazzo}, K., {Frasca}, A., \& {Marilli}, E., 2002.
\newblock {Measuring starspot temperature from line depth ratios. I. The
  method}, {\it \aap\/}, {\bf 394}, 1009--1021.

\bibitem[{Cauley} et~al.(2018){Cauley}, {Kuckein}, {Redfield}, {Shkolnik},
  {Denker}, {Llama}, \& {Verma}]{Cauley2018}
{Cauley}, P.~W., {Kuckein}, C., {Redfield}, S., {Shkolnik}, E.~L., {Denker},
  C., {Llama}, J., \& {Verma}, M., 2018.
\newblock {The Effects of Stellar Activity on Optical High-resolution Exoplanet
  Transmission Spectra}, {\it \aj\/}, {\bf 156}(5), 189.

\bibitem[{Chachan} et~al.(2019){Chachan}, {Knutson}, {Gao}, {Kataria}, {Wong},
  {Henry}, {Benneke}, {Zhang}, {Barstow}, {Bean}, {Mikal-Evans}, {Lewis},
  {Mansfield}, {L{\'o}pez-Morales}, {Nikolov}, {Sing}, \&
  {Wakeford}]{Chachan2019}
{Chachan}, Y., {Knutson}, H.~A., {Gao}, P., {Kataria}, T., {Wong}, I., {Henry},
  G.~W., {Benneke}, B., {Zhang}, M., {Barstow}, J., {Bean}, J.~L.,
  {Mikal-Evans}, T., {Lewis}, N.~K., {Mansfield}, M., {L{\'o}pez-Morales}, M.,
  {Nikolov}, N., {Sing}, D.~K., \& {Wakeford}, H., 2019.
\newblock {A Hubble PanCET Study of HAT-P-11b: A Cloudy Neptune with a Low
  Atmospheric Metallicity}, {\it \aj\/}, {\bf 158}(6), 244.

\bibitem[{Chaplin} et~al.(2011{\natexlab{a}}){Chaplin}, {Kjeldsen}, {Bedding},
  {Christensen-Dalsgaard}, {Gilliland}, {Kawaler}, {Appourchaux}, {Elsworth},
  {Garc{\'\i}a}, {Houdek}, {Karoff}, {Metcalfe}, {Molenda-{\.Z}akowicz},
  {Monteiro}, {Thompson}, {Verner}, {Batalha}, {Borucki}, {Brown}, {Bryson},
  {Christiansen}, {Clarke}, {Jenkins}, {Klaus}, {Koch}, {An}, {Ballot}, {Basu},
  {Benomar}, {Bonanno}, {Broomhall}, {Campante}, {Corsaro}, {Creevey}, {Esch},
  {Gai}, {Gaulme}, {Hale}, {Handberg}, {Hekker}, {Huber}, {Mathur}, {Mosser},
  {New}, {Pinsonneault}, {Pricopi}, {Quirion}, {R{\'e}gulo}, {Roxburgh},
  {Salabert}, {Stello}, \& {Suran}]{Chaplin2011}
{Chaplin}, W.~J., {Kjeldsen}, H., {Bedding}, T.~R., {Christensen-Dalsgaard},
  J., {Gilliland}, R.~L., {Kawaler}, S.~D., {Appourchaux}, T., {Elsworth}, Y.,
  {Garc{\'\i}a}, R.~A., {Houdek}, G., {Karoff}, C., {Metcalfe}, T.~S.,
  {Molenda-{\.Z}akowicz}, J., {Monteiro}, M.~J.~P.~F.~G., {Thompson}, M.~J.,
  {Verner}, G.~A., {Batalha}, N., {Borucki}, W.~J., {Brown}, T.~M., {Bryson},
  S.~T., {Christiansen}, J.~L., {Clarke}, B.~D., {Jenkins}, J.~M., {Klaus},
  T.~C., {Koch}, D., {An}, D., {Ballot}, J., {Basu}, S., {Benomar}, O.,
  {Bonanno}, A., {Broomhall}, A.~M., {Campante}, T.~L., {Corsaro}, E.,
  {Creevey}, O.~L., {Esch}, L., {Gai}, N., {Gaulme}, P., {Hale}, S.~J.,
  {Handberg}, R., {Hekker}, S., {Huber}, D., {Mathur}, S., {Mosser}, B., {New},
  R., {Pinsonneault}, M.~H., {Pricopi}, D., {Quirion}, P.~O., {R{\'e}gulo}, C.,
  {Roxburgh}, I.~W., {Salabert}, D., {Stello}, D., \& {Suran}, M.~D.,
  2011{\natexlab{a}}.
\newblock {Predicting the Detectability of Oscillations in Solar-type Stars
  Observed by Kepler}, {\it \apj\/}, {\bf 732}(1), 54.

\bibitem[{Chaplin} et~al.(2011{\natexlab{b}}){Chaplin}, {Kjeldsen},
  {Christensen-Dalsgaard}, {Basu}, {Miglio}, {Appourchaux}, {Bedding},
  {Elsworth}, {Garc{\'\i}a}, {Gilliland}, {Girardi}, {Houdek}, {Karoff},
  {Kawaler}, {Metcalfe}, {Molenda-{\.Z}akowicz}, {Monteiro}, {Thompson},
  {Verner}, {Ballot}, {Bonanno}, {Brand{\~a}o}, {Broomhall}, {Bruntt},
  {Campante}, {Corsaro}, {Creevey}, {Do{\u{g}}an}, {Esch}, {Gai}, {Gaulme},
  {Hale}, {Handberg}, {Hekker}, {Huber}, {Jim{\'e}nez}, {Mathur}, {Mazumdar},
  {Mosser}, {New}, {Pinsonneault}, {Pricopi}, {Quirion}, {R{\'e}gulo},
  {Salabert}, {Serenelli}, {Silva Aguirre}, {Sousa}, {Stello}, {Stevens},
  {Suran}, {Uytterhoeven}, {White}, {Borucki}, {Brown}, {Jenkins}, {Kinemuchi},
  {Van Cleve}, \& {Klaus}]{Chaplin:2011}
{Chaplin}, W.~J., {Kjeldsen}, H., {Christensen-Dalsgaard}, J., {Basu}, S.,
  {Miglio}, A., {Appourchaux}, T., {Bedding}, T.~R., {Elsworth}, Y.,
  {Garc{\'\i}a}, R.~A., {Gilliland}, R.~L., {Girardi}, L., {Houdek}, G.,
  {Karoff}, C., {Kawaler}, S.~D., {Metcalfe}, T.~S., {Molenda-{\.Z}akowicz},
  J., {Monteiro}, M.~J.~P.~F.~G., {Thompson}, M.~J., {Verner}, G.~A., {Ballot},
  J., {Bonanno}, A., {Brand{\~a}o}, I.~M., {Broomhall}, A.~M., {Bruntt}, H.,
  {Campante}, T.~L., {Corsaro}, E., {Creevey}, O.~L., {Do{\u{g}}an}, G.,
  {Esch}, L., {Gai}, N., {Gaulme}, P., {Hale}, S.~J., {Handberg}, R., {Hekker},
  S., {Huber}, D., {Jim{\'e}nez}, A., {Mathur}, S., {Mazumdar}, A., {Mosser},
  B., {New}, R., {Pinsonneault}, M.~H., {Pricopi}, D., {Quirion}, P.~O.,
  {R{\'e}gulo}, C., {Salabert}, D., {Serenelli}, A.~M., {Silva Aguirre}, V.,
  {Sousa}, S.~G., {Stello}, D., {Stevens}, I.~R., {Suran}, M.~D.,
  {Uytterhoeven}, K., {White}, T.~R., {Borucki}, W.~J., {Brown}, T.~M.,
  {Jenkins}, J.~M., {Kinemuchi}, K., {Van Cleve}, J., \& {Klaus}, T.~C.,
  2011{\natexlab{b}}.
\newblock {Ensemble Asteroseismology of Solar-Type Stars with the NASA Kepler
  Mission}, {\it Science\/}, {\bf 332}(6026), 213.

\bibitem[{Chaplin} et~al.(2015){Chaplin}, {Lund}, {Handberg}, {Basu},
  {Buchhave}, {Campante}, {Davies}, {Huber}, {Latham}, {Latham}, {Serenelli},
  {Antia}, {Appourchaux}, {Ball}, {Benomar}, {Casagrande},
  {Christensen-Dalsgaard}, {Coelho}, {Creevey}, {Elsworth}, {Garc{\'\i}a},
  {Gaulme}, {Hekker}, {Kallinger}, {Karoff}, {Kawaler}, {Kjeldsen},
  {Lundkvist}, {Marcadon}, {Mathur}, {Miglio}, {Mosser}, {R{\'e}gulo},
  {Roxburgh}, {Silva Aguirre}, {Stello}, {Verma}, {White}, {Bedding},
  {Barclay}, {Buzasi}, {Dehuevels}, {Gizon}, {Houdek}, {Howell}, {Salabert}, \&
  {Soderblom}]{Chaplin2015}
{Chaplin}, W.~J., {Lund}, M.~N., {Handberg}, R., {Basu}, S., {Buchhave}, L.~A.,
  {Campante}, T.~L., {Davies}, G.~R., {Huber}, D., {Latham}, D.~W., {Latham},
  C.~A., {Serenelli}, A., {Antia}, H.~M., {Appourchaux}, T., {Ball}, W.~H.,
  {Benomar}, O., {Casagrande}, L., {Christensen-Dalsgaard}, J., {Coelho},
  H.~R., {Creevey}, O.~L., {Elsworth}, Y., {Garc{\'\i}a}, R.~A., {Gaulme}, P.,
  {Hekker}, S., {Kallinger}, T., {Karoff}, C., {Kawaler}, S.~D., {Kjeldsen},
  H., {Lundkvist}, M.~S., {Marcadon}, F., {Mathur}, S., {Miglio}, A., {Mosser},
  B., {R{\'e}gulo}, C., {Roxburgh}, I.~W., {Silva Aguirre}, V., {Stello}, D.,
  {Verma}, K., {White}, T.~R., {Bedding}, T.~R., {Barclay}, T., {Buzasi},
  D.~L., {Dehuevels}, S., {Gizon}, L., {Houdek}, G., {Howell}, S.~B.,
  {Salabert}, D., \& {Soderblom}, D.~R., 2015.
\newblock {Asteroseismology of Solar-Type Stars with K2: Detection of
  Oscillations in C1 Data}, {\it \pasp\/}, {\bf 127}(956), 1038.

\bibitem[{Chapman}(1970)]{Chapman1970}
{Chapman}, G.~A., 1970.
\newblock {An Interference Filter for Observing the Photospheric Network}, {\it
  \solphys\/}, {\bf 13}(1), 78--84.

\bibitem[{Chapman} et~al.(2013){Chapman}, {Cookson}, \&
  {Preminger}]{chapman2013}
{Chapman}, G.~A., {Cookson}, A.~M., \& {Preminger}, D.~G., 2013.
\newblock {Modeling Total Solar Irradiance with San Fernando Observatory
  Ground-Based Photometry: Comparison with ACRIM, PMOD, and RMIB Composites},
  {\it \solphys\/}, {\bf 283}(2), 295--305.

\bibitem[{Charbonneau} et~al.(2000){Charbonneau}, {Brown}, {Latham}, \&
  {Mayor}]{Charbonneau-00}
{Charbonneau}, D., {Brown}, T.~M., {Latham}, D.~W., \& {Mayor}, M., 2000.
\newblock {Detection of Planetary Transits Across a Sun-like Star}, {\it
  \apjl\/}, {\bf 529}, L45--L48.

\bibitem[{Charbonneau} et~al.(2002){Charbonneau}, {Brown}, {Noyes}, \&
  {Gilliland}]{Charbonneau2002}
{Charbonneau}, D., {Brown}, T.~M., {Noyes}, R.~W., \& {Gilliland}, R.~L., 2002.
\newblock {Detection of an Extrasolar Planet Atmosphere}, {\it \apj\/}, {\bf
  568}(1), 377--384.

\bibitem[{Chatzistergos} et~al.(2019{\natexlab{a}}){Chatzistergos}, {Ermolli},
  {Krivova}, \& {Solanki}]{Chatzistergos:2019b}
{Chatzistergos}, T., {Ermolli}, I., {Krivova}, N.~A., \& {Solanki}, S.~K.,
  2019{\natexlab{a}}.
\newblock {Analysis of full disc Ca II K spectroheliograms. II. Towards an
  accurate assessment of long-term variations in plage areas}, {\it \aap\/},
  {\bf 625}, A69.

\bibitem[{Chatzistergos} et~al.(2019{\natexlab{b}}){Chatzistergos}, {Ermolli},
  {Solanki}, {Krivova}, {Giorgi}, \& {Yeo}]{Chatzistergos:2019a}
{Chatzistergos}, T., {Ermolli}, I., {Solanki}, S.~K., {Krivova}, N.~A.,
  {Giorgi}, F., \& {Yeo}, K.~L., 2019{\natexlab{b}}.
\newblock {Recovering the unsigned photospheric magnetic field from Ca II K
  observations}, {\it \aap\/}, {\bf 626}, A114.

\bibitem[{Chatzistergos} et~al.(2020{\natexlab{a}}){Chatzistergos}, {Ermolli},
  {Giorgi}, {Krivova}, \& {Puiu}]{chat2020}
{Chatzistergos}, T., {Ermolli}, I., {Giorgi}, F., {Krivova}, N.~A., \& {Puiu},
  C.~C., 2020{\natexlab{a}}.
\newblock Modelling solar irradiance from ground-based photometric
  observations, {\it J. Space Weather Space Clim.\/}, {\bf 10}, 45.

\bibitem[{Chatzistergos} et~al.(2020{\natexlab{b}}){Chatzistergos}, {Ermolli},
  {Krivova}, {Solanki}, {Banerjee}, {Barata}, {Belik}, {Gafeira}, {Garcia},
  {Hanaoka}, {Hegde}, {Klime{\v{s}}}, {Korokhin}, {Louren{\c{c}}o}, {Malherbe},
  {Marchenko}, {Peixinho}, {Sakurai}, \& {Tlatov}]{Chatzistergos:2020}
{Chatzistergos}, T., {Ermolli}, I., {Krivova}, N.~A., {Solanki}, S.~K.,
  {Banerjee}, D., {Barata}, T., {Belik}, M., {Gafeira}, R., {Garcia}, A.,
  {Hanaoka}, Y., {Hegde}, M., {Klime{\v{s}}}, J., {Korokhin}, V.~V.,
  {Louren{\c{c}}o}, A., {Malherbe}, J.-M., {Marchenko}, G.~P., {Peixinho}, N.,
  {Sakurai}, T., \& {Tlatov}, A.~G., 2020{\natexlab{b}}.
\newblock {Analysis of full-disc Ca II K spectroheliograms. III. Plage area
  composite series covering 1892-2019}, {\it \aap\/}, {\bf 639}, A88.

\bibitem[{Chen} et~al.(2017){Chen}, {Guenther}, {Pall{\'e}}, {Nortmann},
  {Nowak}, {Kunz}, {Parviainen}, \& {Murgas}]{Chen-17}
{Chen}, G., {Guenther}, E.~W., {Pall{\'e}}, E., {Nortmann}, L., {Nowak}, G.,
  {Kunz}, S., {Parviainen}, H., \& {Murgas}, F., 2017.
\newblock {The GTC exoplanet transit spectroscopy survey. V. A
  spectrally-resolved Rayleigh scattering slope in GJ 3470b}, {\it A\&A\/},
  {\bf 600}, A138.

\bibitem[{Chiavassa} et~al.(2009){Chiavassa}, {Plez}, {Josselin}, \&
  {Freytag}]{Chiavassa-09}
{Chiavassa}, A., {Plez}, B., {Josselin}, E., \& {Freytag}, B., 2009.
\newblock {Radiative hydrodynamics simulations of red supergiant stars. I.
  interpretation of interferometric observations}, {\it \aap\/}, {\bf 506}(3),
  1351--1365.

\bibitem[{Chiavassa} et~al.(2017){Chiavassa}, {Caldas}, {Selsis}, {Leconte},
  {Von Paris}, {Bord{\'e}}, {Magic}, {Collet}, \& {Asplund}]{Chiavassa-17}
{Chiavassa}, A., {Caldas}, A., {Selsis}, F., {Leconte}, J., {Von Paris}, P.,
  {Bord{\'e}}, P., {Magic}, Z., {Collet}, R., \& {Asplund}, M., 2017.
\newblock {Measuring stellar granulation during planet transits}, {\it \aap\/},
  {\bf 597}, A94.

\bibitem[{Choudhary} et~al.(2020){Choudhary}, {Cadavid}, {Cookson}, \&
  {Chapman}]{choud2020}
{Choudhary}, D.~P., {Cadavid}, A.~C., {Cookson}, A., \& {Chapman}, G.~A., 2020.
\newblock {Variability in Irradiance and Photometric Indices During the Last
  Two Solar Cycles}, {\it \solphys\/}, {\bf 295}(2), 15.

\bibitem[{Ciceri} et~al.(2013){Ciceri}, {Mancini}, {Southworth}, {Nikolov},
  {Bozza}, {Bruni}, {Calchi Novati}, {D'Ago}, \& {Henning}]{Ciceri-13}
{Ciceri}, S., {Mancini}, L., {Southworth}, J., {Nikolov}, N., {Bozza}, V.,
  {Bruni}, I., {Calchi Novati}, S., {D'Ago}, G., \& {Henning}, T., 2013.
\newblock {Simultaneous follow-up of planetary transits: revised physical
  properties for the planetary systems HAT-P-16 and WASP-21}, {\it \aap\/},
  {\bf 557}, A30.

\bibitem[{Claret}(2009)]{Claret2009}
{Claret}, A., 2009.
\newblock {Does the HD 209458 planetary system pose a challenge to the stellar
  atmosphere models?}, {\it \aap\/}, {\bf 506}(3), 1335--1340.

\bibitem[{Clette}(2021)]{clette2021}
{Clette}, F., 2021.
\newblock {Is the F$_{10.7cm}$ - Sunspot Number relation linear and stable?},
  {\it Journal of Space Weather and Space Climate\/}, {\bf 11}, 2.

\bibitem[{Corsaro} et~al.(2013){Corsaro}, {Fr{\"o}hlich}, {Bonanno}, {Huber},
  {Bedding}, {Benomar}, {De Ridder}, \& {Stello}]{Corsaro2013}
{Corsaro}, E., {Fr{\"o}hlich}, H.~E., {Bonanno}, A., {Huber}, D., {Bedding},
  T.~R., {Benomar}, O., {De Ridder}, J., \& {Stello}, D., 2013.
\newblock {A Bayesian approach to scaling relations for amplitudes of
  solar-like oscillations in Kepler stars}, {\it \mnras\/}, {\bf 430}(3),
  2313--2326.

\bibitem[{Cowan} et~al.(2009){Cowan}, {Agol}, {Meadows}, {Robinson},
  {Livengood}, {Deming}, {Lisse}, {A'Hearn}, {Wellnitz}, {Seager},
  {Charbonneau}, \& {EPOXI Team}]{Cowan2009}
{Cowan}, N.~B., {Agol}, E., {Meadows}, V.~S., {Robinson}, T., {Livengood},
  T.~A., {Deming}, D., {Lisse}, C.~M., {A'Hearn}, M.~F., {Wellnitz}, D.~D.,
  {Seager}, S., {Charbonneau}, D., \& {EPOXI Team}, 2009.
\newblock {Alien Maps of an Ocean-bearing World}, {\it \apj\/}, {\bf 700}(2),
  915--923.

\bibitem[{Cowan} et~al.(2013){Cowan}, {Fuentes}, \& {Haggard}]{Cowan2013}
{Cowan}, N.~B., {Fuentes}, P.~A., \& {Haggard}, H.~M., 2013.
\newblock {Light curves of stars and exoplanets: estimating inclination,
  obliquity and albedo}, {\it \mnras\/}, {\bf 434}(3), 2465--2479.

\bibitem[{Cracchiolo} et~al.(2021){Cracchiolo}, {Micela}, \&
  {Peres}]{Cracchiolo2021}
{Cracchiolo}, G., {Micela}, G., \& {Peres}, G., 2021.
\newblock {Correcting the effect of stellar spots on ARIEL transmission
  spectra}, {\it \mnras\/}, {\bf 501}(2), 1733--1747.

\bibitem[{Cranmer} et~al.(2014){Cranmer}, {Bastien}, {Stassun}, \&
  {Saar}]{Cranmer:2014}
{Cranmer}, S.~R., {Bastien}, F.~A., {Stassun}, K.~G., \& {Saar}, S.~H., 2014.
\newblock {Stellar Granulation as the Source of High-frequency Flicker in
  Kepler Light Curves}, {\it \apj\/}, {\bf 781}(2), 124.

\bibitem[{Crida} et~al.(2018){Crida}, {Ligi}, {Dorn}, \& {Lebreton}]{Crida2018}
{Crida}, A., {Ligi}, R., {Dorn}, C., \& {Lebreton}, Y., 2018.
\newblock {Mass, Radius, and Composition of the Transiting Planet 55 Cnc e:
  Using Interferometry and Correlations}, {\it \apj\/}, {\bf 860}(2), 122.

\bibitem[{Criscuoli}(2019)]{criscuoli2019}
{Criscuoli}, S., 2019.
\newblock {Effects of Continuum Fudging on Non-LTE Synthesis of Stellar
  Spectra. I. Effects on Estimates of UV Continua and Solar Spectral Irradiance
  Variability}, {\it \apj\/}, {\bf 872}(1), 52.

\bibitem[{Criscuoli} \& {Ermolli}(2008)]{criscuoli2008}
{Criscuoli}, S. \& {Ermolli}, I., 2008.
\newblock {Stray-light restoration of full-disk CaII K solar observations: a
  case study}, {\it \aap\/}, {\bf 484}(2), 591--599.

\bibitem[{Criscuoli} et~al.(2017){Criscuoli}, {Norton}, \&
  {Whitney}]{criscuoli2017}
{Criscuoli}, S., {Norton}, A., \& {Whitney}, T., 2017.
\newblock {Photometric Properties of Network and Faculae Derived from HMI Data
  Compensated for Scattered Light}, {\it \apj\/}, {\bf 847}(2), 93.

\bibitem[{Csizmadia} et~al.(2013){Csizmadia}, {Pasternacki}, {Dreyer},
  {Cabrera}, {Erikson}, \& {Rauer}]{Csizmadia-13}
{Csizmadia}, S., {Pasternacki}, T., {Dreyer}, C., {Cabrera}, J., {Erikson}, A.,
  \& {Rauer}, H., 2013.
\newblock {The effect of stellar limb darkening values on the accuracy of the
  planet radii derived from photometric transit observations}, {\it \aap\/},
  {\bf 549}, A9.

\bibitem[{Czesla} et~al.(2009){Czesla}, {Huber}, {Wolter}, {Schr{\"o}ter}, \&
  {Schmitt}]{Czesla-09}
{Czesla}, S., {Huber}, K.~F., {Wolter}, U., {Schr{\"o}ter}, S., \& {Schmitt},
  J.~H.~M.~M., 2009.
\newblock {How stellar activity affects the size estimates of extrasolar
  planets}, {\it \aap\/}, {\bf 505}, 1277--1282.

\bibitem[{Dai} \& {Winn}(2017)]{Dai-2017}
{Dai}, F. \& {Winn}, J.~N., 2017.
\newblock {The Oblique Orbit of WASP-107b from K2 Photometry}, {\it \aj\/},
  {\bf 153}(5), 205.

\bibitem[{Danielski} et~al.(2014){Danielski}, {Deroo}, {Waldmann}, {Hollis},
  {Tinetti}, \& {Swain}]{Danielski2014}
{Danielski}, C., {Deroo}, P., {Waldmann}, I.~P., {Hollis}, M.~D.~J., {Tinetti},
  G., \& {Swain}, M.~R., 2014.
\newblock {0.94-2.42 {\ensuremath{\mu}}m Ground-based Transmission Spectra of
  the Hot Jupiter HD-189733b}, {\it \apj\/}, {\bf 785}(1), 35.

\bibitem[{Danielski} et~al.(2022){Danielski}, {Brucalassi}, {Benatti},
  {Campante}, {Delgado-Mena}, {Rainer}, {Sacco}, {Adibekyan}, {Biazzo},
  {Bossini}, {Bruno}, {Casali}, {Kabath}, {Magrini}, {Micela}, {Morello},
  {Palladino}, {Sanna}, {Sarkar}, {Sousa}, {Tsantaki}, {Turrini}, \& {Van der
  Swaelmen}]{Danielski2022}
{Danielski}, C., {Brucalassi}, A., {Benatti}, S., {Campante}, T.,
  {Delgado-Mena}, E., {Rainer}, M., {Sacco}, G., {Adibekyan}, V., {Biazzo}, K.,
  {Bossini}, D., {Bruno}, G., {Casali}, G., {Kabath}, P., {Magrini}, L.,
  {Micela}, G., {Morello}, G., {Palladino}, P., {Sanna}, N., {Sarkar}, S.,
  {Sousa}, S., {Tsantaki}, M., {Turrini}, D., \& {Van der Swaelmen}, M., 2022.
\newblock {The homogeneous characterisation of Ariel host stars}, {\it
  Experimental Astronomy\/}, {\bf 53}(2), 473--510.

\bibitem[{Danilovic} et~al.(2008){Danilovic}, {Gandorfer}, {Lagg},
  {Sch{\"u}ssler}, {Solanki}, {V{\"o}gler}, {Katsukawa}, \&
  {Tsuneta}]{danilovic+al2008}
{Danilovic}, S., {Gandorfer}, A., {Lagg}, A., {Sch{\"u}ssler}, M., {Solanki},
  S.~K., {V{\"o}gler}, A., {Katsukawa}, Y., \& {Tsuneta}, S., 2008.
\newblock {The intensity contrast of solar granulation: comparing Hinode SP
  results with MHD simulations}, {\it \aap\/}, {\bf 484}, L17--L20.

\bibitem[{Davenport}(2015)]{Davenport2015}
{Davenport}, J., 2015.
\newblock {\it {Spots and Flares: Stellar Activity in the Time Domain Era}\/},
  Ph.D. thesis, University of Washington.

\bibitem[{Davenport}(2016)]{Davenport2016}
{Davenport}, J.~R.~A., 2016.
\newblock {The Kepler Catalog of Stellar Flares}, {\it \apj\/}, {\bf 829}, 23.

\bibitem[{de la Cruz Rodr{\'\i}guez} et~al.(2013){de la Cruz Rodr{\'\i}guez},
  {Rouppe van der Voort}, {Socas-Navarro}, \& {van
  Noort}]{delaCruzRodriguez2013}
{de la Cruz Rodr{\'\i}guez}, J., {Rouppe van der Voort}, L., {Socas-Navarro},
  H., \& {van Noort}, M., 2013.
\newblock {Physical properties of a sunspot chromosphere with umbral flashes},
  {\it \aap\/}, {\bf 556}, A115.

\bibitem[{De Pontieu} et~al.(2006){De Pontieu}, {Carlsson}, {Stein}, {Rouppe
  van der Voort}, {L{\"o}fdahl}, {van Noort}, {Nordlund}, \&
  {Scharmer}]{de_pontieu+al2006}
{De Pontieu}, B., {Carlsson}, M., {Stein}, R., {Rouppe van der Voort}, L.,
  {L{\"o}fdahl}, M., {van Noort}, M., {Nordlund}, {\AA}., \& {Scharmer}, G.,
  2006.
\newblock {Rapid Temporal Variability of Faculae: High-Resolution Observations
  and Modeling}, {\it \apj\/}, {\bf 646}, 1405--1420.

\bibitem[{De Pontieu} et~al.(2014){De Pontieu}, {Title}, {Lemen}, {Kushner},
  {Akin}, {Allard}, {Berger}, {Boerner}, {Cheung}, {Chou}, {Drake}, {Duncan},
  {Freeland}, {Heyman}, {Hoffman}, {Hurlburt}, {Lindgren}, {Mathur}, {Rehse},
  {Sabolish}, {Seguin}, {Schrijver}, {Tarbell}, {W{\"u}lser}, {Wolfson},
  {Yanari}, {Mudge}, {Nguyen-Phuc}, {Timmons}, {van Bezooijen}, {Weingrod},
  {Brookner}, {Butcher}, {Dougherty}, {Eder}, {Knagenhjelm}, {Larsen},
  {Mansir}, {Phan}, {Boyle}, {Cheimets}, {DeLuca}, {Golub}, {Gates}, {Hertz},
  {McKillop}, {Park}, {Perry}, {Podgorski}, {Reeves}, {Saar}, {Testa}, {Tian},
  {Weber}, {Dunn}, {Eccles}, {Jaeggli}, {Kankelborg}, {Mashburn}, {Pust},
  {Springer}, {Carvalho}, {Kleint}, {Marmie}, {Mazmanian}, {Pereira}, {Sawyer},
  {Strong}, {Worden}, {Carlsson}, {Hansteen}, {Leenaarts}, {Wiesmann},
  {Aloise}, {Chu}, {Bush}, {Scherrer}, {Brekke}, {Martinez-Sykora}, {Lites},
  {McIntosh}, {Uitenbroek}, {Okamoto}, {Gummin}, {Auker}, {Jerram}, {Pool}, \&
  {Waltham}]{DePontieu2014}
{De Pontieu}, B., {Title}, A.~M., {Lemen}, J.~R., {Kushner}, G.~D., {Akin},
  D.~J., {Allard}, B., {Berger}, T., {Boerner}, P., {Cheung}, M., {Chou}, C.,
  {Drake}, J.~F., {Duncan}, D.~W., {Freeland}, S., {Heyman}, G.~F., {Hoffman},
  C., {Hurlburt}, N.~E., {Lindgren}, R.~W., {Mathur}, D., {Rehse}, R.,
  {Sabolish}, D., {Seguin}, R., {Schrijver}, C.~J., {Tarbell}, T.~D.,
  {W{\"u}lser}, J.~P., {Wolfson}, C.~J., {Yanari}, C., {Mudge}, J.,
  {Nguyen-Phuc}, N., {Timmons}, R., {van Bezooijen}, R., {Weingrod}, I.,
  {Brookner}, R., {Butcher}, G., {Dougherty}, B., {Eder}, J., {Knagenhjelm},
  V., {Larsen}, S., {Mansir}, D., {Phan}, L., {Boyle}, P., {Cheimets}, P.~N.,
  {DeLuca}, E.~E., {Golub}, L., {Gates}, R., {Hertz}, E., {McKillop}, S.,
  {Park}, S., {Perry}, T., {Podgorski}, W.~A., {Reeves}, K., {Saar}, S.,
  {Testa}, P., {Tian}, H., {Weber}, M., {Dunn}, C., {Eccles}, S., {Jaeggli},
  S.~A., {Kankelborg}, C.~C., {Mashburn}, K., {Pust}, N., {Springer}, L.,
  {Carvalho}, R., {Kleint}, L., {Marmie}, J., {Mazmanian}, E., {Pereira},
  T.~M.~D., {Sawyer}, S., {Strong}, J., {Worden}, S.~P., {Carlsson}, M.,
  {Hansteen}, V.~H., {Leenaarts}, J., {Wiesmann}, M., {Aloise}, J., {Chu},
  K.~C., {Bush}, R.~I., {Scherrer}, P.~H., {Brekke}, P., {Martinez-Sykora}, J.,
  {Lites}, B.~W., {McIntosh}, S.~W., {Uitenbroek}, H., {Okamoto}, T.~J.,
  {Gummin}, M.~A., {Auker}, G., {Jerram}, P., {Pool}, P., \& {Waltham}, N.,
  2014.
\newblock {The Interface Region Imaging Spectrograph (IRIS)}, {\it \solphys\/},
  {\bf 289}(7), 2733--2779.

\bibitem[{de Wijn} et~al.(2009){de Wijn}, {Stenflo}, {Solanki}, \&
  {Tsuneta}]{dewijn2009}
{de Wijn}, A.~G., {Stenflo}, J.~O., {Solanki}, S.~K., \& {Tsuneta}, S., 2009.
\newblock {Small-Scale Solar Magnetic Fields}, {\it \ssr\/}, {\bf 144}(1-4),
  275--315.

\bibitem[{Deeg} et~al.(2001){Deeg}, {Garrido}, \& {Claret}]{Deeg-01}
{Deeg}, H.~J., {Garrido}, R., \& {Claret}, A., 2001.
\newblock {Probing the stellar surface of HD 209458 from multicolor transit
  observations}, {\it New Astronomy\/}, {\bf 6}(2), 51--60.

\bibitem[{Delrez} et~al.(2018){Delrez}, {Gillon}, {Queloz}, {Demory},
  {Almleaky}, {de Wit}, {Jehin}, {Triaud}, {Barkaoui}, {Burdanov}, {Burgasser},
  {Ducrot}, {McCormac}, {Murray}, {Silva Fernandes}, {Sohy}, {Thompson}, {Van
  Grootel}, {Alonso}, {Benkhaldoun}, \& {Rebolo}]{speculoos_Delrez2018}
{Delrez}, L., {Gillon}, M., {Queloz}, D., {Demory}, B.-O., {Almleaky}, Y., {de
  Wit}, J., {Jehin}, E., {Triaud}, A. H.~M.~J., {Barkaoui}, K., {Burdanov}, A.,
  {Burgasser}, A.~J., {Ducrot}, E., {McCormac}, J., {Murray}, C., {Silva
  Fernandes}, C., {Sohy}, S., {Thompson}, S.~J., {Van Grootel}, V., {Alonso},
  R., {Benkhaldoun}, Z., \& {Rebolo}, R., 2018.
\newblock {SPECULOOS: a network of robotic telescopes to hunt for terrestrial
  planets around the nearest ultracool dwarfs}, in {\em Ground-based and
  Airborne Telescopes VII\/}, vol. 10700 of {\bf Society of Photo-Optical
  Instrumentation Engineers (SPIE) Conference Series}, p. 107001I.

\bibitem[{Delrez} et~al.(2022){Delrez}, {Murray}, {Pozuelos}, {Narita},
  {Ducrot}, {Timmermans}, {Watanabe}, {Burgasser}, {Hirano}, {Rackham},
  {Stassun}, {Van Grootel}, {Aganze}, {Cointepas}, {Howell}, {Kaltenegger},
  {Niraula}, {Sebastian}, {Almenara}, {Barkaoui}, {Baycroft}, {Bonfils},
  {Bouchy}, {Burdanov}, {Caldwell}, {Charbonneau}, {Ciardi}, {Collins},
  {Daylan}, {Demory}, {de Wit}, {Dransfield}, {Fajardo-Acosta}, {Fausnaugh},
  {Fukui}, {Furlan}, {Garcia}, {Gnilka}, {G{\'o}mez Maqueo Chew},
  {G{\'o}mez-Mu{\~n}oz}, {G{\"u}nther}, {Harakawa}, {Heng}, {Hooton}, {Hori},
  {Ikoma}, {Jehin}, {Jenkins}, {Kagetani}, {Kawauchi}, {Kimura}, {Kodama},
  {Kotani}, {Krishnamurthy}, {Kudo}, {Kunovac}, {Kusakabe}, {Latham},
  {Littlefield}, {McCormac}, {Melis}, {Mori}, {Murgas}, {Palle}, {Pedersen},
  {Queloz}, {Ricker}, {Sabin}, {Schanche}, {Schroffenegger}, {Seager}, {Shiao},
  {Sohy}, {Standing}, {Tamura}, {Theissen}, {Thompson}, {Triaud}, {Vanderspek},
  {Vievard}, {Wells}, {Winn}, {Zou}, {Z{\'u}{\~n}iga-Fern{\'a}ndez}, \&
  {Gillon}]{Delrez2022}
{Delrez}, L., {Murray}, C.~A., {Pozuelos}, F.~J., {Narita}, N., {Ducrot}, E.,
  {Timmermans}, M., {Watanabe}, N., {Burgasser}, A.~J., {Hirano}, T.,
  {Rackham}, B.~V., {Stassun}, K.~G., {Van Grootel}, V., {Aganze}, C.,
  {Cointepas}, M., {Howell}, S., {Kaltenegger}, L., {Niraula}, P., {Sebastian},
  D., {Almenara}, J.~M., {Barkaoui}, K., {Baycroft}, T.~A., {Bonfils}, X.,
  {Bouchy}, F., {Burdanov}, A., {Caldwell}, D.~A., {Charbonneau}, D., {Ciardi},
  D.~R., {Collins}, K.~A., {Daylan}, T., {Demory}, B.~O., {de Wit}, J.,
  {Dransfield}, G., {Fajardo-Acosta}, S.~B., {Fausnaugh}, M., {Fukui}, A.,
  {Furlan}, E., {Garcia}, L.~J., {Gnilka}, C.~L., {G{\'o}mez Maqueo Chew}, Y.,
  {G{\'o}mez-Mu{\~n}oz}, M.~A., {G{\"u}nther}, M.~N., {Harakawa}, H., {Heng},
  K., {Hooton}, M.~J., {Hori}, Y., {Ikoma}, M., {Jehin}, E., {Jenkins}, J.~M.,
  {Kagetani}, T., {Kawauchi}, K., {Kimura}, T., {Kodama}, T., {Kotani}, T.,
  {Krishnamurthy}, V., {Kudo}, T., {Kunovac}, V., {Kusakabe}, N., {Latham},
  D.~W., {Littlefield}, C., {McCormac}, J., {Melis}, C., {Mori}, M., {Murgas},
  F., {Palle}, E., {Pedersen}, P.~P., {Queloz}, D., {Ricker}, G., {Sabin}, L.,
  {Schanche}, N., {Schroffenegger}, U., {Seager}, S., {Shiao}, B., {Sohy}, S.,
  {Standing}, M.~R., {Tamura}, M., {Theissen}, C.~A., {Thompson}, S.~J.,
  {Triaud}, A.~H.~M.~J., {Vanderspek}, R., {Vievard}, S., {Wells}, R.~D.,
  {Winn}, J.~N., {Zou}, Y., {Z{\'u}{\~n}iga-Fern{\'a}ndez}, S., \& {Gillon},
  M., 2022.
\newblock {Two temperate super-Earths transiting a nearby late-type M dwarf},
  {\it \aap\/}, {\bf 667}, A59.

\bibitem[{Deming} et~al.(2013){Deming}, {Wilkins}, {McCullough}, {Burrows},
  {Fortney}, {Agol}, {Dobbs-Dixon}, {Madhusudhan}, {Crouzet}, {Desert},
  {Gilliland}, {Haynes}, {Knutson}, {Line}, {Magic}, {Mandell}, {Ranjan},
  {Charbonneau}, {Clampin}, {Seager}, \& {Showman}]{Deming2013}
{Deming}, D., {Wilkins}, A., {McCullough}, P., {Burrows}, A., {Fortney}, J.~J.,
  {Agol}, E., {Dobbs-Dixon}, I., {Madhusudhan}, N., {Crouzet}, N., {Desert},
  J.-M., {Gilliland}, R.~L., {Haynes}, K., {Knutson}, H.~A., {Line}, M.,
  {Magic}, Z., {Mandell}, A.~M., {Ranjan}, S., {Charbonneau}, D., {Clampin},
  M., {Seager}, S., \& {Showman}, A.~P., 2013.
\newblock {Infrared Transmission Spectroscopy of the Exoplanets HD 209458b and
  XO-1b Using the Wide Field Camera-3 on the Hubble Space Telescope}, {\it
  \apj\/}, {\bf 774}(2), 95.

\bibitem[{Desai} \& {Giacalone}(2016)]{Desai2016}
{Desai}, M. \& {Giacalone}, J., 2016.
\newblock {Large gradual solar energetic particle events}, {\it Living Reviews
  in Solar Physics\/}, {\bf 13}(1), 3.

\bibitem[{D{\'e}sert} et~al.(2011{\natexlab{a}}){D{\'e}sert}, {Charbonneau},
  {Demory}, {Ballard}, {Carter}, {Fortney}, {Cochran}, {Endl}, {Quinn},
  {Isaacson}, {Fressin}, {Buchhave}, {Latham}, {Knutson}, {Bryson}, {Torres},
  {Rowe}, {Batalha}, {Borucki}, {Brown}, {Caldwell}, {Christiansen}, {Deming},
  {Fabrycky}, {Ford}, {Gilliland}, {Gillon}, {Haas}, {Jenkins}, {Kinemuchi},
  {Koch}, {Lissauer}, {Lucas}, {Mullally}, {MacQueen}, {Marcy}, {Sasselov},
  {Seager}, {Still}, {Tenenbaum}, {Uddin}, \& {Winn}]{Desert-11}
{D{\'e}sert}, J.-M., {Charbonneau}, D., {Demory}, B.-O., {Ballard}, S.,
  {Carter}, J.~A., {Fortney}, J.~J., {Cochran}, W.~D., {Endl}, M., {Quinn},
  S.~N., {Isaacson}, H.~T., {Fressin}, F., {Buchhave}, L.~A., {Latham}, D.~W.,
  {Knutson}, H.~A., {Bryson}, S.~T., {Torres}, G., {Rowe}, J.~F., {Batalha},
  N.~M., {Borucki}, W.~J., {Brown}, T.~M., {Caldwell}, D.~A., {Christiansen},
  J.~L., {Deming}, D., {Fabrycky}, D.~C., {Ford}, E.~B., {Gilliland}, R.~L.,
  {Gillon}, M., {Haas}, M.~R., {Jenkins}, J.~M., {Kinemuchi}, K., {Koch}, D.,
  {Lissauer}, J.~J., {Lucas}, P., {Mullally}, F., {MacQueen}, P.~J., {Marcy},
  G.~W., {Sasselov}, D.~D., {Seager}, S., {Still}, M., {Tenenbaum}, P.,
  {Uddin}, K., \& {Winn}, J.~N., 2011{\natexlab{a}}.
\newblock {The Hot-Jupiter Kepler-17b: Discovery, Obliquity from Stroboscopic
  Starspots, and Atmospheric Characterization}, {\it \apjs\/}, {\bf 197}, 14.

\bibitem[{D{\'e}sert} et~al.(2011{\natexlab{b}}){D{\'e}sert}, {Sing},
  {Vidal-Madjar}, {H{\'e}brard}, {Ehrenreich}, {Lecavelier Des Etangs},
  {Parmentier}, {Ferlet}, \& {Henry}]{Desert2011}
{D{\'e}sert}, J.~M., {Sing}, D., {Vidal-Madjar}, A., {H{\'e}brard}, G.,
  {Ehrenreich}, D., {Lecavelier Des Etangs}, A., {Parmentier}, V., {Ferlet},
  R., \& {Henry}, G.~W., 2011{\natexlab{b}}.
\newblock {Transit spectrophotometry of the exoplanet HD 189733b. II. New
  Spitzer observations at 3.6 {\ensuremath{\mu}}m}, {\it \aap\/}, {\bf 526},
  A12.

\bibitem[{Deutsch}(1958)]{Deutsch1958}
{Deutsch}, A.~J., 1958.
\newblock {Harmonic Analysis of the Periodic Spectrum Variables}, in {\em
  Electromagnetic Phenomena in Cosmical Physics\/}, vol.~6, p. 209.

\bibitem[{Di Folco} et~al.(2004){Di Folco}, {Th{\'e}venin}, {Kervella},
  {Domiciano de Souza}, {Coud{\'e} du Foresto}, {S{\'e}gransan}, \&
  {Morel}]{DiFolco2004}
{Di Folco}, E., {Th{\'e}venin}, F., {Kervella}, P., {Domiciano de Souza}, A.,
  {Coud{\'e} du Foresto}, V., {S{\'e}gransan}, D., \& {Morel}, P., 2004.
\newblock {VLTI near-IR interferometric observations of Vega-like stars. Radius
  and age of {\ensuremath{\alpha}} PsA, {\ensuremath{\beta}} Leo,
  {\ensuremath{\beta}} Pic, {\ensuremath{\epsilon}} Eri and {\ensuremath{\tau}}
  Cet}, {\it \aap\/}, {\bf 426}, 601--617.

\bibitem[{Donati}(1999)]{Donati1999}
{Donati}, J.~F., 1999.
\newblock {Magnetic cycles of HR 1099 and LQ Hydrae}, {\it \mnras\/}, {\bf
  302}(3), 457--481.

\bibitem[{Donati} \& {Collier Cameron}(1997)]{Donati-CollierCameron1997abdor}
{Donati}, J.~F. \& {Collier Cameron}, A., 1997.
\newblock {Differential rotation and magnetic polarity patterns on AB Doradus},
  {\it \mnras\/}, {\bf 291}(1), 1--19.

\bibitem[{Donati} \& {Landstreet}(2009)]{Donati-Landstreet2009}
{Donati}, J.~F. \& {Landstreet}, J.~D., 2009.
\newblock {Magnetic Fields of Nondegenerate Stars}, {\it \araa\/}, {\bf 47}(1),
  333--370.

\bibitem[{Donati} et~al.(1997){Donati}, {Semel}, {Carter}, {Rees}, \& {Collier
  Cameron}]{Donati-97}
{Donati}, J.-F., {Semel}, M., {Carter}, B.~D., {Rees}, D.~E., \& {Collier
  Cameron}, A., 1997.
\newblock {Spectropolarimetric observations of active stars}, {\it \mnras\/},
  {\bf 291}, 658.

\bibitem[{Dorren}(1987)]{Dorren1987}
{Dorren}, J.~D., 1987.
\newblock {A New Formulation of the Starspot Model, and the Consequences of
  Starspot Structure}, {\it \apj\/}, {\bf 320}, 756.

\bibitem[{Doyle} et~al.(2018){Doyle}, {Ramsay}, {Doyle}, {Wu}, \&
  {Scullion}]{Doyle2018}
{Doyle}, L., {Ramsay}, G., {Doyle}, J.~G., {Wu}, K., \& {Scullion}, E., 2018.
\newblock {Investigating the rotational phase of stellar flares on M dwarfs
  using K2 short cadence data}, {\it \mnras\/}, {\bf 480}(2), 2153--2164.

\bibitem[{Dravins} \& {Nordlund}(1990{\natexlab{a}})]{dravins+nordlund1990a}
{Dravins}, D. \& {Nordlund}, {\AA}., 1990{\natexlab{a}}.
\newblock {Stellar Granulation - Part Four - Line Formation in Inhomogeneous
  Stellar Photospheres}, {\it \aap\/}, {\bf 228}, 184.

\bibitem[{Dravins} \& {Nordlund}(1990{\natexlab{b}})]{dravins+nordlund1990b}
{Dravins}, D. \& {Nordlund}, {\AA}., 1990{\natexlab{b}}.
\newblock {Stellar Granulation - Part Five - Synthetic Spectral Lines in Disk
  Integrated Starlight}, {\it \aap\/}, {\bf 228}, 203.

\bibitem[{Ducrot} et~al.(2020){Ducrot}, {Gillon}, {Delrez}, {Agol}, {Rimmer},
  {Turbet}, {G{\"u}nther}, {Demory}, {Triaud}, {Bolmont}, {Burgasser}, {Carey},
  {Ingalls}, {Jehin}, {Leconte}, {Lederer}, {Queloz}, {Raymond}, {Selsis}, {Van
  Grootel}, \& {de Wit}]{Ducrot2020}
{Ducrot}, E., {Gillon}, M., {Delrez}, L., {Agol}, E., {Rimmer}, P., {Turbet},
  M., {G{\"u}nther}, M.~N., {Demory}, B.~O., {Triaud}, A.~H.~M.~J., {Bolmont},
  E., {Burgasser}, A., {Carey}, S.~J., {Ingalls}, J.~G., {Jehin}, E.,
  {Leconte}, J., {Lederer}, S.~M., {Queloz}, D., {Raymond}, S.~N., {Selsis},
  F., {Van Grootel}, V., \& {de Wit}, J., 2020.
\newblock {TRAPPIST-1: Global results of the Spitzer Exploration Science
  Program Red Worlds}, {\it \aap\/}, {\bf 640}, A112.

\bibitem[{Dukes}(1992)]{dukes92}
{Dukes}, Robert~J., J., 1992.
\newblock {Results from the Four College APT.}, in {\em Robotic Telescopes in
  the 1990s\/}, vol. 103 of {\bf Astronomical Society of the Pacific Conference
  Series}, pp. 9--18.

\bibitem[{Dunn} \& {Zirker}(1973)]{Dunn-Zirker1973}
{Dunn}, R.~B. \& {Zirker}, J.~B., 1973.
\newblock {The Solar Filigree}, {\it \solphys\/}, {\bf 33}(2), 281--304.

\bibitem[{Engvold} et~al.(2019){Engvold}, {Vial}, \& {Skumanich}]{Envgold2019}
{Engvold}, O., {Vial}, J.-C., \& {Skumanich}, A., 2019.
\newblock {\it {The Sun as a Guide to Stellar Physics}\/}.

\bibitem[{Ermolli} et~al.(1998){Ermolli}, {Fofi}, {Bernacchia}, {Berrilli},
  {Caccin}, {Egidi}, \& {Florio}]{ermolli1998}
{Ermolli}, I., {Fofi}, M., {Bernacchia}, C., {Berrilli}, F., {Caccin}, B.,
  {Egidi}, A., \& {Florio}, A., 1998.
\newblock {The Prototype Rise-Pspt Instrument Operating in Rome}, {\it
  \solphys\/}, {\bf 177}, 1--10.

\bibitem[{Ermolli} et~al.(2007){Ermolli}, {Criscuoli}, {Centrone}, {Giorgi}, \&
  {Penza}]{ermolli2007}
{Ermolli}, I., {Criscuoli}, S., {Centrone}, M., {Giorgi}, F., \& {Penza}, V.,
  2007.
\newblock {Photometric properties of facular features over the activity cycle},
  {\it \aap\/}, {\bf 465}(1), 305--314.

\bibitem[{Ermolli} et~al.(2010){Ermolli}, {Criscuoli}, {Uitenbroek}, {Giorgi},
  {Rast}, \& {Solanki}]{ermolli2010}
{Ermolli}, I., {Criscuoli}, S., {Uitenbroek}, H., {Giorgi}, F., {Rast}, M.~P.,
  \& {Solanki}, S.~K., 2010.
\newblock {Radiative emission of solar features in the Ca II K line: comparison
  of measurements and models}, {\it \aap\/}, {\bf 523}, A55.

\bibitem[{Ermolli} et~al.(2013){Ermolli}, {Matthes}, {Dudok de Wit}, {Krivova},
  {Tourpali}, {Weber}, {Unruh}, {Gray}, {Langematz}, {Pilewskie}, {Rozanov},
  {Schmutz}, {Shapiro}, {Solanki}, \& {Woods}]{Ermolli:2013}
{Ermolli}, I., {Matthes}, K., {Dudok de Wit}, T., {Krivova}, N.~A., {Tourpali},
  K., {Weber}, M., {Unruh}, Y.~C., {Gray}, L., {Langematz}, U., {Pilewskie},
  P., {Rozanov}, E., {Schmutz}, W., {Shapiro}, A., {Solanki}, S.~K., \&
  {Woods}, T.~N., 2013.
\newblock {Recent variability of the solar spectral irradiance and its impact
  on climate modelling}, {\it Atmospheric Chemistry \& Physics\/}, {\bf 13}(8),
  3945--3977.

\bibitem[{Espinoza} \& {Jord{\'a}n}(2015)]{Espinoza:2015}
{Espinoza}, N. \& {Jord{\'a}n}, A., 2015.
\newblock {Limb darkening and exoplanets: testing stellar model atmospheres and
  identifying biases in transit parameters}, {\it \mnras\/}, {\bf 450}(2),
  1879--1899.

\bibitem[{Espinoza} et~al.(2019){Espinoza}, {Rackham}, {Jord{\'a}n}, {Apai},
  {L{\'o}pez-Morales}, {Osip}, {Grimm}, {Hoeijmakers}, {Wilson}, {Bixel},
  {McGruder}, {Rodler}, {Weaver}, {Lewis}, {Fortney}, \&
  {Fraine}]{Espinoza2019}
{Espinoza}, N., {Rackham}, B.~V., {Jord{\'a}n}, A., {Apai}, D.,
  {L{\'o}pez-Morales}, M., {Osip}, D.~J., {Grimm}, S.~L., {Hoeijmakers}, J.,
  {Wilson}, P.~A., {Bixel}, A., {McGruder}, C., {Rodler}, F., {Weaver}, I.,
  {Lewis}, N.~K., {Fortney}, J.~J., \& {Fraine}, J., 2019.
\newblock {ACCESS: a featureless optical transmission spectrum for WASP-19b
  from Magellan/IMACS}, {\it \mnras\/}, {\bf 482}(2), 2065--2087.

\bibitem[{Faria} et~al.(2016){Faria}, {Haywood}, {Brewer}, {Figueira},
  {Oshagh}, {Santerne}, \& {Santos}]{Faria-16}
{Faria}, J.~P., {Haywood}, R.~D., {Brewer}, B.~J., {Figueira}, P., {Oshagh},
  M., {Santerne}, A., \& {Santos}, N.~C., 2016.
\newblock {Uncovering the planets and stellar activity of CoRoT-7 using only
  radial velocities}, {\it \aap\/}, {\bf 588}, A31.

\bibitem[{Faurobert} et~al.(2020){Faurobert}, {Criscuoli}, {Carbillet}, \&
  {Contursi}]{faurobert2020}
{Faurobert}, M., {Criscuoli}, S., {Carbillet}, M., \& {Contursi}, G., 2020.
\newblock {A new spectroscopic method for measuring the temperature gradient in
  the solar photosphere. Generalized application in magnetized regions}, {\it
  \aap\/}, {\bf 642}, A186.

\bibitem[{Faurobert-Scholl}(1993)]{Faurobert-Scholl1993}
{Faurobert-Scholl}, M., 1993.
\newblock {Investigation of microturbulent magnetic fields in the solar
  photosphere by their Hanle effect in the Sr I 4607 angstroms line.}, {\it
  \aap\/}, {\bf 268}, 765--774.

\bibitem[{Feinstein} et~al.(2019){Feinstein}, {Montet}, {Foreman-Mackey},
  {Bedell}, {Saunders}, {Bean}, {Christiansen}, {Hedges}, {Luger}, {Scolnic},
  \& {Cardoso}]{Feinstein2019}
{Feinstein}, A.~D., {Montet}, B.~T., {Foreman-Mackey}, D., {Bedell}, M.~E.,
  {Saunders}, N., {Bean}, J.~L., {Christiansen}, J.~L., {Hedges}, C., {Luger},
  R., {Scolnic}, D., \& {Cardoso}, J. V. d.~M., 2019.
\newblock {eleanor: An Open-source Tool for Extracting Light Curves from the
  TESS Full-frame Images}, {\it \pasp\/}, {\bf 131}(1003), 094502.

\bibitem[{Feinstein} et~al.(2020){Feinstein}, {Montet}, {Ansdell}, {Nord},
  {Bean}, {G{\"u}nther}, {Gully-Santiago}, \& {Schlieder}]{Feinstein2020}
{Feinstein}, A.~D., {Montet}, B.~T., {Ansdell}, M., {Nord}, B., {Bean}, J.~L.,
  {G{\"u}nther}, M.~N., {Gully-Santiago}, M.~A., \& {Schlieder}, J.~E., 2020.
\newblock {Flare Statistics for Young Stars from a Convolutional Neural Network
  Analysis of TESS Data}, {\it \aj\/}, {\bf 160}(5), 219.

\bibitem[{Feller} et~al.(2020){Feller}, {Gandorfer}, {Iglesias}, {Lagg},
  {Riethm{\"u}ller}, {Solanki}, {Katsukawa}, \& {Kubo}]{SUSI}
{Feller}, A., {Gandorfer}, A., {Iglesias}, F.~A., {Lagg}, A.,
  {Riethm{\"u}ller}, T.~L., {Solanki}, S.~K., {Katsukawa}, Y., \& {Kubo}, M.,
  2020.
\newblock {The SUNRISE UV Spectropolarimeter and imager for SUNRISE III}, in
  {\em Society of Photo-Optical Instrumentation Engineers (SPIE) Conference
  Series\/}, vol. 11447 of {\bf Society of Photo-Optical Instrumentation
  Engineers (SPIE) Conference Series}, p. 11447AK.

\bibitem[{Findeisen} et~al.(2011){Findeisen}, {Hillenbrand}, \&
  {Soderblom}]{Findeisen11}
{Findeisen}, K., {Hillenbrand}, L., \& {Soderblom}, D., 2011.
\newblock {Stellar Activity in the Broadband Ultraviolet}, {\it \aj\/}, {\bf
  142}(1), 23.

\bibitem[{Fligge} et~al.(2000){Fligge}, {Solanki}, \& {Unruh}]{Fligge2000}
{Fligge}, M., {Solanki}, S.~K., \& {Unruh}, Y.~C., 2000.
\newblock {Modelling irradiance variations from the surface distribution of the
  solar magnetic field}, {\it \aap\/}, {\bf 353}, 380--388.

\bibitem[{Fontenla} et~al.(1999){Fontenla}, {White}, {Fox}, {Avrett}, \&
  {Kurucz}]{Fontenla:1999}
{Fontenla}, J., {White}, O.~R., {Fox}, P.~A., {Avrett}, E.~H., \& {Kurucz},
  R.~L., 1999.
\newblock {Calculation of Solar Irradiances. I. Synthesis of the Solar
  Spectrum}, {\it \apj\/}, {\bf 518}(1), 480--499.

\bibitem[{Fontenla} et~al.(1993){Fontenla}, {Avrett}, \&
  {Loeser}]{Fontenla:1993}
{Fontenla}, J.~M., {Avrett}, E.~H., \& {Loeser}, R., 1993.
\newblock {Energy Balance in the Solar Transition Region. III. Helium Emission
  in Hydrostatic, Constant-Abundance Models with Diffusion}, {\it \apj\/}, {\bf
  406}, 319.

\bibitem[{Fontenla} et~al.(2002){Fontenla}, {Avrett}, \&
  {Loeser}]{Fontenla:2002}
{Fontenla}, J.~M., {Avrett}, E.~H., \& {Loeser}, R., 2002.
\newblock {Energy Balance in the Solar Transition Region. IV. Hydrogen and
  Helium Mass Flows with Diffusion}, {\it \apj\/}, {\bf 572}(1), 636--662.

\bibitem[{Fontenla} et~al.(2006){Fontenla}, {Avrett}, {Thuillier}, \&
  {Harder}]{Fontenla:2006}
{Fontenla}, J.~M., {Avrett}, E., {Thuillier}, G., \& {Harder}, J., 2006.
\newblock {Semiempirical Models of the Solar Atmosphere. I. The Quiet- and
  Active Sun Photosphere at Moderate Resolution}, {\it \apj\/}, {\bf 639}(1),
  441--458.

\bibitem[{Fontenla} et~al.(2009){Fontenla}, {Curdt}, {Haberreiter}, {Harder},
  \& {Tian}]{Fontenla:2009}
{Fontenla}, J.~M., {Curdt}, W., {Haberreiter}, M., {Harder}, J., \& {Tian}, H.,
  2009.
\newblock {Semiempirical Models of the Solar Atmosphere. III. Set of Non-LTE
  Models for Far-Ultraviolet/Extreme-Ultraviolet Irradiance Computation}, {\it
  \apj\/}, {\bf 707}(1), 482--502.

\bibitem[{Fossati} et~al.(2015){Fossati}, {Ingrassia}, \& {Lanza}]{Fossati2015}
{Fossati}, L., {Ingrassia}, S., \& {Lanza}, A.~F., 2015.
\newblock {A Bimodal Correlation between Host Star Chromospheric Emission and
  the Surface Gravity of Hot Jupiters}, {\it \apjl\/}, {\bf 812}(2), L35.

\bibitem[{Fraine} et~al.(2014){Fraine}, {Deming}, {Benneke}, {Knutson},
  {Jord{\'a}n}, {Espinoza}, {Madhusudhan}, {Wilkins}, \& {Todorov}]{Fraine2014}
{Fraine}, J., {Deming}, D., {Benneke}, B., {Knutson}, H., {Jord{\'a}n}, A.,
  {Espinoza}, N., {Madhusudhan}, N., {Wilkins}, A., \& {Todorov}, K., 2014.
\newblock {Water vapour absorption in the clear atmosphere of a Neptune-sized
  exoplanet}, {\it \nat\/}, {\bf 513}(7519), 526--529.

\bibitem[{France} et~al.(2016){France}, {Loyd}, {Youngblood}, {Brown},
  {Schneider}, {Hawley}, {Froning}, {Linsky}, {Roberge}, {Buccino},
  {Davenport}, {Fontenla}, {Kaltenegger}, {Kowalski}, {Mauas}, {Miguel},
  {Redfield}, {Rugheimer}, {Tian}, {Vieytes}, {Walkowicz}, \&
  {Weisenburger}]{France2016}
{France}, K., {Loyd}, R.~O.~P., {Youngblood}, A., {Brown}, A., {Schneider},
  P.~C., {Hawley}, S.~L., {Froning}, C.~S., {Linsky}, J.~L., {Roberge}, A.,
  {Buccino}, A.~P., {Davenport}, J. R.~A., {Fontenla}, J.~M., {Kaltenegger},
  L., {Kowalski}, A.~F., {Mauas}, P. J.~D., {Miguel}, Y., {Redfield}, S.,
  {Rugheimer}, S., {Tian}, F., {Vieytes}, M.~C., {Walkowicz}, L.~M., \&
  {Weisenburger}, K.~L., 2016.
\newblock {The MUSCLES Treasury Survey. I. Motivation and Overview}, {\it
  \apj\/}, {\bf 820}(2), 89.

\bibitem[{Freytag} et~al.(2002){Freytag}, {Steffen}, \& {Dorch}]{freytag2002}
{Freytag}, B., {Steffen}, M., \& {Dorch}, B., 2002.
\newblock {Spots on the surface of Betelgeuse -- Results from new 3D stellar
  convection models}, {\it Astronomische Nachrichten\/}, {\bf 323}, 213--219.

\bibitem[{Freytag} et~al.(2012){Freytag}, {Steffen}, {Ludwig},
  {Wedemeyer-B{\"o}hm}, {Schaffenberger}, \& {Steiner}]{Co5bold_freytag12}
{Freytag}, B., {Steffen}, M., {Ludwig}, H.~G., {Wedemeyer-B{\"o}hm}, S.,
  {Schaffenberger}, W., \& {Steiner}, O., 2012.
\newblock {Simulations of stellar convection with CO5BOLD}, {\it Journal of
  Computational Physics\/}, {\bf 231}(3), 919--959.

\bibitem[{Frohlich} et~al.(1997){Frohlich}, {Andersen}, {Appourchaux},
  {Berthomieu}, {Crommelynck}, {Domingo}, {Fichot}, {Finsterle}, {Gomez},
  {Gough}, {Jimenez}, {Leifsen}, {Lombaerts}, {Pap}, {Provost}, {Roca Cortes},
  {Romero}, {Roth}, {Sekii}, {Telljohann}, {Toutain}, \& {Wehrli}]{VIRGO}
{Frohlich}, C., {Andersen}, B.~N., {Appourchaux}, T., {Berthomieu}, G.,
  {Crommelynck}, D.~A., {Domingo}, V., {Fichot}, A., {Finsterle}, W., {Gomez},
  M.~F., {Gough}, D., {Jimenez}, A., {Leifsen}, T., {Lombaerts}, M., {Pap},
  J.~M., {Provost}, J., {Roca Cortes}, T., {Romero}, J., {Roth}, H., {Sekii},
  T., {Telljohann}, U., {Toutain}, T., \& {Wehrli}, C., 1997.
\newblock {First Results from VIRGO, the Experiment for Helioseismology and
  Solar Irradiance Monitoring on SOHO}, {\it \solphys\/}, {\bf 170}(1), 1--25.

\bibitem[{Fuhrmeister} et~al.(2022){Fuhrmeister}, {Czesla}, {Nagel}, {Reiners},
  {Schmitt}, {Jeffers}, {Caballero}, {Shulyak}, {Johnson}, {Zechmeister},
  {Montes}, {L{\'o}pez-Gallifa}, {Ribas}, {Quirrenbach}, {Amado},
  {Galad{\'\i}-Enr{\'\i}quez}, {Hatzes}, {K{\"u}rster}, {Danielski},
  {B{\'e}jar}, {Kaminski}, {Morales}, \& {Zapatero Osorio}]{Fuhrmeister2022}
{Fuhrmeister}, B., {Czesla}, S., {Nagel}, E., {Reiners}, A., {Schmitt},
  J.~H.~M.~M., {Jeffers}, S.~V., {Caballero}, J.~A., {Shulyak}, D., {Johnson},
  E.~N., {Zechmeister}, M., {Montes}, D., {L{\'o}pez-Gallifa}, {\'A}., {Ribas},
  I., {Quirrenbach}, A., {Amado}, P.~J., {Galad{\'\i}-Enr{\'\i}quez}, D.,
  {Hatzes}, A.~P., {K{\"u}rster}, M., {Danielski}, C., {B{\'e}jar}, V.~J.~S.,
  {Kaminski}, A., {Morales}, J.~C., \& {Zapatero Osorio}, M.~R., 2022.
\newblock {The CARMENES search for exoplanets around M dwarfs. Diagnostic
  capabilities of strong K I lines for photosphere and chromosphere}, {\it
  \aap\/}, {\bf 657}, A125.

\bibitem[{Gaidos} et~al.(2023){Gaidos}, {Claytor}, {Dungee}, {Ali}, \&
  {Feiden}]{Gaidos2023}
{Gaidos}, E., {Claytor}, Z., {Dungee}, R., {Ali}, A., \& {Feiden}, G.~A., 2023.
\newblock {The TIME Table: Rotation and ages of cool exoplanet host stars},
  {\it \mnras\/}.

\bibitem[{Gan} et~al.(2022{\natexlab{a}}){Gan}, {Lin}, {Wang}, {Mao},
  {Fouqu{\'e}}, {Fan}, {Bedell}, {Stassun}, {Giacalone}, {Fukui}, {Murgas},
  {Ciardi}, {Howell}, {Collins}, {Shporer}, {Arnold}, {Barclay}, {Charbonneau},
  {Christiansen}, {Crossfield}, {Dressing}, {Elliott}, {Esparza-Borges},
  {Evans}, {Gnilka}, {Gonzales}, {Howard}, {Isogai}, {Kawauchi}, {Kurita},
  {Liu}, {Livingston}, {Matson}, {Narita}, {Palle}, {Parviainen}, {Rackham},
  {Rodriguez}, {Rose}, {Rudat}, {Schlieder}, {Scott}, {Vezie}, {Ricker},
  {Vanderspek}, {Latham}, {Seager}, {Winn}, \& {Jenkins}]{Gan2022_TOI530b}
{Gan}, T., {Lin}, Z., {Wang}, S.~X., {Mao}, S., {Fouqu{\'e}}, P., {Fan}, J.,
  {Bedell}, M., {Stassun}, K.~G., {Giacalone}, S., {Fukui}, A., {Murgas}, F.,
  {Ciardi}, D.~R., {Howell}, S.~B., {Collins}, K.~A., {Shporer}, A., {Arnold},
  L., {Barclay}, T., {Charbonneau}, D., {Christiansen}, J., {Crossfield}, I.
  J.~M., {Dressing}, C.~D., {Elliott}, A., {Esparza-Borges}, E., {Evans}, P.,
  {Gnilka}, C.~L., {Gonzales}, E.~J., {Howard}, A.~W., {Isogai}, K.,
  {Kawauchi}, K., {Kurita}, S., {Liu}, B., {Livingston}, J.~H., {Matson},
  R.~A., {Narita}, N., {Palle}, E., {Parviainen}, H., {Rackham}, B.~V.,
  {Rodriguez}, D.~R., {Rose}, M., {Rudat}, A., {Schlieder}, J.~E., {Scott},
  N.~J., {Vezie}, M., {Ricker}, G.~R., {Vanderspek}, R., {Latham}, D.~W.,
  {Seager}, S., {Winn}, J.~N., \& {Jenkins}, J.~M., 2022{\natexlab{a}}.
\newblock {TOI-530b: a giant planet transiting an M-dwarf detected by TESS},
  {\it \mnras\/}, {\bf 511}(1), 83--99.

\bibitem[{Gan} et~al.(2022{\natexlab{b}}){Gan}, {Soubkiou}, {Wang},
  {Benkhaldoun}, {Mao}, {Artigau}, {Fouqu{\'e}}, {Arnold}, {Giacalone},
  {Theissen}, {Aganze}, {Burgasser}, {Collins}, {Shporer}, {Barkaoui},
  {Ghachoui}, {Howell}, {Lamman}, {Demangeon}, {Burdanov}, {Cadieux},
  {Chouqar}, {Collins}, {Cook}, {Delrez}, {Demory}, {Doyon}, {Dransfield},
  {Dressing}, {Ducrot}, {Fan}, {Garcia}, {Gill}, {Gillon}, {Gnilka}, {G{\'o}mez
  Maqueo Chew}, {G{\"u}nther}, {Henze}, {Huang}, {Jehin}, {Jensen}, {Lin},
  {Manset}, {McCormac}, {Murray}, {Niraula}, {Pedersen}, {Pozuelos}, {Queloz},
  {Rackham}, {Savel}, {Schanche}, {Schwarz}, {Sebastian}, {Thompson},
  {Timmermans}, {Triaud}, {Vezie}, {Wells}, {de Wit}, {Ricker}, {Vanderspek},
  {Latham}, {Seager}, {Winn}, \& {Jenkins}]{Gan2022_TOI2136b}
{Gan}, T., {Soubkiou}, A., {Wang}, S.~X., {Benkhaldoun}, Z., {Mao}, S.,
  {Artigau}, {\'E}., {Fouqu{\'e}}, P., {Arnold}, L., {Giacalone}, S.,
  {Theissen}, C.~A., {Aganze}, C., {Burgasser}, A., {Collins}, K.~A.,
  {Shporer}, A., {Barkaoui}, K., {Ghachoui}, M., {Howell}, S.~B., {Lamman}, C.,
  {Demangeon}, O. D.~S., {Burdanov}, A., {Cadieux}, C., {Chouqar}, J.,
  {Collins}, K.~I., {Cook}, N.~J., {Delrez}, L., {Demory}, B.-O., {Doyon}, R.,
  {Dransfield}, G., {Dressing}, C.~D., {Ducrot}, E., {Fan}, J., {Garcia}, L.,
  {Gill}, H., {Gillon}, M., {Gnilka}, C.~L., {G{\'o}mez Maqueo Chew}, Y.,
  {G{\"u}nther}, M.~N., {Henze}, C.~E., {Huang}, C.~X., {Jehin}, E., {Jensen},
  E. L.~N., {Lin}, Z., {Manset}, N., {McCormac}, J., {Murray}, C.~A.,
  {Niraula}, P., {Pedersen}, P.~P., {Pozuelos}, F.~J., {Queloz}, D., {Rackham},
  B.~V., {Savel}, A.~B., {Schanche}, N., {Schwarz}, R.~P., {Sebastian}, D.,
  {Thompson}, S., {Timmermans}, M., {Triaud}, A. H.~M.~J., {Vezie}, M.,
  {Wells}, R.~D., {de Wit}, J., {Ricker}, G.~R., {Vanderspek}, R., {Latham},
  D.~W., {Seager}, S., {Winn}, J.~N., \& {Jenkins}, J.~M., 2022{\natexlab{b}}.
\newblock {TESS discovery of a sub-Neptune orbiting a mid-M dwarf TOI-2136},
  {\it \mnras\/}, {\bf 514}(3), 4120--4139.

\bibitem[{Garcia} et~al.(2016){Garcia}, {Muterspaugh}, {van Belle}, {Monnier},
  {Stassun}, {Ghasempour}, {Clark}, {Zavala}, {Benson}, {Hutter}, {Schmitt},
  {Baines}, {Jorgensen}, {Strosahl}, {Sanborn}, {Zawicki}, {Sakosky}, \&
  {Swihart}]{Garcia2016}
{Garcia}, E.~V., {Muterspaugh}, M.~W., {van Belle}, G., {Monnier}, J.~D.,
  {Stassun}, K.~G., {Ghasempour}, A., {Clark}, J.~H., {Zavala}, R.~T.,
  {Benson}, J.~A., {Hutter}, D.~J., {Schmitt}, H.~R., {Baines}, E.~K.,
  {Jorgensen}, A.~M., {Strosahl}, S.~G., {Sanborn}, J., {Zawicki}, S.~J.,
  {Sakosky}, M.~F., \& {Swihart}, S., 2016.
\newblock {Vision: A Six-telescope Fiber-fed Visible Light Beam Combiner for
  the Navy Precision Optical Interferometer}, {\it \pasp\/}, {\bf 128}(963),
  055004.

\bibitem[{Gaudi} et~al.(2020){Gaudi}, {Seager}, {Mennesson}, {Kiessling},
  {Warfield}, {Cahoy}, {Clarke}, {Domagal-Goldman}, {Feinberg}, {Guyon},
  {Kasdin}, {Mawet}, {Plavchan}, {Robinson}, {Rogers}, {Scowen}, {Somerville},
  {Stapelfeldt}, {Stark}, {Stern}, {Turnbull}, {Amini}, {Kuan}, {Martin},
  {Morgan}, {Redding}, {Stahl}, {Webb}, {Alvarez-Salazar}, {Arnold}, {Arya},
  {Balasubramanian}, {Baysinger}, {Bell}, {Below}, {Benson}, {Blais}, {Booth},
  {Bourgeois}, {Bradford}, {Brewer}, {Brooks}, {Cady}, {Caldwell}, {Calvet},
  {Carr}, {Chan}, {Cormarkovic}, {Coste}, {Cox}, {Danner}, {Davis}, {Dewell},
  {Dorsett}, {Dunn}, {East}, {Effinger}, {Eng}, {Freebury}, {Garcia}, {Gaskin},
  {Greene}, {Hennessy}, {Hilgemann}, {Hood}, {Holota}, {Howe}, {Huang}, {Hull},
  {Hunt}, {Hurd}, {Johnson}, {Kissil}, {Knight}, {Kolenz}, {Kraus}, {Krist},
  {Li}, {Lisman}, {Mandic}, {Mann}, {Marchen}, {Marrese-Reading}, {McCready},
  {McGown}, {Missun}, {Miyaguchi}, {Moore}, {Nemati}, {Nikzad}, {Nissen},
  {Novicki}, {Perrine}, {Pineda}, {Polanco}, {Putnam}, {Qureshi}, {Richards},
  {Eldorado Riggs}, {Rodgers}, {Rud}, {Saini}, {Scalisi}, {Scharf}, {Schulz},
  {Serabyn}, {Sigrist}, {Sikkia}, {Singleton}, {Shaklan}, {Smith}, {Southerd},
  {Stahl}, {Steeves}, {Sturges}, {Sullivan}, {Tang}, {Taras}, {Tesch},
  {Therrell}, {Tseng}, {Valente}, {Van Buren}, {Villalvazo}, {Warwick}, {Webb},
  {Westerhoff}, {Wofford}, {Wu}, {Woo}, {Wood}, {Ziemer}, {Arney}, {Anderson},
  {Ma{\'\i}z-Apell{\'a}niz}, {Bartlett}, {Belikov}, {Bendek}, {Cenko},
  {Douglas}, {Dulz}, {Evans}, {Faramaz}, {Feng}, {Ferguson}, {Follette},
  {Ford}, {Garc{\'\i}a}, {Geha}, {Gelino}, {G{\"o}tberg}, {Hildebrandt}, {Hu},
  {Jahnke}, {Kennedy}, {Kreidberg}, {Isella}, {Lopez}, {Marchis}, {Macri},
  {Marley}, {Matzko}, {Mazoyer}, {McCandliss}, {Meshkat}, {Mordasini},
  {Morris}, {Nielsen}, {Newman}, {Petigura}, {Postman}, {Reines}, {Roberge},
  {Roederer}, {Ruane}, {Schwieterman}, {Sirbu}, {Spalding}, {Teplitz},
  {Tumlinson}, {Turner}, {Werk}, {Wofford}, {Wyatt}, {Young}, \&
  {Zellem}]{Gaudi2020}
{Gaudi}, B.~S., {Seager}, S., {Mennesson}, B., {Kiessling}, A., {Warfield}, K.,
  {Cahoy}, K., {Clarke}, J.~T., {Domagal-Goldman}, S., {Feinberg}, L., {Guyon},
  O., {Kasdin}, J., {Mawet}, D., {Plavchan}, P., {Robinson}, T., {Rogers}, L.,
  {Scowen}, P., {Somerville}, R., {Stapelfeldt}, K., {Stark}, C., {Stern}, D.,
  {Turnbull}, M., {Amini}, R., {Kuan}, G., {Martin}, S., {Morgan}, R.,
  {Redding}, D., {Stahl}, H.~P., {Webb}, R., {Alvarez-Salazar}, O., {Arnold},
  W.~L., {Arya}, M., {Balasubramanian}, B., {Baysinger}, M., {Bell}, R.,
  {Below}, C., {Benson}, J., {Blais}, L., {Booth}, J., {Bourgeois}, R.,
  {Bradford}, C., {Brewer}, A., {Brooks}, T., {Cady}, E., {Caldwell}, M.,
  {Calvet}, R., {Carr}, S., {Chan}, D., {Cormarkovic}, V., {Coste}, K., {Cox},
  C., {Danner}, R., {Davis}, J., {Dewell}, L., {Dorsett}, L., {Dunn}, D.,
  {East}, M., {Effinger}, M., {Eng}, R., {Freebury}, G., {Garcia}, J.,
  {Gaskin}, J., {Greene}, S., {Hennessy}, J., {Hilgemann}, E., {Hood}, B.,
  {Holota}, W., {Howe}, S., {Huang}, P., {Hull}, T., {Hunt}, R., {Hurd}, K.,
  {Johnson}, S., {Kissil}, A., {Knight}, B., {Kolenz}, D., {Kraus}, O.,
  {Krist}, J., {Li}, M., {Lisman}, D., {Mandic}, M., {Mann}, J., {Marchen}, L.,
  {Marrese-Reading}, C., {McCready}, J., {McGown}, J., {Missun}, J.,
  {Miyaguchi}, A., {Moore}, B., {Nemati}, B., {Nikzad}, S., {Nissen}, J.,
  {Novicki}, M., {Perrine}, T., {Pineda}, C., {Polanco}, O., {Putnam}, D.,
  {Qureshi}, A., {Richards}, M., {Eldorado Riggs}, A.~J., {Rodgers}, M., {Rud},
  M., {Saini}, N., {Scalisi}, D., {Scharf}, D., {Schulz}, K., {Serabyn}, G.,
  {Sigrist}, N., {Sikkia}, G., {Singleton}, A., {Shaklan}, S., {Smith}, S.,
  {Southerd}, B., {Stahl}, M., {Steeves}, J., {Sturges}, B., {Sullivan}, C.,
  {Tang}, H., {Taras}, N., {Tesch}, J., {Therrell}, M., {Tseng}, H., {Valente},
  M., {Van Buren}, D., {Villalvazo}, J., {Warwick}, S., {Webb}, D.,
  {Westerhoff}, T., {Wofford}, R., {Wu}, G., {Woo}, J., {Wood}, M., {Ziemer},
  J., {Arney}, G., {Anderson}, J., {Ma{\'\i}z-Apell{\'a}niz}, J., {Bartlett},
  J., {Belikov}, R., {Bendek}, E., {Cenko}, B., {Douglas}, E., {Dulz}, S.,
  {Evans}, C., {Faramaz}, V., {Feng}, Y.~K., {Ferguson}, H., {Follette}, K.,
  {Ford}, S., {Garc{\'\i}a}, M., {Geha}, M., {Gelino}, D., {G{\"o}tberg}, Y.,
  {Hildebrandt}, S., {Hu}, R., {Jahnke}, K., {Kennedy}, G., {Kreidberg}, L.,
  {Isella}, A., {Lopez}, E., {Marchis}, F., {Macri}, L., {Marley}, M.,
  {Matzko}, W., {Mazoyer}, J., {McCandliss}, S., {Meshkat}, T., {Mordasini},
  C., {Morris}, P., {Nielsen}, E., {Newman}, P., {Petigura}, E., {Postman}, M.,
  {Reines}, A., {Roberge}, A., {Roederer}, I., {Ruane}, G., {Schwieterman}, E.,
  {Sirbu}, D., {Spalding}, C., {Teplitz}, H., {Tumlinson}, J., {Turner}, N.,
  {Werk}, J., {Wofford}, A., {Wyatt}, M., {Young}, A., \& {Zellem}, R., 2020.
\newblock {The Habitable Exoplanet Observatory (HabEx) Mission Concept Study
  Final Report}, {\it arXiv e-prints\/}, p. arXiv:2001.06683.

\bibitem[{Gaulme} et~al.(2016){Gaulme}, {Rowe}, {Bedding}, {Benomar},
  {Corsaro}, {Davies}, {Hale}, {Howe}, {Garcia}, {Huber}, {Jim{\'e}nez},
  {Mathur}, {Mosser}, {Appourchaux}, {Boumier}, {Jackiewicz}, {Leibacher},
  {Schmider}, {Hammel}, {Lissauer}, {Marley}, {Simon}, {Chaplin}, {Elsworth},
  {Guzik}, {Murphy}, \& {Silva Aguirre}]{Gaulme:2016}
{Gaulme}, P., {Rowe}, J.~F., {Bedding}, T.~R., {Benomar}, O., {Corsaro}, E.,
  {Davies}, G.~R., {Hale}, S.~J., {Howe}, R., {Garcia}, R.~A., {Huber}, D.,
  {Jim{\'e}nez}, A., {Mathur}, S., {Mosser}, B., {Appourchaux}, T., {Boumier},
  P., {Jackiewicz}, J., {Leibacher}, J., {Schmider}, F.~X., {Hammel}, H.~B.,
  {Lissauer}, J.~J., {Marley}, M.~S., {Simon}, A.~A., {Chaplin}, W.~J.,
  {Elsworth}, Y., {Guzik}, J.~A., {Murphy}, N., \& {Silva Aguirre}, V., 2016.
\newblock {A Distant Mirror: Solar Oscillations Observed on Neptune by the
  Kepler K2 Mission}, {\it \apjl\/}, {\bf 833}(1), L13.

\bibitem[{Ghosh} et~al.(2016){Ghosh}, {Chatterjee}, {Khan}, {Tripathi},
  {Ramaprakash}, {Banerjee}, {Chordia}, {Gandorfer}, {Krivova}, {Nandy},
  {Rajarshi}, {Solanki}, \& {Sriram}]{Ghosh:2016}
{Ghosh}, A., {Chatterjee}, S., {Khan}, A.~R., {Tripathi}, D., {Ramaprakash},
  A.~N., {Banerjee}, D., {Chordia}, P., {Gandorfer}, A.~M., {Krivova}, N.,
  {Nandy}, D., {Rajarshi}, C., {Solanki}, S.~K., \& {Sriram}, S., 2016.
\newblock {The Solar Ultraviolet Imaging Telescope onboard Aditya-L1}, in {\em
  Space Telescopes and Instrumentation 2016: Ultraviolet to Gamma Ray\/}, vol.
  9905 of {\bf Society of Photo-Optical Instrumentation Engineers (SPIE)
  Conference Series}, p. 990503.

\bibitem[{Gibbs} et~al.(2020){Gibbs}, {Bixel}, {Rackham}, {Apai}, {Schlecker},
  {Espinoza}, {Mancini}, {Chen}, {Henning}, {Gabor}, {Boyle}, {Perez Chavez},
  {Mousseau}, {Dietrich}, {Jay Socia}, {Ip}, {Ngeow}, {Tsai}, {Bhandare},
  {Marian}, {Baehr}, {Brown}, {H{\"a}berle}, {Keppler}, {Molaverdikhani}, \&
  {Sarkis}]{eden_gibbs2020}
{Gibbs}, A., {Bixel}, A., {Rackham}, B.~V., {Apai}, D., {Schlecker}, M.,
  {Espinoza}, N., {Mancini}, L., {Chen}, W.-P., {Henning}, T., {Gabor}, P.,
  {Boyle}, R., {Perez Chavez}, J., {Mousseau}, A., {Dietrich}, J., {Jay Socia},
  Q., {Ip}, W., {Ngeow}, C.-C., {Tsai}, A.-L., {Bhandare}, A., {Marian}, V.,
  {Baehr}, H., {Brown}, S., {H{\"a}berle}, M., {Keppler}, M., {Molaverdikhani},
  K., \& {Sarkis}, P., 2020.
\newblock {EDEN: Sensitivity Analysis and Transiting Planet Detection Limits
  for Nearby Late Red Dwarfs}, {\it \aj\/}, {\bf 159}(4), 169.

\bibitem[{Giles} et~al.(2017){Giles}, {Collier Cameron}, \&
  {Haywood}]{Giles:2017}
{Giles}, H. A.~C., {Collier Cameron}, A., \& {Haywood}, R.~D., 2017.
\newblock {A Kepler study of starspot lifetimes with respect to light-curve
  amplitude and spectral type}, {\it \mnras\/}, {\bf 472}(2), 1618--1627.

\bibitem[{Gizis} et~al.(2015){Gizis}, {Dettman}, {Burgasser}, {Camnasio},
  {Alam}, {Filippazzo}, {Cruz}, {Metchev}, {Berger}, \& {Williams}]{Gizis2015}
{Gizis}, J.~E., {Dettman}, K.~G., {Burgasser}, A.~J., {Camnasio}, S., {Alam},
  M., {Filippazzo}, J.~C., {Cruz}, K.~L., {Metchev}, S., {Berger}, E., \&
  {Williams}, P. K.~G., 2015.
\newblock {Kepler Monitoring of an L Dwarf. II. Clouds with Multi-year
  Lifetimes}, {\it \apj\/}, {\bf 813}(2), 104.

\bibitem[{Gnevyshev}(1938)]{Gnevyshev1938}
{Gnevyshev}, M.~N., 1938.
\newblock {On the nature of solar activity}, {\it Izvestiya Glavnoj
  Astronomicheskoj Observatorii v Pulkove\/}, {\bf 16}, 36--45.

\bibitem[{Goode} et~al.(2010){Goode}, {Coulter}, {Gorceix}, {Yurchyshyn}, \&
  {Cao}]{Goode2010}
{Goode}, P.~R., {Coulter}, R., {Gorceix}, N., {Yurchyshyn}, V., \& {Cao}, W.,
  2010.
\newblock {The NST: First results and some lessons for ATST and EST}, {\it
  Astronomische Nachrichten\/}, {\bf 331}(6), 620.

\bibitem[{Gravity Collaboration} et~al.(2017){Gravity Collaboration}, {Abuter},
  {Accardo}, {Amorim}, {Anugu}, {{\'A}vila}, {Azouaoui}, {Benisty}, {Berger},
  {Blind}, {Bonnet}, {Bourget}, {Brandner}, {Brast}, {Buron}, {Burtscher},
  {Cassaing}, {Chapron}, {Choquet}, {Cl{\'e}net}, {Collin}, {Coud{\'e} Du
  Foresto}, {de Wit}, {de Zeeuw}, {Deen}, {Delplancke-Str{\"o}bele}, {Dembet},
  {Derie}, {Dexter}, {Duvert}, {Ebert}, {Eckart}, {Eisenhauer}, {Esselborn},
  {F{\'e}dou}, {Finger}, {Garcia}, {Garcia Dabo}, {Garcia Lopez}, {Gendron},
  {Genzel}, {Gillessen}, {Gonte}, {Gordo}, {Grould}, {Gr{\"o}zinger}, {Guieu},
  {Haguenauer}, {Hans}, {Haubois}, {Haug}, {Haussmann}, {Henning}, {Hippler},
  {Horrobin}, {Huber}, {Hubert}, {Hubin}, {Hummel}, {Jakob}, {Janssen},
  {Jochum}, {Jocou}, {Kaufer}, {Kellner}, {Kendrew}, {Kern}, {Kervella},
  {Kiekebusch}, {Klein}, {Kok}, {Kolb}, {Kulas}, {Lacour}, {Lapeyr{\`e}re},
  {Lazareff}, {Le Bouquin}, {L{\`e}na}, {Lenzen}, {L{\'e}v{\^e}que}, {Lippa},
  {Magnard}, {Mehrgan}, {Mellein}, {M{\'e}rand}, {Moreno-Ventas}, {Moulin},
  {M{\"u}ller}, {M{\"u}ller}, {Neumann}, {Oberti}, {Ott}, {Pallanca},
  {Panduro}, {Pasquini}, {Paumard}, {Percheron}, {Perraut}, {Perrin},
  {Pfl{\"u}ger}, {Pfuhl}, {Phan Duc}, {Plewa}, {Popovic}, {Rabien},
  {Ram{\'\i}rez}, {Ramos}, {Rau}, {Riquelme}, {Rohloff}, {Rousset},
  {Sanchez-Bermudez}, {Scheithauer}, {Sch{\"o}ller}, {Schuhler}, {Spyromilio},
  {Straubmeier}, {Sturm}, {Suarez}, {Tristram}, {Ventura}, {Vincent},
  {Waisberg}, {Wank}, {Weber}, {Wieprecht}, {Wiest}, {Wiezorrek}, {Wittkowski},
  {Woillez}, {Wolff}, {Yazici}, {Ziegler}, \& {Zins}]{Gravity2017}
{Gravity Collaboration}, {Abuter}, R., {Accardo}, M., {Amorim}, A., {Anugu},
  N., {{\'A}vila}, G., {Azouaoui}, N., {Benisty}, M., {Berger}, J.~P., {Blind},
  N., {Bonnet}, H., {Bourget}, P., {Brandner}, W., {Brast}, R., {Buron}, A.,
  {Burtscher}, L., {Cassaing}, F., {Chapron}, F., {Choquet}, {\'E}.,
  {Cl{\'e}net}, Y., {Collin}, C., {Coud{\'e} Du Foresto}, V., {de Wit}, W., {de
  Zeeuw}, P.~T., {Deen}, C., {Delplancke-Str{\"o}bele}, F., {Dembet}, R.,
  {Derie}, F., {Dexter}, J., {Duvert}, G., {Ebert}, M., {Eckart}, A.,
  {Eisenhauer}, F., {Esselborn}, M., {F{\'e}dou}, P., {Finger}, G., {Garcia},
  P., {Garcia Dabo}, C.~E., {Garcia Lopez}, R., {Gendron}, E., {Genzel}, R.,
  {Gillessen}, S., {Gonte}, F., {Gordo}, P., {Grould}, M., {Gr{\"o}zinger}, U.,
  {Guieu}, S., {Haguenauer}, P., {Hans}, O., {Haubois}, X., {Haug}, M.,
  {Haussmann}, F., {Henning}, T., {Hippler}, S., {Horrobin}, M., {Huber}, A.,
  {Hubert}, Z., {Hubin}, N., {Hummel}, C.~A., {Jakob}, G., {Janssen}, A.,
  {Jochum}, L., {Jocou}, L., {Kaufer}, A., {Kellner}, S., {Kendrew}, S.,
  {Kern}, L., {Kervella}, P., {Kiekebusch}, M., {Klein}, R., {Kok}, Y., {Kolb},
  J., {Kulas}, M., {Lacour}, S., {Lapeyr{\`e}re}, V., {Lazareff}, B., {Le
  Bouquin}, J.~B., {L{\`e}na}, P., {Lenzen}, R., {L{\'e}v{\^e}que}, S.,
  {Lippa}, M., {Magnard}, Y., {Mehrgan}, L., {Mellein}, M., {M{\'e}rand}, A.,
  {Moreno-Ventas}, J., {Moulin}, T., {M{\"u}ller}, E., {M{\"u}ller}, F.,
  {Neumann}, U., {Oberti}, S., {Ott}, T., {Pallanca}, L., {Panduro}, J.,
  {Pasquini}, L., {Paumard}, T., {Percheron}, I., {Perraut}, K., {Perrin}, G.,
  {Pfl{\"u}ger}, A., {Pfuhl}, O., {Phan Duc}, T., {Plewa}, P.~M., {Popovic},
  D., {Rabien}, S., {Ram{\'\i}rez}, A., {Ramos}, J., {Rau}, C., {Riquelme}, M.,
  {Rohloff}, R.~R., {Rousset}, G., {Sanchez-Bermudez}, J., {Scheithauer}, S.,
  {Sch{\"o}ller}, M., {Schuhler}, N., {Spyromilio}, J., {Straubmeier}, C.,
  {Sturm}, E., {Suarez}, M., {Tristram}, K.~R.~W., {Ventura}, N., {Vincent},
  F., {Waisberg}, I., {Wank}, I., {Weber}, J., {Wieprecht}, E., {Wiest}, M.,
  {Wiezorrek}, E., {Wittkowski}, M., {Woillez}, J., {Wolff}, B., {Yazici}, S.,
  {Ziegler}, D., \& {Zins}, G., 2017.
\newblock {First light for GRAVITY: Phase referencing optical interferometry
  for the Very Large Telescope Interferometer}, {\it \aap\/}, {\bf 602}, A94.

\bibitem[{Gray}(2005)]{Gray-05}
{Gray}, D.~F., 2005.
\newblock {\it {The Observation and Analysis of Stellar Photospheres}\/},
  Cambridge University Press.

\bibitem[{Gray} \& {Johanson}(1991)]{GrayJohanson1991}
{Gray}, D.~F. \& {Johanson}, H.~L., 1991.
\newblock {Precise Measurement of Stellar Temperatures Using Line-Depth
  Ratios}, {\it \pasp\/}, {\bf 103}, 439.

\bibitem[{Greene} et~al.(2016){Greene}, {Line}, {Montero}, {Fortney},
  {Lustig-Yaeger}, \& {Luther}]{greene2016}
{Greene}, T.~P., {Line}, M.~R., {Montero}, C., {Fortney}, J.~J.,
  {Lustig-Yaeger}, J., \& {Luther}, K., 2016.
\newblock {Characterizing Transiting Exoplanet Atmospheres with JWST}, {\it
  \apj\/}, {\bf 817}(1), 17.

\bibitem[{Grindlay} et~al.(2009){Grindlay}, {Tang}, {Simcoe}, {Laycock}, {Los},
  {Mink}, {Doane}, \& {Champine}]{Grindlay2009}
{Grindlay}, J., {Tang}, S., {Simcoe}, R., {Laycock}, S., {Los}, E., {Mink}, D.,
  {Doane}, A., \& {Champine}, G., 2009.
\newblock {DASCH to Measure (and preserve) the Harvard Plates: Opening the
  {\ensuremath{\sim}}100-year Time Domain Astronomy Window}, in {\em Preserving
  Astronomy's Photographic Legacy: Current State and the Future of North
  American Astronomical Plates\/}, vol. 410 of {\bf Astronomical Society of the
  Pacific Conference Series}, p. 101.

\bibitem[{Grindlay} et~al.(2012){Grindlay}, {Tang}, {Los}, \&
  {Servillat}]{Grindlay2012}
{Grindlay}, J., {Tang}, S., {Los}, E., \& {Servillat}, M., 2012.
\newblock {Opening the 100-Year Window for Time-Domain Astronomy}, in {\em New
  Horizons in Time Domain Astronomy\/}, vol. 285, pp. 29--34.

\bibitem[{Gudiksen} et~al.(2011){Gudiksen}, {Carlsson}, {Hansteen}, {Hayek},
  {Leenaarts}, \& {Mart{\'\i}nez-Sykora}]{gudiksen2011bifrost}
{Gudiksen}, B.~V., {Carlsson}, M., {Hansteen}, V.~H., {Hayek}, W., {Leenaarts},
  J., \& {Mart{\'\i}nez-Sykora}, J., 2011.
\newblock {The stellar atmosphere simulation code Bifrost. Code description and
  validation}, {\it \aap\/}, {\bf 531}, A154.

\bibitem[{Gully-Santiago} et~al.(2017){Gully-Santiago}, {Herczeg}, {Czekala},
  {Somers}, {Grankin}, {Covey}, {Donati}, {Alencar}, {Hussain}, {Shappee},
  {Mace}, {Lee}, {Holoien}, {Jose}, \& {Liu}]{Gully-Santiago2017}
{Gully-Santiago}, M.~A., {Herczeg}, G.~J., {Czekala}, I., {Somers}, G.,
  {Grankin}, K., {Covey}, K.~R., {Donati}, J.~F., {Alencar}, S. H.~P.,
  {Hussain}, G. A.~J., {Shappee}, B.~J., {Mace}, G.~N., {Lee}, J.-J.,
  {Holoien}, T.~W.~S., {Jose}, J., \& {Liu}, C.-F., 2017.
\newblock {Placing the Spotted T Tauri Star LkCa 4 on an HR Diagram}, {\it
  \apj\/}, {\bf 836}(2), 200.

\bibitem[{G{\"u}nther} et~al.(2020){G{\"u}nther}, {Zhan}, {Seager}, {Rimmer},
  {Ranjan}, {Stassun}, {Oelkers}, {Daylan}, {Newton}, {Kristiansen}, {Olah},
  {Gillen}, {Rappaport}, {Ricker}, {Vanderspek}, {Latham}, {Winn}, {Jenkins},
  {Glidden}, {Fausnaugh}, {Levine}, {Dittmann}, {Quinn}, {Krishnamurthy}, \&
  {Ting}]{Guenther2020}
{G{\"u}nther}, M.~N., {Zhan}, Z., {Seager}, S., {Rimmer}, P.~B., {Ranjan}, S.,
  {Stassun}, K.~G., {Oelkers}, R.~J., {Daylan}, T., {Newton}, E.,
  {Kristiansen}, M.~H., {Olah}, K., {Gillen}, E., {Rappaport}, S., {Ricker},
  G.~R., {Vanderspek}, R.~K., {Latham}, D.~W., {Winn}, J.~N., {Jenkins}, J.~M.,
  {Glidden}, A., {Fausnaugh}, M., {Levine}, A.~M., {Dittmann}, J.~A., {Quinn},
  S.~N., {Krishnamurthy}, A., \& {Ting}, E.~B., 2020.
\newblock {Stellar Flares from the First TESS Data Release: Exploring a New
  Sample of M Dwarfs}, {\it \aj\/}, {\bf 159}(2), 60.

\bibitem[{Hale}(1908)]{Hale:1908}
{Hale}, G.~E., 1908.
\newblock {On the Probable Existence of a Magnetic Field in Sun-Spots}, {\it
  \apj\/}, {\bf 28}, 315.

\bibitem[{Hall} \& {Henry}(1994)]{HallHenry1994}
{Hall}, D.~S. \& {Henry}, G.~W., 1994.
\newblock {The Law of Starspot Lifetimes}, {\it International
  Amateur-Professional Photoelectric Photometry Communications\/}, {\bf 55},
  51.

\bibitem[{Hall} et~al.(2021){Hall}, {Davies}, {van Saders}, {Nielsen}, {Lund},
  {Chaplin}, {Garc{\'\i}a}, {Amard}, {Breimann}, {Khan}, {See}, \&
  {Tayar}]{Hall2021}
{Hall}, O.~J., {Davies}, G.~R., {van Saders}, J., {Nielsen}, M.~B., {Lund},
  M.~N., {Chaplin}, W.~J., {Garc{\'\i}a}, R.~A., {Amard}, L., {Breimann},
  A.~A., {Khan}, S., {See}, V., \& {Tayar}, J., 2021.
\newblock {Weakened magnetic braking supported by asteroseismic rotation rates
  of Kepler dwarfs}, {\it Nature Astronomy\/}.

\bibitem[{Hamann} et~al.(2019){Hamann}, {Montet}, {Fabrycky}, {Agol}, \&
  {Kruse}]{Hamann2019}
{Hamann}, A., {Montet}, B.~T., {Fabrycky}, D.~C., {Agol}, E., \& {Kruse}, E.,
  2019.
\newblock {K2-146: Discovery of Planet c, Precise Masses from Transit Timing,
  and Observed Precession}, {\it \aj\/}, {\bf 158}(3), 133.

\bibitem[{Harvey} \& {Livingston}(1969)]{Harvey:1969}
{Harvey}, J. \& {Livingston}, W., 1969.
\newblock {Magnetograph Measurements with Temperature-Sensitive Lines}, {\it
  \solphys\/}, {\bf 10}(2), 283--293.

\bibitem[{Hathaway}(2015)]{Hathaway2015}
{Hathaway}, D.~H., 2015.
\newblock {The Solar Cycle}, {\it Living Reviews in Solar Physics\/}, {\bf
  12}(1), 4.

\bibitem[{Hawley} \& {Pettersen}(1991)]{Hawley1991}
{Hawley}, S.~L. \& {Pettersen}, B.~R., 1991.
\newblock {The Great Flare of 1985 April 12 on AD Leonis}, {\it \apj\/}, {\bf
  378}, 725.

\bibitem[{Hawley} et~al.(2014){Hawley}, {Davenport}, {Kowalski}, {Wisniewski},
  {Hebb}, {Deitrick}, \& {Hilton}]{Hawley2014}
{Hawley}, S.~L., {Davenport}, J. R.~A., {Kowalski}, A.~F., {Wisniewski}, J.~P.,
  {Hebb}, L., {Deitrick}, R., \& {Hilton}, E.~J., 2014.
\newblock {Kepler Flares. I. Active and Inactive M Dwarfs}, {\it \apj\/}, {\bf
  797}(2), 121.

\bibitem[{Haywood} et~al.(2014){Haywood}, {Collier Cameron}, {Queloz},
  {Barros}, {Deleuil}, {Fares}, {Gillon}, {Lanza}, {Lovis}, {Moutou}, {Pepe},
  {Pollacco}, {Santerne}, {S{\'e}gransan}, \& {Unruh}]{Haywood-14}
{Haywood}, R.~D., {Collier Cameron}, A., {Queloz}, D., {Barros}, S.~C.~C.,
  {Deleuil}, M., {Fares}, R., {Gillon}, M., {Lanza}, A.~F., {Lovis}, C.,
  {Moutou}, C., {Pepe}, F., {Pollacco}, D., {Santerne}, A., {S{\'e}gransan},
  D., \& {Unruh}, Y.~C., 2014.
\newblock {Planets and stellar activity: hide and seek in the CoRoT-7 system},
  {\it \mnras\/}, {\bf 443}, 2517--2531.

\bibitem[{Hedges} et~al.(2021){Hedges}, {Luger}, {Dotson}, {Foreman-Mackey}, \&
  {Barentsen}]{Hedges2021}
{Hedges}, C., {Luger}, R., {Dotson}, J., {Foreman-Mackey}, D., \& {Barentsen},
  G., 2021.
\newblock {Multiwavelength Photometry Derived from Monochromatic Kepler Data},
  {\it \aj\/}, {\bf 161}(2), 95.

\bibitem[{Hellier} et~al.(2015){Hellier}, {Anderson}, {Collier Cameron},
  {Delrez}, {Gillon}, {Jehin}, {Lendl}, {Maxted}, {Pepe}, {Pollacco}, {Queloz},
  {S{\'e}gransan}, {Smalley}, {Smith}, {Southworth}, {Triaud}, {Turner},
  {Udry}, \& {West}]{Hellier-15}
{Hellier}, C., {Anderson}, D.~R., {Collier Cameron}, A., {Delrez}, L.,
  {Gillon}, M., {Jehin}, E., {Lendl}, M., {Maxted}, P.~F.~L., {Pepe}, F.,
  {Pollacco}, D., {Queloz}, D., {S{\'e}gransan}, D., {Smalley}, B., {Smith},
  A.~M.~S., {Southworth}, J., {Triaud}, A.~H.~M.~J., {Turner}, O.~D., {Udry},
  S., \& {West}, R.~G., 2015.
\newblock {Three WASP-South Transiting Exoplanets: WASP-74b, WASP-83b, and
  WASP-89b}, {\it \aj\/}, {\bf 150}(1), 18.

\bibitem[{Henriques} et~al.(2020){Henriques}, {Nelson}, {Rouppe van der Voort},
  \& {Mathioudakis}]{Henriques2020}
{Henriques}, V. M.~J., {Nelson}, C.~J., {Rouppe van der Voort}, L. H.~M., \&
  {Mathioudakis}, M., 2020.
\newblock {Umbral chromospheric fine structure and umbral flashes modelled as
  one: The corrugated umbra}, {\it \aap\/}, {\bf 642}, A215.

\bibitem[{Henry} et~al.(2000){Henry}, {Marcy}, {Butler}, \& {Vogt}]{Henry-00}
{Henry}, G.~W., {Marcy}, G.~W., {Butler}, R.~P., \& {Vogt}, S.~S., 2000.
\newblock {A Transiting ``51 Peg-like'' Planet}, {\it \apjl\/}, {\bf 529},
  L41--L44.

\bibitem[{Henry} et~al.(1996){Henry}, {Soderblom}, {Donahue}, \&
  {Baliunas}]{Henry1996}
{Henry}, T.~J., {Soderblom}, D.~R., {Donahue}, R.~A., \& {Baliunas}, S.~L.,
  1996.
\newblock {A Survey of Ca II H and K Chromospheric Emission in Southern
  Solar-Type Stars}, {\it \aj\/}, {\bf 111}, 439.

\bibitem[{Herbst} et~al.(2021){Herbst}, {Papaioannou}, {Airapetian}, \&
  {Atri}]{herbst2021}
{Herbst}, K., {Papaioannou}, A., {Airapetian}, V.~S., \& {Atri}, D., 2021.
\newblock {From Starspots to Stellar Coronal Mass
  Ejections{\textemdash}Revisiting Empirical Stellar Relations}, {\it \apj\/},
  {\bf 907}(2), 89.

\bibitem[{Herrero} et~al.(2016){Herrero}, {Ribas}, {Jordi}, {Morales},
  {Perger}, \& {Rosich}]{Herrero-16}
{Herrero}, E., {Ribas}, I., {Jordi}, C., {Morales}, J.~C., {Perger}, M., \&
  {Rosich}, A., 2016.
\newblock {Modelling the photosphere of active stars for planet detection and
  characterization}, {\it \aap\/}, {\bf 586}, A131.

\bibitem[{Hirzberger} \& {Wiehr}(2005)]{hirzberger2005}
{Hirzberger}, J. \& {Wiehr}, E., 2005.
\newblock {Solar limb faculae}, {\it \aap\/}, {\bf 438}(3), 1059--1065.

\bibitem[{Hirzberger} et~al.(2010){Hirzberger}, {Feller}, {Riethm{\"u}ller},
  {Sch{\"u}ssler}, {Borrero}, {Afram}, {Unruh}, {Berdyugina}, {Gandorfer},
  {Solanki}, {Barthol}, {Bonet}, {Mart{\'\i}nez Pillet}, {Berkefeld},
  {Kn{\"o}lker}, {Schmidt}, \& {Title}]{Hirzberger2010}
{Hirzberger}, J., {Feller}, A., {Riethm{\"u}ller}, T.~L., {Sch{\"u}ssler}, M.,
  {Borrero}, J.~M., {Afram}, N., {Unruh}, Y.~C., {Berdyugina}, S.~V.,
  {Gandorfer}, A., {Solanki}, S.~K., {Barthol}, P., {Bonet}, J.~A.,
  {Mart{\'\i}nez Pillet}, V., {Berkefeld}, T., {Kn{\"o}lker}, M., {Schmidt},
  W., \& {Title}, A.~M., 2010.
\newblock {Quiet-sun Intensity Contrasts in the Near-ultraviolet as Measured
  from SUNRISE}, {\it \apjl\/}, {\bf 723}(2), L154--L158.

\bibitem[{Holczer} et~al.(2016){Holczer}, {Mazeh}, {Nachmani}, {Jontof-Hutter},
  {Ford}, {Fabrycky}, {Ragozzine}, {Kane}, \& {Steffen}]{Holczer2016}
{Holczer}, T., {Mazeh}, T., {Nachmani}, G., {Jontof-Hutter}, D., {Ford}, E.~B.,
  {Fabrycky}, D., {Ragozzine}, D., {Kane}, M., \& {Steffen}, J.~H., 2016.
\newblock {Transit Timing Observations from Kepler. IX. Catalog of the Full
  Long-cadence Data Set}, {\it \apjs\/}, {\bf 225}, 9.

\bibitem[{Houdek} et~al.(1999){Houdek}, {Balmforth}, {Christensen-Dalsgaard},
  \& {Gough}]{houdek1999a}
{Houdek}, G., {Balmforth}, N.~J., {Christensen-Dalsgaard}, J., \& {Gough},
  D.~O., 1999.
\newblock {Amplitudes of stochastically excited oscillations in main-sequence
  stars}, {\it \aap\/}, {\bf 351}, 582--596.

\bibitem[{Howard} et~al.(2020){Howard}, {Corbett}, {Law}, {Ratzloff},
  {Galliher}, {Glazier}, {Gonzalez}, {Vasquez Soto}, {Fors}, {del Ser}, \&
  {Haislip}]{Howard2020}
{Howard}, W.~S., {Corbett}, H., {Law}, N.~M., {Ratzloff}, J.~K., {Galliher},
  N., {Glazier}, A.~L., {Gonzalez}, R., {Vasquez Soto}, A., {Fors}, O., {del
  Ser}, D., \& {Haislip}, J., 2020.
\newblock {EvryFlare. III. Temperature Evolution and Habitability Impacts of
  Dozens of Superflares Observed Simultaneously by Evryscope and TESS}, {\it
  \apj\/}, {\bf 902}(2), 115.

\bibitem[{Huber} et~al.(2011){Huber}, {Bedding}, {Stello}, {Hekker}, {Mathur},
  {Mosser}, {Verner}, {Bonanno}, {Buzasi}, {Campante}, {Elsworth}, {Hale},
  {Kallinger}, {Silva Aguirre}, {Chaplin}, {De Ridder}, {Garc{\'\i}a},
  {Appourchaux}, {Frandsen}, {Houdek}, {Molenda-{\.Z}akowicz}, {Monteiro},
  {Christensen-Dalsgaard}, {Gilliland}, {Kawaler}, {Kjeldsen}, {Broomhall},
  {Corsaro}, {Salabert}, {Sanderfer}, {Seader}, \& {Smith}]{Huber2011}
{Huber}, D., {Bedding}, T.~R., {Stello}, D., {Hekker}, S., {Mathur}, S.,
  {Mosser}, B., {Verner}, G.~A., {Bonanno}, A., {Buzasi}, D.~L., {Campante},
  T.~L., {Elsworth}, Y.~P., {Hale}, S.~J., {Kallinger}, T., {Silva Aguirre},
  V., {Chaplin}, W.~J., {De Ridder}, J., {Garc{\'\i}a}, R.~A., {Appourchaux},
  T., {Frandsen}, S., {Houdek}, G., {Molenda-{\.Z}akowicz}, J., {Monteiro},
  M.~J.~P.~F.~G., {Christensen-Dalsgaard}, J., {Gilliland}, R.~L., {Kawaler},
  S.~D., {Kjeldsen}, H., {Broomhall}, A.~M., {Corsaro}, E., {Salabert}, D.,
  {Sanderfer}, D.~T., {Seader}, S.~E., \& {Smith}, J.~C., 2011.
\newblock {Testing Scaling Relations for Solar-like Oscillations from the Main
  Sequence to Red Giants Using Kepler Data}, {\it \apj\/}, {\bf 743}(2), 143.

\bibitem[{Huenemoerder} et~al.(1989){Huenemoerder}, {Ramsey}, \&
  {Buzasi}]{Huenemoerder1989}
{Huenemoerder}, D.~P., {Ramsey}, L.~W., \& {Buzasi}, D.~L., 1989.
\newblock {Titanium Oxide Variations in II Pegasi}, {\it \aj\/}, {\bf 98},
  2264.

\bibitem[{Hussain}(2002)]{Hussain2002}
{Hussain}, G.~A.~J., 2002.
\newblock {Starspot lifetimes}, {\it Astronomische Nachrichten\/}, {\bf 323},
  349--356.

\bibitem[{Husser} et~al.(2013){Husser}, {Wende-von Berg}, {Dreizler},
  {Homeier}, {Reiners}, {Barman}, \& {Hauschildt}]{Husser2013}
{Husser}, T.~O., {Wende-von Berg}, S., {Dreizler}, S., {Homeier}, D.,
  {Reiners}, A., {Barman}, T., \& {Hauschildt}, P.~H., 2013.
\newblock {A new extensive library of PHOENIX stellar atmospheres and synthetic
  spectra}, {\it \aap\/}, {\bf 553}, A6.

\bibitem[{Iyer} \& {Line}(2020)]{Iyer2020}
{Iyer}, A.~R. \& {Line}, M.~R., 2020.
\newblock {The Influence of Stellar Contamination on the Interpretation of
  Near-infrared Transmission Spectra of Sub-Neptune Worlds around M-dwarfs},
  {\it \apj\/}, {\bf 889}(2), 78.

\bibitem[{Iyer} et~al.(2023){Iyer}, {Line}, {Muirhead}, {Fortney}, \&
  {Gharib-Nezhad}]{Iyer2023}
{Iyer}, A.~R., {Line}, M.~R., {Muirhead}, P.~S., {Fortney}, J.~J., \&
  {Gharib-Nezhad}, E., 2023.
\newblock {The SPHINX M-dwarf Spectral Grid. I. Benchmarking New Model
  Atmospheres to Derive Fundamental M-dwarf Properties}, {\it \apj\/}, {\bf
  944}(1), 41.

\bibitem[{J{\"a}rvinen} et~al.(2005{\natexlab{a}}){J{\"a}rvinen}, {Berdyugina},
  \& {Strassmeier}]{Jarvinen2005ekdra}
{J{\"a}rvinen}, S.~P., {Berdyugina}, S.~V., \& {Strassmeier}, K.~G.,
  2005{\natexlab{a}}.
\newblock {Spots on EK Draconis. Active longitudes and cycles from long-term
  photometry}, {\it \aap\/}, {\bf 440}(2), 735--741.

\bibitem[{J{\"a}rvinen} et~al.(2005{\natexlab{b}}){J{\"a}rvinen}, {Berdyugina},
  {Tuominen}, {Cutispoto}, \& {Bos}]{Jarvinen2005abdor}
{J{\"a}rvinen}, S.~P., {Berdyugina}, S.~V., {Tuominen}, I., {Cutispoto}, G., \&
  {Bos}, M., 2005{\natexlab{b}}.
\newblock {Magnetic activity in the young solar analog AB Dor. Active
  longitudes and cycles from long-term photometry.}, {\it \aap\/}, {\bf
  432}(2), 657--664.

\bibitem[{J{\"a}rvinen} et~al.(2008){J{\"a}rvinen}, {Korhonen}, {Berdyugina},
  {Ilyin}, {Strassmeier}, {Weber}, {Savanov}, \&
  {Tuominen}]{Jarvinen2008v889her}
{J{\"a}rvinen}, S.~P., {Korhonen}, H., {Berdyugina}, S.~V., {Ilyin}, I.,
  {Strassmeier}, K.~G., {Weber}, M., {Savanov}, I., \& {Tuominen}, I., 2008.
\newblock {Magnetic activity on V889 Herculis. Combining photometry and
  spectroscopy}, {\it \aap\/}, {\bf 488}(3), 1047--1055.

\bibitem[{J{\"a}rvinen} et~al.(2018){J{\"a}rvinen}, {Strassmeier}, {Carroll},
  {Ilyin}, \& {Weber}]{Jarvinen2018ekdra}
{J{\"a}rvinen}, S.~P., {Strassmeier}, K.~G., {Carroll}, T.~A., {Ilyin}, I., \&
  {Weber}, M., 2018.
\newblock {Mapping EK Draconis with PEPSI. Possible evidence for starspot
  penumbrae}, {\it \aap\/}, {\bf 620}, A162.

\bibitem[{Jeffers} et~al.(2007){Jeffers}, {Donati}, \& {Collier
  Cameron}]{Jeffers2007abdor}
{Jeffers}, S.~V., {Donati}, J.~F., \& {Collier Cameron}, A., 2007.
\newblock {Magnetic activity on AB Doradus: temporal evolution of star-spots
  and differential rotation from 1988 to 1994}, {\it \mnras\/}, {\bf 375}(2),
  567--583.

\bibitem[{Johnson} et~al.(2021){Johnson}, {Norris}, {Unruh}, {Solanki},
  {Krivova}, {Witzke}, \& {Shapiro}]{Johnson2021}
{Johnson}, L.~J., {Norris}, C.~M., {Unruh}, Y.~C., {Solanki}, S.~K., {Krivova},
  N., {Witzke}, V., \& {Shapiro}, A.~I., 2021.
\newblock {Forward modelling of Kepler-band variability due to faculae and
  spots}, {\it \mnras\/}, {\bf 504}(4), 4751--4767.

\bibitem[{Jord{\'a}n} et~al.(2013){Jord{\'a}n}, {Espinoza}, {Rabus},
  {Eyheramendy}, {Sing}, {D{\'e}sert}, {Bakos}, {Fortney}, {L{\'o}pez-Morales},
  {Maxted}, {Triaud}, \& {Szentgyorgyi}]{Jordan2013}
{Jord{\'a}n}, A., {Espinoza}, N., {Rabus}, M., {Eyheramendy}, S., {Sing},
  D.~K., {D{\'e}sert}, J.-M., {Bakos}, G.~{\'A}., {Fortney}, J.~J.,
  {L{\'o}pez-Morales}, M., {Maxted}, P. F.~L., {Triaud}, A. H.~M.~J., \&
  {Szentgyorgyi}, A., 2013.
\newblock {A Ground-based Optical Transmission Spectrum of WASP-6b}, {\it
  \apj\/}, {\bf 778}(2), 184.

\bibitem[{Juvan} et~al.(2018){Juvan}, {Lendl}, {Cubillos}, {Fossati},
  {Tregloan-Reed}, {Lammer}, {Guenther}, \& {Hanslmeier}]{Juven-18}
{Juvan}, I.~G., {Lendl}, M., {Cubillos}, P.~E., {Fossati}, L., {Tregloan-Reed},
  J., {Lammer}, H., {Guenther}, E.~W., \& {Hanslmeier}, A., 2018.
\newblock {PyTranSpot: A tool for multiband light curve modeling of planetary
  transits and stellar spots}, {\it \aap\/}, {\bf 610}, A15.

\bibitem[{Kahil} et~al.(2017){Kahil}, {Riethm{\"u}ller}, \&
  {Solanki}]{kahil2017}
{Kahil}, F., {Riethm{\"u}ller}, T.~L., \& {Solanki}, S.~K., 2017.
\newblock {Brightness of Solar Magnetic Elements As a Function of Magnetic Flux
  at High Spatial Resolution}, {\it \apjs\/}, {\bf 229}(1), 12.

\bibitem[{Kallinger} et~al.(2014){Kallinger}, {De Ridder}, {Hekker}, {Mathur},
  {Mosser}, {Gruberbauer}, {Garc{\'\i}a}, {Karoff}, \& {Ballot}]{Kallinger2014}
{Kallinger}, T., {De Ridder}, J., {Hekker}, S., {Mathur}, S., {Mosser}, B.,
  {Gruberbauer}, M., {Garc{\'\i}a}, R.~A., {Karoff}, C., \& {Ballot}, J., 2014.
\newblock {The connection between stellar granulation and oscillation as seen
  by the Kepler mission}, {\it \aap\/}, {\bf 570}, A41.

\bibitem[{Karoff} et~al.(2018){Karoff}, {Metcalfe}, {Santos}, {Montet},
  {Isaacson}, {Witzke}, {Shapiro}, {Mathur}, {Davies}, {Lund}, {Garcia},
  {Brun}, {Salabert}, {Avelino}, {van Saders}, {Egeland}, {Cunha}, {Campante},
  {Chaplin}, {Krivova}, {Solanki}, {Stritzinger}, \& {Knudsen}]{Karoff18}
{Karoff}, C., {Metcalfe}, T.~S., {Santos}, {\^A}. R.~G., {Montet}, B.~T.,
  {Isaacson}, H., {Witzke}, V., {Shapiro}, A.~I., {Mathur}, S., {Davies},
  G.~R., {Lund}, M.~N., {Garcia}, R.~A., {Brun}, A.~S., {Salabert}, D.,
  {Avelino}, P.~P., {van Saders}, J., {Egeland}, R., {Cunha}, M.~S.,
  {Campante}, T.~L., {Chaplin}, W.~J., {Krivova}, N., {Solanki}, S.~K.,
  {Stritzinger}, M., \& {Knudsen}, M.~F., 2018.
\newblock {The Influence of Metallicity on Stellar Differential Rotation and
  Magnetic Activity}, {\it \apj\/}, {\bf 852}(1), 46.

\bibitem[{Katsukawa} et~al.(2007){Katsukawa}, {Berger}, {Ichimoto}, {Lites},
  {Nagata}, {Shimizu}, {Shine}, {Suematsu}, {Tarbell}, {Title}, \&
  {Tsuneta}]{Katsukawa2007}
{Katsukawa}, Y., {Berger}, T.~E., {Ichimoto}, K., {Lites}, B.~W., {Nagata}, S.,
  {Shimizu}, T., {Shine}, R.~A., {Suematsu}, Y., {Tarbell}, T.~D., {Title},
  A.~M., \& {Tsuneta}, S., 2007.
\newblock {Small-Scale Jetlike Features in Penumbral Chromospheres}, {\it
  Science\/}, {\bf 318}(5856), 1594.

\bibitem[{Keller} et~al.(2003){Keller}, {Harvey}, \& {Giampapa}]{Keller2003}
{Keller}, C.~U., {Harvey}, J.~W., \& {Giampapa}, M.~S., 2003.
\newblock {SOLIS: an innovative suite of synoptic instruments}, in {\em
  Innovative Telescopes and Instrumentation for Solar Astrophysics\/}, vol.
  4853 of {\bf Society of Photo-Optical Instrumentation Engineers (SPIE)
  Conference Series}, pp. 194--204.

\bibitem[{Keller} et~al.(2004){Keller}, {Sch{\"u}ssler}, {V{\"o}gler}, \&
  {Zakharov}]{keller+al2004}
{Keller}, C.~U., {Sch{\"u}ssler}, M., {V{\"o}gler}, A., \& {Zakharov}, V.,
  2004.
\newblock {On the Origin of Solar Faculae}, {\it \apjl\/}, {\bf 607}(1),
  L59--L62.

\bibitem[{Kervella} \& {Fouqu{\'e}}(2008)]{Kervella2008}
{Kervella}, P. \& {Fouqu{\'e}}, P., 2008.
\newblock {The angular sizes of dwarf stars and subgiants. Non-linear surface
  brightness relations in BVR\_cI$_{c}$ from interferometry}, {\it \aap\/},
  {\bf 491}(3), 855--858.

\bibitem[{Kervella} et~al.(2004{\natexlab{a}}){Kervella}, {Bersier}, {Mourard},
  {Nardetto}, {Fouqu{\'e}}, \& {Coud{\'e} du Foresto}]{Kervella2004_cepheids}
{Kervella}, P., {Bersier}, D., {Mourard}, D., {Nardetto}, N., {Fouqu{\'e}}, P.,
  \& {Coud{\'e} du Foresto}, V., 2004{\natexlab{a}}.
\newblock {Cepheid distances from infrared long-baseline interferometry. III.
  Calibration of the surface brightness-color relations}, {\it \aap\/}, {\bf
  428}, 587--593.

\bibitem[{Kervella} et~al.(2004{\natexlab{b}}){Kervella}, {Th{\'e}venin}, {Di
  Folco}, \& {S{\'e}gransan}]{Kervella2004_dwarfs}
{Kervella}, P., {Th{\'e}venin}, F., {Di Folco}, E., \& {S{\'e}gransan}, D.,
  2004{\natexlab{b}}.
\newblock {The angular sizes of dwarf stars and subgiants. Surface brightness
  relations calibrated by interferometry}, {\it \aap\/}, {\bf 426}, 297--307.

\bibitem[{Keys} et~al.(2019){Keys}, {Reid}, {Mathioudakis}, {Shelyag},
  {Henriques}, {Hewitt}, {Del Moro}, {Jafarzadeh}, {Jess}, \&
  {Stangalini}]{keys2019}
{Keys}, P.~H., {Reid}, A., {Mathioudakis}, M., {Shelyag}, S., {Henriques}, V.
  M.~J., {Hewitt}, R.~L., {Del Moro}, D., {Jafarzadeh}, S., {Jess}, D.~B., \&
  {Stangalini}, M., 2019.
\newblock {The magnetic properties of photospheric magnetic bright points with
  high-resolution spectropolarimetry}, {\it \mnras\/}, {\bf 488}(1), L53--L58.

\bibitem[{Kiepenheuer}(1953)]{Kiepenheuer1953}
{Kiepenheuer}, K.~O., 1953.
\newblock {Photoelectric Measurements of Solar Magnetic Fields.}, {\it \apj\/},
  {\bf 117}, 447.

\bibitem[{Kiess} et~al.(2014){Kiess}, {Rezaei}, \& {Schmidt}]{Kiess:2014}
{Kiess}, C., {Rezaei}, R., \& {Schmidt}, W., 2014.
\newblock {Properties of sunspot umbrae observed in cycle 24}, {\it \aap\/},
  {\bf 565}, A52.

\bibitem[{Kilcik} et~al.(2020){Kilcik}, {Sarp}, {Yurchyshyn}, {Rozelot}, \&
  {Ozguc}]{Kilcik2020}
{Kilcik}, A., {Sarp}, V., {Yurchyshyn}, V., {Rozelot}, J.-P., \& {Ozguc}, A.,
  2020.
\newblock {Physical Characteristics of Umbral Dots Derived from a
  High-Resolution Observations}, {\it \solphys\/}, {\bf 295}(4), 58.

\bibitem[{Kilpatrick} et~al.(2020){Kilpatrick}, {Kataria}, {Lewis}, {Zellem},
  {Henry}, {Cowan}, {de Wit}, {Fortney}, {Knutson}, {Seager}, {Showman}, \&
  {Tucker}]{Kilpatrick2020}
{Kilpatrick}, B.~M., {Kataria}, T., {Lewis}, N.~K., {Zellem}, R.~T., {Henry},
  G.~W., {Cowan}, N.~B., {de Wit}, J., {Fortney}, J.~J., {Knutson}, H.,
  {Seager}, S., {Showman}, A.~P., \& {Tucker}, G.~S., 2020.
\newblock {Evaluating Climate Variability of the Canonical Hot-Jupiters HD
  189733b and HD 209458b through Multi-epoch Eclipse Observations}, {\it
  \aj\/}, {\bf 159}(2), 51.

\bibitem[{Kiraga} \& {Stepien}(2007)]{2007AcA....57..149K}
{Kiraga}, M. \& {Stepien}, K., 2007.
\newblock {Age-Rotation-Activity Relations for M Dwarf Stars}, {\it \actaa\/},
  {\bf 57}, 149--172.

\bibitem[{Kirk} et~al.(2016){Kirk}, {Wheatley}, {Louden}, {Littlefair},
  {Copperwheat}, {Armstrong}, {Marsh}, \& {Dhillon}]{Kirk-16}
{Kirk}, J., {Wheatley}, P.~J., {Louden}, T., {Littlefair}, S.~P.,
  {Copperwheat}, C.~M., {Armstrong}, D.~J., {Marsh}, T.~R., \& {Dhillon},
  V.~S., 2016.
\newblock {Transmission spectroscopy of the inflated exoplanet WASP-52b, and
  evidence for a bright region on the stellar surface}, {\it \mnras\/}, {\bf
  463}(3), 2922--2931.

\bibitem[{Kirk} et~al.(2021){Kirk}, {Rackham}, {MacDonald},
  {L{\'o}pez-Morales}, {Espinoza}, {Lendl}, {Wilson}, {Osip}, {Wheatley},
  {Skillen}, {Apai}, {Bixel}, {Gibson}, {Jordan}, {Lewis}, {Louden},
  {McGruder}, {Nikolov}, {Rodler}, \& {Weaver}]{Kirk2021}
{Kirk}, J., {Rackham}, B., {MacDonald}, R., {L{\'o}pez-Morales}, M.,
  {Espinoza}, N., {Lendl}, M., {Wilson}, J., {Osip}, D.~J., {Wheatley}, P.~J.,
  {Skillen}, I., {Apai}, D., {Bixel}, A., {Gibson}, N.~P., {Jordan}, A.,
  {Lewis}, N.~K., {Louden}, T., {McGruder}, C.~D., {Nikolov}, N., {Rodler}, F.,
  \& {Weaver}, I.~C., 2021.
\newblock {ACCESS \& LRG-BEASTS: a precise new optical transmission spectrum of
  the ultrahot Jupiter WASP-103b}, {\it arXiv e-prints\/}, p. arXiv:2105.00012.

\bibitem[{Kitai}(1986)]{Kitai1986}
{Kitai}, R., 1986.
\newblock {Photospheric and Chromospheric Umbral Dots in a Decaying Sunspot},
  {\it \solphys\/}, {\bf 104}(2), 287--301.

\bibitem[{Kitai} \& {Muller}(1984)]{Kitai-Muller1984}
{Kitai}, R. \& {Muller}, R., 1984.
\newblock {On the Relation Between Chromospheric and Photospheric Fine
  Structure in an Active Region}, {\it \solphys\/}, {\bf 90}(2), 303--314.

\bibitem[{Kjeldsen} \& {Bedding}(2011)]{kjeldsen2011a}
{Kjeldsen}, H. \& {Bedding}, T.~R., 2011.
\newblock {Amplitudes of solar-like oscillations: a new scaling relation}, {\it
  \aap\/}, {\bf 529}, L8.

\bibitem[{Kleint} et~al.(2010){Kleint}, {Berdyugina}, {Shapiro}, \&
  {Bianda}]{Kleint2010}
{Kleint}, L., {Berdyugina}, S.~V., {Shapiro}, A.~I., \& {Bianda}, M., 2010.
\newblock {Solar turbulent magnetic fields: surprisingly homogeneous
  distribution during the solar minimum}, {\it \aap\/}, {\bf 524}, A37.

\bibitem[{Kleint} et~al.(2011){Kleint}, {Shapiro}, {Berdyugina}, \&
  {Bianda}]{Kleint2011}
{Kleint}, L., {Shapiro}, A.~I., {Berdyugina}, S.~V., \& {Bianda}, M., 2011.
\newblock {Solar turbulent magnetic fields: Non-LTE modeling of the Hanle
  effect in the C$_{2}$ molecule}, {\it \aap\/}, {\bf 536}, A47.

\bibitem[{Knutson} et~al.(2011){Knutson}, {Madhusudhan}, {Cowan},
  {Christiansen}, {Agol}, {Deming}, {D{\'e}sert}, {Charbonneau}, {Henry},
  {Homeier}, {Langton}, {Laughlin}, \& {Seager}]{Knutson2011}
{Knutson}, H.~A., {Madhusudhan}, N., {Cowan}, N.~B., {Christiansen}, J.~L.,
  {Agol}, E., {Deming}, D., {D{\'e}sert}, J.-M., {Charbonneau}, D., {Henry},
  G.~W., {Homeier}, D., {Langton}, J., {Laughlin}, G., \& {Seager}, S., 2011.
\newblock {A Spitzer Transmission Spectrum for the Exoplanet GJ 436b, Evidence
  for Stellar Variability, and Constraints on Dayside Flux Variations}, {\it
  \apj\/}, {\bf 735}(1), 27.

\bibitem[{Knutson} et~al.(2012){Knutson}, {Lewis}, {Fortney}, {Burrows},
  {Showman}, {Cowan}, {Agol}, {Aigrain}, {Charbonneau}, {Deming}, {D{\'e}sert},
  {Henry}, {Langton}, \& {Laughlin}]{knutson12}
{Knutson}, H.~A., {Lewis}, N., {Fortney}, J.~J., {Burrows}, A., {Showman},
  A.~P., {Cowan}, N.~B., {Agol}, E., {Aigrain}, S., {Charbonneau}, D.,
  {Deming}, D., {D{\'e}sert}, J.-M., {Henry}, G.~W., {Langton}, J., \&
  {Laughlin}, G., 2012.
\newblock {3.6 and 4.5 {\ensuremath{\mu}}m Phase Curves and Evidence for
  Non-equilibrium Chemistry in the Atmosphere of Extrasolar Planet HD 189733b},
  {\it \apj\/}, {\bf 754}(1), 22.

\bibitem[{Kobel} et~al.(2011){Kobel}, {Solanki}, \& {Borrero}]{kobel2011}
{Kobel}, P., {Solanki}, S.~K., \& {Borrero}, J.~M., 2011.
\newblock {The continuum intensity as a function of magnetic field. I. Active
  region and quiet Sun magnetic elements}, {\it \aap\/}, {\bf 531}, A112.

\bibitem[{Kochanek} et~al.(2017){Kochanek}, {Shappee}, {Stanek}, {Holoien},
  {Thompson}, {Prieto}, {Dong}, {Shields}, {Will}, {Britt}, {Perzanowski}, \&
  {Pojma{\'n}ski}]{Kochanek2017}
{Kochanek}, C.~S., {Shappee}, B.~J., {Stanek}, K.~Z., {Holoien}, T.~W.~S.,
  {Thompson}, T.~A., {Prieto}, J.~L., {Dong}, S., {Shields}, J.~V., {Will}, D.,
  {Britt}, C., {Perzanowski}, D., \& {Pojma{\'n}ski}, G., 2017.
\newblock {The All-Sky Automated Survey for Supernovae (ASAS-SN) Light Curve
  Server v1.0}, {\it \pasp\/}, {\bf 129}(980), 104502.

\bibitem[{Kopp}(2016)]{2016JSWSC...6A..30K}
{Kopp}, G., 2016.
\newblock {Magnitudes and timescales of total solar irradiance variability},
  {\it Journal of Space Weather and Space Climate\/}, {\bf 6}, A30.

\bibitem[{Kopp}(2021)]{Kopp2021}
{Kopp}, G., 2021.
\newblock {Science Highlights and Final Updates from 17 Years of Total Solar
  Irradiance from the SOlar Radiation and Climate Experiment/Total Irradiance
  Monitor (SORCE/TIM)}, {\it Solar Physics\/}.

\bibitem[{Korhonen} et~al.(2021){Korhonen}, {Roettenbacher}, {Gu}, {Grundahl},
  {Andersen}, {Henry}, {Jessen-Hansen}, {Antoci}, \& {Pall{\'e}}]{Korhonen2021}
{Korhonen}, H., {Roettenbacher}, R.~M., {Gu}, S., {Grundahl}, F., {Andersen},
  M.~F., {Henry}, G.~W., {Jessen-Hansen}, J., {Antoci}, V., \& {Pall{\'e}},
  P.~L., 2021.
\newblock {Observing the changing surface structures of the active K giant
  {\ensuremath{\sigma}} Geminorum with SONG}, {\it \aap\/}, {\bf 646}, A6.

\bibitem[{Kostogryz} \& {Berdyugina}(2015)]{Kostogryz-Berdyugina2015}
{Kostogryz}, N.~M. \& {Berdyugina}, S.~V., 2015.
\newblock {Center-to-limb polarization in continuum spectra of F, G, K stars},
  {\it \aap\/}, {\bf 575}, A89.

\bibitem[{Kostogryz} et~al.(2016){Kostogryz}, {Milic}, {Berdyugina}, \&
  {Hauschildt}]{Kostogryz2016}
{Kostogryz}, N.~M., {Milic}, I., {Berdyugina}, S.~V., \& {Hauschildt}, P.~H.,
  2016.
\newblock {Center-to-limb variation of intensity and polarization in continuum
  spectra of FGK stars for spherical atmospheres}, {\it \aap\/}, {\bf 586},
  A87.

\bibitem[{Kostogryz} et~al.(2017){Kostogryz}, {Yakobchuk}, {Berdyugina}, \&
  {Milic}]{Kostogryz2017}
{Kostogryz}, N.~M., {Yakobchuk}, T.~M., {Berdyugina}, S.~V., \& {Milic}, I.,
  2017.
\newblock {Polarimetry of transiting planets: Differences between
  plane-parallel and spherical host star atmosphere models}, {\it \aap\/}, {\bf
  601}, A6.

\bibitem[{Kov{\'a}ri} et~al.(2007){Kov{\'a}ri}, {Bartus}, {Strassmeier},
  {Vida}, {{\v{S}}vanda}, \& {Ol{\'a}h}]{Kovari2007}
{Kov{\'a}ri}, Z., {Bartus}, J., {Strassmeier}, K.~G., {Vida}, K.,
  {{\v{S}}vanda}, M., \& {Ol{\'a}h}, K., 2007.
\newblock {Anti-solar differential rotation on the active K-giant
  {\ensuremath{\sigma}} Geminorum}, {\it \aap\/}, {\bf 474}(1), 165--168.

\bibitem[{Kowalski} et~al.(2013){Kowalski}, {Hawley}, {Wisniewski}, {Osten},
  {Hilton}, {Holtzman}, {Schmidt}, \& {Davenport}]{Kowalski2013}
{Kowalski}, A.~F., {Hawley}, S.~L., {Wisniewski}, J.~P., {Osten}, R.~A.,
  {Hilton}, E.~J., {Holtzman}, J.~A., {Schmidt}, S.~J., \& {Davenport}, J.
  R.~A., 2013.
\newblock {Time-resolved Properties and Global Trends in dMe Flares from
  Simultaneous Photometry and Spectra}, {\it \apjs\/}, {\bf 207}(1), 15.

\bibitem[{Kreidberg} et~al.(2014){Kreidberg}, {Bean}, {D{\'e}sert}, {Benneke},
  {Deming}, {Stevenson}, {Seager}, {Berta-Thompson}, {Seifahrt}, \&
  {Homeier}]{Kreidberg2014}
{Kreidberg}, L., {Bean}, J.~L., {D{\'e}sert}, J.-M., {Benneke}, B., {Deming},
  D., {Stevenson}, K.~B., {Seager}, S., {Berta-Thompson}, Z., {Seifahrt}, A.,
  \& {Homeier}, D., 2014.
\newblock {Clouds in the atmosphere of the super-Earth exoplanet GJ1214b}, {\it
  \nat\/}, {\bf 505}(7481), 69--72.

\bibitem[{Krivova} et~al.(2003){Krivova}, {Solanki}, {Fligge}, \&
  {Unruh}]{Krivova2003}
{Krivova}, N.~A., {Solanki}, S.~K., {Fligge}, M., \& {Unruh}, Y.~C., 2003.
\newblock {Reconstruction of solar irradiance variations in cycle 23: Is solar
  surface magnetism the cause?}, {\it \aap\/}, {\bf 399}, L1--L4.

\bibitem[{Kron}(1947)]{Kron1947}
{Kron}, G.~E., 1947.
\newblock {The Probable Detecting of Surface Spots on AR Lacertae B}, {\it
  \pasp\/}, {\bf 59}(350), 261.

\bibitem[{Kulow} et~al.(2014){Kulow}, {France}, {Linsky}, \& {Loyd}]{Kulow2014}
{Kulow}, J.~R., {France}, K., {Linsky}, J., \& {Loyd}, R.~O.~P., 2014.
\newblock {Ly{\ensuremath{\alpha}} Transit Spectroscopy and the Neutral
  Hydrogen Tail of the Hot Neptune GJ 436b}, {\it \apj\/}, {\bf 786}(2), 132.

\bibitem[{Kurucz}(1992{\natexlab{a}})]{Kurucz:1992a}
{Kurucz}, R.~L., 1992{\natexlab{a}}.
\newblock {Atomic and Molecular Data for Opacity Calculations}, {\it \rmxaa\/},
  {\bf 23}, 45.

\bibitem[{Kurucz}(1992{\natexlab{b}})]{Kurucz:1992b}
{Kurucz}, R.~L., 1992{\natexlab{b}}.
\newblock {``Finding'' the ``missing'' solar ultraviolet opacity.}, {\it
  \rmxaa\/}, {\bf 23}, 181--186.

\bibitem[{Kurucz}(1992{\natexlab{c}})]{Kurucz:1992c}
{Kurucz}, R.~L., 1992{\natexlab{c}}.
\newblock {Remaining Line Opacity Problems for the Solar Spectrum}, {\it
  \rmxaa\/}, {\bf 23}, 187.

\bibitem[{Kuzmychov} et~al.(2017){Kuzmychov}, {Berdyugina}, \&
  {Harrington}]{Kuzmychov2017}
{Kuzmychov}, O., {Berdyugina}, S.~V., \& {Harrington}, D.~M., 2017.
\newblock {First Spectropolarimetric Measurement of a Brown Dwarf Magnetic
  Field in Molecular Bands}, {\it \apj\/}, {\bf 847}(1), 60.

\bibitem[{Lam} et~al.(2020){Lam}, {Korth}, {Masuda}, {Csizmadia},
  {Eigm{\"u}ller}, {Stef{\'a}nsson}, {Endl}, {Albrecht}, {Robertson}, {Luque},
  {Livingston}, {Hirano}, {Sobrino}, {Barrag{\'a}n}, {Cabrera}, {Carleo},
  {Chaushev}, {Cochran}, {Dai}, {Leon}, {Deeg}, {Erikson}, {Esposito},
  {Fridlund}, {Fukui}, {Gandolfi}, {Georgieva}, {Cuesta}, {Grziwa}, {Guenther},
  {Hatzes}, {Hidalgo}, {Hjorth}, {Kabath}, {Knudstrup}, {Lund}, {Mahadevan},
  {Mathur}, {Rodr{\'\i}guez}, {Murgas}, {Narita}, {Nespral}, {Niraula},
  {Palle}, {P{\"a}tzold}, {Persson}, {Prieto-Arranz}, {Rauer}, {Redfield},
  {Ribas}, {Skarka}, {Smith}, {Subjak}, \& {Van Eylen}]{Lam2020}
{Lam}, K. W.~F., {Korth}, J., {Masuda}, K., {Csizmadia}, S., {Eigm{\"u}ller},
  P., {Stef{\'a}nsson}, G.~K., {Endl}, M., {Albrecht}, S., {Robertson}, P.,
  {Luque}, R., {Livingston}, J.~H., {Hirano}, T., {Sobrino}, R.~A.,
  {Barrag{\'a}n}, O., {Cabrera}, J., {Carleo}, I., {Chaushev}, A., {Cochran},
  W.~D., {Dai}, F., {Leon}, J.~d., {Deeg}, H.~J., {Erikson}, A., {Esposito},
  M., {Fridlund}, M., {Fukui}, A., {Gandolfi}, D., {Georgieva}, I., {Cuesta},
  L.~G., {Grziwa}, S., {Guenther}, E.~W., {Hatzes}, A.~P., {Hidalgo}, D.,
  {Hjorth}, M., {Kabath}, P., {Knudstrup}, E., {Lund}, M.~N., {Mahadevan}, S.,
  {Mathur}, S., {Rodr{\'\i}guez}, P.~M., {Murgas}, F., {Narita}, N., {Nespral},
  D., {Niraula}, P., {Palle}, E., {P{\"a}tzold}, M., {Persson}, C.~M.,
  {Prieto-Arranz}, J., {Rauer}, H., {Redfield}, S., {Ribas}, I., {Skarka}, M.,
  {Smith}, A. M.~S., {Subjak}, J., \& {Van Eylen}, V., 2020.
\newblock {It Takes Two Planets in Resonance to Tango around K2-146}, {\it
  \aj\/}, {\bf 159}(3), 120.

\bibitem[{Lammer} et~al.(2003){Lammer}, {Selsis}, {Ribas}, {Guinan}, {Bauer},
  \& {Weiss}]{Lammer2003}
{Lammer}, H., {Selsis}, F., {Ribas}, I., {Guinan}, E.~F., {Bauer}, S.~J., \&
  {Weiss}, W.~W., 2003.
\newblock {Atmospheric Loss of Exoplanets Resulting from Stellar X-Ray and
  Extreme-Ultraviolet Heating}, {\it \apjl\/}, {\bf 598}(2), L121--L124.

\bibitem[{Langhans} et~al.(2004){Langhans}, {Schmidt}, \&
  {Rimmele}]{langhans2004}
{Langhans}, K., {Schmidt}, W., \& {Rimmele}, T., 2004.
\newblock {Diagnostic spectroscopy of G-band brightenings in the photosphere of
  the sun}, {\it \aap\/}, {\bf 423}, 1147--1157.

\bibitem[{Lanza} et~al.(1998){Lanza}, {Catalano}, {Cutispoto}, {Pagano}, \&
  {Rodono}]{Lanza1998}
{Lanza}, A.~F., {Catalano}, S., {Cutispoto}, G., {Pagano}, I., \& {Rodono}, M.,
  1998.
\newblock {Long-term starspot evolution, activity cycle and orbital period
  variation of AR Lacertae}, {\it \aap\/}, {\bf 332}, 541--560.

\bibitem[{Law} et~al.(2014){Law}, {Morton}, {Baranec}, {Riddle},
  {Ravichandran}, {Ziegler}, {Johnson}, {Tendulkar}, {Bui}, {Burse}, {Das},
  {Dekany}, {Kulkarni}, {Punnadi}, \& {Ramaprakash}]{Law2014}
{Law}, N.~M., {Morton}, T., {Baranec}, C., {Riddle}, R., {Ravichandran}, G.,
  {Ziegler}, C., {Johnson}, J.~A., {Tendulkar}, S.~P., {Bui}, K., {Burse},
  M.~P., {Das}, H.~K., {Dekany}, R.~G., {Kulkarni}, S., {Punnadi}, S., \&
  {Ramaprakash}, A.~N., 2014.
\newblock {Robotic Laser Adaptive Optics Imaging of 715 Kepler Exoplanet
  Candidates Using Robo-AO}, {\it \apj\/}, {\bf 791}(1), 35.

\bibitem[{Le Bouquin} et~al.(2011){Le Bouquin}, {Berger}, {Lazareff}, {Zins},
  {Haguenauer}, {Jocou}, {Kern}, {Millan-Gabet}, {Traub}, {Absil}, {Augereau},
  {Benisty}, {Blind}, {Bonfils}, {Bourget}, {Delboulbe}, {Feautrier},
  {Germain}, {Gitton}, {Gillier}, {Kiekebusch}, {Kluska}, {Knudstrup},
  {Labeye}, {Lizon}, {Monin}, {Magnard}, {Malbet}, {Maurel}, {M{\'e}nard},
  {Micallef}, {Michaud}, {Montagnier}, {Morel}, {Moulin}, {Perraut}, {Popovic},
  {Rabou}, {Rochat}, {Rojas}, {Roussel}, {Roux}, {Stadler}, {Stefl}, {Tatulli},
  \& {Ventura}]{LeBouquin2011}
{Le Bouquin}, J.~B., {Berger}, J.~P., {Lazareff}, B., {Zins}, G., {Haguenauer},
  P., {Jocou}, L., {Kern}, P., {Millan-Gabet}, R., {Traub}, W., {Absil}, O.,
  {Augereau}, J.~C., {Benisty}, M., {Blind}, N., {Bonfils}, X., {Bourget}, P.,
  {Delboulbe}, A., {Feautrier}, P., {Germain}, M., {Gitton}, P., {Gillier}, D.,
  {Kiekebusch}, M., {Kluska}, J., {Knudstrup}, J., {Labeye}, P., {Lizon},
  J.~L., {Monin}, J.~L., {Magnard}, Y., {Malbet}, F., {Maurel}, D.,
  {M{\'e}nard}, F., {Micallef}, M., {Michaud}, L., {Montagnier}, G., {Morel},
  S., {Moulin}, T., {Perraut}, K., {Popovic}, D., {Rabou}, P., {Rochat}, S.,
  {Rojas}, C., {Roussel}, F., {Roux}, A., {Stadler}, E., {Stefl}, S.,
  {Tatulli}, E., \& {Ventura}, N., 2011.
\newblock {PIONIER: a 4-telescope visitor instrument at VLTI}, {\it \aap\/},
  {\bf 535}, A67.

\bibitem[{Leenaarts} \& {Wedemeyer-B{\"o}hm}(2006)]{leenaarts+wedemeyer2006}
{Leenaarts}, J. \& {Wedemeyer-B{\"o}hm}, S., 2006.
\newblock {Time-dependent hydrogen ionisation in 3D simulations of the solar
  chromosphere. Methods and first results}, {\it \aap\/}, {\bf 460}(1),
  301--307.

\bibitem[{Leenaarts} et~al.(2007){Leenaarts}, {Carlsson}, {Hansteen}, \&
  {Rutten}]{leenaarts+al2007}
{Leenaarts}, J., {Carlsson}, M., {Hansteen}, V., \& {Rutten}, R.~J., 2007.
\newblock {Non-equilibrium hydrogen ionization in 2D simulations of the solar
  atmosphere}, {\it \aap\/}, {\bf 473}(2), 625--632.

\bibitem[{Leenaarts} et~al.(2012){Leenaarts}, {Carlsson}, \& {Rouppe van der
  Voort}]{Leenaarts2012}
{Leenaarts}, J., {Carlsson}, M., \& {Rouppe van der Voort}, L., 2012.
\newblock {The Formation of the H{\ensuremath{\alpha}} Line in the Solar
  Chromosphere}, {\it \apj\/}, {\bf 749}(2), 136.

\bibitem[{Lightkurve Collaboration} et~al.(2018){Lightkurve Collaboration},
  {Cardoso}, {Hedges}, {Gully-Santiago}, {Saunders}, {Cody}, {Barclay}, {Hall},
  {Sagear}, {Turtelboom}, {Zhang}, {Tzanidakis}, {Mighell}, {Coughlin}, {Bell},
  {Berta-Thompson}, {Williams}, {Dotson}, \& {Barentsen}]{Lightkurve}
{Lightkurve Collaboration}, {Cardoso}, J.~V.~d.~M., {Hedges}, C.,
  {Gully-Santiago}, M., {Saunders}, N., {Cody}, A.~M., {Barclay}, T., {Hall},
  O., {Sagear}, S., {Turtelboom}, E., {Zhang}, J., {Tzanidakis}, A., {Mighell},
  K., {Coughlin}, J., {Bell}, K., {Berta-Thompson}, Z., {Williams}, P.,
  {Dotson}, J., \& {Barentsen}, G., 2018.
\newblock {Lightkurve: Kepler and TESS time series analysis in Python},
  Astrophysics Source Code Library.

\bibitem[{Linsky}(2017)]{Linsky2017}
{Linsky}, J.~L., 2017.
\newblock {Stellar Model Chromospheres and Spectroscopic Diagnostics}, {\it
  \araa\/}, {\bf 55}(1), 159--211.

\bibitem[{Lites} et~al.(2008){Lites}, {Kubo}, {Socas-Navarro}, {Berger},
  {Frank}, {Shine}, {Tarbell}, {Title}, {Ichimoto}, {Katsukawa}, {Tsuneta},
  {Suematsu}, {Shimizu}, \& {Nagata}]{Lites2008}
{Lites}, B.~W., {Kubo}, M., {Socas-Navarro}, H., {Berger}, T., {Frank}, Z.,
  {Shine}, R., {Tarbell}, T., {Title}, A., {Ichimoto}, K., {Katsukawa}, Y.,
  {Tsuneta}, S., {Suematsu}, Y., {Shimizu}, T., \& {Nagata}, S., 2008.
\newblock {The Horizontal Magnetic Flux of the Quiet-Sun Internetwork as
  Observed with the Hinode Spectro-Polarimeter}, {\it \apj\/}, {\bf 672}(2),
  1237--1253.

\bibitem[{Livingston} \& {Harvey}(1971)]{Livingston-Harvey1971}
{Livingston}, W. \& {Harvey}, J., 1971.
\newblock {The Kitt Peak Magnetograph. Iv: 40-CHANNEL Probe and the Detection
  of Weak Photospheric Fields}, in {\em Solar Magnetic Fields\/}, vol.~43,
  p.~51.

\bibitem[{Livingston} et~al.(2006){Livingston}, {Harvey}, {Malanushenko}, \&
  {Webster}]{Livingston:2006}
{Livingston}, W., {Harvey}, J.~W., {Malanushenko}, O.~V., \& {Webster}, L.,
  2006.
\newblock {Sunspots with the Strongest Magnetic Fields}, {\it \solphys\/}, {\bf
  239}(1-2), 41--68.

\bibitem[{Llama} \& {Shkolnik}(2015)]{Llama2015}
{Llama}, J. \& {Shkolnik}, E.~L., 2015.
\newblock {Transiting the Sun: the Impact of Stellar Activity on X-Ray and
  Ultraviolet Transits}, {\it \apj\/}, {\bf 802}(1), 41.

\bibitem[{Llama} \& {Shkolnik}(2016)]{Llama2016}
{Llama}, J. \& {Shkolnik}, E.~L., 2016.
\newblock {Transiting the Sun. II. The Impact of Stellar Activity on
  Ly{\ensuremath{\alpha}} Transits}, {\it \apj\/}, {\bf 817}(1), 81.

\bibitem[{Ludwig} et~al.(2009){Ludwig}, {Caffau}, {Steffen}, {Freytag},
  {Bonifacio}, \& {Ku{\v{c}}inskas}]{ludwig2009cfist}
{Ludwig}, H.~G., {Caffau}, E., {Steffen}, M., {Freytag}, B., {Bonifacio}, P.,
  \& {Ku{\v{c}}inskas}, A., 2009.
\newblock {The CIFIST 3D model atmosphere grid.}, {\it \memsai\/}, {\bf 80},
  711.

\bibitem[{Luger} et~al.(2017){Luger}, {Sestovic}, {Kruse}, {Grimm}, {Demory},
  {Agol}, {Bolmont}, {Fabrycky}, {Fernandes}, {Van Grootel}, {Burgasser},
  {Gillon}, {Ingalls}, {Jehin}, {Raymond}, {Selsis}, {Triaud}, {Barclay},
  {Barentsen}, {Howell}, {Delrez}, {de Wit}, {Foreman-Mackey}, {Holdsworth},
  {Leconte}, {Lederer}, {Turbet}, {Almleaky}, {Benkhaldoun}, {Magain},
  {Morris}, {Heng}, \& {Queloz}]{Luger-17}
{Luger}, R., {Sestovic}, M., {Kruse}, E., {Grimm}, S.~L., {Demory}, B.-O.,
  {Agol}, E., {Bolmont}, E., {Fabrycky}, D., {Fernandes}, C.~S., {Van Grootel},
  V., {Burgasser}, A., {Gillon}, M., {Ingalls}, J.~G., {Jehin}, E., {Raymond},
  S.~N., {Selsis}, F., {Triaud}, A.~H.~M.~J., {Barclay}, T., {Barentsen}, G.,
  {Howell}, S.~B., {Delrez}, L., {de Wit}, J., {Foreman-Mackey}, D.,
  {Holdsworth}, D.~L., {Leconte}, J., {Lederer}, S., {Turbet}, M., {Almleaky},
  Y., {Benkhaldoun}, Z., {Magain}, P., {Morris}, B.~M., {Heng}, K., \&
  {Queloz}, D., 2017.
\newblock {A seven-planet resonant chain in TRAPPIST-1}, {\it Nature
  Astronomy\/}, {\bf 1}, 0129.

\bibitem[{Luger} et~al.(2021{\natexlab{a}}){Luger}, {Foreman-Mackey}, \&
  {Hedges}]{Luger2021b}
{Luger}, R., {Foreman-Mackey}, D., \& {Hedges}, C., 2021{\natexlab{a}}.
\newblock {Mapping Stellar Surfaces. II. An Interpretable Gaussian Process
  Model for Light Curves}, {\it \aj\/}, {\bf 162}(3), 124.

\bibitem[{Luger} et~al.(2021{\natexlab{b}}){Luger}, {Foreman-Mackey}, {Hedges},
  \& {Hogg}]{Luger2021}
{Luger}, R., {Foreman-Mackey}, D., {Hedges}, C., \& {Hogg}, D.~W.,
  2021{\natexlab{b}}.
\newblock {Mapping stellar surfaces I: Degeneracies in the rotational light
  curve problem}, {\it arXiv e-prints\/}, p. arXiv:2102.00007.

\bibitem[{Lupu} et~al.(2014){Lupu}, {Zahnle}, {Marley}, {Schaefer}, {Fegley},
  {Morley}, {Cahoy}, {Freedman}, \& {Fortney}]{Lupu14}
{Lupu}, R.~E., {Zahnle}, K., {Marley}, M.~S., {Schaefer}, L., {Fegley}, B.,
  {Morley}, C., {Cahoy}, K., {Freedman}, R., \& {Fortney}, J.~J., 2014.
\newblock {The Atmospheres of Earthlike Planets after Giant Impact Events},
  {\it \apj\/}, {\bf 784}, 27.

\bibitem[{MacDonald}(2023)]{MacDonald2023}
{MacDonald}, R.~J., 2023.
\newblock {POSEIDON: A Multidimensional Atmospheric Retrieval Code for
  Exoplanet Spectra}, {\it The Journal of Open Source Software\/}, {\bf 8},
  4873.

\bibitem[{MacDonald} \& {Madhusudhan}(2017)]{MacDonald2017_hd209458b}
{MacDonald}, R.~J. \& {Madhusudhan}, N., 2017.
\newblock {HD 209458b in new light: evidence of nitrogen chemistry, patchy
  clouds and sub-solar water}, {\it \mnras\/}, {\bf 469}(2), 1979--1996.

\bibitem[{Maciejewski} et~al.(2016){Maciejewski}, {Dimitrov}, {Fern{\'a}ndez},
  {Sota}, {Nowak}, {Ohlert}, {Nikolov}, {Bukowiecki}, {Hinse}, {Pall{\'e}},
  {Tingley}, {Kjurkchieva}, {Lee}, \& {Lee}]{Maciejewski:2016}
{Maciejewski}, G., {Dimitrov}, D., {Fern{\'a}ndez}, M., {Sota}, A., {Nowak},
  G., {Ohlert}, J., {Nikolov}, G., {Bukowiecki}, {\L}., {Hinse}, T.~C.,
  {Pall{\'e}}, E., {Tingley}, B., {Kjurkchieva}, D., {Lee}, J.~W., \& {Lee},
  C.~U., 2016.
\newblock {Departure from the constant-period ephemeris for the transiting
  exoplanet WASP-12}, {\it \aap\/}, {\bf 588}, L6.

\bibitem[{Maehara} et~al.(2015){Maehara}, {Shibayama}, {Notsu}, {Notsu},
  {Honda}, {Nogami}, \& {Shibata}]{Maehara2015}
{Maehara}, H., {Shibayama}, T., {Notsu}, Y., {Notsu}, S., {Honda}, S.,
  {Nogami}, D., \& {Shibata}, K., 2015.
\newblock {Statistical properties of superflares on solar-type stars based on
  1-min cadence data}, {\it Earth, Planets, and Space\/}, {\bf 67}, 59.

\bibitem[{Magic} et~al.(2013){Magic}, {Collet}, {Asplund}, {Trampedach},
  {Hayek}, {Chiavassa}, {Stein}, \& {Nordlund}]{magic+al2013a}
{Magic}, Z., {Collet}, R., {Asplund}, M., {Trampedach}, R., {Hayek}, W.,
  {Chiavassa}, A., {Stein}, R.~F., \& {Nordlund}, {\AA}., 2013.
\newblock {The Stagger-grid: A grid of 3D stellar atmosphere models. I. Methods
  and general properties}, {\it \aap\/}, {\bf 557}, A26.

\bibitem[{Magrini} et~al.(2022){Magrini}, {Danielski}, {Bossini}, {Rainer},
  {Turrini}, {Benatti}, {Brucalassi}, {Tsantaki}, {Delgado Mena}, {Sanna},
  {Biazzo}, {Campante}, {Van der Swaelmen}, {Sousa}, {He{\l}miniak}, {Neitzel},
  {Adibekyan}, {Bruno}, \& {Casali}]{Magrini2022}
{Magrini}, L., {Danielski}, C., {Bossini}, D., {Rainer}, M., {Turrini}, D.,
  {Benatti}, S., {Brucalassi}, A., {Tsantaki}, M., {Delgado Mena}, E., {Sanna},
  N., {Biazzo}, K., {Campante}, T.~L., {Van der Swaelmen}, M., {Sousa}, S.~G.,
  {He{\l}miniak}, K.~G., {Neitzel}, A.~W., {Adibekyan}, V., {Bruno}, G., \&
  {Casali}, G., 2022.
\newblock {Ariel stellar characterisation. I. Homogeneous stellar parameters of
  187 FGK planet host stars: Description and validation of the method}, {\it
  \aap\/}, {\bf 663}, A161.

\bibitem[{Mallonn} \& {Strassmeier}(2016)]{Mallonn2016}
{Mallonn}, M. \& {Strassmeier}, K.~G., 2016.
\newblock {Transmission spectroscopy of HAT-P-32b with the LBT: confirmation of
  clouds/hazes in the planetary atmosphere}, {\it \aap\/}, {\bf 590}, A100.

\bibitem[{Mallonn} et~al.(2018){Mallonn}, {Herrero}, {Juvan}, {von Essen},
  {Rosich}, {Ribas}, {Granzer}, {Alexoudi}, \& {Strassmeier}]{Mallonn-18}
{Mallonn}, M., {Herrero}, E., {Juvan}, I.~G., {von Essen}, C., {Rosich}, A.,
  {Ribas}, I., {Granzer}, T., {Alexoudi}, X., \& {Strassmeier}, K.~G., 2018.
\newblock {GJ 1214: Rotation period, starspots, and uncertainty on the optical
  slope of the transmission spectrum}, {\it \aap\/}, {\bf 614}, A35.

\bibitem[{Mancini} et~al.(2013){Mancini}, {Ciceri}, {Chen}, {Tregloan-Reed},
  {Fortney}, {Southworth}, {Tan}, {Burgdorf}, {Calchi Novati}, {Dominik},
  {Fang}, {Finet}, {Gerner}, {Hardis}, {Hinse}, {J{\o}rgensen}, {Liebig},
  {Nikolov}, {Ricci}, {Sch{\"a}fer}, {Sch{\"o}nebeck}, {Skottfelt}, {Wertz},
  {Alsubai}, {Bozza}, {Browne}, {Dodds}, {Gu}, {Harps{\o}e}, {Henning},
  {Hundertmark}, {Jessen-Hansen}, {Kains}, {Kerins}, {Kjeldsen}, {Lund},
  {Lundkvist}, {Madhusudhan}, {Mathiasen}, {Penny}, {Prof}, {Rahvar}, {Sahu},
  {Scarpetta}, {Snodgrass}, \& {Surdej}]{Mancini-13}
{Mancini}, L., {Ciceri}, S., {Chen}, G., {Tregloan-Reed}, J., {Fortney}, J.~J.,
  {Southworth}, J., {Tan}, T.~G., {Burgdorf}, M., {Calchi Novati}, S.,
  {Dominik}, M., {Fang}, X.-S., {Finet}, F., {Gerner}, T., {Hardis}, S.,
  {Hinse}, T.~C., {J{\o}rgensen}, U.~G., {Liebig}, C., {Nikolov}, N., {Ricci},
  D., {Sch{\"a}fer}, S., {Sch{\"o}nebeck}, F., {Skottfelt}, J., {Wertz}, O.,
  {Alsubai}, K.~A., {Bozza}, V., {Browne}, P., {Dodds}, P., {Gu}, S.-H.,
  {Harps{\o}e}, K., {Henning}, T., {Hundertmark}, M., {Jessen-Hansen}, J.,
  {Kains}, N., {Kerins}, E., {Kjeldsen}, H., {Lund}, M.~N., {Lundkvist}, M.,
  {Madhusudhan}, N., {Mathiasen}, M., {Penny}, M.~T., {Prof}, S., {Rahvar}, S.,
  {Sahu}, K., {Scarpetta}, G., {Snodgrass}, C., \& {Surdej}, J., 2013.
\newblock {Physical properties, transmission and emission spectra of the
  WASP-19 planetary system from multi-colour photometry}, {\it MNRAS\/}, {\bf
  436}, 2--18.

\bibitem[{Mancini} et~al.(2014{\natexlab{a}}){Mancini}, {Southworth}, {Ciceri},
  {Calchi Novati}, {Dominik}, {Henning}, {J{\o}rgensen}, {Korhonen}, {Nikolov},
  {Alsubai}, {Bozza}, {Bramich}, {D'Ago}, {Figuera Jaimes}, {Galianni}, {Gu},
  {Harps{\o}e}, {Hinse}, {Hundertmark}, {Juncher}, {Kains}, {Popovas}, {Rabus},
  {Rahvar}, {Skottfelt}, {Snodgrass}, {Street}, {Surdej}, {Tsapras}, {Vilela},
  {Wang}, \& {Wertz}]{Mancini-14}
{Mancini}, L., {Southworth}, J., {Ciceri}, S., {Calchi Novati}, S., {Dominik},
  M., {Henning}, T., {J{\o}rgensen}, U.~G., {Korhonen}, H., {Nikolov}, N.,
  {Alsubai}, K.~A., {Bozza}, V., {Bramich}, D.~M., {D'Ago}, G., {Figuera
  Jaimes}, R., {Galianni}, P., {Gu}, S.-H., {Harps{\o}e}, K., {Hinse}, T.~C.,
  {Hundertmark}, M., {Juncher}, D., {Kains}, N., {Popovas}, A., {Rabus}, M.,
  {Rahvar}, S., {Skottfelt}, J., {Snodgrass}, C., {Street}, R., {Surdej}, J.,
  {Tsapras}, Y., {Vilela}, C., {Wang}, X.-B., \& {Wertz}, O.,
  2014{\natexlab{a}}.
\newblock {Physical properties of the WASP-67 planetary system from
  multi-colour photometry}, {\it \aap\/}, {\bf 568}, A127.

\bibitem[{Mancini} et~al.(2014{\natexlab{b}}){Mancini}, {Southworth}, {Ciceri},
  {Tregloan-Reed}, {Crossfield}, {Nikolov}, {Bruni}, {Zambelli}, \&
  {Henning}]{Mancini-14b}
{Mancini}, L., {Southworth}, J., {Ciceri}, S., {Tregloan-Reed}, J.,
  {Crossfield}, I., {Nikolov}, N., {Bruni}, I., {Zambelli}, R., \& {Henning},
  T., 2014{\natexlab{b}}.
\newblock {Physical properties, star-spot activity, orbital obliquity and
  transmission spectrum of the Qatar-2 planetary system from multicolour
  photometry}, {\it MNRAS\/}, {\bf 443}, 2391--2409.

\bibitem[{Mancini} et~al.(2015){Mancini}, {Esposito}, {Covino}, {Raia},
  {Southworth}, {Tregloan-Reed}, {Biazzo}, {Bonomo}, {Desidera}, {Lanza},
  {Maciejewski}, {Poretti}, {Sozzetti}, {Borsa}, {Bruni}, {Ciceri}, {Claudi},
  {Cosentino}, {Gratton}, {Martinez Fiorenzano}, {Lodato}, {Lorenzi},
  {Marzari}, {Murabito}, {Affer}, {Bignamini}, {Bedin}, {Boccato}, {Damasso},
  {Henning}, {Maggio}, {Micela}, {Molinari}, {Pagano}, {Piotto}, {Rainer},
  {Scandariato}, {Smareglia}, \& {Zanmar Sanchez}]{Mancini-15}
{Mancini}, L., {Esposito}, M., {Covino}, E., {Raia}, G., {Southworth}, J.,
  {Tregloan-Reed}, J., {Biazzo}, K., {Bonomo}, A.~S., {Desidera}, S., {Lanza},
  A.~F., {Maciejewski}, G., {Poretti}, E., {Sozzetti}, A., {Borsa}, F.,
  {Bruni}, I., {Ciceri}, S., {Claudi}, R., {Cosentino}, R., {Gratton}, R.,
  {Martinez Fiorenzano}, A.~F., {Lodato}, G., {Lorenzi}, V., {Marzari}, F.,
  {Murabito}, S., {Affer}, L., {Bignamini}, A., {Bedin}, L.~R., {Boccato}, C.,
  {Damasso}, M., {Henning}, T., {Maggio}, A., {Micela}, G., {Molinari}, E.,
  {Pagano}, I., {Piotto}, G., {Rainer}, M., {Scandariato}, G., {Smareglia}, R.,
  \& {Zanmar Sanchez}, R., 2015.
\newblock {The GAPS Programme with HARPS-N at TNG. VIII. Observations of the
  Rossiter-McLaughlin effect and characterisation of the transiting planetary
  systems HAT-P-36 and WASP-11/HAT-P-10}, {\it \aap\/}, {\bf 579}, A136.

\bibitem[{Mancini} et~al.(2016){Mancini}, {Lillo-Box}, {Southworth}, {Borsato},
  {Gandolfi}, {Ciceri}, {Barrado}, {Brahm}, \& {Henning}]{Mancini-16}
{Mancini}, L., {Lillo-Box}, J., {Southworth}, J., {Borsato}, L., {Gandolfi},
  D., {Ciceri}, S., {Barrado}, D., {Brahm}, R., \& {Henning}, T., 2016.
\newblock {Kepler-539: A young extrasolar system with two giant planets on wide
  orbits and in gravitational interaction}, {\it \aap\/}, {\bf 590}, A112.

\bibitem[{Mancini} et~al.(2017){Mancini}, {Southworth}, {Raia},
  {Tregloan-Reed}, {Molli{\`e}re}, {Bozza}, {Bretton}, {Bruni}, {Ciceri},
  {D'Ago}, {Dominik}, {Hinse}, {Hundertmark}, {J{\o}rgensen}, {Korhonen},
  {Rabus}, {Rahvar}, {Starkey}, {Calchi Novati}, {Figuera Jaimes}, {Henning},
  {Juncher}, {Haugb{\o}lle}, {Kains}, {Popovas}, {Schmidt}, {Skottfelt},
  {Snodgrass}, {Surdej}, \& {Wertz}]{Mancini-17}
{Mancini}, L., {Southworth}, J., {Raia}, G., {Tregloan-Reed}, J.,
  {Molli{\`e}re}, P., {Bozza}, V., {Bretton}, M., {Bruni}, I., {Ciceri}, S.,
  {D'Ago}, G., {Dominik}, M., {Hinse}, T.~C., {Hundertmark}, M.,
  {J{\o}rgensen}, U.~G., {Korhonen}, H., {Rabus}, M., {Rahvar}, S., {Starkey},
  D., {Calchi Novati}, S., {Figuera Jaimes}, R., {Henning}, T., {Juncher}, D.,
  {Haugb{\o}lle}, T., {Kains}, N., {Popovas}, A., {Schmidt}, R.~W.,
  {Skottfelt}, J., {Snodgrass}, C., {Surdej}, J., \& {Wertz}, O., 2017.
\newblock {Orbital alignment and star-spot properties in the WASP-52 planetary
  system}, {\it \mnras\/}, {\bf 465}, 843--857.

\bibitem[{Mandal} et~al.(2021){Mandal}, {Krivova}, {Cameron}, \&
  {Solanki}]{Mandal2021}
{Mandal}, S., {Krivova}, N.~A., {Cameron}, R., \& {Solanki}, S.~K., 2021.
\newblock {On the size distribution of spots within sunspot groups}, {\it
  \aap\/}, {\bf 652}, A9.

\bibitem[{Mann} et~al.(2015){Mann}, {Feiden}, {Gaidos}, {Boyajian}, \& {von
  Braun}]{Mann2015}
{Mann}, A.~W., {Feiden}, G.~A., {Gaidos}, E., {Boyajian}, T., \& {von Braun},
  K., 2015.
\newblock {How to Constrain Your M Dwarf: Measuring Effective Temperature,
  Bolometric Luminosity, Mass, and Radius}, {\it \apj\/}, {\bf 804}(1), 64.

\bibitem[{Mansfield} et~al.(2018){Mansfield}, {Bean}, {Oklop{\v{c}}i{\'c}},
  {Kreidberg}, {D{\'e}sert}, {Kempton}, {Line}, {Fortney}, {Henry}, {Mallonn},
  {Stevenson}, {Dragomir}, {Allart}, \& {Bourrier}]{Mansfield2018}
{Mansfield}, M., {Bean}, J.~L., {Oklop{\v{c}}i{\'c}}, A., {Kreidberg}, L.,
  {D{\'e}sert}, J.-M., {Kempton}, E. M.~R., {Line}, M.~R., {Fortney}, J.~J.,
  {Henry}, G.~W., {Mallonn}, M., {Stevenson}, K.~B., {Dragomir}, D., {Allart},
  R., \& {Bourrier}, V., 2018.
\newblock {Detection of Helium in the Atmosphere of the Exo-Neptune HAT-P-11b},
  {\it \apjl\/}, {\bf 868}(2), L34.

\bibitem[{Marchenko} et~al.(2019){Marchenko}, {Woods}, {DeLand}, {Mauceri},
  {Pilewskie}, \& {Haberreiter}]{Marchenko2019}
{Marchenko}, S.~V., {Woods}, T.~N., {DeLand}, M.~T., {Mauceri}, S.,
  {Pilewskie}, P., \& {Haberreiter}, M., 2019.
\newblock {Improved Aura/OMI Solar Spectral Irradiances: Comparisons With
  Independent Data Sets and Model Predictions}, {\it Earth and Space
  Science\/}, {\bf 6}(12), 2379--2396.

\bibitem[{Marfil} et~al.(2021){Marfil}, {Tabernero}, {Montes}, {Caballero},
  {L{\'a}zaro}, {Gonz{\'a}lez Hern{\'a}ndez}, {Nagel}, {Passegger},
  {Schweitzer}, {Ribas}, {Reiners}, {Quirrenbach}, {Amado}, {Cifuentes},
  {Cort{\'e}s-Contreras}, {Dreizler}, {Duque-Arribas},
  {Galad{\'\i}-Enr{\'\i}quez}, {Henning}, {Jeffers}, {Kaminski}, {K{\"u}rster},
  {Lafarga}, {L{\'o}pez-Gallifa}, {Morales}, {Shan}, \&
  {Zechmeister}]{Marfil2021}
{Marfil}, E., {Tabernero}, H.~M., {Montes}, D., {Caballero}, J.~A.,
  {L{\'a}zaro}, F.~J., {Gonz{\'a}lez Hern{\'a}ndez}, J.~I., {Nagel}, E.,
  {Passegger}, V.~M., {Schweitzer}, A., {Ribas}, I., {Reiners}, A.,
  {Quirrenbach}, A., {Amado}, P.~J., {Cifuentes}, C., {Cort{\'e}s-Contreras},
  M., {Dreizler}, S., {Duque-Arribas}, C., {Galad{\'\i}-Enr{\'\i}quez}, D.,
  {Henning}, T., {Jeffers}, S.~V., {Kaminski}, A., {K{\"u}rster}, M.,
  {Lafarga}, M., {L{\'o}pez-Gallifa}, {\'A}., {Morales}, J.~C., {Shan}, Y., \&
  {Zechmeister}, M., 2021.
\newblock {The CARMENES search for exoplanets around M dwarfs. Stellar
  atmospheric parameters of target stars with SteParSyn}, {\it \aap\/}, {\bf
  656}, A162.

\bibitem[{Martinez} et~al.(2021){Martinez}, {Baron}, {Monnier},
  {Roettenbacher}, \& {Parks}]{Martinez2021}
{Martinez}, A.~O., {Baron}, F.~R., {Monnier}, J.~D., {Roettenbacher}, R.~M., \&
  {Parks}, J.~R., 2021.
\newblock {Dynamical Surface Imaging of {\ensuremath{\lambda}} Andromedae},
  {\it \apj\/}, {\bf 916}(1), 60.

\bibitem[{Mathew} et~al.(2007){Mathew}, {Mart{\'\i}nez Pillet}, {Solanki}, \&
  {Krivova}]{mathew2007}
{Mathew}, S.~K., {Mart{\'\i}nez Pillet}, V., {Solanki}, S.~K., \& {Krivova},
  N.~A., 2007.
\newblock {Properties of sunspots in cycle 23. I. Dependence of brightness on
  sunspot size and cycle phase}, {\it \aap\/}, {\bf 465}(1), 291--304.

\bibitem[{Mathur} et~al.(2011){Mathur}, {Hekker}, {Trampedach}, {Ballot},
  {Kallinger}, {Buzasi}, {Garc{\'\i}a}, {Huber}, {Jim{\'e}nez}, {Mosser},
  {Bedding}, {Elsworth}, {R{\'e}gulo}, {Stello}, {Chaplin}, {De Ridder},
  {Hale}, {Kinemuchi}, {Kjeldsen}, {Mullally}, \& {Thompson}]{Mathur:2011}
{Mathur}, S., {Hekker}, S., {Trampedach}, R., {Ballot}, J., {Kallinger}, T.,
  {Buzasi}, D., {Garc{\'\i}a}, R.~A., {Huber}, D., {Jim{\'e}nez}, A., {Mosser},
  B., {Bedding}, T.~R., {Elsworth}, Y., {R{\'e}gulo}, C., {Stello}, D.,
  {Chaplin}, W.~J., {De Ridder}, J., {Hale}, S.~J., {Kinemuchi}, K.,
  {Kjeldsen}, H., {Mullally}, F., \& {Thompson}, S.~E., 2011.
\newblock {Granulation in Red Giants: Observations by the Kepler Mission and
  Three-dimensional Convection Simulations}, {\it \apj\/}, {\bf 741}(2), 119.

\bibitem[{Matthes} et~al.(2017){Matthes}, {Funke}, {Andersson}, {Barnard},
  {Beer}, {Charbonneau}, {Clilverd}, {Dudok de Wit}, {Haberreiter}, {Hendry},
  {Jackman}, {Kretzschmar}, {Kruschke}, {Kunze}, {Langematz}, {Marsh},
  {Maycock}, {Misios}, {Rodger}, {Scaife}, {Sepp{\"a}l{\"a}}, {Shangguan},
  {Sinnhuber}, {Tourpali}, {Usoskin}, {van de Kamp}, {Verronen}, \&
  {Versick}]{Matthes2017}
{Matthes}, K., {Funke}, B., {Andersson}, M.~E., {Barnard}, L., {Beer}, J.,
  {Charbonneau}, P., {Clilverd}, M.~A., {Dudok de Wit}, T., {Haberreiter}, M.,
  {Hendry}, A., {Jackman}, C.~H., {Kretzschmar}, M., {Kruschke}, T., {Kunze},
  M., {Langematz}, U., {Marsh}, D.~R., {Maycock}, A.~C., {Misios}, S.,
  {Rodger}, C.~J., {Scaife}, A.~A., {Sepp{\"a}l{\"a}}, A., {Shangguan}, M.,
  {Sinnhuber}, M., {Tourpali}, K., {Usoskin}, I., {van de Kamp}, M.,
  {Verronen}, P.~T., \& {Versick}, S., 2017.
\newblock {Solar forcing for CMIP6 (v3.2)}, {\it Geoscientific Model
  Development\/}, {\bf 10}(6), 2247--2302.

\bibitem[{Maxted}(2016)]{Maxted-16}
{Maxted}, P.~F.~L., 2016.
\newblock {ellc: A fast, flexible light curve model for detached eclipsing
  binary stars and transiting exoplanets}, {\it \aap\/}, {\bf 591}, A111.

\bibitem[{Maxted}(2018)]{Maxted2018}
{Maxted}, P.~F.~L., 2018.
\newblock {Comparison of the power-2 limb-darkening law from the STAGGER-grid
  to Kepler light curves of transiting exoplanets}, {\it \aap\/}, {\bf 616},
  A39.

\bibitem[{Maxted} et~al.(2015){Maxted}, {Serenelli}, \&
  {Southworth}]{Maxted2015}
{Maxted}, P.~F.~L., {Serenelli}, A.~M., \& {Southworth}, J., 2015.
\newblock {Comparison of gyrochronological and isochronal age estimates for
  transiting exoplanet host stars}, {\it \aap\/}, {\bf 577}, A90.

\bibitem[{Mayfield} \& {Lawrence}(1985)]{Mayfield1985}
{Mayfield}, E.~B. \& {Lawrence}, J.~K., 1985.
\newblock {The Correlation of Solar Flare Production with Magnetic Energy in
  Active Regions}, {\it \solphys\/}, {\bf 96}(2), 293--305.

\bibitem[{Mayorga} et~al.(2021){Mayorga}, {Lustig-Yaeger}, {May}, {Sotzen},
  {Gonzalez-Quiles}, {Kilpatrick}, {Martin}, {Mandt}, {Stevenson}, \&
  {Izenberg}]{Mayorga2021}
{Mayorga}, L.~C., {Lustig-Yaeger}, J., {May}, E.~M., {Sotzen}, K.~S.,
  {Gonzalez-Quiles}, J., {Kilpatrick}, B.~M., {Martin}, E.~C., {Mandt}, K.,
  {Stevenson}, K.~B., \& {Izenberg}, N.~R., 2021.
\newblock {Transmission Spectroscopy of the Earth-Sun System to Inform the
  Search for Extrasolar Life}, {\it \psj\/}, {\bf 2}(4), 140.

\bibitem[{McCullough} et~al.(2014){McCullough}, {Crouzet}, {Deming}, \&
  {Madhusudhan}]{McCullough2014}
{McCullough}, P.~R., {Crouzet}, N., {Deming}, D., \& {Madhusudhan}, N., 2014.
\newblock {Water Vapor in the Spectrum of the Extrasolar Planet HD 189733b. I.
  The Transit}, {\it \apj\/}, {\bf 791}(1), 55.

\bibitem[{Mehltretter}(1974)]{Mehltretter1974}
{Mehltretter}, J.~P., 1974.
\newblock {Observations of Photospheric Faculae at the Center of the Solar
  Disk}, {\it \solphys\/}, {\bf 38}(1), 43--57.

\bibitem[{Metchev} et~al.(2015){Metchev}, {Heinze}, {Apai}, {Flateau},
  {Radigan}, {Burgasser}, {Marley}, {Artigau}, {Plavchan}, \&
  {Goldman}]{Metchev2015}
{Metchev}, S.~A., {Heinze}, A., {Apai}, D., {Flateau}, D., {Radigan}, J.,
  {Burgasser}, A., {Marley}, M.~S., {Artigau}, {\'E}., {Plavchan}, P., \&
  {Goldman}, B., 2015.
\newblock {Weather on Other Worlds. II. Survey Results: Spots are Ubiquitous on
  L and T Dwarfs}, {\it \apj\/}, {\bf 799}(2), 154.

\bibitem[{Michel} et~al.(2008){Michel}, {Baglin}, {Auvergne}, {Catala},
  {Samadi}, {Baudin}, {Appourchaux}, {Barban}, {Weiss}, {Berthomieu},
  {Boumier}, {Dupret}, {Garcia}, {Fridlund}, {Garrido}, {Goupil}, {Kjeldsen},
  {Lebreton}, {Mosser}, {Grotsch-Noels}, {Janot-Pacheco}, {Provost},
  {Roxburgh}, {Thoul}, {Toutain}, {Tiph{\`e}ne}, {Turck-Chieze}, {Vauclair},
  {Vauclair}, {Aerts}, {Alecian}, {Ballot}, {Charpinet}, {Hubert},
  {Ligni{\`e}res}, {Mathias}, {Monteiro}, {Neiner}, {Poretti}, {Renan de
  Medeiros}, {Ribas}, {Rieutord}, {Roca Cort{\'e}s}, \& {Zwintz}]{Michel2008}
{Michel}, E., {Baglin}, A., {Auvergne}, M., {Catala}, C., {Samadi}, R.,
  {Baudin}, F., {Appourchaux}, T., {Barban}, C., {Weiss}, W.~W., {Berthomieu},
  G., {Boumier}, P., {Dupret}, M.-A., {Garcia}, R.~A., {Fridlund}, M.,
  {Garrido}, R., {Goupil}, M.-J., {Kjeldsen}, H., {Lebreton}, Y., {Mosser}, B.,
  {Grotsch-Noels}, A., {Janot-Pacheco}, E., {Provost}, J., {Roxburgh}, I.~W.,
  {Thoul}, A., {Toutain}, T., {Tiph{\`e}ne}, D., {Turck-Chieze}, S.,
  {Vauclair}, S.~D., {Vauclair}, G.~P., {Aerts}, C., {Alecian}, G., {Ballot},
  J., {Charpinet}, S., {Hubert}, A.-M., {Ligni{\`e}res}, F., {Mathias}, P.,
  {Monteiro}, M. J.~P.~F.~G., {Neiner}, C., {Poretti}, E., {Renan de Medeiros},
  J., {Ribas}, I., {Rieutord}, M.~L., {Roca Cort{\'e}s}, T., \& {Zwintz}, K.,
  2008.
\newblock {CoRoT Measures Solar-Like Oscillations and Granulation in Stars
  Hotter Than the Sun}, {\it Science\/}, {\bf 322}(5901), 558.

\bibitem[{Milbourne} et~al.(2021){Milbourne}, {Phillips}, {Langellier},
  {Mortier}, {Haywood}, {Saar}, {Cegla}, {Collier Cameron}, {Dumusque},
  {Latham}, {Malavolta}, {Maldonado}, {Thompson}, {Vanderburg}, {Watson},
  {Buchhave}, {Cecconi}, {Cosentino}, {Ghedina}, {Gonzalez}, {Lodi},
  {L{\'o}pez-Morales}, {Sozzetti}, \& {Walsworth}]{Milbourne2021}
{Milbourne}, T.~W., {Phillips}, D.~F., {Langellier}, N., {Mortier}, A.,
  {Haywood}, R.~D., {Saar}, S.~H., {Cegla}, H.~M., {Collier Cameron}, A.,
  {Dumusque}, X., {Latham}, D.~W., {Malavolta}, L., {Maldonado}, J.,
  {Thompson}, S., {Vanderburg}, A., {Watson}, C.~A., {Buchhave}, L.~A.,
  {Cecconi}, M., {Cosentino}, R., {Ghedina}, A., {Gonzalez}, M., {Lodi}, M.,
  {L{\'o}pez-Morales}, M., {Sozzetti}, A., \& {Walsworth}, R.~L., 2021.
\newblock {Estimating Magnetic Filling Factors From Simultaneous Spectroscopy
  and Photometry: Disentangling Spots, Plage, and Network}, {\it arXiv
  e-prints\/}, p. arXiv:2105.09113.

\bibitem[{Miles-P{\'a}ez} et~al.(2017){Miles-P{\'a}ez}, {Metchev}, {Heinze}, \&
  {Apai}]{Miles-Paez2017}
{Miles-P{\'a}ez}, P.~A., {Metchev}, S.~A., {Heinze}, A., \& {Apai}, D., 2017.
\newblock {Weather on Other Worlds. IV. H{\ensuremath{\alpha}} Emission and
  Photometric Variability Are Not Correlated in L0-T8 Dwarfs}, {\it \apj\/},
  {\bf 840}(2), 83.

\bibitem[{Millar-Blanchaer} et~al.(2020){Millar-Blanchaer}, {Girard},
  {Karalidi}, {Marley}, {van Holstein}, {Sengupta}, {Mawet}, {Kataria}, {Snik},
  {de Boer}, {Jensen-Clem}, {Vigan}, \& {Hinkley}]{Millar-Blanchaer2020}
{Millar-Blanchaer}, M.~A., {Girard}, J.~H., {Karalidi}, T., {Marley}, M.~S.,
  {van Holstein}, R.~G., {Sengupta}, S., {Mawet}, D., {Kataria}, T., {Snik},
  F., {de Boer}, J., {Jensen-Clem}, R., {Vigan}, A., \& {Hinkley}, S., 2020.
\newblock {Detection of Polarization due to Cloud Bands in the Nearby Luhman 16
  Brown Dwarf Binary}, {\it \apj\/}, {\bf 894}(1), 42.

\bibitem[{Mirtorabi} et~al.(2003){Mirtorabi}, {Wasatonic}, \&
  {Guinan}]{mirtorabi03}
{Mirtorabi}, M.~T., {Wasatonic}, R., \& {Guinan}, E.~F., 2003.
\newblock {Wing Near-Infrared, TiO-Band, and V-Band Photometry of
  Chromospherically Active Star {\ensuremath{\lambda}} Andromedae}, {\it
  \aj\/}, {\bf 125}(6), 3265--3273.

\bibitem[{Miyakawa} et~al.(2021){Miyakawa}, {Hirano}, {Fukui}, {Mann},
  {Gaidos}, \& {Sato}]{Miyakawa21}
{Miyakawa}, K., {Hirano}, T., {Fukui}, A., {Mann}, A.~W., {Gaidos}, E., \&
  {Sato}, B., 2021.
\newblock {Wavelength Dependence of Activity-induced Photometric Variations for
  Young Cool Stars in Hyades}, {\it \aj\/}, {\bf 162}(3), 104.

\bibitem[{Mohler-Fischer} et~al.(2013){Mohler-Fischer}, {Mancini}, {Hartman},
  {Bakos}, {Penev}, {Bayliss}, {Jord{\'a}n}, {Csubry}, {Zhou}, {Rabus},
  {Nikolov}, {Brahm}, {Espinoza}, {Buchhave}, {B{\'e}ky}, {Suc}, {Cs{\'a}k},
  {Henning}, {Wright}, {Tinney}, {Addison}, {Schmidt}, {Noyes}, {Papp},
  {L{\'a}z{\'a}r}, {S{\'a}ri}, \& {Conroy}]{Mohler-Fischer13}
{Mohler-Fischer}, M., {Mancini}, L., {Hartman}, J.~D., {Bakos}, G.~{\'A}.,
  {Penev}, K., {Bayliss}, D., {Jord{\'a}n}, A., {Csubry}, Z., {Zhou}, G.,
  {Rabus}, M., {Nikolov}, N., {Brahm}, R., {Espinoza}, N., {Buchhave}, L.~A.,
  {B{\'e}ky}, B., {Suc}, V., {Cs{\'a}k}, B., {Henning}, T., {Wright}, D.~J.,
  {Tinney}, C.~G., {Addison}, B.~C., {Schmidt}, B., {Noyes}, R.~W., {Papp}, I.,
  {L{\'a}z{\'a}r}, J., {S{\'a}ri}, P., \& {Conroy}, P., 2013.
\newblock {HATS-2b: A transiting extrasolar planet orbiting a K-type star
  showing starspot activity}, {\it A\&A\/}, {\bf 558}, A55.

\bibitem[{Monnier} et~al.(2018){Monnier}, {Ireland}, {Kraus}, {Alonso-Herrero},
  {Bonsor}, {Baron}, {Bayo}, {Berger}, {Boyajian}, {Chiavassa}, {Ciardi},
  {Creech-Eakman}, {de Wit}, {Defr{\`e}re}, {Dong}, {Duch{\^e}ne}, {Espaillat},
  {Gallenne}, {Gandhi}, {Gonzalez}, {Haniff}, {Hoenig}, {Ilee}, {Isella},
  {Jensen}, {Juhasz}, {Kane}, {Kishimoto}, {Kley}, {Kral}, {Kratter},
  {Labadie}, {Lacour}, {Laughlin}, {Le Bouquin}, {Michael}, {Meru},
  {Millan-Gabet}, {Millour}, {Minardi}, {Morbidelli}, {Mordasini}, {Morlok},
  {Mozurkewich}, {Nelson}, {Olofsson}, {Oudmaijer}, {Packham}, {Paladini},
  {Panic}, {Petrov}, {Pope}, {Pott}, {Quiroga-Nunez}, {Ramos Almeida},
  {Raymond}, {Regaly}, {Reynolds}, {Ridgway}, {Rinehart}, {Schreiber}, {Smith},
  {Stassun}, {Surdej}, {ten Brummelaar}, {Tristram}, {Turner}, {Tuthill}, {van
  Belle}, {Vasisht}, {Wallace}, {Weigelt}, {Wishnow}, {Wittkowski}, {Wolf},
  {Young}, {Zhao}, {Zhu}, \& {Z{\'u}{\~n}iga-Fern{\'a}ndez}]{Monnier2018}
{Monnier}, J.~D., {Ireland}, M., {Kraus}, S., {Alonso-Herrero}, A., {Bonsor},
  A., {Baron}, F., {Bayo}, A., {Berger}, J.-P., {Boyajian}, T., {Chiavassa},
  A., {Ciardi}, D., {Creech-Eakman}, M., {de Wit}, W.-J., {Defr{\`e}re}, D.,
  {Dong}, R., {Duch{\^e}ne}, G., {Espaillat}, C., {Gallenne}, A., {Gandhi}, P.,
  {Gonzalez}, J.-F., {Haniff}, C., {Hoenig}, S., {Ilee}, J., {Isella}, A.,
  {Jensen}, E., {Juhasz}, A., {Kane}, S., {Kishimoto}, M., {Kley}, W., {Kral},
  Q., {Kratter}, K., {Labadie}, L., {Lacour}, S., {Laughlin}, G., {Le Bouquin},
  J.-B., {Michael}, E., {Meru}, F., {Millan-Gabet}, R., {Millour}, F.,
  {Minardi}, S., {Morbidelli}, A., {Mordasini}, C., {Morlok}, A.,
  {Mozurkewich}, D., {Nelson}, R., {Olofsson}, J., {Oudmaijer}, R., {Packham},
  C., {Paladini}, C., {Panic}, O., {Petrov}, R., {Pope}, B., {Pott}, J.-U.,
  {Quiroga-Nunez}, L.~H., {Ramos Almeida}, C., {Raymond}, S.~N., {Regaly}, Z.,
  {Reynolds}, M., {Ridgway}, S., {Rinehart}, S., {Schreiber}, M., {Smith}, M.,
  {Stassun}, K., {Surdej}, J., {ten Brummelaar}, T., {Tristram}, K., {Turner},
  N., {Tuthill}, P., {van Belle}, G., {Vasisht}, G., {Wallace}, A., {Weigelt},
  G., {Wishnow}, E., {Wittkowski}, M., {Wolf}, S., {Young}, J., {Zhao}, M.,
  {Zhu}, Z., \& {Z{\'u}{\~n}iga-Fern{\'a}ndez}, S., 2018.
\newblock {Planet formation imager: project update}, in {\em Optical and
  Infrared Interferometry and Imaging VI\/}, vol. 10701 of {\bf Society of
  Photo-Optical Instrumentation Engineers (SPIE) Conference Series}, p.
  1070118.

\bibitem[{Montalto} et~al.(2014){Montalto}, {Bou{\'e}}, {Oshagh}, {Boisse},
  {Bruno}, \& {Santos}]{Montalto-14}
{Montalto}, M., {Bou{\'e}}, G., {Oshagh}, M., {Boisse}, I., {Bruno}, G., \&
  {Santos}, N.~C., 2014.
\newblock {Improvements on analytic modelling of stellar spots}, {\it MNRAS\/},
  {\bf 444}, 1721--1728.

\bibitem[{Montet} et~al.(2017){Montet}, {Tovar}, \&
  {Foreman-Mackey}]{Montet2017}
{Montet}, B.~T., {Tovar}, G., \& {Foreman-Mackey}, D., 2017.
\newblock {Long-term Photometric Variability in Kepler Full-frame Images:
  Magnetic Cycles of Sun-like Stars}, {\it \apj\/}, {\bf 851}(2), 116.

\bibitem[{Morin} et~al.(2008){Morin}, {Donati}, {Forveille}, {Delfosse},
  {Dobler}, {Petit}, {Jardine}, {Collier Cameron}, {Albert}, {Manset},
  {Dintrans}, {Chabrier}, \& {Valenti}]{Morin2008}
{Morin}, J., {Donati}, J.~F., {Forveille}, T., {Delfosse}, X., {Dobler}, W.,
  {Petit}, P., {Jardine}, M.~M., {Collier Cameron}, A., {Albert}, L., {Manset},
  N., {Dintrans}, B., {Chabrier}, G., \& {Valenti}, J.~A., 2008.
\newblock {The stable magnetic field of the fully convective star V374 Peg},
  {\it \mnras\/}, {\bf 384}(1), 77--86.

\bibitem[{Morley} et~al.(2017){Morley}, {Kreidberg}, {Rustamkulov}, {Robinson},
  \& {Fortney}]{Morley_observing_2017}
{Morley}, C.~V., {Kreidberg}, L., {Rustamkulov}, Z., {Robinson}, T., \&
  {Fortney}, J.~J., 2017.
\newblock {Observing the Atmospheres of Known Temperate Earth-sized Planets
  with JWST}, {\it \apj\/}, {\bf 850}(2), 121.

\bibitem[{Morris} et~al.(2017){Morris}, {Hebb}, {Davenport}, {Rohn}, \&
  {Hawley}]{Morris-17}
{Morris}, B.~M., {Hebb}, L., {Davenport}, J.~R.~A., {Rohn}, G., \& {Hawley},
  S.~L., 2017.
\newblock {The Starspots of HAT-P-11: Evidence for a Solar-like Dynamo}, {\it
  \apj\/}, {\bf 846}, 99.

\bibitem[{Morris} et~al.(2018{\natexlab{a}}){Morris}, {Agol}, {Davenport}, \&
  {Hawley}]{Morris2018e}
{Morris}, B.~M., {Agol}, E., {Davenport}, J. R.~A., \& {Hawley}, S.~L.,
  2018{\natexlab{a}}.
\newblock {Possible Bright Starspots on TRAPPIST-1}, {\it \apj\/}, {\bf
  857}(1), 39.

\bibitem[{Morris} et~al.(2018{\natexlab{b}}){Morris}, {Agol}, {Hebb}, \&
  {Hawley}]{Morris2018b}
{Morris}, B.~M., {Agol}, E., {Hebb}, L., \& {Hawley}, S.~L.,
  2018{\natexlab{b}}.
\newblock {Robust Transiting Exoplanet Radii in the Presence of Starspots from
  Ingress and Egress Durations}, {\it \aj\/}, {\bf 156}(3), 91.

\bibitem[{Morris} et~al.(2018{\natexlab{c}}){Morris}, {Agol}, {Hebb}, {Hawley},
  {Gillon}, {Ducrot}, {Delrez}, {Ingalls}, \& {Demory}]{Morris2018d}
{Morris}, B.~M., {Agol}, E., {Hebb}, L., {Hawley}, S.~L., {Gillon}, M.,
  {Ducrot}, E., {Delrez}, L., {Ingalls}, J., \& {Demory}, B.-O.,
  2018{\natexlab{c}}.
\newblock {Non-detection of Contamination by Stellar Activity in the Spitzer
  Transit Light Curves of TRAPPIST-1}, {\it \apjl\/}, {\bf 863}(2), L32.

\bibitem[{Morris} et~al.(2018{\natexlab{d}}){Morris}, {Curtis}, {Douglas},
  {Hawley}, {Ag{\"u}eros}, {Bobra}, \& {Agol}]{Morris2018c}
{Morris}, B.~M., {Curtis}, J.~L., {Douglas}, S.~T., {Hawley}, S.~L.,
  {Ag{\"u}eros}, M.~A., {Bobra}, M.~G., \& {Agol}, E., 2018{\natexlab{d}}.
\newblock {Are Starspots and Plages Co-located on Active G and K Stars?}, {\it
  \aj\/}, {\bf 156}(5), 203.

\bibitem[{Morris} et~al.(2018{\natexlab{e}}){Morris}, {Hawley}, \&
  {Hebb}]{Morris-2018}
{Morris}, B.~M., {Hawley}, S.~L., \& {Hebb}, L., 2018{\natexlab{e}}.
\newblock {Large Starspot Groups on HAT-P-11 in Activity Cycle 1}, {\it
  Research Notes of the American Astronomical Society\/}, {\bf 2}(1), 26.

\bibitem[{Morris} et~al.(2020){Morris}, {Bobra}, {Agol}, {Lee}, \&
  {Hawley}]{Morris2020_plato}
{Morris}, B.~M., {Bobra}, M.~G., {Agol}, E., {Lee}, Y.~J., \& {Hawley}, S.~L.,
  2020.
\newblock {The stellar variability noise floor for transiting exoplanet
  photometry with PLATO}, {\it \mnras\/}, {\bf 493}(4), 5489--5498.

\bibitem[{Mosser} et~al.(2012){Mosser}, {Elsworth}, {Hekker}, {Huber},
  {Kallinger}, {Mathur}, {Belkacem}, {Goupil}, {Samadi}, {Barban}, {Bedding},
  {Chaplin}, {Garc{\'\i}a}, {Stello}, {De Ridder}, {Middour}, {Morris}, \&
  {Quintana}]{Mosser2012}
{Mosser}, B., {Elsworth}, Y., {Hekker}, S., {Huber}, D., {Kallinger}, T.,
  {Mathur}, S., {Belkacem}, K., {Goupil}, M.~J., {Samadi}, R., {Barban}, C.,
  {Bedding}, T.~R., {Chaplin}, W.~J., {Garc{\'\i}a}, R.~A., {Stello}, D., {De
  Ridder}, J., {Middour}, C.~K., {Morris}, R.~L., \& {Quintana}, E.~V., 2012.
\newblock {Characterization of the power excess of solar-like oscillations in
  red giants with Kepler}, {\it \aap\/}, {\bf 537}, A30.

\bibitem[{Mo{\v{c}}nik} et~al.(2016){Mo{\v{c}}nik}, {Clark}, {Anderson},
  {Hellier}, \& {Brown}]{Mocnik-2016}
{Mo{\v{c}}nik}, T., {Clark}, B.~J.~M., {Anderson}, D.~R., {Hellier}, C., \&
  {Brown}, D.~J.~A., 2016.
\newblock {Starspots on WASP-85}, {\it \aj\/}, {\bf 151}(6), 150.

\bibitem[{Mo{\v{c}}nik} et~al.(2017{\natexlab{a}}){Mo{\v{c}}nik}, {Hellier},
  {Anderson}, {Clark}, \& {Southworth}]{Mocnik-2017b}
{Mo{\v{c}}nik}, T., {Hellier}, C., {Anderson}, D.~R., {Clark}, B.~J.~M., \&
  {Southworth}, J., 2017{\natexlab{a}}.
\newblock {Starspots on WASP-107 and pulsations of WASP-118}, {\it \mnras\/},
  {\bf 469}(2), 1622--1629.

\bibitem[{Mo{\v{c}}nik} et~al.(2017{\natexlab{b}}){Mo{\v{c}}nik}, {Southworth},
  \& {Hellier}]{Mocnik-2017a}
{Mo{\v{c}}nik}, T., {Southworth}, J., \& {Hellier}, C., 2017{\natexlab{b}}.
\newblock {Recurring sets of recurring starspot occultations on exoplanet host
  Qatar-2}, {\it \mnras\/}, {\bf 471}(1), 394--403.

\bibitem[{M{\"u}ller} et~al.(2020){M{\"u}ller}, {St. Cyr}, {Zouganelis},
  {Gilbert}, {Marsden}, {Nieves-Chinchilla}, {Antonucci}, {Auch{\`e}re},
  {Berghmans}, {Horbury}, {Howard}, {Krucker}, {Maksimovic}, {Owen}, {Rochus},
  {Rodriguez-Pacheco}, {Romoli}, {Solanki}, {Bruno}, {Carlsson}, {Fludra},
  {Harra}, {Hassler}, {Livi}, {Louarn}, {Peter}, {Sch{\"u}hle}, {Teriaca}, {del
  Toro Iniesta}, {Wimmer-Schweingruber}, {Marsch}, {Velli}, {De Groof},
  {Walsh}, \& {Williams}]{SolarOrbiterMision}
{M{\"u}ller}, D., {St. Cyr}, O.~C., {Zouganelis}, I., {Gilbert}, H.~R.,
  {Marsden}, R., {Nieves-Chinchilla}, T., {Antonucci}, E., {Auch{\`e}re}, F.,
  {Berghmans}, D., {Horbury}, T.~S., {Howard}, R.~A., {Krucker}, S.,
  {Maksimovic}, M., {Owen}, C.~J., {Rochus}, P., {Rodriguez-Pacheco}, J.,
  {Romoli}, M., {Solanki}, S.~K., {Bruno}, R., {Carlsson}, M., {Fludra}, A.,
  {Harra}, L., {Hassler}, D.~M., {Livi}, S., {Louarn}, P., {Peter}, H.,
  {Sch{\"u}hle}, U., {Teriaca}, L., {del Toro Iniesta}, J.~C.,
  {Wimmer-Schweingruber}, R.~F., {Marsch}, E., {Velli}, M., {De Groof}, A.,
  {Walsh}, A., \& {Williams}, D., 2020.
\newblock {The Solar Orbiter mission. Science overview}, {\it \aap\/}, {\bf
  642}, A1.

\bibitem[{Murgas} et~al.(2019){Murgas}, {Chen}, {Pall{\'e}}, {Nortmann}, \&
  {Nowak}]{Murgas2019}
{Murgas}, F., {Chen}, G., {Pall{\'e}}, E., {Nortmann}, L., \& {Nowak}, G.,
  2019.
\newblock {The GTC exoplanet transit spectroscopy survey. X. Stellar spots
  versus Rayleigh scattering: the case of HAT-P-11b}, {\it \aap\/}, {\bf 622},
  A172.

\bibitem[{Murray-Clay} et~al.(2009){Murray-Clay}, {Chiang}, \&
  {Murray}]{Murray-Clay2009}
{Murray-Clay}, R.~A., {Chiang}, E.~I., \& {Murray}, N., 2009.
\newblock {Atmospheric Escape From Hot Jupiters}, {\it \apj\/}, {\bf 693}(1),
  23--42.

\bibitem[{Namekata} et~al.(2019){Namekata}, {Maehara}, {Notsu}, {Toriumi},
  {Hayakawa}, {Ikuta}, {Notsu}, {Honda}, {Nogami}, \& {Shibata}]{Namekate2019}
{Namekata}, K., {Maehara}, H., {Notsu}, Y., {Toriumi}, S., {Hayakawa}, H.,
  {Ikuta}, K., {Notsu}, S., {Honda}, S., {Nogami}, D., \& {Shibata}, K., 2019.
\newblock {Lifetimes and Emergence/Decay Rates of Star Spots on Solar-type
  Stars Estimated by Kepler Data in Comparison with Those of Sunspots}, {\it
  \apj\/}, {\bf 871}(2), 187.

\bibitem[{Namekata} et~al.(2020){Namekata}, {Davenport}, {Morris}, {Hawley},
  {Maehara}, {Notsu}, {Toriumi}, {Ikuta}, {Notsu}, {Honda}, {Nogami}, \&
  {Shibata}]{Namekate2020}
{Namekata}, K., {Davenport}, J. R.~A., {Morris}, B.~M., {Hawley}, S.~L.,
  {Maehara}, H., {Notsu}, Y., {Toriumi}, S., {Ikuta}, K., {Notsu}, S., {Honda},
  S., {Nogami}, D., \& {Shibata}, K., 2020.
\newblock {Temporal Evolution of Spatially Resolved Individual Star Spots on a
  Planet-hosting Solar-type Star: Kepler-17}, {\it \apj\/}, {\bf 891}(2), 103.

\bibitem[{Narukage} et~al.(2016){Narukage}, {McKenzie}, {Ishikawa},
  {Trujillo-Bueno}, {De Pontieu}, {Kubo}, {Ishikawa}, {Kano}, {Suematsu},
  {Yoshida}, {Rachmeler}, {Kobayashi}, {Cirtain}, {Winebarger}, {Asensio
  Ramos}, {del Pino Aleman}, {{\v{S}}t{\k{e}}p{\'a}n}, {Belluzzi},
  {Larruquert}, {Auch{\`e}re}, {Leenaarts}, \& {Carlsson}]{CLASP2}
{Narukage}, N., {McKenzie}, D.~E., {Ishikawa}, R., {Trujillo-Bueno}, J., {De
  Pontieu}, B., {Kubo}, M., {Ishikawa}, S.-n., {Kano}, R., {Suematsu}, Y.,
  {Yoshida}, M., {Rachmeler}, L.~A., {Kobayashi}, K., {Cirtain}, J.~W.,
  {Winebarger}, A.~R., {Asensio Ramos}, A., {del Pino Aleman}, T.,
  {{\v{S}}t{\k{e}}p{\'a}n}, J., {Belluzzi}, L., {Larruquert}, J.~I.,
  {Auch{\`e}re}, F., {Leenaarts}, J., \& {Carlsson}, M. J.~L., 2016.
\newblock {Chromospheric LAyer SpectroPolarimeter (CLASP2)}, in {\em Space
  Telescopes and Instrumentation 2016: Ultraviolet to Gamma Ray\/}, vol. 9905
  of {\bf Society of Photo-Optical Instrumentation Engineers (SPIE) Conference
  Series}, p. 990508.

\bibitem[{Nascimbeni} et~al.(2015){Nascimbeni}, {Mallonn}, {Scandariato},
  {Pagano}, {Piotto}, {Micela}, {Messina}, {Leto}, {Strassmeier}, {Bisogni}, \&
  {Speziali}]{Nascimbeni2015}
{Nascimbeni}, V., {Mallonn}, M., {Scandariato}, G., {Pagano}, I., {Piotto}, G.,
  {Micela}, G., {Messina}, S., {Leto}, G., {Strassmeier}, K.~G., {Bisogni}, S.,
  \& {Speziali}, R., 2015.
\newblock {Large Binocular Telescope view of the atmosphere of GJ1214b}, {\it
  \aap\/}, {\bf 579}, A113.

\bibitem[{Neckel} \& {Labs}(1994)]{Neckel1994}
{Neckel}, H. \& {Labs}, D., 1994.
\newblock {Solar Limb Darkening 1986-1990 Lambda 303-NANOMETERS to
  1099-NANOMETERS}, {\it \solphys\/}, {\bf 153}(1-2), 91--114.

\bibitem[{Neff} et~al.(1995){Neff}, {O'Neal}, \& {Saar}]{Neff1995}
{Neff}, J.~E., {O'Neal}, D., \& {Saar}, S.~H., 1995.
\newblock {Absolute Measurements of Starspot Area and Temperature: II Pegasi in
  1989 October}, {\it \apj\/}, {\bf 452}, 879.

\bibitem[{N{\`e}mec} et~al.(2022){N{\`e}mec}, {Shapiro}, {I{\c{s}}{\i}k},
  {Sowmya}, {Solanki}, {Krivova}, {Cameron}, \& {Gizon}]{Nemec2022}
{N{\`e}mec}, N.~E., {Shapiro}, A.~I., {I{\c{s}}{\i}k}, E., {Sowmya}, K.,
  {Solanki}, S.~K., {Krivova}, N.~A., {Cameron}, R.~H., \& {Gizon}, L., 2022.
\newblock {Faculae Cancel out on the Surfaces of Active Suns}, {\it \apjl\/},
  {\bf 934}(2), L23.

\bibitem[{Netto} \& {Valio}(2020)]{Netto-20}
{Netto}, Y. \& {Valio}, A., 2020.
\newblock {Stellar magnetic activity and the butterfly diagram of Kepler-63},
  {\it \aap\/}, {\bf 635}, A78.

\bibitem[{Neveu-VanMalle} et~al.(2016){Neveu-VanMalle}, {Queloz}, {Anderson},
  {Brown}, {Collier Cameron}, {Delrez}, {D{\'{\i}}az}, {Gillon}, {Hellier},
  {Jehin}, {Lister}, {Pepe}, {Rojo}, {S{\'e}gransan}, {Triaud}, {Turner}, \&
  {Udry}]{Neveu-VanMalle-16}
{Neveu-VanMalle}, M., {Queloz}, D., {Anderson}, D.~R., {Brown}, D.~J.~A.,
  {Collier Cameron}, A., {Delrez}, L., {D{\'{\i}}az}, R.~F., {Gillon}, M.,
  {Hellier}, C., {Jehin}, E., {Lister}, T., {Pepe}, F., {Rojo}, P.,
  {S{\'e}gransan}, D., {Triaud}, A.~H.~M.~J., {Turner}, O.~D., \& {Udry}, S.,
  2016.
\newblock {Hot Jupiters with relatives: discovery of additional planets in
  orbit around WASP-41 and WASP-47}, {\it \aap\/}, {\bf 586}, A93.

\bibitem[{Newman} et~al.(2022){Newman}, {Plavchan}, {Burt}, {Teske}, {Mamajek},
  {Leifer}, {Gaudi}, {Blackwood}, \& {Morgan}]{Newman2022}
{Newman}, P.~D., {Plavchan}, P., {Burt}, J.~A., {Teske}, J., {Mamajek}, E.~E.,
  {Leifer}, .~S., {Gaudi}, B.~S., {Blackwood}, G., \& {Morgan}, R., 2022.
\newblock {Simulations for Planning Next-Generation Exoplanet Radial Velocity
  Surveys}, {\it arXiv e-prints\/}, p. arXiv:2204.13968.

\bibitem[{Newton} et~al.(2016){Newton}, {Irwin}, {Charbonneau},
  {Berta-Thompson}, {Dittmann}, \& {West}]{2016ApJ...821...93N}
{Newton}, E.~R., {Irwin}, J., {Charbonneau}, D., {Berta-Thompson}, Z.~K.,
  {Dittmann}, J.~A., \& {West}, A.~A., 2016.
\newblock {The Rotation and Galactic Kinematics of Mid M Dwarfs in the Solar
  Neighborhood}, {\it \apj\/}, {\bf 821}(2), 93.

\bibitem[{Newton} et~al.(2017){Newton}, {Irwin}, {Charbonneau}, {Berlind},
  {Calkins}, \& {Mink}]{2017ApJ...834...85N}
{Newton}, E.~R., {Irwin}, J., {Charbonneau}, D., {Berlind}, P., {Calkins},
  M.~L., \& {Mink}, J., 2017.
\newblock {The H{\ensuremath{\alpha}} Emission of Nearby M Dwarfs and its
  Relation to Stellar Rotation}, {\it \apj\/}, {\bf 834}(1), 85.

\bibitem[{Newton} et~al.(2018){Newton}, {Mondrik}, {Irwin}, {Winters}, \&
  {Charbonneau}]{Newton2018}
{Newton}, E.~R., {Mondrik}, N., {Irwin}, J., {Winters}, J.~G., \&
  {Charbonneau}, D., 2018.
\newblock {New Rotation Period Measurements for M Dwarfs in the Southern
  Hemisphere: An Abundance of Slowly Rotating, Fully Convective Stars}, {\it
  \aj\/}, {\bf 156}(5), 217.

\bibitem[{Nichols-Fleming} \& {Blackman}(2020)]{NicholsFleming2020}
{Nichols-Fleming}, F. \& {Blackman}, E.~G., 2020.
\newblock {Determination of the star-spot covering fraction as a function of
  stellar age from observational data}, {\it \mnras\/}, {\bf 491}(2),
  2706--2714.

\bibitem[{Nikolov} et~al.(2015){Nikolov}, {Sing}, {Burrows}, {Fortney},
  {Henry}, {Pont}, {Ballester}, {Aigrain}, {Wilson}, {Huitson}, {Gibson},
  {D{\'e}sert}, {Lecavelier Des Etangs}, {Showman}, {Vidal-Madjar}, {Wakeford},
  \& {Zahnle}]{Nikolov2015}
{Nikolov}, N., {Sing}, D.~K., {Burrows}, A.~S., {Fortney}, J.~J., {Henry},
  G.~W., {Pont}, F., {Ballester}, G.~E., {Aigrain}, S., {Wilson}, P.~A.,
  {Huitson}, C.~M., {Gibson}, N.~P., {D{\'e}sert}, J.~M., {Lecavelier Des
  Etangs}, A., {Showman}, A.~P., {Vidal-Madjar}, A., {Wakeford}, H.~R., \&
  {Zahnle}, K., 2015.
\newblock {HST hot-Jupiter transmission spectral survey: haze in the atmosphere
  of WASP-6b}, {\it \mnras\/}, {\bf 447}(1), 463--478.

\bibitem[{Nikolov} et~al.(2016){Nikolov}, {Sing}, {Gibson}, {Fortney}, {Evans},
  {Barstow}, {Kataria}, \& {Wilson}]{Nikolov2016}
{Nikolov}, N., {Sing}, D.~K., {Gibson}, N.~P., {Fortney}, J.~J., {Evans},
  T.~M., {Barstow}, J.~K., {Kataria}, T., \& {Wilson}, P.~A., 2016.
\newblock {VLT FORS2 Comparative Transmission Spectroscopy: Detection of Na in
  the Atmosphere of WASP-39b from the Ground}, {\it \apj\/}, {\bf 832}(2), 191.

\bibitem[{Nordgren} et~al.(1999){Nordgren}, {Germain}, {Benson}, {Mozurkewich},
  {Sudol}, {Elias}, {Hajian}, {White}, {Hutter}, {Johnston}, {Gauss},
  {Armstrong}, {Pauls}, \& {Rickard}]{Nordgren1999}
{Nordgren}, T.~E., {Germain}, M.~E., {Benson}, J.~A., {Mozurkewich}, D.,
  {Sudol}, J.~J., {Elias}, N.~M., I., {Hajian}, A.~R., {White}, N.~M.,
  {Hutter}, D.~J., {Johnston}, K.~J., {Gauss}, F.~S., {Armstrong}, J.~T.,
  {Pauls}, T.~A., \& {Rickard}, L.~J., 1999.
\newblock {Stellar Angular Diameters of Late-Type Giants and Supergiants
  Measured with the Navy Prototype Optical Interferometer}, {\it \aj\/}, {\bf
  118}(6), 3032--3038.

\bibitem[{Nordlund}(1982)]{1982nordlund}
{Nordlund}, A., 1982.
\newblock {Numerical simulations of the solar granulation. I. Basic equations
  and methods.}, {\it \aap\/}, {\bf 107}, 1--10.

\bibitem[{Nordlund}(1984)]{1984nordlund}
{Nordlund}, A., 1984.
\newblock {Modelling of Small-Scale Dynamical Processes: Convection and Wave
  Generation (Keynote)}, in {\em Small-Scale Dynamical Processes in Quiet
  Stellar Atmospheres\/}, p. 181.

\bibitem[{Nordlund}(1985)]{1985SoPh..100..209N}
{Nordlund}, A., 1985.
\newblock {Solar Convection}, {\it \solphys\/}, {\bf 100}, 209.

\bibitem[{Nordlund} \& {Dravins}(1990)]{Nord_Dravins_90A}
{Nordlund}, A. \& {Dravins}, D., 1990.
\newblock {Stellar granulation. III. Hydrodynamic model atmospheres.}, {\it
  \aap\/}, {\bf 228}, 155--183.

\bibitem[{Nordlund} et~al.(2009){Nordlund}, {Stein}, \&
  {Asplund}]{Nordlund2009}
{Nordlund}, {\r{A}}., {Stein}, R.~F., \& {Asplund}, M., 2009.
\newblock {Solar Surface Convection}, {\it Living Reviews in Solar Physics\/},
  {\bf 6}(1), 2.

\bibitem[Norris(2018)]{Norris2018}
Norris, C.~M., 2018.
\newblock {\it {Spectral variability in cool main-sequence stars due to
  small-scale magnetic features}\/}, Ph.D. thesis, Imperial College London.

\bibitem[{Norris} et~al.(2017){Norris}, {Beeck}, {Unruh}, {Solanki}, {Krivova},
  \& {Yeo}]{Norris2017}
{Norris}, C.~M., {Beeck}, B., {Unruh}, Y.~C., {Solanki}, S.~K., {Krivova},
  N.~A., \& {Yeo}, K.~L., 2017.
\newblock {Spectral variability of photospheric radiation due to faculae. I.
  The Sun and Sun-like stars}, {\it \aap\/}, {\bf 605}, A45.

\bibitem[{Notsu} et~al.(2013){Notsu}, {Shibayama}, {Maehara}, {Notsu}, {Nagao},
  {Honda}, {Ishii}, {Nogami}, \& {Shibata}]{Notsu2013}
{Notsu}, Y., {Shibayama}, T., {Maehara}, H., {Notsu}, S., {Nagao}, T., {Honda},
  S., {Ishii}, T.~T., {Nogami}, D., \& {Shibata}, K., 2013.
\newblock {Superflares on Solar-type Stars Observed with Kepler II. Photometric
  Variability of Superflare-generating Stars: A Signature of Stellar Rotation
  and Starspots}, {\it \apj\/}, {\bf 771}(2), 127.

\bibitem[{Notsu} et~al.(2019){Notsu}, {Maehara}, {Honda}, {Hawley},
  {Davenport}, {Namekata}, {Notsu}, {Ikuta}, {Nogami}, \&
  {Shibata}]{Notsu:2019}
{Notsu}, Y., {Maehara}, H., {Honda}, S., {Hawley}, S.~L., {Davenport}, J.
  R.~A., {Namekata}, K., {Notsu}, S., {Ikuta}, K., {Nogami}, D., \& {Shibata},
  K., 2019.
\newblock {Do Kepler Superflare Stars Really Include Slowly Rotating Sun-like
  Stars?{\textemdash}Results Using APO 3.5 m Telescope Spectroscopic
  Observations and Gaia-DR2 Data}, {\it \apj\/}, {\bf 876}(1), 58.

\bibitem[{Noyes} et~al.(1984){Noyes}, {Hartmann}, {Baliunas}, {Duncan}, \&
  {Vaughan}]{Noyes-84}
{Noyes}, R.~W., {Hartmann}, L.~W., {Baliunas}, S.~L., {Duncan}, D.~K., \&
  {Vaughan}, A.~H., 1984.
\newblock {Rotation, convection, and magnetic activity in lower main-sequence
  stars}, {\it \apj\/}, {\bf 279}, 763--777.

\bibitem[{Nutzman} \& {Charbonneau}(2008)]{Mearth_nutzman_charbonneau2008}
{Nutzman}, P. \& {Charbonneau}, D., 2008.
\newblock {Design Considerations for a Ground-Based Transit Search for
  Habitable Planets Orbiting M Dwarfs}, {\it \pasp\/}, {\bf 120}(865), 317.

\bibitem[{Okamoto} et~al.(2021){Okamoto}, {Notsu}, {Maehara}, {Namekata},
  {Honda}, {Ikuta}, {Nogami}, \& {Shibata}]{Okamoto2021}
{Okamoto}, S., {Notsu}, Y., {Maehara}, H., {Namekata}, K., {Honda}, S.,
  {Ikuta}, K., {Nogami}, D., \& {Shibata}, K., 2021.
\newblock {Statistical Properties of Superflares on Solar-type Stars: Results
  Using All of the Kepler Primary Mission Data}, {\it \apj\/}, {\bf 906}(2),
  72.

\bibitem[{Okamoto} \& {Sakurai}(2018)]{Okamoto:2018}
{Okamoto}, T.~J. \& {Sakurai}, T., 2018.
\newblock {Super-strong Magnetic Field in Sunspots}, {\it \apjl\/}, {\bf
  852}(1), L16.

\bibitem[{Ol{\'a}h} et~al.(2009){Ol{\'a}h}, {Koll{\'a}th}, {Granzer},
  {Strassmeier}, {Lanza}, {J{\"a}rvinen}, {Korhonen}, {Baliunas}, {Soon},
  {Messina}, \& {Cutispoto}]{Olah2009}
{Ol{\'a}h}, K., {Koll{\'a}th}, Z., {Granzer}, T., {Strassmeier}, K.~G.,
  {Lanza}, A.~F., {J{\"a}rvinen}, S., {Korhonen}, H., {Baliunas}, S.~L.,
  {Soon}, W., {Messina}, S., \& {Cutispoto}, G., 2009.
\newblock {Multiple and changing cycles of active stars. II. Results}, {\it
  \aap\/}, {\bf 501}(2), 703--713.

\bibitem[{Omodei} et~al.(2018){Omodei}, {Pesce-Rollins}, {Longo}, {Allafort},
  \& {Krucker}]{Omodei2018}
{Omodei}, N., {Pesce-Rollins}, M., {Longo}, F., {Allafort}, A., \& {Krucker},
  S., 2018.
\newblock {Fermi-LAT Observations of the 2017 September 10 Solar Flare}, {\it
  \apjl\/}, {\bf 865}(1), L7.

\bibitem[{O'Neal} et~al.(1998{\natexlab{a}}){O'Neal}, {Saar}, \&
  {Neff}]{ONeal1998}
{O'Neal}, D., {Saar}, S.~H., \& {Neff}, J.~E., 1998{\natexlab{a}}.
\newblock {Spectroscopic Evidence for Nonuniform Starspot Properties on II
  Pegasi}, {\it \apjl\/}, {\bf 501}(1), L73--L76.

\bibitem[{O'Neal} et~al.(1998{\natexlab{b}}){O'Neal}, {Saar}, \&
  {Neff}]{ONeal-98}
{O'Neal}, D., {Saar}, S.~M., \& {Neff}, J.~E., 1998{\natexlab{b}}.
\newblock {Spectroscopic Evidence for Nonuniform Starspot Properties on II
  Pegasi}, {\it \apjl\/}, {\bf 501}, L73.

\bibitem[{Ortiz} et~al.(2002){Ortiz}, {Solanki}, {Domingo}, {Fligge}, \&
  {Sanahuja}]{ortiz2002}
{Ortiz}, A., {Solanki}, S.~K., {Domingo}, V., {Fligge}, M., \& {Sanahuja}, B.,
  2002.
\newblock {On the intensity contrast of solar photospheric faculae and network
  elements}, {\it \aap\/}, {\bf 388}, 1036--1047.

\bibitem[{Osborn} \& {Bayliss}(2020)]{Osborn2020}
{Osborn}, A. \& {Bayliss}, D., 2020.
\newblock {Investigating the planet-metallicity correlation for hot Jupiters},
  {\it \mnras\/}, {\bf 491}(3), 4481--4487.

\bibitem[{Oshagh} et~al.(2013{\natexlab{a}}){Oshagh}, {Boisse}, {Bou{\'e}},
  {Montalto}, {Santos}, {Bonfils}, \& {Haghighipour}]{Oshagh-13a}
{Oshagh}, M., {Boisse}, I., {Bou{\'e}}, G., {Montalto}, M., {Santos}, N.~C.,
  {Bonfils}, X., \& {Haghighipour}, N., 2013{\natexlab{a}}.
\newblock {SOAP-T: a tool to study the light curve and radial velocity of a
  system with a transiting planet and a rotating spotted star}, {\it \aap\/},
  {\bf 549}, A35.

\bibitem[{Oshagh} et~al.(2013{\natexlab{b}}){Oshagh}, {Santos}, {Boisse},
  {Bou{\'e}}, {Montalto}, {Dumusque}, \& {Haghighipour}]{Oshagh-13b}
{Oshagh}, M., {Santos}, N.~C., {Boisse}, I., {Bou{\'e}}, G., {Montalto}, M.,
  {Dumusque}, X., \& {Haghighipour}, N., 2013{\natexlab{b}}.
\newblock {Effect of stellar spots on high-precision transit light-curve}, {\it
  \aap\/}, {\bf 556}, A19.

\bibitem[{Oshagh} et~al.(2014){Oshagh}, {Santos}, {Ehrenreich}, {Haghighipour},
  {Figueira}, {Santerne}, \& {Montalto}]{Oshagh2014}
{Oshagh}, M., {Santos}, N.~C., {Ehrenreich}, D., {Haghighipour}, N.,
  {Figueira}, P., {Santerne}, A., \& {Montalto}, M., 2014.
\newblock {Impact of occultations of stellar active regions on transmission
  spectra. Can occultation of a plage mimic the signature of a blue sky?}, {\it
  \aap\/}, {\bf 568}, A99.

\bibitem[{Oshagh} et~al.(2015){Oshagh}, {Santos}, {Figueira}, {Adibekyan},
  {Santerne}, {Barros}, \& {Lima}]{Oshagh-15}
{Oshagh}, M., {Santos}, N.~C., {Figueira}, P., {Adibekyan}, V.~Z., {Santerne},
  A., {Barros}, S.~C.~C., \& {Lima}, J.~J.~G., 2015.
\newblock {Polar stellar-spots and grazing planetary transits. Possible
  explanation for the low number of discovered grazing planets}, {\it \aap\/},
  {\bf 583}, L1.

\bibitem[{Oshagh} et~al.(2016){Oshagh}, {Dreizler}, {Santos}, {Figueira}, \&
  {Reiners}]{Oshagh-16}
{Oshagh}, M., {Dreizler}, S., {Santos}, N.~C., {Figueira}, P., \& {Reiners},
  A., 2016.
\newblock {Can stellar activity make a planet seem misaligned?}, {\it \aap\/},
  {\bf 593}, A25.

\bibitem[{Oshagh} et~al.(2017){Oshagh}, {Santos}, {Figueira}, {Barros},
  {Donati}, {Adibekyan}, {Faria}, {Watson}, {Cegla}, {Dumusque}, {H{\'e}brard},
  {Demangeon}, {Dreizler}, {Boisse}, {Deleuil}, {Bonfils}, {Pepe}, \&
  {Udry}]{Oshagh2017}
{Oshagh}, M., {Santos}, N.~C., {Figueira}, P., {Barros}, S.~C.~C., {Donati},
  J.~F., {Adibekyan}, V., {Faria}, J.~P., {Watson}, C.~A., {Cegla}, H.~M.,
  {Dumusque}, X., {H{\'e}brard}, E., {Demangeon}, O., {Dreizler}, S., {Boisse},
  I., {Deleuil}, M., {Bonfils}, X., {Pepe}, F., \& {Udry}, S., 2017.
\newblock {Understanding stellar activity-induced radial velocity jitter using
  simultaneous K2 photometry and HARPS RV measurements}, {\it \aap\/}, {\bf
  606}, A107.

\bibitem[{Oshagh} et~al.(2018){Oshagh}, {Triaud}, {Burdanov}, {Figueira},
  {Reiners}, {Santos}, {Faria}, {Boue}, {D{\'{\i}}az}, {Dreizler}, {Boldt},
  {Delrez}, {Ducrot}, {Gillon}, {Guzman Mesa}, {Jehin}, {Khalafinejad}, {Kohl},
  {Serrano}, \& {Udry}]{Oshagh-18}
{Oshagh}, M., {Triaud}, A.~H.~M.~J., {Burdanov}, A., {Figueira}, P., {Reiners},
  A., {Santos}, N.~C., {Faria}, J., {Boue}, G., {D{\'{\i}}az}, R.~F.,
  {Dreizler}, S., {Boldt}, S., {Delrez}, L., {Ducrot}, E., {Gillon}, M.,
  {Guzman Mesa}, A., {Jehin}, E., {Khalafinejad}, S., {Kohl}, S., {Serrano},
  L., \& {Udry}, S., 2018.
\newblock {Activity induced variation in spin-orbit angles as derived from
  Rossiter-McLaughlin measurements}, {\it \aap\/}, {\bf 619}, A150.

\bibitem[{Owen} \& {Jackson}(2012)]{Owen2012}
{Owen}, J.~E. \& {Jackson}, A.~P., 2012.
\newblock {Planetary evaporation by UV \& X-ray radiation: basic
  hydrodynamics}, {\it \mnras\/}, {\bf 425}(4), 2931--2947.

\bibitem[{Owen} \& {Wu}(2013)]{Owen2013}
{Owen}, J.~E. \& {Wu}, Y., 2013.
\newblock {Kepler Planets: A Tale of Evaporation}, {\it \apj\/}, {\bf 775}(2),
  105.

\bibitem[{Owen} \& {Wu}(2017)]{Owen2017}
{Owen}, J.~E. \& {Wu}, Y., 2017.
\newblock {The Evaporation Valley in the Kepler Planets}, {\it \apj\/}, {\bf
  847}(1), 29.

\bibitem[{Pallavicini} et~al.(1981){Pallavicini}, {Golub}, {Rosner}, {Vaiana},
  {Ayres}, \& {Linsky}]{Pallavicini1981}
{Pallavicini}, R., {Golub}, L., {Rosner}, R., {Vaiana}, G.~S., {Ayres}, T., \&
  {Linsky}, J.~L., 1981.
\newblock {Relations among stellar X-ray emission observed from Einstein,
  stellar rotation and bolometric luminosity.}, {\it \apj\/}, {\bf 248},
  279--290.

\bibitem[{Panja} et~al.(2020){Panja}, {Cameron}, \& {Solanki}]{Panja-2020}
{Panja}, M., {Cameron}, R., \& {Solanki}, S.~K., 2020.
\newblock {3D Radiative MHD Simulations of Starspots}, {\it \apj\/}, {\bf
  893}(2), 113.

\bibitem[{Pannetier} et~al.(2020){Pannetier}, {Mourard}, {Berio}, {Cassaing},
  {Allouche}, {Anugu}, {Bailet}, {ten Brummelaar}, {Dejonghe}, {Gies}, {Jocou},
  {Kraus}, {Lacour}, {Lagarde}, {Le Bouquin}, {Lecron}, {Monnier}, {Nardetto},
  {Patru}, {Perraut}, {Petrov}, {Rousseau}, {Stee}, {Sturmann}, \&
  {Sturmann}]{Pannetier2020}
{Pannetier}, C., {Mourard}, D., {Berio}, P., {Cassaing}, F., {Allouche}, F.,
  {Anugu}, N., {Bailet}, C., {ten Brummelaar}, T., {Dejonghe}, J., {Gies}, D.,
  {Jocou}, L., {Kraus}, S., {Lacour}, S., {Lagarde}, S., {Le Bouquin}, J.~B.,
  {Lecron}, D., {Monnier}, J., {Nardetto}, N., {Patru}, F., {Perraut}, K.,
  {Petrov}, R., {Rousseau}, S., {Stee}, P., {Sturmann}, J., \& {Sturmann}, L.,
  2020.
\newblock {Progress of the CHARA/SPICA project}, in {\em Society of
  Photo-Optical Instrumentation Engineers (SPIE) Conference Series\/}, vol.
  11446 of {\bf Society of Photo-Optical Instrumentation Engineers (SPIE)
  Conference Series}, p. 114460T.

\bibitem[{Parker}(1955)]{Parker1955}
{Parker}, E.~N., 1955.
\newblock {Hydromagnetic Dynamo Models.}, {\it \apj\/}, {\bf 122}, 293.

\bibitem[{Parks} et~al.(2021){Parks}, {White}, {Baron}, {Monnier},
  {Kloppenborg}, {Henry}, {Schaefer}, {Che}, {Pedretti}, {Thureau}, {Zhao},
  {ten Brummelaar}, {McAlister}, {Ridgway}, {Turner}, {Sturmann}, \&
  {Sturmann}]{Parks2021}
{Parks}, J.~R., {White}, R.~J., {Baron}, F., {Monnier}, J.~D., {Kloppenborg},
  B., {Henry}, G.~W., {Schaefer}, G., {Che}, X., {Pedretti}, E., {Thureau}, N.,
  {Zhao}, M., {ten Brummelaar}, T., {McAlister}, H., {Ridgway}, S.~T.,
  {Turner}, N., {Sturmann}, J., \& {Sturmann}, L., 2021.
\newblock {Interferometric Imaging of {\ensuremath{\lambda}} Andromedae:
  Evidence of Starspots and Rotation}, {\it \apj\/}, {\bf 913}(1), 54.

\bibitem[{Passegger} et~al.(2018){Passegger}, {Reiners}, {Jeffers}, {Wende-von
  Berg}, {Sch{\"o}fer}, {Caballero}, {Schweitzer}, {Amado}, {B{\'e}jar},
  {Cort{\'e}s-Contreras}, {Hatzes}, {K{\"u}rster}, {Montes}, {Pedraz},
  {Quirrenbach}, {Ribas}, \& {Seifert}]{Passegger2018}
{Passegger}, V.~M., {Reiners}, A., {Jeffers}, S.~V., {Wende-von Berg}, S.,
  {Sch{\"o}fer}, P., {Caballero}, J.~A., {Schweitzer}, A., {Amado}, P.~J.,
  {B{\'e}jar}, V.~J.~S., {Cort{\'e}s-Contreras}, M., {Hatzes}, A.~P.,
  {K{\"u}rster}, M., {Montes}, D., {Pedraz}, S., {Quirrenbach}, A., {Ribas},
  I., \& {Seifert}, W., 2018.
\newblock {The CARMENES search for exoplanets around M dwarfs. Photospheric
  parameters of target stars from high-resolution spectroscopy}, {\it \aap\/},
  {\bf 615}, A6.

\bibitem[{Patra} et~al.(2020){Patra}, {Winn}, {Holman}, {Gillon}, {Burdanov},
  {Jehin}, {Delrez}, {Pozuelos}, {Barkaoui}, {Benkhaldoun}, {Narita}, {Fukui},
  {Kusakabe}, {Kawauchi}, {Terada}, {Bouma}, {Weinberg}, \&
  {Broome}]{Patra:2020}
{Patra}, K.~C., {Winn}, J.~N., {Holman}, M.~J., {Gillon}, M., {Burdanov}, A.,
  {Jehin}, E., {Delrez}, L., {Pozuelos}, F.~J., {Barkaoui}, K., {Benkhaldoun},
  Z., {Narita}, N., {Fukui}, A., {Kusakabe}, N., {Kawauchi}, K., {Terada}, Y.,
  {Bouma}, L.~G., {Weinberg}, N.~N., \& {Broome}, M., 2020.
\newblock {The Continuing Search for Evidence of Tidal Orbital Decay of Hot
  Jupiters}, {\it \aj\/}, {\bf 159}(4), 150.

\bibitem[{Peck} \& {Rast}(2015)]{peck2015}
{Peck}, C.~L. \& {Rast}, M.~P., 2015.
\newblock {Photometric Trends in the Visible Solar Continuum and Their
  Sensitivity to the Center-to-Limb Profile}, {\it \apj\/}, {\bf 808}(2), 192.

\bibitem[{Penza} et~al.(2004){Penza}, {Caccin}, \& {Del Moro}]{penza2004}
{Penza}, V., {Caccin}, B., \& {Del Moro}, D., 2004.
\newblock {The sensitivity of the C I 538.0 nm Fe I 537.9 nm and Ti II 538.1 nm
  lines to solar active regions}, {\it \aap\/}, {\bf 427}, 345--351.

\bibitem[{Pepper} et~al.(2007){Pepper}, {Pogge}, {DePoy}, {Marshall}, {Stanek},
  {Stutz}, {Poindexter}, {Siverd}, {O'Brien}, {Trueblood}, \&
  {Trueblood}]{Pepper2007}
{Pepper}, J., {Pogge}, R.~W., {DePoy}, D.~L., {Marshall}, J.~L., {Stanek},
  K.~Z., {Stutz}, A.~M., {Poindexter}, S., {Siverd}, R., {O'Brien}, T.~P.,
  {Trueblood}, M., \& {Trueblood}, P., 2007.
\newblock {The Kilodegree Extremely Little Telescope (KELT): A Small Robotic
  Telescope for Large-Area Synoptic Surveys}, {\it \pasp\/}, {\bf 119}(858),
  923--935.

\bibitem[{Petrie} et~al.(2021){Petrie}, {Criscuoli}, \& {Bertello}]{petrie2021}
{Petrie}, G., {Criscuoli}, S., \& {Bertello}, L., 2021.
\newblock {Solar Magnetism and Radiation}, in {\em Solar Physics and Solar
  Wind\/}, vol.~1, p.~83.

\bibitem[{Petrovay} \& {van Driel-Gesztelyi}(1997)]{Petrovay1997}
{Petrovay}, K. \& {van Driel-Gesztelyi}, L., 1997.
\newblock {Making Sense of Sunspot Decay. I. Parabolic Decay Law and
  Gnevyshev-Waldmeier Relation}, {\it \solphys\/}, {\bf 176}(2), 249--266.

\bibitem[{Pinhas} et~al.(2018){Pinhas}, {Rackham}, {Madhusudhan}, \&
  {Apai}]{Pinhas2018}
{Pinhas}, A., {Rackham}, B.~V., {Madhusudhan}, N., \& {Apai}, D., 2018.
\newblock {Retrieval of planetary and stellar properties in transmission
  spectroscopy with AURA}, {\it \mnras\/}, {\bf 480}(4), 5314--5331.

\bibitem[{Piskunov} et~al.(1990){Piskunov}, {Tuominen}, \&
  {Vilhu}]{Piskunov1990di}
{Piskunov}, N.~E., {Tuominen}, I., \& {Vilhu}, O., 1990.
\newblock {Surface imaging of late-type stars.}, {\it \aap\/}, {\bf 230},
  363--370.

\bibitem[{Pollacco} et~al.(2006){Pollacco}, {Skillen}, {Collier Cameron},
  {Christian}, {Hellier}, {Irwin}, {Lister}, {Street}, {West}, {Anderson},
  {Clarkson}, {Deeg}, {Enoch}, {Evans}, {Fitzsimmons}, {Haswell}, {Hodgkin},
  {Horne}, {Kane}, {Keenan}, {Maxted}, {Norton}, {Osborne}, {Parley}, {Ryans},
  {Smalley}, {Wheatley}, \& {Wilson}]{Pollacco2006}
{Pollacco}, D.~L., {Skillen}, I., {Collier Cameron}, A., {Christian}, D.~J.,
  {Hellier}, C., {Irwin}, J., {Lister}, T.~A., {Street}, R.~A., {West}, R.~G.,
  {Anderson}, D.~R., {Clarkson}, W.~I., {Deeg}, H., {Enoch}, B., {Evans}, A.,
  {Fitzsimmons}, A., {Haswell}, C.~A., {Hodgkin}, S., {Horne}, K., {Kane},
  S.~R., {Keenan}, F.~P., {Maxted}, P.~F.~L., {Norton}, A.~J., {Osborne}, J.,
  {Parley}, N.~R., {Ryans}, R.~S.~I., {Smalley}, B., {Wheatley}, P.~J., \&
  {Wilson}, D.~M., 2006.
\newblock {The WASP Project and the SuperWASP Cameras}, {\it \pasp\/}, {\bf
  118}(848), 1407--1418.

\bibitem[{Polyansky} et~al.(1997){Polyansky}, {Zobov}, {Viti}, {Tennyson},
  {Bernath}, \& {Wallace}]{Polyansky1997}
{Polyansky}, O.~L., {Zobov}, N.~F., {Viti}, S., {Tennyson}, J., {Bernath},
  P.~F., \& {Wallace}, L., 1997.
\newblock {K-Band Spectrum of Water in Sunspots}, {\it \apjl\/}, {\bf 489},
  L205.

\bibitem[{Pont} et~al.(2008){Pont}, {Knutson}, {Gilliland}, {Moutou}, \&
  {Charbonneau}]{Pont2008}
{Pont}, F., {Knutson}, H., {Gilliland}, R.~L., {Moutou}, C., \& {Charbonneau},
  D., 2008.
\newblock {Detection of atmospheric haze on an extrasolar planet: the 0.55-1.05
  {\ensuremath{\mu}}m transmission spectrum of HD 189733b with the
  HubbleSpaceTelescope}, {\it \mnras\/}, {\bf 385}(1), 109--118.

\bibitem[{Pont} et~al.(2011){Pont}, {Aigrain}, \& {Zucker}]{Pont-11}
{Pont}, F., {Aigrain}, S., \& {Zucker}, S., 2011.
\newblock {Reassessing the radial-velocity evidence for planets around
  CoRoT-7}, {\it \mnras\/}, {\bf 411}, 1953--1962.

\bibitem[{Pont} et~al.(2013){Pont}, {Sing}, {Gibson}, {Aigrain}, {Henry}, \&
  {Husnoo}]{Pont2013}
{Pont}, F., {Sing}, D.~K., {Gibson}, N.~P., {Aigrain}, S., {Henry}, G., \&
  {Husnoo}, N., 2013.
\newblock {The prevalence of dust on the exoplanet HD 189733b from Hubble and
  Spitzer observations}, {\it \mnras\/}, {\bf 432}(4), 2917--2944.

\bibitem[{Pr{\v s}a} \& {Zwitter}(2005)]{Prsa-05}
{Pr{\v s}a}, A. \& {Zwitter}, T., 2005.
\newblock {A Computational Guide to Physics of Eclipsing Binaries. I.
  Demonstrations and Perspectives}, {\it ApJ\/}, {\bf 628}, 426--438.

\bibitem[{Pr{\v s}a} et~al.(2016){Pr{\v s}a}, {Conroy}, {Horvat}, {Pablo},
  {Kochoska}, {Bloemen}, {Giammarco}, {Hambleton}, \& {Degroote}]{Prsa-16}
{Pr{\v s}a}, A., {Conroy}, K.~E., {Horvat}, M., {Pablo}, H., {Kochoska}, A.,
  {Bloemen}, S., {Giammarco}, J., {Hambleton}, K.~M., \& {Degroote}, P., 2016.
\newblock {Physics Of Eclipsing Binaries. II. Toward the Increased Model
  Fidelity}, {\it \apjs\/}, {\bf 227}, 29.

\bibitem[{Quintana} et~al.(2021){Quintana}, {Col{\'o}n}, {Mosby}, {Schlieder},
  {Supsinskas}, {Karburn}, {Dotson}, {Greene}, {Hedges}, {Apai}, {Barclay},
  {Christiansen}, {Espinoza}, {Mullally}, {Gilbert}, {Hoffman}, {Kostov},
  {Lewis}, {Foote}, {Mason}, {Youngblood}, {Morris}, {Newton}, {Pepper},
  {Rackham}, {Rowe}, \& {Stevenson}]{Quintana2021}
{Quintana}, E.~V., {Col{\'o}n}, K.~D., {Mosby}, G., {Schlieder}, J.~E.,
  {Supsinskas}, P., {Karburn}, J., {Dotson}, J.~L., {Greene}, T.~P., {Hedges},
  C., {Apai}, D., {Barclay}, T., {Christiansen}, J.~L., {Espinoza}, N.,
  {Mullally}, S.~E., {Gilbert}, E.~A., {Hoffman}, K., {Kostov}, V.~B., {Lewis},
  N.~K., {Foote}, T.~O., {Mason}, J., {Youngblood}, A., {Morris}, B.~M.,
  {Newton}, E.~R., {Pepper}, J., {Rackham}, B.~V., {Rowe}, J.~F., \&
  {Stevenson}, K., 2021.
\newblock {The Pandora SmallSat: Multiwavelength Characterization of Exoplanets
  and their Host Stars}, {\it arXiv e-prints\/}, p. arXiv:2108.06438.

\bibitem[{Rackham} et~al.(2017){Rackham}, {Espinoza}, {Apai},
  {L{\'o}pez-Morales}, {Jord{\'a}n}, {Osip}, {Lewis}, {Rodler}, {Fraine},
  {Morley}, \& {Fortney}]{Rackham2017}
{Rackham}, B., {Espinoza}, N., {Apai}, D., {L{\'o}pez-Morales}, M.,
  {Jord{\'a}n}, A., {Osip}, D.~J., {Lewis}, N.~K., {Rodler}, F., {Fraine},
  J.~D., {Morley}, C.~V., \& {Fortney}, J.~J., 2017.
\newblock {ACCESS I: An Optical Transmission Spectrum of GJ 1214b Reveals a
  Heterogeneous Stellar Photosphere}, {\it \apj\/}, {\bf 834}(2), 151.

\bibitem[{Rackham} et~al.(2018){Rackham}, {Apai}, \& {Giampapa}]{Rackham2018}
{Rackham}, B.~V., {Apai}, D., \& {Giampapa}, M.~S., 2018.
\newblock {The Transit Light Source Effect: False Spectral Features and
  Incorrect Densities for M-dwarf Transiting Planets}, {\it \apj\/}, {\bf
  853}(2), 122.

\bibitem[{Rackham} et~al.(2019){Rackham}, {Apai}, \& {Giampapa}]{Rackham2019}
{Rackham}, B.~V., {Apai}, D., \& {Giampapa}, M.~S., 2019.
\newblock {The Transit Light Source Effect. II. The Impact of Stellar
  Heterogeneity on Transmission Spectra of Planets Orbiting Broadly Sun-like
  Stars}, {\it \aj\/}, {\bf 157}(3), 96.

\bibitem[{Rackham} et~al.(2022){Rackham}, {Espinoza}, {Berdyugina}, {Korhonen},
  {MacDonald}, {Montet}, {Morris}, {Oshagh}, {Shapiro}, {Unruh}, {Quintana},
  {Zellem}, {Apai}, {Barclay}, {Barstow}, {Bruno}, {Carone}, {Casewell},
  {Cegla}, {Criscuoli}, {Fischer}, {Fournier}, {Giampapa}, {Giles}, {Iyer},
  {Kopp}, {Kostogryz}, {Krivova}, {Mallonn}, {McGruder}, {Molaverdikhani},
  {Newton}, {Panja}, {Peacock}, {Reardon}, {Roettenbacher}, {Scandariato},
  {Solanki}, {Stassun}, {Steiner}, {Stevenson}, {Tregloan-Reed}, {Valio},
  {Wedemeyer}, {Welbanks}, {Yu}, {Alam}, {Davenport}, {Deming}, {Dong},
  {Ducrot}, {Fisher}, {Gilbert}, {Kostov}, {L{\'o}pez-Morales}, {Line},
  {Mo{\v{c}}nik}, {Mullally}, {Paudel}, {Ribas}, \& {Valenti}]{Rackham2022}
{Rackham}, B.~V., {Espinoza}, N., {Berdyugina}, S.~V., {Korhonen}, H.,
  {MacDonald}, R.~J., {Montet}, B.~T., {Morris}, B.~M., {Oshagh}, M.,
  {Shapiro}, A.~I., {Unruh}, Y.~C., {Quintana}, E.~V., {Zellem}, R.~T., {Apai},
  D., {Barclay}, T., {Barstow}, J.~K., {Bruno}, G., {Carone}, L., {Casewell},
  S.~L., {Cegla}, H.~M., {Criscuoli}, S., {Fischer}, C., {Fournier}, D.,
  {Giampapa}, M.~S., {Giles}, H., {Iyer}, A., {Kopp}, G., {Kostogryz}, N.~M.,
  {Krivova}, N., {Mallonn}, M., {McGruder}, C., {Molaverdikhani}, K., {Newton},
  E.~R., {Panja}, M., {Peacock}, S., {Reardon}, K., {Roettenbacher}, R.~M.,
  {Scandariato}, G., {Solanki}, S., {Stassun}, K.~G., {Steiner}, O.,
  {Stevenson}, K.~B., {Tregloan-Reed}, J., {Valio}, A., {Wedemeyer}, S.,
  {Welbanks}, L., {Yu}, J., {Alam}, M.~K., {Davenport}, J. R.~A., {Deming}, D.,
  {Dong}, C., {Ducrot}, E., {Fisher}, C., {Gilbert}, E., {Kostov}, V.,
  {L{\'o}pez-Morales}, M., {Line}, M., {Mo{\v{c}}nik}, T., {Mullally}, S.,
  {Paudel}, R.~R., {Ribas}, I., \& {Valenti}, J.~A., 2022.
\newblock {Final Report for SAG 21: The Effect of Stellar Contamination on
  Space-based Transmission Spectroscopy}, {\it arXiv e-prints\/}, p.
  arXiv:2201.09905.

\bibitem[{Radick} et~al.(1998){Radick}, {Lockwood}, {Skiff}, \&
  {Baliunas}]{Radick1998}
{Radick}, R.~R., {Lockwood}, G.~W., {Skiff}, B.~A., \& {Baliunas}, S.~L., 1998.
\newblock {Patterns of Variation among Sun-like Stars}, {\it \apjs\/}, {\bf
  118}(1), 239--258.

\bibitem[{Radigan} et~al.(2012){Radigan}, {Jayawardhana}, {Lafreni{\`e}re},
  {Artigau}, {Marley}, \& {Saumon}]{Radigan2012}
{Radigan}, J., {Jayawardhana}, R., {Lafreni{\`e}re}, D., {Artigau}, {\'E}.,
  {Marley}, M., \& {Saumon}, D., 2012.
\newblock {Large-amplitude Variations of an L/T Transition Brown Dwarf:
  Multi-wavelength Observations of Patchy, High-contrast Cloud Features}, {\it
  \apj\/}, {\bf 750}(2), 105.

\bibitem[{Radigan} et~al.(2014){Radigan}, {Lafreni{\`e}re}, {Jayawardhana}, \&
  {Artigau}]{Radigan2014}
{Radigan}, J., {Lafreni{\`e}re}, D., {Jayawardhana}, R., \& {Artigau}, E.,
  2014.
\newblock {Strong Brightness Variations Signal Cloudy-to-clear Transition of
  Brown Dwarfs}, {\it \apj\/}, {\bf 793}(2), 75.

\bibitem[{Ragozzine} \& {Wolf}(2009)]{Ragozzine-09}
{Ragozzine}, D. \& {Wolf}, A.~S., 2009.
\newblock {Probing the Interiors of very Hot Jupiters Using Transit Light
  Curves}, {\it \apj\/}, {\bf 698}, 1778--1794.

\bibitem[{Rajpaul} et~al.(2015){Rajpaul}, {Aigrain}, {Osborne}, {Reece}, \&
  {Roberts}]{Rajpaul-15}
{Rajpaul}, V., {Aigrain}, S., {Osborne}, M.~A., {Reece}, S., \& {Roberts}, S.,
  2015.
\newblock {A Gaussian process framework for modelling stellar activity signals
  in radial velocity data}, {\it \mnras\/}, {\bf 452}, 2269--2291.

\bibitem[{Rast} et~al.(1999){Rast}, {Fox}, {Lin}, {Lites}, {Meisner}, \&
  {White}]{rast1999}
{Rast}, M.~P., {Fox}, P.~A., {Lin}, H., {Lites}, B.~W., {Meisner}, R.~W., \&
  {White}, O.~R., 1999.
\newblock {Bright rings around sunspots}, {\it \nat\/}, {\bf 401}(6754),
  678--679.

\bibitem[{Rathcke} et~al.(2021){Rathcke}, {MacDonald}, {Barstow}, {Goyal},
  {Lopez-Morales}, {Mendon{\c{c}}a}, {Sanz-Forcada}, {Henry}, {Sing}, {Alam},
  {Lewis}, {Chubb}, {Taylor}, {Nikolov}, \& {Buchhave}]{Rathcke2021}
{Rathcke}, A.~D., {MacDonald}, R.~J., {Barstow}, J.~K., {Goyal}, J.~M.,
  {Lopez-Morales}, M., {Mendon{\c{c}}a}, J.~M., {Sanz-Forcada}, J., {Henry},
  G.~W., {Sing}, D.~K., {Alam}, M.~K., {Lewis}, N.~K., {Chubb}, K.~L.,
  {Taylor}, J., {Nikolov}, N., \& {Buchhave}, L.~A., 2021.
\newblock {HST PanCET Program: A Complete Near-UV to Infrared Transmission
  Spectrum for the Hot Jupiter WASP-79b}, {\it arXiv e-prints\/}, p.
  arXiv:2104.10688.

\bibitem[{Rauer} et~al.(2014){Rauer}, {Catala}, {Aerts}, {Appourchaux}, {Benz},
  {Brandeker}, {Christensen-Dalsgaard}, {Deleuil}, {Gizon}, {Goupil},
  {G{\"u}del}, {Janot-Pacheco}, {Mas-Hesse}, {Pagano}, {Piotto}, {Pollacco},
  {Santos}, {Smith}, {Su{\'a}rez}, {Szab{\'o}}, {Udry}, {Adibekyan}, {Alibert},
  {Almenara}, {Amaro-Seoane}, {Eiff}, {Asplund}, {Antonello}, {Barnes},
  {Baudin}, {Belkacem}, {Bergemann}, {Bihain}, {Birch}, {Bonfils}, {Boisse},
  {Bonomo}, {Borsa}, {Brand{\~a}o}, {Brocato}, {Brun}, {Burleigh}, {Burston},
  {Cabrera}, {Cassisi}, {Chaplin}, {Charpinet}, {Chiappini}, {Church},
  {Csizmadia}, {Cunha}, {Damasso}, {Davies}, {Deeg}, {D{\'{\i}}az}, {Dreizler},
  {Dreyer}, {Eggenberger}, {Ehrenreich}, {Eigm{\"u}ller}, {Erikson}, {Farmer},
  {Feltzing}, {de Oliveira Fialho}, {Figueira}, {Forveille}, {Fridlund},
  {Garc{\'{\i}}a}, {Giommi}, {Giuffrida}, {Godolt}, {Gomes da Silva},
  {Granzer}, {Grenfell}, {Grotsch-Noels}, {G{\"u}nther}, {Haswell}, {Hatzes},
  {H{\'e}brard}, {Hekker}, {Helled}, {Heng}, {Jenkins}, {Johansen},
  {Khodachenko}, {Kislyakova}, {Kley}, {Kolb}, {Krivova}, {Kupka}, {Lammer},
  {Lanza}, {Lebreton}, {Magrin}, {Marcos-Arenal}, {Marrese}, {Marques},
  {Martins}, {Mathis}, {Mathur}, {Messina}, {Miglio}, {Montalban}, {Montalto},
  {Monteiro}, {Moradi}, {Moravveji}, {Mordasini}, {Morel}, {Mortier},
  {Nascimbeni}, {Nelson}, {Nielsen}, {Noack}, {Norton}, {Ofir}, {Oshagh},
  {Ouazzani}, {P{\'a}pics}, {Parro}, {Petit}, {Plez}, {Poretti}, {Quirrenbach},
  {Ragazzoni}, {Raimondo}, {Rainer}, {Reese}, {Redmer}, {Reffert},
  {Rojas-Ayala}, {Roxburgh}, {Salmon}, {Santerne}, {Schneider}, {Schou},
  {Schuh}, {Schunker}, {Silva-Valio}, {Silvotti}, {Skillen}, {Snellen}, {Sohl},
  {Sousa}, {Sozzetti}, {Stello}, {Strassmeier}, {{\v S}vanda}, {Szab{\'o}},
  {Tkachenko}, {Valencia}, {Van Grootel}, {Vauclair}, {Ventura}, {Wagner},
  {Walton}, {Weingrill}, {Werner}, {Wheatley}, \& {Zwintz}]{Rauer-14}
{Rauer}, H., {Catala}, C., {Aerts}, C., {Appourchaux}, T., {Benz}, W.,
  {Brandeker}, A., {Christensen-Dalsgaard}, J., {Deleuil}, M., {Gizon}, L.,
  {Goupil}, M.-J., {G{\"u}del}, M., {Janot-Pacheco}, E., {Mas-Hesse}, M.,
  {Pagano}, I., {Piotto}, G., {Pollacco}, D., {Santos}, {\.C}., {Smith}, A.,
  {Su{\'a}rez}, J.-C., {Szab{\'o}}, R., {Udry}, S., {Adibekyan}, V., {Alibert},
  Y., {Almenara}, J.-M., {Amaro-Seoane}, P., {Eiff}, M.~A.-v., {Asplund}, M.,
  {Antonello}, E., {Barnes}, S., {Baudin}, F., {Belkacem}, K., {Bergemann}, M.,
  {Bihain}, G., {Birch}, A.~C., {Bonfils}, X., {Boisse}, I., {Bonomo}, A.~S.,
  {Borsa}, F., {Brand{\~a}o}, I.~M., {Brocato}, E., {Brun}, S., {Burleigh}, M.,
  {Burston}, R., {Cabrera}, J., {Cassisi}, S., {Chaplin}, W., {Charpinet}, S.,
  {Chiappini}, C., {Church}, R.~P., {Csizmadia}, S., {Cunha}, M., {Damasso},
  M., {Davies}, M.~B., {Deeg}, H.~J., {D{\'{\i}}az}, R.~F., {Dreizler}, S.,
  {Dreyer}, C., {Eggenberger}, P., {Ehrenreich}, D., {Eigm{\"u}ller}, P.,
  {Erikson}, A., {Farmer}, R., {Feltzing}, S., {de Oliveira Fialho}, F.,
  {Figueira}, P., {Forveille}, T., {Fridlund}, M., {Garc{\'{\i}}a}, R.~A.,
  {Giommi}, P., {Giuffrida}, G., {Godolt}, M., {Gomes da Silva}, J., {Granzer},
  T., {Grenfell}, J.~L., {Grotsch-Noels}, A., {G{\"u}nther}, E., {Haswell},
  C.~A., {Hatzes}, A.~P., {H{\'e}brard}, G., {Hekker}, S., {Helled}, R.,
  {Heng}, K., {Jenkins}, J.~M., {Johansen}, A., {Khodachenko}, M.~L.,
  {Kislyakova}, K.~G., {Kley}, W., {Kolb}, U., {Krivova}, N., {Kupka}, F.,
  {Lammer}, H., {Lanza}, A.~F., {Lebreton}, Y., {Magrin}, D., {Marcos-Arenal},
  P., {Marrese}, P.~M., {Marques}, J.~P., {Martins}, J., {Mathis}, S.,
  {Mathur}, S., {Messina}, S., {Miglio}, A., {Montalban}, J., {Montalto}, M.,
  {Monteiro}, M.~J.~P.~F.~G., {Moradi}, H., {Moravveji}, E., {Mordasini}, C.,
  {Morel}, T., {Mortier}, A., {Nascimbeni}, V., {Nelson}, R.~P., {Nielsen},
  M.~B., {Noack}, L., {Norton}, A.~J., {Ofir}, A., {Oshagh}, M., {Ouazzani},
  R.-M., {P{\'a}pics}, P., {Parro}, V.~C., {Petit}, P., {Plez}, B., {Poretti},
  E., {Quirrenbach}, A., {Ragazzoni}, R., {Raimondo}, G., {Rainer}, M.,
  {Reese}, D.~R., {Redmer}, R., {Reffert}, S., {Rojas-Ayala}, B., {Roxburgh},
  I.~W., {Salmon}, S., {Santerne}, A., {Schneider}, J., {Schou}, J., {Schuh},
  S., {Schunker}, H., {Silva-Valio}, A., {Silvotti}, R., {Skillen}, I.,
  {Snellen}, I., {Sohl}, F., {Sousa}, S.~G., {Sozzetti}, A., {Stello}, D.,
  {Strassmeier}, K.~G., {{\v S}vanda}, M., {Szab{\'o}}, G.~M., {Tkachenko}, A.,
  {Valencia}, D., {Van Grootel}, V., {Vauclair}, S.~D., {Ventura}, P.,
  {Wagner}, F.~W., {Walton}, N.~A., {Weingrill}, J., {Werner}, S.~C.,
  {Wheatley}, P.~J., \& {Zwintz}, K., 2014.
\newblock {The PLATO 2.0 mission}, {\it Experimental Astronomy\/}, {\bf 38},
  249--330.

\bibitem[{Redfield} et~al.(2008){Redfield}, {Endl}, {Cochran}, \&
  {Koesterke}]{Redfield2008}
{Redfield}, S., {Endl}, M., {Cochran}, W.~D., \& {Koesterke}, L., 2008.
\newblock {Sodium Absorption from the Exoplanetary Atmosphere of HD 189733b
  Detected in the Optical Transmission Spectrum}, {\it \apjl\/}, {\bf 673}(1),
  L87.

\bibitem[{Reiners}(2012)]{Reiners2012LRSP}
{Reiners}, A., 2012.
\newblock {Observations of Cool-Star Magnetic Fields}, {\it Living Reviews in
  Solar Physics\/}, {\bf 9}(1), 1.

\bibitem[{Reinhold} et~al.(2013){Reinhold}, {Reiners}, \& {Basri}]{Reinhold-13}
{Reinhold}, T., {Reiners}, A., \& {Basri}, G., 2013.
\newblock {Rotation and differential rotation of active Kepler stars}, {\it
  \aap\/}, {\bf 560}, A4.

\bibitem[Reinhold et~al.(2019)Reinhold, Bell, Kuszlewicz, Hekker, \&
  Shapiro]{reinhold2019transition}
Reinhold, T., Bell, K.~J., Kuszlewicz, J., Hekker, S., \& Shapiro, A.~I., 2019.
\newblock Transition from spot to faculae domination-an alternate explanation
  for the dearth of intermediate kepler rotation periods, {\it Astronomy \&
  Astrophysics\/}, {\bf 621}, A21.

\bibitem[{Reinhold} et~al.(2021){Reinhold}, {Shapiro}, {Witzke}, {N{\`e}mec},
  {I{\c{s}}{\i}k}, \& {Solanki}]{Reinhold2021}
{Reinhold}, T., {Shapiro}, A.~I., {Witzke}, V., {N{\`e}mec}, N.-E.,
  {I{\c{s}}{\i}k}, E., \& {Solanki}, S.~K., 2021.
\newblock {Where Have All the Solar-like Stars Gone? Rotation Period
  Detectability at Various Inclinations and Metallicities}, {\it \apjl\/}, {\bf
  908}(2), L21.

\bibitem[{Rempel}(2012)]{Rempel2012}
{Rempel}, M., 2012.
\newblock {Numerical Sunspot Models: Robustness of Photospheric Velocity and
  Magnetic Field Structure}, {\it \apj\/}, {\bf 750}(1), 62.

\bibitem[{Rempel}(2020)]{rempel2020}
{Rempel}, M., 2020.
\newblock {On the Contribution of Quiet-Sun Magnetism to Solar Irradiance
  Variations: Constraints on Quiet-Sun Variability and Grand-minimum
  Scenarios}, {\it \apj\/}, {\bf 894}(2), 140.

\bibitem[{Rempel} et~al.(2009{\natexlab{a}}){Rempel}, {Sch{\"u}ssler},
  {Cameron}, \& {Kn{\"o}lker}]{Rempel:2009Sci}
{Rempel}, M., {Sch{\"u}ssler}, M., {Cameron}, R.~H., \& {Kn{\"o}lker}, M.,
  2009{\natexlab{a}}.
\newblock {Penumbral Structure and Outflows in Simulated Sunspots}, {\it
  Science\/}, {\bf 325}(5937), 171.

\bibitem[{Rempel} et~al.(2009{\natexlab{b}}){Rempel}, {Sch{\"u}ssler}, \&
  {Kn{\"o}lker}]{rempel09a}
{Rempel}, M., {Sch{\"u}ssler}, M., \& {Kn{\"o}lker}, M., 2009{\natexlab{b}}.
\newblock {Radiative Magnetohydrodynamic Simulation of Sunspot Structure}, {\it
  \apj\/}, {\bf 691}, 640--649.

\bibitem[{Rib{\'a}rik} et~al.(2003){Rib{\'a}rik}, {Ol{\'a}h}, \&
  {Strassmeier}]{Ribarik2003}
{Rib{\'a}rik}, G., {Ol{\'a}h}, K., \& {Strassmeier}, K.~G., 2003.
\newblock {Time-series photometric spot modelling VI. A new computer code and
  its application to 23 years of photometry of the active giant IM Pegasi},
  {\it Astronomische Nachrichten\/}, {\bf 324}(3), 202--214.

\bibitem[{Ricker} et~al.(2014){Ricker}, {Winn}, {Vanderspek}, {Latham},
  {Bakos}, {Bean}, {Berta-Thompson}, {Brown}, {Buchhave}, {Butler}, {Butler},
  {Chaplin}, {Charbonneau}, {Christensen-Dalsgaard}, {Clampin}, {Deming},
  {Doty}, {De Lee}, {Dressing}, {Dunham}, {Endl}, {Fressin}, {Ge}, {Henning},
  {Holman}, {Howard}, {Ida}, {Jenkins}, {Jernigan}, {Johnson}, {Kaltenegger},
  {Kawai}, {Kjeldsen}, {Laughlin}, {Levine}, {Lin}, {Lissauer}, {MacQueen},
  {Marcy}, {McCullough}, {Morton}, {Narita}, {Paegert}, {Palle}, {Pepe},
  {Pepper}, {Quirrenbach}, {Rinehart}, {Sasselov}, {Sato}, {Seager},
  {Sozzetti}, {Stassun}, {Sullivan}, {Szentgyorgyi}, {Torres}, {Udry}, \&
  {Villasenor}]{TESS}
{Ricker}, G.~R., {Winn}, J.~N., {Vanderspek}, R., {Latham}, D.~W., {Bakos},
  G.~{\'A}., {Bean}, J.~L., {Berta-Thompson}, Z.~K., {Brown}, T.~M.,
  {Buchhave}, L., {Butler}, N.~R., {Butler}, R.~P., {Chaplin}, W.~J.,
  {Charbonneau}, D., {Christensen-Dalsgaard}, J., {Clampin}, M., {Deming}, D.,
  {Doty}, J., {De Lee}, N., {Dressing}, C., {Dunham}, E.~W., {Endl}, M.,
  {Fressin}, F., {Ge}, J., {Henning}, T., {Holman}, M.~J., {Howard}, A.~W.,
  {Ida}, S., {Jenkins}, J., {Jernigan}, G., {Johnson}, J.~A., {Kaltenegger},
  L., {Kawai}, N., {Kjeldsen}, H., {Laughlin}, G., {Levine}, A.~M., {Lin}, D.,
  {Lissauer}, J.~J., {MacQueen}, P., {Marcy}, G., {McCullough}, P.~R.,
  {Morton}, T.~D., {Narita}, N., {Paegert}, M., {Palle}, E., {Pepe}, F.,
  {Pepper}, J., {Quirrenbach}, A., {Rinehart}, S.~A., {Sasselov}, D., {Sato},
  B., {Seager}, S., {Sozzetti}, A., {Stassun}, K.~G., {Sullivan}, P.,
  {Szentgyorgyi}, A., {Torres}, G., {Udry}, S., \& {Villasenor}, J., 2014.
\newblock {Transiting Exoplanet Survey Satellite (TESS)}, in {\em Space
  Telescopes and Instrumentation 2014: Optical, Infrared, and Millimeter
  Wave\/}, vol. 9143 of {\bf Society of Photo-Optical Instrumentation Engineers
  (SPIE) Conference Series}, p. 914320.

\bibitem[{Ricker} et~al.(2015){Ricker}, {Winn}, {Vanderspek}, {Latham},
  {Bakos}, {Bean}, {Berta-Thompson}, {Brown}, {Buchhave}, {Butler}, {Butler},
  {Chaplin}, {Charbonneau}, {Christensen-Dalsgaard}, {Clampin}, {Deming},
  {Doty}, {De Lee}, {Dressing}, {Dunham}, {Endl}, {Fressin}, {Ge}, {Henning},
  {Holman}, {Howard}, {Ida}, {Jenkins}, {Jernigan}, {Johnson}, {Kaltenegger},
  {Kawai}, {Kjeldsen}, {Laughlin}, {Levine}, {Lin}, {Lissauer}, {MacQueen},
  {Marcy}, {McCullough}, {Morton}, {Narita}, {Paegert}, {Palle}, {Pepe},
  {Pepper}, {Quirrenbach}, {Rinehart}, {Sasselov}, {Sato}, {Seager},
  {Sozzetti}, {Stassun}, {Sullivan}, {Szentgyorgyi}, {Torres}, {Udry}, \&
  {Villasenor}]{Ricker2015}
{Ricker}, G.~R., {Winn}, J.~N., {Vanderspek}, R., {Latham}, D.~W., {Bakos},
  G.~{\'A}., {Bean}, J.~L., {Berta-Thompson}, Z.~K., {Brown}, T.~M.,
  {Buchhave}, L., {Butler}, N.~R., {Butler}, R.~P., {Chaplin}, W.~J.,
  {Charbonneau}, D., {Christensen-Dalsgaard}, J., {Clampin}, M., {Deming}, D.,
  {Doty}, J., {De Lee}, N., {Dressing}, C., {Dunham}, E.~W., {Endl}, M.,
  {Fressin}, F., {Ge}, J., {Henning}, T., {Holman}, M.~J., {Howard}, A.~W.,
  {Ida}, S., {Jenkins}, J.~M., {Jernigan}, G., {Johnson}, J.~A., {Kaltenegger},
  L., {Kawai}, N., {Kjeldsen}, H., {Laughlin}, G., {Levine}, A.~M., {Lin}, D.,
  {Lissauer}, J.~J., {MacQueen}, P., {Marcy}, G., {McCullough}, P.~R.,
  {Morton}, T.~D., {Narita}, N., {Paegert}, M., {Palle}, E., {Pepe}, F.,
  {Pepper}, J., {Quirrenbach}, A., {Rinehart}, S.~A., {Sasselov}, D., {Sato},
  B., {Seager}, S., {Sozzetti}, A., {Stassun}, K.~G., {Sullivan}, P.,
  {Szentgyorgyi}, A., {Torres}, G., {Udry}, S., \& {Villasenor}, J., 2015.
\newblock {Transiting Exoplanet Survey Satellite (TESS)}, {\it Journal of
  Astronomical Telescopes, Instruments, and Systems\/}, {\bf 1}(1), 014003.

\bibitem[{Riethm{\"u}ller} et~al.(2008){Riethm{\"u}ller}, {Solanki},
  {Zakharov}, \& {Gandorfer}]{Riethmueller:2008}
{Riethm{\"u}ller}, T.~L., {Solanki}, S.~K., {Zakharov}, V., \& {Gandorfer}, A.,
  2008.
\newblock {Brightness, distribution, and evolution of sunspot umbral dots},
  {\it \aap\/}, {\bf 492}(1), 233--243.

\bibitem[{Rimmele} et~al.(2020){Rimmele}, {Warner}, {Keil}, {Goode},
  {Kn{\"o}lker}, {Kuhn}, {Rosner}, {McMullin}, {Casini}, {Lin}, {W{\"o}ger},
  {von der L{\"u}he}, {Tritschler}, {Davey}, {de Wijn}, {Elmore}, {Fehlmann},
  {Harrington}, {Jaeggli}, {Rast}, {Schad}, {Schmidt}, {Mathioudakis},
  {Mickey}, {Anan}, {Beck}, {Marshall}, {Jeffers}, {Oschmann}, {Beard},
  {Berst}, {Cowan}, {Craig}, {Cross}, {Cummings}, {Donnelly}, {de Vanssay},
  {Eigenbrot}, {Ferayorni}, {Foster}, {Galapon}, {Gedrites}, {Gonzales},
  {Goodrich}, {Gregory}, {Guzman}, {Guzzo}, {Hegwer}, {Hubbard}, {Hubbard},
  {Johansson}, {Johnson}, {Liang}, {Liang}, {McQuillen}, {Mayer}, {Newman},
  {Onodera}, {Phelps}, {Puentes}, {Richards}, {Rimmele}, {Sekulic}, {Shimko},
  {Simison}, {Smith}, {Starman}, {Sueoka}, {Summers}, {Szabo}, {Szabo},
  {Wampler}, {Williams}, \& {White}]{DKIST}
{Rimmele}, T.~R., {Warner}, M., {Keil}, S.~L., {Goode}, P.~R., {Kn{\"o}lker},
  M., {Kuhn}, J.~R., {Rosner}, R.~R., {McMullin}, J.~P., {Casini}, R., {Lin},
  H., {W{\"o}ger}, F., {von der L{\"u}he}, O., {Tritschler}, A., {Davey}, A.,
  {de Wijn}, A., {Elmore}, D.~F., {Fehlmann}, A., {Harrington}, D.~M.,
  {Jaeggli}, S.~A., {Rast}, M.~P., {Schad}, T.~A., {Schmidt}, W.,
  {Mathioudakis}, M., {Mickey}, D.~L., {Anan}, T., {Beck}, C., {Marshall},
  H.~K., {Jeffers}, P.~F., {Oschmann}, J.~M., {Beard}, A., {Berst}, D.~C.,
  {Cowan}, B.~A., {Craig}, S.~C., {Cross}, E., {Cummings}, B.~K., {Donnelly},
  C., {de Vanssay}, J.-B., {Eigenbrot}, A.~D., {Ferayorni}, A., {Foster}, C.,
  {Galapon}, C.~A., {Gedrites}, C., {Gonzales}, K., {Goodrich}, B.~D.,
  {Gregory}, B.~S., {Guzman}, S.~S., {Guzzo}, S., {Hegwer}, S., {Hubbard},
  R.~P., {Hubbard}, J.~R., {Johansson}, E.~M., {Johnson}, L.~C., {Liang}, C.,
  {Liang}, M., {McQuillen}, I., {Mayer}, C., {Newman}, K., {Onodera}, B.,
  {Phelps}, L., {Puentes}, M.~M., {Richards}, C., {Rimmele}, L.~M., {Sekulic},
  P., {Shimko}, S.~R., {Simison}, B.~E., {Smith}, B., {Starman}, E., {Sueoka},
  S.~R., {Summers}, R.~T., {Szabo}, A., {Szabo}, L., {Wampler}, S.~B.,
  {Williams}, T.~R., \& {White}, C., 2020.
\newblock {The Daniel K. Inouye Solar Telescope - Observatory Overview}, {\it
  \solphys\/}, {\bf 295}(12), 172.

\bibitem[{Robertson} et~al.(2014){Robertson}, {Mahadevan}, {Endl}, \&
  {Roy}]{Robertson-14}
{Robertson}, P., {Mahadevan}, S., {Endl}, M., \& {Roy}, A., 2014.
\newblock {Stellar activity masquerading as planets in the habitable zone of
  the M dwarf Gliese 581}, {\it Science\/}, {\bf 345}, 440--444.

\bibitem[{Robertson} et~al.(2015){Robertson}, {Roy}, \&
  {Mahadevan}]{Robertson2015}
{Robertson}, P., {Roy}, A., \& {Mahadevan}, S., 2015.
\newblock {Stellar Activity Mimics a Habitable-zone Planet around Kapteyn's
  Star}, {\it \apjl\/}, {\bf 805}(2), L22.

\bibitem[{Robertson} et~al.(2020){Robertson}, {Stefansson}, {Mahadevan},
  {Endl}, {Cochran}, {Beard}, {Bender}, {Diddams}, {Duong}, {Ford}, {Fredrick},
  {Halverson}, {Hearty}, {Holcomb}, {Juan}, {Kanodia}, {Lubin}, {Metcalf},
  {Monson}, {Ninan}, {Palafoutas}, {Ramsey}, {Roy}, {Schwab}, {Terrien}, \&
  {Wright}]{2020ApJ...897..125R}
{Robertson}, P., {Stefansson}, G., {Mahadevan}, S., {Endl}, M., {Cochran},
  W.~D., {Beard}, C., {Bender}, C.~F., {Diddams}, S.~A., {Duong}, N., {Ford},
  E.~B., {Fredrick}, C., {Halverson}, S., {Hearty}, F., {Holcomb}, R., {Juan},
  L., {Kanodia}, S., {Lubin}, J., {Metcalf}, A.~J., {Monson}, A., {Ninan},
  J.~P., {Palafoutas}, J., {Ramsey}, L.~W., {Roy}, A., {Schwab}, C., {Terrien},
  R.~C., \& {Wright}, J.~T., 2020.
\newblock {Persistent Starspot Signals on M Dwarfs: Multiwavelength Doppler
  Observations with the Habitable-zone Planet Finder and Keck/HIRES}, {\it
  \apj\/}, {\bf 897}(2), 125.

\bibitem[{Roettenbacher} \& {Kane}(2017)]{RoettenbacherKane2017}
{Roettenbacher}, R.~M. \& {Kane}, S.~R., 2017.
\newblock {The Stellar Activity of TRAPPIST-1 and Consequences for the
  Planetary Atmospheres}, {\it \apj\/}, {\bf 851}(2), 77.

\bibitem[{Roettenbacher} \& {Vida}(2018)]{Roettenbacher2018}
{Roettenbacher}, R.~M. \& {Vida}, K., 2018.
\newblock {The Connection between Starspots and Flares on Main-sequence Kepler
  Stars}, {\it \apj\/}, {\bf 868}(1), 3.

\bibitem[{Roettenbacher} et~al.(2013){Roettenbacher}, {Monnier}, {Harmon},
  {Barclay}, \& {Still}]{Roettenbacher2013}
{Roettenbacher}, R.~M., {Monnier}, J.~D., {Harmon}, R.~O., {Barclay}, T., \&
  {Still}, M., 2013.
\newblock {Imaging Starspot Evolution on Kepler Target KIC 5110407 Using
  Light-Curve Inversion}, {\it \apj\/}, {\bf 767}(1), 60.

\bibitem[{Roettenbacher} et~al.(2016){Roettenbacher}, {Monnier}, {Korhonen},
  {Aarnio}, {Baron}, {Che}, {Harmon}, {K{\H{o}}v{\'a}ri}, {Kraus}, {Schaefer},
  {Torres}, {Zhao}, {Ten Brummelaar}, {Sturmann}, \&
  {Sturmann}]{Roettenbacher2016}
{Roettenbacher}, R.~M., {Monnier}, J.~D., {Korhonen}, H., {Aarnio}, A.~N.,
  {Baron}, F., {Che}, X., {Harmon}, R.~O., {K{\H{o}}v{\'a}ri}, Z., {Kraus}, S.,
  {Schaefer}, G.~H., {Torres}, G., {Zhao}, M., {Ten Brummelaar}, T.~A.,
  {Sturmann}, J., \& {Sturmann}, L., 2016.
\newblock {No Sun-like dynamo on the active star {\ensuremath{\zeta}}
  Andromedae from starspot asymmetry}, {\it \nat\/}, {\bf 533}(7602), 217--220.

\bibitem[{Roettenbacher} et~al.(2017){Roettenbacher}, {Monnier}, {Korhonen},
  {Harmon}, {Baron}, {Hackman}, {Henry}, {Schaefer}, {Strassmeier}, {Weber}, \&
  {ten Brummelaar}]{Roettenbacher2017}
{Roettenbacher}, R.~M., {Monnier}, J.~D., {Korhonen}, H., {Harmon}, R.~O.,
  {Baron}, F., {Hackman}, T., {Henry}, G.~W., {Schaefer}, G.~H., {Strassmeier},
  K.~G., {Weber}, M., \& {ten Brummelaar}, T.~A., 2017.
\newblock {Contemporaneous Imaging Comparisons of the Spotted Giant
  {\ensuremath{\sigma}} Geminorum Using Interferometric, Spectroscopic, and
  Photometric Data}, {\it \apj\/}, {\bf 849}(2), 120.

\bibitem[{Roettenbacher} et~al.(2022){Roettenbacher}, {Cabot}, {Fischer},
  {Monnier}, {Henry}, {Harmon}, {Korhonen}, {Brewer}, {Llama}, {Petersburg},
  {Zhao}, {Kraus}, {Le Bouquin}, {Anugu}, {Davies}, {Gardner}, {Lanthermann},
  {Schaefer}, {Setterholm}, {Clark}, {Jorstad}, {Kuehn}, \&
  {Levine}]{Roettenbacher2022}
{Roettenbacher}, R.~M., {Cabot}, S. H.~C., {Fischer}, D.~A., {Monnier}, J.~D.,
  {Henry}, G.~W., {Harmon}, R.~O., {Korhonen}, H., {Brewer}, J.~M., {Llama},
  J., {Petersburg}, R.~R., {Zhao}, L.~L., {Kraus}, S., {Le Bouquin}, J.-B.,
  {Anugu}, N., {Davies}, C.~L., {Gardner}, T., {Lanthermann}, C., {Schaefer},
  G., {Setterholm}, B., {Clark}, C.~A., {Jorstad}, S.~G., {Kuehn}, K., \&
  {Levine}, S., 2022.
\newblock {EXPRES. III. Revealing the Stellar Activity Radial Velocity
  Signature of {\ensuremath{\epsilon}} Eridani with Photometry and
  Interferometry}, {\it \aj\/}, {\bf 163}(1), 19.

\bibitem[{Romano} et~al.(2012){Romano}, {Berrilli}, {Criscuoli}, {Del Moro},
  {Ermolli}, {Giorgi}, {Viticchi{\'e}}, \& {Zuccarello}]{romano2012}
{Romano}, P., {Berrilli}, F., {Criscuoli}, S., {Del Moro}, D., {Ermolli}, I.,
  {Giorgi}, F., {Viticchi{\'e}}, B., \& {Zuccarello}, F., 2012.
\newblock {A Comparative Analysis of Photospheric Bright Points in an Active
  Region and in the Quiet Sun}, {\it \solphys\/}, {\bf 280}(2), 407--416.

\bibitem[{Rosich} et~al.(2020){Rosich}, {Herrero}, {Mallonn}, {Ribas},
  {Morales}, {Perger}, {Anglada-Escud{\'e}}, \& {Granzer}]{Rosich-20}
{Rosich}, A., {Herrero}, E., {Mallonn}, M., {Ribas}, I., {Morales}, J.~C.,
  {Perger}, M., {Anglada-Escud{\'e}}, G., \& {Granzer}, T., 2020.
\newblock {Correcting for chromatic stellar activity effects in transits with
  multiband photometric monitoring: application to WASP-52}, {\it \aap\/}, {\bf
  641}, A82.

\bibitem[{Rutten}(2007)]{rutten2007}
{Rutten}, R.~J., 2007.
\newblock {Observing the Solar Chromosphere}, in {\em The Physics of
  Chromospheric Plasmas\/}, vol. 368 of {\bf Astronomical Society of the
  Pacific Conference Series}, p.~27.

\bibitem[{Salhab} et~al.(2018){Salhab}, {Steiner}, {Berdyugina}, {Freytag},
  {Rajaguru}, \& {Steffen}]{co5bold_salhab}
{Salhab}, R.~G., {Steiner}, O., {Berdyugina}, S.~V., {Freytag}, B., {Rajaguru},
  S.~P., \& {Steffen}, M., 2018.
\newblock {Simulation of the small-scale magnetism in main-sequence stellar
  atmospheres}, {\it \aap\/}, {\bf 614}, A78.

\bibitem[{Samadi} et~al.(2010{\natexlab{a}}){Samadi}, {Ludwig}, {Belkacem},
  {Goupil}, {Benomar}, {Mosser}, {Dupret}, {Baudin}, {Appourchaux}, \&
  {Michel}]{samadi2010b}
{Samadi}, R., {Ludwig}, H.~G., {Belkacem}, K., {Goupil}, M.~J., {Benomar}, O.,
  {Mosser}, B., {Dupret}, M.~A., {Baudin}, F., {Appourchaux}, T., \& {Michel},
  E., 2010{\natexlab{a}}.
\newblock {The CoRoT target HD 49933 . II. Comparison of theoretical mode
  amplitudes with observations}, {\it \aap\/}, {\bf 509}, A16.

\bibitem[{Samadi} et~al.(2010{\natexlab{b}}){Samadi}, {Ludwig}, {Belkacem},
  {Goupil}, \& {Dupret}]{samadi2010a}
{Samadi}, R., {Ludwig}, H.~G., {Belkacem}, K., {Goupil}, M.~J., \& {Dupret},
  M.~A., 2010{\natexlab{b}}.
\newblock {The CoRoT target HD 49933 . I. Effect of the metal abundance on the
  mode excitation rates}, {\it \aap\/}, {\bf 509}, A15.

\bibitem[{Samadi} et~al.(2012){Samadi}, {Belkacem}, {Dupret}, {Ludwig},
  {Baudin}, {Caffau}, {Goupil}, \& {Barban}]{samadi2012a}
{Samadi}, R., {Belkacem}, K., {Dupret}, M.~A., {Ludwig}, H.~G., {Baudin}, F.,
  {Caffau}, E., {Goupil}, M.~J., \& {Barban}, C., 2012.
\newblock {Amplitudes of solar-like oscillations in red giant stars. Evidence
  for non-adiabatic effects using CoRoT observations}, {\it \aap\/}, {\bf 543},
  A120.

\bibitem[{S{\'a}nchez Cuberes} et~al.(2002){S{\'a}nchez Cuberes},
  {V{\'a}zquez}, {Bonet}, \& {Sobotka}]{sanchez2002}
{S{\'a}nchez Cuberes}, M., {V{\'a}zquez}, M., {Bonet}, J.~A., \& {Sobotka}, M.,
  2002.
\newblock {Infrared Photometry of Solar Photospheric Structures. II.
  Center-to-Limb Variation of Active Regions}, {\it \apj\/}, {\bf 570}(2),
  886--899.

\bibitem[{S{\'a}nchez Cuberes} et~al.(2003){S{\'a}nchez Cuberes},
  {V{\'a}zquez}, {Bonet}, \& {Sobotka}]{SC:2003}
{S{\'a}nchez Cuberes}, M., {V{\'a}zquez}, M., {Bonet}, J.~A., \& {Sobotka}, M.,
  2003.
\newblock {Centre-to-limb variation of solar granulation in the infrared}, {\it
  \aap\/}, {\bf 397}, 1075--1081.

\bibitem[{Sanchis-Ojeda} \& {Winn}(2011)]{Sanchis-Ojeda-11b}
{Sanchis-Ojeda}, R. \& {Winn}, J.~N., 2011.
\newblock {Starspots, Spin-Orbit Misalignment, and Active Latitudes in the
  HAT-P-11 Exoplanetary System}, {\it \apj\/}, {\bf 743}, 61.

\bibitem[{Sanchis-Ojeda} et~al.(2011){Sanchis-Ojeda}, {Winn}, {Holman},
  {Carter}, {Osip}, \& {Fuentes}]{Sanchis-Ojeda-11a}
{Sanchis-Ojeda}, R., {Winn}, J.~N., {Holman}, M.~J., {Carter}, J.~A., {Osip},
  D.~J., \& {Fuentes}, C.~I., 2011.
\newblock {Starspots and Spin-orbit Alignment in the WASP-4 Exoplanetary
  System}, {\it \apj\/}, {\bf 733}, 127--+.

\bibitem[{Sanchis-Ojeda} et~al.(2012){Sanchis-Ojeda}, {Fabrycky}, {Winn},
  {Barclay}, {Clarke}, {Ford}, {Fortney}, {Geary}, {Holman}, {Howard},
  {Jenkins}, {Koch}, {Lissauer}, {Marcy}, {Mullaly}, {Ragozzine}, {Seader},
  {Still}, \& {Thompson}]{Sanchis-Ojeda-12}
{Sanchis-Ojeda}, R., {Fabrycky}, D.~C., {Winn}, J.~N., {Barclay}, T., {Clarke},
  B.~D., {Ford}, E.~B., {Fortney}, J.~J., {Geary}, J.~C., {Holman}, M.~J.,
  {Howard}, A.~W., {Jenkins}, J.~M., {Koch}, D., {Lissauer}, J.~J., {Marcy},
  J.~W., {Mullaly}, F., {Ragozzine}, D., {Seader}, S.~E., {Still}, M., \&
  {Thompson}, S.~E., 2012.
\newblock {Alignment of the stellar spin with the orbits of a three-planet
  system.}, {\it \nat\/}, {\bf 487}, 449--453.

\bibitem[{Sanchis-Ojeda} et~al.(2013){Sanchis-Ojeda}, {Winn}, {Marcy},
  {Howard}, {Isaacson}, {Johnson}, {Torres}, {Albrecht}, {Campante}, {Chaplin},
  {Davies}, {Lund}, {Carter}, {Dawson}, {Buchhave}, {Everett}, {Fischer},
  {Geary}, {Gilliland}, {Horch}, {Howell}, \& {Latham}]{Sanchis-Ojeda-13}
{Sanchis-Ojeda}, R., {Winn}, J.~N., {Marcy}, G.~W., {Howard}, A.~W.,
  {Isaacson}, H., {Johnson}, J.~A., {Torres}, G., {Albrecht}, S., {Campante},
  T.~L., {Chaplin}, W.~J., {Davies}, G.~R., {Lund}, M.~N., {Carter}, J.~A.,
  {Dawson}, R.~I., {Buchhave}, L.~A., {Everett}, M.~E., {Fischer}, D.~A.,
  {Geary}, J.~C., {Gilliland}, R.~L., {Horch}, E.~P., {Howell}, S.~B., \&
  {Latham}, D.~W., 2013.
\newblock {Kepler-63b: A Giant Planet in a Polar Orbit around a Young Sun-like
  Star}, {\it \apj\/}, {\bf 775}, 54.

\bibitem[{Santos} et~al.(2013){Santos}, {Sousa}, {Mortier}, {Neves},
  {Adibekyan}, {Tsantaki}, {Delgado Mena}, {Bonfils}, {Israelian}, {Mayor}, \&
  {Udry}]{Santos2013}
{Santos}, N.~C., {Sousa}, S.~G., {Mortier}, A., {Neves}, V., {Adibekyan}, V.,
  {Tsantaki}, M., {Delgado Mena}, E., {Bonfils}, X., {Israelian}, G., {Mayor},
  M., \& {Udry}, S., 2013.
\newblock {SWEET-Cat: A catalogue of parameters for Stars With ExoplanETs. I.
  New atmospheric parameters and masses for 48 stars with planets}, {\it
  \aap\/}, {\bf 556}, A150.

\bibitem[{Sarkar} et~al.(2018){Sarkar}, {Argyriou}, {Vandenbussche},
  {Papageorgiou}, \& {Pascale}]{Sarkar:2018}
{Sarkar}, S., {Argyriou}, I., {Vandenbussche}, B., {Papageorgiou}, A., \&
  {Pascale}, E., 2018.
\newblock {Stellar pulsation and granulation as noise sources in exoplanet
  transit spectroscopy in the ARIEL space mission}, {\it \mnras\/}, {\bf
  481}(3), 2871--2877.

\bibitem[{Sayeed} et~al.(2021){Sayeed}, {Huber}, {Wheeler}, \&
  {Ness}]{Sayeed:2021}
{Sayeed}, M., {Huber}, D., {Wheeler}, A., \& {Ness}, M.~K., 2021.
\newblock {Data-Driven Inference of Stellar Surface Gravities for Cool Stars
  from Photometric Light Curves}, in {\em American Astronomical Society Meeting
  Abstracts\/}, vol.~53 of {\bf American Astronomical Society Meeting
  Abstracts}, p. 339.17.

\bibitem[{Scandariato} et~al.(2017){Scandariato}, {Nascimbeni}, {Lanza},
  {Pagano}, {Zanmar Sanchez}, \& {Leto}]{Scandariato2017}
{Scandariato}, G., {Nascimbeni}, V., {Lanza}, A.~F., {Pagano}, I., {Zanmar
  Sanchez}, R., \& {Leto}, G., 2017.
\newblock {TOSC: an algorithm for the tomography of spotted transit chords},
  {\it \aap\/}, {\bf 606}, A134.

\bibitem[{Schad}(2014)]{2014-Schad}
{Schad}, T.~A., 2014.
\newblock {On the Collective Magnetic Field Strength and Vector Structure of
  Dark Umbral Cores Measured by the Hinode Spectropolarimeter}, {\it
  \solphys\/}, {\bf 289}(5), 1477--1498.

\bibitem[{Schanche} et~al.(2022){Schanche}, {Pozuelos}, {G{\"u}nther}, {Wells},
  {Burgasser}, {Chinchilla}, {Delrez}, {Ducrot}, {Garcia}, {G{\'o}mez Maqueo
  Chew}, {Jofr{\'e}}, {Rackham}, {Sebastian}, {Stassun}, {Stern}, {Timmermans},
  {Barkaoui}, {Belinski}, {Benkhaldoun}, {Benz}, {Bieryla}, {Bouchy},
  {Burdanov}, {Charbonneau}, {Christiansen}, {Collins}, {Demory},
  {D{\'e}vora-Pajares}, {de Wit}, {Dragomir}, {Dransfield}, {Furlan},
  {Ghachoui}, {Gillon}, {Gnilka}, {G{\'o}mez-Mu{\~n}oz}, {Guerrero}, {Harris},
  {Heng}, {Henze}, {Hesse}, {Howell}, {Jehin}, {Jenkins}, {Jensen}, {Kunimoto},
  {Latham}, {Lester}, {McLeod}, {Mireles}, {Murray}, {Niraula}, {Pedersen},
  {Queloz}, {Quintana}, {Ricker}, {Rudat}, {Sabin}, {Safonov},
  {Schroffenegger}, {Scott}, {Seager}, {Strakhov}, {Triaud}, {Vanderspek},
  {Vezie}, \& {Winn}]{Schanche2022}
{Schanche}, N., {Pozuelos}, F.~J., {G{\"u}nther}, M.~N., {Wells}, R.~D.,
  {Burgasser}, A.~J., {Chinchilla}, P., {Delrez}, L., {Ducrot}, E., {Garcia},
  L.~J., {G{\'o}mez Maqueo Chew}, Y., {Jofr{\'e}}, E., {Rackham}, B.~V.,
  {Sebastian}, D., {Stassun}, K.~G., {Stern}, D., {Timmermans}, M., {Barkaoui},
  K., {Belinski}, A., {Benkhaldoun}, Z., {Benz}, W., {Bieryla}, A., {Bouchy},
  F., {Burdanov}, A., {Charbonneau}, D., {Christiansen}, J.~L., {Collins},
  K.~A., {Demory}, B.~O., {D{\'e}vora-Pajares}, M., {de Wit}, J., {Dragomir},
  D., {Dransfield}, G., {Furlan}, E., {Ghachoui}, M., {Gillon}, M., {Gnilka},
  C., {G{\'o}mez-Mu{\~n}oz}, M.~A., {Guerrero}, N., {Harris}, M., {Heng}, K.,
  {Henze}, C.~E., {Hesse}, K., {Howell}, S.~B., {Jehin}, E., {Jenkins}, J.,
  {Jensen}, E.~L.~N., {Kunimoto}, M., {Latham}, D.~W., {Lester}, K., {McLeod},
  K.~K., {Mireles}, I., {Murray}, C.~A., {Niraula}, P., {Pedersen}, P.~P.,
  {Queloz}, D., {Quintana}, E.~V., {Ricker}, G., {Rudat}, A., {Sabin}, L.,
  {Safonov}, B., {Schroffenegger}, U., {Scott}, N., {Seager}, S., {Strakhov},
  I., {Triaud}, A.~H.~M.~J., {Vanderspek}, R., {Vezie}, M., \& {Winn}, J.,
  2022.
\newblock {TOI-2257 b: A highly eccentric long-period sub-Neptune transiting a
  nearby M dwarf}, {\it \aap\/}, {\bf 657}, A45.

\bibitem[{Scharmer} et~al.(2003){Scharmer}, {Bjelksjo}, {Korhonen}, {Lindberg},
  \& {Petterson}]{Scharmer2003}
{Scharmer}, G.~B., {Bjelksjo}, K., {Korhonen}, T.~K., {Lindberg}, B., \&
  {Petterson}, B., 2003.
\newblock {The 1-meter Swedish solar telescope}, in {\em Innovative Telescopes
  and Instrumentation for Solar Astrophysics\/}, vol. 4853 of {\bf Society of
  Photo-Optical Instrumentation Engineers (SPIE) Conference Series}, pp.
  341--350.

\bibitem[{Schmidt} et~al.(2014){Schmidt}, {Prieto}, {Stanek}, {Shappee},
  {Morrell}, {Bardalez Gagliuffi}, {Kochanek}, {Jencson}, {Holoien}, {Basu},
  {Beacom}, {Szczygie{\l}}, {Pojmanski}, {Brimacombe}, {Dubberley}, {Elphick},
  {Foale}, {Hawkins}, {Mullins}, {Rosing}, {Ross}, \& {Walker}]{Schmidt2014}
{Schmidt}, S.~J., {Prieto}, J.~L., {Stanek}, K.~Z., {Shappee}, B.~J.,
  {Morrell}, N., {Bardalez Gagliuffi}, D.~C., {Kochanek}, C.~S., {Jencson}, J.,
  {Holoien}, T.~W.~S., {Basu}, U., {Beacom}, J.~F., {Szczygie{\l}}, D.~M.,
  {Pojmanski}, G., {Brimacombe}, J., {Dubberley}, M., {Elphick}, M., {Foale},
  S., {Hawkins}, E., {Mullins}, D., {Rosing}, W., {Ross}, R., \& {Walker}, Z.,
  2014.
\newblock {Characterizing a Dramatic {\ensuremath{\Delta}}V
  \raisebox{-0.5ex}\textasciitilde -9 Flare on an Ultracool Dwarf Found by the
  ASAS-SN Survey}, {\it \apjl\/}, {\bf 781}(2), L24.

\bibitem[{Schmidt} et~al.(2012){Schmidt}, {von der L{\"u}he}, {Volkmer},
  {Denker}, {Solanki}, {Balthasar}, {Bello Gonzalez}, {Berkefeld}, {Collados},
  {Fischer}, {Halbgewachs}, {Heidecke}, {Hofmann}, {Kneer}, {Lagg}, {Nicklas},
  {Popow}, {Puschmann}, {Schmidt}, {Sigwarth}, {Sobotka}, {Soltau}, {Staude},
  {Strassmeier}, \& {Waldmann }]{Schmidt2012}
{Schmidt}, W., {von der L{\"u}he}, O., {Volkmer}, R., {Denker}, C., {Solanki},
  S.~K., {Balthasar}, H., {Bello Gonzalez}, N., {Berkefeld}, T., {Collados},
  M., {Fischer}, A., {Halbgewachs}, C., {Heidecke}, F., {Hofmann}, A., {Kneer},
  F., {Lagg}, A., {Nicklas}, H., {Popow}, E., {Puschmann}, K.~G., {Schmidt},
  D., {Sigwarth}, M., {Sobotka}, M., {Soltau}, D., {Staude}, J., {Strassmeier},
  K.~G., \& {Waldmann }, T.~A., 2012.
\newblock {The 1.5 meter solar telescope GREGOR}, {\it Astronomische
  Nachrichten\/}, {\bf 333}(9), 796.

\bibitem[{Schofield} et~al.(2019){Schofield}, {Chaplin}, {Huber}, {Campante},
  {Davies}, {Miglio}, {Ball}, {Appourchaux}, {Basu}, {Bedding},
  {Christensen-Dalsgaard}, {Creevey}, {Garc{\'\i}a}, {Handberg}, {Kawaler},
  {Kjeldsen}, {Latham}, {Lund}, {Metcalfe}, {Ricker}, {Serenelli}, {Silva
  Aguirre}, {Stello}, \& {Vanderspek}]{Schofield2019}
{Schofield}, M., {Chaplin}, W.~J., {Huber}, D., {Campante}, T.~L., {Davies},
  G.~R., {Miglio}, A., {Ball}, W.~H., {Appourchaux}, T., {Basu}, S., {Bedding},
  T.~R., {Christensen-Dalsgaard}, J., {Creevey}, O., {Garc{\'\i}a}, R.~A.,
  {Handberg}, R., {Kawaler}, S.~D., {Kjeldsen}, H., {Latham}, D.~W., {Lund},
  M.~N., {Metcalfe}, T.~S., {Ricker}, G.~R., {Serenelli}, A., {Silva Aguirre},
  V., {Stello}, D., \& {Vanderspek}, R., 2019.
\newblock {The Asteroseismic Target List for Solar-like Oscillators Observed in
  2 minute Cadence with the Transiting Exoplanet Survey Satellite}, {\it
  \apjs\/}, {\bf 241}(1), 12.

\bibitem[{Schrijver} et~al.(2012){Schrijver}, {Beer}, {Baltensperger},
  {Cliver}, {G{\"u}del}, {Hudson}, {McCracken}, {Osten}, {Peter}, {Soderblom},
  {Usoskin}, \& {Wolff}]{Schrijver:2012}
{Schrijver}, C.~J., {Beer}, J., {Baltensperger}, U., {Cliver}, E.~W.,
  {G{\"u}del}, M., {Hudson}, H.~S., {McCracken}, K.~G., {Osten}, R.~A.,
  {Peter}, T., {Soderblom}, D.~R., {Usoskin}, I.~G., \& {Wolff}, E.~W., 2012.
\newblock {Estimating the frequency of extremely energetic solar events, based
  on solar, stellar, lunar, and terrestrial records}, {\it Journal of
  Geophysical Research (Space Physics)\/}, {\bf 117}(A8), A08103.

\bibitem[{Sch{\"u}ssler} et~al.(2003){Sch{\"u}ssler}, {Shelyag}, {Berdyugina},
  {V{\"o}gler}, \& {Solanki}]{Schuessler2003}
{Sch{\"u}ssler}, M., {Shelyag}, S., {Berdyugina}, S., {V{\"o}gler}, A., \&
  {Solanki}, S.~K., 2003.
\newblock {Why Solar Magnetic Flux Concentrations Are Bright in Molecular
  Bands}, {\it \apjl\/}, {\bf 597}(2), L173--L176.

\bibitem[{Schutte} et~al.(2022){Schutte}, {Hebb}, {Lowry}, {Wisniewski},
  {Hawley}, {Mahadevan}, {Morris}, {Robertson}, {Rohn}, \&
  {Stefansson}]{Schutte2022}
{Schutte}, M.~C., {Hebb}, L., {Lowry}, S., {Wisniewski}, J., {Hawley}, S.~L.,
  {Mahadevan}, S., {Morris}, B.~M., {Robertson}, P., {Rohn}, G., \&
  {Stefansson}, G., 2022.
\newblock {Modeling Stellar Surface Features on a Subgiant Star with an M-dwarf
  Companion}, {\it \aj\/}, {\bf 164}(1), 14.

\bibitem[{Seager} \& {Sasselov}(2000)]{Seager2000}
{Seager}, S. \& {Sasselov}, D.~D., 2000.
\newblock {Theoretical Transmission Spectra during Extrasolar Giant Planet
  Transits}, {\it \apj\/}, {\bf 537}(2), 916--921.

\bibitem[{Sedaghati} et~al.(2015){Sedaghati}, {Boffin}, {Csizmadia}, {Gibson},
  {Kabath}, {Mallonn}, \& {Van den Ancker}]{Sedaghati-15}
{Sedaghati}, E., {Boffin}, H.~M.~J., {Csizmadia}, S., {Gibson}, N., {Kabath},
  P., {Mallonn}, M., \& {Van den Ancker}, M.~E., 2015.
\newblock {Regaining the FORS: optical ground-based transmission spectroscopy
  of the exoplanet WASP-19b with VLT+FORS2}, {\it \aap\/}, {\bf 576}, L11.

\bibitem[See et~al.(2021)See, Roquette, Amard, \& Matt]{see2021photometric}
See, V., Roquette, J., Amard, L., \& Matt, S.~P., 2021.
\newblock Photometric variability as a proxy for magnetic activity and its
  dependence on metallicity, {\it The Astrophysical Journal\/}, {\bf 912}(2),
  127.

\bibitem[{Semel}(1989)]{Semel1989zdi}
{Semel}, M., 1989.
\newblock {Zeeman-Doppler imaging of active stars. I - Basic principles.}, {\it
  \aap\/}, {\bf 225}, 456--466.

\bibitem[{Serrano} et~al.(2020){Serrano}, {Oshagh}, {Cegla}, {Barros},
  {Santos}, {Faria}, \& {Akinsanmi}]{Serrano-20}
{Serrano}, L.~M., {Oshagh}, M., {Cegla}, H.~M., {Barros}, S.~C.~C., {Santos},
  N.~C., {Faria}, J.~P., \& {Akinsanmi}, B., 2020.
\newblock {Can we detect the stellar differential rotation of WASP-7 through
  the Rossiter-McLaughlin observations?}, {\it \mnras\/}, {\bf 493}(4),
  5928--5943.

\bibitem[Shapiro et~al.(2020)Shapiro, Amazo-G{\'o}mez, Krivova, \&
  Solanki]{shapiro2020inflection}
Shapiro, A., Amazo-G{\'o}mez, E., Krivova, N.~A., \& Solanki, S.~K., 2020.
\newblock Inflection point in the power spectrum of stellar brightness
  variations-i. the model, {\it Astronomy \& Astrophysics\/}, {\bf 633}, A32.

\bibitem[{Shapiro} et~al.(2011){Shapiro}, {Fluri}, {Berdyugina}, {Bianda}, \&
  {Ramelli}]{Shapiro2011}
{Shapiro}, A.~I., {Fluri}, D.~M., {Berdyugina}, S.~V., {Bianda}, M., \&
  {Ramelli}, R., 2011.
\newblock {NLTE modeling of Stokes vector center-to-limb variations in the CN
  violet system}, {\it \aap\/}, {\bf 529}, A139.

\bibitem[{Shapiro} et~al.(2014){Shapiro}, {Solanki}, {Krivova}, {Schmutz},
  {Ball}, {Knaack}, {Rozanov}, \& {Unruh}]{Shapiro2014}
{Shapiro}, A.~I., {Solanki}, S.~K., {Krivova}, N.~A., {Schmutz}, W.~K., {Ball},
  W.~T., {Knaack}, R., {Rozanov}, E.~V., \& {Unruh}, Y.~C., 2014.
\newblock {Variability of Sun-like stars: reproducing observed photometric
  trends}, {\it \aap\/}, {\bf 569}, A38.

\bibitem[{Shapiro} et~al.(2016){Shapiro}, {Solanki}, {Krivova}, {Yeo}, \&
  {Schmutz}]{shapiro2016}
{Shapiro}, A.~I., {Solanki}, S.~K., {Krivova}, N.~A., {Yeo}, K.~L., \&
  {Schmutz}, W.~K., 2016.
\newblock {Are solar brightness variations faculae- or spot-dominated?}, {\it
  \aap\/}, {\bf 589}, A46.

\bibitem[{Shapiro} et~al.(2017){Shapiro}, {Solanki}, {Krivova}, {Cameron},
  {Yeo}, \& {Schmutz}]{2017NatAs...1..612S}
{Shapiro}, A.~I., {Solanki}, S.~K., {Krivova}, N.~A., {Cameron}, R.~H., {Yeo},
  K.~L., \& {Schmutz}, W.~K., 2017.
\newblock {The nature of solar brightness variations}, {\it Nature
  Astronomy\/}, {\bf 1}, 612--616.

\bibitem[{Sheeley}(1971)]{Sheeley1971}
{Sheeley}, N.~R., J., 1971.
\newblock {Using CN {\ensuremath{\lambda}} 3883 spectroheliograms to map weak
  photospheric magnetic fields}, {\it \solphys\/}, {\bf 20}(1), 19--25.

\bibitem[{Shimizu}(2009)]{Shimizu2009}
{Shimizu}, T., 2009.
\newblock {HINODE: New Space-borne Observatory for Investigating the Sun}, {\it
  Transactions of Space Technology Japan\/}, {\bf 7}, Tm1--Tm6.

\bibitem[{Shulyak} et~al.(2019){Shulyak}, {Reiners}, {Nagel}, {Tal-Or},
  {Caballero}, {Zechmeister}, {B{\'e}jar}, {Cort{\'e}s-Contreras}, {Martin},
  {Kaminski}, {Ribas}, {Quirrenbach}, {Amado}, {Anglada-Escud{\'e}}, {Bauer},
  {Dreizler}, {Guenther}, {Henning}, {Jeffers}, {K{\"u}rster}, {Lafarga},
  {Montes}, {Morales}, \& {Pedraz}]{Shulyak2019}
{Shulyak}, D., {Reiners}, A., {Nagel}, E., {Tal-Or}, L., {Caballero}, J.~A.,
  {Zechmeister}, M., {B{\'e}jar}, V.~J.~S., {Cort{\'e}s-Contreras}, M.,
  {Martin}, E.~L., {Kaminski}, A., {Ribas}, I., {Quirrenbach}, A., {Amado},
  P.~J., {Anglada-Escud{\'e}}, G., {Bauer}, F.~F., {Dreizler}, S., {Guenther},
  E.~W., {Henning}, T., {Jeffers}, S.~V., {K{\"u}rster}, M., {Lafarga}, M.,
  {Montes}, D., {Morales}, J.~C., \& {Pedraz}, S., 2019.
\newblock {Magnetic fields in M dwarfs from the CARMENES survey}, {\it \aap\/},
  {\bf 626}, A86.

\bibitem[{Silva}(2003)]{Silva-03}
{Silva}, A. V.~R., 2003.
\newblock {Method for Spot Detection on Solar-like Stars}, {\it \apjl\/}, {\bf
  585}(2), L147--L150.

\bibitem[{Silva-Valio} et~al.(2010){Silva-Valio}, {Lanza}, {Alonso}, \&
  {Barge}]{Silva-10}
{Silva-Valio}, A., {Lanza}, A.~F., {Alonso}, R., \& {Barge}, P., 2010.
\newblock {Properties of starspots on CoRoT-2}, {\it A\&A\/}, {\bf 510}, A25.

\bibitem[{Sing}(2010)]{Sing-10}
{Sing}, D.~K., 2010.
\newblock {Stellar limb-darkening coefficients for CoRot and Kepler}, {\it
  \aap\/}, {\bf 510}, A21.

\bibitem[{Sing} et~al.(2009){Sing}, {D{\'e}sert}, {Lecavelier Des Etangs},
  {Ballester}, {Vidal-Madjar}, {Parmentier}, {Hebrard}, \& {Henry}]{Sing2009}
{Sing}, D.~K., {D{\'e}sert}, J.~M., {Lecavelier Des Etangs}, A., {Ballester},
  G.~E., {Vidal-Madjar}, A., {Parmentier}, V., {Hebrard}, G., \& {Henry},
  G.~W., 2009.
\newblock {Transit spectrophotometry of the exoplanet HD 189733b. I. Searching
  for water but finding haze with HST NICMOS}, {\it \aap\/}, {\bf 505}(2),
  891--899.

\bibitem[{Sing} et~al.(2011){Sing}, {Pont}, {Aigrain}, {Charbonneau},
  {D{\'e}sert}, {Gibson}, {Gilliland}, {Hayek}, {Henry}, {Knutson}, {Lecavelier
  Des Etangs}, {Mazeh}, \& {Shporer}]{Sing2011}
{Sing}, D.~K., {Pont}, F., {Aigrain}, S., {Charbonneau}, D., {D{\'e}sert},
  J.~M., {Gibson}, N., {Gilliland}, R., {Hayek}, W., {Henry}, G., {Knutson},
  H., {Lecavelier Des Etangs}, A., {Mazeh}, T., \& {Shporer}, A., 2011.
\newblock {Hubble Space Telescope transmission spectroscopy of the exoplanet HD
  189733b: high-altitude atmospheric haze in the optical and near-ultraviolet
  with STIS}, {\it \mnras\/}, {\bf 416}(2), 1443--1455.

\bibitem[{Sing} et~al.(2015){Sing}, {Wakeford}, {Showman}, {Nikolov},
  {Fortney}, {Burrows}, {Ballester}, {Deming}, {Aigrain}, {D{\'e}sert},
  {Gibson}, {Henry}, {Knutson}, {Lecavelier des Etangs}, {Pont},
  {Vidal-Madjar}, {Williamson}, \& {Wilson}]{Sing2015}
{Sing}, D.~K., {Wakeford}, H.~R., {Showman}, A.~P., {Nikolov}, N., {Fortney},
  J.~J., {Burrows}, A.~S., {Ballester}, G.~E., {Deming}, D., {Aigrain}, S.,
  {D{\'e}sert}, J.~M., {Gibson}, N.~P., {Henry}, G.~W., {Knutson}, H.,
  {Lecavelier des Etangs}, A., {Pont}, F., {Vidal-Madjar}, A., {Williamson},
  M.~W., \& {Wilson}, P.~A., 2015.
\newblock {HST hot-Jupiter transmission spectral survey: detection of potassium
  in WASP-31b along with a cloud deck and Rayleigh scattering}, {\it \mnras\/},
  {\bf 446}(3), 2428--2443.

\bibitem[{Sing} et~al.(2016){Sing}, {Fortney}, {Nikolov}, {Wakeford},
  {Kataria}, {Evans}, {Aigrain}, {Ballester}, {Burrows}, {Deming},
  {D{\'e}sert}, {Gibson}, {Henry}, {Huitson}, {Knutson}, {Lecavelier Des
  Etangs}, {Pont}, {Showman}, {Vidal-Madjar}, {Williamson}, \&
  {Wilson}]{Sing2016}
{Sing}, D.~K., {Fortney}, J.~J., {Nikolov}, N., {Wakeford}, H.~R., {Kataria},
  T., {Evans}, T.~M., {Aigrain}, S., {Ballester}, G.~E., {Burrows}, A.~S.,
  {Deming}, D., {D{\'e}sert}, J.-M., {Gibson}, N.~P., {Henry}, G.~W.,
  {Huitson}, C.~M., {Knutson}, H.~A., {Lecavelier Des Etangs}, A., {Pont}, F.,
  {Showman}, A.~P., {Vidal-Madjar}, A., {Williamson}, M.~H., \& {Wilson},
  P.~A., 2016.
\newblock {A continuum from clear to cloudy hot-Jupiter exoplanets without
  primordial water depletion}, {\it \nat\/}, {\bf 529}(7584), 59--62.

\bibitem[{Siu-Tapia} et~al.(2017){Siu-Tapia}, {Lagg}, {Solanki}, {van Noort},
  \& {Jur{\v{c}}{\'a}k}]{Siu-Tapia:2017}
{Siu-Tapia}, A., {Lagg}, A., {Solanki}, S.~K., {van Noort}, M., \&
  {Jur{\v{c}}{\'a}k}, J., 2017.
\newblock {Normal and counter Evershed flows in the photospheric penumbra of a
  sunspot. SPINOR 2D inversions of Hinode-SOT/SP observations}, {\it \aap\/},
  {\bf 607}, A36.

\bibitem[{Siu-Tapia} et~al.(2019){Siu-Tapia}, {Lagg}, {van Noort}, {Rempel}, \&
  {Solanki}]{Siu-Tapia2019}
{Siu-Tapia}, A., {Lagg}, A., {van Noort}, M., {Rempel}, M., \& {Solanki},
  S.~K., 2019.
\newblock {Superstrong photospheric magnetic fields in sunspot penumbrae}, {\it
  \aap\/}, {\bf 631}, A99.

\bibitem[{Skaf} et~al.(2020){Skaf}, {Bieger}, {Edwards}, {Changeat}, {Morvan},
  {Kiefer}, {Blain}, {Zingales}, {Poveda}, {Al-Refaie}, {Baeyens}, {Gressier},
  {Guilluy}, {Jaziri}, {Modirrousta-Galian}, {Mugnai}, {Pluriel}, {Whiteford},
  {Wright}, {Yip}, {Charnay}, {Leconte}, {Drossart}, {Tsiaras}, {Venot},
  {Waldmann}, \& {Beaulieu}]{Skaf2020}
{Skaf}, N., {Bieger}, M.~F., {Edwards}, B., {Changeat}, Q., {Morvan}, M.,
  {Kiefer}, F., {Blain}, D., {Zingales}, T., {Poveda}, M., {Al-Refaie}, A.,
  {Baeyens}, R., {Gressier}, A., {Guilluy}, G., {Jaziri}, A.~Y.,
  {Modirrousta-Galian}, D., {Mugnai}, L.~V., {Pluriel}, W., {Whiteford}, N.,
  {Wright}, S., {Yip}, K.~H., {Charnay}, B., {Leconte}, J., {Drossart}, P.,
  {Tsiaras}, A., {Venot}, O., {Waldmann}, I., \& {Beaulieu}, J.-P., 2020.
\newblock {ARES. II. Characterizing the Hot Jupiters WASP-127 b, WASP-79 b, and
  WASP-62b with the Hubble Space Telescope}, {\it \aj\/}, {\bf 160}(3), 109.

\bibitem[{Skumanich}(1972)]{Skumanich1972}
{Skumanich}, A., 1972.
\newblock {Time Scales for Ca II Emission Decay, Rotational Braking, and
  Lithium Depletion}, {\it \apj\/}, {\bf 171}, 565.

\bibitem[{Smalley} et~al.(2012){Smalley}, {Anderson}, {Collier-Cameron},
  {Doyle}, {Fumel}, {Gillon}, {Hellier}, {Jehin}, {Lendl}, {Maxted}, {Pepe},
  {Pollacco}, {Queloz}, {S{\'e}gransan}, {Smith}, {Southworth}, {Triaud},
  {Udry}, \& {West}]{Smalley2012}
{Smalley}, B., {Anderson}, D.~R., {Collier-Cameron}, A., {Doyle}, A.~P.,
  {Fumel}, A., {Gillon}, M., {Hellier}, C., {Jehin}, E., {Lendl}, M., {Maxted},
  P.~F.~L., {Pepe}, F., {Pollacco}, D., {Queloz}, D., {S{\'e}gransan}, D.,
  {Smith}, A.~M.~S., {Southworth}, J., {Triaud}, A.~H.~M.~J., {Udry}, S., \&
  {West}, R.~G., 2012.
\newblock {WASP-78b and WASP-79b: two highly-bloated hot Jupiter-mass
  exoplanets orbiting F-type stars in Eridanus}, {\it \aap\/}, {\bf 547}, A61.

\bibitem[{Snellen} et~al.(2008){Snellen}, {Albrecht}, {de Mooij}, \& {Le
  Poole}]{Snellen2008}
{Snellen}, I.~A.~G., {Albrecht}, S., {de Mooij}, E.~J.~W., \& {Le Poole},
  R.~S., 2008.
\newblock {Ground-based detection of sodium in the transmission spectrum of
  exoplanet HD 209458b}, {\it \aap\/}, {\bf 487}(1), 357--362.

\bibitem[{Snellen} et~al.(2010){Snellen}, {de Kok}, {de Mooij}, \&
  {Albrecht}]{Snellen2010}
{Snellen}, I. A.~G., {de Kok}, R.~J., {de Mooij}, E. J.~W., \& {Albrecht}, S.,
  2010.
\newblock {The orbital motion, absolute mass and high-altitude winds of
  exoplanet HD209458b}, {\it \nat\/}, {\bf 465}(7301), 1049--1051.

\bibitem[{Sobotka} \& {Hanslmeier}(2005)]{Sobotka-Hanslmeier2005}
{Sobotka}, M. \& {Hanslmeier}, A., 2005.
\newblock {Photometry of umbral dots}, {\it \aap\/}, {\bf 442}(1), 323--329.

\bibitem[{Sobotka} et~al.(1997{\natexlab{a}}){Sobotka}, {Brandt}, \&
  {Simon}]{Sobotka1997a}
{Sobotka}, M., {Brandt}, P.~N., \& {Simon}, G.~W., 1997{\natexlab{a}}.
\newblock {Fine structure in sunspots. I. Sizes and lifetimes of umbral dots},
  {\it \aap\/}, {\bf 328}, 682--688.

\bibitem[{Sobotka} et~al.(1997{\natexlab{b}}){Sobotka}, {Brandt}, \&
  {Simon}]{Sobotka1997b}
{Sobotka}, M., {Brandt}, P.~N., \& {Simon}, G.~W., 1997{\natexlab{b}}.
\newblock {Fine structure in sunspots. II. Intensity variations and proper
  motions of umbral dots}, {\it \aap\/}, {\bf 328}, 689--694.

\bibitem[{Solanki}(1993)]{solanki1993}
{Solanki}, S.~K., 1993.
\newblock {Smallscale Solar Magnetic Fields - an Overview}, {\it \ssr\/}, {\bf
  63}(1-2), 1--188.

\bibitem[{Solanki}(2003)]{Solanki-03}
{Solanki}, S.~K., 2003.
\newblock {Sunspots: An overview}, {\it \aapr\/}, {\bf 11}, 153--286.

\bibitem[{Solanki} \& {Unruh}(2013)]{2013AN....334..145S}
{Solanki}, S.~K. \& {Unruh}, Y.~C., 2013.
\newblock {Solar irradiance variability}, {\it Astronomische Nachrichten\/},
  {\bf 334}(1-2), 145.

\bibitem[{Solanki} et~al.(2006){Solanki}, {Inhester}, \&
  {Sch{\"u}ssler}]{solanki2006}
{Solanki}, S.~K., {Inhester}, B., \& {Sch{\"u}ssler}, M., 2006.
\newblock {The solar magnetic field}, {\it Reports on Progress in Physics\/},
  {\bf 69}(3), 563--668.

\bibitem[{Sotzen} et~al.(2020){Sotzen}, {Stevenson}, {Sing}, {Kilpatrick},
  {Wakeford}, {Filippazzo}, {Lewis}, {H{\"o}rst}, {L{\'o}pez-Morales}, {Henry},
  {Buchhave}, {Ehrenreich}, {Fraine}, {Garc{\'\i}a Mu{\~n}oz}, {Jayaraman},
  {Lavvas}, {Lecavelier des Etangs}, {Marley}, {Nikolov}, {Rathcke}, \&
  {Sanz-Forcada}]{Sotzen2020}
{Sotzen}, K.~S., {Stevenson}, K.~B., {Sing}, D.~K., {Kilpatrick}, B.~M.,
  {Wakeford}, H.~R., {Filippazzo}, J.~C., {Lewis}, N.~K., {H{\"o}rst}, S.~M.,
  {L{\'o}pez-Morales}, M., {Henry}, G.~W., {Buchhave}, L.~A., {Ehrenreich}, D.,
  {Fraine}, J.~D., {Garc{\'\i}a Mu{\~n}oz}, A., {Jayaraman}, R., {Lavvas}, P.,
  {Lecavelier des Etangs}, A., {Marley}, M.~S., {Nikolov}, N., {Rathcke},
  A.~D., \& {Sanz-Forcada}, J., 2020.
\newblock {Transmission Spectroscopy of WASP-79b from 0.6 to 5.0
  {\ensuremath{\mu}}m}, {\it \aj\/}, {\bf 159}(1), 5.

\bibitem[{Sousa} et~al.(2021){Sousa}, {Adibekyan}, {Delgado-Mena}, {Santos},
  {Rojas-Ayala}, {Soares}, {Legoinha}, {Ulmer-Moll}, {Camacho}, {Barros},
  {Demangeon}, {Hoyer}, {Israelian}, {Mortier}, {Tsantaki}, \&
  {Monteiro}]{Sousa2021}
{Sousa}, S.~G., {Adibekyan}, V., {Delgado-Mena}, E., {Santos}, N.~C.,
  {Rojas-Ayala}, B., {Soares}, B.~M.~T.~B., {Legoinha}, H., {Ulmer-Moll}, S.,
  {Camacho}, J.~D., {Barros}, S.~C.~C., {Demangeon}, O.~D.~S., {Hoyer}, S.,
  {Israelian}, G., {Mortier}, A., {Tsantaki}, M., \& {Monteiro}, M.~A., 2021.
\newblock {SWEET-Cat 2.0: The Cat just got SWEETer. Higher quality spectra and
  precise parallaxes from Gaia eDR3}, {\it \aap\/}, {\bf 656}, A53.

\bibitem[{Southworth}(2011)]{Southworth2011}
{Southworth}, J., 2011.
\newblock {Homogeneous studies of transiting extrasolar planets - IV. Thirty
  systems with space-based light curves}, {\it \mnras\/}, {\bf 417}(3),
  2166--2196.

\bibitem[{Southworth} et~al.(2009){Southworth}, {Hinse}, {Burgdorf}, {Dominik},
  {Hornstrup}, {J{\o}rgensen}, {Liebig}, {Ricci}, \& others.]{Southworth-09}
{Southworth}, J., {Hinse}, T.~C., {Burgdorf}, M.~J., {Dominik}, M.,
  {Hornstrup}, A., {J{\o}rgensen}, U.~G., {Liebig}, C., {Ricci}, D., \&
  others., 2009.
\newblock {High-precision photometry by telescope defocussing - II. The
  transiting planetary system WASP-4}, {\it MNRAS\/}, {\bf 399}, 287--294.

\bibitem[{Southworth} et~al.(2016){Southworth}, {Tregloan-Reed}, {Andersen},
  {Calchi Novati}, {Ciceri}, {Colque}, {D'Ago}, {Dominik}, {Evans}, {Gu},
  {Herrera-Cordova}, {Hinse}, {J{\o}rgensen}, {Juncher}, {Kuffmeier},
  {Mancini}, {Peixinho}, {Popovas}, {Rabus}, {Skottfelt}, {Tronsgaard},
  {Unda-Sanzana}, {Wang}, {Wertz}, {Alsubai}, {Andersen}, {Bozza}, {Bramich},
  {Burgdorf}, {Damerdji}, {Diehl}, {Elyiv}, {Figuera Jaimes}, {Haugb{\o}lle},
  {Hundertmark}, {Kains}, {Kerins}, {Korhonen}, {Liebig}, {Mathiasen}, {Penny},
  {Rahvar}, {Scarpetta}, {Schmidt}, {Snodgrass}, {Starkey}, {Surdej}, {Vilela},
  {von Essen}, \& {Wang}]{Southworth-16}
{Southworth}, J., {Tregloan-Reed}, J., {Andersen}, M.~I., {Calchi Novati}, S.,
  {Ciceri}, S., {Colque}, J.~P., {D'Ago}, G., {Dominik}, M., {Evans}, D.~F.,
  {Gu}, S.-H., {Herrera-Cordova}, A., {Hinse}, T.~C., {J{\o}rgensen}, U.~G.,
  {Juncher}, D., {Kuffmeier}, M., {Mancini}, L., {Peixinho}, N., {Popovas}, A.,
  {Rabus}, M., {Skottfelt}, J., {Tronsgaard}, R., {Unda-Sanzana}, E., {Wang},
  X.-B., {Wertz}, O., {Alsubai}, K.~A., {Andersen}, J.~M., {Bozza}, V.,
  {Bramich}, D.~M., {Burgdorf}, M., {Damerdji}, Y., {Diehl}, C., {Elyiv}, A.,
  {Figuera Jaimes}, R., {Haugb{\o}lle}, T., {Hundertmark}, M., {Kains}, N.,
  {Kerins}, E., {Korhonen}, H., {Liebig}, C., {Mathiasen}, M., {Penny}, M.~T.,
  {Rahvar}, S., {Scarpetta}, G., {Schmidt}, R.~W., {Snodgrass}, C., {Starkey},
  D., {Surdej}, J., {Vilela}, C., {von Essen}, C., \& {Wang}, Y., 2016.
\newblock {High-precision photometry by telescope defocussing - VIII. WASP-22,
  WASP-41, WASP-42 and WASP-55}, {\it \mnras\/}, {\bf 457}, 4205--4217.

\bibitem[{Spake} et~al.(2018){Spake}, {Sing}, {Evans}, {Oklop{\v{c}}i{\'c}},
  {}, {Bourrier}, {Kreidberg}, {Rackham}, {Irwin}, {Ehrenreich}, {Wyttenbach},
  {Wakeford}, {Zhou}, {Chubb}, {Nikolov}, {Goyal}, {Henry}, {Williamson},
  {Blumenthal}, {Anderson}, {Hellier}, {Charbonneau}, {Udry}, \&
  {Madhusudhan}]{Spake2018}
{Spake}, J.~J., {Sing}, D.~K., {Evans}, T.~M., {Oklop{\v{c}}i{\'c}}, {}, A.,
  {Bourrier}, V., {Kreidberg}, L., {Rackham}, B.~V., {Irwin}, J., {Ehrenreich},
  D., {Wyttenbach}, A., {Wakeford}, H.~R., {Zhou}, Y., {Chubb}, K.~L.,
  {Nikolov}, N., {Goyal}, J.~M., {Henry}, G.~W., {Williamson}, M.~H.,
  {Blumenthal}, S., {Anderson}, D.~R., {Hellier}, C., {Charbonneau}, D.,
  {Udry}, S., \& {Madhusudhan}, N., 2018.
\newblock {Helium in the eroding atmosphere of an exoplanet}, {\it \nat\/},
  {\bf 557}(7703), 68--70.

\bibitem[{Stassun} et~al.(2017){Stassun}, {Collins}, \& {Gaudi}]{Stassun:2017}
{Stassun}, K.~G., {Collins}, K.~A., \& {Gaudi}, B.~S., 2017.
\newblock {Accurate Empirical Radii and Masses of Planets and Their Host Stars
  with Gaia Parallaxes}, {\it \aj\/}, {\bf 153}(3), 136.

\bibitem[{Stassun} et~al.(2018){Stassun}, {Corsaro}, {Pepper}, \&
  {Gaudi}]{Stassun:2018}
{Stassun}, K.~G., {Corsaro}, E., {Pepper}, J.~A., \& {Gaudi}, B.~S., 2018.
\newblock {Empirical Accurate Masses and Radii of Single Stars with TESS and
  Gaia}, {\it \aj\/}, {\bf 155}(1), 22.

\bibitem[{Stefansson} et~al.(2018){Stefansson}, {Mahadevan}, {Wisniewski},
  {Li}, {Hebb}, {Morris}, {Halverson}, {Monson}, \& {Robertson}]{Stefansson18}
{Stefansson}, G., {Mahadevan}, S., {Wisniewski}, J., {Li}, Y., {Hebb}, L.,
  {Morris}, B., {Halverson}, S., {Monson}, A., \& {Robertson}, P., 2018.
\newblock {Extreme precision photometry from the ground with beam-shaping
  diffusers for K2, TESS, and beyond}, in {\em Ground-based and Airborne
  Instrumentation for Astronomy VII\/}, vol. 10702 of {\bf Society of
  Photo-Optical Instrumentation Engineers (SPIE) Conference Series}, p.
  1070250.

\bibitem[{Steffen} \& {Holweger}(2002)]{steffen2002}
{Steffen}, M. \& {Holweger}, H., 2002.
\newblock {Line formation in convective stellar atmospheres. I. Granulation
  corrections for solar photospheric abundances}, {\it \aap\/}, {\bf 387},
  258--270.

\bibitem[{Stein}(2012)]{Stein:2012}
{Stein}, R.~F., 2012.
\newblock {Solar Surface Magneto-Convection}, {\it Living Reviews in Solar
  Physics\/}, {\bf 9}(1), 4.

\bibitem[{Stein} \& {Nordlund}(1998)]{Stein_Nordlund98}
{Stein}, R.~F. \& {Nordlund}, {\r{A}}., 1998.
\newblock {Simulations of Solar Granulation. I. General Properties}, {\it
  \apj\/}, {\bf 499}(2), 914--933.

\bibitem[{Steiner}(2005)]{steiner2005}
{Steiner}, O., 2005.
\newblock {Radiative properties of magnetic elements. II. Center to limb
  variation of the appearance of photospheric faculae}, {\it \aap\/}, {\bf
  430}, 691--700.

\bibitem[{Steiner} et~al.(2014){Steiner}, {Salhab}, {Freytag}, {Rajaguru},
  {Schaffenberger}, \& {Steffen}]{Steiner2014}
{Steiner}, O., {Salhab}, R., {Freytag}, B., {Rajaguru}, P., {Schaffenberger},
  W., \& {Steffen}, M., 2014.
\newblock {Properties of small-scale magnetism of stellar atmospheres}, {\it
  \pasj\/}, {\bf 66}, S5.

\bibitem[{Stello} et~al.(2011){Stello}, {Huber}, {Kallinger}, {Basu}, {Mosser},
  {Hekker}, {Mathur}, {Garc{\'\i}a}, {Bedding}, {Kjeldsen}, {Gilliland},
  {Verner}, {Chaplin}, {Benomar}, {Meibom}, {Grundahl}, {Elsworth},
  {Molenda-{\.Z}akowicz}, {Szab{\'o}}, {Christensen-Dalsgaard}, {Tenenbaum},
  {Twicken}, \& {Uddin}]{Stello2011}
{Stello}, D., {Huber}, D., {Kallinger}, T., {Basu}, S., {Mosser}, B., {Hekker},
  S., {Mathur}, S., {Garc{\'\i}a}, R.~A., {Bedding}, T.~R., {Kjeldsen}, H.,
  {Gilliland}, R.~L., {Verner}, G.~A., {Chaplin}, W.~J., {Benomar}, O.,
  {Meibom}, S., {Grundahl}, F., {Elsworth}, Y.~P., {Molenda-{\.Z}akowicz}, J.,
  {Szab{\'o}}, R., {Christensen-Dalsgaard}, J., {Tenenbaum}, P., {Twicken},
  J.~D., \& {Uddin}, K., 2011.
\newblock {Amplitudes of Solar-like Oscillations: Constraints from Red Giants
  in Open Clusters Observed by Kepler}, {\it \apjl\/}, {\bf 737}(1), L10.

\bibitem[{Stello} et~al.(2013){Stello}, {Huber}, {Bedding}, {Benomar},
  {Bildsten}, {Elsworth}, {Gilliland}, {Mosser}, {Paxton}, \&
  {White}]{Stello:2013}
{Stello}, D., {Huber}, D., {Bedding}, T.~R., {Benomar}, O., {Bildsten}, L.,
  {Elsworth}, Y.~P., {Gilliland}, R.~L., {Mosser}, B., {Paxton}, B., \&
  {White}, T.~R., 2013.
\newblock {Asteroseismic Classification of Stellar Populations among 13,000 Red
  Giants Observed by Kepler}, {\it \apjl\/}, {\bf 765}(2), L41.

\bibitem[{Stenflo}(1982)]{Stenflo1982}
{Stenflo}, J.~O., 1982.
\newblock {The Hanle Effect and the Diagnostics of Turbulent Magnetic Fields in
  the Solar Atmosphere}, {\it \solphys\/}, {\bf 80}(2), 209--226.

\bibitem[{Stenflo} et~al.(1998){Stenflo}, {Keller}, \&
  {Gandorfer}]{Stenflo1998}
{Stenflo}, J.~O., {Keller}, C.~U., \& {Gandorfer}, A., 1998.
\newblock {Differential Hanle effect and the spatial variation of turbulent
  magnetic fields on the Sun}, {\it \aap\/}, {\bf 329}, 319--328.

\bibitem[Stephenson(1990)]{Stephenson:1990}
Stephenson, F.~R., 1990.
\newblock Historical evidence concerning the sun: Interpretation of sunspot
  records during the telescopic and pretelescopic eras, {\it Philosophical
  Transactions of the Royal Society of London. Series A, Mathematical and
  Physical Sciences\/}, {\bf 330}(1615), 499--512.

\bibitem[{Strassmeier}(2009)]{Strassmeier-09}
{Strassmeier}, K.~G., 2009.
\newblock {Starspots}, {\it \aapr\/}, {\bf 17}, 251--308.

\bibitem[{Strassmeier} \& {Rice}(1998)]{StrassmeierRice1998}
{Strassmeier}, K.~G. \& {Rice}, J.~B., 1998.
\newblock {Doppler imaging of stellar surface structure. VI. HD 129333 = EK
  Draconis: a stellar analog of the active young Sun}, {\it \aap\/}, {\bf 330},
  685--695.

\bibitem[{Suetterlin}(1998)]{Suetterlin1998}
{Suetterlin}, P., 1998.
\newblock {Properties of solar pores}, {\it \aap\/}, {\bf 333}, 305--312.

\bibitem[{Sulis} et~al.(2020){Sulis}, {Lendl}, {Hofmeister}, {Veronig},
  {Fossati}, {Cubillos}, \& {Van Grootel}]{Sulis:2020}
{Sulis}, S., {Lendl}, M., {Hofmeister}, S., {Veronig}, A., {Fossati}, L.,
  {Cubillos}, P., \& {Van Grootel}, V., 2020.
\newblock {Mitigating flicker noise in high-precision photometry. I.
  Characterization of the noise structure, impact on the inferred transit
  parameters, and predictions for CHEOPS observations}, {\it \aap\/}, {\bf
  636}, A70.

\bibitem[{Sun} et~al.(2017){Sun}, {Gu}, {Wang}, {Collier Cameron}, {Cao},
  {Wang}, {Xiang}, {Hui}, {Kwok}, {Yeung}, {Ng}, \& {Grau Horta}]{Sun-17}
{Sun}, L., {Gu}, S., {Wang}, X., {Collier Cameron}, A., {Cao}, D., {Wang}, Y.,
  {Xiang}, Y., {Hui}, H.-K., {Kwok}, C.-T., {Yeung}, B., {Ng}, E., \& {Grau
  Horta}, F., 2017.
\newblock {Refined System Parameters and TTV Study of Transiting Exoplanetary
  System HAT-P-20}, {\it \aj\/}, {\bf 153}(1), 28.

\bibitem[{Tan} \& {Showman}(2021)]{TanShowman2021}
{Tan}, X. \& {Showman}, A.~P., 2021.
\newblock {Atmospheric circulation of brown dwarfs and directly imaged
  exoplanets driven by cloud radiative feedback: effects of rotation}, {\it
  \mnras\/}, {\bf 502}(1), 678--699.

\bibitem[{Tang} et~al.(2013){Tang}, {Sasselov}, {Grindlay}, {Los}, \&
  {Servillat}]{Tang2013}
{Tang}, S., {Sasselov}, D., {Grindlay}, J., {Los}, E., \& {Servillat}, M.,
  2013.
\newblock {100-year DASCH Light Curves of Kepler Planet-Candidate Host Stars},
  {\it \pasp\/}, {\bf 125}(929), 793.

\bibitem[{ten Brummelaar} et~al.(2005){ten Brummelaar}, {McAlister}, {Ridgway},
  {Bagnuolo}, {Turner}, {Sturmann}, {Sturmann}, {Berger}, {Ogden}, {Cadman},
  {Hartkopf}, {Hopper}, \& {Shure}]{tenBrummelaar2005}
{ten Brummelaar}, T.~A., {McAlister}, H.~A., {Ridgway}, S.~T., {Bagnuolo},
  W.~G., J., {Turner}, N.~H., {Sturmann}, L., {Sturmann}, J., {Berger}, D.~H.,
  {Ogden}, C.~E., {Cadman}, R., {Hartkopf}, W.~I., {Hopper}, C.~H., \& {Shure},
  M.~A., 2005.
\newblock {First Results from the CHARA Array. II. A Description of the
  Instrument}, {\it \apj\/}, {\bf 628}(1), 453--465.

\bibitem[{The LUVOIR Team}(2019)]{TheLUVOIRTeam2019}
{The LUVOIR Team}, 2019.
\newblock {The LUVOIR Mission Concept Study Final Report}, {\it arXiv
  e-prints\/}, p. arXiv:1912.06219.

\bibitem[{Tinetti} et~al.(2007){Tinetti}, {Vidal-Madjar}, {Liang}, {Beaulieu},
  {Yung}, {Carey}, {Barber}, {Tennyson}, {Ribas}, {Allard}, {Ballester},
  {Sing}, \& {Selsis}]{Tinetti2007}
{Tinetti}, G., {Vidal-Madjar}, A., {Liang}, M.-C., {Beaulieu}, J.-P., {Yung},
  Y., {Carey}, S., {Barber}, R.~J., {Tennyson}, J., {Ribas}, I., {Allard}, N.,
  {Ballester}, G.~E., {Sing}, D.~K., \& {Selsis}, F., 2007.
\newblock {Water vapour in the atmosphere of a transiting extrasolar planet},
  {\it \nat\/}, {\bf 448}(7150), 169--171.

\bibitem[{Tinetti} et~al.(2018){Tinetti}, {Drossart}, {Eccleston}, {Hartogh},
  {Heske}, {Leconte}, {Micela}, {Ollivier}, {Pilbratt}, {Puig}, \&
  et~al.]{Tinetti-18}
{Tinetti}, G., {Drossart}, P., {Eccleston}, P., {Hartogh}, P., {Heske}, A.,
  {Leconte}, J., {Micela}, G., {Ollivier}, M., {Pilbratt}, G., {Puig}, L., \&
  et~al., 2018.
\newblock {A chemical survey of exoplanets with ARIEL}, {\it Experimental
  Astronomy\/}, {\bf 46}, 135--209.

\bibitem[{Tinetti} et~al.(2021){Tinetti}, {Eccleston}, {Haswell}, {Lagage},
  {Leconte}, {L{\"u}ftinger}, {Micela}, {Min}, {Pilbratt}, {Puig}, \&
  et~al.]{Tinetti2021}
{Tinetti}, G., {Eccleston}, P., {Haswell}, C., {Lagage}, P.-O., {Leconte}, J.,
  {L{\"u}ftinger}, T., {Micela}, G., {Min}, M., {Pilbratt}, G., {Puig}, L., \&
  et~al., 2021.
\newblock {Ariel: Enabling planetary science across light-years}, {\it arXiv
  e-prints\/}, p. arXiv:2104.04824.

\bibitem[{Toner} \& {Gray}(1988)]{TonerGray1988}
{Toner}, C.~G. \& {Gray}, D.~F., 1988.
\newblock {The Star Patch on the G8 Dwarf chi Bootis A}, {\it \apj\/}, {\bf
  334}, 1008.

\bibitem[{Toner} et~al.(1997){Toner}, {Jefferies}, \& {Duvall}]{toner1997}
{Toner}, C.~G., {Jefferies}, S.~M., \& {Duvall}, T.~L., 1997.
\newblock {Restoration of Long-Exposure Full-Disk Solar Intensity Images}, {\it
  \apj\/}, {\bf 478}(2), 817--827.

\bibitem[{Trampedach} et~al.(2013){Trampedach}, {Asplund}, {Collet},
  {Nordlund}, \& {Stein}]{Trampedach_STAGGER}
{Trampedach}, R., {Asplund}, M., {Collet}, R., {Nordlund}, {\r{A}}., \&
  {Stein}, R.~F., 2013.
\newblock {A Grid of Three-dimensional Stellar Atmosphere Models of Solar
  Metallicity. I. General Properties, Granulation, and Atmospheric Expansion},
  {\it \apj\/}, {\bf 769}(1), 18.

\bibitem[{Tregloan-Reed} \& {Unda-Sanzana}(2019)]{Tregloan-Reed-19}
{Tregloan-Reed}, J. \& {Unda-Sanzana}, E., 2019.
\newblock {Simulations of starspot anomalies within TESS exoplanetary transit
  light curves. I. Detection limits of starspot anomalies in TESS light
  curves}, {\it A\&A\/}, {\bf 630}, A114.

\bibitem[{Tregloan-Reed} \& {Unda-Sanzana}(2021)]{Tregloan-Reed-21}
{Tregloan-Reed}, J. \& {Unda-Sanzana}, E., 2021.
\newblock {Simulations of starspot anomalies within TESS exoplanetary transit
  light curves -- II}, {\it \aap\/}, {\bf 649}, A130.

\bibitem[{Tregloan-Reed} et~al.(2013){Tregloan-Reed}, {Southworth}, \&
  {Tappert}]{Tregloan-Reed-13}
{Tregloan-Reed}, J., {Southworth}, J., \& {Tappert}, C., 2013.
\newblock {Transits and starspots in the WASP-19 planetary system}, {\it
  \mnras\/}, {\bf 428}, 3671--3679.

\bibitem[{Tregloan-Reed} et~al.(2015){Tregloan-Reed}, {Southworth}, {Burgdorf},
  {Novati}, {Dominik}, {Finet}, {J{\o}rgensen}, {Maier}, {Mancini}, {Prof},
  {Ricci}, {Snodgrass}, {Bozza}, {Browne}, {Dodds}, {Gerner}, {Harps{\o}e},
  {Hinse}, {Hundertmark}, {Kains}, {Kerins}, {Liebig}, {Penny}, {Rahvar},
  {Sahu}, {Scarpetta}, {Sch{\"a}fer}, {Sch{\"o}nebeck}, {Skottfelt}, \&
  {Surdej}]{Tregloan-Reed-15}
{Tregloan-Reed}, J., {Southworth}, J., {Burgdorf}, M., {Novati}, S.~C.,
  {Dominik}, M., {Finet}, F., {J{\o}rgensen}, U.~G., {Maier}, G., {Mancini},
  L., {Prof}, S., {Ricci}, D., {Snodgrass}, C., {Bozza}, V., {Browne}, P.,
  {Dodds}, P., {Gerner}, T., {Harps{\o}e}, K., {Hinse}, T.~C., {Hundertmark},
  M., {Kains}, N., {Kerins}, E., {Liebig}, C., {Penny}, M.~T., {Rahvar}, S.,
  {Sahu}, K., {Scarpetta}, G., {Sch{\"a}fer}, S., {Sch{\"o}nebeck}, F.,
  {Skottfelt}, J., \& {Surdej}, J., 2015.
\newblock {Transits and starspots in the WASP-6 planetary system}, {\it
  \mnras\/}, {\bf 450}, 1760--1769.

\bibitem[{Tregloan-Reed} et~al.(2018){Tregloan-Reed}, {Southworth}, {Mancini},
  {Molli{\`e}re}, {Ciceri}, {Bruni}, {Ricci}, {Ayala-Loera}, \&
  {Henning}]{Tregloan-Reed-18}
{Tregloan-Reed}, J., {Southworth}, J., {Mancini}, L., {Molli{\`e}re}, P.,
  {Ciceri}, S., {Bruni}, I., {Ricci}, D., {Ayala-Loera}, C., \& {Henning}, T.,
  2018.
\newblock {Possible detection of a bimodal cloud distribution in the atmosphere
  of HAT-P-32 A b from multiband photometry}, {\it \mnras\/}, {\bf 474},
  5485--5499.

\bibitem[{Tremblay} et~al.(2013){Tremblay}, {Ludwig}, {Freytag}, {Steffen}, \&
  {Caffau}]{tremblay+al2013}
{Tremblay}, P.-E., {Ludwig}, H.-G., {Freytag}, B., {Steffen}, M., \& {Caffau},
  E., 2013.
\newblock {Granulation properties of giants, dwarfs, and white dwarfs from the
  CIFIST 3D model atmosphere grid}, {\it \aap\/}, {\bf 557}, A7.

\bibitem[{Tremblin} et~al.(2020){Tremblin}, {Phillips}, {Emery}, {Baraffe},
  {Lew}, {Apai}, {Biller}, \& {Bonnefoy}]{Tremblin2020}
{Tremblin}, P., {Phillips}, M.~W., {Emery}, A., {Baraffe}, I., {Lew}, B.~W.~P.,
  {Apai}, D., {Biller}, B.~A., \& {Bonnefoy}, M., 2020.
\newblock {Rotational spectral modulation of cloudless atmospheres for L/T
  brown dwarfs and extrasolar giant planets}, {\it \aap\/}, {\bf 643}, A23.

\bibitem[Trotta(2008)]{Trotta2008}
Trotta, R., 2008.
\newblock Bayes in the sky: Bayesian inference and model selection in
  cosmology, {\it Contemporary Physics\/}, {\bf 49}(2), 71--104.

\bibitem[{Trujillo Bueno} et~al.(2004){Trujillo Bueno}, {Shchukina}, \&
  {Asensio Ramos}]{TrujilloBueno2004}
{Trujillo Bueno}, J., {Shchukina}, N., \& {Asensio Ramos}, A., 2004.
\newblock {A substantial amount of hidden magnetic energy in the quiet Sun},
  {\it \nat\/}, {\bf 430}(6997), 326--329.

\bibitem[{Tsuneta} et~al.(2008){Tsuneta}, {Ichimoto}, {Katsukawa}, {Nagata},
  {Otsubo}, {Shimizu}, {Suematsu}, {Nakagiri}, {Noguchi}, {Tarbell}, {Title},
  {Shine}, {Rosenberg}, {Hoffmann}, {Jurcevich}, {Kushner}, {Levay}, {Lites},
  {Elmore}, {Matsushita}, {Kawaguchi}, {Saito}, {Mikami}, {Hill}, \&
  {Owens}]{tsuneta+al2008}
{Tsuneta}, S., {Ichimoto}, K., {Katsukawa}, Y., {Nagata}, S., {Otsubo}, M.,
  {Shimizu}, T., {Suematsu}, Y., {Nakagiri}, M., {Noguchi}, M., {Tarbell}, T.,
  {Title}, A., {Shine}, R., {Rosenberg}, W., {Hoffmann}, C., {Jurcevich}, B.,
  {Kushner}, G., {Levay}, M., {Lites}, B., {Elmore}, D., {Matsushita}, T.,
  {Kawaguchi}, N., {Saito}, H., {Mikami}, I., {Hill}, L.~D., \& {Owens}, J.~K.,
  2008.
\newblock {The Solar Optical Telescope for the Hinode Mission: An Overview},
  {\it \solphys\/}, {\bf 249}, 167--196.

\bibitem[{Turova}(1984)]{Turova1984}
{Turova}, I.~P., 1984.
\newblock {On Umbral Flashes in Different Sunspot Groups}, {\it \solphys\/},
  {\bf 91}(1), 51--54.

\bibitem[{Unruh} et~al.(1999){Unruh}, {Solanki}, \& {Fligge}]{unruh1999}
{Unruh}, Y.~C., {Solanki}, S.~K., \& {Fligge}, M., 1999.
\newblock {The spectral dependence of facular contrast and solar irradiance
  variations}, {\it \aap\/}, {\bf 345}, 635--642.

\bibitem[{Usoskin} et~al.(2021){Usoskin}, {Solanki}, {Krivova}, {Hofer},
  {Kovaltsov}, {Wacker}, {Brehm}, \& {Kromer}]{Usoskin:2021}
{Usoskin}, I.~G., {Solanki}, S.~K., {Krivova}, N.~A., {Hofer}, B., {Kovaltsov},
  G.~A., {Wacker}, L., {Brehm}, N., \& {Kromer}, B., 2021.
\newblock {Solar cyclic activity over the last millennium reconstructed from
  annual $^{14}$C data}, {\it \aap\/}, {\bf 649}, A141.

\bibitem[{Valio} et~al.(2017){Valio}, {Estrela}, {Netto}, {Bravo}, \& {de
  Medeiros}]{Valio-17}
{Valio}, A., {Estrela}, R., {Netto}, Y., {Bravo}, J.~P., \& {de Medeiros},
  J.~R., 2017.
\newblock {Activity and Rotation of Kepler-17}, {\it \apj\/}, {\bf 835}(2),
  294.

\bibitem[{Valio} et~al.(2020){Valio}, {Spagiari}, {Marengoni}, \&
  {Selhorst}]{Valio2020_sunspots}
{Valio}, A., {Spagiari}, E., {Marengoni}, M., \& {Selhorst}, C.~L., 2020.
\newblock {Correlations of Sunspot Physical Characteristics during Solar Cycle
  23}, {\it \solphys\/}, {\bf 295}(9), 120.

\bibitem[{van Belle} et~al.(2020{\natexlab{a}}){van Belle}, {Clark},
  {Armstrong}, {Baines}, {Martinez}, {Restaino}, {Schmitt}, {Jorgensen},
  {Mozurkewich}, {Pugh}, {Clark}, {Green}, {Jones}, {Kingsley}, {Kurtz}, \&
  {Schilperoort}]{vanBelle2020}
{van Belle}, G., {Clark}, J., {Armstrong}, J.~T., {Baines}, E., {Martinez}, T.,
  {Restaino}, S., {Schmitt}, H., {Jorgensen}, A., {Mozurkewich}, D., {Pugh},
  T., {Clark}, W., {Green}, N., {Jones}, K., {Kingsley}, B., {Kurtz}, P., \&
  {Schilperoort}, A., 2020{\natexlab{a}}.
\newblock {The Navy Precision Optical Interferometer: two years of development
  towards large-aperture observations}, in {\em Society of Photo-Optical
  Instrumentation Engineers (SPIE) Conference Series\/}, vol. 11446 of {\bf
  Society of Photo-Optical Instrumentation Engineers (SPIE) Conference Series},
  p. 1144608.

\bibitem[{van Belle} et~al.(2020{\natexlab{b}}){van Belle}, {Hillsberry},
  {Kloske}, {Kugler}, {Patan{\'e}}, {Paul-Gin}, {Piness}, {Riley}, {Schomer},
  {Snyder}, \& {Tobiassen}]{vanBelle2020b}
{van Belle}, G.~T., {Hillsberry}, D., {Kloske}, J., {Kugler}, J., {Patan{\'e}},
  S., {Paul-Gin}, N., {Piness}, J., {Riley}, D., {Schomer}, J., {Snyder}, M.,
  \& {Tobiassen}, T., 2020{\natexlab{b}}.
\newblock {Optimast structurally connected interferometry enabled by in-space
  robotic manufacturing and assembly}, in {\em Society of Photo-Optical
  Instrumentation Engineers (SPIE) Conference Series\/}, vol. 11446 of {\bf
  Society of Photo-Optical Instrumentation Engineers (SPIE) Conference Series},
  p. 114462K.

\bibitem[{van Noort} et~al.(2013){van Noort}, {Lagg}, {Tiwari}, \&
  {Solanki}]{vanNoort:2013}
{van Noort}, M., {Lagg}, A., {Tiwari}, S.~K., \& {Solanki}, S.~K., 2013.
\newblock {Peripheral downflows in sunspot penumbrae}, {\it \aap\/}, {\bf 557},
  A24.

\bibitem[{van Saders} et~al.(2016){van Saders}, {Ceillier}, {Metcalfe}, {Silva
  Aguirre}, {Pinsonneault}, {Garc{\'\i}a}, {Mathur}, \&
  {Davies}]{vanSaders2016}
{van Saders}, J.~L., {Ceillier}, T., {Metcalfe}, T.~S., {Silva Aguirre}, V.,
  {Pinsonneault}, M.~H., {Garc{\'\i}a}, R.~A., {Mathur}, S., \& {Davies},
  G.~R., 2016.
\newblock {Weakened magnetic braking as the origin of anomalously rapid
  rotation in old field stars}, {\it \nat\/}, {\bf 529}(7585), 181--184.

\bibitem[{Vaughan} \& {Preston}(1980)]{Vaughan1980}
{Vaughan}, A.~H. \& {Preston}, G.~W., 1980.
\newblock {A survey of chromospheric CA II H and K emission in field stars of
  the solar neighborhood.}, {\it \pasp\/}, {\bf 92}, 385--391.

\bibitem[{Vaughan} \& {Zirin}(1968)]{Vaughan1968}
{Vaughan}, Arthur~H., J. \& {Zirin}, H., 1968.
\newblock {The Helium Line {\ensuremath{\lambda}} 10830 {\r{A}} in Late-Type
  Stars}, {\it \apj\/}, {\bf 152}, 123.

\bibitem[{Vernazza} et~al.(1981){Vernazza}, {Avrett}, \&
  {Loeser}]{Vernazza:1981}
{Vernazza}, J.~E., {Avrett}, E.~H., \& {Loeser}, R., 1981.
\newblock {Structure of the solar chromosphere. III. Models of the EUV
  brightness components of the quiet sun.}, {\it \apjs\/}, {\bf 45}, 635--725.

\bibitem[{Vida} et~al.(2007){Vida}, {Kov{\'a}ri}, {{\v{S}}vanda}, {Ol{\'a}h},
  {Strassmeier}, \& {Bartus}]{Vida2007}
{Vida}, K., {Kov{\'a}ri}, Z., {{\v{S}}vanda}, M., {Ol{\'a}h}, K.,
  {Strassmeier}, K.~G., \& {Bartus}, J., 2007.
\newblock {Anti-solar differential rotation and surface flow pattern on UZ
  Libr{\ae}}, {\it Astronomische Nachrichten\/}, {\bf 328}(10), 1078.

\bibitem[{Vida} et~al.(2016){Vida}, {Kriskovics}, {Ol{\'a}h}, {Leitzinger},
  {Odert}, {K{\H{o}}v{\'a}ri}, {Korhonen}, {Greimel}, {Robb}, {Cs{\'a}k}, \&
  {Kov{\'a}cs}]{Vida2016}
{Vida}, K., {Kriskovics}, L., {Ol{\'a}h}, K., {Leitzinger}, M., {Odert}, P.,
  {K{\H{o}}v{\'a}ri}, Z., {Korhonen}, H., {Greimel}, R., {Robb}, R.,
  {Cs{\'a}k}, B., \& {Kov{\'a}cs}, J., 2016.
\newblock {Investigating magnetic activity in very stable stellar magnetic
  fields. Long-term photometric and spectroscopic study of the fully convective
  M4 dwarf V374 Pegasi}, {\it \aap\/}, {\bf 590}, A11.

\bibitem[{Vida} et~al.(2017){Vida}, {K{\H{o}}v{\'a}ri}, {P{\'a}l}, {Ol{\'a}h},
  \& {Kriskovics}]{Vida2017}
{Vida}, K., {K{\H{o}}v{\'a}ri}, Z., {P{\'a}l}, A., {Ol{\'a}h}, K., \&
  {Kriskovics}, L., 2017.
\newblock {Frequent Flaring in the TRAPPIST-1 System{\textemdash}Unsuited for
  Life?}, {\it \apj\/}, {\bf 841}(2), 124.

\bibitem[{Vidal-Madjar} et~al.(2004){Vidal-Madjar}, {D{\'e}sert}, {Lecavelier
  des Etangs}, {H{\'e}brard}, {Ballester}, {Ehrenreich}, {Ferlet}, {McConnell},
  {Mayor}, \& {Parkinson}]{Vidal-Madjar2004}
{Vidal-Madjar}, A., {D{\'e}sert}, J.~M., {Lecavelier des Etangs}, A.,
  {H{\'e}brard}, G., {Ballester}, G.~E., {Ehrenreich}, D., {Ferlet}, R.,
  {McConnell}, J.~C., {Mayor}, M., \& {Parkinson}, C.~D., 2004.
\newblock {Detection of Oxygen and Carbon in the Hydrodynamically Escaping
  Atmosphere of the Extrasolar Planet HD 209458b}, {\it \apjl\/}, {\bf 604}(1),
  L69--L72.

\bibitem[{Vinyoles} et~al.(2017){Vinyoles}, {Serenelli}, {Villante}, {Basu},
  {Bergstr{\"o}m}, {Gonzalez-Garcia}, {Maltoni}, {Pe{\~n}a-Garay}, \&
  {Song}]{Vinyoles2017}
{Vinyoles}, N., {Serenelli}, A.~M., {Villante}, F.~L., {Basu}, S.,
  {Bergstr{\"o}m}, J., {Gonzalez-Garcia}, M.~C., {Maltoni}, M.,
  {Pe{\~n}a-Garay}, C., \& {Song}, N., 2017.
\newblock {A New Generation of Standard Solar Models}, {\it \apj\/}, {\bf
  835}(2), 202.

\bibitem[{Viticchi{\'e}} et~al.(2010){Viticchi{\'e}}, {Del Moro}, {Criscuoli},
  \& {Berrilli}]{viticchie2010}
{Viticchi{\'e}}, B., {Del Moro}, D., {Criscuoli}, S., \& {Berrilli}, F., 2010.
\newblock {Imaging Spectropolarimetry with IBIS. II. On the Fine Structure of
  G-band Bright Features}, {\it \apj\/}, {\bf 723}(1), 787--796.

\bibitem[{V{\"o}gler} et~al.(2005){V{\"o}gler}, {Shelyag}, {Sch{\"u}ssler},
  {Cattaneo}, {Emonet}, \& {Linde}]{Vogler2005}
{V{\"o}gler}, A., {Shelyag}, S., {Sch{\"u}ssler}, M., {Cattaneo}, F., {Emonet},
  T., \& {Linde}, T., 2005.
\newblock {Simulations of magneto-convection in the solar photosphere.
  Equations, methods, and results of the MURaM code}, {\it \aap\/}, {\bf 429},
  335--351.

\bibitem[{Vogt}(1979)]{Vogt1979}
{Vogt}, S.~S., 1979.
\newblock {A spectroscopic and photometric study of the star spot on HD
  224085.}, {\it \pasp\/}, {\bf 91}, 616.

\bibitem[{Vogt} \& {Penrod}(1983)]{VogtPenrod1983}
{Vogt}, S.~S. \& {Penrod}, G.~D., 1983.
\newblock {Doppler imaging of spotted stars : application to the RS Canum
  Venaticorum star HR 1099.}, {\it \pasp\/}, {\bf 95}, 565--576.

\bibitem[{Vogt} et~al.(1987){Vogt}, {Penrod}, \& {Hatzes}]{Vogt1987}
{Vogt}, S.~S., {Penrod}, G.~D., \& {Hatzes}, A.~P., 1987.
\newblock {Doppler Images of Rotating Stars Using Maximum Entropy Image
  Reconstruction}, {\it \apj\/}, {\bf 321}, 496.

\bibitem[{Vogt} et~al.(1999){Vogt}, {Hatzes}, {Misch}, \&
  {K{\"u}rster}]{Vogt1999hr1099}
{Vogt}, S.~S., {Hatzes}, A.~P., {Misch}, A.~A., \& {K{\"u}rster}, M., 1999.
\newblock {Doppler Imagery of the Spotted RS Canum Venaticorum Star HR 1099
  (V711 Tauri) from 1981 to 1992}, {\it \apjs\/}, {\bf 121}(2), 547--589.

\bibitem[{Vrard} et~al.(2018){Vrard}, {Kallinger}, {Mosser}, {Barban},
  {Baudin}, {Belkacem}, \& {Cunha}]{Vrard2018a}
{Vrard}, M., {Kallinger}, T., {Mosser}, B., {Barban}, C., {Baudin}, F.,
  {Belkacem}, K., \& {Cunha}, M.~S., 2018.
\newblock {Amplitude and lifetime of radial modes in red giant star spectra
  observed by Kepler}, {\it \aap\/}, {\bf 616}, A94.

\bibitem[{Wakeford} et~al.(2019){Wakeford}, {Lewis}, {Fowler}, {Bruno},
  {Wilson}, {Moran}, {Valenti}, {Batalha}, {Filippazzo}, {Bourrier},
  {H{\"o}rst}, {Lederer}, \& {de Wit}]{Wakeford2019}
{Wakeford}, H.~R., {Lewis}, N.~K., {Fowler}, J., {Bruno}, G., {Wilson}, T.~J.,
  {Moran}, S.~E., {Valenti}, J., {Batalha}, N.~E., {Filippazzo}, J.,
  {Bourrier}, V., {H{\"o}rst}, S.~M., {Lederer}, S.~M., \& {de Wit}, J., 2019.
\newblock {Disentangling the Planet from the Star in Late-Type M Dwarfs: A Case
  Study of TRAPPIST-1g}, {\it \aj\/}, {\bf 157}(1), 11.

\bibitem[{Waldmeier}(1955)]{Waldmeier1955}
{Waldmeier}, M., 1955.
\newblock {\it {Ergebnisse und Probleme der Sonnenforschung.}\/}.

\bibitem[{Walton} et~al.(2003){Walton}, {Preminger}, \& {Chapman}]{walton2003}
{Walton}, S.~R., {Preminger}, D.~G., \& {Chapman}, G.~A., 2003.
\newblock {A Statistical Analysis of the Characteristics of Sunspots and
  Faculae}, {\it \solphys\/}, {\bf 213}(2), 301--317.

\bibitem[{Watanabe}(2014)]{Watanabe2014}
{Watanabe}, H., 2014.
\newblock {Observations of umbral dots and their physical models}, {\it
  \pasj\/}, {\bf 66}, S1.

\bibitem[{Watson} et~al.(2006){Watson}, {Henden}, \& {Price}]{watson06}
{Watson}, C.~L., {Henden}, A.~A., \& {Price}, A., 2006.
\newblock {The International Variable Star Index (VSX)}, {\it Society for
  Astronomical Sciences Annual Symposium\/}, {\bf 25}, 47.

\bibitem[{Watson} et~al.(2014){Watson}, {Penn}, \& {Livingston}]{Watson:2014}
{Watson}, F.~T., {Penn}, M.~J., \& {Livingston}, W., 2014.
\newblock {A Multi-instrument Analysis of Sunspot Umbrae}, {\it \apj\/}, {\bf
  787}(1), 22.

\bibitem[{Wedemeyer} et~al.(2013){Wedemeyer}, {Ludwig}, \&
  {Steiner}]{Wedemeyer2013}
{Wedemeyer}, S., {Ludwig}, H.~G., \& {Steiner}, O., 2013.
\newblock {Three-dimensional magnetohydrodynamic simulations of M-dwarf
  chromospheres}, {\it Astronomische Nachrichten\/}, {\bf 334}(1-2), 137--140.

\bibitem[{Wedemeyer-B{\"o}hm} \& {Rouppe van der Voort}(2009)]{wed2009}
{Wedemeyer-B{\"o}hm}, S. \& {Rouppe van der Voort}, L., 2009.
\newblock {On the continuum intensity distribution of the solar photosphere},
  {\it \aap\/}, {\bf 503}(1), 225--239.

\bibitem[{Wells} et~al.(2021){Wells}, {Rackham}, {Schanche}, {Petrucci},
  {G{\'o}mez Maqueo Chew}, {Demory}, {Burgasser}, {Burn}, {Pozuelos},
  {G{\"u}nther}, {Sabin}, {Schroffenegger}, {G{\'o}mez-Mu{\~n}oz}, {Stassun},
  {Van Grootel}, {Howell}, {Sebastian}, {Triaud}, {Apai}, {Plauchu-Frayn},
  {Guerrero}, {Guill{\'e}n}, {Landa}, {Melgoza}, {Montalvo}, {Serrano},
  {Riesgo}, {Barkaoui}, {Bixel}, {Burdanov}, {Chen}, {Chinchilla}, {Collins},
  {Daylan}, {de Wit}, {Delrez}, {D{\'e}vora-Pajares}, {Dietrich}, {Dransfield},
  {Ducrot}, {Fausnaugh}, {Furlan}, {Gabor}, {Gan}, {Garcia}, {Ghachoui},
  {Giacalone}, {Gibbs}, {Gillon}, {Gnilka}, {Gore}, {Guerrero}, {Henning},
  {Hesse}, {Jehin}, {Jenkins}, {Latham}, {Lester}, {McCormac}, {Murray},
  {Niraula}, {Pedersen}, {Queloz}, {Ricker}, {Rodriguez}, {Schroeder},
  {Schwarz}, {Scott}, {Seager}, {Theissen}, {Thompson}, {Timmermans},
  {Twicken}, \& {Winn}]{Wells2021}
{Wells}, R.~D., {Rackham}, B.~V., {Schanche}, N., {Petrucci}, R., {G{\'o}mez
  Maqueo Chew}, Y., {Demory}, B.~O., {Burgasser}, A.~J., {Burn}, R.,
  {Pozuelos}, F.~J., {G{\"u}nther}, M.~N., {Sabin}, L., {Schroffenegger}, U.,
  {G{\'o}mez-Mu{\~n}oz}, M.~A., {Stassun}, K.~G., {Van Grootel}, V., {Howell},
  S.~B., {Sebastian}, D., {Triaud}, A.~H.~M.~J., {Apai}, D., {Plauchu-Frayn},
  I., {Guerrero}, C.~A., {Guill{\'e}n}, P.~F., {Landa}, A., {Melgoza}, G.,
  {Montalvo}, F., {Serrano}, H., {Riesgo}, H., {Barkaoui}, K., {Bixel}, A.,
  {Burdanov}, A., {Chen}, W.~P., {Chinchilla}, P., {Collins}, K.~A., {Daylan},
  T., {de Wit}, J., {Delrez}, L., {D{\'e}vora-Pajares}, M., {Dietrich}, J.,
  {Dransfield}, G., {Ducrot}, E., {Fausnaugh}, M., {Furlan}, E., {Gabor}, P.,
  {Gan}, T., {Garcia}, L., {Ghachoui}, M., {Giacalone}, S., {Gibbs}, A.~B.,
  {Gillon}, M., {Gnilka}, C., {Gore}, R., {Guerrero}, N., {Henning}, T.,
  {Hesse}, K., {Jehin}, E., {Jenkins}, J.~M., {Latham}, D.~W., {Lester}, K.,
  {McCormac}, J., {Murray}, C.~A., {Niraula}, P., {Pedersen}, P.~P., {Queloz},
  D., {Ricker}, G., {Rodriguez}, D.~R., {Schroeder}, A., {Schwarz}, R.~P.,
  {Scott}, N., {Seager}, S., {Theissen}, C.~A., {Thompson}, S., {Timmermans},
  M., {Twicken}, J.~D., \& {Winn}, J.~N., 2021.
\newblock {A large sub-Neptune transiting the thick-disk M4 V TOI-2406}, {\it
  \aap\/}, {\bf 653}, A97.

\bibitem[{Wende} et~al.(2009){Wende}, {Reiners}, \& {Ludwig}]{Wende2009}
{Wende}, S., {Reiners}, A., \& {Ludwig}, H.~G., 2009.
\newblock {3D simulations of M star atmosphere velocities and their influence
  on molecular FeH lines}, {\it \aap\/}, {\bf 508}(3), 1429--1442.

\bibitem[{Wheatley} et~al.(2018){Wheatley}, {West}, {Goad}, {Jenkins},
  {Pollacco}, {Queloz}, {Rauer}, {Udry}, {Watson}, {Chazelas}, {Eigm{\"u}ller},
  {Lambert}, {Genolet}, {McCormac}, {Walker}, {Armstrong}, {Bayliss}, {Bento},
  {Bouchy}, {Burleigh}, {Cabrera}, {Casewell}, {Chaushev}, {Chote},
  {Csizmadia}, {Erikson}, {Faedi}, {Foxell}, {G{\"a}nsicke}, {Gillen},
  {Grange}, {G{\"u}nther}, {Hodgkin}, {Jackman}, {Jord{\'a}n}, {Louden},
  {Metrailler}, {Moyano}, {Nielsen}, {Osborn}, {Poppenhaeger}, {Raddi},
  {Raynard}, {Smith}, {Soto}, \& {Titz-Weider}]{NGTS_weathley2018}
{Wheatley}, P.~J., {West}, R.~G., {Goad}, M.~R., {Jenkins}, J.~S., {Pollacco},
  D.~L., {Queloz}, D., {Rauer}, H., {Udry}, S., {Watson}, C.~A., {Chazelas},
  B., {Eigm{\"u}ller}, P., {Lambert}, G., {Genolet}, L., {McCormac}, J.,
  {Walker}, S., {Armstrong}, D.~J., {Bayliss}, D., {Bento}, J., {Bouchy}, F.,
  {Burleigh}, M.~R., {Cabrera}, J., {Casewell}, S.~L., {Chaushev}, A., {Chote},
  P., {Csizmadia}, S., {Erikson}, A., {Faedi}, F., {Foxell}, E.,
  {G{\"a}nsicke}, B.~T., {Gillen}, E., {Grange}, A., {G{\"u}nther}, M.~N.,
  {Hodgkin}, S.~T., {Jackman}, J., {Jord{\'a}n}, A., {Louden}, T.,
  {Metrailler}, L., {Moyano}, M., {Nielsen}, L.~D., {Osborn}, H.~P.,
  {Poppenhaeger}, K., {Raddi}, R., {Raynard}, L., {Smith}, A. M.~S., {Soto},
  M., \& {Titz-Weider}, R., 2018.
\newblock {The Next Generation Transit Survey (NGTS)}, {\it \mnras\/}, {\bf
  475}(4), 4476--4493.

\bibitem[{White} et~al.(2018){White}, {Huber}, {Mann}, {Casagrande},
  {Grunblatt}, {Justesen}, {Silva Aguirre}, {Bedding}, {Ireland}, {Schaefer},
  \& {Tuthill}]{White2018}
{White}, T.~R., {Huber}, D., {Mann}, A.~W., {Casagrande}, L., {Grunblatt},
  S.~K., {Justesen}, A.~B., {Silva Aguirre}, V., {Bedding}, T.~R., {Ireland},
  M.~J., {Schaefer}, G.~H., \& {Tuthill}, P.~G., 2018.
\newblock {Interferometric diameters of five evolved intermediate-mass
  planet-hosting stars measured with PAVO at the CHARA Array}, {\it \mnras\/},
  {\bf 477}(4), 4403--4413.

\bibitem[{Wilson}(1979)]{Wilson-79}
{Wilson}, R.~E., 1979.
\newblock {Eccentric orbit generalization and simultaneous solution of binary
  star light and velocity curves}, {\it \apj\/}, {\bf 234}, 1054--1066.

\bibitem[{Wilson}(1990)]{Wilson-90}
{Wilson}, R.~E., 1990.
\newblock {Accuracy and efficiency in the binary star reflection effect}, {\it
  \apj\/}, {\bf 356}, 613--622.

\bibitem[{Wilson}(2012)]{Wilson-12}
{Wilson}, R.~E., 2012.
\newblock {Spotted Star Light Curves with Enhanced Precision}, {\it \aj\/},
  {\bf 144}, 73.

\bibitem[{Wilson} \& {Devinney}(1971)]{Wilson-71}
{Wilson}, R.~E. \& {Devinney}, E.~J., 1971.
\newblock {Realization of Accurate Close-Binary Light Curves: Application to MR
  Cygni}, {\it ApJ\/}, {\bf 166}, 605--+.

\bibitem[{Winn} et~al.(2010){Winn}, {Fabrycky}, {Albrecht}, \&
  {Johnson}]{Winn2010}
{Winn}, J.~N., {Fabrycky}, D., {Albrecht}, S., \& {Johnson}, J.~A., 2010.
\newblock {Hot Stars with Hot Jupiters Have High Obliquities}, {\it \apjl\/},
  {\bf 718}, L145--L149.

\bibitem[{Witzke} et~al.(2018){Witzke}, {Shapiro}, {Solanki}, {Krivova}, \&
  {Schmutz}]{witzke2018}
{Witzke}, V., {Shapiro}, A.~I., {Solanki}, S.~K., {Krivova}, N.~A., \&
  {Schmutz}, W., 2018.
\newblock {From solar to stellar brightness variations. The effect of
  metallicity}, {\it \aap\/}, {\bf 619}, A146.

\bibitem[{Witzke} et~al.(2020){Witzke}, {Reinhold}, {Shapiro}, {Krivova}, \&
  {Solanki}]{Witzke2020}
{Witzke}, V., {Reinhold}, T., {Shapiro}, A.~I., {Krivova}, N.~A., \& {Solanki},
  S.~K., 2020.
\newblock {Effect of metallicity on the detectability of rotational periods in
  solar-like stars}, {\it \aap\/}, {\bf 634}, L9.

\bibitem[{Witzke} et~al.(2022){Witzke}, {Shapiro}, {Kostogryz}, {Cameron},
  {Rackham}, {Seager}, {Solanki}, \& {Unruh}]{Witzke2022}
{Witzke}, V., {Shapiro}, A.~I., {Kostogryz}, N.~M., {Cameron}, R., {Rackham},
  B.~V., {Seager}, S., {Solanki}, S.~K., \& {Unruh}, Y.~C., 2022.
\newblock {Can 1D Radiative-equilibrium Models of Faculae Be Used for
  Calculating Contamination of Transmission Spectra?}, {\it \apjl\/}, {\bf
  941}(2), L35.

\bibitem[{Witzke} et~al.(2023){Witzke}, {Duehnen}, {Shapiro}, {Przybylski},
  {Bhatia}, {Cameron}, \& {Solanki}]{Witzke2023}
{Witzke}, V., {Duehnen}, H.~B., {Shapiro}, A.~I., {Przybylski}, D., {Bhatia},
  T.~S., {Cameron}, R., \& {Solanki}, S.~K., 2023.
\newblock {Small-scale dynamo in cool stars. II. The effect of metallicity},
  {\it \aap\/}, {\bf 669}, A157.

\bibitem[{W{\"o}hl}(1971)]{Wohl1971}
{W{\"o}hl}, H., 1971.
\newblock {On Molecules in Sunspots}, {\it \solphys\/}, {\bf 16}(2), 362--372.

\bibitem[{W{\"o}hl} et~al.(1970){W{\"o}hl}, {Wittmann}, \&
  {Schr{\"o}ter}]{Wohl1970}
{W{\"o}hl}, H., {Wittmann}, A., \& {Schr{\"o}ter}, E.~H., 1970.
\newblock {A complete photoelectric sunspot spectrum: An atlas from 3900 8000
  {\r{A}}}, {\it \solphys\/}, {\bf 13}(1), 104--117.

\bibitem[{Wootten} \& {Thompson}(2009)]{Wootten2009}
{Wootten}, A. \& {Thompson}, A.~R., 2009.
\newblock {The Atacama Large Millimeter/Submillimeter Array}, {\it IEEE
  Proceedings\/}, {\bf 97}(8), 1463--1471.

\bibitem[{Wright} et~al.(2018){Wright}, {Newton}, {Williams}, {Drake}, \&
  {Yadav}]{2018MNRAS.479.2351W}
{Wright}, N.~J., {Newton}, E.~R., {Williams}, P. K.~G., {Drake}, J.~J., \&
  {Yadav}, R.~K., 2018.
\newblock {The stellar rotation-activity relationship in fully convective M
  dwarfs}, {\it \mnras\/}, {\bf 479}(2), 2351--2360.

\bibitem[{Yadav} \& {Mathew}(2018)]{Yadav:2018}
{Yadav}, R. \& {Mathew}, S.~K., 2018.
\newblock {Physical Properties of Umbral Dots Observed in Sunspots: A Hinode
  Observation}, {\it \solphys\/}, {\bf 293}(4), 54.

\bibitem[{Yadav} et~al.(2015){Yadav}, {Christensen}, {Morin}, {Gastine},
  {Reiners}, {Poppenhaeger}, \& {Wolk}]{Yadav2015dynamo}
{Yadav}, R.~K., {Christensen}, U.~R., {Morin}, J., {Gastine}, T., {Reiners},
  A., {Poppenhaeger}, K., \& {Wolk}, S.~J., 2015.
\newblock {Explaining the Coexistence of Large-scale and Small-scale Magnetic
  Fields in Fully Convective Stars}, {\it \apjl\/}, {\bf 813}(2), L31.

\bibitem[{Yeo} et~al.(2013){Yeo}, {Solanki}, \& {Krivova}]{yeo2013}
{Yeo}, K.~L., {Solanki}, S.~K., \& {Krivova}, N.~A., 2013.
\newblock {Intensity contrast of solar network and faculae}, {\it \aap\/}, {\bf
  550}, A95.

\bibitem[{Yeo} et~al.(2014{\natexlab{a}}){Yeo}, {Feller}, {Solanki},
  {Couvidat}, {Danilovic}, \& {Krivova}]{yeo+al2014}
{Yeo}, K.~L., {Feller}, A., {Solanki}, S.~K., {Couvidat}, S., {Danilovic}, S.,
  \& {Krivova}, N.~A., 2014{\natexlab{a}}.
\newblock {Point spread function of SDO/HMI and the effects of stray light
  correction on the apparent properties of solar surface phenomena}, {\it
  \aap\/}, {\bf 561}, A22.

\bibitem[{Yeo} et~al.(2014{\natexlab{b}}){Yeo}, {Krivova}, {Solanki}, \&
  {Glassmeier}]{Yeo2014_SATIRE}
{Yeo}, K.~L., {Krivova}, N.~A., {Solanki}, S.~K., \& {Glassmeier}, K.~H.,
  2014{\natexlab{b}}.
\newblock {Reconstruction of total and spectral solar irradiance from 1974 to
  2013 based on KPVT, SoHO/MDI, and SDO/HMI observations}, {\it \aap\/}, {\bf
  570}, A85.

\bibitem[Yeo et~al.(2020)Yeo, Solanki, \& Krivova]{yeo2020faculae}
Yeo, K.~L., Solanki, S.~K., \& Krivova, N.~A., 2020.
\newblock How faculae and network relate to sunspots, and the implications for
  solar and stellar brightness variations, {\it Astronomy \& Astrophysics\/},
  {\bf 639}, A139.

\bibitem[{Yeo} et~al.(2020){Yeo}, {Solanki}, {Krivova}, {Rempel}, {Anusha},
  {Shapiro}, {Tagirov}, \& {Witzke}]{Yeo2020}
{Yeo}, K.~L., {Solanki}, S.~K., {Krivova}, N.~A., {Rempel}, M., {Anusha},
  L.~S., {Shapiro}, A.~I., {Tagirov}, R.~V., \& {Witzke}, V., 2020.
\newblock {The Dimmest State of the Sun}, {\it \grl\/}, {\bf 47}(19), e90243.

\bibitem[{Yu} et~al.(2018){Yu}, {Huber}, {Bedding}, {Stello}, {Hon}, {Murphy},
  \& {Khanna}]{Yu2018a}
{Yu}, J., {Huber}, D., {Bedding}, T.~R., {Stello}, D., {Hon}, M., {Murphy},
  S.~J., \& {Khanna}, S., 2018.
\newblock {Asteroseismology of 16,000 Kepler Red Giants: Global Oscillation
  Parameters, Masses, and Radii}, {\it \apjs\/}, {\bf 236}(2), 42.

\bibitem[{Zakharov} et~al.(2005){Zakharov}, {Gandorfer}, {Solanki}, \&
  {L{\"o}fdahl}]{zakharov2005}
{Zakharov}, V., {Gandorfer}, A., {Solanki}, S.~K., \& {L{\"o}fdahl}, M., 2005.
\newblock {A comparative study of the contrast of solar magnetic elements in CN
  and CH}, {\it \aap\/}, {\bf 437}(3), L43--L46.

\bibitem[{Zaleski} et~al.(2019){Zaleski}, {Valio}, {Marsden}, \&
  {Carter}]{Zaleski-19}
{Zaleski}, S.~M., {Valio}, A., {Marsden}, S.~C., \& {Carter}, B.~D., 2019.
\newblock {Differential rotation of Kepler-71 via transit photometry mapping of
  faculae and starspots}, {\it \mnras\/}, {\bf 484}(1), 618--630.

\bibitem[{Zaleski} et~al.(2020){Zaleski}, {Valio}, {Carter}, \&
  {Marsden}]{Zaleski-20}
{Zaleski}, S.~M., {Valio}, A., {Carter}, B.~D., \& {Marsden}, S.~C., 2020.
\newblock {Activity and differential rotation of the early M dwarf Kepler-45
  from transit mapping}, {\it \mnras\/}, {\bf 492}(4), 5141--5151.

\bibitem[{Zellem} et~al.(2010){Zellem}, {Guinan}, {Messina}, {Lanza},
  {Wasatonic}, \& {McCook}]{Zellem2010}
{Zellem}, R., {Guinan}, E.~F., {Messina}, S., {Lanza}, A.~F., {Wasatonic}, R.,
  \& {McCook}, G.~P., 2010.
\newblock {Multiband Photometry of the Chromospherically Active \& Spotted
  Binary System IM Peg{\textemdash}the Guide Star for the Gravity Probe B
  Mission}, {\it \pasp\/}, {\bf 122}(892), 670.

\bibitem[{Zellem} et~al.(2017){Zellem}, {Swain}, {Roudier}, {Shkolnik},
  {Creech-Eakman}, {Ciardi}, {Line}, {Iyer}, {Bryden}, {Llama}, \&
  {Fahy}]{Zellem2017}
{Zellem}, R.~T., {Swain}, M.~R., {Roudier}, G., {Shkolnik}, E.~L.,
  {Creech-Eakman}, M.~J., {Ciardi}, D.~R., {Line}, M.~R., {Iyer}, A.~R.,
  {Bryden}, G., {Llama}, J., \& {Fahy}, K.~A., 2017.
\newblock {Forecasting the Impact of Stellar Activity on Transiting Exoplanet
  Spectra}, {\it \apj\/}, {\bf 844}(1), 27.

\bibitem[{Zellem} et~al.(2019){Zellem}, {Swain}, {Cowan}, {Bryden}, {Komacek},
  {Colavita}, {Ardila}, {Roudier}, {Fortney}, {Bean}, {Line}, {Griffith},
  {Shkolnik}, {Kreidberg}, {Moses}, {Showman}, {Stevenson}, {Wong}, {Chapman},
  {Ciardi}, {Howard}, {Kataria}, {Kempton}, {Latham}, {Mahadevan},
  {Mel{\'e}ndez}, \& {Parmentier}]{Zellem2019}
{Zellem}, R.~T., {Swain}, M.~R., {Cowan}, N.~B., {Bryden}, G., {Komacek},
  T.~D., {Colavita}, M., {Ardila}, D., {Roudier}, G.~M., {Fortney}, J.~J.,
  {Bean}, J., {Line}, M.~R., {Griffith}, C.~A., {Shkolnik}, E.~L., {Kreidberg},
  L., {Moses}, J.~I., {Showman}, A.~P., {Stevenson}, K.~B., {Wong}, A.,
  {Chapman}, J.~W., {Ciardi}, D.~R., {Howard}, A.~W., {Kataria}, T., {Kempton},
  E. M.~R., {Latham}, D., {Mahadevan}, S., {Mel{\'e}ndez}, J., \& {Parmentier},
  V., 2019.
\newblock {Constraining Exoplanet Metallicities and Aerosols with the
  Contribution to ARIEL Spectroscopy of Exoplanets (CASE)}, {\it \pasp\/}, {\bf
  131}(1003), 094401.

\bibitem[{Zellem} et~al.(2020){Zellem}, {Pearson}, {Blaser}, {Fowler},
  {Ciardi}, {Biferno}, {Massey}, {Marchis}, {Baer}, {Ball}, {Chasin}, {Conley},
  {Dixon}, {Fletcher}, {Hernandez}, {Nair}, {Perian}, {Sienkiewicz}, {Tock},
  {Vijayakumar}, {Swain}, {Roudier}, {Bryden}, {Conti}, {Hill}, {Hergenrother},
  {Dussault}, {Kane}, {Fitzgerald}, {Boyce}, {Peticolas}, {Gee}, {Cominsky},
  {Zimmerman-Brachman}, {Smith}, {Creech-Eakman}, {Engelke}, {Iturralde},
  {Dragomir}, {Jovanovic}, {Lawton}, {Arbouch}, {Kuchner}, \&
  {Malvache}]{zellem20}
{Zellem}, R.~T., {Pearson}, K.~A., {Blaser}, E., {Fowler}, M., {Ciardi}, D.~R.,
  {Biferno}, A., {Massey}, B., {Marchis}, F., {Baer}, R., {Ball}, C., {Chasin},
  M., {Conley}, M., {Dixon}, S., {Fletcher}, E., {Hernandez}, S., {Nair}, S.,
  {Perian}, Q., {Sienkiewicz}, F., {Tock}, K., {Vijayakumar}, V., {Swain},
  M.~R., {Roudier}, G.~M., {Bryden}, G., {Conti}, D.~M., {Hill}, D.~H.,
  {Hergenrother}, C.~W., {Dussault}, M., {Kane}, S.~R., {Fitzgerald}, M.,
  {Boyce}, P., {Peticolas}, L., {Gee}, W., {Cominsky}, L.,
  {Zimmerman-Brachman}, R., {Smith}, D., {Creech-Eakman}, M.~J., {Engelke}, J.,
  {Iturralde}, A., {Dragomir}, D., {Jovanovic}, N., {Lawton}, B., {Arbouch},
  E., {Kuchner}, M., \& {Malvache}, A., 2020.
\newblock {Utilizing Small Telescopes Operated by Citizen Scientists for
  Transiting Exoplanet Follow-up}, {\it \pasp\/}, {\bf 132}(1011), 054401.

\bibitem[{Zhang} et~al.(2019){Zhang}, {Chachan}, {Kempton}, \&
  {Knutson}]{Zhang2019_platon}
{Zhang}, M., {Chachan}, Y., {Kempton}, E. M.~R., \& {Knutson}, H.~A., 2019.
\newblock {Forward Modeling and Retrievals with PLATON, a Fast Open-source
  Tool}, {\it \pasp\/}, {\bf 131}(997), 034501.

\bibitem[{Zhang} \& {Showman}(2014)]{Zhang2014}
{Zhang}, X. \& {Showman}, A.~P., 2014.
\newblock {Atmospheric Circulation of Brown Dwarfs: Jets, Vortices, and Time
  Variability}, {\it \apjl\/}, {\bf 788}(1), L6.

\bibitem[{Zhang} et~al.(2018){Zhang}, {Zhou}, {Rackham}, \& {Apai}]{Zhang2018}
{Zhang}, Z., {Zhou}, Y., {Rackham}, B.~V., \& {Apai}, D., 2018.
\newblock {The Near-infrared Transmission Spectra of TRAPPIST-1 Planets b, c,
  d, e, f, and g and Stellar Contamination in Multi-epoch Transit Spectra},
  {\it \aj\/}, {\bf 156}(4), 178.

\bibitem[{Ziegler} et~al.(2017){Ziegler}, {Law}, {Morton}, {Baranec}, {Riddle},
  {Atkinson}, {Baker}, {Roberts}, \& {Ciardi}]{Ziegler2017}
{Ziegler}, C., {Law}, N.~M., {Morton}, T., {Baranec}, C., {Riddle}, R.,
  {Atkinson}, D., {Baker}, A., {Roberts}, S., \& {Ciardi}, D.~R., 2017.
\newblock {Robo-AO Kepler Planetary Candidate Survey. III. Adaptive Optics
  Imaging of 1629 Kepler Exoplanet Candidate Host Stars}, {\it \aj\/}, {\bf
  153}(2), 66.

\bibitem[{Ziegler} et~al.(2020){Ziegler}, {Tokovinin}, {Brice{\~n}o}, {Mang},
  {Law}, \& {Mann}]{Ziegler2020}
{Ziegler}, C., {Tokovinin}, A., {Brice{\~n}o}, C., {Mang}, J., {Law}, N., \&
  {Mann}, A.~W., 2020.
\newblock {SOAR TESS Survey. I. Sculpting of TESS Planetary Systems by Stellar
  Companions}, {\it \aj\/}, {\bf 159}(1), 19.

\end{thebibliography}




\clearpage

\appendix

\section{Acronyms Used}
\label{S:Acronyms}

For convenience, \autoref{tab:acronyms} lists the acronyms and abbreviations used in this review.

\begin{table*}
    \caption{Acronyms and abbreviations used in this review.}
    \label{tab:acronyms}
    \centering
    \begin{tabular}{ll}
        \hline\hline
        Acronym or Abbreviation & Definition \\
        \hline
        3D & Three-dimensional \\
        ALMA & Atacama Large Millimeter/submillimeter Array \\
        ASAS-SN & All Sky Automated Survey for SuperNovae \\
        BFI & Broadband Filter Imager \\
        CASE & Contribution to Ariel Spectroscopy of Exoplanets \\
        CHARA & Center for High Angular Resolution Astronomy \\
        CLASP & Chromospheric Lyman-Alpha Spectro-Polarimeter \\
        CO5BOLD & COnservative COde for the COmputation of COmpressible COnvection in a BOx of L Dimensions, L=2,3 \\
        CoRoT & Convection, Rotation, and planetary Transits \\
        CLV & center-to-limb variation \\
        DASCH & Digital Access to a Sky Century @ Harvard \\
        DKIST & Daniel K. Inouye Solar Telescope \\
        DI & Doppler imaging \\
        EUV & extreme ultraviolet \\
        ExoFOP & Exoplanet Follow-up Observing Program \\
        ExoPAG & Exoplanet Exploration Program Analysis Group \\
        GOES & Geostationary Operational Environmental Satellites \\
        HabEx & Habitable Exoplanet Observatory \\
        HARPS-N & High Accuracy Radial velocity Planet Searcher for the Northern hemisphere \\
        HD & hydrodynamic \\
        HST & Hubble Space Telescope \\
        IRIS & Interface Region Imaging Spectrograph \\
        ISS & International Space Station \\
        JWST & James Webb Space Telescope \\
        KELT & Kilodegree Extremely Little Telescope \\
        Kepler & Kepler Space Telescope \\
        LTE & local thermodynamic equilibrium \\
        LUVOIR & Large UV/Optical/IR Surveyor \\
        MCMC & Markov chain Monte Carlo \\
        MHD & magnetohydrodynamic \\
        MHS & millionths of the hemisphere of the Sun \\
        MIRC-X & Michigan InfraRed Combiner-eXeter \\
        MURaM & Max-Planck-Institute for Aeronomy/ University of Chicago Radiation Magneto-hydrodynamics code \\
        NASA &  National Aeronautics and Space Administration \\
        NIR & near infrared \\
        NPOI & Navy Precision Optical Interferometer \\
        NSF & National Science Foundation \\
        OMI & Ozone Monitoring Experiment \\
        PCH & photospheric and chromospheric heterogeneities \\
        PIONIER & Precision Integrated Optics Near Infrared ExpeRiment \\
        ppm & parts per million \\
        ppt & parts per thousand \\
        PLATO & PLAnetary Transit and Oscillations \\
        PSPT & Precision Solar Photometric Telescope \\
        RMS & root mean square \\
        RV & radial velocity \\
        SAG21 & Study Analysis Group 21 \\
        SATIRE & Spectral And Total Irradiance REconstruction \\
        SDO & Solar Dynamics Observatory \\
        SED & spectral energy distribution \\
        SOHO & Solar and Heliospheric Observatory \\
        SOLIS & Synoptic Long-term Investigations of the Sun \\
        SORCE & Solar Radiation and Climate Experiment \\
        SOT & Solar Optical Telescope \\
        SPICA & Stellar Parameters and Images with a Cophased Array \\
        Spitzer & Spitzer Space Telescope \\
        SST & Swedish Solar Tower \\
        SUIT & Solar Ultraviolet Imaging Telescope \\
        SUVI & Solar Ultraviolet Imager \\
        TESS & Transiting Exoplanet Survey Satellite \\
        TOI & TESS Object of Interest \\
        TIM & Total Irradiance Monitor \\
        TLSE & transit light source effect \\
        \hline
    \end{tabular}
\end{table*}

\begin{table*}
    \contcaption{Acronyms and abbreviations used in this review.}
    \centering
    \begin{tabular}{ll}
        \hline\hline
        Acronym or Abbreviation & Definition \\
        \hline
        TSI & total solar irradiance \\
        UV & ultraviolet \\
        VIRGO & Variability of solar IRradiance and Gravity Oscillations \\
        VISION & Visible Imaging System for Interferometric Observations \\
        VLTI & Very Large Telescope Interferometer \\
        WASP & Wide Angle Search for Planets \\
        ZDI & Zeeman--Doppler imaging \\
        \hline
    \end{tabular}
\end{table*}


\bsp	
\label{lastpage}
\end{document}